\documentclass[phd,lfcs,twoside, singlespacing, logo, listintoc, timesfont]{infthesis}

\usepackage[OT1]{fontenc}
\usepackage{cmbright}
\usepackage{graphicx, color, graphpap}
\usepackage[most]{tcolorbox}
\usepackage{thmtools, thm-restate}
\usepackage{amsmath}
\usepackage{enumitem}
\usepackage{amssymb}
\usepackage{amsthm}

\usepackage{pstricks}
\usepackage{float}
\usepackage[colorlinks=true, citecolor=blue, linkcolor=cyan, pagebackref]{hyperref}
\usepackage[T1]{fontenc}
\usepackage{bbm}
\usepackage{dsfont}
\usepackage[algo2e,linesnumbered,ruled,vlined]{algorithm2e}
\SetKwInput{kwInit}{Init}
\usepackage{mathtools}
\usepackage{cancel}
\DeclarePairedDelimiter\ceil{\lceil}{\rceil}

\usetikzlibrary{positioning, scopes,svg.path,shapes.geometric,shadows}
\usepackage{xcolor}
\usepackage[caption=false]{subfig}
\usepackage{xfrac}

\usepackage{hyperref}
\usepackage{lipsum}
\usepackage{extarrows}
\usepackage{braket}
\usepackage{caption}
\usepackage{dirtytalk}
\usepackage{float}
\usepackage{authblk}
\usepackage{enumitem}
\usepackage{tikz}
\usepackage{xpatch}
\usepackage{qcircuit}
\usepackage{algpseudocode}
\usepackage{algorithm}
\usepackage{algpseudocode}
\usepackage{sidecap, caption}
\usepackage{siunitx}
\usepackage{pifont} 
\usepackage{tabularx}
\usepackage{multirow}
\usepackage{slashbox}

\usepackage{wrapfig}
\usepackage{url}

\usepackage{breakurl}
\usepackage{relsize}

\usepackage[numbers]{natbib}

\MHInternalSyntaxOn
\def\MT_extended_eqref:n #1{
  \protected@write\@auxout{}
  {\string\MT@newlabel{#1}}
  \textup{\let\df@label\@empty\MT_prev_tagform:n {\ref{#1}}}
}
\MHInternalSyntaxOff

\makeatletter
\newtheorem*{rep@theorem}{\rep@title}
\newcommand{\newreptheorem}[2]{%
\newenvironment{rep#1}[1]{%
 \def\rep@title{#2 \ref{##1}}%
 \begin{rep@theorem}}%
 {\end{rep@theorem}}}
\makeatother

\newreptheorem{theorem}{Theorem}
\newreptheorem{lemma}{Lemma}

\makeatletter
\renewcommand{\@chapapp}{}
\newenvironment{chapquote}[2][2em]
  {\setlength{\@tempdima}{#1}%
   \def\chapquote@author{#2}%
   \parshape 1 \@tempdima \dimexpr\textwidth-2\@tempdima\relax%
   \itshape}
  {\par\normalfont\hfill--\ \chapquote@author\hspace*{\@tempdima}\par\bigskip}
\makeatother

\DeclareUnicodeCharacter{2212}{-}

\newcommand{\corrref}[1]{\hyperref[#1]{Corollary~\ref{#1}}}
\newcommand{\defref}[1]{\hyperref[#1]{Definition~\ref{#1}}}
\newcommand{\secref}[1]{\hyperref[#1]{Sec.~\ref{#1}}}
\newcommand{\chapref}[1]{\hyperref[#1]{Chapter~\ref{#1}}}
\newcommand{\appref}[1]{\hyperref[#1]{Appendix~\ref{#1}}}
\newcommand{\thmref}[1]{\hyperref[#1]{Theorem~\ref{#1}}}
\newcommand{\lemref}[1]{\hyperref[#1]{Lemma~\ref{#1}}}
\newcommand{\figref}[1]{\hyperref[#1]{Fig.~\ref{#1}}}
\renewcommand{\eqref}[1]{\hyperref[#1]{Eq. (\ref{#1})}}
\newcommand{\algoref}[1]{\hyperref[#1]{Algorithm~\ref{#1}}}
\newcommand{\attref}[1]{\hyperref[#1]{Attack~\ref{#1}}}
\newcommand{\gameref}[1]{\hyperref[#1]{Game~\ref{#1}}}
\newcommand{\tableref}[1]{\hyperref[#1]{Table~\ref{#1}}}
\newcommand{\protoref}[1]{\hyperref[#1]{Protocol~\ref{#1}}}
\newcommand{\constref}[1]{\hyperref[#1]{Construction~\ref{#1}}}

\newcommand{\sq}{\textsf{sq}}

\renewcommand{\geq}{\geqslant}
\renewcommand{\leq}{\leqslant}

\newcommand{\mbraket}[2]{\bra{#1}#2\rangle}

\newcommand{\norm}[2][]{#1| \! #1| #2 #1| \! #1|}

\newcommand{\dtr}{d_{\text{Tr}}}
\newcommand{\dba}{\Theta_{\text{BA}}}
\newcommand{\dfs}{d_{\text{FS}}}
\newcommand{\RNum}[1]{\fontfamily{lmtt}\selectfont\uppercase\expandafter{\romannumeral #1\relax}}

\newtheorem{requirement}{Requirement}
\newtheorem{example}{Example}
\newtheorem{theorem}{Theorem}
\newtheorem{lemma}{Lemma}
\newtheorem{corollary}{Corollary}

\theoremstyle{definition}
\newtheorem{definition}{Definition}

\newtheorem{game}{Game}
\newtheorem{construction}{Construction}
\newtheorem{protocol}{Protocol}
\newtheorem{attack}{Attack}

\newcommand{\ES}{\mathcal{S}}
\newcommand{\T}{\mathcal{T}}
\newcommand{\Td}{\mathcal{T}^{ideal}_{\delta}}
\newcommand{\A}{\mathcal{A}}
\newcommand{\B}{\mathcal{B}}
\newcommand{\C}{\mathcal{C}}
\newcommand{\D}{\mathcal{D}}
\newcommand{\E}{\mathcal{E}}
\newcommand{\F}{\mathcal{F}}
\newcommand{\Ler}{\mathcal{L}}
\newcommand{\K}{\mathcal{K}}
\newcommand{\X}{\mathcal{X}}
\newcommand{\Y}{\mathcal{Y}}

\newcommand{\M}{\mathcal{M}}
\newcommand{\V}{\mathcal{V}}
\newcommand{\Pv}{\mathcal{P}}
\newcommand{\nul}{\mathsf{null}}
\newcommand{\Hil}{\mathcal{H}}
\newcommand{\HilD}{\mathcal{H}^D}
\newcommand{\Hild}{\mathcal{H}^d}
\newcommand{\negl}{negl}
\newcommand{\nonnegl}{non\text{-}\negl}
\newcommand{\Ue}{\mathrm{U_{\mathcal{E}}}}
\newcommand{\U}{\mathrm{U}}
\newcommand{\UqPUF}{\mathrm{UqPUF}}
\newcommand{\qPUF}{\mathrm{qPUF}}

\newcommand{\sw}{\text{SWAP}}
\newcommand{\err}{\texttt{Err}}

\newcommand{\QE}{\hyperref[sec:prelim-qe]{\textcolor{cyan}{QE}}}
\newcommand{\fm}[1]{f^{-1}(#1)}
\newcommand{\Uf}{U_f}
\newcommand{\Ufi}{U_{f^{-1}}}

\newcommand{\Ora}{\mathcal{O}}
\newcommand{\eO}{\mathcal{O}^{\mathcal{E}}}
\newcommand{\reO}{R \mathcal{O}^{\mathcal{E}}}
\newcommand{\vO}{\mathcal{O}^{\mathcal{V}}}
\newcommand{\HilR}{\mathcal{H}^{\mathcal{R}}}

\newcommand{\Conc}{\mathfrak{C}}

\newcommand{\euf}{1\text{-}qGEU}
\newcommand{\eufm}{$\mu$\text{-}qGEU}

\newcommand{\suf}{1\text{-}qGSU}
\newcommand{\sufm}{$\mu$\text{-}qGSU}

\newcommand{\uuf}{qGUU}
\newcommand{\bz}{BZ}
\newcommand{\bu}{BU}
\newcommand{\qprf}{qPRF}
\newcommand{\prf}{PRF}
\newcommand{\pru}{PRU}

\newcommand{\GCM}[2]{\mathcal{G}^{#1}_{#2}}

\newcommand{\Gn}[1]{\mathcal{G}^{#1}}

\newcommand{\qEx}{\mathsf{qEx}}
\newcommand{\qSel}{\mathsf{qSel}}
\newcommand{\qUni}{\mathsf{qUni}}
\newcommand{\Be}{\mathcal{B}_{\epsilon}}

\newcommand{\PEX}{\mathsf{PEX}}
\newcommand{\QPEX}{\mathsf{QPEX}}
\newcommand{\QMQ}{\mathsf{QMQ}}
\newcommand{\inp}{\mathrm{in}}
\newcommand{\out}{\mathrm{out}}

\newcommand{\Gnn}[2]{\mathcal{G}^{#1}_{#2}}

\newcommand{\Sin}{S_{in}}
\newcommand{\Sout}{S_{out}}
\newcommand{\Hilin}{\mathcal{H}_{in}}
\newcommand{\Hilout}{\mathcal{H}_{out}}

\newcommand{\Hildperp}{\mathcal{H}^{d^{\perp}}}
\newcommand{\Hildperpo}{\mathcal{H}^{d^{\perp}}_{out}}
\newcommand{\din}{d_{in}}

\newcommand{\HildIn}{\Hil^{d_{in}}}
\newcommand{\Hildout}{\Hil^{d}_{out}}
\newcommand{\HildOut}{\Hil^{d_{out}}}

\newcommand{\kpo}{\ket{\psi^{out}}}
\newcommand{\po}{\psi^{out}}

\newcommand{\UqPUFid}{\UqPUF_{\mathbf{id}}}

\newcommand{\qPUFGen}{\mathrm{QGen}}
\newcommand{\qPUFEval}{\mathrm{QEval}}
\newcommand{\id}{\mathbf{id}}
\newcommand{\qPUFidi}{\qPUF_{\id_i}}
\newcommand{\qPUFidj}{\qPUF_{\id_j}}
\newcommand{\qPUFid}{\qPUF_{\id}}
\newcommand{\rhoin}{\rho_{in}}
\newcommand{\rhoout}{\rho_{out}}
\newcommand{\sigmain}{\sigma_{in}}
\newcommand{\sigmaout}{\sigma_{out}}
\newcommand{\psiin}{\psi_{in}}
\newcommand{\psiout}{\psi_{out}}
\newcommand{\notmu}{\not\in_{\mu}}

\newcommand{\Gnef}{\mathcal{G}^{eff}_{\qUni}}

\newcommand{\Uset}{\mathcal{U}}

\newcommand{\kc}{\ket{\phi^c_i}}
\newcommand{\kr}{\ket{\phi^r_i}}
\newcommand{\krm}{\ket{\phi^r_i}^{\otimes M}}
\newcommand{\kp}{\ket{\phi^p_i}}
\newcommand{\kb}{\ket{\phi^b_i}}

\newcommand{\hrv}{\textcolor{BlueViolet}{hrv-id}}
\newcommand{\lrv}{\textcolor{BlueViolet}{lrv-id}}
\newcommand{\hrvs}{\textcolor{BlueViolet}{hrv-id-swap}}
\newcommand{\hrvg}{\textcolor{BlueViolet}{hrv-id-gswap}}
\newcommand{\ke}{\ket{\phi^{\mathcal{E}}_i}}
\newcommand{\Ti}{\mathcal{T}_{ideal}}
\newcommand{\R}{\mathcal{R}}
\newcommand{\gen}{\mathrm{Gen}}
\newcommand{\Gnh}{\mathcal{G}^{HPUF}}
\newcommand{\Gnhl}{\mathcal{G}^{HLPUF}}
\newcommand{\Gnre}{\mathcal{G}_{re}}
\newcommand{\Gea}{\mathcal{G}^{\E_{f_1}}}
\newcommand{\Geb}{\mathcal{G}^{\E_{f_2}}}

\newcommand{\Gel}{\mathcal{G}^{\E^L_f}}
\newcommand{\Aad}{\mathcal{A}_{ad}}
\newcommand{\Ana}{\mathcal{A}_{weak}}
\newcommand{\ver}{\texttt{Ver}}
\newcommand{\forge}{\text{forge}}
\newcommand{\fail}{\text{fail}}
\newcommand{\extract}{\text{extract}}
\newcommand{\classical}{\text{classic}}
\newcommand{\quantum}{\text{quant}}
\usetikzlibrary{matrix,shapes,arrows,positioning,chains,arrows.meta,calc,fit}

\newcommand{\Dcrit}{D_{\text{crit}}}

\newcommand*{\guess}{\text{guess}}

\newcommand{\IBM}{\mathsf{QCIBM}}

\newcommand{\paramtheta}{\boldsymbol{\theta}}

\DeclareMathOperator{\Lbs}{\textsf{L}}
\DeclareMathOperator{\Gbs}{\textsf{G}}
\DeclareMathOperator{\Cbs}{\textsf{C}}

\newcommand{\Ansatze}{\text{Ans\"{a}tze}}
\newcommand{\Ansatz}{\text{ansatz}}
\newcommand{\ansatze}{\text{ans\"{a}tze}}

\newtheoremstyle{example}{\topsep}{\topsep}%
{}
{}
{\bfseries}
{:}
{   }
{\thmname{#1}\thmnumber{ #2}}
\theoremstyle{example}

\newcommand{\RY}{\mathsf{R}_y}
\newcommand{\CNOT}{\mathsf{CNOT}}
\newcommand{\CZ}{\mathsf{CZ}}

\newcommand{\SWAP}{\mathsf{SWAP}}

\DeclareMathOperator{\VQC}{\textsf{VarQlone}}

\newtheorem*{theorem*}{Theorem}

\def\orcid#1{\kern -0.4em\href{https://orcid.org/#1}{\includegraphics[keepaspectratio,width=0.7em]{orcid_logo.pdf}}}

\DeclareMathOperator*{\argmin}{arg\,min}

\newcommand{\opt}{\mathrm{opt}}

\renewcommand{\epsilon}{\varepsilon}

\newcommand{\tr}{\mathsf{Tr}}
\newcommand{\Tr}{\mathsf{Tr}}

\definecolor{ForestGreen}{RGB}{34, 139, 34}
\definecolor{DeepSkyBlue}{RGB}{0, 191,255}
\definecolor{Lavender}{RGB}{230, 230, 250}

\long\def\ca#1\cb{} 
\apptocmd{\sloppy}{\hbadness 10000\relax}{}{}
\newcommand{\computerfont}[1]{{\fontfamily{cmtt}\selectfont #1}}

\def\orcid#1{\kern -0.4em\href{https://orcid.org/#1}{\includegraphics[keepaspectratio,width=0.7em]{images/orcid_logo.pdf}}}

\newtcolorbox{defbox}{colback=white,colframe=blue!45!black}
\newtcolorbox{thmbox}{colback=violet!3!white,colframe=violet!75!black}
\newtcolorbox{gamebox}[1]{colback=gray!5!white,colframe=darkgray,title=#1}
\newtcolorbox{corrbox}{colback=orange!5!white,colframe=orange!75!black}
\newtcolorbox{lembox}{colback=purple!2!white,colframe=purple!85!black}
\newtcolorbox{constbox}{colback=green!2!white,colframe=green!85!black}
\DefineNamedColor{named}{GreenYellow}   {cmyk}{0.15,0,0.69,0}
\DefineNamedColor{named}{Yellow}        {cmyk}{0,0,1,0}
\DefineNamedColor{named}{Goldenrod}     {cmyk}{0,0.10,0.84,0}
\DefineNamedColor{named}{Dandelion}     {cmyk}{0,0.29,0.84,0}
\DefineNamedColor{named}{Apricot}       {cmyk}{0,0.32,0.52,0}
\DefineNamedColor{named}{Peach}         {cmyk}{0,0.50,0.70,0}
\DefineNamedColor{named}{Melon}         {cmyk}{0,0.46,0.50,0}
\DefineNamedColor{named}{YellowOrange}  {cmyk}{0,0.42,1,0}
\DefineNamedColor{named}{Orange}        {cmyk}{0,0.61,0.87,0}
\DefineNamedColor{named}{BurntOrange}   {cmyk}{0,0.51,1,0}
\DefineNamedColor{named}{Bittersweet}   {cmyk}{0,0.75,1,0.24}
\DefineNamedColor{named}{RedOrange}     {cmyk}{0,0.77,0.87,0}
\DefineNamedColor{named}{Mahogany}      {cmyk}{0,0.85,0.87,0.35}
\DefineNamedColor{named}{Maroon}        {cmyk}{0,0.87,0.68,0.32}
\DefineNamedColor{named}{BrickRed}      {cmyk}{0,0.89,0.94,0.28}
\DefineNamedColor{named}{Red}           {cmyk}{0,1,1,0}
\DefineNamedColor{named}{OrangeRed}     {cmyk}{0,1,0.50,0}
\DefineNamedColor{named}{RubineRed}     {cmyk}{0,1,0.13,0}
\DefineNamedColor{named}{WildStrawberry}{cmyk}{0,0.96,0.39,0}
\DefineNamedColor{named}{Salmon}        {cmyk}{0,0.53,0.38,0}
\DefineNamedColor{named}{CarnationPink} {cmyk}{0,0.63,0,0}
\DefineNamedColor{named}{Magenta}       {cmyk}{0,1,0,0}
\DefineNamedColor{named}{VioletRed}     {cmyk}{0,0.81,0,0}
\DefineNamedColor{named}{Rhodamine}     {cmyk}{0,0.82,0,0}
\DefineNamedColor{named}{Mulberry}      {cmyk}{0.34,0.90,0,0.02}
\DefineNamedColor{named}{RedViolet}     {cmyk}{0.07,0.90,0,0.34}
\DefineNamedColor{named}{Fuchsia}       {cmyk}{0.47,0.91,0,0.08}
\DefineNamedColor{named}{Lavender}      {cmyk}{0,0.48,0,0}
\DefineNamedColor{named}{Thistle}       {cmyk}{0.12,0.59,0,0}
\DefineNamedColor{named}{Orchid}        {cmyk}{0.32,0.64,0,0}
\DefineNamedColor{named}{DarkOrchid}    {cmyk}{0.40,0.80,0.20,0}
\DefineNamedColor{named}{Purple}        {cmyk}{0.45,0.86,0,0}
\DefineNamedColor{named}{Plum}          {cmyk}{0.50,1,0,0}
\DefineNamedColor{named}{Violet}        {cmyk}{0.79,0.88,0,0}
\DefineNamedColor{named}{RoyalPurple}   {cmyk}{0.75,0.90,0,0}
\DefineNamedColor{named}{BlueViolet}    {cmyk}{0.86,0.91,0,0.04}
\DefineNamedColor{named}{Periwinkle}    {cmyk}{0.57,0.55,0,0}
\DefineNamedColor{named}{CadetBlue}     {cmyk}{0.62,0.57,0.23,0}
\DefineNamedColor{named}{CornflowerBlue}{cmyk}{0.65,0.13,0,0}
\DefineNamedColor{named}{MidnightBlue}  {cmyk}{0.98,0.13,0,0.43}
\DefineNamedColor{named}{NavyBlue}      {cmyk}{0.94,0.54,0,0}
\DefineNamedColor{named}{RoyalBlue}     {cmyk}{1,0.50,0,0}
\DefineNamedColor{named}{Blue}          {cmyk}{1,1,0,0}
\DefineNamedColor{named}{Cerulean}      {cmyk}{0.94,0.11,0,0}
\DefineNamedColor{named}{Cyan}          {cmyk}{1,0,0,0}
\DefineNamedColor{named}{ProcessBlue}   {cmyk}{0.96,0,0,0}
\DefineNamedColor{named}{SkyBlue}       {cmyk}{0.62,0,0.12,0}
\DefineNamedColor{named}{Turquoise}     {cmyk}{0.85,0,0.20,0}
\DefineNamedColor{named}{TealBlue}      {cmyk}{0.86,0,0.34,0.02}
\DefineNamedColor{named}{Aquamarine}    {cmyk}{0.82,0,0.30,0}
\DefineNamedColor{named}{BlueGreen}     {cmyk}{0.85,0,0.33,0}
\DefineNamedColor{named}{Emerald}       {cmyk}{1,0,0.50,0}
\DefineNamedColor{named}{JungleGreen}   {cmyk}{0.99,0,0.52,0}
\DefineNamedColor{named}{SeaGreen}      {cmyk}{0.69,0,0.50,0}
\DefineNamedColor{named}{Green}         {cmyk}{1,0,1,0}
\DefineNamedColor{named}{ForestGreen}   {cmyk}{0.91,0,0.88,0.12}
\DefineNamedColor{named}{PineGreen}     {cmyk}{0.92,0,0.59,0.25}
\DefineNamedColor{named}{LimeGreen}     {cmyk}{0.50,0,1,0}
\DefineNamedColor{named}{YellowGreen}   {cmyk}{0.44,0,0.74,0}
\DefineNamedColor{named}{SpringGreen}   {cmyk}{0.26,0,0.76,0}
\DefineNamedColor{named}{OliveGreen}    {cmyk}{0.64,0,0.95,0.40}
\DefineNamedColor{named}{RawSienna}     {cmyk}{0,0.72,1,0.45}
\DefineNamedColor{named}{Sepia}         {cmyk}{0,0.83,1,0.70}
\DefineNamedColor{named}{Brown}         {cmyk}{0,0.81,1,0.60}
\DefineNamedColor{named}{Tan}           {cmyk}{0.14,0.42,0.56,0}
\DefineNamedColor{named}{Gray}          {cmyk}{0,0,0,0.50}
\DefineNamedColor{named}{Black}         {cmyk}{0,0,0,1}
\DefineNamedColor{named}{White}         {cmyk}{0,0,0,0}

\title{Unclonability and Quantum Cryptanalysis: From Foundations to Applications}
\author{Mina Doosti}

\abstract{
The impossibility of creating perfect identical copies of unknown quantum systems is a fundamental concept in quantum theory and one of the main non-classical properties of quantum information. This limitation imposed by quantum mechanics, famously known as the no-cloning theorem, has played a central role in quantum cryptography as a key component in the security of quantum protocols. In this thesis, we look at Unclonability in a broader context in physics and computer science and more specifically through the lens of cryptography, learnability and hardware assumptions. We introduce new notions of unclonability in the quantum world, namely quantum physical unclonability, and study the relationship with cryptographic properties and assumptions such as unforgeability, randomness and pseudorandomness. The purpose of this study is to bring new insights into the field of quantum cryptanalysis and into the notion of unclonability itself. We also discuss applications of this new type of unclonability as a cryptographic resource for designing provably secure quantum protocols.

First, we study the unclonability of quantum processes and unitaries in relation to their learnability and unpredictability. The instinctive idea of unpredictability from a cryptographic perspective is formally captured by the notion of unforgeability. Intuitively, unforgeability means that an adversary should not be able to produce the
output of an unknown function or process from a limited number of input-output samples of it. Even though this notion is almost easily formalized in classical cryptography, translating it to the quantum world against a quantum adversary has been proven challenging. One of our contributions is to define a new unified framework to analyse the unforgeability property for both classical and quantum schemes in the quantum setting. This new framework is designed in such a way that can be readily related to the novel notions of unclonability that we will define in the following chapters. Another question that we try to address here is "What is the fundamental property that leads to unclonability?" In attempting to answer this question, we dig into the relationship between unforgeability and learnability, which motivates us to repurpose some learning tools as a new cryptanalysis toolkit. We introduce a new class of quantum attacks based on the concept of `emulation' and learning algorithms, breaking new ground for more sophisticated and complicated algorithms for quantum cryptanalysis. 

Second, we formally represent, for the first time, the notion of physical unclonability in the quantum world by introducing Quantum Physical Unclonable Functions (qPUF) as the quantum analogue of Physical Unclonable Functions (PUF). PUF is a hardware assumption introduced previously in the literature of hardware security, as physical devices with unique behaviour, due to manufacturing imperfections and natural uncontrollable disturbances that make them essentially hard to reproduce. We deliver the mathematical model for qPUFs, and we formally study their main desired cryptographic property, namely unforgeability, using our previously defined unforgeability framework. In light of these new techniques, we show several possibility and impossibility results regarding the unforgeability of qPUFs. We will also discuss how the quantum version of physical unclonability relates to randomness and unknownness in the quantum world, exploring further the extended notion of unclonability.

Third, we dive deeper into the connection between physical unclonability and related hardware assumptions with quantum pseudorandomness. Like unclonability in quantum information, pseudorandomness is also a fundamental concept in cryptography and complexity. We uncover a deep connection between Pseudorandom Unitaries (PRU) and quantum physical unclonable functions by proving that both qPUFs and the PRU can be constructed from each other. We also provide a novel route towards realising quantum pseudorandomness, distinct from computational assumptions. 

Next, we propose new applications of unclonability in quantum communication, using the notion of physical unclonability as a new resource to achieve provably secure quantum protocols against quantum adversaries. We propose several protocols for mutual entity identification in a client-server or quantum network setting. Authentication and identification are building-block tasks for quantum networks, and our protocols can provide new resource-efficient applications for quantum communications. The proposed protocols use different quantum and hybrid (quantum-classical) PUF constructions and quantum resources, which we compare and attempt in reducing, as much as possible throughout the various works we present. Specifically, our hybrid construction can provide quantum security using limited quantum communication resources that cause our protocols to be implementable and practical in the near term. 

Finally, we present a new practical cryptanalysis technique concerning the problem of approximate cloning of quantum states. We propose variational quantum cloning ($\VQC$), a quantum machine learning-based cryptanalysis algorithm which allows an adversary to obtain optimal (approximate) cloning strategies with short depth quantum circuits, trained using the hybrid classical-quantum technique. This approach enables the end-to-end discovery of hardware efficient quantum circuits to clone specific families of quantum states, which has applications in the foundations and cryptography. In particular, we use a cloning-based attack on two quantum coin-flipping protocols and show that our algorithm can improve near term attacks on these protocols, using approximate quantum cloning as a resource. Throughout this work, we demonstrate how the power of quantum learning tools as attacks on one hand, and the power of quantum unclonability as a security resource, on the other hand, fight against each other to break and ensure security in the near term quantum era.
}

\begin{document}

\begin{preliminary}

\maketitle

\standarddeclaration

\clearpage
\section*{Lay summary} \label{sec:lay_summary}

\begin{small}
One of the most routine tasks we do almost every day on our computers is copying a file. A computer file contains information in the form of a string of zeros and ones. But, what if instead of a normal file, data was encoded inside a tiny physical system? In fact, a system from the subatomic world where different rules of physic would apply to it. The set of rules in that scale is known as the theory of quantum mechanics, and quantum mechanics says that if you have a `quantum file', it is forbidden to copy it! This fundamental rule of physics called the `no-cloning theorem', or unclonability, while seemingly very limiting, is a very convenient property of nature. It allows us to conceal information and share it securely. Controlling quantum systems for the secure transmission of information and similar cryptographic tasks is called quantum cryptography. 

On the other hand, we can also control these subatomic systems to perform computation, leading to physical devices known as quantum computers that perform computational tasks in a fundamentally different way from any `classical' computer. Despite applications in many areas, such as solving some mathematical problems, optimizing some operations, and simulating complex molecules, quantum computers do not bring good news for our cryptosystems. For the same reason that they are efficient in solving some mathematical problems, they can break many of today's cryptosystems as they are based on the assumption that solving those problems would take too long time to be feasible. 

Even given the long and challenging technological road ahead of building quantum computers, there has been incredible progress in recent years that has brought the idea of quantum computing to reality. Today's quantum computers, although `small' in scale and `low' in quality, can perform interesting tasks even now. Plus, we believe that sooner or later, we will get to the regime where quantum computers will surpass the limit of computation for any classical computers. Thus we need to be prepared for the threats that they will bring on.

The study of cryptography, in the near future, where we can both exploit quantum systems in our favour and will be at risk due to their computational power, is the art and science called quantum cryptanalysis. To master this art, one needs to understand the strengths and limitations of quantum systems. As such, the unclonability will be at the heart of it.

In this thesis, we study quantum unclonability beyond its usual scope. We explore other forms of natural unclonability that not only, are fundamentally connected to no-cloning, but can also be exploited for cryptography. An example of unclonable objects is optical devices that are particularly unique since their formation or manufacturing processes involve factors that we cannot control. As a result, they become physical devices that are not reproducible. This uniqueness makes them also a physical key. These physically unclonable objects can be modelled in the regime of quantum mechanics. A major part of this thesis includes the comprehensive study of them in the quantum regime, their several interesting properties, and finally, their applications in cryptography.

We also explore the relationship between unclonability and learning, that is how efficiently one can learn a quantum or classical system. In this research area, we use different tools from other fields of physics and computer science, such as machine learning. Specifically, we show that we can make a quantum machine to learn how to efficiently create an 'almost' satisfactory copy of a quantum system. This machine-learning algorithm can be used to attack the security of protocols. These attack analyses give a better perspective on the security of cryptosystems with current and future quantum technology and help us design our systems more securely. 

\end{small}
\clearpage

\section*{Publications and manuscripts}
The author has contributed to the following papers and publications in the course of her doctoral program.\\

\hspace{-1.1em}\textbf{Fully or partially included in this thesis:}
\begin{enumerate}
    \item {\color{Violet} Quantum physical unclonable functions: Possibilities and impossibilities. Quantum 5 (2021)}\cite{arapinis_quantum_2021}\\
    Myrto Arapinis, Mahshid Delavar, \textbf{Mina Doosti}, and Elham Kashefi\\
    \href{https://quantum-journal.org/papers/q-2021-06-15-475/}{DOI:10.22331/q-2021-06-15-475}
    \item {\color{Violet} On the connection between quantum pseudorandomness and quantum hardware assumptions. Quantum Science and Technology 7.3 (2022)}\cite{doosti_connection_2022}\\
    \textbf{Mina Doosti}, Niraj Kumar, Elham Kashefi, and Kaushik Chakraborty\\
    \href{https://iopscience.iop.org/article/10.1088/2058-9565/ac66fb}{DOI:10.1088/2058-9565/ac66fb}
    \item {\color{Violet}Client-server identification protocols with quantum PUF. ACM Transactions on Quantum Computing 2.3 (2021)}\cite{doosti_client-server_2021}\\
    \textbf{Mina Doosti}, Niraj Kumar, Mahshid Delavar, and Elham Kashefi\\
    \href{https://dl.acm.org/doi/abs/10.1145/3484197}{DOI:10.1145/3484197}
    \item {\color{Violet}Progress toward practical quantum cryptanalysis by variational quantum cloning. Physical Review A 105.4 (2022)}\cite{coyle_progress_2022}\\
    Brian Coyle, \textbf{Mina Doosti}, Elham Kashefi, and Niraj Kumar\\
    \href{https://journals.aps.org/pra/abstract/10.1103/PhysRevA.105.042604}{DOI:10.1103/PhysRevA.105.042604}
    \item {\color{Violet} A Unified Framework For Quantum Unforgeability. arXiv preprint (2021)}\cite{doosti_unified_2021}\\
    \textbf{Mina Doosti}, Mahshid Delavar, Elham Kashefi, and Myrto Arapinis\\
    \href{https://arxiv.org/abs/2103.13994}{Arxiv:2103.13994}
    \item {\color{Violet} Quantum Lock: A Provable Quantum Communication Advantage. arXiv preprint (2021)}\cite{chakraborty_quantum_2021}\\
    Kaushik Chakraborty, \textbf{Mina Doosti}, Yao Ma, Myrto Arapinis, and Elham Kashefi\\
    \href{https://arxiv.org/abs/2110.09469}{Arxiv:2110.09469}\\
    
    \hspace{-1.5em}\textbf{Excluded from this thesis:}
    \item {\color{Violet} Differential Privacy Amplification in Quantum and Quantum-inspired Algorithms. arXiv preprint (2022) / workshop paper (SRML)}\cite{angrisani_differential_2022}\\
    Armando Angrisani, \textbf{Mina Doosti}, Elham Kashefi \\
    \href{https://arxiv.org/abs/2203.03604}{Arxiv:2203.03604}
    
\end{enumerate}

    

\clearpage

\begin{acknowledgements}
I do not believe that PhD is a one-man (or, in this case, one-woman) journey, or at least it has not been so for me. If I have reached the centre of this spiral, it has been the result of many fortunate 'environmental factors' and 'lucky interactions' with amazing people I met along the way (just like a lucky quantum state that has finally collapsed to the right state). So I do not intend to keep this acknowledgement short, by no means. 

First and foremost, I have to thank my supervisors, Elham Kashefi and Myrto Arapinis, the two brilliant, enthusiastic, inspiring and, certainly lovely women I have had the honour to spend my PhD under their supervision. I can never be thankful enough for their non-stop support, their patience, their insightful pieces of advice, and for being much more than just advisors to me. Starting my PhD and coming from a physics background, I have been an outsider to computer science and cryptography. This thesis is the product of their tireless effort in transforming me into the hybrid creature that I am now. Elham, who never came short on thrilling ideas, whose energy always awed me, and who taught me how to be an independent researcher. Myrto, who patiently educated me to appreciate 'formal' mathematical frameworks and taught me how to think like a cryptographer, and, was there for me in all the ups and downs. I also greatly thank Anne Broadbent, and Petros Wallden, my PhD examiners for their valuable comments on this thesis.

Although I cannot only thank Petros for being my examiner, a special thank goes to him, as he is one of the most intelligent men I have met during my PhD, who never withheld me his time, someone whom I could always go to with my questions about almost any topic, and whom I learned a lot from, and with whom I have had countless hours of fruitful discussions, and pleasant conversations. My PhD would have not been the same without him. I should also thank other faculties of the quantum group at the University of Edinburgh: Raul Garcia-Patron Sanchez and Chris Heunen, as I greatly benefited from their wisdom and insight during my time as a PhD student. I appreciate having the opportunity to study and research in such a generous environment. While almost half of my PhD collided with the covid pandemic and was spent at home, in the other half, I gathered many good memories from my time at the Informatic Forum and all the people there. I also would like to thank Marc Kaplan, my mentor at VeriQloud, for his support and mentorship.

Then, of course, I should thank the comrades, the members of our amazing quantum group over these years, a group that not even a global pandemic could diminish its merit: First, Alex Cojocaru, who is like my brother, who knows how much I am thankful for his kindness, his friendship and his help all the way to the end, and I thank him also for giving me feedback on this thesis. Mahshid Delavar and Meisam Tarabkhah, for being such all-in friends, and in a way, my family here. Niraj Kumar, who not only was a mentor to me but a fantastic colleague and friend. Brian Coyle, as he truly deserves his title of "the most efficient man in the group", as he is a brilliant researcher and caring friend whom I have learned a lot from. Kaushik Chakraborty, my smart, patient and wonderful colleague, office-mate and friend, whom I greatly enjoyed working with. Atul Mantri, who despite his short time in the group, was one of the most memorable persons, as he is clever in his work and humour equally. Ellen Derbyshire, for the unique kindness and company she always offered me. Yao Ma, my awesome colleague and friend with all the exciting memories we made together. Rawad Mehzer, an equally clever and deep man whom I always enjoyed having a conversation with. Daniel Mills, a character that I will always respect and never forget. James Mills, with whom I spent memorable times. Ieva Cepaite, a cheerful and free soul. And I also thank Ross Grassie, Sima Bahrani, Theodoros Kapourniotis, Debasis Sadhukhan, other amazing postdocs that I had the pleasure to know in the group and discuss with, as well as Pablo Andr\'es-Martinez and Nuiok Dicaire from the extended Edinburgh quantum group.

And that is not all, since our group is a highly non-local one, with half of the entangled pair in Paris. I should extend my thanks to Dominik Leichtle, L\'eo Colisson, Ulysse Chabaud, Shraddha Singh, Armando Angrisani, Jonas Landman, Pierre-Emmanuel Emariau, Luka Music, Raja Yehia, Damian Markham, Harold Ollivier, Fred Groshans, Eleni Diamanti with all of whom I have great memories and interactions.

I am perhaps one of the luckiest human beings in having friends whose friendship expands over space-time. I start with my friends in Iran: to Armita, my best friend, the ever-present being in my life despite the distance, and my companion through all the pains and joys, and she knows what else. To Alma and Aria, because thinking about them (and our memories of 225) has always been the most heart-warming thought. To Laaya, Nima, Babak, Alireza and Ghazal, as they all have their own special place for me. I also thank Azadeh, our ever-supportive and great friend. Then I thank the lovely people I have met here: Mohammad and Lucy, for all the pleasing times we have spent in these years, and to S\'ego, for being such a beautiful person who she is and for all the music we played together and all the things we shared. And finally, among friends, I thank my dearest friend, who knows who he is, and although he might find it dull if he ever reads this page, I can't go on without thanking him as he has been such a great friend to me during the toughest parts of this. 

I give my sincere thanks to my mom and dad, my biggest supporters, who never ceased to encourage me during this time, even from the other side of the world, and I thank them for everything they have done for me in my life and for enduring the best and worst of me.

And as I will keep the best for the last, I thank Ramin, my love, my partner in life and crime, and my closest, most reliable company. Without him and his endless support, his love, and his beautiful mind, I probably would not be where I am now, let alone finishing this thesis and this period. Although no gratitude would do justice to what he has been for me all these years, I am glad that I have this chance to properly thank him.
\end{acknowledgements}

\dedication{\begin{large}To Ramin, my love, \\ And to every being who seeks knowledge and beauty.\end{large}}

{
\hypersetup{linkcolor=blue}
\tableofcontents
}
\listoffigures

\end{preliminary}


\chapter{Introduction}\label{chap:intro}
\begin{chapquote}{Richard Feynman}
``See that the imagination of nature is far, far greater than the imagination of man.''
\end{chapquote}

One of the most impactful scientific revolutions of the 20th century was the development of quantum mechanics. Revolutionary discoveries about the behaviour of light and matter in the late 19th and early 20th centuries, not explainable by existing knowledge of physics by that time, or what we call today \emph{classical physics}, led to a need for a completely new theory. Arguably, the most important among these was the illumination of the nature of light by Planck in 1900 and the photoelectric effect between 1883 and 1900. These physical phenomena that classical physics has still to date failed to explain have created one of those mutually horrifying and exciting situations in science: the sparks of \emph{maybe we got it all wrong!}

Fanning the flames of this idea led to the birth of the main concepts of quantum theory and eventually to a well-established formalisation of quantum mechanics as we know it today. Quantum theory has changed our mindset and understanding of nature to a great degree. Even though it has gracefully and even surprisingly explained those phenomena which classical physics fell short in describing, it did so at the cost of being counter-intuitive and odd to our \emph{classical} minds. It comes with predictions about nature, not as it appears in our everyday life, but as Richard Feynman famously put it ``nature, ... she is absurd'' \cite{richard_phillips_feynman_qed_1988}. Quantum mechanics left our human mind with questions and mysteries about \emph{probabilistic nature of observation}, \emph{non-locality}, \emph{unclonability} (which is the core topic of this thesis), and many more to ponder about over the years.

Attempting to unravel some of the mysteries of quantum mechanics led to the appearance of quantum information theory \cite{nielsen_quantum_2010}, which takes an information theory approach to study quantum systems. This new field, together with the rise of quantum computation, has engaged physicists, mathematicians, and computer scientists with new fundamental questions about the concept of computation and the differences between classical and quantum versions of it \cite{aaronson_complexity-theoretic_2016}. Quantum information, in its simplest form, starts with the idea of considering discrete quantum systems as carriers of information and treating them as quantum versions of the binary systems we use for the storage and manipulation of information. However, the field expands extensively beyond this humble foundation and incorporates the full framework of information theory and many other powerful mathematical tools such as probability theory, group theory, representation theory and so on. Using all this powerful machinery, quantum information sheds a light on complicated problems of dealing with quantum properties of nature. Thanks to quantum information, we have now developed a much better understanding of these problems to the point that most of them are no longer mysteries, even if still strange. The idea of quantum computing, on the other hand, emerged from the idea of using physical quantum mechanical systems to simulate themselves. A task that seemed to be too hard to simulate using classical digital computers. This idea was introduced by Feynman\footnote{Although not all the credit should be given to Mr Feynman! The idea of quantum computing in other forms, was also mentioned by Yuri Manin \cite{manin_computable_1980} and Paul Benioff \cite{benioff_computer_1980} in 1980.} in 1981, at a conference where he talks about the difficulties of such simulations and asks ``Can you do it with a \emph{new kind} of computer? A quantum computer?''\cite{preskill_quantum_2021}. 

To realise Feynman's groundbreaking idea would require us to understand this \emph{new kind} of computation and to eventually acquire the ability to control quantum systems for performing our desired computational or simulation task. Obtaining this ability, as Feynman has predicted ``doesn't look easy'', and almost 40 years of relentless research (from his talk) has proved to be truly the case. Notwithstanding the unresolved challenges of controlling quantum systems, there has been remarkable progress in this area, especially in recent years. One of the main challenges is to achieve quantum computers able to perform useful computational tasks, outside the reach of classical computers, which requires a considerably large scale and a high level of control over such systems. In 2019 the Google AI Quantum group announced their quantum computer, with 53 working qubits, has surpassed this limit and has achieved what is famously known as \emph{quantum supremacy} or \emph{quantum advantage} in the realm of computation \cite{arute_quantum_2019}. Despite the scepticism and critics about this result \cite{kalai_argument_2021,horner_what_2021,rinott_statistical_2021}, it is an undeniable indication of an important fact: quantum computers are no longer just an idea, and we have entered a new era. More specifically, this new era is named NISQ, standing for \emph{Noisy Intermediate-Scale Quantum} devices~\cite{preskill_quantum_2018}. The NISQ devices provide on the orders of 10s to 100 noisy qubits, but they can exploit the quantum behaviours of light or matter to execute \emph{some limited} quantum programs. Although limited, they can provide a \emph{laboratory} for theoretical research in the field of quantum computation and quantum information that was not possible until very recently.

The future large-scale quantum computers, on the other hand, are believed to have significant quantum advantages over classical computers for some specific problems, which are not limited to the simulation of physical systems. The range of problems we hope to be able to solve more efficiently with quantum computers extends to decision problems, search problems and learning problems. The ability to solve these problems, not just faster, but perhaps \emph{in a different way}, using quantum properties, has already and will continue to impact many areas of physics, computer science, chemistry \cite{lanyon_towards_2010,cao_quantum_2019,mcardle_quantum_2020}, biology \cite{emani_quantum_2021,cao_potential_2018,outeiral_prospects_2021}, and even linguistics \cite{heunen_quantum_2013,coecke_picturing_2018,meichanetzidis_grammar-aware_2020,meichanetzidis_quantum_2021}. One of the fields that has been hugely affected by quantum computing and quantum information is \emph{cryptography}. Quantum mechanics has a rather fascinating and somewhat contradictory relationship with cryptography. When quantum steps into the realm of cryptography, not only does it threaten security, but it may as well enhance it! But most definitely, it has changed the way one can look at cryptography, as it has done with computation. Let us first start with the negative side of the story.

It is well-known that Shor’s quantum algorithm \cite{shor_algorithms_1994} menaces many widely-used cryptographic schemes that are based on the mathematical hardness assumptions of factoring and the discrete logarithm problems (such as RSA). A sufficiently large, fault-tolerant and universal quantum computer will be able to run this algorithm and solve these problems efficiently \cite{wallden_cyber_2019}. Furthermore, Shor’s algorithm is not the only quantum algorithm that can be used as an attack on classical cryptography schemes. Other famous algorithms, such as Grover or Simon, have also been used for this purpose \cite{brassard_quantum_1998,kaplan_quantum_2016,jordan_quantum_2018,leander_grover_2017,grassl_applying_2016,alkhzaimi_cryptanalysis_2013}. Generally speaking, a quantum computer can be seen as a powerful computational resource in the hands of an attacker. Yet, this extra computational power is not the only aspect that can raise an issue. An adversary who has been given the possibility to exploit non-classical properties of quantum data may as well, have other advantages. For example, a quantum adversary can also use entanglement to extract crucial information from a system or use the power of quantum superposition while interacting with cryptosystems. Hence a very central question to ask is \emph{What are the possible advantages of an attacker equipped with quantum capabilities?} Answering this question, in full generality, is brutally challenging and requires a profound understanding of the underlying assumptions (both computational and physical) of cryptographic schemes, as well as the new potential ways for these quantum capabilities to be exploited to break them. Nevertheless, partially addressing this question is one of the central ideas of this thesis.

On the bright side, quantum mechanics and its odd properties provide us with a new way of achieving cryptographic functionalities, or that is to say, a fundamentally distinct type of cryptography: one that is based on the laws of quantum mechanics and the limitations that it imposes on the adversary. The field of research that studies this direction is known as \emph{quantum cryptography}. The most well-known problem studied in this field is \emph{Quantum Key Distribution (QKD)}, a protocol that enables two remote parties to establish a secure key by relying on the characteristics of quantum mechanics \cite{bennett_quantum_2014} in the presence of the most possibly powerful quantum adversary. Perhaps one of the most intriguing aspects of QKD is that, under a carefully specified set of assumptions and requirements about the underlying physical systems, it provably achieves the strongest known level of security without any computational assumptions. QKD, however, is not the only example of what quantum cryptography can bring to the table. For example, the wide range of capabilities that quantum features equip us with has motivated the construction of networks where the nodes are armed with the ability to transmit, store and manipulate small quantities of quantum information\footnote{Often referred to as \emph{quantum communication}.}\cite{wehner_quantum_2018}. These quantum networks (also called a quantum internet) can enable applications fundamentally impossible for purely classical networks and systems, such as quantum money, quantum multi-party computations, and delegated quantum computing. For a better overview of these functionalities and developed protocols, we refer the reader to the \emph{quantum protocol zoo}~\cite{veriqloud_quantum_2019}, an open repository for quantum protocols.

In quantum cryptanalysis\footnote{In cryptography, the term `cryptanalysis' is commonly used for referring to the study of attacks on cryptosystems and often at a practical or implementation level. However, in this thesis, we use this term in a slightly different sense to address both cases of breaking cryptosystems and designing secure ones, or more generally for \emph{analysing} the cryptographic properties of a system}, which is the other main topic of this thesis, we walk a thin line between these two sides, trying to win the everlasting battle between making and breaking cryptosystems in a world (probably not too far away in the future) where both sides can make the most of quantum mechanical systems and computers. As such, quantum cryptanalysis encompasses many subfields of quantum sciences from quantum computing and quantum information to the foundations of quantum mechanics, to better manipulate them for designing secure systems. Furthermore, it has even merged with relatively younger fields, such as quantum machine learning and quantum learning theory, since they provide a new ground for cryptanalysis. Also, given that quantum technology is usually more expensive and resource-intensive than the usual classical computers and existing systems, another balance to maintain in quantum cryptanalysis is between the security guarantees and the required quantum resources. Maintaining this balance, although challenging, is what makes the design of such quantum systems and protocols thought-provoking and theoretically satisfying.

As mentioned above, a key factor for building efficient and secure quantum cryptosystems in the presence of a quantum attacker is to deeply understand the fundamental and non-classical aspects of quantum systems in general. In this spirit, we can ask: \emph{What are the key elements of quantum security?} or in other words, \emph{What is it that leads to the security (of primitives or protocols) in the regime of quantum mechanics?} Depending on the required functionality or protocol, different quantum features have been used, such as entanglement and non-locality, the probabilistic nature of measurements, the indistinguishability of quantum states and conjugate coding, and most of all, \emph{unclonability}.

The unclonability of the quantum state is one of the most exploited and most common non-classical features in any cryptographic functionality that uses quantum systems. This fundamental limitation of quantum mechanics that forbids creating perfect copies of unknown quantum systems is one of the most central properties of quantum mechanics. Maybe that is why it is inherent in almost all quantum protocols and functionalities, even if not consciously employed. It is perhaps safe to say that unclonability is a \emph{resource} for achieving quantum security. However, many questions still remain regarding unclonability, such as \emph{Is the no-cloning of quantum states the only existing form of unclonability?} If not, \emph{what are the other notions of unclonability?} and \emph{how can we relate them to cryptographic properties?} Or maybe we can ask more fundamental questions such as \emph{Is there any deeper level to the concept of unclonability?} These are some of the questions that we try to tackle in this thesis as we continue to uncover the relationship between the broader notion of unclonability and quantum cryptanalysis. Finally, we aim to use the tools and concepts that we develop and gather along the way for practical applications.

\section{Thesis overview}
We give a brief summary of our contributions and the structure of the thesis. We exclude \chapref{chap:prelim}, which includes the preliminaries and background materials for the various tools we used in this thesis.

\begin{itemize}
    \item \chapref{chap:unf-tools}: We start from the foundations while focusing on three primary notions: unclonability, unforgeability and learnability. We first lay an argument about the relationship between unclonability and unknownness, a concept that we formally define for unitary transformations, and then we bring unclonability into a greater regime which encompasses both cryptography and learning theory. This chapter serves as a roadmap for the rest of the thesis, where we focus on different aspects of each of the concepts we will discuss. Moreover, we introduce two main contributions which we will widely employ in the rest of the thesis. The first one is a new class of quantum attacks based on the concept of \emph{emulation}, and the next one is a new unified framework for quantum unforgeability. Within this framework, we enclose the notion of unforgeability for both quantum and classical primitives, and we also provide a hierarchy of definitions. Finally, as a case study of our framework, several impossibility results are given, and some quantum-secure constructions have been introduced. The content of this chapter is the combination of two papers, \href{https://quantum-journal.org/papers/q-2021-06-15-475/}{\color{Violet}{Quantum physical unclonable functions: possibilities and impossibilities. Quantum 5 (2021)}}\cite{arapinis_quantum_2021} and \href{https://arxiv.org/abs/2103.13994}{\color{Violet}{A Unified Framework For Quantum Unforgeability." arXiv preprint arXiv:2103.13994 (2021)}}\cite{doosti_unified_2021}, and some unpublished results which were excluded from the mentioned papers.
    
    \item \chapref{chap:qpuf}: This chapter focuses on defining the notion of \emph{quantum Physical Unclonable Functions (qPUF)} and studying its cryptographic properties using the unforgeability framework that has been introduced in the previous chapter. PUFs are a concept borrowed from the hardware security literature. However, as we will see in this chapter, defining a quantum counterpart is not a straightforward translation to the quantum regime but a rather more fundamental generalisation of the notion of physical unclonability. Here, we answer one of the questions we have asked before, \emph{i.e.} we confirm the existence of other forms of unclonability, not unrelated to the unclonability of quantum states in quantum mechanics, while this relationship is more lucid in the context of quantum PUF compared to classical ones. We focus on the unitary subclass of quantum PUFs, and we prove general no-go results about the unforgeability property of any such primitives. Finally, we formally prove that a large class of them can satisfy a level of quantum unforgeability powerful enough to make them strong and useful hardware tokens for cryptography. This chapter is based on the paper \href{https://quantum-journal.org/papers/q-2021-06-15-475/}{\color{Violet}{Quantum physical unclonable functions: Possibilities and impossibilities. Quantum 5 (2021)}}\cite{arapinis_quantum_2021} as a part of a collaboration with \emph{Mahshid Delavar}, \emph{Myrto Arapinis} and \emph{Elham Kashefi}.
    
    \item \chapref{chap:pr-connection}: This chapter is concerned with quantum pseudorandomness and its relationship with physical unclonability and, more generally, quantum hardware assumptions. Pseudorandomness is also one of the most rudimentary building blocks of modern cryptography as it provides the randomness required for cryptographic schemes in an efficient manner. In the quantum world, quantum pseudorandomness has been recently introduced by \cite{shacham_pseudorandom_2018} via the notions of pseudorandom quantum states and pseudorandom unitaries. The pseudorandom quantum objects provide an efficient and computational form of perfect uniform randomness over the Hilbert spaces known as Haar-randomness. In this chapter, we first study the connection between pseudorandom quantum states and unforgeability. We prove that using quantum pseudorandomness will allow the same level of security guarantee for unforgeability while improving efficiency. Then we delve into the relationship between quantum physical unclonability and pseudorandom unitaries, and we show that they are closely connected to the point that they can be derived from each other in terms of functionality. We also show that, interestingly, considering some assumptions over the family of qPUFs, even without assuming the full extent of quantum physical unclonability, will lead to quantum pseudorandom objects. This chapter is the result of a collaboration between \emph{Kaushik Chakraborty}, \emph{Niraj Kumar} and \emph{Elham Kashefi}, published in \href{https://iopscience.iop.org/article/10.1088/2058-9565/ac66fb}{\color{Violet}{On the connection between quantum pseudorandomness and quantum hardware assumptions. Quantum Science and Technology 7.3 (2022)}}\cite{doosti_connection_2022}.
    
    \item \chapref{chap:application}: This chapter which includes three main results from three projects is dedicated to applications of quantum physical unclonability and quantum-enhanced physical unclonable functions. In the first part of the chapter, we introduce two new identification protocols based on qPUFs as we have defined and studied in earlier chapters. Our protocols include client-server scenarios: that is, in the first one, a quantum server intends to identify a low-resource client who only owns a qPUF device, and in the second protocol, the client identifies a quantum server with a qPUF device, while we manage to delegate the quantum verification to the server as well such that the client only needs to run a classical verification test. Amid the security proof of these two protocols lies one of our leading arguments earlier concerning unclonability, since we will see how quantum physical unclonability serves as a resource for these protocols to achieve exponential security in only a polynomial number of rounds of quantum communication. We also thoroughly discuss the role of differed quantum testing algorithms as our verification subroutines and compare them, which can be of interest even outside the scope of the presented protocols and more generally for other quantum communication protocols. Furthermore, to provide sufficient theoretical ground and benchmark for the experimental realisation of these protocols in the future, we give a resource analysis in terms of quantum memory, quantum communication and quantum computation resources. In the second part of the chapter, we will show that the results we have demonstrated in \chapref{chap:pr-connection} bring on efficiency and practicality to our qPUF-based protocols. Finally, in the last part of this chapter, we introduce a new quantum-enhanced PUF construction that combines classical physical unclonability with quantum communication and combines the best of both worlds. This construction, although weaker than a full quantum PUF, allows for an efficient identification protocol that is implementable with today's technology, for instance, the existing QKD infrastructure, while it achieves a high level of security against quantum adversaries. This application also shows a particular \emph{provable} advantage of quantum communication vs classical ones, as we will discuss through different properties that the protocol achieves. To prove the security of our construction and protocol, we use many tools and previous results from quantum information theory, including entropic uncertainty relations. The first part of the chapter is based on the work done in collaboration with \emph{Niraj Kumar}, \emph{Mahshid Delavar} and \emph{Elham Kashefi}, which resulted in this publication \href{https://dl.acm.org/doi/abs/10.1145/3484197}{\color{Violet}{Client-server identification protocols with quantum puf." ACM Transactions on Quantum Computing 2.3 (2021)}}\cite{doosti_client-server_2021}. The second part is from a small section of the paper mentioned before \cite{doosti_connection_2022}, while it was more appropriate to be included in this chapter. Lastly, the third part of this chapter is from a collaboration with \emph{Kaushik Chakraborty}, \emph{Yao Ma}, \emph{Myrto Arapinis}, and \emph{Elham Kashefi}, resulted in the paper \href{https://arxiv.org/abs/2110.09469}{\color{Violet}{Quantum Lock: A Provable Quantum Communication Advantage. arXiv preprint arXiv:2110.09469 (2021)}}\cite{chakraborty_quantum_2021}. We note that this last paper is included partially for more coherence and brevity of the chapter. 
    
    \item \chapref{chap:varqlone}: Finally, we turn to another aspect of the relation between quantum unclonability and quantum cryptanalysis, this time from a machine learning perspective. This chapter introduces a new cryptanalysis toolkit and method based on approximate quantum cloning and variational algorithms. We introduce our machine learning algorithm, $\VQC$, that can efficiently learn to (approximately) clone quantum states of a specified family optimally and in a hardware-friendly manner. This algorithm can have several applications in the context of quantum foundation and quantum computing, specifically since it can be run on NISQ devices, as we have done so. However, in this thesis, we are particularly interested in its application for cryptanalysis. For this purpose, we take two classes of quantum protocols, QKD and quantum coin-flipping, for case studies, and we relate their security to cloning-based attacks based on $\VQC$. We argue that cryptanalysis in this new fashion, even for protocols that have been information-theoretically proven secure like QKD, is beneficial since it allows for benchmarking the state of the art of the current technology with the state of the art of sophisticated attacks that are also implementable on current hardware. Besides the relevance in the application, this chapter allows us to come back to the foundations. In the course of this chapter, we connect the security of certain classes of quantum protocols to specific classes of approximate cloning. This type of cloning-based cryptanalysis brings us one step closer to understanding the role of unclonability as a source of security in quantum cryptography. We even offer several theoretical guarantees and results on the specifications of this algorithm that could be of interest to the quantum machine learning community. The content of this chapter is from a collaboration with \emph{Brian Coyle}, \emph{Niraj Kumar} and \emph{Elham Kashefi} and was published in \href{https://journals.aps.org/pra/abstract/10.1103/PhysRevA.105.042604}{\color{Violet}{Progress toward practical quantum cryptanalysis by variational quantum cloning." Physical Review A 105.4 (2022)}}\cite{coyle_progress_2022}. Again, since the context of the research done in this paper is wider than the focus and interest of this thesis, we have only included the most relevant parts and main contributions of the current author.
\end{itemize}
\chapter{Preliminaries} \label{chap:prelim}
\begin{chapquote}{Nightwish, The Greatest Show on Earth}
Man, he took his time in the sun\\
Had a dream to understand\\
A single grain of sand\\
He gave birth to poetry\\
But one day'll cease to be\\
Greet the last light of the library
\end{chapquote}

\noindent We start with a general remark regarding this chapter. This chapter attempts to cover all the necessary backgrounds, topics, concepts and tools used in this thesis. Some sections mainly provide general knowledge on the subject, and others, briefly introduce, sometimes in more detail, the definitions or tools used later on in the thesis. Throughout this preliminary chapter, whenever a specific notion or tool is introduced, we navigate the reader to the part of the thesis it is employed. However, for a reader with familiarity with the general topics covered here, we suggest skipping this chapter and returning to each subsection when referred to subsequently in the following chapters.

\section{Quantum information and quantum computing}\label{sec:prelim-qi}
Let us begin the chapter by giving some background on quantum information and quantum computing. We assume some familiarity with quantum mechanics and although it will not be necessary for understanding the content of this thesis, it is encouraged for enjoying it. We also assume familiarity with linear algebra.

\subsection{Quantum states and Hilbert space}\label{sec:prelim-qstate}
The concept of \emph{quantum states} and where they live, which is called \emph{state space} comes from the first postulate of quantum mechanics. According to this postulate, any isolated physical system can be described (or be associated with) a \emph{vector}, in a complex vector space with an inner product, known as \emph{Hilbert space}. This vector, that completely describes the physical system, or sometimes the physical quantum mechanical property of a physical system, is called the \emph{state} of the system and is a normalised unit vector in the Hilbert space \cite{nielsen_quantum_2010}\footnote{There is an alternative version of the first postulate of quantum mechanics that has an additional statement that the other way is also assumed, meaning that giving a Hilbert space, where a physical system is described in it, any vector of the Hilbert space is also a potential state of a physical system. Sometimes this is inherent in the evolution postulate, however, it is interesting to think about it as the first postulate as well.}. We denote a Hilbert space of dimension $d$ by $\Hil$ or sometimes $\Hild$. Any $d$-dimensional Hilbert space is equipped with a set of $d$ orthonormal vectors called a basis. The most important quantum systems in quantum information are 2-level quantum systems or quantum states living in a 2-dimensional Hilbert space. We call this special state a \emph{qubit}. The following set of vectors is a complete basis for a qubit, referred to as the computational bases:
\begin{equation}
    \ket{0}= \begin{bmatrix}
1 \\ 0
\end{bmatrix}\quad \quad \ket{1}=\begin{bmatrix}
0 \\ 1
\end{bmatrix}
\end{equation}
Here $\ket{.}$ (called `ket') or $\bra{.}$ (called `bra') are known as \emph{Dirac notation} and are the most common notation in quantum. Moreover, for any Hilbert space $\Hil$ we can define a dual space denoted by $\Hil^*$, where for any $\ket{\psi} \in \Hil$, there exists a dual $\bra{\psi} \in \Hil^*$, that is its complex conjugate, such that $\mbraket{\psi}{\psi} = 1$. Also the inner product between two vectors $\ket{\psi}, \ket{\phi} \in \Hil$ is shown in the Dirac notation as $\mbraket{\phi}{\psi}$.

Any qubit state can be written as a linear combination of the basis for instance: 
$\ket{\psi}= \alpha \ket{0} + \beta \ket{1}$
where $|\alpha|^2 +|\beta|^2 =1$ for any $\alpha, \beta \in \mathbb{C}$, since the state should be normalised. This linear combination of other quantum states (for instance any basis state) is called a \emph{superposition} of those quantum states. According to quantum mechanics, any normalised superposition of the states is also a valid member of the Hilbert space, due to linearity and hence is another valid quantum state. The coefficients $\alpha$ and $\beta$ are also called \emph{amplitudes} and are complex numbers. One of the most useful superpositions, in the equal-weight superposition of the computational basis like the following:

\begin{equation}
    \ket{+} = \frac{1}{\sqrt{2}}(\ket{0} + \ket{1}), \quad \quad \ket{-} = \frac{1}{\sqrt{2}}(\ket{0} - \ket{1})
\end{equation}
one can see that the state $\ket{+}/\ket{-}$ are also orthonormal and hence form another basis for qubit Hilbert space. This basis is called \emph{plus-minus basis} or \emph{X-basis} (we will see why in \ref{sec:prelim-quantum-gates}). The generalisation of such uniform superposition basis states in higher dimension is also known as \emph{Fourier basis}.

\subsection{Mixed states and density matrices}\label{sec:prelim-mixed-state}
Now, let us give a more general and complementary formalism for describing qubits and quantum states. When we describe a quantum system by a vector in the Hilbert space, we \emph{deterministically} describe its state (however, as we will see, the process of revealing the state is itself probabilistic), in this case, we call it a \emph{pure} quantum state. Nevertheless, not all the systems in nature are like that. Some systems, are in fact, a probability distribution over different pure quantum states. We call these states \emph{mixed states}, and we can represent them as follows:
\begin{equation}
    \rho :=\sum_{s} p_{s}\ket{\psi_s}\bra{\psi_s}
\end{equation}
where $\ket{\psi_s}\bra{\psi_s}$ is the outer product of the two vectors. The above representation means that we prepare a pure state $\ket{\psi_s}$ with probability $p_s$. Also, one can see that $\rho$ is no longer a vector, but a matrix. This kind of matrices, called \emph{density matrices} \cite{fano_description_1957} are a more general way of describing all quantum systems, including the pure one, since if we have only one probability $p=1$, then $\rho = \ket{\psi}\bra{\psi}$, which is a pure state and the density matrix equivalent of $\ket{\psi}$.

In operator language, a density operator for a system is a positive semi-definite, Hermitian operator of trace one ($\tr(\rho) = 1$) acting on the Hilbert space of the system. We denote the set of all the density matrices associated with the Hilbert space $\Hil$, as $\ES(\Hil)$. Geometrically, this is a convex set. Also a pure state always satisfy $\tr(\rho^2) = 1$, while as for a mixed state $\tr(\rho^2) < 1$. This is a good criterion for checking the purity of a density matrix.

\subsubsection{Bloch sphere}\label{sec:prelim-bloch}
\begin{figure}[ht!]
    \centering
    \includegraphics[width=0.35\textwidth]{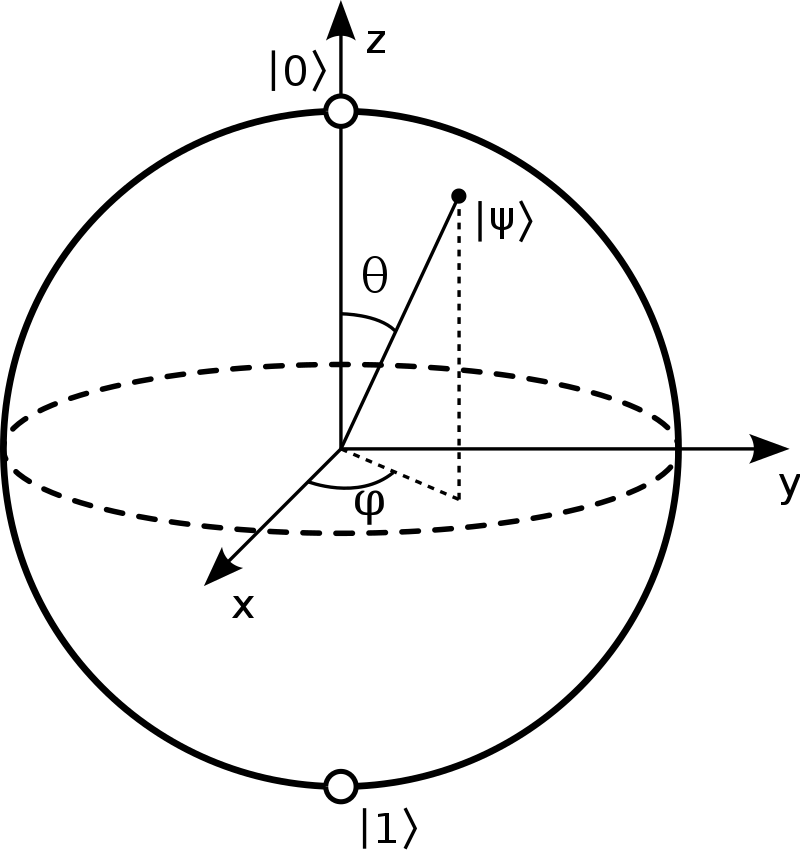}
    \caption{The Bloch sphere (figure from:~\cite{wikipedia_bloch_2022})}
    \label{fig:bloch}
\end{figure}

For the 2-dimensional space of qubits, there is a simple and pleasant geometrical representation for all the possible pure and mixed states. It is described as a unit sphere, called \emph{Bloch sphere}, as shown in \figref{fig:bloch}.

The surface of the Bloch sphere represents the pure state of a qubit and all the points inside of the sphere represent the set of the mixed states. A pure qubit state can be described in terms of the angles associated with its Bloch vector, as can be seen in the figure. For $0 \leq \theta \leq \pi$ and $0 \leq \phi < 2\pi$, the state of a qubit $ \ket{\psi}$ is described as:

\begin{equation}\label{eq:prelim-bloch-sphere-state}
    \ket{\psi} = \cos{\frac{\theta}{2}}\ket{0} + e^{i\phi} \sin{\frac{\theta}{2}} \ket{1}
\end{equation}

\noindent A general qubit density matrix, can be written as \cite{brus_lectures_2006}:

\begin{equation}
    \rho = \frac{1}{2} (\mathbb{I} + x_1 X + x_2 Y + x_3 Z)
\end{equation}
where $\mathbb{I}$ is the identity matrix and X, Y and Z are the following matrices known as \emph{Pauli matrices}:
\begin{equation}\label{eq:prelim-pauli-matrices}
    X = \begin{pmatrix}
    0 & 1\\
    1 & 0
    \end{pmatrix}, \quad
    Y = \begin{pmatrix}
    0 & -i\\
    i & 0
    \end{pmatrix}, \quad
    Z = \begin{pmatrix}
    1 & 0\\
    0 & -1
    \end{pmatrix}
\end{equation}

Thus, states of single qubits are characterized by a vector $(x1, x2, x3) \in \mathbb{R}^3$ taken from the unit
ball, which is the Bloch sphere. 
\subsubsection{Composition of quantum systems and entanglement}\label{sec:prelim-tensor}
The very original formulation of quantum mechanics talks about a single quantum system described by a vector in Hilbert space. However, if we want to describe the joint state of two quantum systems $\Hil_1$ and $\Hil_2$, we need to deal with their composition. The most common framework for the composition of quantum states is tensor product composition. In other words, the composite system $\ket{\psi_{1,2}}$ is described as a unit vector in the tensor product of the Hilbert spaces $\Hil_1 \otimes \Hil_2$.\footnote{This is usually considered as an (extended) axioms of the quantum mechanics, however, it is also possible not to assume it, and to derive it instead from general composition rules and physical evidence~\cite{aerts_physical_1978}.} 

If a quantum state can be written as the tensor product of all its subsystems, we say that the state is \emph{separable}, for example: $\ket{\psi_{AB}} = \ket{\psi_A}\otimes\ket{\psi_B}$. Nevertheless, not all the states in $\Hil_A \otimes \Hil_B$ can be written as such. The states that cannot be described in this tensor product form are called \emph{entangled} states, and physically, they contain some non-classical correlation known as \emph{entanglement}. The following defines the general definition of separable and entangled states for bipartite mixed states: 

\begin{defbox}
\begin{definition}[Separable and entangled mixed states \cite{brus_lectures_2006}]\label{def:prelim-entg}
A mixed state $\rho_{AB}$ is separable if and only if it can be represented as a convex combination of the product of projectors on local states in the form of the following equation. Otherwise, the mixed state is said to be entangled.
\begin{equation}\label{eq:prelim-separable}
    \rho_{AB} = \sum^K_{i=1} p_i \ket{e_i}\bra{e_i} \otimes \ket{f_i}\bra{f_i}
\end{equation}
where $\ket{e_i}$ and $\ket{f_i}$ are a basis for subsystem $A$ and $B$ respectively. 
\end{definition}
\end{defbox}

We also note that the states in \eqref{eq:prelim-separable} describe the most general state of a class of states called LOCC, meaning the most general states that two parties, Alice and Bob can prepare using only \emph{local operation} and \emph{classical communication}.

Another advantage of the density matrix formalism is that it allows us to describe the quantum states of the subsystems of a joint quantum system, even if the state is not separable \cite{nielsen_quantum_2010}. For this we can take the partial trace of the other subsystem, to obtain the subsystems of interest, for instance:
\begin{equation}
    \rho_A = \tr_B(\rho_{AB}), \quad \text{and}, \quad \rho_B = \tr_A(\rho_{AB})
\end{equation}
where $\tr_B$ meaning taking the trace over subsystem $B$, using the basis of this subspace. $\rho_A$ and $\rho_B$ are called a \emph{reduced density matrix} of the system.

Now, let us introduce the most important entangled bipartite states in quantum information, also known as \emph{Bell states}, which are as follows:
\begin{equation}
\begin{split}
    & \ket{\Phi^{\pm}} := \frac{1}{\sqrt{2}}(\ket{00} \pm \ket{11})\\
    & \ket{\Psi^{\pm}} := \frac{1}{\sqrt{2}}(\ket{01} \pm \ket{10})
\end{split}
\end{equation}
where $\ket{00} = \ket{0}_A \otimes \ket{0}_B$ (and similarly for the rest of the basis). These states contain the maximum amount of entanglement between all the pure bipartite states. Furthermore, they have an interesting property that their reduced density matrices are the state $\frac{\mathbb{I}}{2}$, which is a state known as \emph{maximally mixed} state.\footnote{In general maximally mixed state for Hilbert space of dimension $d$ is given as $\frac{\mathbb{I}_d}{d}$.} For instance, we have:
\begin{equation}
    \rho_A = \rho_B = \tr_A(\ket{\Phi^+}\bra{\Phi^+}_{AB}) = \frac{\mathbb{I}}{2}
\end{equation}
Looking at the reduced density matrix of joint quantum systems can in general give information about the amount of entanglement contained in these systems.

\subsection{Quantum operations and measurements}\label{sec:prelim-channel-povm}
So far we gave a brief introduction to quantum systems and some of their properties. Now it is time to talk about how quantum systems evolve and transform into other quantum systems. 

The first form of quantum operation that we know, according to postulates of quantum mechanics, are unitary operators. A unitary matrix $U$, can transform a pure and mixed quantum state as follows:
\begin{equation}
    \ket{\psi'} = U\ket{\psi} \quad \text{and}, \quad
    \rho' = U \rho U^{\dagger}
\end{equation}
Recalling the Bloch sphere, the unitary operation of any pure qubit state is equivalent to a rotation of the vector on the surface of the Bloch sphere (up to a phase factor). 
However, unitary matrices are not the most general form of quantum operations. General quantum transformations are \emph{Completely Positive Trace Preserving (CPTP or CPT)} maps which include also unitary matrices. These operations are also called a \emph{quantum channel} and can map a general density matrix $\rho \in \Hil$ to another density matrix $\rho' \in \Hil'$ (where $\Hil'$ is often the same as $\Hil$, but not necessarily), as follows:
\begin{equation}
    \E: \ES(\Hil) \rightarrow \ES(\Hil'), \quad \quad \rho' = \E(\rho)
\end{equation}
Quantum channels can take the following general from:
\begin{equation}
    \E(\rho) = \sum_k E_k \rho E^{\dagger}_k
\end{equation}
where $E_k$ are operators that should satisfy the following criterion for the overall operation $\E$ to be trace-preserving:
\begin{equation}
    \sum_k E^{\dagger}_k E_k = \mathbb{I}
\end{equation}
This representation of quantum channels is called \emph{operator-sum} formalism \cite{nielsen_quantum_2010} and the decomposition of a quantum channel into such operators is also sometimes called \emph{Kraus decomposition}. There also exists an alternative way of representing quantum channels from the point of view of system-environment interaction. This point of view is very interesting since it shows that all the operations can eventually be described via a unitary operation on a larger or expanded Hilbert space which also includes the environment. Let the quantum state $\rho$ be entangled with a system $\ket{E}$ that describes the environment. If a unitary operation is applied to the joint state of the system-environment, the operation that is applied to the system alone can be described as follows:
\begin{equation}
    \rho' = \E(\rho) = \tr_E [U(\rho \otimes \ket{E}\bra{E})U^{\dagger}]
\end{equation}
This operation is no longer a unitary but a CPTP map. We should also note that this later interpretation gives us a good intuition and toolkit to study the effect of noise on the quantum system in the same way as we describe any other transformations of them. A quantum \emph{noise} is also described as a CPTP map and be studied with the same mathematical toolkits theoretically. Commonly we define specific classes of quantum channels that model the most common errors and noisy behaviour that happens to the actual devices. The most famous ones are \emph{bit-flip noise channel}, \emph{phase-flip noise channel}, \emph{Pauli noise channel}, \emph{depolarising noise channel}, \emph{dephasing noise channel}, and \emph{amplitude-damping noise channel} \cite{brus_lectures_2006,nielsen_quantum_2010}.

\subsubsection{Measurements}
We now introduce one of the most central types of operations in quantum mechanics \emph{i.e.} measurements. Measurement operators offer a mathematical formalism for studying the process of observation and extracting the real values for the physical properties of a quantum system. These values are in some sense \emph{the classical information} of the system given by expectation values of a Hermitian observable. Quantum measurements are described as a set of linear operators $\{M_m\}$ acting on the state where the index $m$ refers to each measurement outcome. If $\ket{\psi}$ is the pure quantum state before the measurement, then the probability of obtaining result $m$, and the state of the system after the measurement is given as follows \cite{nielsen_quantum_2010}:
\begin{equation}
\begin{split}
& p(m) = Pr[\text{obtaining result } m]=\bra{\psi}M_m^{\dagger}M_m\ket{\psi} \\
    & \ket{\psi_m} = \frac{M_m \ket{\psi}}{\sqrt{\bra{\psi}M_m^{\dagger}M_m\ket{\psi}}}
\end{split}
\end{equation}
Thus quantum measurements are probabilistic in nature, and since the probability should be preserved over the full set of measurement, they also satisfy the following completeness equation: 
\begin{equation}
    \sum_m \bra{\psi}M_m^{\dagger}M_m\ket{\psi} = \sum_m p(m) = 1
\end{equation}

This probability rule for quantum systems, also known as \emph{Born's rule}, for a general mixed system is given as follows:
\begin{equation}
    p(m) = Pr[\text{obtaining result } m]= \tr[M_m \rho M_m^{\dagger}]
\end{equation}

The first type of measurement, and a very useful one, are \emph{projective measurements}, which are given by a set of projective operators. A simple example is a set $\{M_0,M_1\}$ where $M_0 = \ket{0}\bra{0}$ and $M_1 = \ket{1}\bra{1}$. This is a qubit measurement which projects everything in the $Z$ basis of the Bloch sphere, and it is also famously known as \emph{measurement in the computational basis}.\footnote{The computational basis measurement can be easily generalised to any dimension by only making the projective operator from the computational basis of that Hilbert space.}

The most general class of measurement in the quantum world is given by a mathematical formalism known as \emph{Positive Operator-Valued Measure (POVM)}. A POVM is described as a set of positive operators $\{E_m\}$ satisfying the relation $\sum_m E_m = \mathbb{I}$, and they obey the same Born's rule as we described earlier. This class also includes the projective measurements, however, one difference between the projective measurements and a POVM non-projective one is that the cardinality of the set of POVM measurements over a Hilbert space can be larger than the dimension as opposed to the projective ones. For instance, for qubits, we have seen that the full set of computational basis measurements, includes 2 projectors (which is the same for measuring on any arbitrary basis), but one can define the following valid set of POVM measurements on a qubit:
\begin{equation}
    \begin{split}
        & E_1 = (2 - \sqrt{2})\ket{1}\bra{1}, \quad \quad E_2 = (2 - \sqrt{2})\ket{-}\bra{-}\\
        & E_3 = \mathbb{I} - E_1 - E_2
    \end{split}
\end{equation}

The POVM can be interpreted physically in different ways. The first one is when we are applying a projective measurement on the joint state of our system within a larger system or correlated with another system. In this case, although the measurement on the larger Hilbert space is projective, the resulting measurement on the main systems that we are interested in is not, and it's instead a POVM. This scenario is similar to the case we have discussed before regarding the CPTP maps and there is a good reason for this resemblance due to the fact that POVMs are also CPTP maps and can be described in that formalism. Another way of physically interpreting the POVM is when the real measurement devices are not perfect and instead of performing a perfect projection, they perform a combination of a projection and another quantum operation. In that case, again the measurement can be mathematically described as a POVM. This last point is very important in distinguishing quantum states from each other as we will see in Section~\ref{sec:prelim-distinguish-quantum-test}.

\subsection{Distance measures}\label{sec:prelim-distances}
Distance measures are mathematical tools for comparing systems on different aspects, for instance, their information quantity. In the classical world, these comparisons are often straightforward. As an example, to compare classical bit strings, we can simply check their equality. But one can also define a better and more fine-grained distance for classical information, for example, by counting the number of places where two bitstrings are different. This distance is called \emph{Hamming distance} in classical information theory and gives a good measure for quantifying classical information in many cases. But how about quantum information. As we know by now, the quantum information lives inside the state of a qubit, that is, a vector in a continuous vector space. And more importantly, while revealing this information (measurement) we are dealing with a probabilistic process. Hence comparing and quantifying the distance between quantum information and generally quantum systems are more tricky! Fortunately, the mathematical background of quantum mechanics and quantum information is strong enough to handle this more complicated situation, and a large variety of quantum distance measures have been defined in the literature, each of which, is useful for different problems that we face in this field \cite{nielsen_quantum_2010,brus_lectures_2006,ghosh_quantum_2018,modi_unified_2010,maciejewski_exploring_2022,bera_quantum_2018,gilchrist_distance_2005}. Here we only introduce the very few most relevant distance measures for this thesis.

But before, that let us start with a classical distance, which has been incorporated in the quantum regime very similarly. This is the case when we want to compare two probability distributions $\{p(x)\}$ and $\{q(x)\}$. One very common and quite intuitive way of defining a notion of distance for them is as follows:
\begin{equation}
    d(p(x),q(x)) := d_{\ell_1}(p(x),q(x)) = \frac{1}{2} \sum_x |p(x) - q(x)|
\end{equation}
This distance is called \emph{trace distance} or \emph{$\ell_1$-norm}. And the first quantum distance that we introduce is the generalisation of this distance. The trace distance is defined as follows:

\begin{defbox}
\begin{definition}[Trace distance]\label{def:prelim-trace-distance}
For any general quantum state $\rho$ and $\sigma$, their trace distance is defined as:
\begin{equation}
    \dtr(\rho, \sigma) := \frac{1}{2} \tr|\rho - \sigma| = \frac{1}{2} \sum_i |\lambda^{\rho\sigma}_i|
\end{equation}
where  $|\rho - \sigma| := \sqrt{(\rho - \sigma)^{\dagger}(\rho - \sigma)}$ is defined as the positive square root of the matrix, and $\lambda^{\rho\sigma}_i$ are eigenvalues of the Hermitian, but not necessarily positive, matrix $(\rho -\sigma)$.
\end{definition}
\end{defbox}

The second measure of distance that is widely used all over quantum information, and this thesis included, is \emph{fidelity}. Although fidelity is not a \emph{metric} on the space of density matrices \cite{nielsen_quantum_2010}, it is one of the most useful measures of the `closeness' of two quantum states. One of the reasons is that fidelity has an operational meaning: intuitively it expresses the probability that one state will pass a test to identify as the other one. The fidelity between two quantum states in the most general case is known as \emph{Uhlmann's fidelity} and is defined as follows:

\begin{defbox}
\begin{definition}[Fidelity]\label{def:prelim-fidelity}
For any general quantum state $\rho$ and $\sigma$, their Uhlmann's fidelity is defined as:
\begin{equation}
    F(\rho, \sigma) :=  \left(\tr\left[\sqrt{\sqrt{\rho}\sigma\sqrt{\rho}}\right]\right)^2
\end{equation}
which is equal to the squared overlap $F(\ket{\psi_{\rho}},\ket{\psi_{\sigma}}) := |\mbraket{\psi_{\rho}}{\psi_{\sigma}}|^2$, for two pure quantum states $\ket{\psi_{\rho}}$ and $\ket{\psi_{\sigma}}$.
\end{definition}
\end{defbox}

The third quantum distance that we introduce, is closely related to the fidelity. In fact, we first define a geometrical metric in the space of quantum states, known as \emph{Bures angle}, as follows:
\begin{equation}\label{eq:prelim-bures-angle}
     \dba(\rho, \sigma) := \arccos{\sqrt{F(\rho, \sigma)}}
\end{equation}

The Bures angle is itself a distance but it is also associated with another important metric in the quantum information, known as \emph{Bures distance} which is also the quantum equivalent of \emph{Fubini-Study} metric \cite{study_kurzeste_1905}.

\begin{defbox}
\begin{definition}[Fubini-Study/Bures distance]\label{def:prelim-fubini-distance}
For any general quantum state $\rho$ and $\sigma$, their Bures/Fubini-Study distance is defined as follows:
\begin{equation}
    \dfs(\rho, \sigma) :=  \left(2(1 - \sqrt{F(\rho, \sigma)})\right)^{\frac{1}{2}} = \sqrt{2(1 - \cos{\dba})}
\end{equation}
\end{definition}
\end{defbox}

There are several important and useful properties and features of these distances which we need to cover for the purpose of this thesis. First, we need to note that all these distances, including the trace distance and fidelity, should be preserved under the unitary evolution of quantum states. This is because unitaries, preserve the inner product and hence should also preserve any notion of distance that we define over the space of density matrices. Thus we have the following central relations \cite{nielsen_quantum_2010}:

\begin{equation}
    \begin{split}
        \dtr(U\rho U^{\dagger}, U\sigma U^{\dagger}) & = \dtr(\rho, \sigma) \\
        F(U\rho U^{\dagger}, U\sigma U^{\dagger}) & = F(\rho, \sigma)
    \end{split}
\end{equation}

But how about the distance between quantum states, after a non-unitary general quantum channel is applied to them? It can be shown that quantum channels are \emph{contractive}, meaning that they decrease the distance between quantum states. This is captured in the following theorem in terms of trace distance:

\begin{thmbox}
\begin{theorem}\label{th:prelim-trce-dist-contractive}[Contractivity of quantum channels \cite{nielsen_quantum_2010}]
Suppose $\E$ is a CPTP map. Let $\rho$ and $\sigma$ be any two density operators. We have:
\begin{equation}
    \dtr(\E(\rho),\E(\sigma)) \leq \dtr(\rho,\sigma)
\end{equation}
\end{theorem}
\end{thmbox}

The same result can be reformulated in terms of fidelity leading to the fact that $F(\E(\rho),\E(\sigma)) \geq F(\rho,\sigma)$ under any CPTP operation, which is usually referred to as \emph{Monotonicity of the fidelity}\cite{nielsen_quantum_2010}.

Another property of trace distance that will come in very handy in our proofs is what is known as \emph{strong convexity} and is stated as follows:

\begin{thmbox}
\begin{theorem}\label{th:prelim-trce-dist-convexity}[Strong convexity of the trace distance \cite{nielsen_quantum_2010}]
Let $\{p_i\}$ and $\{q_i\}$ be two probability distributions over the same index set, and $\rho_i$ and $\sigma_i$ be density matrices associated with the same index set. Then the trace distance satisfies the following:
\begin{equation}
    \dtr\left(\sum_i p_i\rho_i,\sum_i q_i\sigma_i\right) \leq d_{\ell_1}(p_i,q_i) + \sum_i \dtr(\rho_i,\sigma_i)
\end{equation}
where $d_{\ell_1}(p_i,q_i)$ is the classical trace-distance between the two probability distribution.
\end{theorem}
\end{thmbox}

Similarly, we have \emph{strong concavity} for fidelity:

\begin{thmbox}
\begin{theorem}\label{th:prelim-fidelity-concavity}[Strong concavity of fidelity \cite{nielsen_quantum_2010}]
Let $\{p_i\}$ and $\{q_i\}$ be two probability distributions over the same index set, and $\rho_i$ and $\sigma_i$ be density matrices associated with the same index set. Then the trace distance satisfies the following:
\begin{equation}
    F\left(\sum_i p_i\rho_i,\sum_i q_i\sigma_i\right) \geq \sum_i \sqrt{p_i q_i} F(\rho_i,\sigma_i)
\end{equation}
\end{theorem}
\end{thmbox}

which also results in the weaker version called \emph{concavity} of the fidelity:
\begin{equation}
    F(\sum_i p_i\rho_i,\sigma) \geq \sum_i p_i F(\rho_i,\sigma)
\end{equation}

Finally, it is important to be able to translate between fidelity and trace distance. The relation between the two, is given via the following inequality:
\begin{equation}\label{eq:prelim-trace-dist-fidelity}
    1 - \sqrt{F(\rho ,\sigma)} \leq \dtr(\rho ,\sigma) \leq \sqrt{1-F(\rho ,\sigma )}
\end{equation}

One can also define distances and norms on the operator space. These distances are called \emph{operator norms}. The first important example is the \emph{operator infinity norm} or $\ell^{\infty}$-norm. In general, an operator norm $\ell^\infty$ is defined on a Banach space\footnote{Banach space is a complete normed vector space. That is, a Banach space is a vector space with a metric that allows the computation of vector length and distance between vectors, and is complete in the sense that a `Cauchy sequence' of vectors always converges to a well-defined limit that is within the space \cite{allan_introduction_2011}} or a bounded sequence of elements or vectors of that space as follows:
\begin{equation}
\label{eq:infnorm-op}
    ||x||_{\infty} = \sup_n |x_n|, 
\end{equation}
The operator norm on the Hilbert space is defined over the space of bounded linear operators as,
\begin{equation}
\label{eq:infnorm-hilbert}
    ||O||_{\infty} = \sup ||Ox|| : \forall ||x|| \leq 1, 
\end{equation}
We also note that for the operator norms, $||.||_1$ is the dual norm of $||.||_{\infty}$ \cite{hiai_contraction_2016}.

The final distance measures that we introduce, which is particularly beneficial distance in the quantum setting, is a distance called \emph{diamond norm}, defined as follows: 
\begin{defbox}
\begin{definition}[Diamond norm]\label{def:prelim-diamond-norm}
For any two CPTP map (quantum channel) $\Lambda_1$, $\Lambda_2$, their diamond norm is defined as,
\begin{equation}
    \parallel \Lambda_1 - \Lambda_2 \parallel_{\diamond} := \underset{\rho}{\max} (\parallel(\Lambda_1 \otimes \mathbb{I})[\rho] - (\Lambda_2 \otimes \mathbb{I})[\rho]\parallel_1) 
\end{equation}
where $ \parallel.\parallel_1$ is the \emph{$\ell_1$-norm}, and the maximum has been taken over all the density matrices $\rho$.
\end{definition}
\end{defbox}

Operationally diamond norm quantifies the maximum probability of distinguishing operation $\Lambda_1$ from $\Lambda_2$ in a single-use, and it is a sensible measure to quantify the difference between unitary operators or other quantum channels.

\subsection{Entropic uncertainty relations}\label{sec:prelim-uncert}
In this section, we introduce a more advanced but very useful toolkit in quantum information that is also related to cryptography. It consists of a mathematical framework and several inequalities known as \emph{conditional entropies} or \emph{entropic uncertainty relations}, which have been used in the formal security proofs of several quantum protocols, specifically, the Quantum Key Distribution (QKD) protocols \cite{renner_security_2008,tomamichel_largely_2017}.\footnote{For this subsection we assume some familiarity with the concept of QKD protocols since we refer to it several times. However, the details of the protocol are not compulsory for understanding the tools that we introduce here. We do not intend to give a background on the QKD protocol(s) as it will make this preliminary even longer than it is. We refer the readers to \cite{nielsen_quantum_2010} for a general overview of the protocol and to \cite{tomamichel_largely_2017} for a comprehensive and advanced description and security proofs.} We have mostly exploited the content of this section in \chapref{chap:application}, Section~\ref{sec:application-hpuf-security-proofs} and \ref{sec:application-hlpuf-reusability} and \chapref{chap:varqlone} Section~\ref{sec:varqlone-phasecov-crypt}.

But first, we need to introduce the notion of entropy in quantum and classical information. In classical information theory, the entropy for a random variable is defined as follows, and it is called \emph{Shannon’s entropy}.

\begin{defbox}
\begin{definition}[Shannon’s entropy]\label{def:prelim-shannon-ent}
Let $X$ be a discrete random variable on a finite set $\X=\{x_1,\dots,x_n\}$, with probability distribution function $p(x)=Pr(X=x)$. The entropy $H(X)$ of $X$ is defined as:
\begin{equation}
    h(X) = - \sum_{x\in \X} p(x) \log p(x) = \mathbb{E}[- \log p(X)]
\end{equation}
while we consider $0 \log 0 = 0$ and logarithm is usually taken to the base $2$, in which case the entropy is measured in bits.
\end{definition}
\end{defbox}

Intuitively, this measure quantifies the amount of `information' (or on the other side `uncertainty') in a system. As we mentioned, quantum states also include a certain degree of classical information, which we can extract through the probabilistic procedure of measurements. As a result, we can also assign entropy to quantum states. The quantum version of Shannon's entropy is called \emph{Von Neumann entropy}\footnote{In this thesis, we mostly use $H(.)$ to denote the general notion of entropy which can in some cases refer to Shannon's entropy and in some others to Von Neumann entropy. However, if we specifically want to emphasise Shannon's entropy, we use the notation $h(.)$} which is defined as follows:

\begin{defbox}
\begin{definition}[Von Neumann entropy]\label{def:prelim-von-neumann-ent}
For a quantum-mechanical system described by a density matrix $\rho$, the von Neumann entropy is
\begin{equation}
   S(\rho) = -\tr(\rho \log \rho) = -\sum_i \lambda_i \log_2 (\lambda_i)
\end{equation}
where $\lambda_i$ are the eigenvalues of $\rho$, we consider $0 \log 0 = 0$ and logarithm is taken to the base $2$ or $e$. Furthermore $S(\rho)$ is zero if and only if $\rho$ represents a pure state.
\end{definition}
\end{defbox}

We are now ready to talk about uncertainty in quantum mechanics, in terms of entropy. Heisenberg's uncertainty principle is one of the most important fundamental properties of quantum mechanics which is mathematically speaking due to the non-commuting property of observables, like Pauli $X$ and $Z$. Reformulating these relations in terms of entropic quantities has been very useful and the most well-known uncertainty relation for these operators was given by Deutsch~\cite{deutsch_uncertainty_1983} and later improved~\cite{maassen_generalized_1988}. The relation, for $X$ and $Z$ observables, is given as follows:

\begin{equation}\label{eq:prelim-uncert-xz}
  H(X) + H(Z) \geq \log_2 (\frac{1}{c})  
\end{equation}

where $c$ denotes the maximum overlap between any two eigenvectors of $X$ and $Z$. Since the entropy is defined with respect to a random variable, we need to see what are our random variables here. First, we consider a quantum system $A$ where the state is described with the density matrix $\rho_A$ on a finite-dimensional Hilbert space. We then assume a measurement is performed on a $X$ and $Z$ basis as projective operators that project the state into the subspace spanned by those bases. Thus the random variables are defined via measurements of the observers $X$ and $Y$. In the most general case, the measurements are a set of POVM operators on system $A$ denoted as $\{M^x\}_x$ and $\{N^z\}_z$ satisfying the Born rule for obtaining outcomes $x$ and $z$ to be as follows:
\begin{equation}
  P_X(x) = \tr[\rho_A M^x] \quad , \quad  P_Z(z) = \tr[\rho_A N^z]
\end{equation}

In this case, the \eqref{eq:prelim-uncert-xz} still gives the generalised uncertainty relation with the difference that the $c$ is defined as follows:

\begin{equation}
  c = \max_{x,z} c_{zx}, \quad \text{and} \quad c_{xz} = \parallel \sqrt{M^x}\sqrt{N^z} \parallel^2
\end{equation}

where $\parallel.\parallel$ denotes the operator norm (or infinity norm) defined in Section \ref{sec:prelim-distances}. The above uncertainty relation can be extended to conditional entropy as well in the context of \emph{guessing games} as has been defined in \cite{coles_entropic_2017}. Assume two parties, Alice and Bob, where Bob prepares a state $\rho_A$ and Alice randomly performs the $X$ and $Z$ measurements leading to a bit $K$. Then Bob wants to guess $K$ given the basis choice $R=\{0,1\}$. The conditional Shannon entropy is defined as follows:

\begin{equation}
  H(K|R) := H(KR) - H(R)
\end{equation}
 Thus one can get the same uncertainty relation with the conditional entropy as:
 \begin{equation}
  H(K|R=0) + H(K|R=1) \geq \log_2(\frac{1}{c})
\end{equation}
Similar, to the classical case, for a bipartite system $\rho_{AB}$ the conditional Von Neumann entropy is defined as follows:

\begin{equation}
  H(A|B) := H(\rho_{AB}) - H(\rho_B)
\end{equation}

Furthermore, this can be generalised to any tripartite quantum system with state $\rho_{ABC}$. An interesting property here is an inequality referred to as \emph{data processing inequality}~\cite{coles_entropic_2017} which states that the uncertainty of $A$ conditioned on some system $B$ never goes down if $B$ performs a quantum channel on the system. In other words for any tripartite system $\rho_{ABC}$ where system $C$ will perform a quantum operation on the quantum state to extract some information, we have the following:

\begin{equation}\label{eq:prelim-data-processing-ineq}
  H(A|BC) \leq H(A|B)
\end{equation}

Given the above inequality leads to the general uncertainty relations between any tripartite system. One of the most common scenarios is when we have two honest parties, Alice and Bob, and an \emph{eavesdropper} or \emph{adversary} called Eve, for example in the QKD protocol. In this case, the following entropic inequality holds:
\begin{equation}
  H(K|ER) + H(K|BR) \geq \log_2(\frac{1}{c})
\end{equation}

Where $K$ is the measurement output and $R$ is the basis bit. This imposes a fundamental bound on the uncertainty in terms of von Neumann entropy, in other words, the amount of information that an eavesdropper can extract from the joint quantum systems shared between the three parties, is fundamentally bounded by quantum mechanics. These inequalities can also be extended to the case where $n$ bits are encoded in $n$ quantum states where $R^n$ and $K^n$ are bit-strings denoting the basis random choices for the qubits and measurement outputs respectively, and $B^n$ denotes Bob's bit-string. Also, $E$ denotes Eve's system which is a general quantum system operating on $n$-qubit messages and any arbitrary local system. We have the following inequality:\footnote{This is the main result we will use in Section~\ref{sec:application-hpuf-security-proofs} and Section~\ref{sec:application-hlpuf-reusability} for our security proof.}
\begin{equation}\label{eq:prelim-n-fold-entropic-unc}
  H(K^n|ER^n) + H(K^n|B^nR^n) \geq n \log_2(\frac{1}{c})
\end{equation}

The amount of information shared between joint quantum systems can also be defined in terms of other informatic quantities such as \emph{mutual information} or \emph{accessible information}. Again, let us discuss these quantities in a two party scenario. Consider a scenario where Alice prepares a pure quantum state drawn from the ensemble $\{p_y,|\psi_y\rangle\}$ with the density matrix $\rho_{AB}$, where
\begin{equation}
    \label{eq:prelim-denst-alice-bob}
    \rho_{AB} = \sum_{y}p_y|y\rangle_A\langle y| \otimes |\psi_y\rangle_B \langle \psi_y|.
\end{equation}

Bob knows the ensemble i.e., the mixed state $\rho_{AB}$, but not the particular state that Alice chose. He wants to acquire as much information as possible about $y$. Bob collects his information by performing a generalized measurement, the POVM ${M_{\tilde y}}$. Bob's state is of the form $\rho_B = \tr_A(\rho_{AB})$ as it is the subsystem of the larger density matrix. If Alice's preparation choice was $y$, Bob will obtain the measurement outcome $\tilde y$ with conditional probability $p(\tilde y|y) = \bra{\psi_y}M_{\tilde y}\ket{\psi_y}$. For this kind of classical-quantum state $\rho_{AB}$, the amount of information Bob can extract from this measurement is given by a quantity called \emph{mutual information (MI)} $I(Y;\tilde Y)_{\rho}$ between $Y,\tilde Y$ which is defined as follows.
\begin{equation}\label{eq:prelim-mutual-inf}
    I(Y :\tilde Y) :=  h(Y) - h(\tilde Y|Y),
\end{equation}



Nonetheless, all the entropic quantities that we have discussed so far, work well in the asymptotic limit, while other similar quantities are more suited to capture the finite-size systems. \emph{Min- and max-entropy} are the notion first proposed by Renner \cite{renner_security_2008} as the natural generalizations of what was known as \emph{conditional Rényi entropies}~\cite{renyi_measures_1961} to the quantum setting. The definition is as follows:

\begin{defbox}
\begin{definition}[Min- and max- entropy \cite{konig_operational_2009}]\label{eq:prelim-min-max-ent}
Let $\rho = \rho_{AB}$ be a bipartite density operator. The min-entropy of $A$ conditioned on $B$ is defined by:
\begin{equation}
    H_{\text{min}}(A|B)_{\rho} := - \inf D_{\infty}(\rho_{AB} \parallel \mathbb{I}_A \otimes \sigma_B)
\end{equation}
where the infimum ranges over all normalized density operators $\sigma_B$ on subsystem $B$ and $D_{\infty}(.\parallel .)$ is defined as follows:
\begin{equation}
    D_{\infty}(\tau \parallel \tau') := \inf\{\lambda \in \mathbb{R}: \tau \leq 2^{\lambda}\tau'\}
\end{equation}
and the max-entropy is defined as:
\begin{equation}
    H_{\text{max}}(A|B)_{\rho} := - H_{\text{min}}(A|C)_{\rho}
\end{equation}
where the min-entropy is evaluated for a purification $\rho_{ABC}$ of $\rho_{AB}$.
\end{definition}
\end{defbox}

The above entropies can be then parameterised by a parameter $\varepsilon \geq 0$ called the \emph{smoothness} parameter. The smooth version of the min and max entropies can be defined as follows:

\begin{defbox}
\begin{definition}[$\varepsilon$-smooth min/max entropy \cite{konig_operational_2009}]\label{eq:prelim-smooth-min-max-ent}
Let $\rho = \rho_{AB}$ be a bipartite density operator and let $\varepsilon \geq$ be a parameter. The $\varepsilon$-smooth min/max entropy of $A$ conditioned on $B$ are defined by:
\begin{equation}
\begin{split}
    H^{\varepsilon}_{\text{min}}(A|B)_{\rho} & := \underset{\rho'}{\sup} H_{\text{min}}(A|B)_{\rho'},\\
    H^{\varepsilon}_{\text{max}}(A|B)_{\rho} & := \underset{\rho'}{\inf} H_{\text{max}}(A|B)_{\rho'}
\end{split}
\end{equation}
where the supremum ranges over all density operators $\rho' = \rho'_{AB}$ which are $\varepsilon$-close to $\rho$.\footnote{The $\varepsilon$-closeness can be defined with respect to both trace distance and fidelity. However usually defining them in terms of fidelity is more suitable since it is invariant under purification.}
\end{definition}
\end{defbox}

\noindent The final relevant tool of information theory that we need to introduce in this section is called quantum \emph{Asymptotic Equipartition Property (AEP)} defined in \cite{renner_security_2008}. This is the quantum equivalent of classical (AEP) that roughly speaking, talks about the probability of typical sets occurring in a random or stochastic process when having a series of many random variables. Here we need a special case of quantum AEP for a $n$-fold quantum-classical system, which we represent as the following theorem.

\begin{thmbox}
\begin{theorem}[quantum AEP for quantum-classical states~\cite{renner_security_2008}]\label{th:prelim-quantum-aep}
Let $\rho_{XB} \in \ES(\Hil_X\otimes \Hil_B)$ be density operator that is classical on $\Hil_X$ and let $N \in \mathbb{N}$. Then, for any $\varepsilon \geq 0$ we have,
\begin{equation}
   \frac{1}{N} H^{\varepsilon}_{min}(\rho_{XB}^{\otimes N}|\rho_{B}^{\otimes N}) \geq H(\rho_{XB}) - H(\rho_B) - \eta
\end{equation}
where $\eta := (2H_{max}(\rho_X)+3)\sqrt{\frac{\log(\frac{1}{\varepsilon})}{N} + 1}$, is a function of the smoothing parameter $\epsilon$ and $N$.
\end{theorem}
\end{thmbox}

\subsection{Quantum computing}\label{sec:prelim-quantum-computing}
We have so far given a brief and general background on quantum information. In this section, we will glance over the necessary tools and concepts from quantum computing that we will require for the rest of this thesis. Let us begin with this question: \emph{What do we need to make a quantum computer?} This question was answered by DiVincenzo in \cite{divincenzo_physical_2000}, where certain criteria have been proposed for constructing a quantum computer. The following are the seven proposed criteria (the last two are necessary for \emph{quantum communication}).

\begin{defbox}
\textbf{DiVincenzo criteria}~\cite{divincenzo_physical_2000}
\begin{enumerate}
    \item A scalable physical system with well-characterized qubit
    \item The ability to initialize the state of the qubits to a simple fiducial state (quantum state preparation)
    \item Long relevant decoherence times
    \item A qubit-specific measurement capability
    \item A \emph{universal} set of quantum gates
    \item[*] The ability to interconvert stationary and flying qubits
    \item[*] The ability to faithfully transmit flying qubits between specified locations
\end{enumerate}
\end{defbox}

In the previous sections, we have covered the first four criteria since we have introduced qubits and their transformations (as well as the notion of noise) and measurements. In this section, we will focus on the fifth one while we introduce quantum computation and quantum gates. To see why we need quantum gates, we need to have a look at different models of computation, especially in the quantum world.

\subsubsection{Different models of quantum computation}
The first abstract classical model for computation was a \emph{Turing machine (TM)}. A Turing machine contains four main elements \cite{nielsen_quantum_2010}: (a) a program, (b) a finite state control, co-ordinating the other operations of the machine; (c) a tape, (which is like a memory); and (d) a read-write tape-head, which points to the position on the tape which is currently readable or writable. This simple system is capable of capturing any classical algorithm. The \emph{Quantum Turing machine (QTM)} gives the same type of abstraction for quantum computing. Quantum Turing Machine was firstly proposed by Paul Benioff in 1980 and 1982 \cite{benioff_computer_1980,benioff_quantum_1982}, and then further formalised by David Deutsch in 1985 \cite{deutsch_quantum_1985} while the alternative model which we will talk about has also been introduced. Similar to the classical case, a QTM has also a finite set of states $Q = \{q_0,q_1,\dots\}$, a finite set of input and working alphabet and an infinite quantum tape that models the quantum memory and a single `head'. QTM is usually initialised at a state $\ket{\psi(0)}$ and will perform the computation by applying unitary transformation to the state \emph{i.e.} at every step $\ket{\psi(i + 1)} = U \ket{\psi(i)}$. Finally, the process of reading will include quantum measurements as one can expect. Although the intuitive notion of QTM is quite simple, the formal definition is rather complicated and hence we skip introducing it here.\footnote{Moreover this is not the model of computation that we use in this thesis.}

The most common model of quantum computation (and the one used in this thesis) is the \emph{quantum circuit model}, which is a quantum generalisation of the classical \emph{circuit model}. Classically,  a circuit consists of several inputs and outputs (bits), wires which describe these systems, and several logical gates \cite{nielsen_quantum_2010}. A logical gate is a binary function $f:\{0,1\}^m \rightarrow \{0,1\}^n$ for example, \texttt{AND}, \texttt{OR} or \texttt{NOT} gates. In the quantum circuit model, on the other hand, our inputs are qubit (or more generally quantum states), and our logical gates are unitary transformations. In the classical circuit model, to be able to perform \emph{any} classical computation we need a \emph{universal gate set}. The quantum circuit model is no different. A set of quantum gate $\mathcal{G} = \{G_i\}$ is universal if \emph{any general $n$-qubit unitary operation} can be approximated using a quantum circuits that uses this gate set, with an arbitrary accuracy \cite{nielsen_quantum_2010}.

The set of universal quantum gates is not \emph{unique} and different options have been proposed with different theoretical and most importantly implementational advantages and disadvantages for implementing over different types of hardware \cite{chow_universal_2012}. All of them, however, require some \emph{sing-qubit} gates and some \emph{entangling gates}. In the next section, we introduce some of the most widely used quantum gates.

To conclude this section, let us briefly mention another model of quantum computing known as \emph{Measurement-Based Quantum Computing (MBQC)} introduced in \cite{raussendorf_one-way_2001}. The reason for this naming is that in this model, the initial resource is an entangled state in a form of a graph or cluster (called \emph{graph state} and \emph{cluster state}), and each operation is performed by applying a measurement. This technique is also known as \emph{gate teleportation} \cite{gottesman_demonstrating_1999}. Moreover, MBQC has been shown to be equivalent to the circuit model. We do not go into further details about this model, since we will not use it in the thesis. 

\subsubsection{Quantum gates}\label{sec:prelim-quantum-gates}
The first set of single-qubit quantum gates that we introduce is \emph{Pauli} gates represented by the Pauli matrices we have seen in \eqref{eq:prelim-pauli-matrices}. We first note that the computational basis, are the eigenvectors of $Z$, and the plus-minus basis, are the eigenvectors of the Pauli matrix $X$. The eigenvectors of the $Y$ operator are also very similar to the $X$ and are given as follows:
\begin{equation}
    \ket{+i} = \frac{1}{\sqrt{2}}(\ket{0} + i\ket{1}), \quad \quad \ket{-i} = \frac{1}{\sqrt{2}}(\ket{0} - i\ket{1})
\end{equation}
Also one can easily check the action of $X$, $Y$ and $Z$ gate on the computational basis, by applying their matrix on the basis vectors.
\begin{equation}
\begin{split}
    & X\ket{0} = \ket{1}, \quad X\ket{1} = \ket{0} \\
    & Y\ket{0} = -i\ket{1}, \quad Y\ket{1} = i\ket{0} \\
    & Z\ket{0} = \ket{0}, \quad Z\ket{1} = -\ket{1} \\
\end{split}
\end{equation}
The $X$ gate is the equivalent of the classical `bit-flip' gate, and the $Z$ gate is a `Phase gate'. The $Y$ gate is the combination of both since $Y=iXZ$.

The next gate is called \emph{Hadamard gate}, denoted as $H$ that acts as follows on the computational basis:
\begin{equation}
\begin{split}
    H\ket{0}=\ket{+} & = \frac{1}{\sqrt{2}}(\ket{0} + \ket{1})\\
    H\ket{1}=\ket{-} & = \frac{1}{\sqrt{2}}(\ket{0} - \ket{1})
\end{split}
\end{equation}
The Hadamard gate in fact transforms the computational basis to plus-minus basis. Also, the Hadamard gate creates the symmetric superposition of computational basis even in higher dimension if it is applied as a tensor product form $H^{\otimes n}$ over $n$ qubits. The matrix representation of $H$ is as follows:
\begin{equation}
    H = \frac{1}{\sqrt{2}}\begin{pmatrix}
    1 & 1\\
    1 & -1
    \end{pmatrix}
\end{equation}
As mentioned before, the unitary transformation of a qubit is equivalent (up to a global phase) to rotation on the Bloch sphere, hence one can define the general following rotation single-qubit gates \cite{nielsen_quantum_2010}:
\begin{equation}
\begin{split}
    R_X (\theta) & := e^{-i\theta/2 X} = \begin{pmatrix}
    \cos{\frac{\theta}{2}} & -i\sin{\frac{\theta}{2}}\\
    -i\cos{\frac{\theta}{2}} & \cos{\frac{\theta}{2}}
    \end{pmatrix} \\
    R_Y (\theta) & := e^{-i\theta/2 Y} = \begin{pmatrix}
    \cos{\frac{\theta}{2}} & -\sin{\frac{\theta}{2}}\\
    -\cos{\frac{\theta}{2}} & \cos{\frac{\theta}{2}}
    \end{pmatrix} \\
    R_Z (\theta) & := e^{-i\theta/2 Z} = \begin{pmatrix}
    e^{-i\theta/2} & 0\\
    0 & e^{i\theta/2}
    \end{pmatrix}
\end{split}
\end{equation}

The most useful 2-qubit gates are \text{CNOT} (also called \text{CX} or controlled-X) and \texttt{CZ} (or controlled-Z) gates. The importance of these gates is that they can create entanglement. Let us first give the matrix representation of these gates:
\begin{equation}
    \text{CNOT} = \begin{pmatrix}
    1 & 0 & 0 & 0 \\ 0 & 1 & 0 & 0 \\ 0 & 0 & 0 & 1 \\  0 & 0 & 1 & 0
    \end{pmatrix}, \quad \quad
    \text{CZ} = \begin{pmatrix}
    1 & 0 & 0 & 0 \\ 0 & 1 & 0 & 0 \\ 0 & 0 & 1 & 0 \\  0 & 0 & 0 & -1
    \end{pmatrix}
\end{equation}
In the above gates, the first qubit acts as a `control' and the second qubit as a `target'. \text{CNOT} applies a bit-flip or $X$ gate on the second qubit if the control qubit is $\ket{1}$ (and does nothing if it is $\ket{0}$), and similarly, \text{CZ} applies a $Z$ gate on the second qubit conditioned on the first one being $\ket{1}$. We also note that the \text{CNOT} together with the set of all single-qubit unitary gates form a universal gate set.

Other general \emph{controlled-gates} can be also defined similarly as follows:
\begin{equation}
    \text{CU}_{12} = \ket{0}\bra{0}_1 \otimes \mathbb{I}_1 + \ket{1}\bra{1}_2 \otimes U_2
\end{equation}
where $U$ is conditionally applied on the second qubit.

Finally, a quantum gate that we will use throughout this thesis is another 2-qubit (or generally multi-qubit gate) gate known as the $\SWAP$ gate. The $\SWAP$ gate on two quantum states with arbitrary dimensions acts as follows:
\begin{equation}
    \SWAP\ket{\psi}\ket{\phi} = \ket{\phi}\ket{\psi}
\end{equation}
This gate swaps between the Hilbert space of two quantum states. The qubit $\SWAP$ gate can be built from three CNOT gates, and is given with the following matrix:
\begin{equation}
    \SWAP = \begin{pmatrix}
    1 & 0 & 0 & 0 \\ 0 & 0 & 1 & 0 \\ 0 & 1 & 0 & 0 \\  0 & 0 & 0 & 1
    \end{pmatrix}
\end{equation}

As a final remark, we note that any single-qubit gate may be approximated to
arbitrary accuracy using a finite set of gates \cite{nielsen_quantum_2010}. This is the result of one of the most important theorems in quantum computing, namely \emph{Solovay–Kitaev theorem}~\cite{kitaev_quantum_1997}. More precisely, the Solovay–Kitaev theorem states that for any single-qubit gate $U$ and any $\epsilon \geq 0$, it is possible to approximate $U$ to a precision $\epsilon$ using $\Theta(log^c(1/\epsilon))$ gates from a fixed finite set, where $c$ is a small constant approximately equal to 2.

\section{Distinguishability and verification of quantum states}\label{sec:prelim-distinguish-quantum-test}
An important difference between qubits and classical bits (and generally quantum and classical states) is that it is impossible to obtain the exact classical description\footnote{Here by `classical description' we mean the value of the amplitudes in a specific basis with arbitrary precision} of a single given copy of a quantum system. This important limitation imposed by quantum mechanics is closely related to \emph{no-cloning theorem}, which we will thoroughly introduce in Section~\ref{sec:prelim-cloning}, and we will also further discuss this fundamental connection in Section~\ref{sec:unclone-unknown}. However, having access to copies of the same quantum system allows for the extraction of the state's description. As a result, there exists a bound on how well one can derive the classical description of quantum states depending on their dimension and the number of available copies. This problem is known as the problem of \emph{state estimation} in quantum information \cite{hradil_quantum-state_1997,paris_quantum_2004}. Due to the experimental relevance, in addition to quantum information, this problem has been also widely studied in quantum optics \cite{paris_quantum_2004,doherty_feedback_1999,fischer_quantum-state_2000,okamoto_experimental_2012}.

Another tightly related yet different problem in quantum information is the problem of \emph{state discrimination}. State discrimination refers to the task of distinguishing an unknown (pure or mixed) state $\rho$ in a known set of states. More precisely, given an $d$-level quantum system $\rho$ be one of the states from the set $\{q_i,\rho_i\}^N_{i=1}$, that the ensemble of states $\rho_i$s, each happening with probability $q_i$, the goal is to determine $\rho$ is which one of the states of the set, by performing the best possible POVM, leading to the minimum error discrimination probability.

For the case of two mixed states, the best probability of discrimination is given by a famous bound in quantum information theory known as \emph{Holevo-Helstrom} bound:
\begin{equation}\label{eq:prelim-holevo-holstrom-bound}
    Pr^{opt}_{\text{guess}} = \frac{1}{2} + \frac{1}{2}\norm{q_1 \rho_1 - q_2 \rho_2}_1 = \frac{1}{2} + \dtr(q_1 \rho_1 - q_2 \rho_2)
\end{equation}

Also, for two pure quantum states $\ket{\psi_1}$ and $\ket{\psi_2}$, there exists a general optimal strategy for state discrimination with projective measurements. This result which we represent in the following theorem is an indirect consequence of \emph{Neumark's theorem} (or Naimark's theorem) for general POVMs \cite{bae_quantum_2015}, and hence sometimes called Neumark's measurements.

\begin{thmbox}
\begin{theorem}\label{th:prelim-neumark-disc}
The best discrimination strategy for two pure state $\ket{\psi_1}$ and $\ket{\psi_2}$ with projective measurements $\{\ket{v_1},\ket{v_2}\}$, where
$\ket{v_1}$ and $\ket{v_2}$ are in the span of $\ket{\psi_1}$ and $\ket{\psi_2} $such that $\mbraket{v_1}{v_2} = 0$, is when they are symmetric with respect to the angle bisector of $\ket{\psi_1}$ and $\ket{\psi_2}$, and $\ket{v_i}$ is closer to $\ket{\psi_i}$ for $i = 1, 2$. On outcome "$\ket{v_i}$", we guess $\ket{\psi_i}$. Moreover, let the angle between $\ket{\psi_1}$ and $\ket{\psi_2}$ be defined as: $\theta = \arccos{|\mbraket{\psi_1}{\psi_2}|^2}$. Then the success probability of this strategy is given by:
\begin{equation}\label{eq:prelim-neumark-measurement-succ}
    Pr_{\text{succ}} = |\mbraket{\psi_1}{v_1}|^2 = \cos^2{(\frac{\pi/2 - \theta}{2})} = \frac{1}{2} + \frac{1}{2}\sin{\theta}
\end{equation}
\end{theorem}
\end{thmbox}

One can check that this optimal probability, can be obtained from Holevo-Helstrom bound in \eqref{eq:prelim-holevo-holstrom-bound} as a special case.

We can also assume another quantum state discrimination scenario where we do not get any false results, while we allow the measurement outcome to be inconclusive. This means that if the measurement outcome indicates one of the states, we know, for sure, that it's the correct state. Although sometimes the measurement's outcome is: `I don't know'! In the literature of quantum computing, this scenario is known as \emph{unambiguous state discrimination} \cite{ivanovic_how_1987,dieks_overlap_1988,peres_how_1988}. We note that this problem is particularly of interest from an experimental point of view and while dealing with imperfect measurement devices and observers \cite{bergou_quantum_2007,mohseni_optical_2004,bergou_discrimination_2010}. In this problem again, the goal is to find the best set of POVM measurements for this problem. Then minimise the probability of an inconclusive outcome, finding the optimal POVMs. For the case of two pure quantum states with equal prior probability, the following optimal strategy has been given and proved optimal in \cite{ivanovic_how_1987,dieks_overlap_1988,peres_how_1988}.

\begin{thmbox}
\begin{theorem}\label{th:prelim-unamb-disc}
The best strategy to unambiguously discriminate two pure states $\ket{\psi_1}$ and $\ket{\psi_2}$ is to carry out the POVM $\{E_0, E_1, E_2\}$ where we guess $\ket{\psi_2}$ if the outcome is 1, we guess $\ket{\psi_1}$ if the outcome is 2, and the discrimination is inconclusive if the outcome is 0. The optimal probabilities for this case are given as follows for the angle between $\ket{\psi_1}$ and $\ket{\psi_2}$ is $\theta$.
\begin{equation}\label{eq:prelim-unamb-disc-succ}
\begin{split}
    & Pr[\text{outcome 1}] = tr(E_1 \rho) = 1 - \cos{\theta} \\
    & Pr[\text{outcome 2}] = tr(E_2 \rho) = 0 \\
    & Pr[\text{outcome 0}] = tr(E_0 \rho) = \cos{\theta}
\end{split}
\end{equation}
\end{theorem}
\end{thmbox}

Nevertheless, this is the simplest case of unambiguous state discrimination and the problem has been generalised to $N$ linearly independent states in \cite{chefles_optimum_1998}, and to mixed states in \cite{bergou_optimal_2006,rudolph_unambiguous_2003} and the reader can also find one of the most recent developments on this topic in \cite{karimi_optimal_2021}.

Note that, in general, the discrimination problem is directly related to the trace distance between the given quantum states. More generally, distinguishing between different quantum states is related to their \emph{quantum distance}, quantified by distance measures. We have introduced some of them in Section~\ref{sec:prelim-distances}. While possibly the most common distance measure for this purpose is the trace distance, here we want to give a general definition of distinguishability with Uhlmann fidelity. This definition is one of the most acquainted definitions in the thesis.

\begin{defbox}
\begin{definition}[$\mu$-distinguishability] \label{def:prelim-dist}
Let $0 \leq \mu \leq 1$ be the distinguishability threshold parameter. We say two quantum states $\rho$ and $\sigma$ are $\mu$-distinguishable if $0\le F(\rho,\sigma) \leq 1-\mu$.
\end{definition}
\end{defbox}
Note that two quantum states, $\rho$ and $\sigma$, are \emph{completely distinguishable} or 1-distinguishable ($\mu=1$), if $F(\rho,\sigma)= 0$.

One can also define the $\nu$-indistinguishability in the same manner:
\begin{defbox}
\begin{definition}[$\nu$-indistinguishability] \label{def:prelim-indist}
Let $0 \leq \nu \leq 1$ the indistinguishability threshold parameter. We say two quantum states $\rho$ and $\sigma$ are $\nu$-indistinguishable if $\nu \le F(\rho,\sigma) \le 1$.
\end{definition}
\end{defbox}

\subsection{Verifying quantum states}\label{sec:prelim-swap-gswap}
Due to the impossibility of perfectly distinguishing quantum states, checking the equality of two completely unknown states is a non-trivial task. The task of \emph{equality testing} is a simple but an extremely important task and a building block for lots of complicated quantum protocols \cite{buhrman_quantum_2001, barenco_stabilization_1997,xu_experimental_2015}. The objective is to test whether two \emph{unknown} quantum states are the same. First, we introduce the most well-known quantum algorithm for equality testing. The content of this section is widely used throughout the thesis in all the chapters (perhaps less used in \chapref{chap:varqlone}). Thus, familiarity with the notions and notations used here is crucial.

\subsubsection{SWAP test}\label{sec:prelim-swap}
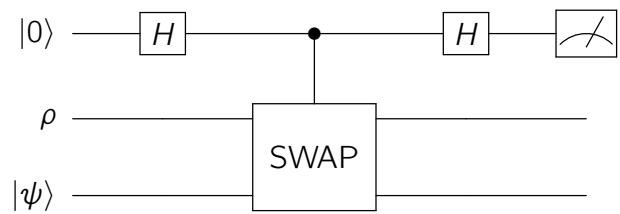
\begin{figure}[ht!]
    \centering
    \[    \Qcircuit @C=2em @R=1.4em {
       & \lstick{\ket{0}} & \gate{H} & \ctrl{1} & \gate{H} & \meter \\
       & \lstick{\rho} & \qw & \multigate{1}{\text{SWAP}} & \qw & \qw \\
       & \lstick{\ket{\psi}} & \qw & \ghost{\text{SWAP}} & \qw & \qw
    }\]
    \caption{The SWAP test circuit}
    \label{fig:swap}
\end{figure}

Given a single copy of two unknown quantum states $\rho$ and $\sigma$, is there a simple test to optimally determine whether the two states are equal or not? This question was answered affirmatively by Buhrman et al. \cite{buhrman_quantum_2001} when they provided a test called the SWAP test. This test was initially used by the authors to prove an exponential separation between classical and quantum resources in the simultaneous message passing model. Since then it has been used as a standard tool in the design of various quantum algorithms \cite{buhrman_nonlocality_2010,kumar_efficient_2017}. A SWAP test circuit takes as an input the two unknown quantum states $\rho$ and $\sigma$ and attaches an ancilla $\ket{0}$. A Hadamard gate is applied to the ancilla followed by the control-SWAP gate and again a Hadamard on the ancilla qubit. Finally, the ancilla is measured in the computational basis and we conclude that the two states are equal if the measurement outcome is `0' (labelled accept). \figref{fig:swap} illustrates this test in the special case when the state $\sigma$ is a pure state and shown by $\ket\psi$.

It can be shown that the probability the SWAP test accepts the states $\rho$ and $\sigma$ is \cite{kobayashi_quantum_2003},
\begin{equation}
    \text{Pr}[\text{SWAP accept}] = \frac{1}{2} + \frac{1}{2}\text{Tr}(\rho\sigma)
\end{equation}

In the special case of when at least one of the states (let's say $\sigma$) is a pure state $\sigma = \ket{\psi}\bra{\psi}$, the probability of acceptance is,
\begin{equation}\label{eq:swapaccept}
     \text{Pr}[\text{SWAP accept}] = \frac{1}{2} + \frac{1}{2} |\bra{\psi}\rho\ket{\psi}| = \frac{1}{2} + \frac{1}{2}F(\rho, \ket{\psi}\bra{\psi})
\end{equation}

Thus when at least one of the two states is a pure state, the acceptance probability is related to the fidelity between the states. Implying that if the states are the same, the acceptance probability is 1. However, when the states are different, the SWAP test accepting the states implies an error. The error in the SWAP test, when the states are not identical (also called the one-sided error), is $\text{Pr}[\text{accept}]$. Nevertheless, this error can be brought down to any desired error $\epsilon > 0$ by running multiple instances of the SWAP test circuit. Let $M$ be the number of copies of both input states. Then the number of instances, required to bring down the error probability to a desired $\epsilon$ is,

\begin{equation}
\begin{split}
    \text{Pr}[\text{SWAP error}] & = \prod^{M}_{j=1}\text{Pr}[\text{SWAP accept}]_j = (\frac{1}{2} + \frac{1}{2}F)^M = \epsilon \\
    & \Rightarrow M(\log(1+F)-1) = \log(\epsilon) \Rightarrow M\approx \mathcal{O}(\log(1/\epsilon))
\end{split}
\end{equation}
where $F = F(\rho, \ket{\psi}\bra{\psi}) = \bra{\psi}\rho\ket{\psi}$ and we use the fact that fidelity is independent of $\epsilon$.

\noindent Now let us introduce a generalisation of this equality test.

\subsubsection{Generalised SWAP test}\label{sec:gswap}
\begin{figure}[ht!]
\includegraphics[scale=0.35]{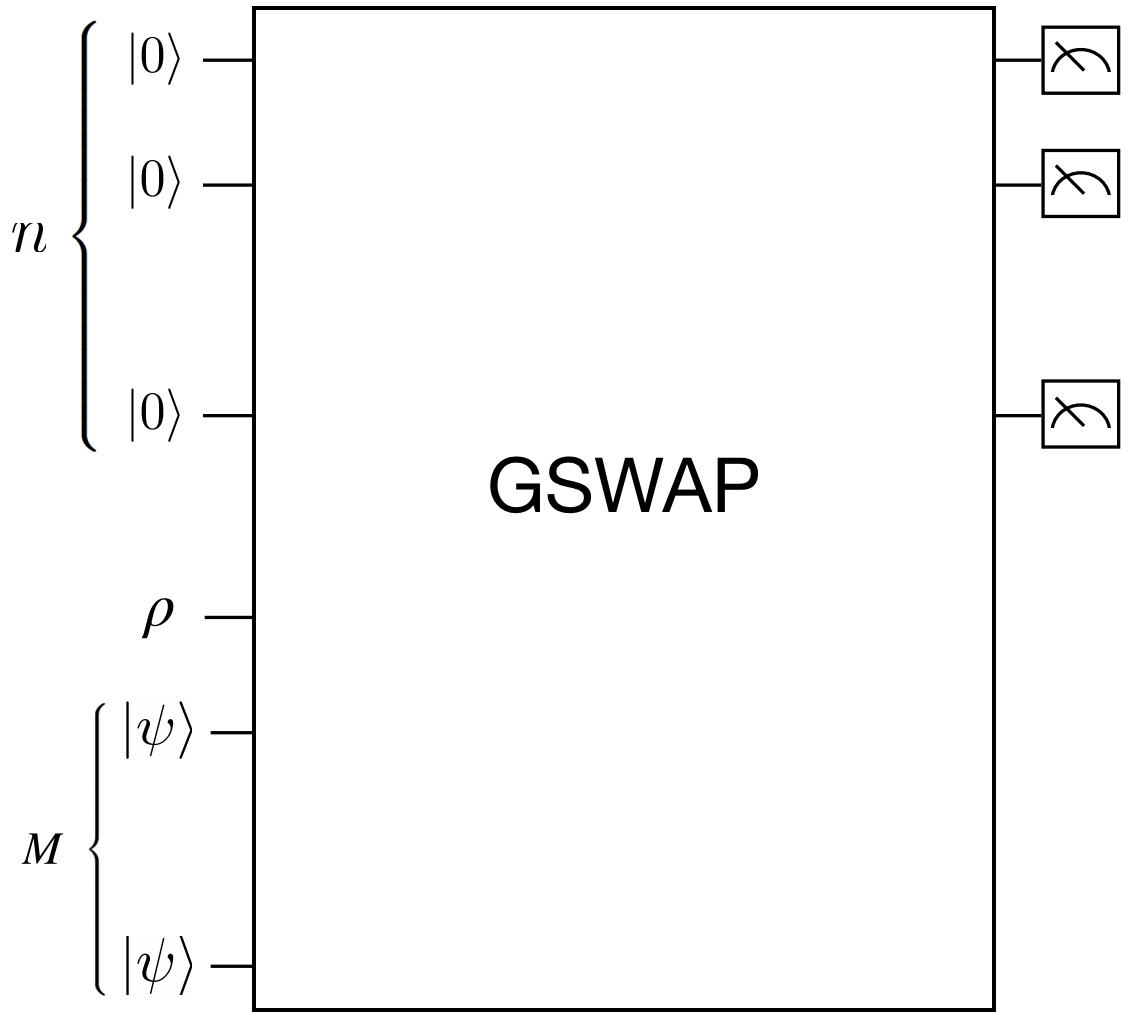}
    \centering
    \caption[The circuit of GSWAP]{GSWAP: A generalisation of the SWAP test with a single copy of $\rho$ and $M$ copies of $\ket{\psi}$. The circuit also inputs $n = \ceil[\big]{\log M+1}$ ancilla qubits in the state $\ket{0}$. At the end of the circuit, the ancilla states are measured in the computational basis.}
    \label{fig:gswap}
\end{figure}

The above SWAP test is optimal in Equality testing (in a single instance) of two unknown quantum states when one has a single copy of the two states. However, there are certain quantum protocols where one has access to multiple copies of one unknown state $\ket{\psi}$ and only a single copy of the other unknown state $\rho$ and the objective is to provide an optimal Equality testing circuit. Considering this scenario, Chabaud et al. \cite{chabaud_optimal_2018} provided an efficient construction of such a circuit, a generalised SWAP (GSWAP) test circuit. A GSWAP circuit takes as an input a single copy of $\rho$, M copies of $\ket{\psi}$ and $\ceil{\log M+1}$ copies of the ancilla qubit $\ket{0}$. The generalised circuit is then run on the inputs, and the ancilla qubits are measured in the computational basis. \figref{fig:gswap} is a generic illustration of such a circuit. For more details on the circuit refer to the original work \cite{chabaud_optimal_2018}.

It can be shown that the probability the GWAP circuit accepts two quantum states $\rho$ and $\ket{\psi}$ is,
\begin{equation}
     \text{Pr}[\text{GSWAP accept}] = \frac{1}{M+1} + \frac{M}{M+1} \bra{\psi}\rho\ket{\psi} = \frac{1}{M+1} + \frac{M}{M+1}F
     \label{eq:gswap}
\end{equation}
where $F = F(\rho, \ket{\psi}\bra{\psi})$. We note that in the special case of $M=1$, the GSWAP test reduces to the SWAP test. Also in a single instance, GSWAP provides a better Equality test compared to the SWAP test since it reduces the one-sided error probability. In the limit $M \rightarrow \infty$, we obtain the optimal acceptance probability of $\text{Pr}[\text{accept}] = F = \bra{\psi}\rho\ket{\psi}$. Another important feature of GSWAP is that it can achieve any desired success probability $\epsilon (\geqslant F)$ in just a single instance which is impossible to achieve using SWAP circuit. However, the number of copies required is exponentially more than the number of instances that the SWAP circuit has to run to achieve the same error probability,
\begin{equation}
\begin{split}
    \text{Pr}[\text{GSWAP error}] & = \text{Pr}[\text{GSWAP accept}] = \frac{1}{M+1} + \frac{M}{M+1}F = \epsilon \\
    & \Rightarrow M\approx \mathcal{O}(1/\epsilon)
\end{split}
\label{eq:gswaperror}
\end{equation}

\subsubsection{Abstract quantum test}\label{sec:prelim-abstract-test}
We can also abstract the notion of quantum equality testing of quantum states. We introduce our own abstract version of such test algorithms by defining the necessary conditions for a general quantum test.
\begin{defbox}
\begin{definition}[Quantum Testing Algorithm]\label{def:test} Let $\rho^{\otimes \kappa_1}$ and $\sigma^{\otimes \kappa_2}$ be $\kappa_1$ and $\kappa_2$ copies of two quantum states $\rho$ and $\sigma$, respectively. A Quantum Testing algorithm $\T$ is a quantum algorithm that takes as input the tuple ($\rho^{\otimes \kappa_1}$,$\sigma^{\otimes \kappa_2}$) and accepts $\rho$ and $\sigma$ as equal (outputs 1) with the following probability 
\begin{equation*}
\small
    \mathrm{Pr}[1 \leftarrow \T(\rho^{\otimes \kappa_1}, \sigma^{\otimes \kappa_2})] = 1 - \mathrm{Pr}[0 \leftarrow \T(\rho^{\otimes \kappa_1}, \sigma^{\otimes \kappa_2})] = f(\kappa_1,\kappa_2, F(\rho, \sigma))
\end{equation*}
where $F(\rho, \sigma)$ is the fidelity and $f(\kappa_1,\kappa_2, F(\rho, \sigma))$ satisfies the following limits:
\begin{equation}
 \begin{cases}
    \lim_{F(\rho, \sigma) \rightarrow 1}f(\kappa_1,\kappa_2, F(\rho, \sigma)) = 1  & \quad \forall\:(\kappa_1,\kappa_2)\\
    \lim_{\kappa_1=1,\kappa_2 \rightarrow \infty}f(\kappa_1,\kappa_2, F(\rho, \sigma)) = F(\rho, \sigma)\\
    \lim_{\kappa_1 \rightarrow \infty,\kappa_2 = 1}f(\kappa_1,\kappa_2, F(\rho, \sigma)) = F(\rho, \sigma)\\
    \lim_{F(\rho, \sigma) \rightarrow 0}f(\kappa_1,\kappa_2, F(\rho, \sigma)) = \err(\kappa_1, \kappa_2)
  \end{cases} 
\end{equation}
with $\err(\kappa_1,\kappa_2)$ characterising the statistical error of the test algorithm.
\end{definition}
\end{defbox}
As an example, for the GSWAP test where $\kappa_1 = 1$ and $\kappa_2 = M$, we obtain from \eqref{eq:gswaperror} that the probability of acceptance in the limit ${F(\rho, \ket{\psi}\bra{\psi}) \rightarrow 1}$ is 1, while it is $\frac{1}{M+1}$ in the limit ${F(\rho, \ket{\psi}\bra{\psi}) \rightarrow 0}$. It can be inferred from the above definition that the quantum test can be idealized by forcing the $\err(\kappa_1,\kappa_2)$ to be zero for any given number of copies. We discuss this last point later in \chapref{chap:qpuf}, when we introduce an ideal version of such abstract tests.

\section{Quantum cloning}\label{sec:prelim-cloning}
In this section, we introduce one of the core concepts of this thesis: the no-cloning theorem and quantum cloning. This section (specifically subsection~\ref{sec:prelim-other-clonings}) is essential for \chapref{chap:varqlone} and not mostly used in other chapters, except for the general notion of no-cloning. First, we discuss the impossibility of \emph{perfectly} cloning quantum states via the no-cloning theorem and then we discuss how we can step out of this limitation on the quantum world and go beyond this impossibility. 

The no-cloning theorem states that it is not possible to perfectly clone an \emph{unknown} quantum state. The proof of this theorem is most commonly known to be the work of Wootters and Zurek \cite{wootters_single_1982}, and also independently done by Dieks \cite{dieks_communication_1982} in 1982. Although it seems that it has originally been discovered in 1970 by Park \cite{park_concept_1970}. One formulation of the no-cloning theorem is as follows.

\begin{thmbox}
\begin{theorem}[The no-cloning theorem]\label{th:prelim-no-cloning}
There exists no unitary transformation $U_c$ that performs the following operation on an arbitrary unknown state $\ket{\psi} \in \Hil$, a blank (or reference) state $\ket{0} \in \Hil$ of the same Hilbert space and any arbitrary ancillary state $\ket{a}$:
\begin{equation}\label{eq:prelim-no-cloning}
    \ket{\psi}\ket{0}\ket{a} \overset{U_c}{\rightarrow} \ket{\psi}\ket{\psi}\ket{a_{\psi}} 
\end{equation}
\end{theorem}
\end{thmbox}

There exist several proofs of this theorem. Here we give a simple one, similar to what can be found in \cite{brus_lectures_2006}. The proof is by contradiction. Let us assume that such a unitary $U_c$ exists. We note that all the states $\ket{\psi}$, $\ket{0}$ and $\ket{a}$ (whose dimension does not need to be specified) are normalized. Since the state $\ket{\psi}$ is arbitrary, the unitary should be working similarly for any state. Now we assume two non-orthogonal input states $\ket{\psi}$ and $\ket{\phi}$, for which the cloning transformation should be as follows.

\begin{equation}\label{eq:prelim-no-cloning-proof}
\begin{split}
    & U_c (\ket{\psi}\ket{0}\ket{a}) =  \ket{\psi}\ket{\psi}\ket{a_{\psi}} \\
    & U_c (\ket{\phi}\ket{0}\ket{a}) =  \ket{\phi}\ket{\psi}\ket{a_{\phi}}
\end{split}
\end{equation}
where $\ket{a_{\psi}}$ and $\ket{a_{\phi}}$ represent the output states of the ancilla after the cloning transformation. Regardless of the dimension of the ancillary states, we have that $|\mbraket{a_{\psi}}{a_{\phi}}| \leq 1$. Also, the unitary transformation preserves the inner product. Now, let us inner product both sides of the \eqref{eq:prelim-no-cloning-proof}, which leads to the following:
\begin{equation}
    \mbraket{\psi}{\phi} = \mbraket{\psi}{\phi}^2 \mbraket{a_{\psi}}{a_{\phi}} \Rightarrow  \mbraket{\psi}{\phi} = \frac{1}{\mbraket{a_{\psi}}{a_{\phi}}}
\end{equation}
which can clearly be never satisfied and hence the contradiction has been shown. There are two cases where one cannot reach such contradictions. The first one is the trivial case of $\mbraket{\psi}{\phi} = 1$, and the other one is the two states are known to be orthogonal \emph{i.e.} $\mbraket{\psi}{\phi} = 0$. The latter case is intuitively very insightful since it corresponds to cloning a classical bit, which we know is possible. We will dig further into this in \chapref{chap:unf-tools}. Moreover, we note that here, no assumption has been made on the unitary, and the no-cloning has been only the result of the unitarity of the transformations in quantum mechanics.

It is also worth mentioning that no-cloning is such a fundamental aspect of nature that it extends to other areas of physics. In fact, cloning is also impossible if we consider the impossibility of superluminal signalling to be held, which we do, due to special relativity. This no-go theorem is known as \emph{no-signalling} \cite{gisin_quantum_1998,navez_cloning_2003} and it is known that if perfect cloning could be possible, it would lead to a contradiction with the fact that no signal can travel faster than the speed of light. No-cloning, is also deeply connected to many other no-go results in quantum information, such as no-broadcasting theorem~\cite{barnum_noncommuting_1996,piani_no-local-broadcasting_2008}, no-deleting theorem~\cite{kumar_pati_impossibility_2000}, no-superposing theorem~\cite{oszmaniec_creating_2016,doosti_universal_2017}.
\subsection{Cloning beyond the no-cloning theorem}\label{sec:prelim-other-clonings}
Well, \emph{the show must go on}! The no-cloning theorem has not been the end of the road for this part of quantum information. On the contrary, the beginning of a rich field of research. To go around the no-cloning limitation, we must lower our expectations from the cloning machine! Remember that we expected the cloning machine to be \emph{deterministic} (meaning that we always want to get the clones with probability 1) and \emph{exact} (which means that the output states should be both perfect copies of the initial state). It turned out that by relaxing each of these conditions, quantum cloning can be made possible. We call the transformation that achieves such tasks a quantum cloning machine (QCM). 

Relaxing the first condition leads to a class of quantum cloning known as \emph{probabilistic cloning}, originated by these works \cite{duan_two_1997,duan_probabilistic_1998}. A probabilistic cloning machine produces \emph{perfect} clones, but only \emph{some times}, i.e. it succeeds with a certain probability. As one can guess, there are information-theoretic upper bounds on this success probability. In this thesis, we do not focus on probabilistic cloning, although we refer the interested readers to these reviews on quantum cloning, including the probabilistic cloning \cite{scarani_quantum_2005,fan_quantum_2014,brus_lectures_2006}. 

The second condition, on the other hand, was historically the first one to be relaxed by Buzek and Hillery in \cite{buzek_quantum_1996}, leading to the field of \emph{Approximate quantum cloning}. In approximate cloning, the operation is deterministic, however, we allow the clones to be not perfect, i.e. have some distance from the original quantum state. The most important property of an approximate cloner is the quality of the output clones, which is measured by the fidelity between the clone and the original state. The next important factor is the family of states we require the cloning machine to clone. Each family of states leads to a class of approximate cloning machines. For instance, if the family of states we consider is \emph{all} the possible states of a Hilbert space, the cloner is called \emph{universal quantum cloner}. However, this set can be made more restricted, meaning that more prior information of the initial states is known, which in turn will lead to cloners with higher optimal fidelity as we will see.

Another property of a cloning machine is \emph{symmetry}. When a cloning machine is symmetric, it means that both of the outputted clones should be the same, relative to the comparison measure. This symmetry is considered when the fidelity is to be optimised, which means that most of the time the symmetric cloners are optimised with respect to \emph{local fidelity}, i.e. the fidelity of each clone state. While we can also have \emph{asymmetric cloners} \cite{cerf_pauli_2000,cerf_asymmetric_2000,iblisdir_multipartite_2005,durt_economical_2005} where the fidelity of one clone is higher than the other one. The quality (especially in the asymmetric case) of the cloner can also be measured with respect to the joint state of both of the clones in comparison to two perfect clones. This fidelity measured is referred to as \emph{global fidelity}. In \chapref{chap:varqlone}, where we introduce a machine learning algorithm for the task of cloning, this distinction between global and local fidelity becomes a subtle and important factor.

Finally, the generalisation of the cloning problem is when we have $M > 1$ copies of the input state and we require to create $N > M$ approximate clones (known as $N\rightarrow M$ cloning). In this case enforcing \emph{symmetry} would correspond to
\begin{equation}
F^j_{\Lbs} = F^k_{\Lbs}, \qquad \forall j, k \in \{1, \dots N\}.
\end{equation}
Where $F$ denotes the fidelity. Now let us briefly discuss three main classes of cloning machines.

\subsubsection{Universal quantum cloning}
The earliest result in approximate cloning was a universal symmetric cloning machine (UQCM) \cite{buzek_quantum_1996}. A UQCM is completely agnostic about the input state and is aimed to clone any given quantum state of a Hilbert space with a given dimension. For the case of a qubit, where we want to clone all the qubits on the Bloch sphere, it has been shown that this cloner can achieve the optimal cloning fidelity of $ 5/6 \approx 0.8333$. This optimal cloner, takes as input an initially unknown state, a blank state and an ancillary qubit, and maps them to a 2-qubit state (the ancillary state is traced out) where the fidelity of each subsystem, or in other words, each reduced density matrices, is optimised to the maximum value of fidelity, leading to the optimal local qubit fidelity of $F_{\mathsf{L}, \textrm{opt}}^{\text{U},1} = F_{\mathsf{L}, \textrm{opt}}^{\text{U},2} = 5/6$.

Now, in the generalised case, we can provide multiple ($M$) copies of a state to the cloner and request $N$ output approximate clones which is referred to as $M\rightarrow N$ cloning~\cite{gisin_optimal_1997, bruss_optimal_1998}. Generalizing the universal cloning fidelity ($F_{\mathsf{L}, \textrm{opt}}^{\text{U}, j}) := F_{\mathsf{L}, \textrm{opt}}^{\text{U}, j}(1, 2))$ to the $M\rightarrow N$ scenario, the optimal local fidelity becomes:
\begin{equation} \label{eq:prelim-mton_universal_optimal_fidelity}
    F^{\textrm{U}, j}_{\Lbs, \textrm{opt}}(M, N) = F^{\textrm{U}}_{\Lbs, \textrm{opt}}(M, N) = \frac{MN + M + N}{N(M+2)}
\end{equation}
Here $N - M$ ancilla qubits are used to assist, so the initial state is $\ket{\psi_A}^{\otimes M}\otimes\ket{0}^{\otimes N-M}$. We also note that in the limit $M\rightarrow \infty$, an optimal cloning machine becomes equivalent to a quantum state estimation machine~\cite{scarani_quantum_2005} for universal cloning. We will further discuss the intuitive meaning and relevance of this important result for our purpose in \chapref{chap:unf-tools}. In the context of cryptography, $M\rightarrow N$ cloning can be also modelled as having $N$ adversaries, $E_1\dots E_N$ who receive $M$ copies of the state to be cloned.

\subsubsection{Phase-covariant cloning}\label{sec:prelim-cloning-phase-cov}
Phase-covariant (introduced by \cite{brus_phase-covariant_2000}) states are equatorial states of Bloch sphere. It is common to choose the $\mathsf{X} - \mathsf{Y}$ plane to describe such states:
\begin{equation} \label{eq:prelim-xy-phasecov-states}
    \ket{\psi_{xy}(\eta)} = \frac{1}{\sqrt{2}}\left(\ket{0} + e^{i\eta}\ket{1}\right)
\end{equation}
It is known that a phase-covariant cloning machine (PCQCM) that clones one equatorial qubit to two clones has the optimal local fidelity $F_{\mathsf{L},  \text{opt}}^{\text{PC}} \approx 0.85 > 5/6$, which is notably higher than the universal case.

These states are relevant since they are used in BB84 QKD protocols and also in universal blind quantum computation protocols~\cite{bennett_quantum_2014, broadbent_universal_2009}. Here in the context of QKD, we have the following scenario: Assume that Alice wishes to transmit quantum information to Bob (here the information is the bits of the key) but the channel is subject to an eavesdropper, Eve, who wishes to adversarially gain knowledge on the message sent by Alice. 

\begin{figure}[ht]
    \centering
\includegraphics[width=0.6\textwidth]{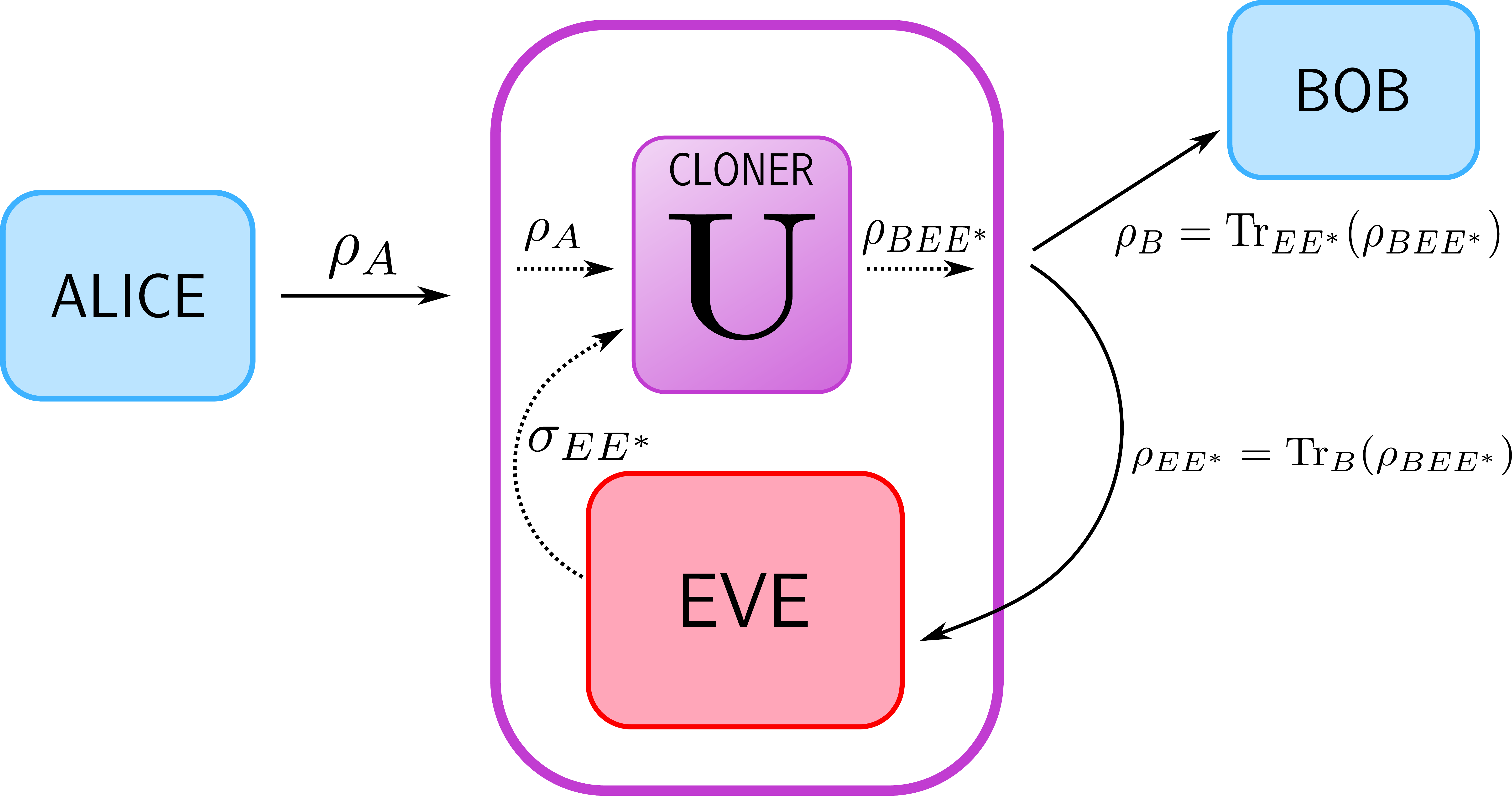}
\caption[Cartoon illustration of a cloning-based eavesdropping attack by Eve, trying to clone the state, $\rho_A$, Alice sends to Bob]{Cartoon illustration of an eavesdropping attack by Eve, trying to clone the state, $\rho_A$, Alice sends to Bob. Eve injects a `blank' state (which can be a specific state or an arbitrary state depending on the scenario, which ends up as a clone of $\rho_A$) and an ancillary system ($E^*$). She then applies the cloning unitary, $U$. We can assume the cloner is manufactured by Eve to give her the greatest advantage. The state Bob receives will be the partial trace over Eve's subsystems, $\rho_{B} = \tr_{EE^*}(\rho_{BEE^*})$, and Eve's clone will be $\rho_E$, where $\rho_{BEE^*}$ is the full output state from the QCM.} 
\label{fig:cloning}
\end{figure}

We illustrate this in \figref{fig:cloning}, for a single Eve. In this picture, Alice ($A$) sends a quantum state\footnote{This will typically be a pure state, $\rho_A := \ket{\psi}\bra{\psi}_A$, but it can also be generalised to include mixed states, in which the task is referred to as \emph{broadcasting}\cite{barnum_noncommuting_1996,chen_mixed_2007,dang_optimal_2007}. The \emph{no-broadcasting} theorem is a generalisation of the no-cloning theorem in this setting.}, $\rho_A$, to Bob ($B$). A cloning based attack strategy for Eve ($E$) could be to try and clone Alice's state, producing a second (approximate) copy which she can use later in her attack, with some ancillary register ($E^*$).

\begin{figure}[ht]
    \centering
\includegraphics[width=0.85\columnwidth]{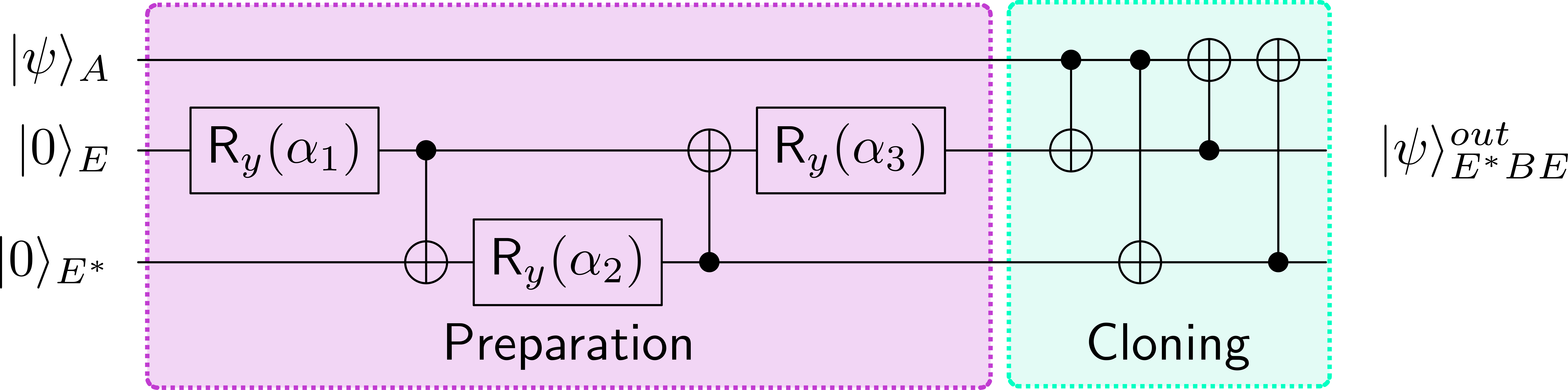}
\caption{Ideal cloning circuit for universal and phase covariant cloning. The \textsf{Preparation} circuit prepares Eve's system to receive the cloned states, while the \textsf{Cloning} circuit transfers information. Notice that the output registers which contain the two clones of $\ket{\psi}_A$ to Bob and Eve in this circuit are registers $2$ and $3$ respectively.} \label{fig:qubit_cloning_ideal_circ}
\end{figure}

Interestingly, the cloning of phase-covariant states can be accomplished in an \emph{economical} manner, meaning without needing an ancilla system for Eve, $E^*$\cite{niu_two-qubit_1999}. However, as noted in~\cite{scarani_quantum_2005}, removing the ancilla is useful to reduce resources if one is \emph{only} interested in performing cloning, but if Eve wishes to attack Alice and Bob's communication, it is more beneficial to apply an ancilla-based attack. Intuitively, this is because the ancilla also contains information about the input state which Eve can extract.
Of interest to our purposes, is an explicit quantum circuit which implements the cloning transformation. A unified circuit~\cite{buzek_quantum_1997, fan_quantum_2014, fan_quantum_2001} for the above cases (universal and $\mathsf{X} - \mathsf{Y}$ phase covariant) can be seen in \figref{fig:qubit_cloning_ideal_circ}.  The parameters of the circuit, $\boldsymbol{\alpha} = \{\alpha_1, \alpha_2, \alpha_3\}$, are given by the family of states the circuit is built for~\cite{buzek_quantum_1997, fan_quantum_2014, fan_quantum_2001}.

For phase-covariant cloning of $\mathsf{X} - \mathsf{Y}$ states, we explicitly have the following optimal angles:
\begin{equation}
    \begin{split}
        & \alpha^{XY}_1 = \alpha^{XY}_3 = \arcsin{\sqrt{(\frac{1}{2} - \frac{1}{2\sqrt{3}})}} \approx 0.477, \\
        & \alpha^{XY}_2 = - \arcsin{\sqrt{(\frac{1}{2} - \frac{\sqrt{3}}{4})}} \approx -0.261
    \end{split}
\end{equation}

For completeness, let us also have a look at the optimal local fidelity for the general $M \rightarrow N$ case, which has been studied in \cite{fan_quantum_2001,dariano_optimal_2003}. There is no unique expression for the optimal local fidelity as a function of $N$ and $M$. However, for the
$1 \rightarrow N$ case the optimal phase covariant fidelity is given by \cite{dariano_optimal_2003} as follows depending on $N$ being odd or even:
\begin{equation}\label{eq:prelim-local_optimal_phase-cov_fidelity_1toM}
F_{\mathsf{L},  \text{opt}}^{\text{PC}}(1,N) = \begin{cases}
        \frac{1}{2}(1 + \frac{N+1}{2N})  & \quad \text{odd } N\\
        \frac{1}{2}(1 + \frac{\sqrt{N(N+2)}}{2N})  & \quad \text{even } N
    \end{cases}
\end{equation}

\subsubsection{State-dependent cloning with fixed overlap}\label{sec:prelim-state-dep-cloning}
Now we introduce another class of cloning machines where we aim to clone two non-orthogonal unknown quantum states with a known fixed overlap\footnote{This cloning is originally referred to as `state-dependent cloning', so we herein use this term referring to this scenario.}. This was one of the original scenarios studied in the realm of approximate cloning \cite{brus_optimal_1998} but is difficult to tackle analytically. Let us consider the simplest case first, where one considers two states of the type:
\begin{equation}\label{eq:prelim-state_dependent_cloning_states}
\begin{split}
    \ket{\psi_1} = \cos\phi\ket{0} +\sin\phi\ket{1}\\
    \ket{\psi_2} = \sin\phi\ket{0} +\cos\phi\ket{1} 
\end{split}
\end{equation}%
which have a fixed overlap, $s = \mbraket{\psi_1}{\psi_2} = \sin 2\phi$. It has been shown in~\cite{brus_optimal_1998} that the optimal \emph{local fidelity} for this scenario is the following:
\begin{equation} \label{eq:prelim-local_optimal_non_ortho_fidelity_1to2}
\begin{split}
    &F^{\textrm{FO}, j}_{\mathsf{L}, \textrm{opt}}  = \frac{1}{2} + \frac{\sqrt{2}}{32 s}(1+s)(3−3s+\sqrt{1−2s+ 9s^2})\\
    & \times\sqrt{−1 + 2s + 3s^2 + (1−s)\sqrt{1−2s+ 9s^2}}, ~j \in \{1,2\}
\end{split}
\end{equation}
It can be shown that the \emph{minimum} value for this expression is achieved when $s=\frac{1}{2}$ and gives $F^{\textrm{FO}, j}_{\mathsf{L}, \textrm{opt}} \approx 0.987$, which is much better than the symmetric phase-covariant cloner.

Let us also have a look at the global fidelity in the general $M \rightarrow N$ case for this cloning machine which is given as \cite{brus_lectures_2006,brus_optimal_1998}:

\begin{equation} \label{eq:prelim-global_optimal_non_ortho_fidelity_1to2}
    F^{\textrm{FO}}_{\mathsf{G}, \textrm{opt}}  = \frac{1}{2}(1 + s^{M+N} + \sqrt{1 - s^{2M}}\sqrt{1 - s^{2N}})
\end{equation}

Interestingly, it can be shown that the state-dependent quantum cloning machine (SDQCM) which achieves this optimal \emph{global} fidelity, does not saturate the optimal \emph{local} fidelity. Computing the local fidelity for the globally optimized SDQCM gives~\cite{brus_lectures_2006}:
\begin{equation}\label{eq:prelim-state_dep_local_fidelity_from_global}
\begin{split}
    F_{\Lbs, *}^{\mathrm{FO}, j}(M, N) = \frac{1}{4}(
    \frac{1+s^M}{1+s^N}\left[1+s^2+2s^N\right] & + \frac{1-s^M}{1-s^N}\left[1+s^2-2s^N\right] \\ & +
     2\frac{1-s^{2M}}{1-s^{2N}}\left[1-s^2\right]
    ) \quad \quad \forall j 
\end{split}    
\end{equation}
In contrast, computing the optimal local fidelity for this scenario~\cite{brus_optimal_1998} (for $1\rightarrow 2$ cloning) is:
\begin{equation} \label{eq:prelim-local_optimal_non_ortho_fidelity_1to2_s}
\begin{split}
    F^{\textrm{FO}, j}_{\mathsf{L}, \textrm{opt}} = & \frac{1}{2} + \frac{\sqrt{2}}{32 s}(1+s)\left(3−3s+\sqrt{1−2s+ 9s^2}\right) \\
    & \times \sqrt{−1 + 2s + 3s^2 + (1−s) \sqrt{1−2s+ 9s^2}}, ~\forall j
\end{split}
\end{equation}
It can be shown that the \emph{minimum} value for this expression is achieved when $s=\frac{1}{2}$ and gives $F^{\textrm{FO}, j}_{\mathsf{L}, \textrm{opt}} \approx 0.987$, which is also much better than the symmetric phase-covariant cloner. Nevertheless, comparing \eqref{eq:prelim-local_optimal_non_ortho_fidelity_1to2} and \eqref{eq:prelim-state_dep_local_fidelity_from_global} reveals that $F_{\Lbs, *}^{\mathrm{FO}, j}(1,2)$ is actually a \emph{lower} bound for the optimal local fidelity, $F^{\textrm{FO}, j}_{\mathsf{L}, \textrm{opt}} $ in \eqref{eq:prelim-local_optimal_non_ortho_fidelity_1to2}. This point is crucially relevant for us and we will go back to it in \chapref{chap:varqlone} where we will use this scenario as a case study for quantum coin flipping protocols and for the design of our variational cloning algorithm.

The state-dependent cloning has been studied concerning the security of QKD. Although, as we have discussed optimal cloning-based attacks for the BB84 protocol are given with the optimal phase covariant cloner. However, this type of cloning has not been widely used for the study of other cryptographic protocols, which is one of our main contributions in  \chapref{chap:varqlone}. We also note that, interestingly, state-dependent cloning has been recently used to demonstrate advantages related to quantum contextuality~\cite{lostaglio_contextual_2020}.
\clearpage
\section{Haar measure and random matrix theory}\label{sec:prelim-haar-random}
In this section, we introduce some mathematical background for a concept that is the building block of quantum randomness. The notion of the Haar measure is a particularly important one,  and consequently, we have also used it in almost all the chapters. However, other random matrix theory toolkits introduced in this section are used in \chapref{chap:pr-connection}. In mathematics, the \emph{Haar measure} assigns an `invariant volume' to subsets of a locally compact topological group. This measure was introduced by Haar in 1933 \cite{haar_massbegriff_1933}, though its special case for Lie groups had been introduced earlier by Hurwitz as \emph{invariant integral} \cite{diaconis_hurwitz_2017}. This measure has been used in many fields such as group theory, representation theory, random matrix theory, ergodic theory and quantum information. Let us first introduce the mathematical definitions and then give an intuition on its application in quantum information.

A Haar measure is a non-zero measure on any locally compact group $G$ such that $\mu: G \rightarrow [0,\infty)$ 
such that for all $X \subset G$ and $x \in G$ we have the following translation invariance property for $\mu(X) = \int_{x \in G} d\mu(x)$:
\begin{equation}
    \mu(xX) = \mu(Xx) = \mu(X)
\end{equation}
In particular, the Haar measure $d\mu(U)$ can be defined for a unitary group $U(d)$. Sampling unitaries from Haar measure on $U(d)$ is equivalent to geometrically uniform sampling from unitary groups of that dimension.

\begin{figure}[ht]
    \centering
\includegraphics[width=1.0\textwidth]{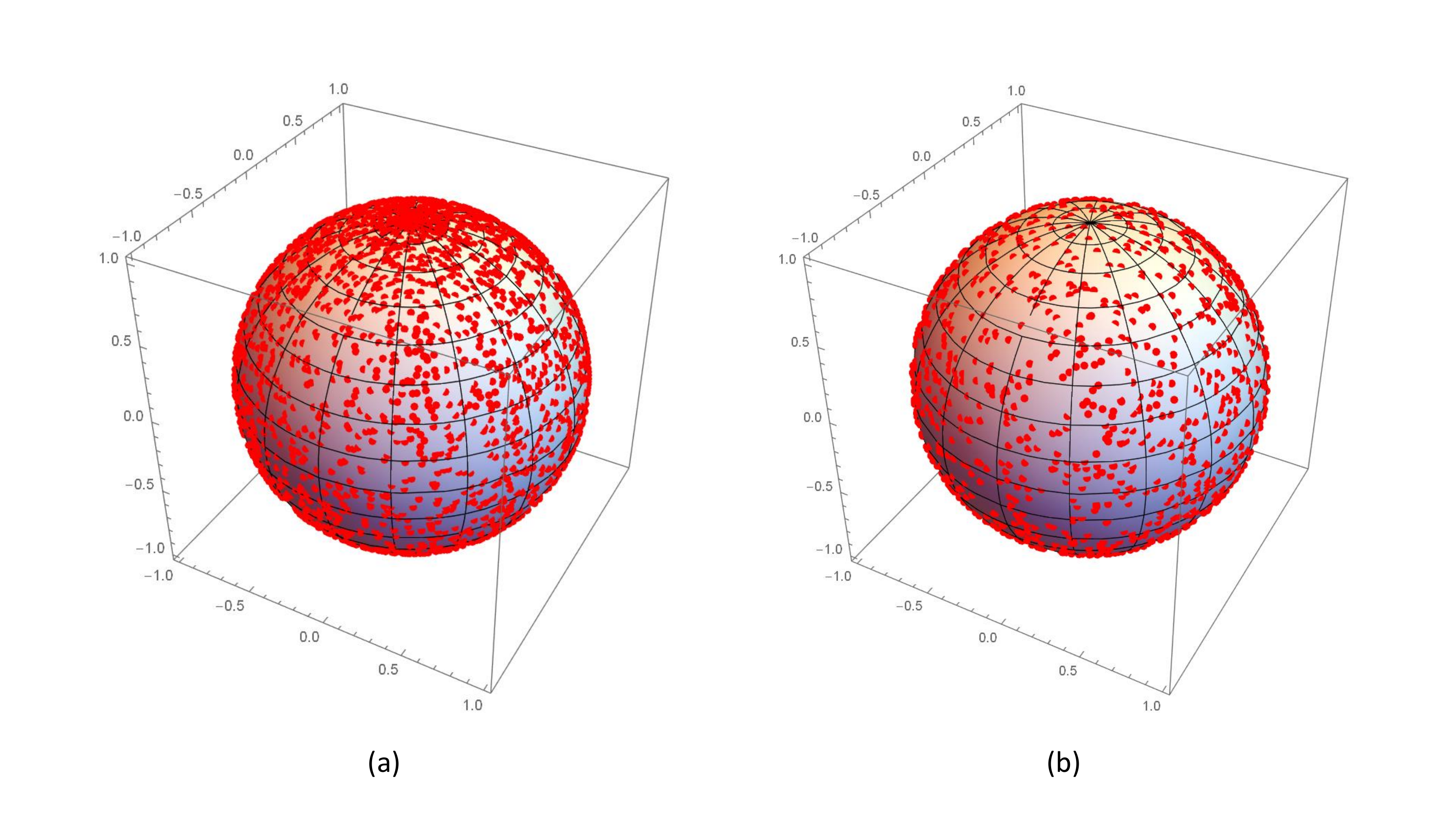}
    \caption[Haar measure and non-uniform sampling over the Bloch sphere]{Haar measure and non-uniform sampling over the Bloch sphere. Figure (a) shows the case where the qubits have been sampled through a uniform sampling of the qubit parameters. This sampling leads to a geometrically non-uniform sampling over the sphere. Figure (b) shows sampling qubits according to Haar measure over $SU(2)$ which leads to a uniform sampling over the Bloch sphere.
    }
    \label{fig:haar-bloch}
\end{figure}

Let us take the 2-dimensional Hilbert space as an example. We recall that pure qubit states can be represented as a vector or a point on the surface of the Bloch sphere. Assume that we want to pick uniformly random qubits on the surface of the Bloch sphere. Since every point on the sphere is parameterised by $\theta$ and $\phi$ according to \eqref{eq:prelim-bloch-sphere-state}, one approach would be to uniformly sample values for these two parameters and the result will be random qubits on the Bloch sphere. Although, as it is shown in \figref{fig:haar-bloch}, by doing this, the resulting vectors will not be uniformly distributed on the surface of the Bloch sphere and they will be more concentrated around the poles. On the contrary, if one samples the vectors uniformly at random according to the Haar measure over $SU(2)$, they will be distributed uniformly over the Bloch sphere (\figref{fig:haar-bloch} (b)). In practice, however, sampling from the Haar measure requires exponential (in $d$) resources~\cite{knill_approximation_1995}.

We are also interested in characterising the properties of the eigenvalues of Haar-random unitary matrices and their distributions. Problems of this sort have been widely studied in the field of \emph{random matrix theory}. Here we introduce some of the important results in this field that we will use later on in \chapref{chap:pr-connection}.

The first result that we need, is known as \emph{Weyl density formula} or \emph{Weyl integration formula}, and is stated as follows:

\begin{lembox}
\begin{lemma}[Weyl integration formula on $U(n)$~\cite{meckes_random_2019}]
Let $\{e^{i\theta_j}\}^n_{j=1}$ be the eigenvalues of $n\times n$ random unitary matrix. The unordered eigenvalues of a random unitary matrix have the following eigenvalue density
\begin{equation}
    \frac{1}{n!(2\pi)^n} \prod_{1\leq j < k \leq n} |e^{i\theta_j} - e^{i\theta_k}|^2
\end{equation}
with respect to $d\theta_1\dots d\theta_n$ on $(2\pi)^n$. That is, for any $g:U(n)\rightarrow R$ with
\begin{equation*}
    g(U) = g(VUV^*) \quad \text{for any } U,V \in U(n), 
\end{equation*}
(i.e., $g$ is a class function), if $U$ is Haar-distributed on $U(n)$, then
\begin{equation}
    \mathbb{E}[g(U)] = \frac{1}{n!(2\pi)^n} \int_{[0,2\pi)^n} \tilde{g}(\theta_1,\dots,\theta_n) \prod_{1\leq j < k \leq n} |e^{i\theta_j} - e^{i\theta_k}|^2 d\theta_1\dots d\theta_n
\end{equation}
where $\tilde{g}: [0,2\pi)^n \rightarrow \mathbb{R}$ is the (necessarily symmetric) expression of $g(U)$ as a
function of the eigenvalues of $U$.
\end{lemma}
\end{lembox}

As discussed in~\cite{meckes_random_2019}, one consequence of the above lemma is that the eigenvalues of random unitary matrices want to spread out. For any given pair of eigenvalues labelled by $(j,k)$, $|e^{i\theta_j} - e^{i\theta_k}|^2$ is zero if $\theta_j = \theta_k$, and is 4 if $\theta_j = \theta_k + \pi$ (and in that neighborhood if they are roughly antipodal). This produces the
effect alternatively known as `eigenvalue repulsion'.

Another important tool in the study of the eigenvalues of random matrices is the \emph{empirical spectral measure} defined as,
\begin{equation}
     \tilde{\mu} = \frac{1}{n}\sum^n_{j=1} \delta_{e^{i\theta_j}}
\end{equation}
where $e^{i\theta_j}$ are the eigenvalues of the unitary matrix and $\delta$ is the probability distribution function over the eigenvalues. The empirical spectral measure is a probability measure to encode the ensemble of eigenvalues which puts equal mass at each of the eigenvalues of $U$. This encoding is very useful for representing the spreading of the eigenvalues on the complex unit circle denoted by $\mathbb{S}^1 \subseteq \mathbb{C}$.

Next, we need the following important theorem by Diaconis-Shashahani~\cite{diaconis_eigenvalues_1994}, that shows the convergence of the eigenvalues of the Haar-random matrices to the uniform distribution over the unit circle:

\begin{thmbox}
\begin{theorem}[\cite{diaconis_eigenvalues_1994}]\label{th:prelim-diaconis-shah}
Let $U$ be uniformly chosen from Haar-measure in $U(d)$, Let $\nu$ be the uniform distribution on $\mathbb{S}^1$. Then as $d \rightarrow \infty$, the $\tilde{\mu}_U$ converges, weakly in probability, to $\nu$:
\begin{equation}
    \tilde{\mu}_U \overset{d\rightarrow \infty}{\longrightarrow} \nu
\end{equation}
\end{theorem}
\end{thmbox}

Finally, we introduce the following result by Wieand~\cite{wieand_eigenvalue_2002} which is very useful in working with the statistics of the eigenvalues of random unitaries.

\begin{thmbox}
\begin{theorem}[\cite{wieand_eigenvalue_2002}]\label{th:prelim-wieand}
Let $U$ be a unitary matrix chosen from Haar measure in $U(d)$, and let $\{e^{i\theta_1},\dots,e^{i\theta_d}\}$ be the eigenvalues of $U$. Fix a finite number of intervals on the unit circle $I_1 =(e^{i\theta_{1j}} , e^{i\theta_{1l}}),\dots,I_m =(e^{i\theta_{mj}} , e^{i\theta_{ml}})$. Define the random variables $N_{\theta_1},\dots,N_{\theta_m}$ to be the number of eigenvalues in each arc defined by the intervals. In the limit of large $d$, the mean and variance of $N_{\theta_k}$ are as follows:
\begin{equation}
    \mathbb{E}_d[N_{\theta_k}] = \frac{d(\theta_{kj} - \theta_{kl})}{2\pi}
\end{equation}
and 
\begin{equation}
     Var(N_{\theta_k}) = \frac{1}{\pi^2}(\log(d) + 1 + \gamma + \log|2\sin(\frac{\theta_{kj} - \theta_{kl}}{2})|) + o(1).
\end{equation}
where $\gamma \approx 0.577$ is the Euler's constant.
\end{theorem}
\end{thmbox}

This theorem, gives a concrete formula for calculating the expectation value and variance of the random variable that represents the number of eigenvalues of a random unitary matrix, in each arc of the unit circle and hence can be used to study the distribution of eigenvalues of random matrices.
\section{Quantum cryptography}\label{sec:prelim-quntum-crypto}
Now we focus on another field of research that we tightly connected with this thesis, namely \emph{quantum cryptography}. Quantum cryptography is almost as old as quantum computing itself and studies different cryptographic problems that involve, in several ways, quantum mechanical systems. These quantum systems can be employed by honest parties to perform a cryptographic task, or else can be exploited by an adversary, or a dishonest party, trying to attack the system. We have already seen an example of a secure protocol (\emph{i.e.} QKD) where parties have some limited quantum capabilities like preparing and measuring qubit states in a specific basis. We will introduce another example of such protocols in this chapter that achieves a functionality which is impossible with only `classical' cryptography (see \ref{sec:prelim-coin-flip}). Yet another sub-field of quantum cryptography is \emph{post-quantum} cryptography, dedicated to studying the security of `classical' systems against quantum adversaries and the design of quantum-secure cryptographic schemes. In this field, the most vital aspect of a quantum adversary is its quantum computing capability, especially after the discovery of algorithms such as Shor's and Grover's where there exists (a potentially) significant \emph{quantum-speedup} \cite{wallden_cyber_2019}. The main idea here is to keep the cryptographic schemes classical while designing them based on assumptions and mathematical problems that are also hard for quantum computers to solve. Due to the current technological challenges and inefficiency of quantum systems, as well as the incompatibility of many of the quantum protocols with today's existing classical cryptosystems, this idea is perhaps the most popular discipline today for achieving security in the quantum world \cite{wallden_cyber_2019,buchmann_post-quantum_2016}. Among the existing attempts in this field to guarantee `quantum-resistant' with classical cryptographic schemes, one of the most successful ones is \emph{lattice-based cryptography} \cite{micciancio_lattice-based_2009}. However, there is a full spectrum between going towards fully quantum systems and keeping them fully classical. Numerous work has been done in this area, which is also of particular interest in this thesis. Therefore, in this section, we will introduce a handpick of concepts and protocols from different branches of quantum cryptography, which are either essential for the results we will establish later or will help the reader with an improved understanding of future topics. However, the introduction we give here is by no means exhaustive.

Before going over the more technical materials, let us settle on a few basic notations and terminologies that we will widely throughout the thesis.

We start with the notion of \emph{security parameter}. The \textbf{security parameter}, which we denote as $\lambda$ in this thesis, is a parameter that quantifies the security level of the systems, or in other words, the complexity of the problem based on which the cryptographic scheme has been designed. Roughly speaking, the security parameter measures how `hard' it is for an attacker (which we call adversary from now on) to break the cryptographic scheme. As such, the objective is to design cryptographic schemes that \emph{for any} adversary, the success probability of the adversary is `small' relative to this parameter.

Now let us focus on the word \emph{`small'} in the previous sentence and try to formalise that. The smallness, in the world of cryptography, is usually formalised via a concept known as \emph{negligible function}, defined as follows:

\begin{defbox}
\begin{definition}[Negligible function]\label{def:prelim-negl}
A function $\epsilon: \mathbb{N} \rightarrow \mathbb{R}$ is a \emph{negligible function}, if for every positive integer $c$, (or equivalently every polynomial function of the security parameter $poly(.)$) there exists an integer $N_c > 0$ (or $N_{poly} > 0$) such that for all $x > N_c$ ($x > N_{poly}$) the following holds:
\begin{equation}
    |\epsilon(x)| < \frac{1}{x^c} \quad \quad \left(\text{or } |\epsilon(x)| < \frac{1}{poly(x)}\right)
\end{equation}
\end{definition}
\end{defbox}

Thus, we require the success probability of the adversary to be a negligible function of the security parameter $\lambda$, which we denote as either $\negl(\lambda)$ or $\epsilon(\lambda)$. We refer to \cite{katz_introduction_2020} for properties of the negligible functions. 

Another terminology that we need to introduce, is the notion of \emph{One-Way Function (OWF)}. An OWF is a function that is `easy' to compute on every input, but `hard' to invert given the image of a random input. OWFs are usually considered as a \emph{computational assumption} on cryptography, referring to the hardness complexity in the definition. More formally, an OWF is defined as follows \cite{katz_introduction_2020}:

\begin{defbox}
\begin{definition}[One-way function (OWF)]\label{def:prelim-owf}
A function $f : \{0, 1\}^n \rightarrow \{0,1\}^l$ is a one-way function if:
\begin{enumerate}
    \item $f$ can be evaluated in polynomial time on every input.
    \item for every \emph{probabilistic polynomial time (PPT)} algorithm $\A$ there exists a negligible function $\epsilon$ such that:
    \begin{equation}
        \underset{X \leftarrow \{0,1\}^n}{Pr}[\A(f(X),1^n) \in f^{-1}(f(X))] \leq \epsilon(n) \quad \forall n.
    \end{equation}
\end{enumerate}
\end{definition}
\end{defbox}

\noindent We also note that one-way-ness can be also defined over a family of functions. OWFs are an important part of modern cryptography, hence we refer the interested reader to \cite{katz_introduction_2020} for more information about the topic. Now, let us introduce another important terminology in the next section.

\subsection{Formal frameworks for cryptanalysis}\label{sec:prelim-gamebase-uc-ac}
Security proofs in cryptography are usually given in a formal mathematical framework that allows for careful analysis of the cryptographic tasks. The most widely-used framework in modern cryptography is the \emph{game-based framework}. In the game-based framework, the cryptographic definition (or task) is formalised as a \emph{game} played between the adversary (In this thesis, we usually denote the adversary as $\A$) and a hypothetical honest party called \emph{challenger} (we usually denote the challenger by $\C$). Both parties can be (and usually are) probabilistic algorithms and processes. The game should be defined in a way that captures the task of interest, and it is defined over a probability space. The security then is proved by performing a probabilistic analysis and measured in the success probability of the adversary in winning the game. This model is quite popular in cryptography as it is relatively easy to understand and utilise, while it is powerful enough to capture the security of any cryptographic primitive. Moreover, the framework can also be translated in the quantum regime, as has been previously done by several works such as \cite{boneh_quantum-secure_2013,boneh_secure_2013,alagic_quantum-access-secure_2020}. In this thesis, we will also use this framework for our security proofs and will define our quantum definitions based on this paradigm.

However, for completeness, we will also briefly mention other security paradigms. The other famous security framework is \emph{simulation paradigm} \cite{goldreich_how_1986}.\footnote{Also called \emph{simulation-based framework} and \emph{real world/ideal world paradigm}} Here the security is captured by comparing two scenarios (or two worlds): An ideal scenario which is secure by definition, and a real-world, where an adversary can interact with the real execution of the protocol. In this regime, the protocol is secure, if any adversary in the real model cannot do much better than if it was involved in the ideal model or, in other words, the two scenarios are \emph{indistinguishable}.

The simulation-based framework, however, investigates the security in a stand-alone model, \emph{i.e.} when we consider the execution of that protocol only, as a single instance. However, more complicated cryptographic protocols are often composed of smaller sub-protocols and components and one requires to ensure the security of the whole system is preserved if \emph{secure protocols} are being composed together. Although this is not only non-trivial, there are several examples where this is not the case, \emph{i.e.} composing secure protocols leads to a non-secure system \cite{katz_introduction_2020}. A cryptographic framework that addresses this problem has been introduced by Canetti \cite{canetti_universally_2001} and is called \emph{Universal Composability (UC)} framework. This framework is closer in nature to the simulation paradigm. Despite being very powerful, often proving the security of protocols in this manner is more complicated, and there are considerably fewer cryptographic schemes that have been proven \emph{composabily secure}. Another similar framework that also captures the composability issue into account is a framework introduced by Maurer~\cite{maurer_abstract_2005} and is called \emph{Abstract Cryptography (AC)} framework. Both of these frameworks have also been extended to quantum setting \cite{muller-quade_composability_2009,maurer_indifferentiability_2016}.

\subsection{Adversarial models in the quantum world}\label{sec:prelim-adversarial-models}
In this section, we will introduce different adversarial models in the quantum world, specifically the ones that we will mostly encounter in this thesis. Some of the adversaries that we discuss are more common in the quantum information literature (for instance against quantum protocols such as QKD). While some others are, generally speaking, translations of different classes of classical adversaries into the quantum world, and as a result, mostly adopted from cryptography literature. As in this thesis, we deal with various quantum adversaries we attempt to give a general and coherent overview of them in this section.

Let us first set up the definition of an adversary! An adversary is an algorithm, defined as a polynomial-time uniform family of quantum circuits, that can be either deterministic or probabilistic (the probabilistic ones are more common to consider due to generality), and their goal is to perform a task that leads to breaking a protocol, or more generally a cryptosystem, under specific assumptions. Hence, each class of adversaries is usually characterised by the set of assumptions that we consider for such algorithms.

In this thesis, we are interested in \emph{quantum adversaries}, i.e. adversaries who also possess quantum capabilities, in addition to their usual classical computational power. As a result, the first assumption on our adversaries of interest is \emph{quantum mechanics} itself! Thus, we assume that a quantum adversary is subject to the laws of quantum mechanics, which we also assume to be \emph{correct} and \emph{complete}\footnote{Although one might consider quantum mechanics as an established model that describes the nature, which is true to some extent and precision as all of the theories in physics, its correctness and completeness is still an assumption we (happily) carry along with ourselves throughout this thesis (and generally in quantum cryptography and quantum information). Despite debates on the \emph{completeness} of quantum theory, as discussed in \cite{portmann_security_2021}, this is still a very justifiable assumption to make}.

If no additional assumptions have been made on the adversary, which inherently means that the adversary's computational power is \emph{unbounded}, the adversary is often called \emph{unbounded quantum adversary}. The security that is achieved against such adversaries is called \emph{information-theoretic security}, as opposed to \emph{computational security} where there are assumptions on the computational capabilities \cite{portmann_security_2021}. This is the strongest known notion of security that the marriage of quantum mechanics and information theory has been made possible \cite{broadbent_quantum_2016}. The security of many quantum protocols, such as QKD, quantum money, quantum coin-flipping and so on, has been studied in this security model. The adversary sometimes appears in the form of an eavesdropper (usually called Eve) who wants to access some encoded information that is being exchanged through a channel controlled by this adversary (like in the case of QKD). In some other protocols, the adversary plays the role of a malicious party in a protocol who wants to cheat or deviate from the honest behaviour (like in the case of quantum money or quantum coin-flipping). 

Nevertheless, this notion is usually too strong, and almost none of the classical cryptosystems that we have can resist unbounded quantum adversaries \cite{wallden_cyber_2019, mosca_cybersecurity_2018}. Thus a more common and standard adversarial model is the class of \emph{quantum polynomial-time} (QPT) adversaries. Here the computational power of the adversary has been limited to polynomial time (in the security parameter). The QPT adversary is, in fact, an \emph{efficient} quantum adversary that, in terms of complexity theory, is allowed to run polynomial-time uniform family of quantum circuits.
Despite being computationally bounded, this class of adversary is still very powerful,  and it is known that many existing cryptosystems based on computational hardness assumptions are still broken against this adversary as well, due to the existence of efficient quantum algorithms such as Shor's algorithm \cite{shor_algorithms_1994}, and Grover's algorithm \cite{grover_fast_1996} that the adversary can exploit\cite{wallden_cyber_2019,song_note_2014}. A QPT adversary can also be given oracle access to the classical or quantum primitive. The oracle model is a common cryptographic technique that is widely used in security proofs since it facilitates modelling adversarial behaviours where some information about the scheme is gathered (in some earlier stages or by interacting with the scheme and observing its properties). We discuss such oracles in the next section as well in Section~\ref{sec:unf-quantum-oracles} in \chapref{chap:unf-tools}. Since a QPT adversary is polynomial bounded, it is also bounded in the oracle model to a polynomial number of queries to the given oracle \footnote{In general, in the field of algorithm complexity, time complexity and query complexity has been considered separately in many cases, while as in cryptography we usually consider the QPT adversary to be polynomial in both.}. The quantum adversaries that have this somewhat quantum communicative access to the primitives are also sometimes called \emph{online} quantum adversaries, while an \emph{offline} quantum adversary can only have classical information of the primitive, and later on, use a quantum algorithm together with the classical data to break the cryptosystem. The study of the security of classical cryptosystems against \emph{offline} quantum adversaries, is famously known as post-quantum cryptography \cite{buchmann_post-quantum_2016,bernstein_post-quantum_2017,song_note_2014}. Moreover, the more technical term for this security model is the \emph{standard security model}, while as when the oracle access to the primitive is considered quantum, the term \emph{quantum security model} is used \cite{boneh_quantum-secure_2013,boneh_secure_2013,gagliardoni_semantic_2016}. One of the key elements of the quantum security model is the superposition queries, that is the adversary can query many classical values in one quantum query in the form of a superposition of those states. Superposition queries enables a broader range of non-trivial attacks~\cite{kaplan_quantum_2016,santoli_using_2017,boneh_quantum-secure_2013,gagliardoni_semantic_2016} that are not possible in the classical regime, or the standard security model.

Although the quantum security model might seem too strong to be considered for classical schemes, there are well-justified reasons for considering it. First of all, we note that for a quantum scheme, the natural model to consider \textit{is} the quantum security model, since any type of interaction with the primitives will be via quantum states and since classical primitives can also be generalised as quantum ones, this model is theoretically more general and hence interesting. Moreover, in the future, one can consider a world where classical computers have been replaced with quantum ones and hence even the classical routines and cryptographic algorithms are being run on a quantum computer. This scenario argument has been also given by Boneh and Zhandry in \cite{boneh_secure_2013}. Nonetheless, from a more practical point of view, one argument against this model for classical primitives is that a possible countermeasure against \emph{superposition attacks} is to forbid any kind of quantum access to the oracle through measurements. However, in such a setting the security relies on the physical implementation of the measurement tool which itself could be potentially exploited by a quantum adversary. Thus, and as it has previously been advocated in~\cite{boneh_secure_2013,boneh_quantum-secure_2013,kaplan_quantum_2016,alagic_quantum-access-secure_2020}, providing security guarantees in the quantum security model is crucial even practically.

Once the oracle's access to the primitive is considered, the adversarial models can be further categorised based on the assumptions of the access level of the adversary. These classes of adversaries are usually the adaptation of the usual classical models in the quantum regime. As usual, we start with the strongest case. The strongest access level is when an adversary can directly access the oracle and query any arbitrary quantum state of their choice. Also, the queries can be issued \emph{adaptively}, meaning that the adversary can choose the next query depending on the responses of the oracle to previous queries. In classical cryptography, this attack model is called \textit{Chosen Message Attack (CMA)} model. The quantum analogue of this model has been introduced by Boneh and Zhandry \cite{boneh_secure_2013}, and called \emph{Quantum Chosen Message Attack (qCMA)}\footnote{When used for encryption scheme, there are several attack models associated with this security level. If the adversary can choose the plaintext arbitrarily, it is referred to as \emph{(quantum) chosen-plaintext attack ((q)CPA)} and if the adversary is allowed to choose the ciphertext, \emph{(quantum) chosen ciphertext attack ((q)CCA)}. Also, due to the complications that exist in the quantum setting and different notions of oracles, there are several definitions for this security level in the literature \cite{gagliardoni_semantic_2016,chevalier_security_2020,gagliardoni_quantum_2021,carstens_relationships_2021}}. This is one of the main quantum adversarial models that we use in this thesis. However, we will carefully define our version of qCMA within our given security game in \chapref{chap:unf-tools}.

Another important note that is worth mentioning here is that although in CMA/qCMA models the queries are \emph{adaptive}, the term `adaptive' is also used in the literature for another security level, where the adversary has been given an extra learning phase, usually after receiving the main message. This level of adaptiveness is meaningful for some definitions for instance for encryption schemes one can consider such a model for ciphertext, known as CCA2. This model has also been brought into the quantum world \cite{chevalier_security_2020}. We will also briefly mention this model for our case studies. 

Next, the adversary can be weakened if we restrict the direct access to the oracle and instead the adversary has access to a random set of queries, chosen from a certain distribution\footnote{Which is usually the uniform distribution over all the possible set of messages}. This is often referred to as \emph{Random Message Attack (RMA)} and can also be translated in the quantum setting when the set of queries are quantum input and output samples of the quantum oracle. Sometimes this type of adversaries is also called \emph{non-adaptive} or \emph{weak} adversaries. As we will see in \chapref{chap:unf-tools}, and also later in \chapref{chap:application}, this adversarial model has close connections with the models that are considered in learning theory.

Bounding the computational power of the quantum adversary is not the only option to go to a weaker quantum adversarial model. It is also possible to remain in the information-theoretic security regime while making instead, some reasonable assumptions about the storage capabilities of the adversary. Making an assumption about the adversary's capability in storing quantum data is a technologically sensible assumption due to the difficulty of building quantum memories, despite the latest efforts and progress \cite{lvovsky_optical_2009,wang_efficient_2019,bradley_ten-qubit_2019,gyongyosi_optimizing_2020,lago-rivera_telecom-heralded_2021,bouillard_quantum_2019,wallucks_quantum_2020,dennis_topological_2002}. This has given rise to a quantum adversarial model known as \emph{bounded quantum-storage model}. This model has been introduced by \cite{damgard_cryptography_2005} and inspired from its classical counterparts \cite{brickell_protocols_1993,goos_unconditional_1997}. In this model, we assume that a quantum adversary can only store a limited number of qubits, yet it is computationally unbounded. It is common for the protocols in this model to assume no quantum memory for the
honest parties, while the adversaries can only store a small fraction
of the qubits sent in the protocol by assumption. Several protocols that are impossible to achieve in the unbounded model have proven to be secure in the quantum bounded storage model, such as oblivious transfer \cite{damgard_cryptography_2005} or bit commitment \cite{unruh_concurrent_2011}. Moreover, another realistic assumption to be made about the quantum memories is that they are noisy. This assumption has been considered in \cite{wehner_cryptography_2008}, leading to a model called \emph{noisy quantum-storage model}. In this thesis, we do not consider the memory-restricted adversarial models, but an enthusiastic reader can find further readings in \cite{broadbent_quantum_2016,portmann_security_2021}.

As the final note, since entanglement is also a precious quantum resource, one can also consider models where quantum adversaries are restricted in using this resource. This model has been studied in \cite{buhrman_position-based_2014} in the context of position-based cryptography, where it has been assumed that the adversaries cannot share entanglement.

\subsection{Quantum accessible oracles for classical functions}\label{sec:prelim-oracles}
In the previous section, we have discussed different adversarial models and the role of oracles in them. In this section, we define quantum oracles for classical primitives. These oracles are also called \emph{quantum accessible oracles}. 

A quantum oracle is a unitary transformation $\bold{\Ora}$ over a $D$-dimensional Hilbert space that can be queried with quantum states. The quantum oracle can grant quantum access to the evaluation transformation of a classical or quantum primitive \emph{i. e.} a classical function. The quantum accessible oracle gives, in fact, a reversible quantum implementation of that function. The first way of doing so is what is referred to as \emph{the standard oracle}~\cite{kashefi_comparison_2002,boneh_secure_2013,boneh_quantum-secure_2013,gagliardoni_semantic_2016,gagliardoni_quantum_2021,chevalier_security_2020}.

In the standard quantum-query model, the adversary $\A$ has black-box access to a reversible version of $f$, which is a classical-polynomial-time computable deterministic or randomised function of the evaluation $\E$, through an oracle $\reO_f$ which is a unitary transformation. The evaluation oracle can be represented as:
\begin{equation}\label{eq:prelim-standard-oracle}
    \reO_f: \sum_{m,y} \alpha_{m,y} \ket{r}_{\Ora}\ket{m,y} \rightarrow \sum_{m,y} \alpha_{m,y}\ket{r}_{\Ora}\ket{m, y \oplus f(m;r)}
\end{equation}
Here $m$ is the message, and $y$ is the ancillary system required for unitarity. In general, the standard oracle can also capture randomised evaluations with a randomness $r$ picked from $\R \subseteq \{0,1\}^l$ as the randomness space, although in this case, the oracle may not be a unitary transformation. The unitary representation of the standard oracle has been introduced in several works such as~\cite{gagliardoni_semantic_2016,gagliardoni_quantum_2021,chevalier_security_2020} with slightly different approaches that lead to an equivalent adversary's state, which is a completely mixed density matrix with respect to the randomness subspace. Nevertheless, in this thesis, to emphasise that the adversary cannot gain access to the internal randomness register of the oracle directly and avoid some potential artificial entanglement attacks, we opt for the approach of~\cite{gagliardoni_quantum_2021} and consider the randomness as an internal state of the oracle which is re-initiated for each query with a new classical value $r$. This choice is also due to the fact that the oracle needs to output the randomness register as a separable state, otherwise, an unwanted entanglement will be created between the adversary's output state and the internal register of the oracle, as also mentioned in~\cite{gagliardoni_quantum_2021}. Moreover, if the primitive requires that the randomness is returned to the adversary for each query (as a classical bit-string or a function of $r$), it can be recorded in the adversary's auxiliary state $y$ that can be extended to also capture the randomness space. An example of such construction will be introduced later in Section~\ref{sec:unf-results}. Finally, we specify that for deterministic primitives (denoted by $\eO_f$) the structure is similar, except that the randomness register is not used.

Now let us introduce another type of quantum accessible oracles. First, let's assume that the function $f: \{0,1\}^n \rightarrow \{0,1\}^n$ is a bijection. In this case, the following transformation is a unitary:
\begin{equation}\label{eq:prelim-minimal-oracle-1}
    \Ora_f: \sum_{m} \alpha_{m}\ket{m} \rightarrow \sum_{m} \alpha_{m}\ket{f(m)}
\end{equation}
This can be generalised for non-length-preserving functions as the following transformation:
\begin{equation}\label{eq:prelim-minimal-oracle}
    \Ora_f: \sum_{m,y} \alpha_{m,y}\ket{m,y} \rightarrow \sum_{m,y} \alpha_{m,y}\ket{\phi_{m,y}}
\end{equation}
where the length of the ancillary register $|y| = |f(m)| - |m|$ and $\phi_{m,0} = f(m)$ for every $m$. This type of oracles are called \emph{minimal oracles}, and as one can see this is closer to a general unitary, and hence they can consider to be a more powerful oracle than the standard oracle. One main difference between minimal and standard oracle is that if the function $f$ is an encryption scheme $Enc_k(.)$, then the adjoint of the minimal encryption oracle is the decryption oracle i.e. $\Ora^{\dagger}_{Enc} = \Ora_{Dec}$, while as this is not the case for the standard oracle \cite{gagliardoni_semantic_2016}.

Another type of quantum oracles is the \emph{Fourier oracle} or \emph{Fourier phase oracle} which is defined as follows \cite{kashefi_comparison_2002,zhandry_how_2019}:
\begin{equation}\label{eq:prelim-fourier-oracle}
    \text{Fourier}\eO_f: \frac{1}{\sqrt{2^{m2^n}}}\sum_{m,y}\ket{m,y} \rightarrow \frac{1}{\sqrt{2^{m2^n}}} \sum_{m,y} e^{2\pi i f(m).y / 2^n}\ket{m,y}
\end{equation}

It has been shown that the Fourier oracle and standard oracle are equivalent \cite{kashefi_comparison_2002}. Also in \cite{zhandry_how_2019} sophisticated techniques have been developed to record the adversary's queries made to the oracle, which is a very challenging task in the quantum regime when the queries are quantum, due to properties such as unclonability. Being able to record quantum queries is needed in some of the proof techniques and reductions in cryptography, and it is specifically relevant in the quantum random oracle model (QROM) \cite{boneh_random_2011}. This issue has been addressed in \cite{zhandry_how_2019} by introducing a new type of oracle called \emph{compressed oracle}, which can be both defined as a standard or Fourier oracle. We avoid introducing this technique here since we do not use oracle recording techniques in this thesis.  
\subsection{Classical pseudorandomness}\label{sec:prelim-classic-pseudorandom}
Randomness is perhaps one of the most crucial elements in modern cryptography. However, it is folklore knowledge that achieving true randomness in the classical world is practically impossible. That is why the concept of \emph{pseudorandomness} has been introduced in cryptography as an efficient and practical approximation of truly random objects. Generally, a pseudorandom object, should not be distinguishable from its truly random counterpart. To formally define this concept, first, we need to ask the following question: `Indistinguishable to what?' Since this is a cryptographic concept, let's consider an adversarial scenario. It is rather obvious that an unbounded adversary who can cover all the possible objects of the set can always make this distinction, then pseudorandomness is an \emph{computational} security assumption. Thus the answer to that question is to an efficient or computationally bounded adversary (or, more generally, distinguisher). The  pseudorandomness can be generally defined as follows \cite{katz_introduction_2020}:

\begin{defbox}
\begin{definition}[Pseudorandomness]\label{def:prelim-pseudorandom}
Let $\D$ be a distribution over $n$-bit strings.$\D$ is $(t,\epsilon)$-pseudorandom if for all adversaries $\A$ running in time at most $t$, $\A$ cannot distinguish $\D$ with a uniformly random distribution $U_n$ over the $n$-bit. In other words, the following holds:
\begin{equation}\label{eq:prelim-pseudorandom}
    |\underset{x \leftarrow \D}{Pr}[\A(x) = 1] - \underset{x \leftarrow U_n}{Pr}[\A(x) = 1]| \leq \epsilon
\end{equation}
\end{definition}
\end{defbox}

Nevertheless, in the asymptotic case, the $\epsilon$ needs to be a negligible function in the security parameter and the pseudorandomness has been usually defined over a finite family of objects with a specified distribution. Now, we can define the first important pseudorandom object in modern cryptography, the Pseudorandom Generators (PRG):

\begin{defbox}
\begin{definition}[PRG~\cite{katz_introduction_2020}]\label{def:prelim-prg}
Let $\ell$ be a polynomial and let $G$ be a deterministic polynomial-time algorithm such that for any $n$ and any input $x \in \{0,1\}^n$, the result $G(x)$ is a string of length $\ell(n)$. We say that $G$ is a Pseudorandom Generator (PRG) if the following conditions hold:
\begin{itemize}
    \item \textbf{(Expansion:)} For every $n$ it holds that $\ell(n) > n$
    \item \textbf{(Pseudorandomness:)} For any PPT algorithm $\A$, there is a negligible function $\negl$ such that:
    \begin{equation}\label{eq:prelim-prg}
    |{Pr}[\A(G(x)) = 1] - {Pr}[\A(r) = 1]| \leq \negl(n) 
\end{equation}
\end{itemize}
where both probabilities are taken over  the randomness of $\A$, the first one over uniform choice of $x \in \{0, 1\}^n
$ and, and the second one, over uniform
choice of $r \in \{0,1\}^{\ell(n)}$.
\end{definition}
\end{defbox}

Thus a PRG is an efficient and deterministic algorithm that expands a \emph{short} uniformly random seed into a longer pseudorandom string.

Next, we define \emph{Pseudorandom Functions (PRF)} which are a family (usually keyed-family) of functions that are computationally indistinguishable from the set of uniformly random functions from the same domain and range. PRFs are formally defined as follows:

\begin{defbox}
\begin{definition}\label{def:prelim-prf}[Pseudorandom Functions (PRF)]
Let $\K,\X,\Y$ be the keyspace, the domain and range, all implicitly depending on the security parameter $\lambda$. A keyed family of functions $\{PRF_k: \X \rightarrow \Y\}_{k\in \K}$ is a pseudorandom function (PRF) if for any polynomial-time (PPT) algorithm $\A$, $PRF_k$ with a  random $k \leftarrow \K$ is indistinguishable from a truly random function $f \leftarrow \Y^{\X}$ in the sense that:
\begin{equation}
|\underset{k \leftarrow \K}{Pr}[\A^{PRF_k}(1^{\lambda})=1] -\underset{f \leftarrow \Y^{\X}}{Pr}[\A^{f}(1^{\lambda})=1]| = \negl(\lambda).
\end{equation}
\end{definition}
\end{defbox}

PRFs can also be equivalently defined in a game-based fashion as an indistinguishability game \cite{katz_introduction_2020}. At the beginning of the game, an honest challenger flips a coin and selects to be in a random or pseudorandom world. Then according to the selected world, the challenger picks either a function $f$ from the truly random family of functions; or picks a random key, and consequently, a pseudorandom function $F_k$. Then every time the adversary issues a query, the challenger responds with $f$ or $F_k$, depending on the random bit $b$. The adversary's objective is then to guess $b$, or in other words, distinguish between truly random and pseudorandom worlds.

PRFs are extremely practical tools for cryptography, and many classically secure constructions are based on them. It is also known that under the assumption of OWF, a PRF family can be constructed.

One can also translate this concept to the quantum setting, where the adversary/distinguisher is quantum. This notion is referred to as \emph{quantum-secure Pseudorandom Functions (qPRF)} and is defined as follows:

\begin{defbox}
\begin{definition}\label{def:prelim-qprf}[quantum-secure Pseudorandom Functions (qPRF) \cite{zhandry_how_2012}]
Let $\K,\X,\Y$ be the keyspace, the domain and range, all implicitly depending on the security parameter $\lambda$. A keyed family of functions $\{PRF_k: \X \rightarrow \Y\}_{k\in \K}$ is a quantum-secure pseudorandom function (qPRF) if for any polynomial-time quantum oracle algorithm $\A$, $PRF_k$ with a  random $k \leftarrow \K$ is indistinguishable from a truly random function $f \leftarrow \Y^{\X}$ in the sense that:
\begin{equation}
|\underset{k \leftarrow \K}{Pr}[\A^{PRF_k}(1^{\lambda})=1] -\underset{f \leftarrow \Y^{\X}}{Pr}[\A^{f}(1^{\lambda})=1]| = \negl(\lambda).
\end{equation}
\end{definition}
\end{defbox}

Similarly, it has been proven that qPRFs can exist under the assumption of quantum-secure OWFs \cite{zhandry_how_2012}.

\subsection{Quantum pseudorandomness}\label{sec:prelim-quanutm-pseudorandom}
The definitions of this sections are directly used in \chapref{chap:unf-tools}, Section~\ref{sec:sel-unf-randomized} and throughout \chapref{chap:pr-connection}.
Like what we have seen in the previous section, \emph{quantum} pseudorandom objects can also be defined. The notion of quantum pseudorandomness has been defined for the first time in \cite{shacham_pseudorandom_2018}, by introducing \emph{Pseudorandom Quantum States (PRS)} and \emph{Pseudorandom Unitaries (PRU)} as a computational version of true quantum randomness. In Section~\ref{sec:prelim-haar-random} we have introduced the Haar measure as a measure for perfect and uniform randomness over the quantum states and unitary transformations. Informally, pseudorandom states/unitaries are a set of states/unitaries that are computationally indistinguishable from Haar-random states/unitaries to a quantum polynomial-time adversary.

\noindent More formally, PRS is defined as follows:

\begin{defbox}
\begin{definition}\label{def:prelim-prs}[Pseudorandom Quantum States (PRS)~\cite{shacham_pseudorandom_2018}]
Let $\Hil$ be a Hilbert space and $\K$ the key space. $\Hil$ and $\K$ depend on the security parameter $\lambda$. A keyed family of quantum states $\{\ket{\phi_k}\in S(\Hil)\}_{k\in\K}$ is \textit{pseudorandom}, if the following two conditions hold:
\begin{itemize}
    \item \textbf{Efficient generation}. There is an efficient quantum algorithm $G$ which generates the state $\ket{\phi_k}$ on input $k$. That is, for all $k\in\K, G(k) = \ket{\phi_k}$.
    \item \textbf{Pseudorandomness}. Any polynomially many copies of $\ket{\phi_k}$ with the same random $k\in\K$ is computationally indistinguishable from the same number of copies of a Haar random state. More precisely, for any efficient quantum algorithm $\A$ and any $m\in poly(\lambda)$, 
\end{itemize}
\begin{equation}
    |\underset{k \leftarrow \K}{Pr}[\A(\ket{\phi_k}^{\otimes m})=1] - \underset{\ket{\psi} \leftarrow \mu}{Pr}[\A(\ket{\psi}^{\otimes m})=1]| = \negl(\lambda).
\end{equation}
where $\mu$ is the Haar measure on $S(\Hil)$.
\end{definition}
\end{defbox}

\noindent And PRU that are the quantum equivalent of PRFs are also defined as follows:

\begin{defbox}
\begin{definition}\label{def:prelim-pru}[Pseudorandom Unitary Operators (PRU)~\cite{shacham_pseudorandom_2018}]
A family of unitary operators $\{U_k \in \mathcal{U}(\Hil)\}_{k \in \mathcal{K}}$ is a pseudorandom unitary if two conditions hold:
\begin{itemize}
    \item \textbf{Efficient computation}. There is an efficient quantum algorithm $Q$ such that for all $k$ and any state $\ket{\psi} \in S(\Hil), Q(k,\ket{\psi}) = U_k\ket{\psi}$.
    \item \textbf{Pseudorandomness}. $U_k$ with a random key $k$ is computationally indistinguishable from a Haar random unitary operator. More precisely, for any efficient quantum algorithm $\A$ that makes at most polynomially many queries to the oracle:  
\end{itemize}
\begin{equation}
    |\underset{k \leftarrow \K}{Pr}[\A^{U_k}(1^{\lambda})=1] - \underset{U \leftarrow \mu}{Pr}[\A^U(1^{\lambda})=1]| = \negl(\lambda).
\end{equation}
where $\mu$ is the Haar measure on $S(\Hil)$. Note that here we focus on the Pseudorandomness condition of the PRU definition.
\end{definition}
\end{defbox}

Another approximation of Haar-randomness in quantum information is the notion of \emph{t-designs}. Although these objects are also often called `pseudorandom' in the mathematical physics literature, they are analogous to t-wise independent random variables in theoretical computer science \cite{shacham_pseudorandom_2018}. Quantum state and unitary t-designs are informally approximating the Haar measure up to $t$-th order polynomials or tensor products. Thus one way of defining the quantum states t-design is as follows:
\begin{equation}
    \sum_i p_i (\ket{\phi_i}\bra{\phi_i})^{\otimes t} = \int_{\mu_{\psi}} (\ket{\psi}\bra{\psi})^{\otimes t} d\mu_{\psi}
\end{equation}
where $p_i$ is the probability of each state $\ket{\phi_i}$ and the integration in the right hand side is over Haar measure \cite{dankert_exact_2009,emerson_scalable_2005}. Similarly, a unitary t-design can be defined as follows:
\begin{equation}
    \sum_i p_i U_i^{\otimes t}\rho (U_i^{\otimes t})^{\dagger} = \int_{\text{Haar}} U^{\otimes t}\rho (U^{\otimes t})^{\dagger} dU
\end{equation}
On the right-hand side of the equation is the expectation for the t-fold tensor product of Haar measure is also denoted by $\mathbb{E}^t_H (\rho)$. Also, a t-design can be defined approximately. The $\epsilon$-approximate t-design has been introduced by Brandão, Harrow and Horodecki \cite{brandao_efficient_2016} as follows:

\begin{defbox}
\begin{definition}[$\epsilon$-approximate t-design]\label{def:prelim-epsilon-approx-tdesing}
We say a family of unitary with distribution $\D$ given as the set $\{p_i, U_i\}$ forms an $\epsilon$-approximate t-design if the following holds:
\begin{equation}
    (1 - \epsilon)\mathbb{E}^t_H (\rho) \leq \sum_i p_i U_i^{\otimes t}\rho (U_i^{\otimes t})^{\dagger} \leq (1 + \epsilon)\mathbb{E}^t_H (\rho) \quad \quad \forall \rho \in \ES(\Hil^{\otimes t})
\end{equation}
\end{definition}
\end{defbox}

We conclude this section by mentioning some of the numerous applications of t-designs in different areas of quantum computing and quantum information, including quantum supremacy \cite{bouland_quantum_2019}, verification and benchmarking \cite{hashagen_real_2018,nakata_quantum_2021,eisert_quantum_2020}, the physics of blackholes \cite{hayden_black_2007}, cryptography \cite{alagic_quantum_2017,shacham_pseudorandom_2018}, and machine learning \cite{mcclean_barren_2018}.

\subsection{Unforgeability}\label{sec:prelim-unf}
Unforgeability is the desired security property for many primitives such as Message Authentication Codes (MACs) and digital signatures. Informally, unforgeability ensures that an adversary cannot produce valid input-output pairs of the evaluation function of the primitive with only limited access to its oracle, or in other words, from a previously learnt set of input and outputs of the function. The unforgeability of a classical primitive can be studied against classical or quantum adversaries in the different adversarial models that we have introduced in Section~\ref{sec:prelim-adversarial-models}. In this section, we first introduce different levels of classical unforgeability, and then we also give some of the proposals for translating this notion to the quantum setting. Later in \chapref{chap:unf-tools}, we generalise the quantum unforgeability inside a formal and unified framework. Thus this section is mostly relevant for \chapref{chap:unf-tools}, Section~\ref{sec:unf-framework}. Unforgeability is also a central security property for quantum schemes such as quantum money, as we will further discuss in that chapter.

\subsubsection{Classical Unforgeability}\label{sec:prelim-classical-unf}
Goldwasser et al.~\cite{goldwasser_digital_1988} define different notions of unforgeability for digital signatures. They consider various types of attacks including CMA where the adversary is allowed access to the signing oracle on a list of messages of their choice. They define \emph{existential forgery} as the attack where the adversary can forge a valid signature for at least one new message; and the notion of \emph{selective forgery} as an attack where the adversary can forge a valid signature with non-negligible probability for a particular message chosen by the adversary prior to accessing the signing oracle.

An et al. \cite{an_security_2002} define a slightly stronger notion of unforgeability called \emph{strong unforgeability} that requires the adversary not only to be unable to generate a valid signature on a `new' message but also to be unable to generate even a valid `new' signature on an already signed message. \emph{Strong Existential Unforgeability (SEUf)}, also called \emph{strong unforgeability}, has formally been defined in \cite{boneh_strongly_2006} by Boneh et al.

Bellare \emph{et al.}~\cite{bellare_xor_1995} define the notion of Strong Existential Unforgeability under chosen message and chosen verification queries attack (SEUF-CMVA) for message authentication codes (MACs). In both of these attack models, the adversary is allowed a chosen message oracle access, as defined for digital signatures in~\cite{goldwasser_digital_1988}. Although in the later attack model for message authentication codes, the experiment also allows verifying queries through oracle access. This model is justified for MACs as unlike digital signatures, where the verification algorithm is public, the adversary cannot run the verification algorithm on their own. \emph{(Weak) Existential Unforgeability (EUf) under chosen message attacks} is a natural definition for MACs defined by Bellare et al.~\cite{bellare_security_2000} and comes by extending the one for digital signatures~\cite{goldwasser_digital_1988}.

Moreover, Dodis et al.~\cite{dodis_message_2012} define the notion of \emph{selective unforgeability under adaptive chosen message and chosen verification queries (SelUF-CMVA)}.

A yet weaker notion called \emph{universal unforgeability} requires the adversary to produce a fresh tag for a uniformly random message given as a challenge to the adversary~\cite{alwen_key-indistinguishable_2014}. This notion, again, can be considered against both attack models: chosen message and chosen verification query attack (UniUF-CMVA) and chosen message attack (UniUF-CMA). Table~\ref{table:cunf} summarizes all these different classical notions of unforgeability.

\begin{table}[h]
\centering
\captionsetup{font=small}
\resizebox{0.6\textwidth}{!}{
\begin{tabular}{|c|c|c|}
\hline
\backslashbox{Def. level}{Attack Model} & CMVA & CMA \\ \hline
SEUf (strong) & - & \cite{an_security_2002,boneh_strongly_2006,bellare_xor_1995} \\ \hline
EUf (weak) & \cite{dodis_message_2012,bellare_power_2004} & \cite{boneh_strongly_2006,bellare_security_2000} \\ \hline
SelUf (selective) & - & \cite{dodis_message_2012} \\ \hline
UniUf (universal) & \cite{alwen_key-indistinguishable_2014} & \cite{alwen_key-indistinguishable_2014} \\ \hline 
\end{tabular}}
\caption{Classical unforgeability definitions from strongest to weakest. \textit{CMVA} - adaptive chosen message queries and limited access to the verification oracle. \textit{CMA} - (adaptive) chosen message attacks. In the cases marked with ``-", no definition has been proposed yet to the best of our knowledge.}
\label{table:cunf}
\end{table}

\subsubsection{Unforgeability in the quantum world}\label{sec:prelim-quantum-unf}
In the quantum regime, the definition of unforgeability defined by Boneh and Zhandry~\cite{boneh_secure_2013,boneh_quantum-secure_2013} (denoted by \bz), is described as a quantum analogue of \emph{strong existential unforgeability} and it is in the chosen message attack (CMA) model. The formal definition of \bz\ (EUF-qCMA) for digital signatures is as follows:

\begin{defbox}
\begin{definition}\label{def:bz}[\bz\ or (EUF-qCMA) \cite{boneh_quantum-secure_2013}]
A system S (Sign/Mac), is existentially unforgeable under a quantum chosen message attack (EUF-qCMA) if no adversary after issuing $q$ quantum chosen message queries, can generate $q+1$ valid classical message-tag pairs with non-negligible probability in the security parameter.
\end{definition}
\end{defbox}

Another definition of unforgeability against quantum adversaries called \emph{blind unforgeability} was proposed in~\cite{alagic_quantum-access-secure_2020}. This more recent definition aims to capture some attacks that are not captured by \bz. This notion defines an algorithm to be forgeable if there exists an adversary who can use access to a `partially blinded' oracle to validate responses of the messages that are in the blinded region and hence only respond to the queries that are not in this region.
A blinded operation for a function $f: X \rightarrow Y$ and a subset of messages $B \subseteq X$ is defined as:
\begin{equation}
    Bf(x)= 
    \begin{cases}
    \perp, & \text{if } x \in B\\
    f(x), & \text{otherwise}
    \end{cases}
\end{equation}
Where in particular for the definition of unforgeability, the elements of X are placed in B independently at random with a particular probability $\epsilon$, denoted by $B_{\epsilon}$. Then the security game of unforgeability has been defined as follows with the adversary having access to the blinded oracle.

\begin{defbox}
\begin{definition}\label{def:prelim-bu}[\cite{alagic_quantum-access-secure_2020}(Def.$4 \& 5$)]
Let $\Pi = (KeyGen,Mac,Ver)$ be a MAC with message set $X$. Let $\A$ be an algorithm, and $\epsilon: \mathbb{N} \rightarrow \mathbb{R}_{\geq 0}$ an efficiently computable function. The blind forgery experiment $BlindForge_{\A,\Pi}(n,\epsilon)$ proceeds as follows:
\begin{enumerate}
    \item Generate key: $k \leftarrow KeyGen(1^n)$
    \item Generate blinding: select $B_{\epsilon}\subseteq X$ by placing each $m$ into $B_{\epsilon}$ independently with probability $\epsilon(n)$.
    \item Produce forgery: $(m,t) \leftarrow \A^{B_{\epsilon}{MAC}_k}(1^n)$.
    \item Outcome: output 1 if $Ver_k(m,t) = acc$ and $m \in B_{\epsilon}$ ; otherwise output 0.
\end{enumerate}
From this game blind-unforgeability is defined as follows.\\
A MAC scheme $\Pi$ is blind-unforgeable (BU) if for every polynomial-time uniform adversary $(\A,\epsilon)$
\[
    Pr[BlindForge_{\A,\Pi}(n,\epsilon(n)) = 1] \leq \negl(n).
\]
and the probability is taken over the choice of key, the choice of blinding set, and any internal randomness of the adversary.
\end{definition}
\end{defbox}

Thus, in this definition, a forgery happens if the adversary can produce a valid tag for a message within the blinded region. We refer to this definition of unforgeability as \bu. This definition imposes the challenge to be orthogonal to the previously queried messages.

We also recall the following useful theorem from \cite{alagic_quantum-access-secure_2020}:

\begin{thmbox}
\begin{theorem}\label{th:prelim-bu}[from~\cite{alagic_quantum-access-secure_2020}]
Let $\A$ be a QPT such that $supp(\A)\cap R = \emptyset$\footnote{Here $supp(\A)$ denotes the support of $\A$ that is defined as follows. Let $\A$ have oracle access to a classical function $f:\{0,1\}^n \rightarrow \{0,1\}^m$. Let $\ket{\psi_i}$ be the state of the the query $i$ or equivalently the intermediate state after applying $U_i$ in the sequence of $\Ora U_q\Ora\dots U_1$ on an initial state $\ket{0}_{XYZ}$ where $X$ denotes the input registers. Then $supp(\A)$ is defined to be the set of input strings $x$ such that there exists a function $f$ with the respective oracle such that $\mbraket{x}{\psi_i}_X \neq 0$ for at least one of the queries.} for some $R \neq \emptyset$. Let $\mathtt{MAC}$ be a MAC, and suppose $A^{\mathtt{MAC}_k}(1^n)$ outputs a valid pair $(m,\mathtt{Mac}_k(m))$ with $m \in R$ with non-negligible probability. Then $\mathtt{MAC}$ is not BU-secure.
\end{theorem}
\end{thmbox}

In addition to these two main definitions, another definition for quantum unforgeability has been given in \cite{garg_new_2017} for \emph{one-time} unforgeable schemes, which we will skip representing it due to the lack of generality and since it is less relevant for this thesis. Another related and interesting work is the study of \emph{non-malleability} and its relation to authentication in the quantum regime which has been studied in \cite{alagic_quantum_2017}.

\subsection{Coin-flipping}\label{sec:prelim-coin-flip}
In this section, we introduce \emph{coin-flipping} which is a cryptographic task that allows two mutually distrustful parties to agree on a common random bit. We particularly need the familiarity with this cryptographic functionality for \chapref{chap:varqlone} (Section~\ref{sec:varqlone-statedep-cloning-crypt} and \ref{sec:varqlone-numerical-results-statedep}). This task has been first introduced by Blum~\cite{blum_coin_1983} and has been motivated in the following scenario: Alice and Bob need to agree on the output of a coin-flip over the phone for an important decision. However, they don't trust each other. The formal definition of a coin-flipping task is given as follows:

\begin{defbox}
\begin{definition}[(Strong) coin-flipping]\label{def:prelim-coin-flip}
The task of coin flipping consists of two mutually distrustful players, Alice and Bob, and the goal is for both players to output the same random bit $c \in \{0,1\}$ such that the following properties hold:
\begin{itemize}
    \item \textbf{Correctness:} if both Alice and Bob are honest then $b$ is uniformly distributed: $p(c = 0)  = p(c = 1) = 1/2$.
    \item \textbf{$\epsilon$-secure:} neither player can force $p(c = 0) \geq  1/2 + \epsilon$ or $p(c = 1) \geq 1/2 + \epsilon$, where $p(c)$ is the probability that the honest player outputs a value $c$.
\end{itemize}
The smallest $\epsilon$ for which a protocol is $\epsilon$-secure is called the \textbf{bias}.
\end{definition}
\end{defbox}

It has also been shown in \cite{blum_coin_1983} that unconditionally secure coin-flipping is impossible in the classical world, meaning that no classical coin-flipping protocol is secure, or no value of $\epsilon < 1/2$ can be achieved for security. Nevertheless, coin-flipping with computational assumptions is possible since there exists a coin-flipping protocol, assuming perfectly secure OWF exists. Also, Cleve \cite{cleve_limits_1986,cleve_martingales_1993} extended the computational coin-flipping into $r$-rounds and showed an upper bound of $\Omega(1/r)$ for any two-party $r$-round coin-flipping protocol as the bias.

Historically, the first quantum coin-flipping protocol was introduced by~\cite{mayers_unconditionally_1999} which has conjectured to achieve arbitrary small bias, although a full security proof has not been given in the paper. In \chapref{chap:varqlone} we will introduce this protocol and as one of our contributions, we show how it can be broken with cloning-based cryptanalysis.

Later a quantum coin-flipping protocol has been introduced by  Aharonov et al. \cite{aharonov_quantum_2000} which provably achieves the bias $0.42$ in an information-theoretic way. We will also closely study this protocol in \chapref{chap:varqlone}, so we avoid repetition in here.

Another quantum coin-flipping protocol with qutrits has been introduced by Ambainis in \cite{ambainis_new_2004} which achieves a better bias than Aharonov's protocols with $0.25$ bias. Given this improvement, an interesting question was whether one we achieve arbitrary small bias for coin-flipping using quantum information. This question has been answered negatively by Kitaev when given the following bound for the bias of any strong coin-flipping protocols:
\begin{equation}\label{eq:prelim-kitaev-bound-strong-coinflip}
    \epsilon^{qcf}_{min} = \frac{\sqrt{2} - 1}{2} \approx 0.207.
\end{equation}
Hence perfectly secure quantum coin-flipping is also impossible.

The requirement on the coin-flipping can we weakened if the choice of Alice and Bob is predetermined, say Alice always wants to bias the coin towards $0$, and Bob is vice versa. This leads to the notion of \emph{weak coin-flipping}. Weak coin-flipping has been studied in the literature for many years and it has been shown in 2007 \cite{mochon_quantum_2007} that that weak quantum coin flipping, with arbitrarily small (but non-zero) bias, is possible. Although the protocol that achieved that arbitrary small bias is complicated and scales exponentially in $1/\epsilon$ in the number of rounds. The long-standing problem of the weak coin-flipping has been finally solved by Arora et al. in 2019 \cite{arora_quantum_2019}.

\section{Quantum and classical learning}\label{sec:prelim-learning}
In this section, we enter the last field of research where we have adopted and exploited many of our toolkits and the concepts we have used in this thesis and which is rather different from the other fields of research we have talked about so far. Here, we will talk about `learning' in a broad context which includes learning theory, machine learning and some areas of quantum information processing such as tomography. However, as each of these subjects has a very rich literature on its own, and entering each of them with enough precision and care requires a separate thesis, our introduction here will be very brief and we will mostly focus on the tools that we have directly used in the thesis. 

Learning is the act of acquiring knowledge, but generally speaking, in physics and computer science, there are two types of learning. Either we want to learn a `system' or an unknown property or feature of the system by interacting with it, or we want the `machines' and computers to be able to do something of a similar nature. However, later we need to teach the machines to `learn' first, which means adopting a methodical approach to leverage `data' to improve the performance of a learning task. The first case is a problem usually studied in physics (and generally through experiments and simulations), and the second is a sub-field of computer science known as \emph{learning theory}\footnote{Also called computational learning theory} and \emph{machine learning}. Although seemingly very different, the latest progress in the field of machine learning, as well as different approaches to simulating physical systems, has brought these fields closer together. An example of this is the applications of machine learning in particle physics \cite{radovic_machine_2018,andreassen_junipr_2019}. On the other hand, physics has also inspired machine learning models, for instance in development of \emph{Boltzmann machine}~\cite{sherrington_solvable_1975} or \emph{Born machine}~\cite{cheng_information_2018,coyle_born_2020}.

On the other hand, the potential advantage of quantum computing in bringing speedup to some problems has motivated researchers to design quantum machine learning techniques and algorithms. One of the earliest examples (perhaps the earliest one) was an algorithm developed for solving linear equations and similar problems in matrix algebra by Harrow, Hasidim, and Lloyd \cite{harrow_quantum_2009}, where exponential speedup for some operations has been shown. However, the field of \emph{Quantum Machine Learning (QML)} has expanded fast over the very few years that have been past since its birth. For a great review of the field we refer the reader to \cite{schuld_introduction_2015,martin-guerrero_quantum_2022}. 

We start on the physics side, explaining the existing notions of learning in quantum information processing, and we finish by introducing some of the tools that we require from classical and quantum learning theory and quantum machine learning.

\subsection{Quantum state and process tomography}\label{sec:prelim-tomography}
Quantum state tomography and quantum process tomography are the process of determining the state of a quantum state or describing quantum dynamics and are of great importance to quantum computation and quantum information, both theoretically and experimentally. However, they are both challenging and resource-intensive tasks \cite{bisio_optimal_2009}.

In Section~\ref{sec:prelim-distinguish-quantum-test}, we have explained the reason behind the difficulty of determining and distinguishing quantum states. Quantum process tomography is even more challenging since we aim to fully characterise the dynamics of a quantum system. For example, we want to characterise a quantum gate (a unitary transformation), or a quantum channel. Both of these tasks have applications in the verification and benchmarking of quantum systems and quantum computers.

Let us start with state tomography. The easiest way (and clearly the most inefficient) way to perform process tomography, is to simply use the Born's rule. This method also sometimes called \emph{linear inversion} involves preparing (or acquiring in any way) many copies of an unknown state $\rho$, and repeatedly performing projective or POVM measurements, in order to extract the expectation values of the probability, of obtaining a histogram of the observations of the measurements, and finally, the amplitudes to fully describe the state. However, this method is highly infeasible and requires asymptotically many copies and measurements to achieve good precision. This duo to the fact that to reconstruct a $d$-dimensional density matrix, one needs to determine $d^2 − 1$ independent parameters and requires many measurements for each. (where $d$ is already exponential in the input size \emph{i.e.} the number of qubits). A better approach is to restrict the domain of the density matrices to a more `likely' space. This method is called \emph{Maximum Likelihood Estimation (MLE)} and involves searching for the density matrix that maximises the likelihood of giving the experimental results. The `likelihood' is a probability function assigned to the observable that would most likely detect the state. Using this method a complexity of $O(d^4)$ has been achieved in the general cases \cite{paris_quantum_2004,qi_quantum_2013}. Yet another method is to use Bayesian estimations such as \emph{Bayesian Mean Estimation (BME)} for this task \cite{blume-kohout_optimal_2010}. This method requires reasonable prior information about the systems however it can achieve a minimum estimation probability of $\frac{1}{N + d}$, with $N$ being the number of observables.

Exploiting machine learning techniques in recent years has enabled tremendous improvements in this field. For instance a general complexity of $O(d^3)$ has also been achieved using neural networks \cite{xu_neural_2018}, and the results have been improved in the following works \cite{araya-polo_deep-learning_2018,torlai_neural-network_2018,rymarczyk_logistic_2019}.

Finally, the most recent breakthrough in this field has been made by Huang et. al in \cite{huang_predicting_2020}, where they have shown that even though the best-known techniques for full state tomography are still exponential (and believed to remain so), one can still extract many useful properties of a quantum state, efficiently and without requiring exponential copies of that state. This discovery has a great deal of significance in very different areas of quantum information and quantum computing, as we will discuss further in the future chapters. 

Going back to quantum process tomography, we first note that this process is closely related to quantum state tomography. Let us give a simple example. Assume you are given an unknown unitary gate (a single-qubit gate, for instance) and want to extract the full unitary matrix. To do so, you need to know the action of the unitary on a full set of basis (say, the computational basis). Thus you will prepare states in the computational basis and apply the unitary to them. However, the output state of the gate, $U\ket{\psi}$, is unknown and can be any state on the Bloch sphere in our specific example. Thus one needs to measure the state and repeat the process many times to get a good approximation of the action of $U$, only on the computational basis. Then for the state $\ket{1}$, the same process needs to be repeated. Nevertheless, similar to state tomography, this is the most naive and inefficient way. A variety of different disciplines exist here such as \emph{Standard quantum-process tomography (SQPT)}, \emph{ancilla-assisted process tomography (AAPT)}, and \emph{direct characterization of quantum dynamics (DCQD)}. We refer the reader to \cite{mohseni_quantum-process_2008} for a review of these different strategies and their required resources.

A point worth mentioning here is that for process tomography (maybe unlike state tomography), the full characterisation of the dynamics might not be needed for many problems and applications. An example of this is \emph{randomised benchmarking} which is a method for testing the quality and capability of quantum hardware by estimating the average error rates of the gates~\cite{emerson_scalable_2005}. The other example is the idea of \emph{quantum emulation}. Since quantum emulation will have particular importance in our work, we will introduce it separately in the next section.

\subsection{Quantum Emulation}\label{sec:prelim-qe}
We now describe the concept of \emph{Quantum Emulation (QE)} and an algorithm called \emph{universal quantum emulator} developed by Marvian and Lloyd in \cite{marvian_universal_2016}.\footnote{Although this may not be the only possible algorithm for the purpose of quantum emulation, is the only one we are aware of by the time of writing this thesis, and hence the content of this section are mainly based on the mentioned work.} The \emph{quantum emulation algorithm} is a quantum process learning tool that can outperform the existing approaches based on quantum tomography \cite{dariano_quantum_2001}. Generally speaking, the goal of the quantum emulator is to mimic the action of an unknown unitary transformation on an unknown input quantum state\footnote{The algorithm can be applied to any unknown state, however, the high fidelity performance is achieved when the target state is in the span of the data hence, not fully unknown.} while having access to some `data' in the form of input-output samples of the unitary\footnote{The term `emulation', however, has been used in other meanings both in physics and cryptography. While throughout this thesis whenever we use the term, we refer to emulation in the context used in this section.}. We return to this algorithm in \chapref{chap:unf-tools} Section~\ref{sec:unf-qe-revisited} to further analyse the algorithm and repurpose it. 

Before diving into the details of this algorithm, we make a small remark on the difference between `emulation' and `simulation'.

\subsubsection{Emulation vs simulation for a quantum process}\label{sec:emulation-simulation}
Emulating a quantum process, in the sense that is introduced here, is different from process tomography since the goal is not to extract the full description of a quantum process but instead to `learn' enough from it to be able to mimic its behaviour. However, when one defines the notion of emulation in this manner, it suggests a similarity with another notion, that is also particularly important for physical systems \emph{i.e.} `simulation'. Especially because the `unknown unitary' that we are trying to emulate here, can correspond to the Hamiltonian of a physical system. Despite similarities, here we want to emphasise that the notion of simulation and emulation for a quantum process have crucial differences. 

In simulating a quantum dynamics, for instance, a unitary, we usually start from an initial state of a quantum system (for example, a many-body system of particles), and we use abstractly speaking a \emph{`simulation machine'} that produces a `simulation' or a mathematical approximation that mimics the required property of the system. Sometimes the method consists of letting the system evolve with a stochastic process \cite{berry_simulating_2015}.\footnote{This is usually referred to as quantum random walk} The key point here, is that the simulation of the system needs to completely (however with some approximation) obey the same dynamics and transformations, otherwise, it will not be useful to study the original system. 

\begin{figure}[h!]
   \centering
     \includegraphics[width=0.9\textwidth]{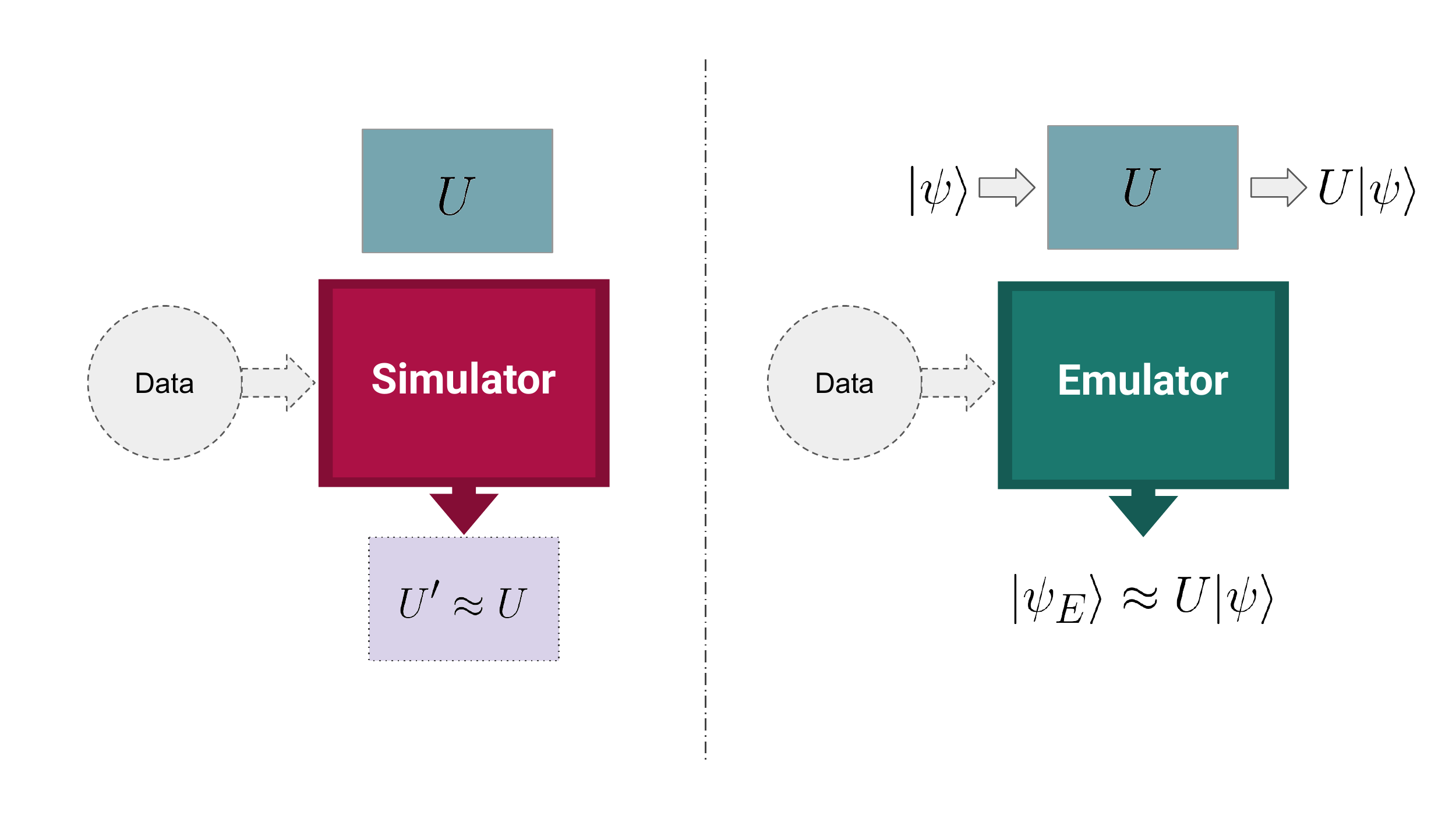}
     \caption[Quanutm emulation vs. quantum simulation]{Illustration of the contrast between the notions of emulation and simulation. Particularly regarding a quantum process, denoted by a unitary $U$. An emulator, as opposed to a simulator, does not necessarily recreate the same dynamics but instead mimics the action of the unitary on a new quantum state. }\label{fig:simulate-emulate}
 \end{figure}

An emulator, on the other hand, is not trying to completely recreate the transformation. Instead, it outputs what that transformation is `supposed to do' on a new quantum state. Although it mimics the system's behaviour in an input-output manner, it does not need to obey the same dynamics. In that sense, an emulation algorithm is closer to a machine learning process, as we will also discuss later in \chapref{chap:unf-tools}. This distinction between simulation and emulation has been illustrated in \figref{fig:simulate-emulate}. With these in mind, we now introduce the quantum emulation algorithm.

\subsubsection{QE: the circuit and description of the algorithm}
\begin{figure}[ht]
    \centering
\includegraphics[width=1.0\columnwidth]{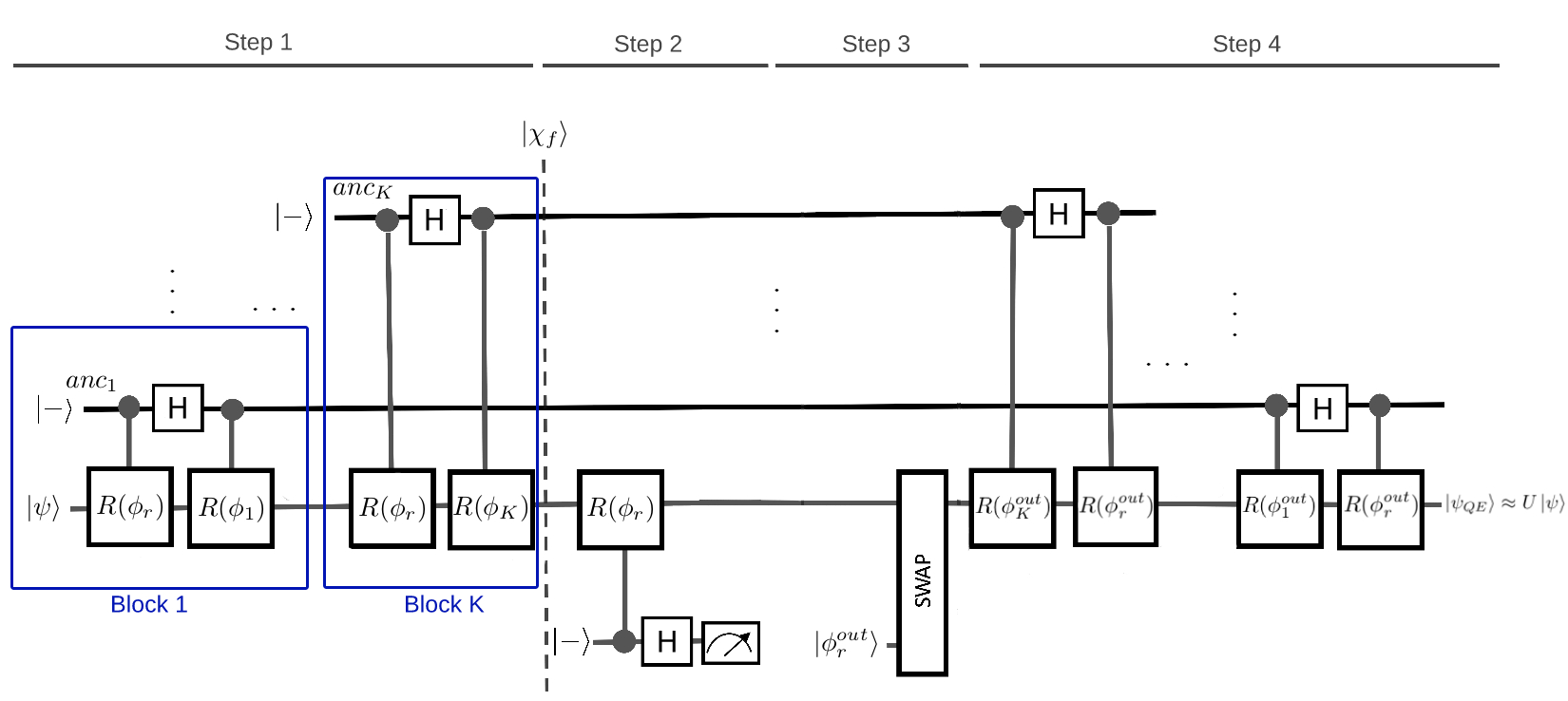}
    \caption[The circuit of the quantum emulation algorithm.]{
    The circuit of the quantum emulation algorithm. $\ket{\phi_r}$ is the reference state and $\ket{\phi^{out}_r}$ is the output of the reference state. $R(*)$ gates are controlled-reflection gates. In each block of Step 1, a reflection around the reference and another sample state is being performed.
    }
    \label{fig:qe}
\end{figure}

The circuit of the quantum emulation algorithm is depicted in \figref{fig:qe} (recreated from \cite{marvian_universal_2016}) and works as follows: Let $\U$ be a unitary transformation on a D-dimensional Hilbert space $\HilD$, $\Sin = \{\ket{\phi_i}; i = 1, ..., K\}$ be a sample of input states and $\Sout = \{\ket{\phi^{out}_i}; i = 1, ..., K\}$ the set of corresponding outputs, i.e $\ket{\phi^{out}_i} = \U\ket{\phi_i}$. Also, let $d$ be the dimension of the Hilbert space $\Hild$ spanned by $\Sin$ and $\ket{\psi}$, a challenge state. The goal of the algorithm is to find the output of $\U$ on $\ket{\psi}$, that is $\U\ket{\psi}$.

The main building blocks of the algorithm are controlled-reflection gates described as:
\begin{equation}
    R_c(\phi) = \ket{0}\bra{0} \otimes \mathbb{I} + \ket{1}\bra{1} \otimes e^{i\pi\ket{\phi}\bra{\phi}}
\end{equation}
A controlled-reflection gate acts as the identity ($\mathbb{I}$) if the control qubit is $\ket{0}$, and as $R(\phi) = e^{i\pi\ket{\phi}\bra{\phi}} = \mathbb{I} - 2\ket{\phi}\bra{\phi}$ if the control qubit is $\ket{1}$.
The circuit also uses Hadamard and SWAP gates and consists of four stages.

\noindent\textbf{Stage 1.} $K$ number of sample states and a specific number of ancillary qubits are chosen and used through the algorithm. We assume the algorithm uses all of the states in $\Sin$. The ancillary systems are all qubits prepared at $\ket{-}$. Let $\ket{\phi_r} \in \Sin$ be considered as the reference state. This state can be chosen at random or according to some distribution. The first step consists of $K-1$ blocks wherein each block the following gates run on the state of the system and an ancilla:
    \begin{equation}\label{eq:qe-block}
        W(i) = R_c(\phi_i) H R_c(\phi_r).
    \end{equation}
    In each block represented by \eqref{eq:qe-block}, a controlled-reflection around the reference state $\ket{\phi_r}$ is performed on $\ket{\psi}$ with the control qubit being on the $\ket{-}$ ancillary state. Then a Hadamard gate (H) runs on the ancilla followed by another controlled-reflection around the sample state $\ket{\phi_i}$. This repeats for each of the $K$ states in $\Sin$, such that the input state is being entangled with the ancillas, and also it is being projected into the subspace $\Hild$. By doing this, the information of $\ket{\psi}$ is encoded in the coefficients of the general entangled state. This information is the overlap of $\ket{\psi}$ with all the sample inputs. By reflecting around the reference state in each block, the main state is pushed to $\ket{\phi_r}$ and the probability of finding the system at the reference state increases. The overall state of the circuit after Stage 1 is:
    \begin{equation}
        [W(K)...W(1)]\ket{\psi}\ket{-}^{\otimes K} \approx \ket{\phi_r}\ket{\Omega(anc)}
    \end{equation}
    where $\ket{\Omega(anc)}$ is the entangled state of $K$ ancillary qubits. The approximation comes from the fact that the state is not only projected on the reference quantum state but is also projected on other sample quantum states with some probability. We present a more precise formula in the next subsection.\\
    
\noindent\textbf{Stage 2.} In this stage, first a reflection around $\ket{\phi_r}$ is performed and after applying a Hadamard gate on an extra ancilla, that ancilla is measured in the computational basis $\{\ket{0}, \ket{1}\}$. Based on the output of the measurement, one can decide whether the first step was successful (i.e. the output of the measurement is 0) or not. If the first step is successful, the main state has been pushed to the reference state. In this case, the algorithm proceeds with Stage 3. If the output is 1, it implies that the projection was unsuccessful and that the input state remains almost unchanged. In this case, either the algorithm aborts or it goes back to the first stage and picks a new state as the reference. This stage has a post-selection role which can be skipped, to output a mixed state of two possible outputs.\\
    
\noindent\textbf{Stage 3.} The main state is swapped with $\ket{\phi^{out}_r} = \U\ket{\phi_r}$ that is the output of the reference state. This is done by employing a SWAP gate. At this point, the overall state of the system is:
\begin{equation}
(\mathrm{SWAP}\otimes I^{\otimes K}) \ket{\phi^{out}_r}\ket{\phi_r}\ket{\Omega(anc)} = \ket{\phi_r}\ket{\phi^{out}_r}\ket{\Omega(anc)}.
\end{equation}
By tracing out the first qubit, the state of the system becomes $\ket{\phi^{out}_r}\ket{\Omega(anc)}$.\\
    
\noindent\textbf{Stage 4.} The last stage is very similar to the first one except that all blocks are run in reverse order, and the reflection gates are made from corresponding output quantum states. The action of stage 4 is equivalent to:
\begin{equation}
W^{out}(i) = R_c(\phi^{out}_i) H R_c(\phi^{out}_r) = (\U \otimes I)W(i)(\U^{\dagger} \otimes \mathbb{I}).
\end{equation}
After repeating this gate for all the output samples, $\U$ is applied to the projected components of $\ket{\psi}$, and by restoring the information of $\ket{\psi}$ from the ancilla, the input state approaches $\U\ket{\psi}$. The overall output state of the circuit at the end of this stage is:
\begin{equation}
[W^{out}(1)...W^{out}(K)]\ket{\phi^{out}_r}\ket{\Omega(anc)} \approx \U\ket{\psi}\ket{-}^{\otimes K}
\end{equation}
where equality is obtained whenever the success probability of Stage 2 is equal to 1. 

The property of interest to measure the success or quality of the emulation algorithm is the fidelity of the output state $\ket{\psi_{QE}}$ (the output state of QE on the main register) and the intended output $\U\ket{\psi}$. In the original paper, the fidelity analysis is first provided for ideal controlled-reflection gates and later a protocol is presented to implement them efficiently using a technique called \emph{quantum principal component analysis} introduced by Lloyd, Mohseni and Rebentrost \cite{lloyd_quantum_2014}.\footnote{For the purpose of this thesis as we are more interested in the theoretical bounds for the fidelity, all the gates including the controlled-reflection gates are assumed to be ideal keeping in mind that the implementation is possible.} We now recall the following central theorem from \cite{marvian_universal_2016}: 

\begin{thmbox}
\begin{theorem}\label{th:qe-fidel}\text{\cite{marvian_universal_2016}}
Let $\E_{\U}$ be the quantum channel that describes the overall effect of the algorithm presented above. Then for any input state $\rho$, the Uhlmann fidelity of $\E_{\U}(\rho)$ and the desired state $\U\rho\U^{\dagger}$ satisfies:
\begin{equation}\label{eq:qe-fidel}
    F(\rho_{QE}, \U\rho\U^{\dagger}) \geq F(\E_{\U}(\rho), \U\rho\U^{\dagger}) \geq \sqrt{P_{succ-stage1}}
\end{equation}
where $\rho_{QE} = \ket{\psi_{QE}}\bra{\psi_{QE}}$ is the main output state(tracing out the ancillas) when the post-selection in Stage 2 has been performed. $\E_{\U}(\rho)$ is the output of the whole circuit without the post-selection measurement in Stage 2 and $P_{succ-stage1}$ is the success probability of Stage 1.
\end{theorem}
\end{thmbox}

We point out that this algorithm with ideal controlled-reflection gates performs the emulation task with $K$ total blocks and arbitrary precision $\epsilon$ given that,
\begin{equation}\label{eq:prelim-qe-blocks-complex}
    T \geq \frac{d \times \log{(8d\epsilon^{-2})}}{1 - |\lambda_{\D}|}
\end{equation}
where $\lambda_{\D}$ is the eigenvalue of the overall channel with the second largest magnitude. Thus the algorithm (both sample and run time) complexity, in this case, is polynomial in $d$. Nevertheless, the spectral gap of the channel, namely $1 - |\lambda_{\D}|$, plays an important role as well, according to the \eqref{eq:prelim-qe-blocks-complex}, which is not very clear and easy to determine. This is why in \chapref{chap:unf-tools} where we use this algorithm, we do not use these complexity results. Instead, we perform fidelity analysis based on the output of the algorithm. 

Finally, in the imperfect setting and given the algorithm for approximating the controlled reflection gates, the overall algorithm can be implemented using the $N_{tot}$ number of samples, and in time $t_{tot}$ with precision $\epsilon > 0$, as follows:
\begin{equation}\label{eq:prelim-qe-runtume-sample}
    N_{tot} = \mathcal{\tilde{O}}\left( \frac{d^2 \times \epsilon^{-1}}{(1 - |\lambda_{\D}|)^2} \right), \quad \quad t_{tot} = \mathcal{\tilde{O}}(N_{tot} \times \log D)
\end{equation}
where $\mathcal{\tilde{O}}$ suppresses more slowly-growing terms.

\subsection{Learning theory}\label{sec:prelim-learning-theory}
In this section we discuss learning theory and we mainly use the definitions provided here in \chapref{chap:unf-tools}, Section~\ref{sec:unf-relation-uuf-pac}. Learning theory is a theoretical subfield of machine learning or \emph{computational learning} that provides the mathematical framework for studying and quantifying learning problems as well as the design and analysis of algorithms to solve them. The `learner' is a classical/quantum algorithm (either deterministic or probabilistic), but the target of learning, can be very different objects, including functions, quantum states, quantum processes, distributions, etc. Also, depending on the learning task, one might not need to fully learn all the characteristics or a full description of the target, but perhaps some specific properties. Often the main question here is \emph{how efficient} can the learning target be learned. This efficiency is measured commonly in time (time complexity) or the size of the sample data (query complexity) \cite{angluin_computational_1992}. Another essential element of learning is data. We need data to learn from or train our machines to perform the desired task. The dataset, in learning theory, is also called  \emph{learning data}, or\emph{training set}. Let of start by defining this sample data:

\begin{defbox}
\begin{definition}[Learning/sample/training dataset]\label{def:prelim-sample-data}
Let $\T$ be the learning object and let $\X$ be the domain or instance space of $\T$, and $\Y$ be the label set (for instance the range, if $\T$ is a function). The learning/sample/training data is a set $S = \{(x_i,y_i)\}^K_{i=1} \subseteq (\X \times \Y)^K$, where $K = |S|$ is the size of the set. Furthermore, the datapoints $x_i$ may have been chosen according a distribution $\D$.
\end{definition}
\end{defbox}

Note that we have defined the above definition in a general way, such that it can be applied to different learning targets and different classical or quantum data. For instance, for a fully quantum data we have $\rho_i,\sigma_i \in \Hil$ and hence $\X = \Y = \Hil$.

Also, the learning problems are usually categorised in two types in the literature of learning theory, as follows \cite{kearns_introduction_1994}:
\begin{itemize}
    \item \textbf{Unsupervised learning:} we require the learner to discover hidden structures in a set of unlabeled data points.
    
    \item \textbf{Supervised learning:} we want to learn a property or make a prediction based on a labelled dataset.
\end{itemize}

The next terminology that we need to introduce here is the notions of \emph{concept class} and \emph{hypothesis}. A `concept' is a specific sample of the learning object. It is most commonly used for Boolean functions, so a concept is a specific $f:\X \rightarrow \Y$. A `concept class' is the set of all the learning objects of interest \emph{e.g.} for a function that would be a family of functions $\F$ such that $f \in \F$. A `hypothesis' $h$, is the learner's output or guess for the function $f$. One ideally wants $h$ to be as close as possible to $f$, for the learning process to be successful. We can also define a `hypothesis class' $H$ where all the $h \in H$, and by restricting the learner to choose $h$ from $H$, we can bias the learner towards a particular set of solutions.

Before introducing the first formal learning definition for functions, we need to introduce one last concept. In Section~\ref{sec:prelim-oracles} we have introduced oracles and their importance in cryptography. Here as well, we can assume the data has been obtained via interacting with an oracle. However, this is an specific oracle called \emph{example oracle} or $\PEX$ \cite{servedio_equivalences_2004,arunachalam_guest_2017}. An example oracle gets a concept $f$, a distribution $\D$, and when queried, outputs a sample data point as defined in \defref{def:prelim-sample-data}:
\begin{equation}
    \PEX(f,\D) \rightarrow [(x,f(x)) : (x \leftarrow \D)]
\end{equation}

Now, we are ready to introduce the definition of \emph{Probably Approximately Correct (PAC)} learning, introduced first by Valiant \cite{valiant_theory_1984}. The name refers to the learner not being required to learn a function \emph{exactly} ($h(x) = f(x)$), but rather approximately with a high probability. The definition has been given in different ways with slight variations in the literature, however, for our purpose, we choose the one closer to \cite{servedio_equivalences_2004,arunachalam_guest_2017}. 

\begin{defbox}
\begin{definition}[PAC-learnability]\label{def:prelim-pac-learn}
A concept class $\F$ is $(\epsilon,\delta)$-PAC learnable, if a learning algorithm $\Ler$, given access to a $\PEX$ oracle, can generate a hypothesis $h$, for all distributions $\D$, and for any concept $f \in \F$, such that $h$ is an $\epsilon$-approximation of $f$ under $\D$, with at least $1-\delta$ probability. \emph{i.e.}
\begin{equation}
    Pr[h \leftarrow \Ler^{\PEX} : \underset{x \in \D}{Pr}[h(x) \neq f(x)] \leq \epsilon] \geq 1 - \delta
\end{equation}
\end{definition}
\end{defbox}

Similar to what has been discussed in Section~\ref{sec:prelim-oracles}, the classical oracles can be translated to quantum setting, for modelling quantum access and to be able to leverage the quantum properties of a data that is encoded in quantum states, such as superposition. In this case, as well, the example oracle has been defined in the quantum setting in \cite{bshouty_learning_1995}. A \emph{Quantum Example Oracle ($\QPEX$)} (also called \emph{quantum random access oracle}), for a classical function (or concept) $f$, over a distribution $\D$ outputs the following state when queried:

\begin{equation}\label{eq:prelim-qpex-oracle}
    \QPEX(f,\D) \rightarrow \ket{\psi_{EX}} := \sum_x \sqrt{\D(x)} \ket{x, f(x)}
\end{equation}

Another way of translating the dataset queries into the quantum world is by \emph{Quantum Membership Query ($\QMQ$)} oracles \cite{servedio_equivalences_2004}, that is much closer to the back-box oracles we have discussed in Section~\ref{sec:prelim-oracles}. A QMQ for a concept $f$, operates as follows:

\begin{equation}\label{eq:prelim-qmq-oracle}
    \QMQ_f: \ket{x,b,y} \rightarrow \ket{x,b\oplus f(x), y}
\end{equation}
which can also take as input superposition state of classical inputs. However, the $\QMQ$ is mostly used in the context of \emph{exact} learning instead of PAC-learning. Using the $\QPEX$, one can define a quantum variant of PAC-learnability\footnote{Although it is better to call this quantum-assisted PAC-learning or PAC-learning with $\QPEX$ oracle, to make a distinction with PAC-learning of quantum objects, for instance in this work \cite{padakandla_pac_2022-1}} as follows:

\begin{defbox}
\begin{definition}[PAC-learnability with $\QPEX$]\label{def:prelim-pac-learn-qpex}
We say a concept class $\F$ is $(\epsilon,\delta)$-quantum-PAC learnable or $(\epsilon,\delta)$-PAC learnable with $\QPEX$, if it is PAC-learnable according to \defref{def:prelim-pac-learn}, with the difference that the learner has been given oracle access to $\QPEX$.
\end{definition}
\end{defbox}

We also point out that this notion of PAC-learnability is quite strong since it is over all the possible distributions. However, one can be interested in the learnability of a concept class over a specific distribution, for instance, a uniform distribution. We refer to this case as \emph{PAC-learnability over D}, where we fix a distribution $D$ from a larger set or all the possible distributions $\D$. We come back to this point in \chapref{chap:unf-tools}, Section~\ref{sec:unf-relation-uuf-pac}.

We conclude this section by mentioning that, as shown in \cite{servedio_equivalences_2004}, the notions of classical and quantum PAC-learning are equivalent, although not in terms of efficiency. More precisely, if a concept class $\F$ is quantum-PAC-learnable according to \defref{def:prelim-pac-learn-qpex}, it is also classically PAC-learnable, while there exists a gap in terms of \emph{efficient} PAC-learnability between quantum and classical case.

\subsection{Quantum machine learning and variational algorithms}\label{sec:prelim-qml}
We have reached now the final section of the background and preliminary materials. In this section, we will give a very high-level introduction to \emph{Quanutm Machine Learning (QML)}. This section is only needed for \chapref{chap:varqlone}. As mentioned before, the QML is only about a decade old. However, due to the huge background brought into the field from classical machine learning and the particular interest and attention of the researchers in this field, it has grown quickly compared to its age. Here we only focus on a sub-filed quantum machine learning that we will need for this thesis, namely \emph{Variational Quantum Algorithms (VQA)}. For a comprehensive review of quantum machine learning we refer the enthusiastic reader to \cite{schuld_introduction_2015,biamonte_quantum_2017,martin-guerrero_quantum_2022,schuld_prospects_2018}.

We also note that the term \emph{quantum} in QML, refers to different classes of problems. In the first class, the learning algorithm is classical but the data is quantum (for instance, using neural networks to analyse measurement statistics from a quantum experiment), in the second one the learning algorithm is quantum, but the data is classical. However, we encode them in quantum states to enable the quantum algorithms to run on them (for instance \cite{harrow_quantum_2009}). And lastly, both data and learning algorithms are quantum (we will see an example of this case in \chapref{chap:varqlone}). For a better overview of each of these subfields and to see examples of each case in the NISQ era, we refer the reader to this thesis \cite{coyle_machine_2022}. Let us begin by introducing VQC.

\subsubsection{Variational quantum algorithms}\label{sec:prelim-vqc}
Generally speaking, a variational quantum algorithm is, in fact, a hybrid quantum-classical algorithm where the classical part is usually in charge of optimising parameters of a quantum object (a parameterised quantum state or a parameterised unitary circuit). The quantum part is the part of the algorithm that deals with interacting with a quantum input and producing a parametrised output such as $\rho(\boldsymbol{\theta})$, while $\boldsymbol{\theta}$ is the hyper-parameter that will be optimised \cite{coyle_machine_2022}.

Since the first part of the systems that interact with the quantum data, has a quantum nature, and the outputted result are quantum states (while the classical part works with classical data), at some point, the data needs to be measured. This process is defined through a set of observable $\{ O_i\}^K_{i=1}$, producing the set of measurement outcomes, or expectation values $\{ \langle O_i \rangle_{\boldsymbol{\theta}}\}$, passed on to the classical optimiser. As one can guess, VQAs are quite heuristic methods, but lately, several works have been done to provide theoretical frameworks and guarantees for them, including \cite{mcclean_theory_2016}. One of the key components of VQAs is the \emph{cost function} which defines the learning problem of interest, and we will briefly introduce it in what follows. 

\subsubsection{Cost function}\label{sec:prelim-qml-cost}
A cost function is a function of the parameters of our model, which quantifies the quality of the learning algorithm. We define the cost function as follows:

\begin{defbox}
\begin{definition}[Cost function]\label{def:prelim-cost-function}
Let a learning problem $\T$ be parameterised and characterised by the hyperparamaeter $\boldsymbol{\theta}$, a \emph{cost function} $C(\boldsymbol{\theta})$ is a function of $\boldsymbol{\theta}$, which a learner $\Ler$, attempts to find its global minimum in order to solve the learning task $\T$.
\end{definition}
\end{defbox}

According to \cite{cerezo_variational_2021}, a good cost function needs satisfy four main properties. \textbf{Faithfulness:} meaning that its minimum point should be a solution for the problem of interest in the parameter landscape; \textbf{efficient computability:} meaning that it should be reasonably feasible to estimate $C$ in polynomial time. \textbf{trainability (or efficient optimisation property):} meaning that the cost function should be optimisable efficiently, \emph{i.e.} it should be differentiable for calculating the gradients, or navigatable in the parameter landscape. And finally, it should have \textbf{operational meaning:} that is, a smaller cost function should correspond to closeness to the solution and quality of the learning.

Finding a suitable cost function for the learning problem of interest is one of the most crucial parts of a VQA, and it would contribute significantly to the success and efficiency of the VQA algorithm. 

Finally, a cost function can be \emph{local} or \emph{global} \cite{coyle_machine_2022,cerezo_cost_2021}. A local cost function corresponds to a local observable and a global cost function to a global one respectively. This distinction about the locality becomes very important in the quantum case, unlike classical cost functions, since local and global observables may differ in important quantum qualities such as entanglement. In \chapref{chap:varqlone}, we will see that this property will become very relevant and theoretically interesting to study for our problem.

\subsubsection{\Ansatze}\label{sec:prelim-qml-ansatz}
Another essential part of a VQA is its ansatz. This is the part of VQA that creates the parameterised states. Therefore, the form of the ansatz is relevant in the geometrical properties and the landscape of $\boldsymbol{\theta}$ \cite{cerezo_variational_2021}. More precisely, from an initial state $\ket{\psi_0}$ which is the input of the algorithm, the ansatz creates the parameterised state as follows:

\begin{equation}
    \ket{\psi(\boldsymbol{\theta})} = U(\boldsymbol{\theta})\ket{\psi_0} = U_L(\theta_L)\dots U_2(\theta_2)U_1(\theta_1) \ket{\psi_0}
\end{equation}

As it is clear from the above equation, the parameterisation can be made by applying a series of parameterised unitaries $U(\boldsymbol{\theta})$, that is also referred to as \emph{parameterised quantum circuits}. The structure of the \ansatze can be tailored to the problem (\emph{problem-inspired ansatz}), or it can be generic (which is also called \emph{problem-agnostic ansatz}) \cite{cerezo_variational_2021}. One advantage of adopting problem-agnostic \ansatze is that they can be tailored to the specific hardware instead, or in other words, be made \emph{hardware-native} or \emph{hardware-efficient}. This type will be particularly desirable for NISQ devices where a limited set of native gates are available, such as Rigetti or $\IBM$ quantum machines. Finally, hardware-efficient \ansatze aim to reduce circuit depth in VQAs, which is yet another important point for NISQ machines.

\subsubsection{Optimisation techniques}\label{sec:prelim-vqc-opt}
Finally, the last part of VQA is the classical optimisation method used to minimise the cost function. This part is essentially a classical machine learning subroutine. Hence, different classical optimisation methods can be used depending on the problem and training structure. One of the most exploited optimisation methods that are of particular interest for VQAs is the \emph{gradient-based optimisation} technique. We will also use the same method for our purpose in this thesis. 

The gradient of a parameterised cost function in the parametrised landscape is given by the following important result, famously known as the \emph{parameter shift rule}. It states the following \cite{cerezo_variational_2021,schuld_evaluating_2019,mari_estimating_2021}:

\begin{thmbox}
\begin{theorem}[Parameter-shift rule]\label{th:perlim-param-shift-rule}
Let $C(\boldsymbol{\theta})$ be the cost function, generated using an ansatz of the form $U(\theta_i) = e^{-i \theta_i/2 \sigma_i}$ where $\sigma_i$ are the Pauli operators, the gradient of $C$ with respect to parameter $\theta_i$ is given as follows:
\begin{equation}
\begin{split}
     \frac{\partial C(\boldsymbol{\theta})}{\partial \theta_i} & := \sum_k \frac{1}{2} (\tr[O_k U^{\dagger}(\boldsymbol{\theta_i^+})\rho_k U^{\dagger}(\boldsymbol{\theta_i^+})] - \tr[O_k U^{\dagger}(\boldsymbol{\theta_i^-})\rho_k U^{\dagger}(\boldsymbol{\theta_i^-})]) \\
     & =  \frac{1}{2}[C(\boldsymbol{\theta}_i^+) - C(\boldsymbol{\theta_i^-})]
\end{split}
\end{equation}
where $\boldsymbol{\theta}_i^{\pm} = (\theta_1,\dots,\theta_i\pm\frac{\pi}{2},\dots,\theta_L)$
\end{theorem}
\end{thmbox}

In general the above theorem can be generalised to a shift $\alpha$, instead of $\frac{\pi}{2}$, in that case the coefficient will be $\frac{1}{2\sin{\alpha}}$, instead of $\frac{1}{2}$ \cite{cerezo_variational_2021}. Generally speaking, the above theorem shows that one can evaluate the gradient by shifting the parameter by some amount $\alpha$, which makes the calculation of the cost function more efficient.
\chapter{Unclonability, Unforgeability and Learnability} \label{chap:unf-tools}
\begin{chapquote}{Maryam Mirzakhani}
``I like crossing the imaginary boundaries people set up between different fields - it's very refreshing.''
\end{chapquote}
\section{Introduction}\label{sec:intro-chap3}
As mentioned, unclonability is one of the pillars of quantum information and perhaps one of the most fundamental sources of security in quantum cryptography. In this chapter, we make an effort to shine new lights on the meaning of unclonability by bringing it into a broader context. We try to understand a generalisation of no-cloning via two other equally fascinating notions: \emph{Unforgeability} and \emph{Learnability}. The former is a cryptographic notion which we have introduced in Section~\ref{sec:prelim-unf}, and the latter is the subject of study in learning theory and machine learning, which we have briefly discussed in Section~\ref{sec:prelim-learning}. We first set the scene for the rest of the thesis by expressing the intuitive meaning and connections among these concepts. As we move forward in the chapter, we attempt to formalise some of the presented ideas while also introducing the tools and definitions that we will need to establish our results in the upcoming chapters. In fact, in this chapter, we sketch a big picture, of which we manage to paint only some tiny sections in detail. However, we believe that the general and intuitive overview represents the idea that binds different chapters of this thesis together and will hopefully initiate further thought-provoking questions in this area of research.

Apart from introducing the concepts and notions of interest, we also discuss two main contributions in this chapter. The first one is a new cryptanalysis tool, namely quantum emulation, which serves as a new attack model, as well as a learning tool which builds a bridge between cryptography and learning and leads to several no-go results that we will introduce in this chapter and \chapref{chap:qpuf}.

The second contribution is a new framework that formalises the notion of unforgeability in the quantum world. This framework generalises unforgeability for both quantum and classical primitives and provides us with an accurate yet intuitive tool to study the notion of unclonability from a cryptographic point of view in our future chapters. We also show several case studies as well as no-go and positive results in this framework to demonstrate the broader applicability of the framework in cryptography, and outside the coverage of the cases studied in connection to unclonability.

\subsection{Structure of the chapter}
In Sections~\ref{sec:unclone-unknown} and \ref{sec:unclone-different-learning} we mostly provide intuitions on the relationships between several concepts including unclonability, unknown transformations, learnability, unforgeability and so on. These two sections can be viewed as extensions of the introduction since we introduce a few novel results. Nevertheless, the reader may find novelty in some of the arguments and more importantly, the way they have been described. In Section~\ref{sec:unf-qe-revisited}, we revisit the quantum emulation algorithm that we introduced in Section~\ref{sec:prelim-qe} (\chapref{chap:prelim}), and we give several attack examples based on this algorithm in~\ref{sec:qe-attacks}. In Section~\ref{sec:unf-framework}, we introduce our generalised quantum framework for unforgeability and finally, in Section~\ref{sec:unf-results}, we discuss the relevance of our framework through several no-go results, examples and new secure constructions. 

\section{Unclonability and Unknownness}\label{sec:unclone-unknown}
Let us start with this rudimentary question: \emph{Why quantum states are unclonable?} A straightforward answer to this question, and one often found in quantum mechanics and quantum computing textbooks is something along these lines: \emph{Due to the unitarity of quantum mechanics.} Even though this answer is correct (as we have also seen in the proof of the no-cloning theorem in \chapref{chap:prelim}), we want to start this section by pondering whether perhaps there is a more fundamental answer to this question. To dive into this deeper level, we first need to ask the question in a manner that captures the more precise statement of the no-cloning theorem: \emph{Why `unknown' quantum states are unclonable?}

Allow us to discuss why the word `unknown' bares such a great deal of significance here. First, a `known' quantum state, \emph{i.e.} a state which we precisely know its classical description, \emph{is} perfectly clonable merely because by knowing the classical description, one can prepare as many copies as desired of that state. In fact, one can see the classical description as a recipe for making a unitary operation that generates that state over and over from a fixed state (for instance the computational basis $\ket{0}$). The second point is, as we have seen in Section~\ref{sec:prelim-cloning}, \emph{not all} quantum states are unclonable! A known set of orthogonal quantum states is \emph{clonable}. For the qubit case, if, for example, the state $\ket{\psi} \in \{\ket{0},\ket{1}\}$ (but let's say we don't know which one) then there is a simple cloning machine using the $\CNOT$ gate as follows: 
\begin{equation}\label{eq:clonable-state}
    \ket{\psi}\ket{0} \overset{\CNOT}{\longrightarrow} \ket{\psi}\ket{\psi}
\end{equation}

\begin{figure}[ht!]
\includegraphics[scale=0.35]{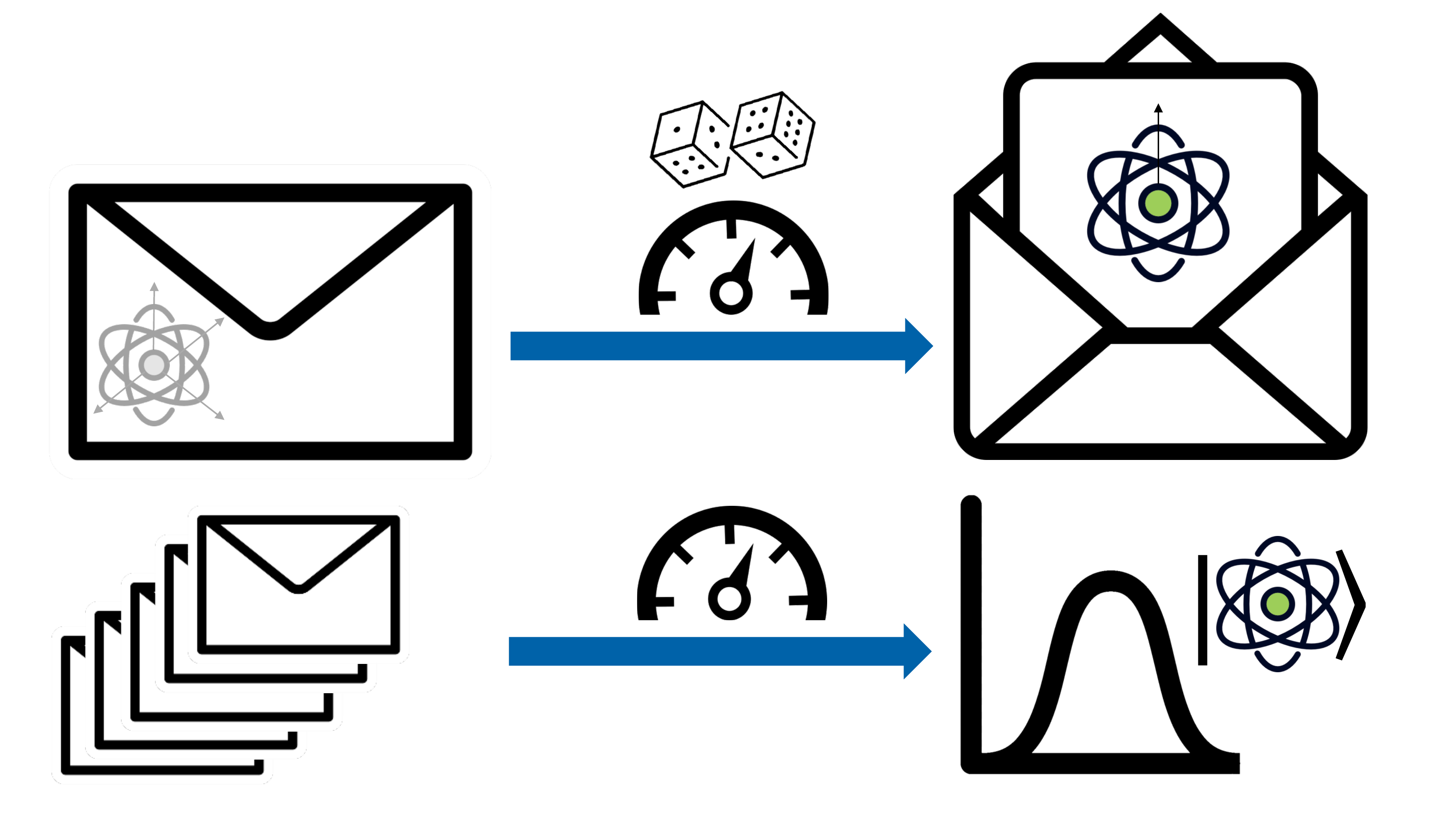}
\centering
\caption[Illustration of `unknownness' property of quantum states and measurement]{Illustration of `unknownness' property of quantum states and act of measurement as a probabilistic and destructive operation. Here the quantum object, can be seen as a closed envelope that contains information (for instance about its spin, polarization, etc.). While to know this information one needs to open the envelope (measurement) which leads to a probabilistic outcome, and the envelope cannot be closed again, illustrating the destructiveness of the measurement. Moreover having multiple copies of the quantum states allows for extracting the classical description of the state in terms of the statistics of the measurement outcomes.}
\label{fig:state-measurment}
\end{figure}

However, a set of orthogonal quantum states with a known basis, is not a piece of quantum information at all, but rather a classical one, even though carried by a quantum system. The reason is that one can always measure the quantum system on the given basis and \emph{deterministically} know the state of the system. We note that in both examples, we are still in the regime of quantum mechanics, where all the transformations between pure quantum states are unitary, and we are still talking about a quantum mechanical property of a \emph{physical} system, such as the spin of an electron. Therefore, it seems that unclonability, is an aspect of the \emph{quantum information} that we believe to exist in a quantum mechanical system, not the system itself, or in other words, the \emph{lack of information} about a quantum mechanical system, in the realm where we have uncertainty.

This `unknownness' for quantum states comes from the two most fundamental qualities of measurement in quantum mechanics: Any measurement is inherently \emph{probabilistic}, and it is usually \emph{destructive}. That is, a quantum system, carries some information about its own physical property \footnote{Whether this physical properties are `carried' by the system in reality, or even whether such states bare existential reality, is a matter of philosophical and scientific debates to date. Despite the author's personal affection for the subject, it is very far from the topic of this thesis, and hence we avoid entering that realm, and we note that the sentence has been merely used with an illustrative purpose. We refer enthusiastic readers to some interesting references \cite{hardy_are_2013,pusey_reality_2012,lewis_distinct_2012,brown_reality_2019,omnes_quantum_2002,adlam_quantum_2016}} (for instance, about its spin, polarization, etc.), while the physics would not allow us to know that information with certainty and without \emph{leaving a mark} on the object, that is in a non-reversible way. We illustrate this in \figref{fig:state-measurment}. On the other hand, imagine a world where perfect cloning of these unknown states is possible. In this world, an observer who owns such a cloning machine, could make many copies of the quantum state and start measuring them one by one. This observer does not need to worry about destrying the states via measurement since they can keep increasing their knowledge of the quantum system, to the point of certainty, only forming a single copy\footnote{However, we need to emphasise that this does not mean that the uncertainty which exists due to quantum measurements disappears by having the classical deception of a quantum state}. Therefore, a perfect cloner would provide a \emph{free source of information} for a quantum system, which is nonphysical.

Finally, we emphasise that as the amount of a priori information about the quantum state increases, cloning becomes more and more feasible. We recall the concept of approximate cloning that we have introduced in Section~\ref{sec:prelim-cloning}. As we have seen, restricting the family of states to be cloned to a specific family, and therefore have \emph{partial prior information} about the state (for instance, phase-covariant or state-dependent cloning as opposed to universal cloning), lead to cloning machines that produce clones with higher optimal fidelity. We will go back to this point in \chapref{chap:varqlone} where we use cloning machines for cryptanalysis. 

As we move along this thesis, we aim to show that the unknownness that is profoundly connected to the unclonability of quantum states, manifests itself in other forms of unclonability. 

\subsection{From unclonability of quantum states to unclonability of transformations}\label{sec:unclone-states-unitaries}
Now, we discuss another notion of quantum unclonability which is the unclonability of quantum transformations. Here, by quantum transformation, we mean either a unitary transformation or more generally, a CPTP map. First, we need to clarify what it means to clone a quantum process.

There are several ways to capture the unclonability of quantum transformations. The first one is as defined by Chiribella et. al. in \cite{chiribella_optimal_2008} where cloning a transformation $\Lambda$ means exploiting a single use of $\Lambda$ inside a quantum circuit, to perform the transformation $\Lambda\otimes\Lambda$ on bipartite states. In this context, a more general no-cloning theorem exists which subsumes the no-cloning of quantum states as a special case: Two black boxes $\mathcal{O}_1$ and $\mathcal{O}_2$ cannot be perfectly cloned by a single-use, unless they are the same, or they are perfectly distinguishable. Here the cloning is intertwined with the task of discrimination between two black boxes via a single-use and is characterised by the minimum of the worst-case error probability for discrimination.\footnote{The clonable cases then are when this discrimination probability is $p=0$ (perfect discrimination) or $p=\frac{1}{2}$ (no discrimination).} An interesting consequence of this theorem is that not only quantum black-box operations, like unitary gates, are unclonable by a single use, but also classical transformations such as permutations of classical registers, are also unclonable by this definition. Let us highlight two key ingredients in this notion of unclonability: \emph{black-box} and \emph{single-use}. One can easily relate the `black-box' to the concept of `unknownness' that we have been discussing so far. But the notion of `single-use' here reveals a yet more interesting fact about the unclonability of quantum transformations, which shows an elementary relation between quantum cloning and quantum learning, as has been also noted in the paper. The reason is that the approximate version of the cloner that clones black boxes is allowed to slightly learn the transformation (up to a single interaction). We will elaborate on this point further and try to portray this connection more clearly in Section \ref{sec:unclone-learning}.

Another way one may look at the unclonability of quantum transformations is as a pre-processing step for the unclonability of quantum states. Assume we have again a black-box or unknown unitary $U$, which we use as a generator for an unknown state $\ket{\psi}$ \emph{i.e.} we always input a known state (take state $\ket{0}$ as an example) and receive a state $\ket{\psi} = U\ket{0}$. Now, since the unitary is fully unknown, the output state is also unknown, therefore unclonable. Thus the unclonability of quantum states can also be viewed as the unclonability of its unknown generator. Nonetheless, in this case too, if the unitary is used repeatedly, the generated copies could be used to learn the state, as well as the unitary itself. Generalising this notion, we can think of the task of cloning a quantum transformation, as a general operation while having multiple time access to a fully unknown unitary, which can generate similar outputs of that unitary on different states (\emph{e.g.}: on the full or partial set of bases). We pursue the understanding of this more general notion of unclonability, and its relation to cryptography.

Our first contribution to this end is to formalise the concept of unknown or black-box unitary in a way that would best suit our purpose. We introduce the notion of \emph{Unknown Unitary (UU)}, which we will use throughout the thesis. Intuitively, an unknown unitary is a unitary that we have no prior information about prior to any interactions with it, \emph{i.e.} querying it. However, we formalise this `lack of prior information as a distinguishability problem. Let us elaborate on this: What we mean by knowing nothing about a unitary, in fact, means that the unitary can be any unitary matrix from the family of \emph{all possible unitaries} that can operate over a Hilbert space (or the associated linear operator space) of a certain dimension. In other words, if this set was finite, the probability that we could guess which unitary has been selected from the set would be the completely random guessing probability over the uniform distribution, which would depend on the cardinality of the set. However, this set is infinite, but as we have seen in Section \ref{sec:prelim-haar-random}, the Haar measure, can describe a uniform measure of randomness over the space of quantum states and unitary transformations. Therefore, we can define the `unknownness' of a unitary in terms of distinguishability from the Haar measure, as follows:

\begin{defbox}
\begin{definition}[Unknown Unitary (UU)]\label{def:unk-uni} Let $U^u$ be a family of unitary transformation where each $U \in U^u$ is a unitary over a $D$-dimensional Hilbert space, $\HilD$. Also, let $\lambda$ be a parameter related to $D$.\footnote{In cryptography, since this parameter is usually related to the security, it is called the security parameter.} We say $U^u$ is a family of Unknown Unitaries (UU), if for all the quantum polynomial-time algorithms $\A$, there exists a negligible function $\epsilon(\lambda) = \negl(\lambda)$ such that the difference between the probability of estimating the output of any $U \in U^u$ on any randomly picked state $\ket{\psi}\in\HilD$ and the probability of estimating the output of a Haar random unitary operator on the same state is bounded by $\epsilon(\lambda)$:
\begin{equation}
\small
        \underset{\ket{\psi}}{\mathbb{E}}\left [\left|\underset{U \leftarrow U^u}{Pr}[F(\A(\ket{\psi}),U\ket{\psi}) \geq \delta(\lambda)] - \underset{U_{\mu} \leftarrow \mu}{Pr}[F(\A(\ket{\psi}),U_{\mu}\ket{\psi})
        \geq \delta(\lambda)] \right|\right ] \leq \epsilon(\lambda).
\end{equation}
where $\mu$ denotes the Haar measure, $F$ denotes fidelity, $\delta(\lambda) = \nonnegl(\lambda)$ is non-negligible value in $\lambda$.
\end{definition}
\end{defbox}

In Section~\ref{sec:prelim-quanutm-pseudorandom}, we introduced the notion of quantum pseudorandomness and particularly pseudorandom quantum unitaries (PRUs). The definition of UU is, in its essence, very similar to that of PRUs. The reason is that, as discussed above, we have intuitively defined the unknownness with respect to \emph{perfect randomness} in the quantum world, which is the Haar measure. Quantum pseudorandomness is an \emph{approximation} of Haar measure or, in another perspective, a computational version of Haar measure. Additionally, we note that UU is a weaker notion, than PRU and can be considered as a \emph{single-shot pseudorandomness}. We further explore this beautiful relationship between randomness/pseudorandomness, unknownness and unclonability in \chapref{chap:pr-connection}. As a final remark to conclude this section, we note that the relationship between unclonability and pseudorandomness has been discussed in \cite{shacham_pseudorandom_2018} as well, where the authors have demonstrated a cryptographic variant of the no-cloning theorem. 

\section{Unclonability and different notions of learning}\label{sec:unclone-learning}
In the previous section, we discussed the intuitive connection between unclonability and the lack of knowledge about a quantum mechanical object such as a quantum state or a quantum transformation, which is what we referred to as \emph{unknownness}. As anything `unknown' can eventually become `known' through the process of interacting and learning,\footnote{Although this may sound like a philosophical statement, it is also a scientific one! From a physics point of view, it simply refers to the fact that any quantum operation can be learned asymptotically.} our next favourite concept to study in this chapter would be `learning'. Here, we look at learning from two perspectives: learning theory and cryptography. We seek to find similarities in what can be called learning in these two different views, which will help us better connect them for the rest of the chapter and the following chapters.

\subsection{Learning, forging and emulation}\label{sec:unclone-different-learning}
First, we start with learning. The term \emph{learning} can correspond to numerous different definitions and meanings in different contexts covering purely fundamental to entirely practical spectrum. We are interested in learning in its most theoretical and fundamental sense, as it is often studied in learning theory. Therefore we consider learning as the operation of an algorithm (either deterministic or probabilistic) that targets to learn/predict an object (a function, quantum state, quantum process, distribution, etc.) from a given set of data, usually known as \emph{learning data}, or sometimes \emph{training set}, that includes information about that object. Moving into the quantum realm, both the learning algorithm and the sample data can be quantum, which has given rise to the massive field of quantum learning theory and quantum machine learning. Our principal interest lies in quantum objects such as quantum states and quantum operations.\footnote{However, classical functions can also be represented via a quantum unitary (Section~\ref{sec:prelim-oracles}). Thus, they too are subjects of our investigation.} We recall from Section~\ref{sec:prelim-learning-theory} that the main question regarding learning is \emph{How efficiently can we learn the object?} Where the question is usually answered in time complexity or sample complexity. In particular, the separation between classical and quantum algorithms in terms of the two above efficiency factors stands among the most challenging problems in the field.

Given this very general description of the learning task, the first notion of learning that we discuss is \emph{learning an unknown quantum state}, which usually means learning the \emph{classical description} or similar properties and characteristics of that state. Here the learning data is multiple physical copies of the same unknown state. The learning algorithm measures these copies and then post-processes the measurement outcomes to extract the classical description or the respective property.\footnote{This process is also known as \emph{state tomography} \cite{nielsen_quantum_2010,bisio_optimal_2009}, discussed in Section~\ref{sec:prelim-tomography}.} In general, extracting a complete classical description with arbitrary precision requires an exponential number of samples. However, Huang \emph{et.al.} introduced a technique known as \emph{classical shadow}~\cite{huang_predicting_2020}, which allows the efficient learning of many properties of a quantum state. Another note worth mentioning here is that there is a well-known result in quantum information showing an equivalence between optimal universal $N \rightarrow \infty$ quantum cloning of pure states and optimal state estimation devices taking as input $N$ copies of an unknown pure state \cite{scarani_quantum_2005,gisin_optimal_1997}. More importantly, bounds on optimal cloning can be derived from this equivalence. This result establishes the deep connection between unclonability and unlearnability of quantum states.

The next notion is perhaps the most well-studied concept in learning theory: function learnability. We recall from Section~\ref{sec:prelim-learning-theory}, the scenario where given a family of functions, $\F$, we want to efficiently learn a representation of any function, $f \in \F$. The task of learning $f$ can be considered to be \emph{exact} or \emph{approximate} where the latter, the function is learned approximately but with high probability, which is denoted as PAC-learning (\defref{def:prelim-pac-learn}). In this learning model, the sample data is sample pairs $(x,f(x))$ where $x$ has been sampled according to some distribution $\mathcal{D}$, given to the learning algorithm via an oracle, which outputs one such pair on each call. In the quantum version of these learning notions, one can also allow the oracle to give quantum access to the underlying classical function, producing superposition queries.

Next, assume a weaker notion of learning, where we do not expect the learner to be able to learn \emph{all} the functions in a family (or a concept class), but instead given a function $f$, selected from a family $\F$, the learning algorithm needs to output a correct new pair $(x,h(x))$, from the set of inputs and outputs of $f$ (which can also be given via interacting with an oracle). The validity of the pair is tested with a verification algorithm, which in many cases (but not all the times), checks whether $h(x) = f(x)$ or not. This scenario is a very well-known scenario in cryptography, known as \emph{forgery}. In fact, in cryptography, one would ideally want to avoid all such cases where an efficient forger can exist for a function $f$. More precisely, in classical cryptography, the function $f$ is usually selected from a keyed-family of functions (where the key is sampled uniformly at random) and the input $x$ can be either selected from a distribution or by the forger itself arbitrarily. Here to ensure the security of a cryptographic scheme which employs function $f$, we require that no adversary is able to produce such valid pairs. In other words, the function $f$ should be not easily learnable/predictable from a limited set of samples. Here the adversary, which is a probabilistic algorithm, runs an efficient learning process for this specific instance of a learning problem that we call forgery. In Section~\ref{sec:unf-quantum-oracles} we will further discuss this relationship and the specifications of the oracles.

Finally, learning a quantum transformation is yet another notion of learning, that shares similarities with the above. The most conventional notion is known as \emph{quantum process tomography} in the literature. In process tomography, given an unknown unitary or quantum channel, we are interested in learning the classical description, or characteristics of the quantum process by interacting with it with many quantum states and measuring the outcomes. It is not hard to see that without any prior information about the unitary or channel, this task is highly inefficient~\cite{nielsen_quantum_2010,mohseni_quantum-process_2008}, given the fact that even learning a good approximation of an unknown quantum state is quite resource-intensive. Nevertheless, this is not the only notion of learning one can imagine for quantum processes. In Section~\ref{sec:emulation-simulation}, we introduced the notion of \emph{emulation} and how it differs from \emph{simulation}. As the simulation is probably closer to the concept of tomography (and even PAC-learning to some degree), we argue that emulation is very close to the notion of forging, where here, the emulation will forge an unknown unitary instead of a classical function. We recall that an emulation algorithm aims to produce a close approximation of the output of a unitary $U$, on a given state $\ket{\psi}$, from a set of input and output samples, which is similar to the forgery scenario presented above.

We keep these intuitions in mind as we move forward, and we focus on some of these notions of learning such as emulation (Section~\ref{sec:unf-qe-revisited}).

\subsection{Unforgeability and unclonability}\label{sec:unclone-unforge}
Here, we look at the relation between unforgeability and unclonability. These two properties become particularly related for \emph{quantum tokens} like quantum money \cite{wiesner_conjugate_1983,aaronson_quantum_2012}, quantum coin \cite{gavinsky_quantum_2012}, and more generally what is known as \emph{unforgeable tokens}. These are unique objects which can be produced and verified by an honest party, but no untrusted party can generate such valid tokens. Considering that the generators of such objects are classical functions or quantum processes, the role of the adversary would be to \emph{forge} or learn them, as discussed in the previous section, and in the cryptography world, resisting forgery attacks is captured by the property of \emph{unforgeability}. Despite its simple intuitive meaning, unforgeability is not very easy and trivial to capture formally, especially in the quantum world. In Section~\ref{sec:prelim-unf}, we discussed different classical definitions of unforgeability and some candidates for unforgeability definitions in the quantum world. We will also present our formal framework for quantum unforgeability later, in this chapter. But for now, let us see how this property is related to unclonability. 

Let us look at it through a very simple example regarding quantum money. Assume a bank (or the mint) producing some notes, \emph{i.e.} the physical objects used as money. Each note has a unique serial number attached to it which provides a basis for the verification of the note when the user wants to use it for a transaction~\cite{veriqloud_quantum_2019}. These notes are distributed among untrusted users, who are willing to create more notes than they originally had in their possession. Thus unforgeability is an important property for a note, meaning that the dishonest party cannot come up with another unique serial number that would be also valid, and hence pass the verification. But on the other hand, an easy way of creating more notes is to simply copy the whole note, since the new one will also have a valid serial number. As a result, cloning a token is a simple but applicable forging attack.

In the classical world, nothing prevents a user with sufficient resources to forge the note physically. However, in 1983, Wiesner first introduced the idea of quantum money, based on the no-cloning theorem \cite{wiesner_conjugate_1983}. Assuming your token to be a physical quantum system described by the quantum state $\ket{\$_{\texttt{serial}}}$ with the serial number $\texttt{serial}$, the unclonability of these states leads to the unforgeability of the quantum money scheme. This unforgeability stems from quantum mechanics, without any extra assumption. On the contrary, if a perfect quantum cloning machine could exist, no such quantum systems could satisfy unforgeability, irrespective of the scheme. 

Wiesner's quantum money is not the only cryptographic primitives where unforgeability and unclonability are related. Another example is a public-key quantum money scheme known as \emph{Quantum Lightning} \cite{zhandry_quantum_2021}. Similar to Wiesner's quantum money, quantum lightning also relies on the unclonability of quantum states to prevent forgery and duplication attacks. This scheme uses a one-time signature, and as discussed in the preliminaries, the main security property of a signature scheme is unforgeability. Moreover, the public-key quantum money functionality has often been seen as a computational or cryptographic variant of the no-cloning theorem.

Another relevant cryptographic functionality that is worth mentioning in this section, is \emph{Quantum copy-protection}, initially proposed by Aaronson \cite{aaronson_quantum_2009}, and that has been developed throughout several works recently~\cite{ananth_secure_2021,aaronson_new_2021,broadbent_secure_2021,sattath_uncloneable_2022}. The idea behind quantum copy-protection is to use quantum unclonability to achieve programs that cannot be copied. This functionality has many interesting applications such as \emph{Secure software leasing} \cite{ananth_secure_2021,broadbent_secure_2021}. Assume a program $f$ is given not as a classical function but instead, in the form of a quantum state $\ket{\psi_f}$, such that from this state, $f$ can be computed on any arbitrary input, yet it is infeasible to copy it or convert it into two other arbitrary states from which you can still compute $f$. Despite the clear connection between this functionality and the unclonability of quantum states, quantum copy protection is a stronger and more demanding requirement than simple unclonability. Also, quantum copy-protection does not directly link with unforgeability to the best of our knowledge. However, we argue that it is related to another concept that we have been discussing in this section, \emph{i.e.} \emph{learnability}. Aaronson shows that any `learnable' program can be copy-protected. Here learnability means that the output of the function cannot be predicted from input and output behaviour, which is very similar to unforgeability. Thus it would not be surprising that `forgeable' functions could also not be copy-protected. Nonetheless, as there are different definitions and levels of unforgeability, there are also several definitions for learnability in this context. Some of them have been introduced in~\cite{aaronson_new_2021}. More surprisingly, a recent result \cite{ananth_secure_2021} shows, that under certain computational assumptions, even certain unlearnable programs cannot be copy-protected. Therefore, the connection between learnability (as well as unforgeability) and copy-protection is still an intriguing open problem. We conjecture that the categorisation of copy-protectable functions from an unforgeability point of view can also be insightful. 

Finally, we note that there are plenty of other cryptographic primitives of this sort, where unclonability is a core aspect which enables features that are infeasible classically. However, the careful study of every one of them, including the above examples, is a research field on its own and indeed, outside the scope of this thesis. Our goal in this section was to sketch the existence of this relation via some well-known examples to be able to highlight it later in the other forms of unclonability, such as the one we present in \chapref{chap:qpuf}. 

\subsection{Unforgeability and learnability with quantum oracles}\label{sec:unf-quantum-oracles}
As we have discussed, unforgeability has a close and intuitive connection with the notion of learning. On the other hand, unforgeability is a cryptographic property we require for many different primitives and cryptographic schemes. We can therefore ask: 
\begin{center}
    \emph{What does it mean to learn a cryptographic primitive?}
\end{center}

First, we note that the core element of a cryptographic primitive is a classical function (or equivalently a quantum process in the quantum case) which maps the domain to the range, where the specific properties of this function (or operation) often are employed to achieve different cryptographic functionalities. In this context, learning a primitive means either learning the full underlying function (or quantum process) or alternatively learning the desired property. This learning process as we discussed in Section~\ref{sec:unclone-different-learning}, can be used by an adversary who tries to break the cryptosystem. For this learning algorithm to be able to function, one needs the \emph{learning data}. In the cryptography literature, similar to the learning theory, this learning data is usually obtained by interacting with an oracle that models an interactive platform for the full and perfect implementation of the evaluation function.

Here, we are mainly interested in quantum algorithms (and hence quantum adversaries) and the notion of quantum unclonability. Thus the relevant case for us is when the interaction with the primitive is also realised quantumly. For quantum primitives, this is a natural model to consider, as we are dealing with a quantum process that produces quantum outputs and often takes quantum inputs as well. For classical primitives on the other hand, according to the different adversarial models that we have introduced in Section~\ref{sec:prelim-adversarial-models}, this brings us to the \emph{quantum security regime}, where the interaction with the primitive is also considered to be quantum. 

In the preliminaries (Section~\ref{sec:prelim-oracles}), we have already discussed how a quantum accessible oracle can be defined for a classical function. Here, we recall that notion and we also present the same model for quantum primitives. In this way, we can study both classes of primitives in an analogous model, \emph{i.e.} the quantum oracle model, which we will use throughout the chapter and the thesis. We also discuss how learning from these quantum oracles, can essentially link to unforgeability or other similar cryptographic properties of interest. 

\subsubsection{Quantum oracle for classical vs quantum primitives}
Let us first recall the \emph{standard quantum oracle} for classical primitives which we introduced in Section~\ref{sec:prelim-oracles}. The standard oracle is a black-box unitary of a reversible version of a classical-polynomial-time computable function $f$, which can represent a deterministic or randomised primitive, defined as follows:
\begin{equation}
    R\Ora_f: \sum_{m,y} \alpha_{m,y} \ket{r}_{\Ora}\ket{m,y} \rightarrow \sum_{m,y} \alpha_{m,y}\ket{r}_{\Ora}\ket{m, y \oplus f(m;r)}
\end{equation}

Having access to this quantum operation, a quantum adversary $\A$ can get the outcome of the function $f$ for several classical inputs $m$ in one query, in a superposition form over the message set of their choice. However since the primitive can be randomised, the value of the function in each execution can also depend on the random value. This scenario is modelled via the randomised version of the oracle or the randomised oracle as shown above. In the case of deterministic primitives, the adversary gets full query access to the unitary that maps all the computational basis to another combination of computational basis that represents the range of the function. In the randomised case, on the other hand, the adversary usually gets a more \emph{limited} access to an extended unitary transformation over the joint Hilbert space of the function and the randomisation, or in other words, to the full Hilbert space of all the possible outcomes of all the functions, parameterised with the random value. In Section~\ref{sec:prelim-oracles} we have briefly discussed some of the interesting cases and questions that arise regarding which level of access to the extended unitary can be permitted for the quantum adversary. Nonetheless, the key point that we want to emphasise is that the information about the primitives is accessible to the adversary via the \emph{quantum} input and outputs of this unitary, which is, initially unknown to the adversary. 

For quantum primitives, the modelling of this scenario is even more evident. Quantum schemes often work with families of quantum states which are usually unknown (or partially unknown) quantum states from the adversary's point of view, or similarly, an unknown quantum operation. In both cases, one can model the primitive with the evaluation between quantum states and thus can define the oracle as a general unitary transformation for a deterministic primitive, as follows:
\begin{equation}\label{eq:qoracle-quantum-det}
    \Ora_U: \sum_{i} \alpha_{i} \ket{b_i} \overset{U}{\rightarrow} \sum_{i} \beta_{i}\ket{b_i}
\end{equation}
Here $\{\ket{b_i}\}$ are a basis (not necessary computational basis) for $\HilD$, the Hilbert space that the unitary operates upon. We note that quantum primitives can perform an arbitrary rotation of the bases. The analogue of this type of oracles for classical primitives, are \emph{type-2} oracles (also called \emph{minimal oracles})\cite{gagliardoni_semantic_2016,gagliardoni_quantum_2021}. We also note that in the non-randomised case, this oracle generalises the standard quantum oracle. In other words, the class of all the possible standard oracles of a certain dimension are a subclass of all the quantum oracles of the form \eqref{eq:qoracle-quantum-det} over the same dimension. 

A randomised quantum primitive can also be defined similarly to the classical case. Here we give an abstract notation of a general randomised quantum primitive, but we further clarify the realisation of such oracles in the upcoming sections. We denote a general randomised unitary oracle for quantum primitives as follows:
\begin{equation}\label{qoracle-quantum-rand}
\Ora_U: \sum_{i} \alpha_{i} \ket{r}_{\Ora} \ket{m_i} \overset{U}{\rightarrow} \sum_{i} \beta_{i}(r)\ket{r}_{\Ora}\ket{m_i}
\end{equation}
Hence a $\Ora_U$ is a unitary over the joint space of the oracle's randomness register and the main input state, which consists of a family of smaller unitaries parameterised by a random internal parameter $r$.

Now, back to the problem of learning, we can see that in both cases, the learning data is a set of input and output quantum states $\{(\rho^{in},\rho^{out})_i\}^q_{i=1}$\footnote{In general, these output states can be entangled across different queries. In some cases, where the query to the oracle was performed adaptively and sequentially, the number of quantum states and the mathematical structure of this quantum learning data might be different. However, we will argue that this generality can still hold. For additional technical discussion, see Section~\ref{sec:unf-game}} of an unknown unitary that is provided to the learning algorithm from interacting $q$-times with the respective quantum oracle. Let us point out a few aspects of this analogy between learning and cryptographic problems in the oracle model. 

First, depending on whether the access to the oracle is direct or indirect via another agent or honest party, different quantum adversarial models emerge. From another perspective, the learning data can, in general, be sampled from a distribution, which can also impose restrictions or relax conditions on the learning problem. Second, the number of queries $q$ (either in the worst case or average case) specifies the \emph{query complexity} of the learning algorithm but also translates to the power of the quantum adversary in the cryptography language. Third, it is clear that in both of these models, any such unknown unitaries are eventually `learnable' if the number of queries to the oracle is unbounded. The same situation in cryptography is commonly known as \emph{brute-force attacks}. The lower the query complexity, the higher the efficiency of the algorithm and the more they become interesting attacks on cryptosystems. Therefore, at the intersection of cryptography and learning theory, we are usually interested in learning algorithms with polynomial \footnote{In a given parameter that either quantifies the resource, or the security, or in some cases both. For instance, the number of input qubits of the algorithm.} query complexity. Many of the learning algorithms, such as process tomography~\cite{dariano_quantum_2001,paris_quantum_2004,hayashi_asymptotic_2005,kaznady_numerical_2009,bisio_optimal_2009,odonnell_efficient_2016}, or certain quantum machine learning algorithms~\cite{biamonte_quantum_2017,wang_quantum_2021,arunachalam_optimal_2017,padakandla_pac_2022} do not have such polynomial-size query complexity and thus despite being very useful in other areas, are not usually compelling toolkits for cryptanalysis. However, in recent years, there has been significant progress in the development of \emph{efficient} algorithms for learning quantum states and quantum processes \cite{marvian_universal_2016,aaronson_shadow_2020,huang_predicting_2020}. We believe these algorithms are powerful yet fairly unexplored tools while studying problems in the domain of cryptography. An example of this sort of learning algorithm is a technique called \emph{shadow tomography} introduced by Aaronson~\cite{aaronson_shadow_2020}, which has also been studied in the context of quantum money.

In what follows, we study one candidate of such efficient learning algorithms. The algorithm that we study is an algorithm for quantum emulation which aims to generate the output of an unknown unitary to an unknown quantum state from learning samples of that unitary. We have sought to uncover the relevance of such methods in relation to unclonability and the cryptographic properties of unclonable objects. From the next section on, we use these two main ingredients to understand unclonability in a broader context: \emph{quantum emulation} as a learning tool and \emph{unforgeability} as a cryptographic characterisation. In order to make our intuitive arguments more precise, we will need to have a closer look at both of them.

\section{Universal quantum emulator revisited}\label{sec:unf-qe-revisited}
In the previous section, we discussed different notions of \emph{learning} and their relation to unclonability, as well as cryptographic concepts like unforgeability. We have also seen that emulating an unknown quantum transformation is a way of \emph{learning} with close proximity to \emph{forging} that process. Previously in Section~\ref{sec:prelim-qe} of \chapref{chap:prelim}, we have seen a quantum algorithm from \cite{marvian_universal_2016} that performs the task of emulating an unknown quantum unitary on an unknown input quantum state by having some of the input-output samples of the unitary. We have seen some properties of the universal quantum emulator algorithm, such as efficiency and complexity results. Re-purposing the algorithm from its original target in the context of tomography, here we take a new look at this algorithm, and more generally emulation, from a cryptanalysis perspective. In other words, since it seems the objectives of emulation and unforgeability are in the complete opposite direction of each other, we propose emulation as a general attack model against the notion of unforgeability in general. To make this statement more formal, we will first require a formal definition of \emph{unforgeability} itself, for which we need to wait until the next section. While in this section, we revisit the quantum emulation algorithm from an adversarial point of view and as a cryptanalysis toolkit. For this purpose, we need to provide a new fidelity analysis of the algorithm exploiting a specific asymmetry in the algorithm that can be used effectively in adversarial scenarios. We then show a few examples of this new class of quantum attacks, which we call \emph{quantum emulation attacks}. 

\subsection{Output fidelity analysis}\label{sec:unf-qe-fid}
We are interested in the fidelity of the output state $\ket{\psi_{QE}}$ of the algorithm and the intended output $\U\ket{\psi}$ to estimate the success. Here we are more interested in the explicit form of the output states and the theoretical bounds for the fidelity, rather than the complexity analysis provided in \cite{marvian_universal_2016}. We note that for our calculations, all the gates including the controlled-reflection gates are assumed to be ideal keeping in mind that the implementation is possible with the technique of \emph{quantum principal component analysis} developed in \cite{lloyd_quantum_2014}, as also mentioned in \cite{marvian_universal_2016}. We recall from Section~\ref{sec:prelim-qe} that the fidelity of the output is related to the success probability of the first stage of the algorithm in the following way: 

\begin{equation}\label{eq:qunftools-qe-fidel}
    F(\rho_{QE}, \U\rho\U^{\dagger}) \geq F(\E_{\U}(\rho), \U\rho\U^{\dagger}) \geq \sqrt{P_{succ-stage1}}
\end{equation}

Also, from the proof of \thmref{th:qe-fidel} from \cite{marvian_universal_2016}, it can be seen that the success probability of Stage 1 is calculated as follows:
\begin{equation}\label{eq:qunftools-qe-ps}
P_{succ-stage1} = |\bra{\phi_r}\tr_{anc}(\ket{\chi_f}\bra{\chi_f})\ket{\phi_r}|^2
\end{equation}

\noindent where $\ket{\chi_f}$ is the final state of the circuit after Stage 1 and $\tr_{anc}(\cdot)$ computes the reduced density matrix by tracing out the ancillas. The overlap of the resulting state and the reference state equals the success probability of Stage 1. Now we only use \eqref{eq:qunftools-qe-ps} for our analysis henceforward. For this section, we need a more precise and concrete expression for the output fidelity.

Here we point an important observation about the algorithm. The fidelity of the output state of the circuit highly depends on the choice of the reference state \eqref{eq:qunftools-qe-ps} such that it may increase or decrease the success probability of the adversary in different security models as we will discuss. We establish the following recursive relation for the state of the circuit after the $i$-th block of Stage 1, in terms of the previous state:
\begin{equation} \label{eq:qe-recur}
        \ket{\chi_i} = \frac{1}{2}[(I - R(\phi_r))\ket{\chi_{i-1}}\ket{0} + R(\phi_i)(\mathbb{I} + R(\phi_r))\ket{\chi_{i-1}}\ket{1}].
\end{equation}

Now by using this relation, we can prove the following theorem:
\begin{thmbox}
\begin{theorem}\label{th:qe-fins}
Let $\ket{\chi_K}$ be the output state of $K$-th block of the circuit  (\figref{fig:qe}). Let $\ket{\psi}$ be the input state of the circuit, $\ket{\phi_r}$ the reference state and $\ket{\phi_i}$ other sample states. We have:
\begin{equation}\label{eq:qe-fins}
\begin{split}
    \ket{\chi_K} &= \mbraket{\phi_r}{\psi} \ket{\phi_r}\ket{0}^{\otimes K} + \ket{\psi}\ket{1}^{\otimes K} - \mbraket{\phi_r}{\psi} \ket{\phi_r}\ket{1}^{\otimes K} \\
    & + \sum^{K}_{i=1}\sum^{i}_{j=0} [f_{ij} 2^{l_{ij}} |\mbraket{\phi_r}{\psi}|^{x_{ij}} |\mbraket{\phi_i}{\psi}|^{y_{ij}} |\mbraket{\phi_r}{\phi_i}|^{z_{ij}}]\ket{\phi_r}\ket{q_{anc}(i,j)} \\
    & + \sum^{K}_{i=1}\sum^{i}_{j=0} [g_{ij} 2^{l'_{ij}} |\mbraket{\phi_r}{\psi}|^{x'_{ij}} |\mbraket{\phi_i}{\psi}|^{y'_{ij}} |\mbraket{\phi_r}{\phi_i}|^{z'_{ij}}]\ket{\phi_i}\ket{q'_{anc}(i,j)}
\end{split}
\end{equation} 
where $l_{ij}$, $x_{ij}$, $y_{ij}$, $z_{ij}$, $l'_{ij}$, $x'_{ij}$, $y'_{ij}$ and $z'_{ij}$ are integer values indicating the power of the terms of the coefficient. Note that $f_{ij}$ and $g_{ij}$ can be 0, 1 or -1 and $q_{anc}(i,j)$ and $q'_{anc}(i,j)$ output a computational basis of $K$ qubits (other than $\ket{0}^{\otimes K}$).
\end{theorem}
\end{thmbox}
\noindent We give an induction proof of this theorem in \appref{app:qe-final-state-proof}.

Having a precise expression for $\ket{\chi_f}$ from \thmref{th:qe-fins}, one can calculate $P_{succ-stage1}$ of \eqref{eq:qunftools-qe-ps} by tracing out all the ancillary systems from the density matrix of $\ket{\chi_f}\bra{\chi_f}$. Also, now it is clear that if $\ket\psi$ is orthogonal to $\Hild$, the only term remaining in \eqref{eq:qe-fins} is $\ket{\psi}\ket{1}^{\otimes K}$. So, the input state remains unchanged after the first stage and $P_{succ-step1}=0$. For states projected in the subspace spanned by $\Sin$, the overall channel describing the quantum emulation algorithm has always a fixed point inside the subspace \cite{marvian_universal_2016}. Hence, Stage 1 is successful with probability close to 1 by assuming the gates to be ideal.

Let us see two simple examples of the above theorem, which we will use in the future. First, assume that we have only two sample states: one reference state $\ket{\phi_r}$, and another sample state $\ket{\phi_1}$, and their respective output states. Trying to run a quantum emulation algorithm with this database, will lead to an emulation algorithm that has only one block in the first stage. In this case, the output state $\ket{\chi_1}$ of the circuit after the first stage is given as a function of the following overlaps,
\begin{equation}
    \mbraket{\phi_r}{\psi} = \alpha, \quad \mbraket{\phi_1}{\psi} = \beta, \quad \mbraket{\phi_r}{\phi_1} = \gamma,
\end{equation}
as follows:
\begin{equation}\label{eq:qe-one-block}
    \ket{\chi_1} = \alpha\ket{\phi_r}\ket{0} + \ket{\psi}\ket{1} - \alpha\ket{\phi_r}\ket{1} -2\beta \ket{\phi_1}\ket{1} + 2\alpha\gamma \ket{\phi_1}\ket{1}
\end{equation}

\noindent Expanding this two two-blocks emulation, using the same formula, we have:
\begin{equation}\label{eq:qe-two-block}
\begin{split}
    \ket{\chi_2} & = \alpha\ket{\phi_r}\ket{00} + 2\gamma_1(\alpha\gamma_1 - \beta_1)\ket{\phi_r}\ket{01} + \ket{\psi}\ket{11} \\
    & + (2\beta_1\gamma_1 - \alpha - 2\alpha\gamma_1^2)\ket{\phi_r}\ket{11} + 2(\alpha\gamma_1 - \beta_1)\ket{\phi_1}\ket{11} \\
    & + 2(\alpha\gamma_2 - \beta_2 + 2\beta_1\delta - 2\alpha\delta\gamma_1 - 2\alpha\beta_1\gamma_1\gamma_2 + 2\alpha\gamma_1^2\gamma_2)\ket{\phi_2}\ket{11}
\end{split}
\end{equation}
where the coefficients are given by the following pair overlaps:
\begin{equation}\label{eq:qe-two-block-2}
\begin{split}
    & \mbraket{\phi_r}{\psi} = \alpha, \quad ~\mbraket{\phi_1}{\psi} = \beta_1, \quad \mbraket{\phi_2}{\psi} = \beta_2\\
    & \mbraket{\phi_1}{\phi_r} = \gamma_1, \quad \mbraket{\phi_2}{\phi_r} = \gamma_2, \quad \mbraket{\phi_1}{\phi_2} = \delta .
\end{split}
\end{equation}

Although calculating these explicit forms seems repetitive, having them provides the ability to optimise the output fidelity with respect to the choice of states. We underline that for the initial purpose of the algorithm, this was not relevant as the sample states were assumed to be chosen randomly, while in some adversarial models, specifically when giving oracle access to the unknown target unitary, the adversary has the advantage of selecting the best possible states. Hence, having the explicit forms in terms of overlaps is a significant step towards using quantum emulation as a cryptanalysis toolkit.  

\subsection{Quantum Emulation Attacks}\label{sec:qe-attacks}
Now, let us see how we can use this algorithm with the new given picture and results in creative ways. The first quantum emulation attack that we present is the one that we will use several times in this thesis to show non-trivial impossibility results and also happens to be the simplest emulation attack. Here we present the most general format, without explicitly specifying the model or formal game in which it is used. Later in Section~\ref{sec:unf-results} and also \chapref{chap:qpuf} we will go back to this attack and use it in a formal way within game-based security models. The other two attacks we give in this section, mostly serve as toy examples to demonstrate the possibility of different attacks one can build given a small-size quantum emulator.

\subsubsection{One-block quantum emulation attack}\label{sec:unf-qe-one-block}
Imagine the scenario where an adversary $\A$ who has access to the following samples of an unknown unitary $U$:
\begin{equation}
    \{\ket{\phi_1},\ket{\phi_r}\}, \quad
    \{\ket{\phi^{out}_1},\ket{\phi^{out}_r}\}
\end{equation}
and tries to closely approximate the output $U\ket{\psi}$ for a target state $\ket{\psi}$. We assume the adversary $\A$, has the ability to select $\ket{\phi_1}$ and $\ket{\phi_r}$ and interact with the unitary to obtain the respective outcome. 

Without loss of generality assume the case that $\ket{\phi_1}$ is a computational basis and the target state $\ket{\psi}$ is another computational basis (or more generally any states such that $\mbraket{\psi}{\phi_1} = 0$). Now the question will be how to choose the best $\ket{\phi_r}$ for this emulation. From \thmref{th:qe-fins} and \eqref{eq:qe-fidel} we already know that the reference state should have some overlap with the target state for the emulation to be successful in this case. For finding a general result, we parameterise the choice of the reference state in the amplitudes of the reference state, as follows:

\begin{equation}
    \ket{\phi_r} = \sqrt{1-\alpha^2}\ket{\phi_1} + \alpha \ket{\psi}
\end{equation}

Now $\A$ can run a one-block quantum emulation algorithm. We can directly use \eqref{eq:qe-one-block}, noting that $\mbraket{\phi_1}{\psi} = 0$ and $|\mbraket{\psi}{\phi_r}|^2 = \alpha^2$ and $|\mbraket{\phi_1}{\phi_r}|^2 = 1 - \alpha^2$
\begin{equation}
    \ket{\chi_1} = \alpha \ket{\phi_r}\ket{0} + \ket{\psi}\ket{1} - \alpha\ket{\phi_r}\ket{1} +2\alpha(\sqrt{1-\alpha^2})\ket{\phi_1}\ket{1}.
\end{equation}
By calculating $\ket{\chi_1}\bra{\chi_1}$, tracing out the ancillary systems and using \thmref{th:qe-fidel}, we can bound the fidelity of the output state of the emulator, denoted by $\ket{\psi_{QE}}$, and the target which is $U\ket{\psi}$ as follows:

\begin{equation}\label{eq:qe-one-block-attack-fid}
    F(\ket{\psi_{QE}}\bra{\psi_{QE}}, U^{\dagger} \ket{\psi}\bra{\psi} \U) \geq \alpha^2[1+4(1-\alpha^2)^2]
\end{equation}

We can see that this fidelity, is a considerable value as long as $\alpha$ is not too small (in a cryptographic sense, the fidelity is a non-negligible function of the security parameter as long as $\alpha$ is not a negligible function). A trivial case is where fidelity is 1, for $\alpha = 1$, which means the reference state and the target state are the same. But there is also a non-trivial case for the fidelity to become unity, and that happens for $\alpha = \frac{1}{\sqrt{2}}$. This is when the reference state is a uniform superposition of the target state and the other sample. 

Thus, we can see that this freedom of carefully choosing the sample states of an unknown unitary, enables an adversary to perform a successful emulation with high fidelity. This is important in the context of cryptography as the unknown unitary can be a quantum oracle of a classical primitive, and the output of the emulation can leak important information about the underlying function. In fact, as we will see later, this sort of attack directly connects to the unforgeability property.


\subsubsection{Quantum emulation attack on the inverse function}\label{sec:unf-qe-attack-inverse-f}
Let us see another example. Assume that we have a classical function $f$ that is efficiently computable but hard or inefficient to invert (for instance a one-way function). However, we assume that the function is invertible \emph{i.e.} the $f^{-1}$ exists. Our goal is to show some information about the inverse of $f$ that can be extracted, with only having a standard quantum oracle access to $\Uf$, using a small quantum emulation attack.

Let the following unitary be the standard quantum oracle for classical function $f$:
\begin{equation}
    \Uf: \sum_{x,t}\alpha_{x,t}\ket{x, t} \rightarrow \sum_{x,t}\alpha_{x,t}\ket{x, f(x) \oplus t}
\end{equation}
Since $f$ is easy to compute the unitary $\Uf$ is also efficiently implemented since we have assumed the inefficiency of computing the inverse implies that in general, an efficient implementation of $\Ufi$ does not exists, or in other words, we can assume that $\Uf^{\dagger} \neq \Ufi$ and having access to $\Uf$ and its complex conjugate does not lead to a trivial implementation of $\Ufi$.
Nonetheless, the standard oracle form of $\Ufi$ takes the following form:
\begin{equation}
    \Ufi: \sum_{y,t}\alpha_{y,t}\ket{y, t} \rightarrow \sum_{y,t}\alpha_{y,t}\ket{y, \fm{y} \oplus t}
\end{equation}
where $y = f(x)$ and thus $\ket{y, \fm{y} \oplus t} = \ket{f(x), \fm{f(x)}=x \oplus t}$.

We show that an adversary $\A$ can emulate $\Ufi$ and extract information about $f^{-1}$ without having any oracle access to $\Ufi$, and by only querying $\Uf$\footnote{We note that this is a forgery attack on such functions, however it clearly does not break the one-wayness property since to break this property, one needs a pre-image of a random $y$ should be found.}.

First, we show the analysis of the emulation's output if the sample states from $\Ufi$ were available. Then we propose a method to translate the queries of $\Uf$ to the queries of $\Ufi$ which are required for the emulation attack.

\begin{description}
    \item[Sample states]
    $\A$ needs the following queries from the unknown unitary $\Ufi$:\\
    \begin{equation}
        \{\ket{\phi_1} = \ket{y_1,0},  \quad \ket{\phi_r} = \frac{1}{\sqrt{2}}(\ket{y_1,0} + \ket{y_k,0})\}
    \end{equation}
    \noindent and their respective outputs:
    \begin{equation}
    \begin{split}
        \{ & \ket{\phi_1^{out}} = \ket{y_1, \fm{y_1}} = \ket{f(x_1), x_1},\\
        & \ket{\phi_r^{out}} = \frac{1}{\sqrt{2}}(\ket{y_1, \fm{y_1}} + \ket{y_k, \fm{y_k}}) = \frac{1}{\sqrt{2}}(\ket{f(x_1), x_1} + \ket{f(x_k), x_k)}\}
    \end{split}
    \end{equation}
    
    where $y_1 = f(x_1)$ and $y_k = f(x_k)$ are classical outputs of the function $f$. Also note that $\ket{\phi_1}$, as well as $\ket{\phi_1^{out}}$, are a computational basis of $\HilD$ over which $\Uf$ and $\Ufi$ operate. Thus, this query is equivalent to a classical query.\\
    
    \item[Target state] Let the target state of emulation, and the expected outcome be $\ket{y_k, 0}$, and $\ket{y_k, x_k}$ respectively, where the second register $\ket{x_k}$ is the desired output of $\Ufi$ (or the pre-image of $y_k$).
\end{description}

\noindent Given the one-block emulation attack from the previous section, we know that the output fidelity for this case ($\alpha - \frac{1}{\sqrt{2}}$) is equal to 1. Therefore $\A$ could perfectly extract the $x_k$ with probability one if having access to $\Ufi$ oracle to be able to get the above samples. Now let us see how one can obtain the required samples, by interacting with $\Uf$ instead.\\

\noindent\textbf{Translating queries of $\Uf$ to queries of $\Ufi$:}\\

\noindent Let the following be the sample states of $\A$ after querying $\Uf$:
\begin{equation}
\begin{split}
& \ket{\phi^{u}_1}=\ket{x_1,0}, \quad  \ket{\phi^{u,out}_1}=\ket{x_1,f(x_1)} \\ 
& \ket{\phi^{u}_2}= \frac{1}{\sqrt{2}}(\ket{x_1,0} + \ket{x_k,0}), \quad \ket{\phi^{u,out}_2}=\frac{1}{\sqrt{2}}(\ket{x_1,f(x_1)} + \ket{x_k,f(x_k)})
\end{split}
\end{equation}

Our goal is to make the transformation in a non-destructive way, instead of simply measuring the states which would lead them to collapse. Obviously, $\ket{\phi_1}$ can be prepared classically and the output, $\ket{\phi^{out}_1}$ can be easily obtained from $\ket{\phi^{u,out}_1}$ by simply performing a SWAP gate between first and second part of the register \emph{i.e.} $\ket{\phi^{out}_1} = \sw_{12} \ket{\phi^{u,out}_1}$. Similarly, $\ket{\phi^{out}_r}$, can be obtained by swapping the first and second register of $\ket{\phi^{u,out}_2}$, \emph{i.e.} $\ket{\phi^{out}_{r}} = \sw_{12} \ket{\phi^{u,out}_2}$.

It remains to obtain $\ket{\phi_r}$ from $\ket{\phi^{u, out}_2}$ which is more complicated as there is an entanglement between the registers that need to be removed. We give a small sample algorithm, \algoref{alg:ufm-from-uf}, to perform this task for the qubit case, although it is easily generalizable to n-qubits as well.

\begin{algorithm}[ht!]
\SetAlgoLined
\textbf{Description:} Translating $\ket{\phi^{u,out}_2}$ to $\ket{\phi_r}$: We assume that $\ket{x_1}$, $\ket{x_k}$ and respectively $\ket{y_1}$ and $\ket{y_k}$ are qubit. If $x_1 \neq x_k$, then the state $\ket{\phi^{u,out}_2}$ is an entangled state which can be written as: $\ket{\phi^{u,out}_2} = \frac{1}{\sqrt{2}}(\ket{0,y_1} + \ket{1,y_k})$. The algorithm proceeds as follows:

\begin{itemize}
    \item Add an ancillary qubit $\ket{0}$, and perform a SWAP gate in the first qubit and the ancillary qubit leading to:
    \begin{equation*}
    \begin{split}
        \sw_{a1}\ket{0_a}\ket{\phi^{u,out}_{2}} & = \sw (\frac{1}{\sqrt{2}}(\ket{0}\ket{0}\ket{y_1}) + \ket{0}\ket{1}\ket{y_k})) \\
        & = \frac{1}{\sqrt{2}}(\ket{0}\ket{0}\ket{y_1}) + \ket{1}\ket{0}\ket{y_k})
    \end{split}
    \end{equation*}
    
    \item Rewrite the first qubit in $\ket{\pm}$ basis:
    \begin{equation*}
      \ket{+}(\frac{\ket{0}\ket{y_1}) + \ket{0}\ket{y_k}}{2}) + \ket{-}(\frac{\ket{0}\ket{y_1}) - \ket{0}\ket{y_k}}{2})
    \end{equation*}
    
    \item Measure the first qubit in $\ket{\pm}$ basis.
    \item One of the two outcomes $\frac{\ket{0}\ket{y_1}) + \ket{0}\ket{y_k}}{\sqrt{2}}$ or $\frac{\ket{0}\ket{y_1}) - \ket{0}\ket{y_k}}{\sqrt{2}}$ are outputed with probability $\frac{1}{2}$.
    
    \item Apply SWAP gate on two registers, leading to the following states: 
    \begin{equation*}
      \frac{1}{\sqrt{2}}(\ket{y_1}\ket{0}) + \ket{y_k}\ket{0}) \quad \text{or} \quad \frac{1}{\sqrt{2}}(\ket{y_1}\ket{0}) - \ket{y_k}\ket{0})
    \end{equation*}
\end{itemize}
 \caption{Mini sample converter algorithm}\label{alg:ufm-from-uf}
\end{algorithm}

Note that, even though the algorithm's outcome is probabilistic, both of the output states have the desired superposition form, albeit with different relative phases\footnote{There is a simple way, however to achieve a deterministic outcome. We can add an additional step to correct the output conditioned on the measurement outcome, as it is usually done in MBQC.}. This relative phase, will not affect the success probability of the emulation algorithm. In other words, emulation samples can be obtained deterministically, from the given states. This case is similar to the technique used in \cite{doosti_universal_2017} where superposition is created with the desired weight but with different phases.

Examples of this section, although simple, showcase the applicability of quantum algorithms, such as quantum emulation, that exploit the power of quantum data (quantum queries) as new attacks on cryptosystems. As shown through these toy examples, many such algorithms, when used in an adversarial scenario, can be adjusted to perform even stronger attacks, compared to the cases made for their original purposes, such as tomography or general learning. Hence the study of this class of algorithms from a cryptanalysis perspective bares outstanding importance in the field.


\section{A unified framework for quantum
unforgeability}\label{sec:unf-framework}
After establishing the intuitive relations between unclonability, unforgeability and learning, and introducing the emulation cryptanalysis toolkit, now it is time to formalize those intuitions in the form of a formal framework for unforgeability as a cryptographic property for quantum and classical primitives. 
The game-based security framework is a standard model for formally defining security properties of cryptographic primitives such as encryption algorithms, digital signature schemes, physical unclonable functions or quantum money~\cite{boneh_quantum-secure_2013,gagliardoni_semantic_2016,alagic_quantum-access-secure_2020,armknecht_towards_2016,soukharev_post-quantum_2016,aaronson_quantum_2012,aaronson_quantum_2009}. Classical cryptographic primitives have also widely been studied in a quantum game-based framework, where parties are quantum (are able to run quantum circuits)~\cite{boneh_quantum-secure_2013,gagliardoni_semantic_2016,alagic_quantum-access-secure_2020,soukharev_post-quantum_2016}. Inspired by these works, we generalise the quantum game-based framework to formalize quantum unforgeability, in a way that it is compatible with the notion of learnability and unclonability and is useful for the rest of the thesis. Additionally, our framework unifies different levels of unforgeability as well as capturing quantum and classical primitives. First, we show the abstract and formal version of the definition and then we show how it can naturally cater for quantum primitives and different adversarial levels.

\subsection{Framework and Formal definitions}\label{sec:unf-game}
Let $\F = (\ES, \E, \V)$ be a classical or quantum primitive with $\ES$, $\E$, and $\V$ being the setup, evaluation, and verification algorithms respectively. We specify unforgeability as a game between a challenger $\C$ (that models the honest parties) and an adversary $\A$ (that captures the corrupted parties). The adversary's goal is to \emph{closely approximate} the output of the evaluation algorithm $\E$ on a \emph{new challenge} such that it passes the verification with high probability. As we work in the quantum regime, we give adversary full quantum query access to the primitive, either classical or quantum. For classical primitives, as we assume the adversary has quantum oracle access to the primitive, we adopt the technique of quantum oracles defined in~\cite{boneh_quantum-secure_2013,boneh_random_2011} for formalizing quantum query-response interaction between the adversary and the challenger.

The security game considered here consists of several phases. 
First, $\C$ runs the setup algorithm $\mathcal{S}$ to generate the parameters required throughout the game and instantiates the evaluation oracle $\eO$, the verification oracle $\vO$, and the message space $\M$. The learning phase defines the threat model. For now, we only consider the quantum equivalent of the chosen-message attack model for coherence and simplicity. Nevertheless, we show the extension to other attack models and types of adversaries, such as weak adversaries, in Section~\ref{sec:unf-weak-adaptive-adv}. The challenge phase determines the security notion captured by the game. The formal specification of our quantum games is presented in \gameref{game:unf-full}. But let us first go informally over each phase of the game and clarify the differences between quantum and classical primitives.

\begin{itemize}[leftmargin=*,label={}]
\item{\bf Setup.} In the setup phase, $\C$ generates the parameters required in subsequent phases by running the setup algorithm of the primitive $\F$ on input $\lambda$ (the security parameter), and the oracles are being instantiated accordingly.\\
\noindent\textit{Quantum case:} For quantum primitives, the evaluation oracle is defined according to \eqref{qoracle-quantum-rand} for deterministic and randomised primitives respectively and the verification oracle implements a quantum test algorithm as defined in \defref{def:test}.

\item{\bf Learning phase.}
In the learning phase, the adversary interacts with the evaluation oracle. For now, we only focus on chosen-message attack (cma) security. $\A$ requires the oracle evaluation on any input state $\rho^{in}_i$. The oracle evaluations are handled by $\C$ who issues the requests on $\rho^{in}_i$ to $\eO$ and forwards to $\A$ the respectively received outputs $\rho^{out}_i$, where $i = \{1,\dots, q=poly(\lambda)\}$. We also note that $\A$ can have an internal register $\sigma$ and we allow for creating entanglement between $\A$'s register and output queries. Specifically for classical primitives, each $\rho^{in}_i = \ket{\phi^{in}_i}\bra{\phi^{in}_i}$ where $\ket{\phi^{in}_i} = \sum_{m_i,y_i}\ket{m_i, y_i}$ is usually a pure state with $m_i$ being the message and $y_i$ the ancillary system. If the queries are being generated by $\A$, in most cases it can be assumed that they have the classical information underlying them, while output queries need to be considered as unknown quantum states to the adversary. In Section~\ref{sec:unf-weak-adaptive-adv} we also represent the model for \emph{weak adversary} which is the quantum equivalent of random-message attack (rma) in the classical world, as well as an alternative way of capturing adaptive adversaries. We also note that this phase for quantum primitives is similar to the classical ones, where $\{\rho^{in}_i\}^q_{i=1}$ represents input chosen message queries and $\{\rho^{out}_i\}^q_{i=1}$ is the respective outputs after the interaction with the oracle sent to $\A$ by the challenger.

\item{\bf Challenge phase.} In this phase, the challenge that the adversary has to respond to is chosen in three different ways, each corresponding to a specific level of unforgeability. Similar to classical notions of unforgeability, the strongest notion is \emph{existential unforgeability} denoted by $\qEx$ in the game, and whereby the adversary picks the message for which it will produce a forgery. On the other hand, in \emph{selective unforgeability}, denoted  $\qSel$, the adversary picks the challenge but needs to commit to it before interacting with the oracle. Hence in \gameref{game:unf-full} the selective challenge phase happens before the learning phase. A further way of weakening the unforgeability notion is when the challenge message is chosen by the challenger $\C$ uniformly at random from the set of all the messages.
In any case, a classical message $m \in \M$ is selected (for classical primitives) where $\M$ is the set of classical messages.\\
\noindent\emph{Quantum case:} If the primitive is quantum, the main difference is that $\M = \HilD$ (or $\M = \ES(\HilD)$ for density operators) is a Hilbert space and $m = \ket{\psi_{m}} \in \HilD$ (or equivalently, $m = \rho_{m} \in \ES(\HilD)$) is a quantum challenge in the $D$-dimensional Hilbert space. In the $\qUni$ challenge phase where the message is chosen by the challenger $\C$ uniformly at random from the set of all the messages, for quantum primitives it should be selected uniformly according to the Haar measure over $\HilD$. We also need to mention that for $\qSel$ challenge phases, $\A$ is required to submit the (efficient) classical description of the quantum state $\rho_m$. This is a technicality related to the verification phase, as it allows the challenger to prepare the required number of copies of the correct output for the most general form of verification. 

We impose different conditions on the challenge phases which will be formalized later in the guessing phase. These conditions prevent the adversary from mounting trivial attacks.

\item{\bf Guess phase.} In this phase, the adversary submits their forgery $t$ for the challenge $m$. They win the game if the output pair $(m,t)$ passes the verification algorithm with high probability. In addition, for $\qSel$, the message $m$ should be the same as the message submitted in the challenge phase. Here the condition in the challenge phase that we have mentioned is formally checked. The quantum challenge phase needs to be carefully specified to avoid capturing trivial attacks such as sending one of the previously learnt states as the challenge of the adversary. As a result, we have introduced the notation $m \notmu \rho^{in}$ denoting  $\mu$-distinguishability from all the input learning phase states. When $m$ is a classical bit-string the same condition should hold for the quantum encoding of $m$ into a computational basis \emph{i.e.} $\ket{m}$ (or $\ket{m, 0}$). Note that the case $\mu=1$ implies the challenge quantum state has no overlap with any of the quantum states queried in the learning phase. 

\noindent\emph{Important note:} We emphasise, that we do not specify how the challenger could check whether the adversary meets the condition or not. Implementing this check is not crucial for our security analysis, where we only need to be able to characterise the instances that might present a security violation. The key point to note is that this can effectively be checked given a run against a given adversary. Indeed, then $\rho^{in}_i$ and $\rho^{out}_i$ can be characterised by the probability analysis allowing proofs of security or exhibition of attacks.\footnote{This argument is a matter of debate in different areas of cryptography. Some researchers believe imposing any condition within the formal game needs to be done via an efficient and specified process while others, including the author, believe that the conditions only need to specify instances with calculatable probability for the purpose of the proofs. A similar case has been discussed in \cite{cortier_sok_2016} (Section VI) regarding the definitions of verifiability in e-voting protocols, where some very widely-accepted definitions such as the one proposed in \cite{kiayias_demos-2_2015,kiayias_end--end_2015} have verifying subroutines that do not necessarily run in polynomial time. However, we note that imposing such conditions in the definition leads to the fact that extra care is needed when definitions such as this one are used to prove security. Since some reductions may not carry over if the conditions are not being executed, the security proofs can be more complicated. As we will see in what follows, this has been taken into account in our security proofs.}

Regarding the verification oracle, for classical primitives the forgery pair $(m,t)$ is classical and the verification oracle $\vO_f$ runs the classical verification algorithm $\V = \mathtt{Ver}(k,m,t,r)$. Here $r$ is the randomness if the primitive is randomised.

\noindent\emph{Quantum case:} This phase is similar to the classical case. Here, it can be seen that this is the most natural way of characterising a forgery for quantum primitives since the difference between quantum states is usually measured by their indistinguishability and with quantum distance measures. The main difference in this phase is the difference between the classical and quantum verification procedures. The verification is fairly straightforward for classical primitives since the equality test can be easily performed whereas for quantum primitives, both message and forgery are quantum states, and the verification oracle $\vO_U$ should call a quantum test algorithm $\T$ that checks the equality of quantum states as in \defref{def:test}. Note that the challenger can prepare copies of correct outputs locally.
\end{itemize}

\begin{gamebox}{Formal definition of Generalised Quanutm Unforgeability (qGU)}
\begin{game}\label{game:unf-full}
Formal definition of the quantum games $\GCM{\F}{q, c,\mu}$($\lambda,\A$) where $\lambda$ is the security parameter, $q$ the number of queries issued to the evaluation oracle in the learning phase, $\mu$ the overlap allowed between the challenge and previously queries messages, and $c$ the level of unforgeability.\underline{\emph{The game $\GCM{\F}{q, c,\mu}$($\lambda,\A$)}\footnote{$c\in\{\qEx, \qSel, \qUni\};\ 0 < \mu\le 1$.}}
\medskip\\
{\bf Setup phase:}
\vspace{-0.5em}
\begin{itemize}
\setlength\itemsep{-0.1em}
\item $\texttt{param} \leftarrow \ES(\lambda)$
\item The oracles $\eO$ and $\vO$ and the message space $\M$ are instantiated given $\texttt{param}$.
\end{itemize}
{\bf Selective challenge phase:}
\vspace{-0.5em}
\begin{itemize}
\item if $c=\qSel$: $\A$ picks $m \in \M$ and sends it to $\C$.
\end{itemize}
{\bf First learning phase:}
\vspace{-0.5em}
\begin{itemize}
\setlength\itemsep{-0.1em}
    \item $\A$ (adaptively) issues queries $\rho^{in}_1, \dots, \rho^{in}_q$ (where $q=poly(\lambda)$) to $\C$. To each query $\rho^{in}_i$ the challenger $\C$ queries $\eO$ on $\rho^{in}_i$, and forwards the received respective output $\rho^{out}_i$ to $\A$. The adversary can also have an internal register $\sigma$ which may be entangled with the output queries.
\end{itemize}
{\bf Challenge phase:}
\vspace{-0.5em}
\begin{itemize}
\setlength\itemsep{-0.2em}
\item if $c=\qEx$: $\A$ picks $m \in \M$ and sends it to $\C$.
\item if $c=\qUni$: $\C$ picks $m \overset{\$}{\leftarrow} \M$ uniformly at random and sends $m$ to $\A$
\end{itemize}
{\bf Second learning phase:}
As the \emph{first learning phase}

{\bf Guess phase:}
\vspace{-0.5em}
\begin{itemize}
\setlength\itemsep{-0.1em}
        \item if $c=\qEx$ OR $c=\qSel$: continue if $m \notmu \rho^{in}$\footnote{$\notmu \rho^{in}$ denotes at least $\mu$-distinguishability from all the $\rho^{in}_{i}$. For the classical message $m \in \{0,1\}^n$, the condition should hold for $\ket{m}$, \emph{i.e.} the quantum encoding of $m$ in computational basis.}, else aborts.
        \item $\A$ generates the forgery $t$, and outputs to $\C$ the pair:\\
        $(m,t) \leftarrow \A(\{\rho_i^{in},\rho_i^{out}\}^q_{i=1},\sigma)$
        \item $\C$ queries the verification oracle: $b \leftarrow \vO (m,t)$
        \item $\C$ outputs $b$
\end{itemize} 
\medskip{}
\end{game}
\end{gamebox}

We omit the parameter $q$ when we consider arbitrarily polynomially many queries to the evaluation oracle issued by $\A$. We can now formally define \emph{Existential}, \emph{Selective} and \emph{Universal Unforgeability} of primitives as instances of our game as follows.

\begin{defbox}
\begin{definition}[\eufm]\label{def:euf-qCMA} A cryptographic primitive $\F$ provides $\mu$-quantum existential unforgeability if the probability of any QPT adversary $\A$ of winning the game $\GCM{\F}{\qEx, \mu}(\lambda, \A)$ is at most negligible in the security parameter,
\begin{equation}
Pr[1\leftarrow \GCM{\F}{\qEx, \mu}(\lambda, \A)] \leq \negl(\lambda).
\end{equation}
\end{definition}
\end{defbox}

We also define a stronger security notion for existential unforgeability which considers any overlap $\mu$.

\begin{defbox}
\begin{definition}[qGEU]\label{def:euf-without-mu} A cryptographic primitive $\F$ provides quantum existential unforgeability if it provides $\mu$-quantum existential unforgeability for all non-negligible $\mu$.
\end{definition}
\end{defbox}

\begin{defbox}
\begin{definition}[\sufm]\label{def:sel-qCMA} A cryptographic primitive $\F$ provides $\mu$-quantum selective unforgeability if for any $q$ the advantage of any QPT adversary $\A$ of winning the game $\GCM{\F}{q, \qSel, \mu}(\lambda, \A)$ over $P_{ov}(q,\mu)$ is at most negligible in the security parameter,
\begin{equation}
Pr[1\leftarrow \GCM{\F}{q, \qSel, \mu}(\lambda, \A)] \leq P_{ov}(q, \mu) + \negl(\lambda).
\end{equation}
We call $P_{ov}(q, \mu)$ the ``overlap probability" describing the probability for trivial attacks via the overlap allowed by the parameter $\mu$.\footnote{Note that by definition $\A$ can always achieve $P_{ov}(q, \mu)$, hence $\A$'s winning probability is always lower-bounded by this value.}
\end{definition}
\end{defbox}

The need for allowing an adversary to win with probability $P_{ov}(q, \mu)$ is similar to the classical definitions where the adversary is required to boost the success probability from some trivial value such as a random guess. Here, by allowing the adversary to create an overlap between the learning phase space and challenge, some unavoidable attacks exist which are independent of the actual primitive at hand, and as such needs to be extracted to characterise the gap between trivial and effective adversaries and hence precisely define a proper distance-based definition.

\begin{defbox}
\begin{definition}[$P_{ov}$ for classical primitives]\label{def:pov-standard-orc}
For all $q$ and for all $\mu$,
for a classical primitive where the evaluation oracle is a standard oracle $\eO_f$, the overlap probability for $q$-query games is equal to $P_{ov}(q, \mu) = 1 - \mu^q$.
\end{definition}
\end{defbox}

The expression $P_{ov}(q, \mu) = 1 - \mu^q$ that is chosen in the above definition for the overlap probability for classical primitives, is the probability of a trivial attack performed via simply measuring the superposition queries. A straightforward calculation of this measurement probability for $q$ queries with the same degree of overlap leads to the expression $1 - \mu^q$.

A similar notion can be defined for quantum primitives. In this case, it is clear that the adversary's success probability in finding the output by measurement strategy is almost zero and hence defining the $P_{ov}$ as defined by \defref{def:pov-standard-orc} leads to zero overlap probability. However, in this case, as well, there is another scenario that may lead to trivial attacks, which is due to the error produced by the quantum test algorithm in distinguishing the states with certain overlap. An example of this is the SWAP test which has a one-sided error of $\frac{1}{2}$ even for perfectly distinguishable states. This is a fundamental difference between the quantum world and classical primitives where equality can be checked deterministically. To have a general characterisation of $P_{ov}$ for quantum primitives, this probability needs to be defined with respect to the test algorithm as follows. 

\begin{defbox}
\begin{definition}[$P_{ov}$ for quantum primitives]\label{def:pov-quantum}
Let $\rho_{max}$ be the input learning phase query with the maximum overlap with the challenge state $\ket{\psi}$, allowed by the $\mu$-distinguishability condition. Let the $\eO_U$ be the unitary oracle for the quantum primitive applying $\Ue$ to the quantum inputs and let $\vO$ implement a quantum test algorithm $\T$. Then $\rho^{out}_{max} = \Ue \rho_{max} \Ue^{\dagger}$ is the output of the query from the oracle and $\rho^{out} = \ket{\psi^{out}}\bra{\psi^{out}} = \Ue \ket{\psi}\bra{\psi} \Ue^{\dagger}$ is the correct output of the challenge $\ket{\psi}$. We define the $P_{ov}$ as the error probability of the test algorithm $\T$ on distinguishing $\rho^{out}_{max}$ and $\rho^{out}$ as follows:
\begin{equation}
    P_{ov} = Pr[1 \leftarrow \T((\rho^{out}_{max})^{\otimes \kappa}, (\rho^{out})^{\otimes \kappa})]
\end{equation}
\end{definition}
\end{defbox}

This definition also implies an intuitive and practical approach to determine the desired $\mu < 1$ for quantum primitives, as it states that for any specific quantum primitive or the protocols based on that primitive, the $\mu$ should not allow for the above overlap attacks with a probability larger than the required security threshold. Nevertheless, if one assumes a reasonably good quantum test algorithm, this probability for quantum primitives is usually less than the classical ones due to quantum state distinguishability and lack of adversary's knowledge over the transformation of the output bases.\\

\noindent When selective unforgeability holds for any overlap $\mu$ we say that the primitive is quantum selective unforgeable.

\begin{defbox}
\begin{definition}[qGSU]\label{def:suf-without-mu} A cryptographic primitive $\F$ provides quantum selective unforgeability if it provides $\mu$-quantum selective unforgeability for all non-negligible $\mu$.
\end{definition}
\end{defbox}

Now we give yet a weaker definition, namely \emph{Universal Unforgeability}. Note that the $\mu$-distinguishability condition is not necessary for universal unforgeability, as the challenge is chosen by the challenger, independently of the adversary's queries and the probability is taken over all the choices of the challenge state hence it is no longer meaningful to count for possible overlaps as trivial attacks.

\begin{defbox}
\begin{definition}[\uuf]\label{def:uni-qCMA}
 A cryptographic primitive $\F$ is quantum universally unforgeable if the probability of any QPT adversary $\A$ of winning the game $\GCM{\F}{\qUni}(\lambda, \A)$ is negligible in the security parameter $\lambda$,
\begin{equation}
Pr[1\leftarrow \GCM{\F}{\qUni}(\lambda, \A)] \leq \negl(\lambda).
\end{equation}
\end{definition}
\end{defbox}

\subsection{Hierarchy and relationship to other definitions}\label{sec:unf-hier-other-defs}
\begin{figure}[h!]
   \centering
     \includegraphics[width=1
     \textwidth]{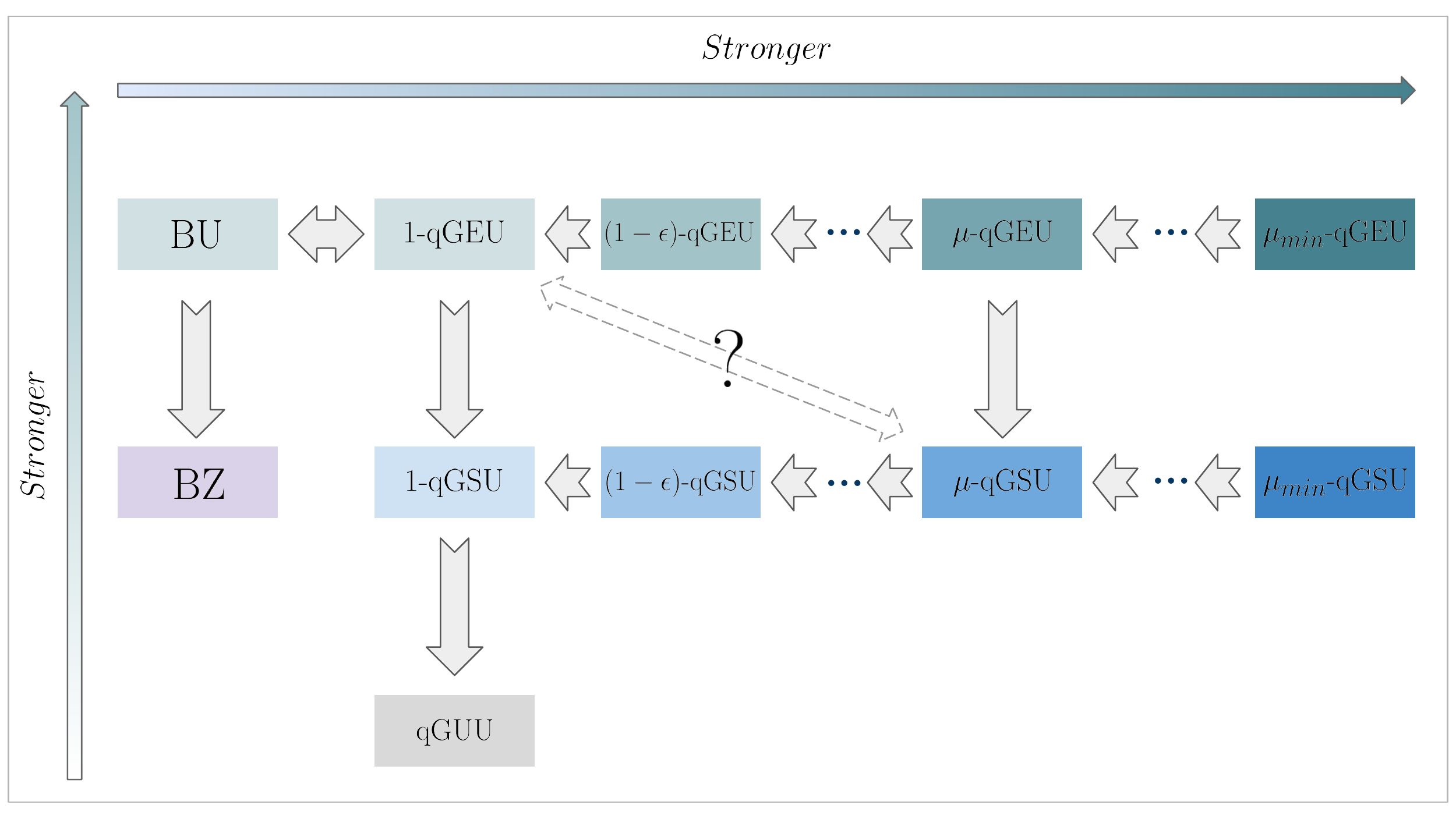}
     \caption[Relationship between different definitions of \emph{Generalised Quantum Unforgeability}, \bu\ and \bz]{Relationship between different definitions of \emph{Generalised Quantum Unforgeability}, \bu\ and \bz. From down to up and left to right the definitions become stronger. $\epsilon = \epsilon(\lambda)$ is a negligible function in the security parameter and $\mu_{min} = \nonnegl(\lambda)$ is the smallest valid degree for $\mu$. It is unknown whether \sufm\ with smaller $\mu$, implies \eufm\ with bigger $\mu$. }\label{fig:unf-hierarchy}
 \end{figure}

In this section, we formally establish the hierarchy between the different levels of Generalised Unforgeability captured by our framework. Furthermore, for completeness, we also investigate how our definitions formally relate to the previously proposed ones for classical primitives. In particular, we show this relationship between \euf\ and the definitions of \bz\ and \bu\ introduced in Section~\ref{sec:prelim-quantum-unf} in the preliminaries. In \figref{fig:unf-hierarchy}, we map out the results presented in this section.

First, we establish the relationship between different instances of our game-based definition. We show that as expected for both existential and selective unforgeability, the definitions become stronger when decreasing the $\mu$ parameter from 1 and hence \eufm\ implies \euf. 

\begin{thmbox}
\begin{theorem}\label{th:mu-smaller-stronger}
If $\mu_1 \leq \mu_2$ then $\mu_1$-qGEU (resp. $\mu_1$-qGSU) implies $\mu_2$-qGEU (resp. $\mu_2$-qGSU)
\end{theorem}
\end{thmbox}
\begin{proof}
The proof is straightforward for qGEU. Let $\A$ win against $\mu_2$-qGEU, Let $\A'$ be the adversary who wants to attack $\mu_1$-qGEU. $\A'$ queries the same learning phase queries as $\A$ and then calls $\A$. Since $\mu_1 \leq \mu_2$ any two states that are $\mu_2$-distinguishable are also $\mu_1$-distinguishable, then the challenge of $\A$ will necessarily satisfy the condition for $\mu_1$-qGEU. Then $\A'$ can also win the game with non-negligible probability. For $\mu$-qGSU the distinguishability argument is similar, although there is also the $P_{ov}$ probability that is a function of $\mu$. Thus we need to show the following:
\begin{equation*}
    Pr[1\leftarrow \GCM{\F}{\qSel, \mu_2}(\lambda, \A)] - P_{ov}(\mu_2) \geq  Pr[1\leftarrow \GCM{\F}{\qSel, \mu_1}(\lambda, \A)] - P_{ov}(\mu_1)
\end{equation*}
Which is also equivalent to showing the following statement:
\begin{equation*}
    Pr[1\leftarrow \GCM{\F}{\qSel, \mu_2}(\lambda, \A)] - Pr[1\leftarrow \GCM{\F}{\qSel, \mu_1}(\lambda, \A)] \geq P_{ov}(\mu_2) - P_{ov}(\mu_1)
\end{equation*}
The LHS of the inequality is always positive due to the above distinguishability argument, and the $P_{ov}$ is always a non-increasing function of $\mu$ for both types of primitives (also the negligible factor is omitted from both sides). Take the $P_{ov}$ for the classical primitives for instance, which is equal to $1-\mu^q$. Therefore, the RHS of the inequality will be equal to $\mu_1^q - \mu_2^q$ which is always a non-positive value as $\mu_1 \leq \mu_2$. Then the above inequality holds and the theorem has been proved. 
\end{proof}

Furthermore, it is easy to observe that for any given $\mu$, \eufm\ implies \sufm. This is due to the fact that if the adversary wins the game by committing to their favourite message before the learning phase, they will necessarily win when picking the message after the learning phase.

Universal unforgeability is also intuitively weaker than existential unforgeability similarly to their classical counterpart. The same thing holds despite the winning condition for these two instances being very different. In universal unforgeability, the adversary wins only if they win the game on average over all the randomly picked messages. In our case, we are only interested in QPT adversaries, and as the universal definition is not parameterised by $\mu$, it is not evident whether $\uuf$ is weaker than \sufm. The following theorem formally establishes the implication. We prove the theorem for \suf\ which, in turn, implies \sufm\ for any $\mu$. 

\begin{thmbox}
\begin{theorem}\label{th:suf-uuf}
\sufm\ implies \uuf.
\end{theorem}
\end{thmbox}
\begin{proof}[Proof Sketch]
The full proof can be found in the \appref{app:proof-suf-uuf}. Here we present the key ideas of the proof. We show if there exists an adversary $\A$ that wins the \uuf\ game then \euf\ (\suf) also breaks and the implication to \eufm\ (\sufm) is straightforward. First, we show that the distinguishability condition for $\mu=1$ can be satisfied. Thus we write the winning probability of $\A$ as the combination of probabilities of winning for the selected message being orthogonal to the learning phase or not:
\begin{equation}
\begin{split}
    \underset{x \in \M}{Pr}[1\leftarrow \A(x)] & = \underset{x \in \M'}{Pr}[1\leftarrow \A(x)]Pr[x \in \M'] + \underset{x \not\in \M'}{Pr}[1\leftarrow \A(x)]Pr[x \not\in \M'] \\
    & = \nonnegl(\lambda)
\end{split}
\end{equation}
where $\M'$ is the set of all the challenges with no overlap with the learning-phase states. By calculating this probability we show that $\underset{x \in \M'}{Pr}[1\leftarrow \A(x)]$ is also non-negligible. In the second part of the proof we show that as long as the previous average probability holds, we can always construct an efficient adversary $\A'$ that uses $\A$ to win the selective unforgeability game. We prove this by partitioning the space of $\M'$ into equal polynomial-size subspaces and show that if the average probability over $\M'$ is non-negligible, then $\A'$ can always win the \euf\ game by randomly picking one of the subsets to pick the message from, as there will exist at least one message that allows $\A$ to win the game with non-negligible probability. As a result, $\A'$ wins the game with non-negligible probability.
\end{proof}

Now, we show an equivalence between an instance of our existential unforgeability definition and \bu. Since this result is mainly for the sake of completeness and will not be directly related to the rest of the thesis, again we give a proof sketch here, and we give the full proof in the \appref{app:proof-1gu-bu-equivalent}.

\begin{thmbox}
\begin{theorem}\label{th:1guf-bu}
\euf\ is equivalent to BU.
\end{theorem}
\end{thmbox}
\begin{proof}[Proof Sketch]
We show that \euf\ implies BU and vice versa. First, we show that if a scheme is not BU unforgeable against a QPT adversary then it is not \euf\ unforgeable either. Let $\A$ be a QPT adversary who forges a scheme $\F = (\ES, \E, \V)$ with message set $\M = \{0,1\}^n$ in the BU definition. Following the definition of BU, if  $\A$ can win against the \bu\ game, there exists a non-empty set $\Be$ for which $\A$ interacts with the blinded oracle associated with it and outputs a pair $(m^*,t^*)$ where $t^* = f(m^*)$ (where $f$ is the classical function of the evaluation $\E$) such that $\V = Ver_k(m^*,t^*) = acc$, and also the $m^* \in \Be$ with non-negligible probability in $\lambda = poly(n)$. By rewriting a general query state to the blinding oracle in orthogonal and non-orthogonal sub-spaces to the main forgery state, we can show that there exists a unitary non-blinding oracle that generates equivalent queries for this scenario. We then show that this new unitary oracle can be queried equivalently by an adversary who satisfies all the conditions of \euf and therefore can also generate a forgery that passes the test algorithm with also non-negligible probability. Hence we have shown that \euf\ implies \bu.

For the other way of implication, we show that a QPT adversary $\A$ who wins the \euf, has also a non-empty support, $supp(\A)\cap R = \emptyset$, for some $R \neq \emptyset$, and can output a valid pair $(m^*,f(m^*))$ with $m^* \in R$ with non-negligible probability. Intuitively, this is due to the orthogonality condition that is required to be satisfied in the \euf\ game between the learning subspace and the forgery state. According to the \thmref{th:prelim-bu} in \chapref{chap:prelim}, this implies that the primitives against such an adversary are not BU-secure. This concludes the proof.
\end{proof}

From the above theorem and the equivalence of \bu\ and \bz\ against classical adversaries, we derive the following corollary.

\begin{corrbox}
\begin{corollary}
$\euf \equiv \bu \equiv \bz$ against classical adversaries.
\end{corollary}
\end{corrbox}

\subsection{Unforgeability against weak vs adaptive adversaries}\label{sec:unf-weak-adaptive-adv}
Another variant of the unforgeability definition appears when we weaken the adversary in choosing the queries in the learning phase freely and adaptively. One way of imposing such limitations on the adversary is to assume that the adversary has no direct oracle access and instead has only access to a random set of queries (random input-output samples from the oracle) being selected at random from a specific distribution by the honest party. This attack model is commonly called the \emph{random message attack} model in classical cryptography. Moreover, in the practical sense, this type of limited adversary represents \emph{network adversarial model}, \emph{i.e.}, when the adversary has only access to the communication channel. Network adversaries can get the input and output samples of a primitive, only by intercepting the messages that are exchanged between the two honest parties during a protocol.

The adversary we have considered so far, which we refer to as \emph{adaptive adversary}, can query the evaluation function of the primitive through oracle access adaptively, in the sense that not only the queries have been chosen by the adversary, but they can depend on the previously received responses in general. We also introduce an alternative way of capturing full adaptive quantum adversaries in the \appref{app:alt-adaptive-adv}. On the other hand, a weak non-adaptive adversary, cannot choose the queries and instead receives $q$ queries (where $q = poly(\lambda)$) of $\E$, in the form of input and output pair. Most commonly, the queries are being picked at random from a uniform distribution by an honest party, but more generally, one can consider the case where queries are being selected from a distribution $\delta_{\mathcal{D}}$ over the message space.

In what follows, we restate a new instance of our \gameref{game:unf-full}, which captures weak adversaries as well as adaptive ones. We also note that weak adversaries are mostly of interest regarding universal unforgeability, thus we only restate the game for universal unforgeability. 

\begin{gamebox}{Universal Unforgeability with weak and adaptive adversaries}
\begin{game}\label{game:uni-unf-weak-adv}
Formal definition of the quantum games $\GCM{\F}{q}$($\lambda,\A$) where $\lambda$ is the security parameter, $q$ the number of queries issued to the evaluation oracle in the learning phase.
\medskip\\
{\bf Setup phase:}
\vspace{-0.5em}
\begin{itemize}
\setlength\itemsep{-0.1em}
\item same as \gameref{game:unf-full}
\end{itemize}
{\bf Learning phase:}
\vspace{-0.5em}
\begin{itemize}
\setlength\itemsep{-0.1em}
            \item If the adversary is adaptive, $\A = \Aad$ (same as \gameref{game:unf-full}):
            \begin{itemize}
                \item $\Aad$ selects any desired query $c_i \in \M$, and issues to $\C$ (up to $q$ queries).
                \item $\C$ queries $\eO$ on $c_i$, and sends the respective output $r_i$ back to $\Aad$.
            \end{itemize}
            \item If the adversary is weak (non-adaptive), $\A = \Ana$:
            \begin{itemize}
                \item $\C$ selects a challenge $c_i \in \M$ uniformly at random from $\M$ and independent of $i$.
                \item $\C$ queries the $\eO$ on $c_i$ and produces the response $r_i = \E(c_i)$.
                \item $\C$ issues to $\Ana$ the set of random challenges and their respective responses $\{(c_i,r_i)\}^q_{i=1}$.
            \end{itemize}
\end{itemize}
{\bf Challenge phase:}
\vspace{-0.5em}
\begin{itemize}
\setlength\itemsep{-0.2em}
\item same as \gameref{game:unf-full} for universal challenge phase ($\qUni$)
\end{itemize}

{\bf Guess phase:}
\vspace{-0.5em}
\begin{itemize}
\setlength\itemsep{-0.1em}
    \item same as \gameref{game:unf-full}
\end{itemize} 
\medskip{}
\end{game}
\end{gamebox}

Note that in the above game all the queries $c_i$ and the responses $r_i$ have been abstracted for simplicity, but can capture both quantum and classical queries. We will widely use this variant on the unforgeability game later in \chapref{chap:application}, Section~\ref{sec:application-hpuf-practical}. 

\noindent Let us define universal unforgeability with weak adversaries as follows:

\begin{defbox}
\begin{definition}[Universal Unforgeability against weak Adversary]\label{def:uni-unf-weak}
A cryptographic primitive $\F$ is quantum universally unforgeable against a `weak (non-adaptive) adversary' if the success probability of any weak QPT adversary $\Ana$ in winning the game $\Gn{\F}(\Ana, \lambda)$ is at most a negligible function, $\epsilon(\lambda)$, in the security parameter.
\begin{equation}
    Pr[1\leftarrow \Gn{\F}(\Ana, \lambda)] \leq \epsilon(\lambda)    
\end{equation}
\end{definition}
\end{defbox}

\subsubsection{A note on weak vs strong unforgeability}\label{sec:weak-strong-unf}
We also note that there is another way of characterising the strength of the unforgeability definition in the literature.
We have formally defined our different instances of unforgeability as a quantum analogue of \emph{weak unforgeability}. However, the same definition albeit with a small modification can be applied to capture \emph{strong unforgeability}. First, we note that the difference between strong and weak unforgeability is only relevant to randomised primitives. For non-randomised primitives, these definitions are equivalent. In the classical world, for strong unforgeability, it is sufficient for the adversary to output a new pair to win the game and hence the adversary is allowed to pick one of the learning phase messages as the challenge and produce a new output with fresh randomness. In our definition, it is sufficient to expand the $\mu$-distinguishability condition to the overall input of the oracle including the randomness, \emph{i.e.} adversary's challenge state $\ket{r^*}\bra{r^*}\otimes\rho_m$ needs to be $\mu$-distinguishable from all the learning phase states with their randomness registers which can be written as $\ket{r_i}\bra{r_i}\otimes\rho^{in}_i$. Once again for $\mu=1$, this will capture the same definition as is expected.
\section{Applications of qGU: possibility and impossibility results}\label{sec:unf-results}
In this section, we study the unforgeability of general classical and quantum primitives under the lens of our generalised unforgeability framework. We start with the strongest level of unforgeability in our framework, \emph{i.e.} existential unforgeability, and we try to give examples for both classical and quantum primitives, all the way to the weakest unforgeability notion. We will see how this framework allows us to establish general possibility and impossibility results on different levels. It will also help us design non-trivial cryptographic primitives that satisfy a high level of quantum unforgeability against any quantum adversaries.

\subsection{Generalised existentially unforgeable schemes}
\label{sec:ex-unf}
In this section, we turn our attention to \euf. First, we show a general and intuitive, yet important no-go result for \eufm\ that is, no classical primitive (deterministic or randomized) can satisfy this level of unforgeability for any $\mu \neq 1$. This result states that \euf, (which is also equivalent to \bu\ according to \thmref{th:1guf-bu}), is the strongest notion of existential unforgeability that any classical primitive can possibly achieve.

\begin{thmbox}
\begin{theorem}[No classical primitive $\F$ is \eufm~ secure]\label{th:ex-qCM} For any classical primitive $\F$ and for any $\mu$ such that $\mu \leq 1 - \frac{1}{D}$, where $D$ is the dimension of the Hilbert space on which the evaluation oracle operates, there exists a QPT adversary $\A$ such that
\begin{equation}
    Pr[1\leftarrow \GCM{\F}{\qEx, \mu}(\lambda, \A)] = \nonnegl(\lambda).
\end{equation}
\end{theorem}
\end{thmbox}

\begin{proof}
There exists a simple superposition attack that breaks \eufm. Let $\A$ issue only one query which is the uniform superposition of all the inputs, which leads to an output of the form $\frac{1}{\sqrt{2^n}}\sum_{m} \ket{r}_{\Ora}\ket{m, f(m;r)}$, where we have taken $D = 2^n$. Then by measuring the first part of the register in the computational basis, the state will collapse to one of the basis and the adversary is able to produce a valid message-tag pair for a classical message with a negligible overlap with the learning phase. Hence $\A$ can always win the game for any $\mu \leq 1 - \frac{1}{2^n}$.
\end{proof}

Nevertheless, it is still possible to have schemes that are \euf\ secure through the following positive result:
\begin{thmbox}
\begin{theorem}\label{th:qprf-1euf} $\qprf$s are \euf\ (\suf) unforgeable.
\end{theorem}
\end{thmbox}
\begin{proof}
This is a straightforward result via equivalence of \euf\ to \bu ~and a corollary  from~\cite{alagic_quantum-access-secure_2020}, where it is shown that $\qprf$s are $\bu$ secure.
\end{proof}

Although the general no-go result for classical primitives does not directly apply to quantum primitives, in \chapref{chap:qpuf}, we show that most quantum primitives that we are interested in do not satisfy this definition either. However, it will not be surprising given the general no-go result we provide for selective unforgeability. 

Another interesting positive result that we can demonstrate, is the peer of \thmref{th:qprf-1euf} for quantum primitives. We show that pseudorandom unitaries (\pru), which are the quantum counterpart of pseudorandom functions in the quantum world, can also satisfy \euf\ (and \suf). 

\begin{thmbox}
\begin{theorem}\label{th:suf-pru}
$\pru$ quantum primitives are \suf\ (\euf) secure.
\end{theorem}
\end{thmbox}

\begin{proof}
We prove this by contradiction. Let $\A$ be an adversary who wins the $\suf$ game with non-negligible probability (Note that here $P_{ov} = 0$). $\A$ selects a message $m$ before (or after) the learning phase and then outputs the respective $t$ such that it passes the verification test with non-negligible probability. Also by definition of \suf, $m \notmu \rho^{in}$ for $\mu = 1$ and hence the message $\rho_m$ is completely orthogonal to  all $\rho^{in}_i$. Now we construct an adversary $\A'$ who is playing the \pru\ game. Let $\A'$ first query all the learning phase states of $\A$ and then also issue one more query which is $\rho_{m}$. Then $\A'$ calls $\A$ and receives the input-output pair of $(m,t)$ such that $\rho_t$ is non-negligible close to the actual output, \emph{i.e.}
\begin{equation}
    F(\rho_t, \Ue\rho_m\Ue^{\dagger}) = \nonnegl(\lambda)
\end{equation}
Now $\A'$ can use this last query as a distinguisher between PRU and a unitary picked from the Haar measure since $\A'$ can estimate the output with non-negligible fidelity if the $U_k$ had been picked from the family. Let $\A'$ runs a quantum equality test as described in \defref{def:test} on the $U_k \ket{\psi}$ obtained in the learning phase and $\rho_t$. Also note that if $U$ is picked from the Haar measure family, the probability of producing the output is negligible by definition. Thus whenever the test shows equality, $\A'$ can conclude that the unitary has been picked from PRU. Thus for $\A'$, we have:
\begin{equation}
    \underset{U \leftarrow U_k}{Pr}[\A'^{U}(1^{\lambda})=1] - \underset{U_{\mu} \leftarrow \mu}{Pr}[\A'^{U_{\mu}}(1^{\lambda})=1]=\nonnegl(\lambda)
\end{equation}
Which is a contradiction and the proof is complete.
\end{proof}

\subsection{Generalised selectively unforgeable schemes}
\label{sec:sel-unf}
In this section, we establish results for \sufm\ which restricts the adversary in two ways. First, by requiring the adversary to commit to the challenge before the learning phase, we prevent the adversary from picking any post-measurement state as their forgery challenge. Second, by subtracting the probability of any potential trivial attack, especially for classical primitives, from the winning probability of the game, we make the probability bounds tighter for the adversary. We also discuss why defining unforgeability in such a way leads to non-trivial results and establishes a separation between randomised and non-randomised constructions, therefore motivating the usefulness of the given definition. 

\subsubsection{Non-randomised schemes}
Let us start with non-randomised schemes. To establish our result, we now take advantage of our proposed cryptanalysis toolkit, namely the \emph{quantum emulation attack} (QEA), which we introduced earlier on in Section~\ref{sec:unf-qe-revisited}. Here we only show this no-go result for classical non-randomised primitives to avoid repetitions, but the same result holds for quantum constructions. 

\begin{thmbox}
\begin{theorem}[No classical (or quantum) non-randomised primitive $\F$ is \sufm\ secure]\label{th:sel-qCM} For any classical/quantum deterministic primitive $\F$ and for any $\mu$, in the range $\frac{1}{4} + \nonnegl(\lambda)\leq \mu \leq 1-\nonnegl(\lambda)$, there exists an effective QPT adversary $\A$ such that
\begin{equation}
Pr[1\leftarrow \GCM{\F}{q(\lambda), \qSel, \mu}(\lambda, \A)] - P_{ov}(q(\lambda), \mu) = \nonnegl(\lambda).
\end{equation}
\end{theorem}
\end{thmbox}

\begin{proof}
We show the proof for classical primitives but the same attack and results also holds for quantum primitives. We show that there exists a QPT adversary $\A$ who can win the game with non-negligible probability for any $\mu$ except when it is negligibly close to 0 or 1.
The attack is the one-block emulation attack from Section~\ref{sec:unf-qe-one-block} with the following setting. First $\A$ picks any two messages $m, m' \in \M$ and sets $m$ as the challenge. Then $\A$ queries the states $\ket{\phi_1} = \ket{m', 0}$ and $\ket{\phi_r} = \sqrt{1-\gamma^2}\ket{m', 0} + \gamma \ket{m, 0}$ by interacting with $\eO_f$,
where $\gamma$ is a real value such that $0 \leq \gamma \leq \sqrt{1-\mu}$ and such that the distinguishability condition of the \sufm\ game is satisfied. After the learning phase, $\A$'s output state can be written as $\sigma_{out} = \ket{\phi^{out}_1}\otimes\ket{\phi^{out}_r}$ where $\ket{\phi^{out}_1} = \Ue \ket{\phi_1}$ and $\ket{\phi^{out}_r} = \Ue \ket{\phi_r}$. Followed by the fidelity analysis given in Section~\ref{sec:unf-qe-one-block}, we show that the success probability of $\A$ in producing the output of $m$ \emph{i.e.} $f(m)$ is bounded by $\gamma^2(1 + 4(1-\gamma^2)^2)$. This is because we have: $\mbraket{\phi_1}{\psi} = 0$ and $|\mbraket{\psi}{\phi_r}|^2 = \gamma^2$ and $|\mbraket{\phi_1}{\phi_r}|^2 = 1 - \gamma^2$, which gives us the following bound on the fidelity:
\begin{equation}
    F(\ket{\omega}\bra{\omega}, \Ue^{\dagger} \ket{\psi}\bra{\psi} \Ue) \geq \gamma^2(1+4(1-\gamma^2)^2)
\end{equation}

In general, $\gamma^2$ which is the overlap between the challenge state and the learning phase state can be as large as $1 - \mu$ allowed by the definition, thus we set the maximum allowed value of overlap which is $\gamma = \gamma_{max} = \sqrt{1 - \mu}$. Now we need to also determine $P_{ov}$ and to show whether the adversary can boost the success probability by a non-negligible value. Here one of the queries is orthogonal to the challenge and there is only one query ($\ket{\phi_r}$) with overlap, thus according to \thmref{def:pov-standard-orc} we have $P_{ov}(2, \mu) = 1 - \mu^2$. As a result
\begin{equation}
\begin{split}
    Pr[1\leftarrow \GCM{\F}{\qSel, \mu}(\lambda, \A)] - P_{ov}(2, \mu) & = (1 - \mu)[1+4(1 - (1 - \mu))^2] - (1 - \mu^2)\\
    & = \mu(1-\mu)(4\mu - 1)
\end{split}
\end{equation}
Since $\frac{1}{4} + \nonnegl(\lambda) \leq \mu \leq 1 - \nonnegl(\lambda)$, then all the terms are non-negligible in the security parameter and this concludes the proof.
\end{proof}

The above theorem has a direct consequence which we represent as the following corollary:

\begin{corrbox}
\begin{corollary}
No deterministic classical or quantum primitive $\F$ is qGSU (\defref{def:suf-without-mu}) secure.
\end{corollary}
\end{corrbox}

Let us now discuss the intuitive meaning of it. First, we note that despite the above no-go theorem, \qprf s still provide \suf\ security (\thmref{th:qprf-1euf}). However, this no-go result shows a fundamental vulnerability of any non-randomised classical primitive against forgeries, since the only way to ensure the security of primitives against such effective attacks is to guarantee that the adversary's forgery message is orthogonal to their learning subspace. Practically this guarantee can only be given by relying on the device implementation, which is arguably in contradiction with the whole motivation of obtaining security against more powerful quantum adversaries, to begin with~\cite{boneh_quantum-secure_2013}. Let us consider a non-randomised MAC scheme such as HMAC and NMAC. According to \thmref{th:sel-qCM}, these schemes do not satisfy existential nor selective unforgeability except for $\mu = 1$ and hence are always vulnerable against more powerful quantum adversaries implementing superposition attacks. Nevertheless, one might argue that such definitions might be too strong, and the proposed attack might not demonstrate an intuitive forgery. To better demonstrate this potential vulnerability, let us show a slightly different example from what is used in the proof of the above theorem to argue there are instances of the game and attacks that can demonstrate an intuitive forgery situation.

\begin{example}\label{example:qea-three-states}
Let $\A$'s state after the learning phase be $\sigma_{in} = \ket{\phi^{in}_1}\otimes\ket{\phi^{in}_r}^{\otimes 2}$ and $\sigma_{out} = \ket{\phi^{out}_1}\otimes\ket{\phi^{out}_r}^{\otimes 2}$ where the query states have been chosen as follows:
\begin{equation}
    \ket{\phi_1} = \ket{m_1, 0} \quad \ket{\phi_r} = \delta \ket{m_1, 0} + \gamma \ket{m_2, 0} + \gamma \ket{m_3, 0}
\end{equation}
Where due to normalisation $|\delta|^2 + 2|\gamma|^2 = 1$, although we pick the $\delta = \sqrt{1-2\gamma^2}$ and $\gamma$ to be real values for simplicity, thus $\gamma^2 \leq \frac{1}{2}$. Also note that $\A$ has two identical copies of $\ket{\phi^{out}_r}$. The attack consists of running two separate emulations for $\ket{m_2, 0}$ and $\ket{m_3, 0}$.

Let $\ket{\phi_r}$ be the reference state for the emulation, and the target state to be $\ket{\psi} = \ket{m_2, 0}$ or $\ket{\psi} = \ket{m_3, 0}$. Note that as $\ket{\phi_1} = \ket{m_1, 0}$ is orthogonal to both states and the reference state is symmetric with respect to them, the emulation's fidelity will be the same for both these states. Relying on Theorem~\ref{th:qe-fins}, the output state of the QE algorithm with only one block will be:
\begin{equation}
\begin{split}
     \ket{\chi_f} =& \mbraket{\phi_r}{\psi} \ket{\phi_r}\ket{0} + \ket{\psi}\ket{1} - \mbraket{\phi_r}{\psi}\ket{\phi_r}\ket{1}-2\mbraket{\phi_1}{\psi}\ket{\phi_1}\ket{1} \\ & +2\mbraket{\phi_r}{\psi}\mbraket{\phi_r}{\phi_1}\ket{\phi_1}\ket{1}.
\end{split}
\end{equation}
Note that $|\mbraket{\phi_1}{\psi}| = 0$ and $|\mbraket{\psi}{\phi_r}|^2 = \gamma^2$ and $|\mbraket{\phi_1}{\phi_r}|^2 = 1 - 2\gamma^2$. Then according to \thmref{th:qe-fidel}, the fidelity of the emulation for both states is:
\begin{equation}
    F(\ket{\omega}\bra{\omega}, \Ue \ket{\psi}\bra{\psi} \Ue^{\dagger}) \geq \gamma^2(1+4(1-2\gamma^2)^2)
\end{equation}

Now we need to compare this probability with the $P_{ov}$ probability which is $P_{ov}(3, \mu) = 1-\mu^3$ since the size of the learning phase includes 3 queries. We write the effective success probability of the adversary as: 
\begin{equation}\label{eq:prob-win-selective-two-messages}
\begin{split}
    Pr_{forge}[\A(m_2)] & = Pr_{forge}[\A(m_3)] = Pr[1\leftarrow \GCM{\F}{3, \qSel, \mu}(\lambda, \A)] - P_{ov}(3, \mu) \\
    & = \gamma^4(1+4(1-2\gamma^2)^2)^2 - (1-\mu^3)
\end{split}
\end{equation}

Finally, we do a functional analysis of the above probability to see in which cases it becomes non-negligible. First, we note that the success probability of the emulation attack is not greater than the trivial success probability for all the values of $\mu$ which shows that if we allow for too much overlap, the trivial attack already has a very high probability which is higher than the emulation's fidelity in this case. Next, since the highest allowed overlap is achieved when $1 - \mu = \gamma^2$, we substitute the variable $\mu$ with $1 - \gamma^2$ to find the degrees of $\mu$ for which an effective adversary exists. Hence we rewrite the winning probability of the \eqref{eq:prob-win-selective-two-messages} as follows:
\begin{equation}
\begin{split}
    Pr_{forge}[\A(m_2 \vee m_3)] & = \gamma^2(1+4(1-2\gamma^2)^2) - (1 - (1-\gamma^2)^3)) \\
    & = \gamma^2(2 - 5 \gamma^2 + 3 \gamma^4)
\end{split}
\end{equation}
Noting that the valid range for $\gamma$ is $0\leq \gamma \leq \frac{\sqrt{2}}{2}$, we plot the above function as it is shown in \figref{fig:example-prob-plot} and we can see that there is exist a valid range for $\mu$ such that the above forgery attack happens with non-negligible probability.

But more importantly, now having access to two copies of the reference state, the adversary can actually run the emulation attack twice, and produce the outputs of both $m_2$ and $m_3$ at the same time, with non-negligible probability. Thus for these values of $\mu$, we have presented an adversary who can produce effective forgery for three classical messages $m_1$, $m_2$ and $m_3$ (Note that the first learning phase query is $\ket{m_1, 0}$ which is basically a classical query and as a result, $\A$ will always have the output for $m_1$) from a classical query, and two copies of the same quantum state which shows an intuitive forgery, especially that the presented attack is independent of the size of the messages and the dimensionality of the Hilbert space of the oracle. This sort of attack cannot be captured in the definitions of unforgeability that count the queries, such as \bz. Nevertheless, our approach to defining the notion of unforgeability is capable of showing such vulnerabilities against strong quantum adversaries.
\begin{SCfigure}
     \includegraphics[width=0.50\textwidth]{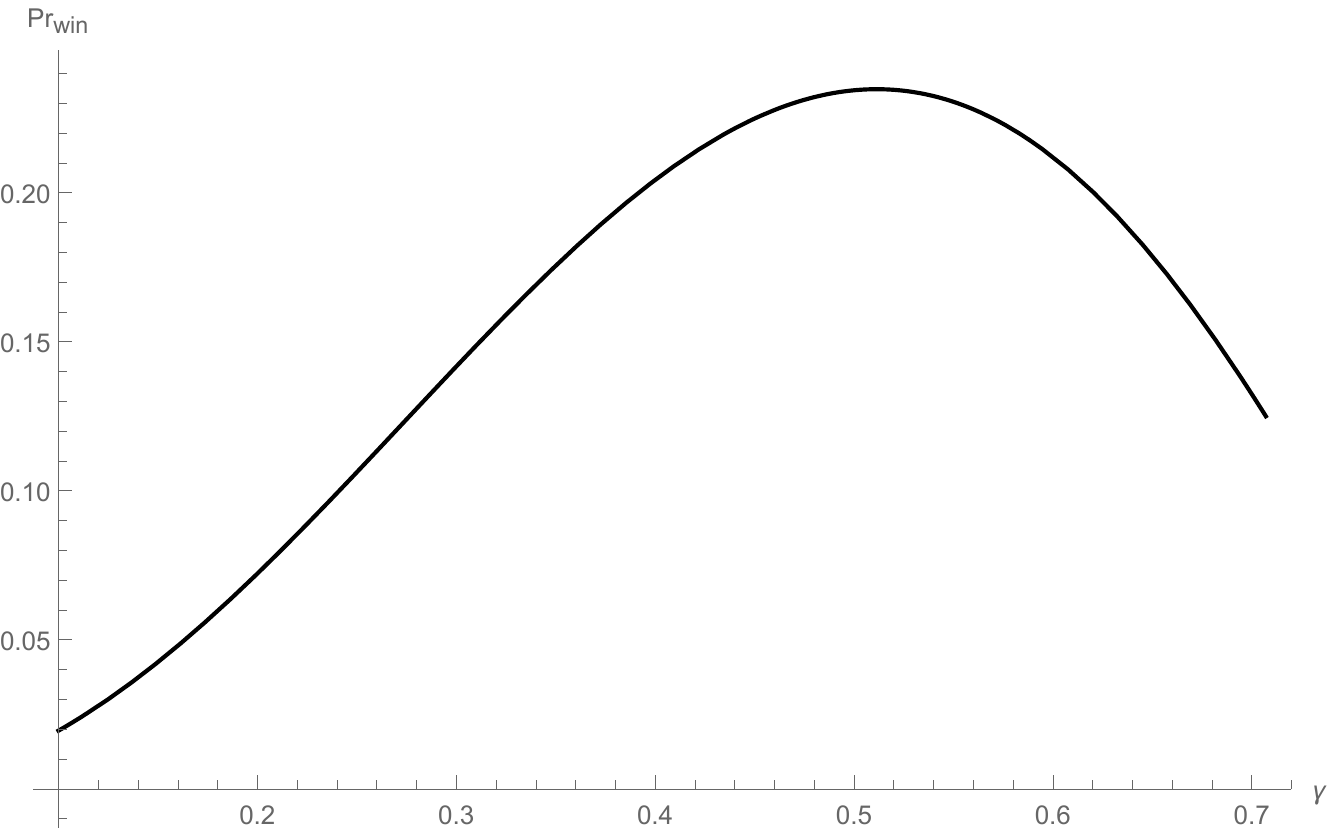}
     \caption[The winning probability of $\A$ to forge classical messages $\{m_2,m_3\}$ with the emulation attack.]{The winning probability of $\A$ to forge classical messages $\{m_2,m_3\}$ with the emulation attack. $\gamma$ represents the overlap between the learning phase query and the target message.}\label{fig:example-prob-plot}
\end{SCfigure}
\end{example}

\subsubsection{Randomised schemes:}\label{sec:sel-unf-randomized}
We have seen so far that by letting quantum adversaries exploit the power of superposition queries, they can mount effective attacks to break selective unforgeability in almost all the cases (most valid ranges of $\mu$). A relevant question here would be whether there exists any scheme that can satisfy this quite strong level of unforgeability. Since it is an impossibility for non-randomised primitives, the only possible road ahead would be to employ randomisation of the primitive. In this section, we explore how to defend against general superposition adversaries. We show that this task is possible via randomisation. Concretely, we present randomised constructions for both classical and quantum cases, which satisfy qGSU, \emph{i.e.} \sufm\ for any $\mu$. The key ingredient that allows this construction to be secure is that the randomisation has been used in an effective way such that the adversary is prevented from creating a known subspace for a specific unitary, even though they can query the challenge message in superposition. First, we start with classical primitives.

Let us first define the desired characteristic for the family of the classical functions used in our construction.

\begin{defbox}
\begin{definition}[Inter-function independent family:]\label{def:computational-function-pairwise} Let $F_k: \K \times \X \rightarrow \Y$ be a keyed family of functions with domain $\X$ and range $\Y$, where $\X=\{0,1\}^n$ and $\Y=\{0,1\}^m$. We say $F_k$ is an \emph{inter-function (pairwise) independent family} if for any efficient PPT adversary $\A$ and any two functions $F(k, .)$ and $F(k', .)$ picked uniformly at random from $F_k$, the probability of $\A$ finding an $x \in \X$ such that $F(k, x) = F(k', x)$, is negligible in the security parameter, \emph{i.e.} the following condition should hold:
\begin{equation}\label{eq:computational-function-pairwise}
    \underset{k, k' \leftarrow \K}{Pr}[x \leftarrow \A(1^{\lambda}) \wedge F(k, x) = F(k', x)] = \negl(\lambda)
\end{equation}
\end{definition}
\end{defbox}

\noindent The next step, is to show that a \prf\ family satisfies the above condition.

\begin{lembox}
\begin{lemma}\label{lemma:prf-inter-function-ind}
A \prf\ is an inter-function independent family.
\end{lemma}
\end{lembox}

\begin{proof}
We want to show that any two randomly selected functions from a PRF family, satisfy the required pairwise-independency property of \defref{def:computational-function-pairwise}. Let $F_k: \K \times \X \rightarrow \Y$ be a PRF family of functions where $|\X| = 2^n$ and $|\Y| = 2^m$. We want to show that there is no efficient adversary that can find an $x$ such that $F(k, x) = F(k', x)$ for any two different, randomly picked keys $k, k'$.
We prove by contradiction. We assume that $F_k$ is a PRF but there exist an efficient adversary $\A$ that can find at least one $x \in \X$ such that for any two randomly picked functions from $F_k$ we have:
\begin{equation}\label{eq:prf-pairwise-false}
    \underset{k, k' \leftarrow \K}{Pr}[x \leftarrow \A(1^{\lambda}) \wedge F(k, x) = F(k', x)] = \nonnegl(\lambda).
\end{equation}

Now we construct a new family of functions from $F_k$ which is a PRF. Let $F'_{k,k'}: \K^2 \times \X \rightarrow \Y$ be constructed as follows:
\begin{equation}
    F'((k,k'), x) = F(k, x) \oplus F(k', x)
\end{equation}
It is a well-known example in the literature that if $F_k$ is a PRF, then $F'_{k,k'}$ is also a PRF.
Now we show that if the \eqref{eq:prf-pairwise-false} holds, then there also exist an adversary who can distinguish $F'((k,k'), x)$ form truly random function. Let $\A'$ query the same $x'$ that has been found by $\A$. If $\A'$ queries $F'((k,k'), x)$, since $F(k, x') = F(k', x')$ with non-negligible probability, then the queries to $F'((k,k'), x)$ on $x'$ should return $0^m$. On the other hand the queries to the truly random function will return random bit-strings. As a results, $\A'$ can distinguish $F'((k,k'), x)$ from a truly random function which is a contradiction and hence we have proved that \prf\ satisfies the \defref{def:computational-function-pairwise}.
\end{proof}

We can now give our construction based on \prf s or more generally, based on any family of classical functions satisfying the \defref{def:computational-function-pairwise}.

\begin{constbox}
\begin{construction}\label{const:classical-random-const-prf}
Let $F: \K \times \X \rightarrow \Y$ be a \prf\ (or any other family satisfying \defref{def:computational-function-pairwise}). Let $\R = \K = \{0,1\}^l$ be the randomness space. And let $\lambda$ be the security parameter and $l$ be polynomial in $\lambda$. The construction is defined by the following key generation algorithm, keyed evaluation algorithm, and keyed verification algorithm:
\begin{itemize}
    \item{\bf Key generation:} The secret key is picked uniformly at random from $\K$: $k \xleftarrow{\text{\$}} \K$
    \item{\bf Evaluation:} The evaluation under key $k$ on input $m$ picks randomness $r$ and applies $F(k\oplus r, \cdot)$ to $m$. Note that when responding to a quantum query, the same randomness is used for all the states of the superposition.
    \begin{itemize}
        \setlength\itemsep{-0.2em}
        \item On input $m\in \X$:
        \item $r \xleftarrow{\text{\$}} \R$
        \item Return $F(k \oplus r, m)||r$ 
    \end{itemize}
    \item{\bf Verification:} The verification under key $k$ of a pair $(m, (t, r))$, runs the evaluation algorithm on $m$ under $k$ with randomness $r$, and checks equality with $t$.
    \begin{itemize}
    \setlength\itemsep{-0.2em}
        \item On input $(m, (t, r))\in \X \times (\Y \times \R)$:
        \item If $F(k \oplus r, m) = t$ return $\top$, otherwise return $\bot$
    \end{itemize}
\end{itemize}
\end{construction}
\end{constbox}

\noindent We show that this construction satisfies \sufm\ security. 

\begin{thmbox}
\begin{theorem}\label{th:classical-random-const-prf-secure}
\constref{const:classical-random-const-prf} is qGSU secure.
\end{theorem}
\end{thmbox}

\begin{proof}
We prove by contraposition. Let us assume there exists a QPT adversary $\A$ who plays the \sufm\ game where the evaluation is according to \constref{const:classical-random-const-prf} and wins with non-negligible probability in the security parameter \emph{i.e.} $\A$ wins the game by producing a valid tag $t^*$ for their selected message $m^*$ and randomness $r^*$ with the following probability:
\begin{equation}\label{eq:adv-prob-wining-suf}
    Pr[1\leftarrow \GCM{\F}{q, \qSel, \mu}(\lambda, \A)] - P_{ov}(q_r, \mu) = \nonnegl(\lambda)
\end{equation}
Where the verification algorithm checks if $F(k \oplus r^*,m^*) = t^*$. We introduce the following games:
\begin{itemize}
    \item \texttt{Game 0.} This game is the \sufm\ for \constref{const:classical-random-const-prf}, where $F(k \oplus r, .)$ is picked from $F$.
    \item \texttt{Game 1.} This game is similar to \texttt{Game 0}, except that $\A$ needs to produce forgery for a $r^*$ which is one of the previously received random values of $\{r_i\}^q_{i=1}$ in the learning phase.
\end{itemize}

First, it is straightforward that the probability of the adversary winning \sufm\ in \texttt{Game 0}, is at most negligibly higher than winning \texttt{Game 1}. Since $r_i$ in both cases have been picked independently and uniformly at random and the probability of producing a forgery for a specific function with no query is negligible. Thus \texttt{Game 0} and \texttt{Game 1} are indistinguishable.

Now we recall the quantum oracle for this randomised construction. Let $\reO_c$ be the random oracle for both games:
\begin{equation}
    \reO_c: \sum_{m,y} \alpha_{m,y} \ket{r}_{\Ora}\ket{m,y} \rightarrow \sum_{m,y} \alpha_{m,y}\ket{r}_{\Ora}\ket{m, y \oplus (F(k \oplus r,m) || r)}
\end{equation}

Note that in each query a new function has been picked from $F$, but it is the same for all the messages in the superposition for that query.

Now we use the inter-function (pairwise) independent property of the family $F$.
The construction requires the $F$ to be a \prf\ family which is inter-function independent according to \defref{def:computational-function-pairwise}, for two randomly selected keys. Now we need to also show that $F(k\oplus r, .)$ is a \prf\ as well, with a key $k$ and any randomly selected randomness $r$, and as a result, we can use the inter-function independent property. This is clearly the case as the key $k$ and any randomness $r$ have been picked independently at random and if there exists a non-negligible advantage for the adversary to distinguish a $F(k\oplus r, .)$ from a truly random function for a value of $r$, there also exists an equivalent non-negligible advantage to distinguish a $F(k', .)$ where $k' = k \oplus r$ is a key selected uniformly at random. This is still the case even if the value $r$ becomes public after the experiment. This is in contrast with the assumption that the family is PRF, hence we conclude that $F(k\oplus r, .)$ is a \prf. Now we can rely on the \lemref{lemma:prf-inter-function-ind} that $F(k\oplus r, .)$ also satisfies the inter-function independent property and the following holds for each of the two functions drawn in any of the two queries:
\begin{equation}\label{eq:inter-func-inside-proof}
    \underset{i,j (i\neq j)}{Pr}[x \leftarrow \A(1^{\lambda}) \wedge F(k \oplus r_i, x) = F(k \oplus r_j, x)] = \negl(\lambda)
\end{equation}

As a result, we show that the adversary can at most span a one-dimensional subspace of each $U_{k\oplus r}$. To show this we will calculate the probability of $\A$ in spanning at least a 2-dimensional common subspace from two different queries. This means that $\A$ needs to find at least two bases mapping to the same 2-dimensional subspace in the output Hilbert space. Moreover, we exclude that part of $\A$'s register that contains the classical value of the randomness to only capture the Hilbert space of each $U_{k\oplus r}$. Thus let the input bases be denoted by $\ket{b} = \ket{m, z}$ where $z$ is a subset of $y$ excluding the space for the randomness, for a specific $m$. Let $\ket{e_i} = U_{k\oplus r_i}\ket{b} = \ket{z \oplus F(k \oplus r_i, m)}$ and $\ket{e_j} = U_{k\oplus r_i}\ket{b} = \ket{z \oplus F(k \oplus r_j, m)}$ be the output states from two different queries. For these output bases to have some overlap, the two functions $F(k \oplus r_i,.)$ and $F(k \oplus r_j, .)$ need to return the same classical output with high probability. Although from \eqref{eq:inter-func-inside-proof}, we have that the probability of finding such inputs that leads to a common basis is negligible:

\begin{equation}
\begin{split}
    & \underset{i,j (i\neq j)}{Pr}[\{\ket{e_i}, \ket{e_j}\} \leftarrow \A(1^{\lambda}) \wedge \mbraket{e_i}{e_j} \neq 0] \\
    & \quad\quad = \underset{i,j (i\neq j)}{Pr}[\ket{b} \leftarrow \A(1^{\lambda}) \wedge \bra{b}U^{\dagger}_{k\oplus r_i} U_{k\oplus r_j}\ket{b} \neq 0] \\
    & \quad\quad = \underset{i,j (i\neq j)}{Pr}[\ket{b} \leftarrow \A(1^{\lambda}) \wedge \mbraket{z \oplus F(k \oplus r_i, m)}{z \oplus F(k \oplus r_j, m)} \neq 0] \\
    & \quad\quad = \underset{i,j (i\neq j)}{Pr}[m \leftarrow \A(1^{\lambda}) \wedge F(k \oplus r_i, m) = F(k \oplus r_j, m)] = \negl(\lambda)
\end{split}
\end{equation}
This means that finding an even 2-dimensional common subspace between the different unitaries of the set is computationally hard for $\A$. Also since unitaries are distance preserving operators, this property holds for any sets of orthonormal basis, not necessarily the computational basis. As a result, by selecting a uniformly random function for each query, we have shown that no more than a one-dimensional subspace can be spanned for each specific unitary. 

Now we calculate the upper-bound of $\A$'s probability from a single query to a fixed unitary $U_{k \oplus r^*}$ which we denote by $U^*$ for simplicity. We recall that this query should be $\mu$-distinguishable with the quantum encoding of $m^*$. Without loss of generality, let us write $\A$'s selected query for $r^*$ as follows:
\begin{equation}
\begin{split}
        & \ket{\phi_{r^*}} = \alpha\ket{m^*, z, 0} + \beta\ket{\Omega}\ket{0},\\
        & \ket{\phi^{out}_{r^*}} = (\alpha\ket{m^*, z \oplus F(k \oplus r^*, m^*)} + \beta U^*\ket{\Omega})\ket{r^*}
\end{split}
\end{equation}
where $\ket{\Omega}$ is a normalised state that includes a superposition of a set of messages $m \neq m^*$ and as a result, $\mbraket{m^*,z}{\Omega} = 0$ and $\A$ sets the second part of the register to $0$, such that the output randomness is a separable state and it can be excluded in the rest of the proof. Due to the fact that $U^*$ is unitary, we know that $\bra{m^*, z\oplus F(k \oplus r^*, m^*)}U^*\ket{\Omega} = 0$ and hence the probability of outputting  $F(k \oplus r^*, m^*)||r^*$ from $\ket{\phi^{out}_{r^*}}$ is at most the probability of measuring it in the computation basis which is $|\alpha|^2$. This probability is maximum when $|\alpha| = |\alpha_{max}|$ which is when $\A$ uses the maximum allowed overlap of size $\sqrt{1-\mu}$. Hence we have:
\begin{equation}
    Pr[1\leftarrow \GCM{\F}{q_r,\qSel, \mu}(\lambda, \A)] \leq 1 - \mu
\end{equation}
But on the other hand we have $P_{ov}(1, \mu) = 1-\mu$, which is the lower bound for $P_{ov}(q, \mu)$, and also since there is only one query to each function selected by each $r$, and \eqref{eq:adv-prob-wining-suf} states that this probability is negligibly higher than $1 - \mu$. Thus we have reached a contradiction that concludes our proof.
\end{proof}

We point out that for this construction to be secure, we did not need to use quantum secure PRFs (\qprf) as an assumption, and the PRF assumption plus the randomisation would bring the quantum security as a byproduct. This is in contrast with most quantum-secure unforgeable schemes in the quantum world~\cite{boneh_quantum-secure_2013,zhandry_secure_2015,alagic_quantum-access-secure_2020}. Nevertheless, \qprf s can also be used in \constref{const:classical-random-const-prf}.

Now, we shall study the same problem for quantum primitives. Similar to the classical constructions, for quantum primitives too, we can use randomisation to effectively secure them. The main idea is to select a new unitary transformation for each query using a classical randomness register. In this case, we need to clarify how such randomised quantum oracles can be implemented in a way that the overall transformation remains a specific unitary. 

By recalling the abstract representation of the randomised quantum oracle that we gave in Section~\ref{sec:unf-quantum-oracles}, the input state $\ket{\psi_b} = \sum_i \alpha_i\ket{b_i}$ (where $\{\ket{b_i}\}$ is a set of orthonormal bases) is mapped to a state $U(r)\ket{\psi_b} = \sum_i \beta_i(r)\ket{b_i}$ where $U(r)$ depends on the randomness and is different for each query \emph{i.e.} the oracle uses its internal register $\ket{r}_{\Ora}$ to activate different $U(r)$ unitaries. However, for many constructions this randomness value $r$ or a function of it like $g(r)$, will be necessary for verification and hence need to also be outputted. On the other hand, the register $\ket{r}_{\Ora}$ is the internal register of the oracle re-initiated for each query and some problems may arise if the adversary gets access to this register (see \ref{sec:unf-quantum-oracles}), thus in order to be able to output this value we expand the query space and we allow the input queries to be $\ket{\mathtt{0}}\otimes\ket{\psi_b}$. We formulate the oracle as follows:
\begin{equation}
\reO_U: \ket{r}_{\Ora}\otimes \ket{\mathtt{0}}\otimes\ket{\psi_b} \rightarrow [\mathcal{I}\otimes\mathcal{I}\otimes U(r)]\ket{r}_{\Ora}\ket{r}\ket{\psi_b}
\end{equation}

\begin{figure}[ht]
   \centering
     \includegraphics[width=1\textwidth, height=0.75\textwidth]{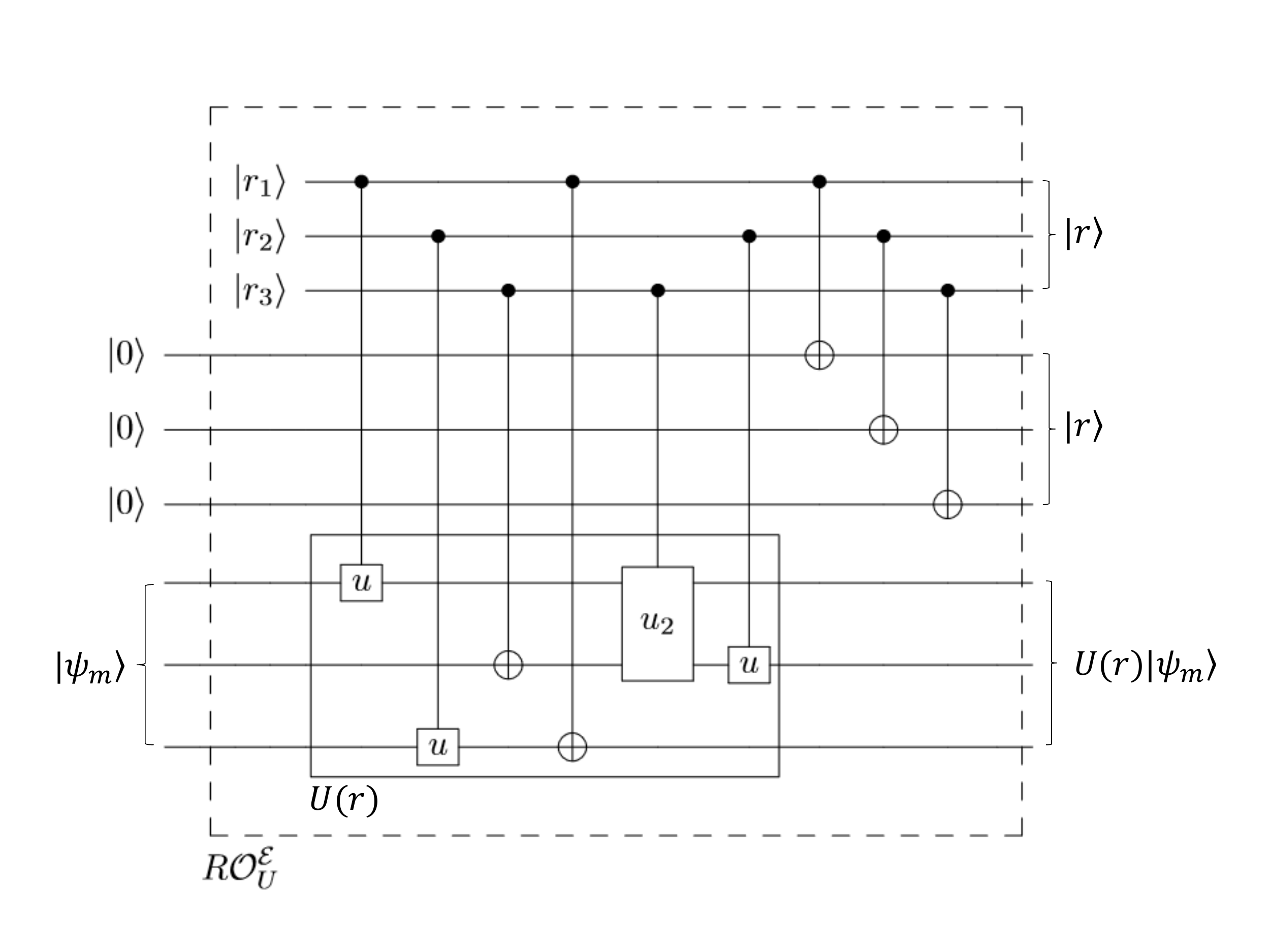}
     \caption[A sample circuit for randomised quantum oracle for quantum primitives]{A sample circuit for randomised quantum oracle for quantum primitives. On each input query $\ket{\mathtt{0}}\ket{\psi_m}$, a new randomness is initialised and the random unitary $U(r)$ acts on $\ket{\psi_m}$. The random unitary $U(r)$ consists of single and 2-qubit unitary gates selected at random in the setup phase, from a gate set required to construct any unitary $U(r)$ in the family $\mathcal{U}$ specified by the construction. These single and two-qubit gates are controlled by the randomness values $\ket{r} = \ket{r_1, r_2, r_3}$. In the last step, the classical value of randomness is recorded in the ancillary qubits of the query to be returned for verification.}\label{fig:qrandomised-oracle-circuit}
 \end{figure}

Note that for the purpose of our construction, in what follows, we assume that the ancillary state is initiated as a separable state $\ket{\mathtt{0}}$ for simplicity, although if the adversary's ancillary register has not been initiated to zero, the randomness can be XORed to that value. The above oracle can be realised in several different ways but for a better demonstration, we give an explicit example in the circuit model, shown in \figref{fig:qrandomised-oracle-circuit}. The input to the unitary evaluation of the oracle consists of two parts; one part includes the query and the second part is the internal randomness register which is initiated to a new value or equivalently to a new basis, for each query. This part in general acts as control qubits for the gates in the other part of the register that leads to applying a new overall unitary on the main query state. We note that the randomness register itself will remain untouched throughout the evaluation and finally its value is recorded in the $\ket{\mathtt{0}}$ part of the input query. Here, $\ket{r}_{\Ora}$ is always on the computational basis. We also emphasize that for our construction we do not use, nor need to use, any explicit construction for the randomised oracle and we only rely on the specified assumption.

As follows from the above discussion, in quantum primitives with such randomised oracles, the security lies in the assumptions we consider on the family of $U(r)$s generated for each $r$. For instance, it is intuitive that a primitive where $U(r)$ are Haar random unitaries can be secure since the overall adversary's state after issuing polynomial queries to the oracle is almost indistinguishable from a totally mixed state. However, this assumption might be too strong. Hence we give a construction based on \pru s which is also the quantum analogue of \prf s that we used in our previous classical construction.  

\begin{constbox}
\begin{construction}\label{const:quanum-random-pru}
Let $\F=(\ES, \E, \V)$ be a quantum primitive with the evaluation unitary $\Ue: \HilR\otimes\HilD \rightarrow \HilR\otimes\HilD$ where $D$ is the overall dimension of the query and $\HilR$ is a $2^l$ dimensional Hilbert space for the randomness. And let $\lambda$ be the security parameter and $l$ and $\log(D)$ be polynomial in $\lambda$. Also, let $\mathcal{U}_{\pru} = \{U_r\}^{L}_{r=0}$ be a \pru\ family with a cardinality $L$ to be at least $2^l$. The construction is defined as follows:
\begin{itemize}
    \item{\bf Setup:} The required parameters \texttt{param} is generated to instantiate the oracles.
    \item{\bf Evaluation:} The evaluation picks randomness $r \xleftarrow{\text{\$}} \R$ uniformly, initialises the randomness register to $\ket{r}_{\Ora}$ and applies the following unitary, on each input query $\ket{\psi_b} = \sum_i \alpha_i\ket{b_i}$ where each $U(r) = U_r \in \mathcal{U}_{\pru}$
    \begin{equation}\label{eq:random-quantum-oracle-const3}
        \reO_U: \ket{r}_{\Ora}\ket{\mathtt{0}}\ket{\psi_b} \overset{U_{\mathcal{E}}}{\rightarrow} [\mathcal{I}\otimes\mathcal{I}\otimes U(r)]\ket{r}_{\Ora}\ket{r}\ket{\psi_b}
    \end{equation}
    \item{\bf Verification:} The verification oracle calls a quantum test algorithm $\T$ as defined in \defref{def:test} on $U(r)\ket{\psi_b}\bra{\psi_b}U(r)^{\dagger}$ and the tag state $\rho_t$:
    \begin{itemize}
    \setlength\itemsep{-0.2em}
        \item If $F(\rho_t, U(r)\ket{\psi_b}\bra{\psi_b}U(r)^{\dagger}) = 1 - \negl(\lambda)$ return $\top$ with a probability $1 - \negl(\lambda)$
        \item and $Pr[1 \leftarrow \T[(\Ue\rho_{\delta}\Ue^{\dagger})^{\otimes \kappa_1}, (\Ue \rho_m \Ue^{\dagger})^{\otimes \kappa_2}]] = \negl(\lambda)$ for any state $\rho_{\delta}$ with $\delta^2$-indistinguishable from $\rho_m$.
    \end{itemize}
\end{itemize}
\end{construction}
\end{constbox} 

\noindent Now let us prove the selective unforgeability of the above construction. 

\begin{thmbox}
\begin{theorem}
\constref{const:quanum-random-pru} is \sufm\ secure for any $\mu \geq 1-\delta^2$.
\end{theorem}
\end{thmbox}

\begin{proof}
We prove by contraposition. Let $\A$ be a QPT adversary who plays the \sufm\ game where the evaluation oracle is as shown in the \eqref{eq:random-quantum-oracle-const3}, and wins with non-negligible probability in the security parameter \emph{i.e.} $\A$, wins the game by producing a valid tag $\rho_t$ for their selected message $\rho_m$ and randomness $r^*$ with the following probability, after interacting with the oracle in the learning phase:
\begin{equation}
    Pr[1\leftarrow \GCM{\F}{\qSel, \mu}(\lambda, \A)] - P_{ov} = \nonnegl(\lambda)
\end{equation}
Where the $P_{ov} = Pr[1 \leftarrow \T(\rho^{out}_{max})^{\otimes \kappa_1}, (\Ue \rho_m \Ue^{\dagger})^{\otimes \kappa_2}]$ according to \defref{def:pov-quantum}, and $\rho^{out}_{max}$ is query with maximum allowed overlap from $\mu$-distinguishability condition. Since the construction implies that $P_{ov} = \negl(\lambda)$, this means:
\begin{equation}\label{eq:const3-adversary-wins-probability}
    Pr[1\leftarrow \GCM{\F}{\qSel, \mu}(\lambda, \A)] = \nonnegl(\lambda)
\end{equation}
Consequently, $\A$ can produce an output $\rho_t$ with non-negligible fidelity with the actual output $\U(r^*)\rho_m\U(r^*)^{\dagger}$, for a $\U_{r^*} \in \mathcal{U}_{\pru}$.
Now we consider two cases. Either $r^*$ is one of the randomnesses that $\A$ has received during the learning phase, which means $\A$ can closely approximate the output of a random unitary $U(r^*)$ from a single query, or $r^*$ is a new randomness value, for a new random unitary $U(r^*)$ where $\A$ has no query on it. We will show that each case leads to a contradiction.

First, we show that $\A$'s output state after the learning phase, \emph{i.e.} $\sigma_{out}$ cannot include more than a one-dimensional subspace of each of the $U(r)$ unitaries. To cover a subspace with a dimension of at least two, $\A$ needs to find a common output basis from two different queries. On the other hand, we note that as shown in~\cite{shacham_pseudorandom_2018}, any PRUs are generators of PRS that are a family of quantum states computationally indistinguishable from Haar measure. Hence the joint output states $\sigma_{out}$ is also indistinguishable from Haar random states for $\A$ who is a QPT adversary. Now if $\A$ can find a common output subspace, it means that there are at least two states, corresponding to the bases of the 2-dimensional subspace, that are indistinguishable (or $0$-distinguishable according to \defref{def:prelim-dist}), and hence $\A$ can use those queries to distinguish the distribution of states $\sigma_{out}$ and a Haar random distribution which contradicts the fact that the oracle will generate a PRS set of states after $q$ queries.
Now we show that each case will lead to a contradiction. We start with the second case where if $\A$ produces an indistinguishable (concerning $\T$) output for a random unitary with no query, then $\A$ can perform the learning phase locally without any interaction with the oracle and hence produce the output of any unitary picked from a family indistinguishable to Haar measure, which is a clear contradiction.
For the first case, relying on the previous argument, we rewrite the learning phase states of the $\A$ after $q$ queries, as follows:
\begin{equation}
    \sigma_{in} = \ket{\phi_{r^*}}\bra{\phi_{r^*}} \otimes \sigma^{q-1}_{in}, \quad \sigma_{out} = U_{r^*}\ket{\phi_{r^*}}\bra{\phi_{r^*}}U_{r^*}^{\dagger} \otimes \sigma^{q-1}_{out}
\end{equation}
where $\ket{\phi_{r^*}}$ is the query associated to $U_{r^*}$ for which $\A$ produces a forgery and $\sigma^{q-1}_{in}$ and $\sigma^{q-1}_{out}$ are the input and output states of the remaining $q-1$ query respectively. We note that $\sigma^{q-1}_{out}$ consists of $q-1$ quantum states with a distribution $\delta$ over a $D'$-dimensional Hilbert space s.t. $\delta$ is Haar-indistinguishable. Furthermore, the ancillary register where the $r$ is encoded consists of $q$ independent random values. Now let us construct an adversary $\A'$ who is a \pru\ distinguisher. Let $\A'$ interact with a unitary $U$ either selected from $\mathcal{U}_{\pru}$ or from Haar measure, and query a state $\ket{\phi_{r^*}}$ as described above, and returns $U\ket{\phi_{r^*}}$ together with an ancillary register $\ket{r}$ where $r$ picked uniformly at random. Then $\A'$ also locally creates $q-1$ Haar-random states and returns to $\A$ as the $ \sigma^{q-1}_{out}$. Then $\A'$ also queries $\rho_m$ from the oracle. Now $\A'$ uses the same test algorithm $\T$ to check the output of $\A$ \emph{i.e.} $\rho_t$ with the the oracle's output for the last query which is $U\rho_m U^{\dagger}$. From \eqref{eq:const3-adversary-wins-probability}, we know that this probability is non-negligible, while as for a Haar random unitary the probability is negligible, thus can conclude that 
\begin{equation}
    |\underset{r \leftarrow \R}{Pr}[\A'^{U_r}(1^{\lambda})=1] - \underset{U \leftarrow Haar}{Pr}[\A'^U(1^{\lambda})=1]| = \nonnegl(\lambda).
\end{equation}
which is a contradiction and the theorem has been proved.
\end{proof}

We conclude that even though generalised quantum selective unforgeability is too strong to be attained by deterministic schemes, one can come up with randomised constructions that satisfy even this strong level of unforgeability in the quantum world.  

\subsection{Generalised universally unforgeable schemes}
We now draw our attention to the weakest notion of unforgeability in the hierarchy of our definitions and provide results for the universal unforgeability of different schemes. We recall that here the adversary receives a challenge picked by the challenger uniformly at random from the full message space. We need to emphasise that universal unforgeability is the most useful notion of unforgeability for our purpose, despite being the weakest. From now on and throughout the thesis we will mainly use this definition and explore its close relation to unclonability in the same context as we have discussed in this chapter. We will also study the universal unforgeability of different primitives and protocols in future chapters.

But for now, for the sake of completeness and to complete the investigation of unforgeable schemes in our framework, we give two straightforward results for classical and quantum primitives.

\begin{corrbox}
\begin{corollary}\label{cor:qprf-uuf}
\qprf s are \uuf\ secure.
\end{corollary}
\end{corrbox}
\begin{proof}
This is a direct implication of \thmref{th:qprf-1euf} where we have proved that \qprf s are \suf\ secure and \thmref{th:suf-uuf} showing that \suf\ implies \uuf.
\end{proof}

We can also show that quantum \pru\ primitives are generally \uuf\ secure. 

\begin{corrbox}
\begin{corollary}
Deterministic quantum primitives based on \pru\ are \uuf\ secure.  
\end{corollary}
\end{corrbox}
\begin{proof}
From \thmref{th:qprf-1euf} we know that \pru\ primitives are \suf\ secure. Also from \thmref{th:suf-uuf}, we have shown that \uuf\ is weaker than \suf. Thus any \pru\ primitive is \uuf\ secure.
\end{proof}

In \gameref{game:unf-full}, we have also introduced a second learning phase, after the challenge phase to capture universal unforgeability against stronger adaptive attack models. Here we also give a general no-go result for \uuf\ security of quantum primitives against such adversaries. This attack model is stronger than the usual chosen-message attack considered for universal unforgeability and is particularly interesting for quantum primitives. This is because for a quantum primitive, the adversary receives an unknown quantum state from the challenger and enabling the second learning phase does not lead to a trivial attack. We call this attack model, an adaptive-universal attack (\emph{aua}). Nevertheless, we can show that a quantum adversary who can use entanglement can break the \uuf\ security of any deterministic primitive if the second learning phase is allowed. We show this specific instance of the game as $\GCM{\F}{\qUni-aua, \mu}(\lambda, \A)$ and we note that again this instance should be parameterised with $\mu$ since a trivial attack can be mount if $\A$ tries to query the challenge phase again in the second learning phase. We present the result in the following theorem. However, we leave the proof for \appref{app:uni-adaptive}.

\begin{thmbox}
\begin{theorem}[No quantum non-randomised primitive $\F$ is aua-\uuf\ secure]\label{th:unf-uni-aua} For any deterministic quantum primitive $\F$ and for any $\mu$ such that $0 \leq \mu \leq 1-\nonnegl(\lambda))$, there exists a QPT adversary $\A$ such that
\begin{equation}
Pr[1\leftarrow \GCM{\F}{\qUni-aua, \mu}(\lambda, \A)] = \nonnegl(\lambda).
\end{equation}
\end{theorem}
\end{thmbox}

\subsubsection{Relationship between universal unforgeability and learnability}\label{sec:unf-relation-uuf-pac}
Finally, we show a connection between universal unforgeability and the notion of function learnability, which we discussed in sections \ref{sec:prelim-learning-theory} and \ref{sec:unclone-different-learning}. More precisely, we show that universal unforgeability implies unlearnability in the PAC-learning setting. To do so, first, we need to clarify some technical remarks concerning the universal unforgeability to be able to link it with PAC-learnability.

First, we note that in \gameref{game:unf-full} the learning phase is characterised by interaction with a general oracle $\eO$, which is included in the primitive $\F$. Here, to establish our result we assume that the oracle for the primitives of interest is a \emph{quantum example oracle} (QPEX) as defined in \eqref{eq:prelim-qpex-oracle}. However, unlike \defref{def:prelim-pac-learn-qpex}, in order to capture the chosen-message attack model that we consider in \gameref{game:unf-full}, we assume that the adversary gets to choose the distribution $\D$ (but not the selected function $f$ from the concept class $\Conc$). We argue that since the QPEX returns a quantum state of the superposition of all the inputs $m$ with uniform weight over the distribution $\D$, for this type of oracles, choosing the distribution will be the equivalent of choosing the input quantum state (or its efficient classical description) of the oracle and receiving the respective quantum output. Therefore, we do not need to make a significant change in the learning phase of our game in order to capture this scenario. Also, in the $\qUni$ challenge phase, the message is not chosen uniformly, but from the distribution $\D$. We refer to this variant of universal unforgeability as \emph{universal unforgeability under distribution $\D$}. Now, we can establish the following theorem:

\begin{thmbox}
\begin{theorem}\label{th:unf-uniunf-learnability}
Any family of universally unforgeable functions $\Conc$, over distribution $\D$, is not PAC-learnable over $\D$.
\end{theorem}
\end{thmbox}
\begin{proof}
Let $f \in \Conc$ be the evaluation function of a primitive $\F$ that is universally unforgeable under distribution $\D$, then by the definition of universal unforgeability, for any QPT adversary $\A$ who can make up to polynomial copy to the oracle, we have:
\begin{equation}
\underset{m\in\D}{Pr}[1\leftarrow \GCM{\F}{\qUni}(\lambda, \A)] \leq \negl(\lambda).
\end{equation}
Let the verification algorithm check the equality of adversary's forgery, \emph{i.e.} $t = h(m)$ with precision $\epsilon$, that is the verification algorithm will pass the forgery if the following holds: 
\begin{equation}
\underset{m\in\D}{Pr}[h(m) \neq f(m)] \leq \epsilon.
\end{equation}
Thus we can rewrite the universal unforgeability of $f$ as follows:
\begin{equation}\label{eq:unf-uniD-learning}
\underset{m\in\D}{Pr}[\underset{m\in\D}{\mathbb{E}}[h(m) \neq f(m)] \leq \epsilon] \leq \negl(\lambda).
\end{equation}
Now we assume a learner $\A_p$. We note that according to the definition of PAC-learnability with a QPEX oracle, the adversary gets samples from an unknown distribution $\D$. While as in the universal unforgeability, the adversary gets to choose a new desired distribution $\D_i$ for every query, where $i$ denotes the index number of the query. We note that for each selected function $f \in \Conc$, the learner $\A_p$, is weaker than $\A$. We denote the hypothesis of $\A_p$ as $h_p$, we then have:
\begin{equation}
\underset{m\in\D}{Pr}[\underset{m\in\D}{\mathbb{E}}[h(m) \neq f(m)] \leq \epsilon] \leq \underset{m\in\D}{Pr}[\underset{m\in\D}{\mathbb{E}}[h_p(m) \neq f(m)]
\end{equation}
From \eqref{eq:unf-uniD-learning} we have that the success probability $\A$ is bounded by a negligible value, thus, the probability of learner $\A_p$, in successfully outputting a hypothesis $h_p(m)$ such that $\underset{m\in\D}{Pr}[h_p(m) \neq f(m)] \leq \epsilon$ is also $\negl(\lambda)$. For $\Conc$ to be PAC-learnable, this probability needs to be $1-\delta$ for every function $f \in \Conc$, while here the $\delta$ can only be negligibly close to 1. This concludes that $\Conc$ is not PAC-learnable.
\end{proof}

We have shown a link between PAC-learning (with QPEX oracle) and quantum universal unforgeability. One can see from the above result that even if one of the functions in the family (concept class) is universally unforgeable, it is enough to show that the family is \emph{not} PAC-learnable since PAC-learning requires the learner to learn \emph{all} the functions in the concept class with the specified conditions.

\section{Discussion and conclusions}
We have seen in this chapter, how unclonability and especially the unclonability of quantum operations is related to the lack of information, which we characterise with the notion of unknownness. Looking at unclonability from this angle allowed us to expand our horizons into the realm of quantum randomness, cryptography and learning theory. We have discussed the connection between unclonability and unforgeability and between unforgeability and different other notions of learning. We have also talked about emulation as a learning mechanism that can be used as a new class of attacks. On the same note, we have studied a quantum emulation algorithm and developed some simple attacks based on the freshly provided analysis of the algorithm.
More importantly, we have developed a universal and generalised framework for unforgeability in the quantum world. Unforgeability will become one of the principal components of this thesis and we will use the definitions and results of our framework in all the remaining chapters (except \chapref{chap:varqlone}). Additionally, our case studies on different quantum and classical primitives in this chapter have shown that the generalised quantum unforgeability has shown the applicability of our framework and has also led us to propose both quantum and classical primitives that are secure against powerful quantum adversaries in a strong quantum security model. The first interesting future direction to this work would be to construct efficient and practical constructions for selective and universal unforgeability. These constructions can serve as quantum-secure MACs. Also, building efficient randomised oracles for quantum primitives using random quantum circuits or t-designs is an interesting future research direction. 

We have also discussed the close relation between unforgeability and unclonability in the context of quantum money. A potentially attractive application of our framework would be for quantum money schemes. A question that can be of interest is whether one can design quantum money schemes with different levels of unforgeability, as captured in our framework, and to what extent they can be practical? 

As our final contributions, we have formalised our intuitive arguments about the correspondence between unforgeability and learning theory by showing a connection between PAC-learning with a quantum example oracle and a slightly different variant of universal unforgeability. Since in some cases proving universal unforgeability might be easier than PAC-learnability, this result can potentially give some criteria or cryptographic measures for checking the learnability of concept classes. Moreover, we conjecture that a similar result can potentially be obtained for quantum primitives, using the definitions of fully quantum PAC-learnability that exist in the literature, such as \cite{heidari_theoretical_2021,padakandla_pac_2022}. Nonetheless, we leave the investigation of this problem as future work. 

Lastly, the link between cryptography and learning theory, especially in the quantum world and in the presence of fascinating phenomena such as unclonability, is an appealing and, in some ways fundamental area of research that we could only slightly touch upon in this chapter. Exploiting more novel quantum learning techniques such as shadow tomography and classical shadow \cite{aaronson_shadow_2020,huang_predicting_2020} for cryptanalysis would be the next step in this line of research.

\chapter{Quantum Physical Unclonable Functions} \label{chap:qpuf}
\begin{chapquote}{Marcus Tullius Cicero}
``In everything truth surpasses the imitation and copy.''
\end{chapquote}
\section{Introduction}\label{sec:qpuf-intro}
In the previous chapter we discussed cryptographic properties that are related to unclonability and we defined a new security framework, as well as attack tools to expand the study of the unclonability of quantum processes from a cryptographic point of view. In this chapter, we introduce a different form of unclonability which is, neither restricted to quantum systems nor originated from quantum mechanics. Yet, this notion of unclonability is also a natural property of certain physical systems which emerges from uncontrollable imperfections, randomness and physical disorders. We refer to this type of unclonability as \emph{Physical Unclonability} adopted from the term \emph{Physical Unclonable Function (PUF)} originally introduced in hardware security. We bring the physical unclonability to the quantum world, and we study it as an abstract mathematical notion concerning the previously introduced concepts in cryptography and quantum information.

\begin{figure}[ht!]
\includegraphics[scale=0.50]{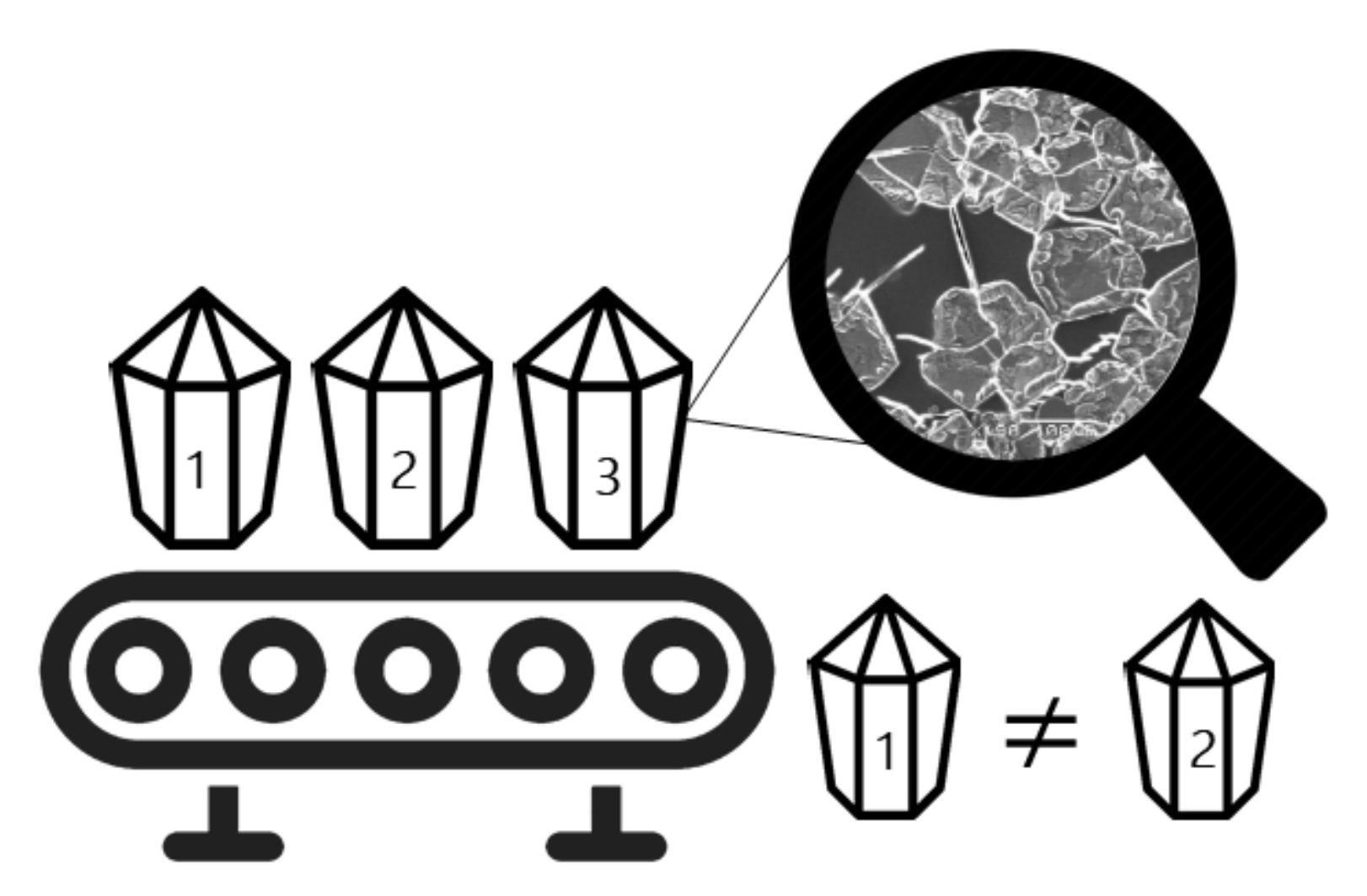}
\centering
\caption[Illustration of the concept of a physical unclonable function]{Illustration of the concept of a physical unclonable function showing that the underlying microscopic structure and the randomness appearing in the manufacturing process make each PUF device a unique one.}
\label{fig:puf-concept}
\end{figure}

But first, let us intuitively describe physical unclonability. Imagine a factory that produces crystals for optical laboratories. The manufacturer intends to produce many copies of the same crystal over and over, with common specific optical properties such as scattering factors. Nevertheless, often no matter how good the product line is and how accurate the devices are, no two crystals produced by this factory are exact clones of each other. This is due to the fact that many uncontrollable parameters and physical randomnesses are involved in the process of crystal formation. Hence on some small scale, the internal structures of two crystals, even though sharing the same large-scale properties, are very different (\figref{fig:puf-concept}), which makes each crystal unique at that level. Now, if such unique features can be somehow detected (for instance, by shining light on the crystal, which results in producing a unique scattering pattern), one can use each of such crystals as a physical key. This key has a notable feature that not even the key-maker can have a copy of. Our example crystal or any similar physical device for that matter is called a physical(ly) unclonable function or a physical unclonable key.

Thus PUFs are hardware structures designed to utilize this random and uncontrollable physical disorder that appears in any physical device during the manufacturing process. The behaviour of a PUF is usually equivalent to a set of Challenge-Response Pairs (CRPs) which are extracted through physically querying the PUF and measuring its responses (In our previous example, the optical parameters of the light are the challenge and the scattering pattern produced by the crystal is the response). The PUF's responses depend on its physical features and are assumed to be fundamentally unpredictable, \emph{i.e.} even the manufacturer of the PUF, with access to many CRPs, cannot easily predict the response to a new challenge \cite{ruhrmair_pufs_2014}. This property makes PUFs different from other hardware tokens in the sense that the manufacturer of a hardware token is usually completely aware of the behaviour of the token they have built \cite{brzuska_physically_2011}. 

In classical cryptography, physical unclonability is often considered as a \emph{hardware assumption}. Considering hardware assumptions in cryptography, originated from an impossibility result by  
Canetti and Fischlin \cite{goos_universally_2001} on the impossibility of achieving secure cryptographic protocols without any setup assumptions. This result has motivated a rich line of research investigating the advantages of making hardware assumptions in protocol design. The idea was first introduced by Katz in \cite{katz_universally_2007} and attracted the attention of researchers and developers as it adopts physical assumptions and eliminates the need to trust a designated party or to rely on computational assumptions. Among different hardware assumptions, PUFs have hugely impacted the field \cite{badrinarayanan_unconditional_2017}.

So far, the cryptographic literature has mainly considered what we will call classical PUFs (or cPUFs/CPUF). This includes, on an abstract level, the physical systems modelled by a classical function and restricted to classical CRPs. Most common cPUFs are electronic devices such as Arbiter PUFs \cite{gassend_silicon_2002}, Ring-Oscillator based PUFs \cite{suh_physical_2007,delavar_ring_2016} and SRAM PUFs \cite{guajardo_fpga_2007}. Optical PUFs were also introduced as cPUFs by Pappu \emph{et al.} \cite{pappu_physical_2002}. For a comprehensive overview of existing PUF structures, we refer the reader to \cite{maes_physically_2013,halak_physically_2018}. Even though cPUF as an unclonable token is highly appealing and of interest for several real-world applications \cite{chang_retrospective_2017,herder_physical_2014,delavar_puf-based_2017,ameri_provably_2019,marchand_implementation_2018,liu_xor-based_2019,mukhopadhyay_pufs_2016}, most cPUFs suffer from two major problems. First, most cPUFs generate only a finite and usually very limited number of completely independent CRPs \cite{chang_retrospective_2017} which is not ideal for many of the mentioned applications. Second, most of them are vulnerable against different attacks like side-channel \cite{tebelmann_side-channel_2019,chang_retrospective_2017} and machine-learning \cite{ganji_strong_2016,ruhrmair_pufs_2014,ruhrmair_modeling_2010,khalafalla_pufs_2019}. In other words, they are not as unpredictable and unclonable as they were initially assumed to be. The aforementioned shortcomings of classical PUFs, on one hand, and their importance as a hardware security primitive in practice, on the other hand, call for investigating other potential physical unclonability in other areas of physics and cryptography. The quantum realm specifically, is one of the best areas to look for such phenomena for several reasons. First, the fundamental unclonability of quantum systems brings forward a potential advantage for achieving a stronger notion of physical unclonability. Second, from the point of view of physics, usually many random disorders that lead to physical unclonability, happen on the atomic and subatomic scale, ruled by the laws of quantum mechanics. Thus having a framework for the study of PUFs as a quantum objects seems to be much more informative. Third, quantum operations can usually generalise classical operations, and if defined carefully, the quantum analogue can encompass classical PUFs, leading to a better understanding of physical unclonability in general. And finally, from a cryptanalysis point, the recent advances in quantum technologies give rise to the question that whether quantum technologies can boost the security of cPUFs or if they, on the contrary, threaten their security. As we will argue later, some more promising classical PUFs such as optical PUFs, \emph{are} quantum devices and can be attacked by a quantum adversary who exploits the power of quantum states and quantum algorithms. Hence to achieve any PUF-based application in the quantum era, the security needs to be properly analysed in a setting that includes quantum adversaries. To conclude, this is one of the few areas of research that lies in the intersection of fundamental physics and cryptography, and human curiosity calls for its theoretical exploration. 

In the current chapter, we address the general and formal treatment of PUFs in the quantum world by defining quantum PUFs (qPUFs) as a quantum token/process that can be challenged with quantum states and output quantum states as a response. Our mathematical framework for qPUFs as a new quantum primitive is inspired by the theoretical literature of classical PUF, while we take into account the full capabilities of a quantum adversary. Similar to cPUF, not any function and process can be considered as a PUF and several requirements need to be satisfied. We identify the requirements a qPUF needs to meet to provide the main security property required for most of the qPUF-based applications, that is \emph{unforgeability}\footnote{\emph{Unpredictability} and \emph{unclonability} are other equivalent terms for this notion used often in the literature.}. One of the main breakthroughs here is to show that the requirements of qPUFs are more restricted than their classical counterparts to achieve the same functionality promises, \emph{i.e.} for qPUFs, the unpredictability is satisfied on a more fundamental level and under fewer assumptions. However, it is worth mentioning that in this chapter we do not focus on the practical constructions for qPUF as designing and implementing concrete qPUFs that satisfy our proposed level of security, remains a challenging task that we will slightly touch upon in the last chapter.

\subsection{Structure of the chapter}
We begin by giving a short background on classical PUF, in Section~\ref{sec:qpuf-classic-puf}. Then in Section~\ref{sec:qpuf-defs} we define qPUFs as general quantum channels and formalize the standard requirements of robustness, uniqueness and collision-resistance for qPUFs guided by their classical counterparts. We will show that given all the requirements, black-box unitary transformations are perfect candidates for qPUFs. We then formally define the notion of Unitary Quantum PUF (UqPUF). We also discuss the importance of the notion of  \emph{unknownness}, as defined in \defref{def:unk-uni} in \chapref{chap:unf-tools}, as the minimal assumption that leads to the unclonability of qPUFs.

In Section \ref{sec:qPUF-crypt}, we use our unforgeability game-based framework to study the security or unforgeability of general qPUFs. Also using the quantum emulation attacks and techniques that we have introduced in the previous chapter, we demonstrate successful attacks on qPUFs for some security levels. This leads to a general impossibility result for qPUFs. In doing so we establish several possibility and impossibility results. On the other hand, we formally prove that any qPUF provides \emph{quantum universal unforgeability}, \emph{i.e.} no QPT adversary can, on average, generate the response of a qPUF to random challenges. This is the main possibility result of this chapter, which shows a promising direction for research on quantum PUFs. 

We conclude the chapter with a discussion and conclusion in Section~\ref{sec:qpuf-disc}. More specifically, we discuss the relevance of our definitions and security framework for other related types of PUF, including classical PUFs. We will argue how our proposed attacks can threaten the security of some of the existing PUF proposals and, we suggest a solution for making them secure by employing our framework and results.

\subsection{Related works}
The concept of Physical Unclonable Functions was first introduced by Pappu \emph{et al.} \cite{pappu_physical_2002} in 2001, devising the first implementation of an Optical PUF. Optical PUFs were subsequently improved to generate an independent number of CRPs~\cite{mesaritakis_physical_2018}.  

More recently, the concept of \emph{quantum read-out of PUF (QR-PUF)} was introduced in \cite{skoric_quantum_2010} to exploit the no-cloning feature of quantum states to solve the spoofing problem in remote device identification protocols. The QR-PUF-based identification protocol has been implemented in \cite{goorden_quantum-secure_2014}. In addition to the security analysis of this protocol against intercept-resend attack in \cite{skoric_quantum_2010}, its security has also been analysed against other special types of attacks targeting extracting information from an unknown challenge state \cite{skoric_security_2013,yao_quantum_2016}. In another work, \cite{nikolopoulos_continuous-variable_2017}, the continuous variable encoding is exploited to implement another practical QR-PUF based identification protocol. The security of this protocol has also been analysed only against prepare-and-resend attacks \cite{nikolopoulos_continuous-variable_2018,fladung_intercept-resend_2019}. Moreover, some other applications of QR-PUFs have been introduced in \cite{skoric_authenticated_2017} and \cite{uppu_asymmetric_2019}. However, all these prior similar works can be considered special cases of qPUFs and in a restricted security setting. (for more discussion, see Section~\ref{sec:qpuf-disc})

In another independent and parallel recent work, Gianfelici et al. have presented a common theoretical framework for both cPUFs and QR-PUFs \cite{gianfelici_theoretical_2020}. They quantitatively characterise the PUF properties, particularly robustness and unclonability. They also introduce a generic PUF-based identification scheme and parameterise its security based on the experimental implementation of PUF.

\section{Background on classical Physical Unclonable Functions} \label{sec:qpuf-classic-puf}
In this section, we briefly present the formal definition of PUFs as found in the classical literature \cite{armknecht_towards_2016,ruhrmair_puf_2014,brzuska_physically_2011}. Let a $\mathcal{D}$-family be a set of physical devices generated through the same manufacturing process. Due to unavoidable variations during manufacturing, each device has some unique features that are not easily clonable. A PUF is an operation making these features observable and measurable by the holder of the device. 

As in \cite{armknecht_towards_2016,brzuska_physically_2011}, we formalize the manufacturing process of a PUF by defining the $\mathrm{Gen}$ algorithm that takes the security parameter $\lambda$ as input and generates a PUF with an identifier $\mathbf{id}$. Note that each time the $\mathrm{Gen}$ algorithm is run, a new PUF with new $\mathbf{id}$ is built. So, we have:
\begin{equation}
    \mathbf{id} \leftarrow \mathrm{Gen}(\lambda).
\end{equation}
Also, we define the $\mathrm{Eval}$ algorithm that takes a challenge $x$ and a PUF $\mathbf{id}$ as inputs and generates the corresponding response $y_\mathbf{id}$ as output:
\begin{equation}
    y_{\mathbf{id}} \leftarrow \mathrm{Eval}(\mathbf{id},x).
\end{equation}
Due to variations in the environmental conditions, for any given PUF with the identifier $\mathbf{id}$ (Let us call it $\mathrm{PUF}_{\mathbf{id}}$ from now on for a more intuitive notation), the $\mathrm{Eval}$ algorithm may generate a different response to the same challenge $x$. It is required that this noise be bounded as follows; if $\mathrm{Eval}(\mathbf{id},x)$ is run several times, the maximum distance between the corresponding responses should at most be $\delta_r$. This requirement is termed the \emph{robustness requirement}.

Consider a family of $\mathrm{PUF}$ generated by the same $\mathrm{Gen}$ algorithm, and assume the algorithm $\mathrm{Eval}$ is run on all of them with a single challenge $x$. To be able to distinguish each $\mathrm{PUF}_\mathbf{id}$, it is required that the minimum distance between the corresponding responses be at least $\delta_u$. This requirement is termed the \emph{uniqueness requirement}.

The other requirement considered in \cite{armknecht_towards_2016} is \emph{collision-resistance}. This imposes that whenever the $\mathrm{Eval}$ algorithm is run on $\mathrm{PUF}_\mathbf{id}$ with different challenges, the minimum distance between the different responses must be at least $\delta_c$.
The parameters $\delta_r$, $\delta_u$, $\delta_c$ are determined by the security parameter $\lambda$. Robustness, uniqueness and collision-resistance are crucial for correctness of cryptographic schemes built on top of PUFs. The conditions $\delta_r \le \delta_u$ and $\delta_r \le \delta_c$ must be satisfied to allow for distinguishing different challenges and PUFs~\cite{armknecht_towards_2016}.

According to the above, a $(\lambda,\delta_r,\delta_u,\delta_c)$-PUF is defined as a pair of algorithms: $\mathrm{Gen}$ and $\mathrm{Eval}$ that provides the robustness, uniqueness and collision-resistance requirements. We call a $(\lambda,\delta_r,\delta_u,\delta_c)$-PUF a Classical PUF (cPUF), if the $\mathrm{Eval}$ algorithm runs on classical information such as bit strings.
We also recall that a cPUF's $\mathrm{Eval}$ as a classical function $f:\{0,1\}^n \rightarrow \{0,1\}^m$, can be represented as a unitary transformation as follows (see Section~\ref{sec:prelim-oracles} in the preliminaries):
\begin{equation}
    \forall x\in \{0,1\}^n, \forall y\in \{0,1\}^m: U_f\ket{x,y}:=\ket{x,f(x)\oplus y}
\end{equation}
and thus if physically possible, a quantum adversary can query $U_f$ on any desired quantum states such as the superposition of all the classical inputs.

\section{Quantum Physical Unclonable Functions}\label{sec:qpuf-defs}
In this section, we define a general notion for quantum PUFs. We consider a set of quantum devices that have been created through the same manufacturing process.
These devices produce a general quantum state when challenged with a quantum state. Similar to the previously presented classical setting, we formalize the manufacturing process of qPUFs by defining a $\qPUFGen$ algorithm:
\begin{equation}\label{eq:qPUFGen}
    \mathbf{id} \leftarrow \qPUFGen(\lambda)
\end{equation}
where $\mathbf{id}$ is the identifier of $\qPUFid$ and $\lambda$ the security parameter. 

We also need to define the $\qPUFEval$ algorithm mapping any input quantum state $\rhoin\in\ES(\HildIn)$ to an output quantum state $\rhoout\in\ES(\HildOut)$ where $\HildIn$ and $\HildOut$ are the domain and range Hilbert spaces of $\qPUFid$, denoted as:
\begin{equation}\label{eq:qPUFEval}
       \rhoout \leftarrow \qPUFEval(\qPUFid,\rhoin).
\end{equation}
 
For now, we allow $\qPUFEval$ to be a general trace-preserving quantum map. We have:
\begin{equation}
    \rhoout = \Lambda_{\id}(\rhoin)
\end{equation}

Apart from these common algorithms that are analogue to the classical setting, we also require qPUFs as a primitive, to include an efficient test algorithm $\T$ as we have formally define in \defref{def:test}  to test the equality between two unknown quantum states.
We will also need the concept of quantum state distinguishability, which can be defined with different quantum distance measures such as trace distance or fidelity. Here we use the fidelity-based definitions of \defref{def:prelim-dist} and \defref{def:prelim-indist}. We can now define a \emph{Quantum Physical Unclonable Function (qPUF)} as follows. 

\begin{defbox}
\begin{definition}[Quantum Physical Unclonable Function]\label{def:qPUF}
Let $\lambda$ be the security parameter, and $\delta_r, \delta_u, \delta_c\in[0,1]$ the robustness, uniqueness and collision resistance thresholds. A ($\lambda,\delta_r,\delta_u,\delta_c$)-qPUF includes the algorithms: $\qPUFGen$, $\qPUFEval$ and $\T$ satisfying Requirements \ref{def:robust}, \ref{def:uniq}, and \ref{def:col-res}\footnote{In Requirements~\ref{def:robust} and \ref{def:col-res} the probabilities have been taken over the states of the domain Hilbert space, picked from any arbitrary distribution. In Requirement~\ref{def:uniq} the probability is over the family of CPTP maps between same input and output Hilbert spaces picked from an arbitrary distribution.}.
\end{definition}
\end{defbox}

\begin{requirement}[\textbf{$\delta_r$-Robustness}]\label{def:robust}\footnote{We should note that this requirement is satisfied for any qPUF, by definition, due to the contractivity of quantum channels, as we have defined the evolution algorithm as a CPTP map. However, since this is a crucial requirement for classical PUFs and an important property required for PUFs in general, we have decided to include it as a requirement for the completeness of the framework and for comparison's sake. Also, one might use a framework similar to the one presented in this chapter but with a PUF that is not necessarily a CPTP map, in which case the requirement is not always satisfied and needs to be checked.} For any $\qPUFid$ generated through $\qPUFGen(\lambda)$ and evaluated using $\qPUFEval$ on any two input states $\rhoin$ and $\sigmain$ that are $\delta_r$-indistinguishable, the corresponding output quantum states $\rhoout$ and $\sigmaout$ are also $\delta_r$-indistinguishable with overwhelming probability,
\begin{equation}
 \mathrm{Pr}[\delta_r\le F(\rhoout, \sigmaout)\le 1] = 1-\negl(\lambda).
\end{equation}
\end{requirement}

\begin{requirement}[\textbf{$\delta_u$-Uniqueness}]\label{def:uniq} For any two $\qPUF$s generated by the $\qPUFGen$ algorithm, \emph{i.e.} $\qPUFidi$ and $\qPUFidj$, the corresponding CPTP map models, \emph{i.e.} $\Lambda_i$ and $\Lambda_j$ are $\delta_u$-distinguishable with overwhelming probability, 
\begin{equation}
\mathrm{Pr}[\ \parallel (\Lambda_i - \Lambda_j)_{i\neq j}\parallel_\diamond \ge \delta_{u} \ ]=1-\negl(\lambda).
\end{equation}
\end{requirement}

\begin{requirement}
[\textbf{$\delta_c$-Collision-Resistance (Strong)}]\label{def:col-res} For any $\qPUFid$ generated by $\qPUFGen(\lambda)$ and evaluated by $\qPUFEval$ on any two input states $\rhoin$ and $\sigmain$ that are $\delta_c$-distinguishable, the corresponding output states $\rhoout$ and $\sigmaout$ are also $\delta_c$-distinguishable with overwhelming probability,\footnote{A weaker variant of Collision-Resistance, with separate input/output bound can be also defined in a similar fashion where the responses generated by $\qPUFEval$ on any two $\delta^i_c$-distinguishable input states $\rhoin$ and $\sigmain$, should be at least $\delta^o_c$-distinguishable. In fact, if $\delta^i_c = \delta^o_c = \delta_c$ we call the requirement a strong collision-resistance. Note that this equality holds up to a negligible value in the security parameter, \emph{i.e.} if $\delta^i_c = \delta^o_c \pm \negl(\lambda)$, the strong collision-resistance requirement has still been satisfied. If $\delta^o_c < \delta^i_c$ (the difference is non-negligible) then this is referred to as weak collision-resistance.}
\begin{equation}
\mathrm{Pr}[0\le F(\rhoout, \sigmaout)\le 1-\delta_c] = 1-\negl(\lambda).
\end{equation}
\end{requirement} 

In many PUF-based applications such as authentication and identification, it is necessary that there be a clear distinction between different qPUF instances generated by the same $\qPUFGen$ algorithm running on the same parameters $\lambda$ \cite{armknecht_towards_2016}. To this end, the following conditions need to be satisfied: $\delta_c \leq 1-\delta_r$ and $\delta_u \leq 1-\delta_r$. We also note that Requirements~\ref{def:robust} and \ref{def:col-res} impose conditions on the evaluation algorithm of each of the qPUFs in the family while Requirement~\ref{def:uniq} is a property of the family of physical unclonable functions. In the majority of this chapter, we are interested in the security properties of each one of such functions in the family thus our results mostly concern the $\qPUFEval$ algorithm.

We note that if qPUF is a general noisy quantum channel, the $\delta_c$ parameter can allow for some specific noise models. More specifically, the weak-collision resistance parameter \emph{i.e.} the ratio of $\delta^o_c / \delta^i_c$ is directly related to the channel parameters of the qPUF evaluation. Since we are interested in the cryptographic properties of qPUFs, and the collision-resistance is an important requirement for security, we choose the strong collision-resistance as the main requirement for quantum PUFs. We specify that the strong collision-resistance parameter can allow for noisy PUF evaluation under the coherent noise models. Such noise models preserve distances between the input and output states of the qPUF and this property makes them suitable candidates for quantum PUF. Also, it has been shown in~\cite{greenbaum_modeling_2018} that a general noise can be modelled as a combination of coherent and incoherent noises. In other words, only the class of noise model with a close to zero incoherent factor can be considered to satisfy the $\delta_c$ (strong) collision resistance. Hence for the rest of this work, aiming to formalise the first general security framework, we consider this restricted noise setting that allows for an ideal qPUF and we leave further investigation that would depend on particular constructions for future works.

We have initially allowed for any CPTP map as $\qPUFEval$ algorithm. Now, we let the $\qPUFEval$ algorithm be a channel with the same dimension of domain and range Hilbert space, \emph{i.e.} $d_{in}=d_{out}$. We show that under this assumption, only unitary transformations and CPTP maps that are highly close to unitary class, can simultaneously provide the (strong)collision-resistance and robustness requirements of qPUFs. 

\begin{thmbox}
\begin{theorem}\label{theorem:non-unitary}
Let $\E(\rho)$ be a completely positive and trace-preserving map described as follows:
\begin{equation}\label{eq:non-unitary-puf}
    \E(\rho) = (1-\epsilon)U \rho U^{\dagger} + \epsilon \Tilde{\E}(\rho) 
\end{equation}
where $U$ is a unitary transformation, $\Tilde{\E}$ is an arbitrary (non-negligibly) contractive channel and $0 \leq \epsilon \leq 1$. Then $\E(\rho)$ is a valid qPUF's evaluation algorithm (with equal domain and range dimensionality) for any $\lambda$ and $\delta_c$ (up to a negligible factor), if and only if $\epsilon = \negl(\lambda)$.
\end{theorem}
\end{thmbox}

\begin{proof}
First, we recall the contractive property of trace-preserving operations \cite{nielsen_quantum_2010}, the robustness is trivially satisfied.
Hence the robustness is generally satisfied. As a result, the proof of the theorem reduces to proving for collision-resistance. Let $\rho$ and $\delta$ be two $\delta_c$-distinguishable challenge with fidelity $F(\rho, \sigma) \leq 1-\delta_c$. Again with the above argument the fidelity of the outputs cannot be smaller than $F(\rho, \sigma)$. Thus the $\delta_c$ requirement is satisfied if the fidelity of the response density matrices are equal up to a negligible value.

Now let $\rho_1 = U\rho U^{\dagger}$, $\sigma_1 = U\sigma U^{\dagger}$, $\rho_2 = \Tilde{\E}(\rho)$, and $\sigma_2 = \Tilde{\E}(\sigma)$. We use the joint concavity of the fidelity \cite{nielsen_quantum_2010} to obtain the following relation for the channel's output fidelity:
\begin{equation}\label{eq:fidelity-joint-concavity}
\begin{split}
        F(\E(\rho),\E(\sigma)) & = F((1-\epsilon)\rho_1 + \epsilon \rho_2, (1-\epsilon)\sigma_1 + \epsilon \sigma_2) \\
        & \geq (1-\epsilon)F(\rho_1,\sigma_1) + \epsilon F(\rho_2,\sigma_2)
\end{split}
\end{equation}
Since the first part of the channel is unitary which is distance preserving, we have $F(\rho_1,\sigma_1) = F(\rho,\sigma)$. Also due to contractivity we know that $F(\rho_2,\sigma_2) \geq F(\rho,\sigma)$. We then have:
\begin{equation}
    F(\E(\rho),\E(\sigma)) - F(\rho,\sigma) \geq \epsilon (F(\rho_2,\sigma_2)-F(\rho,\sigma))
\end{equation}
Now since the channel $\Tilde{\E}$ is non-negligibly contractive, the value $F(\rho_2,\sigma_2)-F(\rho,\sigma)$ is not necessarily negligible and in order for the LHS of ~\eqref{eq:fidelity-joint-concavity} to be always negligible, $\epsilon$ has to be negligible. So we have proved that CPTP maps of the form~\eqref{eq:non-unitary-puf} can be $\delta_c$ collision resistance qPUFs only if $\epsilon = \negl(\lambda)$.

Now we show that all channels of the form of~\eqref{eq:non-unitary-puf} where $\epsilon$ is negligible satisfy the strong collision resistance property up to a negligible value. To show that we recall the relation between fidelity and trace distance, that is $\dtr(\rho, \sigma) \leq \sqrt{1 - F(\rho,\sigma)}$. We use this inequality to relate the distance between the states $\E(\rho)$ and $\E(\sigma)$ and the original distance between $\rho$ and $\sigma$. By subtracting both sides, we get the following inequality:
\begin{equation}
\begin{split}
        F(\E(\rho),\E(\sigma)) - F(\rho,\sigma) & \leq \dtr^2(\rho, \sigma) - \dtr^2(\E(\rho),\E(\sigma))\\
        & \leq (\dtr(\rho, \sigma) - \dtr(\E(\rho),\E(\sigma)))(\dtr(\rho, \sigma) + \dtr(\E(\rho),\E(\sigma)))\\
        & \leq 2(\dtr(\rho, \sigma) - \dtr(\E(\rho),\E(\sigma)))
\end{split}
\end{equation}

Next, we show the following inequality stating that the difference between the trace distance of the input and output for channels described as~\eqref{eq:non-unitary-puf}, is bounded by $\epsilon \dtr(\rho, \sigma)$,
\begin{equation}
    \dtr(\rho, \sigma) - \dtr(\E(\rho), \E(\sigma)) \leq \epsilon \dtr(\rho, \sigma)
\end{equation}

First, we note that the first part of the channel $\E$, which outputs density matrix $U \rho U^{\dagger}$ with probability $(1-\epsilon)^2$, is a unitary and preserves the distance. As a result, for a fixed value of $\epsilon$ and any fixed arbitrary states $\rho$ and $\sigma$, the difference between the trace distances of the output of $\E$ and the input states increases as $\Tilde{\E}$ becomes more contractive. As the maximum contractivity of $\Tilde{\E}$ occurs when $\Tilde{\E} = \frac{I}{d}$, then the maximum difference between the output and input trace distances is satisfied for this instance of the channel. Let $\E'(\rho) = (1-\epsilon)U \rho U^{\dagger} + \epsilon \frac{I}{d}$. Then for a fixed $\epsilon$ we will have:
\begin{equation}\label{eq:trdistance-difference-depolazing}
    \dtr(\rho, \sigma) - \dtr(\E(\rho), \E(\sigma)) \leq \dtr(\rho, \sigma) - \dtr(\E'(\rho), \E'(\sigma))
\end{equation}

Now we calculate $\dtr(\E'(\rho), \E'(\sigma))$ using the definition of the trace distance:
\begin{equation}
\begin{split}
        \dtr(\E'(\rho), \E'(\sigma)) &= \frac{1}{2}\tr[|\E'(\rho), \E'(\sigma)|] \\
        & =  \frac{1}{2}\tr[|(1-\epsilon)U \rho U^{\dagger} + \epsilon \frac{I}{d} - (1-\epsilon)U \sigma U^{\dagger} - \epsilon \frac{I}{d}|] \\
        & = (1-\epsilon)(\frac{1}{2}\tr[|U \rho U^{\dagger} - U \sigma U^{\dagger}|]) \\
        & = (1-\epsilon)\dtr(U \rho U^{\dagger},U \sigma U^{\dagger})\\
        & = (1-\epsilon)\dtr(\rho, \sigma)
\end{split}
\end{equation}
Substituting this back to \eqref{eq:trdistance-difference-depolazing}, we get
\begin{equation}
       \dtr(\rho, \sigma) - \dtr(\E(\rho), \E(\sigma)) \leq \dtr(\rho, \sigma) - (1-\epsilon)\dtr(\rho, \sigma) = \epsilon \dtr(\rho, \sigma)
\end{equation}
Thus we have:
\begin{equation}
    F(\E(\rho),\E(\sigma)) - F(\rho,\sigma) \leq 2\epsilon \dtr(\rho, \sigma)
\end{equation}
Now if and only if $\epsilon = \negl(\lambda)$ and since $0 \leq \dtr(\rho, \sigma) \leq 1$, we conclude that the difference between the fidelity is also negligible and hence the $\delta_c$ collision-resistance is satisfied up to a negligible value, and the proof is complete.
\end{proof}

The above theorem shows that only unitary or more generally, \emph{$\epsilon$-disturbed unitary maps} where $\epsilon$ is small, are suitable candidates for qPUFs, especially when strong collision resistance is required. In the rest of the chapter, we choose the $\qPUFEval$ algorithm to be a unitary map.
We call this type of qPUFs, Unitary qPUFs (or simply UqPUFs) and formally define them in \defref{def:uqPUF}. Nevertheless, we believe studying more general non-unitary qPUFs will be interesting future research directions in this field (see Section~\ref{sec:qpuf-disc}).

\begin{figure}
\includegraphics[scale=0.55]{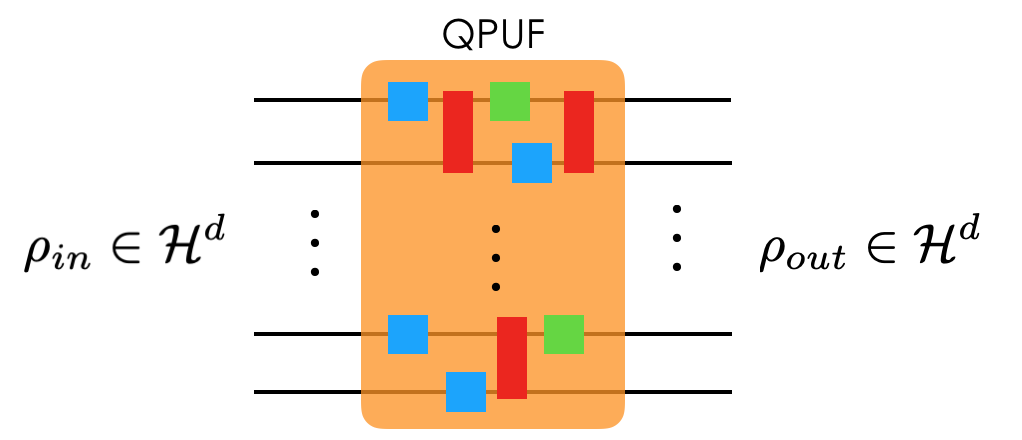}
\centering
\caption[Illustration of qPUF as a unitary operation with input and output quantum states]{Illustration of qPUF as a unitary operation with input and output quantum states in $\mathcal{H}^d$. The blue and green boxes are single-qubit gates, and the red boxes demonstrate two-qubit and entangling gates. These are the abstract building blocks for the d-dimensional UqPUF, while to a QPT adversary, the unitary and the inherent structure is initially unknown.}
\label{fig:qpuf}
\end{figure}

So far, we concluded that in terms of the mathematical model, unitary quantum transformations are best suited to describe qPUFs. Now it is time to formalize the main hardware assumption of our qPUFs. We recall that in the classical setting it is assumed that the PUF behaviour is unknown even to the manufacturer. We also require UqPUF transformations to be initially unknown or in other words, behave as a unitary black-box of it which is exponentially hard to recover the full description. In the previous chapter, we have discussed the notion of \emph{unknownness} and its relation to the unclonability and learnability of quantum processes. Now we invoke the same definition to formalize the hardware assumption of physical unclonability by requiring the unitary matrix of a qPUF to satisfy \emph{unknownness} according to \defref{def:unk-uni} from \chapref{chap:unf-tools}, which formalises single-shot indistinguishability of the unitary from the family of Haar-random unitaries. An illustration of a unitary qPUF is given in \figref{fig:qpuf}. Let us formally define Unitary qPUFs (UqPUFs).

\begin{defbox}
\begin{definition}[Unitary qPUF (UqPUF)]\label{def:uqPUF} A Unitary $\qPUF$ $((\lambda,\delta_r)-\UqPUF)$ is a $(\lambda,\delta_r)-\qPUF$ where the $\qPUFEval$ algorithm is modelled by an unknown unitary transformation $\U_\id$ over a $D$-dimensional Hilbert space, $\HilD$ according to \defref{def:unk-uni}, such that for any quantum challenge $\rhoin$ the respective response $\rhoout$ is given as follows,
\begin{equation}
    \rhoout = \qPUFEval(\UqPUFid, \rhoin) = \U_\id\rhoin\U_\id^{\dagger}.
\end{equation}
\end{definition}
\end{defbox}

For simplicity and practical reasons, usually, the challenge is a pure quantum state denoted as $\ket{\psiin}$, and the response of a UqPUF is simply given by $\ket\psiout = U_{\id}\ket{\psiin}$. Also, due to the distance-preserving property of UqPUFs, we drop $\delta_r$ from the notation and simply characterise UqPUF as $\lambda$-UqPUFs.

There are a couple of notes that are worth mentioning concerning this requirement. First, from the theoretical point of view, this requirement is a minimal and pre-challenge assumption, and considerably weaker than the assumptions needed for classical PUFs. A common requirement needed for classical PUF is \emph{min-entropy} that informally captures the minimum extractable information about a cPUF from subsets of CRPs \cite{armknecht_towards_2016}. This requirement is morally closely related to the unpredictability, or unforgeability of PUFs. Nevertheless, in the quantum setting, we aim to characterize the unpredictability as a byproduct of rather simpler hardware assumptions. Our proposed requirement intuitively requires the information of a UqPUF to be obtained only through querying it.
One straightforward construction for UqPUF is to sample a unitary from a Haar-random Unitary family, but we believe there are more efficient ways to do this sampling~\cite{dankert_exact_2009,ambainis_quantum_2007} (see also Section~\ref{sec:qpuf-disc}, for more discussion about subsequent works and constructions of quantum PUF).

From a construction point of view, this condition may not seem easily achievable, but again practically, it is a reasonable assumption considering limited fabrication capabilities or the fact that simulating an arbitrary unitary on a quantum computer is not technologically easy due to noise and accumulated errors in each gate, even when the structure of the unitary is known. Moreover, there are promising constructions such as the family of optical schemes implemented using crystals or optical scattering media~\cite{nikolopoulos_continuous-variable_2017}, where usually even the manufacturer does not know the underlying unitary unless querying it. On the other hand, in gate-based construction, one cannot avoid the fact that the manufacturer knows the underlying unitary. Hence this type of construction cannot provide security against an adversarial manufacturer. Nevertheless, if predicting the evolution of a quantum state is difficult this is enough for security under the usual PUF assumptions. As a result, such devices are still useful and practical for many applications as they can still provide security against any malicious adversary other than the manufacturer. The security framework that we will propose, on the other hand, covers both adversarial models where the manufacturer could be trusted or not.

The final deserving remark, before we move to the cryptanalysis of UqPUFs is that they also satisfy another natural notion of unclonability, known as no-cloning of unitary transformation \cite{chiribella_optimal_2008}, discussed in Section~\ref{sec:unclone-states-unitaries}. We recall that under this notion, two black-box unitary transformations $\mathcal{O}_1$ and $\mathcal{O}_2$ cannot be perfectly cloned by a single use, apart from the trivial cases of perfect distinguishability or when $\mathcal{O}_1 = \mathcal{O}_2$. Thus, two UqPUFs, as long as they correspond to different black-box unitaries, satisfied by the uniqueness requirement and our proposed assumption, are unclonable by quantum mechanics via a single use. Specifically, in the following section, we show how this unclonability property, can be expanded to the multiple-shot case by introducing the formal notion of unforgeability for quantum PUFs.

\section{Cryptanalysis of Quantum Physical Unclonable Functions}\label{sec:qPUF-crypt}

Using the tools and framework that we have established in the previous chapter, for the study of unclonability and unpredictability via the cryptographic notion of unforgeability, we can now formally define this security notion for quantum PUFs and study the extent to which this property is satisfied for general UqPUFs as we have defined them. Other than the fundamental relationship that we aim to establish between unclonability and unforgeability, in terms of cryptographic applications, the security of most PUF-based protocols relies on the unforgeability of PUFs \cite{armknecht_towards_2016}. In the context of PUF, unforgeability informally means that given a subset of challenge-response pairs of the target PUF, the probability of correctly guessing a new challenge-response pair shall be considerably small.

In the literature of classical cryptography and hardware security, the unforgeability of PUFs as classical functions is often studied in a game-based framework~\cite{armknecht_towards_2016,boneh_secure_2013,delavar_puf-based_2017}. However, there have been studies of PUFs in the UC framework as well \cite{brzuska_physically_2011,ostrovsky_universally_2013}.

We recall the definition of unforgeability from the unified framework that we have defined in \chapref{chap:unf-tools} and \gameref{game:unf-full}. Since the framework is defined to capture both quantum and classical primitives, we can easily adapt it for qPUFs. Here we only elaborate on what each of the stages means in the context of qPUFs.

In the \textbf{setup phase}, the necessary public and private parameters and functions are shared between the adversary and the challenger and the qPUF is generated. 

The \textbf{learning phase} models the knowledge that the adversary can gain over a qPUF through queries. We consider chosen-input attacks model where the quantum adversary can choose any arbitrary (and potentially adaptive) query from the domain Hilbert space. Due to the quantum nature of queries, and to be able to fully characterize adversary's database, they have to prepare two copies of each challenge query, keep one in their database, and send the other one to the challenger. 

The \textbf{challenge phase} captures the intended security notion. For qPUFs, we consider two types of challenge phase: existential and universal\footnote{This level of security is usually known as `selective unforgeability' in the context of PUFs. Nevertheless, to avoid confusions, with the similar term used in \chapref{chap:unf-tools} and for consistency, here we keep the term of universal unforgeability} as defined before. In the universal case, since we are in the regime of quantum unforgeability for quantum schemes, the uniform selection of the challenge is equivalent to choosing the challenge uniformly at random according to the Haar measure.

Finally, in the \textbf{guess phase}, the adversary outputs his guess of the response corresponding to the challenge chosen in the challenge phase. The equality of the adversary's response to the correct response is being tested by a quantum test algorithm as we have abstracted in \defref{def:test}. The adversary wins the game if the output of the test algorithm is 1. \gameref{game:qpuf-unf} is adapted directly from \gameref{game:unf-full} and formalises the unforgeability of qPUFs.

\begin{gamebox}{Formal game-based unforgeability of qPUF}
\begin{game}\label{game:qpuf-unf}
Let $\qPUF=(\qPUFGen,\qPUFEval,\T)$ and $\T$ be defined as \defref{def:qPUF} and \defref{def:test}, respectively. We define the following game $\Gn{\qPUF}{c}{\mu}(\A,\lambda)$ running between an adversary $\A$ and a challenger $\C$:
\begin{itemize}[leftmargin=*]
    \item [] \textbf{Setup.} The challenger $\C$    
    runs $\qPUFGen(\lambda)$ to build an instance of the qPUF family, $\qPUFid$. Then, $\C$ reveals to the adversary $\A$, the domain and range Hilbert space of $\qPUFid$ respectively denoted by $\Hilin$ and $\Hilout$
    as well as the identifier of $\qPUFid$, $\mathbf{id}$. The challenger initialises two empty databases, $\Sin$ and $\Sout$ and shares them with $\A$.
    \item [] \textbf{Learning.} For $i=1:k$
        \begin{itemize}
            \item $\A$ prepares two copies of a quantum state $\rho_i \in \ES(\Hilin)$, appends one to $\Sin$ and sends the other to $\C$;
            \item $\C$ runs $\qPUFEval(\qPUFid,\rho_i)$ and sends $\rho_i^{out}$, to $\A$;
            \item $\A$ appends $\rho_i^{out}$ to $\Sout$.
        \end{itemize} 
    \item [] \textbf{Challenge.}
    \begin{itemize}
        \item If $c=\qEx$: $\A$ picks a quantum state $\rho^*\in\ES(\Hilin)$ at least $\mu$-distinguishable from all the states in $\Sin$ and sends it to $\C$;
        \item If $c=\qUni$: $\C$ chooses a quantum state $\rho^*$ uniformly at random from the Haar-measure over the Hilbert space $\Hilin$. The challenger keeps copies of $\rho^*$ if necessary and sends one copy of $\rho^*$ to $\A$.
    \end{itemize}
    \item [] \textbf{Guess.} 
    \begin{itemize}
        \item $\A$ sends his guess $\rho'$ to~ $\C$;
        \item $\C$ runs $\qPUFEval(\qPUFid,\rho^*)$, and gets $\rho_{out}$;
        \item $\C$ runs the test algorithm $b\leftarrow \mathcal{T}(\rho_{out},\rho')$
        where $b\in\{0,1\}$ and outputs $b$. The adversary wins the game if $b=1$.\footnote{Note that the learning phase queries include any general separable or entangled state.
        }
    \end{itemize}
\end{itemize}
\end{game}
\end{gamebox}

We follow the same definitions of \emph{quantum existential unforgeability} defined in \defref{def:euf-qCMA} and \emph{quantum universal unforgeability} in \defref{def:uni-qCMA} for qPUFs based on the above game. Nevertheless, for the sake of completeness in the study of physical unclonability, we add here another level of security, which is against an exponential (or computationally unbounded) adversary instead of the usual QPT adversary considered in the unforgeability framework. We call this \emph{quantum exponential unforgeability}, \emph{quantum existential unforgeability}, and we define it as follows:

\begin{defbox}
\begin{definition}[Quantum Exponential Unforgeability]\label{def:QunconUf}
A $\qPUF$ provides quantum exponential unforgeability if the success probability of any \emph{exponential} adversary $\A$ in winning the game $\Gn{\qPUF}{\qEx}{\mu}(\A, \lambda)$ or $\Gnn{\qPUF}{\qUni}(\A, \lambda)$ is negligible in $\lambda$
\begin{equation}
    Pr[1\leftarrow \Gnn{\qPUF}{\qEx/\qUni}(\A, \lambda)] = \negl(\lambda)    
\end{equation}
\end{definition}
\end{defbox}

\subsection{Impossibility of exponential unforgeability for UqPUFs}
After formalizing all the security games and definitions, it is time to derive general possibility and impossibility results regarding the quantum unforgeability of UqPUFs. 

We start with the most powerful setting which is against the exponential quantum adversary. In the classical setting, cPUFs can be fully described by the finite set of CRPs, and this suffices for breaking unforgeability. More precisely, an unbounded or exponential adversary can extract the entire set of CRPs by querying the target cPUF with all possible challenges \cite{chang_retrospective_2017}. If the challenges are n-bit strings, the number of possible challenges is $2^n$. However, in the quantum setting, a UqPUF can generate an infinite number of quantum challenge-response pairs such that extracting all of them is hard, even for exponential adversaries. This point, combined with limitations such as no-cloning and the limits on state estimation \cite{bruss_optimal_1998}, raise the question whether UqPUFs could satisfy unforgeability against exponential adversaries. Nevertheless, we answer this question negatively by proving that no UqPUF provides quantum exponential unforgeability as defined in \defref{def:QunconUf}.

\begin{thmbox}
\begin{theorem}\label{th:no-qPUF-unbound}
\textbf{(No UqPUF provides quantum exponential unforgeability)} For any $\lambda$-UqPUF,  there exists an exponential quantum adversary $\A$ such that
\begin{equation}
Pr[1\leftarrow \Gnn{\UqPUF}{\qEx/\qUni}(\lambda, \A)] = \nonnegl(\lambda)    
\end{equation}
\end{theorem}
\end{thmbox}

\begin{proof}
The key idea of the proof is based on complexity analysis of unitary tomography and implementation of a general unitary by single and double qubit gates, since for an exponential quantum adversary, it will be feasible to extract the unitary matrix by tomography and then build the extracted unitary by general gate decomposition method. By using the Solovay-Kitaev theorem \cite{nielsen_quantum_2010}, we then show that the adversary can build the unitary matrix of the UqPUF performing on $n$-qubits, within an arbitrarily small distance $\epsilon$ using $O(n^2 4^n \log^c(n^2 4^n))$ gates and hence win the game with any test algorithm $\T$.
Let $\UqPUF_\id$ operate on $n$-qubit input-output pairs where $n=\log(D)$. In the learning phase, $\A$ selects a complete set of orthonormal basis of $\HilD$ denoted as $\{\ket{b_i}\}^{2^n}_{i=1}$ and queries $\UqPUF_\id$ with each base $2^n$ times. So, the total number of queries in the learning phase is $k_1=2^{2n}$.

Then, $\A$ runs a \emph{unitary tomography} algorithm to extract the mathematical description of the unknown unitary transformation corresponding to the $\UqPUF_\id$, say $\U_\id$. It has been shown \cite{nielsen_quantum_2010} that the complexity of this algorithm is $\mathcal{O}(2^{2n})$ for n-qubit input-output pairs. This is feasible for an exponential adversary. It is clear that once the mathematical description of the unitary is extracted, $\A$ can simply calculate the response of the unitary to any known quantum challenge. Nonetheless, we want to show the exponential adversary wins the weaker notion of the security, \emph{i.e.} quantum universal unforgeability, where they have only one copy of the challenge state.

To win the game with the universal challenge phase, the adversary needs to implement the unitary. It is known that any unitary transformation over $\Hil^{2^n}$ requires $O(2^{2n})$ two-level unitary operations or $O(n^2 2^{2n})$ single qubit and CNOT gates \cite{nielsen_quantum_2010} to be implemented. However, according to Solovay-Kitaev theorem \cite{nielsen_quantum_2010}, to implement a unitary with an accuracy $\epsilon$ using any circuit consisting of $m$ single qubit and CNOT gates, $O(m \log^c(m/c))$ gates from the discrete set are required where $c$ is a constant approximately equal to 2. Thus, an arbitrary unitary performing on $n$-qubit can be approximately implemented within an arbitrarily small distance $\epsilon$ using $O(n^2 4^n \log^c(n^2 4^n))$ gates.

Finally, $\A$ implements the unitary $\U'_\id$ with error $\epsilon$. Let $\A$ get the challenge state $\ket{\psi}$ in the $\qUni$ Challenge phase. The adversary queries $\U'_\id$ with $\ket\psi$ and gets $\ket\omega=\U'_\id\ket\psi$ as output. Since the $\epsilon$ can be arbitrary small, then $F(\U_\id\ket\psi,\U'_\id\ket\psi)\ge 1-\negl(\lambda)$. So, $\A$'s output $\ket\omega$ passes any test algorithm $\T(\kpo^{\otimes \kappa_1}, \ket{\omega}^{\otimes \kappa_2})$ with probability close to 1. Again, an unbounded adversary wins the game $\Gnn{\UqPUF}{\qUni}(\lambda, \A)$ with probability 1. Also, since the existential challenge phase is a stronger definition, if $\A$ wins in game $\Gnn{\UqPUF}{\qUni}(\lambda, \A)$ they will also when $\Gn{\UqPUF}{\qEx}{\mu}(\lambda, \A)$ as well. Therefore we have:
\begin{equation}
    Pr[1\leftarrow \Gnn{\UqPUF}{\qEx/\qUni}(\lambda, \A)] = 1.
\end{equation}
that concludes the proof.
\end{proof}

We note that this result is expected as any qPUF (same as a classical PUF) can, in principle, be simulated with enough computational resources. Therefore no physical unclonability exists against an adversary with an unbounded quantum power since there is no level of \emph{unknownness}, as in an exponentially powerful setting. We point out that similar relation exists in the relation between asymptotic state estimation and approximate quantum cloning (see Section~\ref{sec:prelim-cloning} and Section~\ref{sec:unclone-unknown}). That is why the reasonable and achievable security model is usually against a qPUF in the hands of the adversary for a limited time or limited query such as QPT adversaries. It is also worth mentioning that from an engineering point of view, limiting the adversary to a certain number of queries on a hardware level can depend on the construction and, it might be possible in some qPUF implementations, while might not be feasible with some others. While this is an interesting problem to consider in qPUF implementations, from a cryptanalysis point, our given security analysis against a quantum adversary who is given polynomial time in the security parameter of the qPUF is independent of the construction.

\subsection{Impossibility of existential unforgeability for UqPUFs}
Exploiting the quantum emulation tools introduced in \chapref{chap:unf-tools} for cryptanalysis, we now turn to quantum existential unforgeability and show that no UqPUF provides quantum existential unforgeability for any $\mu\neq 1$ as defined in \defref{def:euf-qCMA}. Note that the case $\mu=1$ corresponds to the existential challenge state being orthogonal to all the queried states in the learning phase. With $\mu=1$, the adversary is prevented from taking advantage of its quantum access to the qPUF to win the game.

\begin{thmbox}
\begin{theorem}\label{th:noeuf-qexunf}
\textbf{(No UqPUF provides quantum existential unforgeability)} For any $\lambda$-UqPUF, and $0\leq \mu \leq 1-\nonnegl(\lambda)$, there exits a QPT adversary $\A$ such that
\begin{equation}
    Pr[1\leftarrow \Gn{\UqPUF}{\qEx}{\mu}(\lambda, \A)] = \nonnegl(\lambda).    
\end{equation}
\end{theorem}
\end{thmbox}

\begin{proof}
We show there is a QPT adversary $\A$ who wins the game $\Gn{\UqPUF}{\qEx}{\mu}(\lambda,\A)$ with non-negligible probability in $\lambda$. We use a similar emulation attack presented in Section~\ref{sec:unf-qe-one-block}, which uses only one block of emulation algorithm. 
The learning phase queries are as follows where $\ket{\phi_1}$ can be any quantum state in $\HilD$, and $\ket{\phi_3}$ is any orthogonal state to $\ket{\phi_1}$ in the domain Hilbert space:
\begin{equation}
    \ket{\phi_2}=\begin{cases}
\frac{1}{\sqrt{2}}(\ket{\phi_1} + \ket{\phi_3}) & if~0 \leq \mu \leq \frac{1}{2}\\
\sqrt{\mu}\ket{\phi_1} + \sqrt{1-\mu}\ket{\phi_3} & if~ \frac{1}{2} < \mu \leq 1-\nonnegl(\lambda)
\end{cases}
\end{equation}
Then, $\A$ sets $\ket{\phi_3}$ as his chosen challenge in the existential challenge phase. Note that $\ket{\phi_3}$ satisfies the $\mu$-distinguishability condition with both $\ket{\phi_1}$ and $\ket{\phi_2}$. In the guess phase, to estimate the output of UqPUF to $\ket{\phi_3}$, the adversary $\A$ runs the \QE\ with the reference state $\ket{\phi_r}=\ket{\phi_2}$. 

Relying on \thmref{th:qe-fins}, the output state of Stage 1 of the \QE\ algorithm is:
\begin{equation}
\begin{split}
    \ket{\chi_f} &= \mbraket{\phi_2}{\phi_3} \ket{\phi_2}\ket{0} + \ket{\phi_3}\ket{1} - \mbraket{\phi_2}{\phi_3}\ket{\phi_2}\ket{1}\\
    & -2\mbraket{\phi_1}{\phi_3}\ket{\phi_1}\ket{1} +2\mbraket{\phi_2}{\phi_3}\mbraket{\phi_2}{\phi_1}\ket{\phi_1}\ket{1}.
\end{split}
\end{equation}
Having $\bra{\phi_1}\phi_3\rangle = 0$ and we setting $\bra{\phi_2}\phi_3\rangle = \alpha$ and $\bra{\phi_2}\phi_1\rangle = \beta$, the final fidelity in terms of the success probability of Stage 1 is given as follows according to \eqref{eq:qe-one-block-attack-fid}
\begin{equation}
    P_{succ-stage1} = |\alpha^2(1+4\alpha^2\beta^2)|^2.
\end{equation}
We have different choices for the reference state depending on the distinguishability parameter $\mu$. For cases where the adversary is allowed to produce a new state with at least overlap half with all the states in the learning phase, by choosing the uniform superposition of the states where $\alpha = \beta = \frac{1}{\sqrt{2}}$, the output fidelity will be:
\begin{equation}
    F(\ket{\phi_3^{out'}}\bra{\phi_3^{out'}}, \ket{\phi_3^{out}}\bra{\phi_3^{out}}) \geq \sqrt{P_{succ-stage1}} = 1.
\end{equation}
where $\ket{\phi_3^{out'}}$ and $\ket{\phi_3^{out}}$ are the output of the \QE\ algorithm and UqPUF to $\ket{\phi_3}$, respectively. According to the calculated fidelity, these two states are completely indistinguishable So, the success probability of $\A$ for any test according to \defref{def:test} is:
\begin{equation}
        Pr[1\leftarrow \Gn{\UqPUF}{\qEx}{\mu}(\lambda, \A)] = Pr[1\leftarrow\T(\ket{\psi^{out}}, \ket{\omega})] = 1
\end{equation}
which is the optimal choice of the reference. On the other hand, for the cases where the adversary is restricted to produce a challenge more than half distinguishable, we can still create a superposed state with $\alpha = \sqrt{1-\mu}$ and $\beta = \sqrt{\mu}$ and end up with the following fidelity of the emulation by setting $\mu=1-\nonnegl(\lambda)$
\begin{equation}
\begin{split}
    F(\ket{\phi_3^{out'}}\bra{\phi_3^{out'}}, \ket{\phi_3^{out}}\bra{\phi_3^{out}}) & \geq |\alpha^2(1+4\alpha^2\beta^2)| \\ &= |(1-\mu)(1 + 4\mu(1-\mu))| \\&= \nonnegl(\lambda).
\end{split}
\end{equation}
Thus for any $\frac{1}{2} < \mu \leq 1 -\nonnegl(\lambda)$:
\begin{equation}
        Pr[1\leftarrow \Gn{\UqPUF}{\qEx}{\mu}(\lambda, \A)] = Pr[1\leftarrow\T(\ket{\phi_3^{out}}, \ket{\phi_3^{out'}})] = \nonnegl(\lambda)
\end{equation}
And the proof is complete.
\end{proof}

This theorem implies that the adversary can always generate the correct response to his chosen challenge provided that he can query it in superposition with other quantum states during the learning phase in terms of the parameter $\mu$. Note that since output quantum states in the learning phase are unknown to the adversary, the more straightforward strategy of superposing the learnt output quantum states cannot be efficiently performed. More precisely, the adversary cannot prepare the precise target superposition of the output states that are completely unknown~\cite{oszmaniec_creating_2016,doosti_universal_2017}. Therefore, the proposed attack in the above proof is general yet non-trivial.

\subsection{Universal unforgeability of UqPUFs}
We now show a positive result for the security of UqPUFs in general by further relaxing the level of security and considering quantum universal unforgeability. We will show that any UqPUF can provide this notion. This result also establishes the relationship between unforgeability, unknownness and physical unclonability in the quantum regime. 
Furthermore, note that in most PUF-based applications, the universal unforgeability is sufficient. We will discuss this further in \chapref{chap:application}. 

We start by proving the following lemma which is a crucial step towards our proof. The lemma establishes the average probability of any state in $\HilD$ to be projected in a subspace $\Hild$ where $d\leq D$. Based on this lemma, we calculate the probability of a state chosen uniformly at random from $\HilD$ (according to Haar-measure) to belong in the orthogonal subspace to the adversary's subspace. We recall that since quantum emulation is successful with a high probability when the target state has enough overlap with the sample subspace, the idea is to exploit the randomness involved in selecting the target state to prevent quantum emulation or similar attacks.

\begin{lembox}
\begin{lemma}\label{lem:qpuf-avg-projection-prob}
    Let $\HilD$ be a $D$-dimensional Hilbert space and $\Hild$ a subspace of $\HilD$ with dimension $d$. Also, let $\Pi_d$ be a projector, projecting any quantum state in $\HilD$ into $\Hild$. The average probability that any state, chosen uniformly at random from $\HilD$ from a Haar distribution, to be projected into $\Hild$ is equal to $\frac{d}{D}$
    \begin{equation}
        \underset{{\ket{\psi},\Pi_d}}{Pr}[|\bra{\psi}\Pi_d\ket{\psi}| = 1] = \frac{d}{D}
    \end{equation}
\end{lemma}
\end{lembox}
\begin{proof}
    The proof is mainly based on the symmetry of the Hilbert space and the fact that the probability of falling into each subspace is equal for any state uniformly picked at random.
    
    Note that any state $\ket{\psi} \in \HilD$ can be written in terms of the orthonormal bases of $\HilD$ denoted by $\ket{b_i}$, as follows:
\begin{equation}
    \ket{\psi} = \sum_{i=0}^{D-1} \alpha_i\ket{b_i} \quad \text{with} \quad \sum_{i=0}^{D-1} |\alpha_i|^2 = 1
\end{equation}
where $\alpha_i$ are complex coefficients. A projection into a smaller subspace consists of choosing $d$ bases of $\HilD$ in the form of $\sum_{j=0}^{d-1} \ket{b_j}\bra{b_j} $. Without loss of generality, we can assume $D = md$ where $m$ is an integer. This assumption is always correct for qubit spaces. This means that the larger Hilbert space can be divided into $m$ smaller subspaces each with dimension $d$. Let $\{\ket{e_i}\}_{i=0}^{d-1}$ be a subset of $\HilD$ which makes a complete set of bases for one of the $d$-dimensional subspaces. A projector projects $\ket{\psi}$ into one of the subspaces. As $\ket{\psi}$ has been picked at random and the subspaces are symmetric, the probability of falling into each subspace is the same and equal to $\frac{1}{m}$ which is $\frac{d}{D}$. Otherwise either the sum of all probabilities would not be 1 or the $\ket\psi$ has not been picked uniformly at random from $\HilD$. This shows that on average the probability of projecting a state $\psi$ is $\frac{d}{D}$. This can also be seen by the fact that the sum of all projectors in a complete set of projectors is equal to one. In this case, we have
\begin{equation}
    \sum_{i=0}^{D-1} \Pi_i = \mathbb{I}
\end{equation}
By sandwiching $\ket{\psi}$ on both sides we have:
\begin{equation}
    \sum_{i=0}^{D-1} \bra{\psi}\Pi_i\ket{\psi} = 1.
\end{equation}
Each $\bra{\psi}\Pi_i\ket{\psi}$ is itself equal to $\sum_{j=0}^{d-1} |\mbraket{\psi}{d_{ij}}|^2$ where $\ket{d_{ij}}$s are the bases associated to the subspace that the projector $\Pi_i$ projects into. This corresponds to all the permutations of $d$ number of the coefficient $|\alpha_i|^2$ which will be $\frac{1}{d}$ on average. Since we have $\sum_{i=0}^{D-1} \frac{Pr_{\Pi_i}}{d} = 1$, we can conclude that the average probability for all the projectors will be $\frac{d}{D}$.
\end{proof}

We need another small technical toolkit which allows us to derive our next result. We define another abstraction of the test algorithm of \defref{def:test} in direct relation to fidelity. We formalize the ideal test $\Td$ as follows:

\begin{defbox}
\begin{definition}[$\Td$ Test Algorithm]\label{def:test-delta} We call a test algorithm according to \defref{def:test}, a $\Td$ Test Algorithm when for any two state $\ket{\psi}$ and $\ket{\phi}$ the test responds as follows:
\begin{equation}
 \Td = \begin{cases}
    1  & \:F(\ket{\psi}, \ket{\phi})\geq \delta\\
    0  & \:otherwise
  \end{cases} 
\end{equation}
\end{definition}
\end{defbox}

To establish our positive result, we first present a preliminary theorem which demonstrates the unforgeability of the UqPUF considering an ideal test algorithm which asymptotically satisfies the fidelity as defined in \defref{def:test-delta}.

\begin{thmbox}
\begin{theorem}\label{th:qpuf-uni-unf-fid}
For any unitary qPUF characterised by $\UqPUF=(\qPUFGen,\qPUFEval,\Td)$, with dimension $D$, and any non-zero $\delta$, the success probability of any QPT adversary $\A$ in the game $\Gnn{\UqPUF}{\qUni}(\lambda, \A)$ is bounded as follows:
\begin{equation}
    Pr[1\leftarrow \Gnn{\UqPUF}{\qUni}(\lambda, \A)] \leq \frac{d+1}{D}
\end{equation}
where $0\leq d \leq D-1$ is the dimension of the largest subspace of $\HilD$ that can be spanned by $\A$'s sample database, during learning phase.
\end{theorem}
\end{thmbox}

\begin{proof}
Let $\A$ be a QPT adversary playing the game $\Gnn{\UqPUF}{\qUni}(\lambda, \A)$ where $\UqPUF$ is defined over $\HilD$. Let $\Sin$ and $\Sout$ be the input and output database of the adversary after the learning phase respectively, both with size $k$. Also, Let $\Hild$ be the $d$-dimensional Hilbert space spanned by elements of $\Sin$ where $d \leq k$ and $\Hildout$ be the Hilbert space spanned by elements of $\Sout$ with the same dimension. $\A$ receives an unknown pure quantum state $\ket{\psi}$ as a challenge in the $\qUni$ challenge phase and tries to output a state $\rho_{\omega}$ as close as possible to $\kpo$. For the simplicity in the proof, we assume the adversary's forgery state is either a pure state or has a purification in the form of $\ket{\omega}$. We are interested in calculating the average probability for the fidelity of $\A$'s output state $\ket{\omega}$ and $\kpo$ be larger or equal to $\delta$. We calculate this probability on average over all the possible states chosen uniformly at random according to the Haar measure over $\HilD$.
\begin{equation}
Pr[1\leftarrow\Gnn{\UqPUF}{\qUni}(\lambda, \A)] = \underset{\ket{\psi}\in\HilD}{Pr}[F(\ket\omega,\kpo)\geq\delta]
\end{equation}
We aim to show, that for any $\delta \neq 0$, the success probability of $\A$ is negligible in $\lambda$. 

Following the game definition, as the adversary selects states of the learning phase, the classical description of these states is usually known for them while the corresponding responses are unknown quantum states. Let $\A'$ be an adversary who also receives the classical description of the outputs or the complete set of bases of $\Hild$ and $\Hildout$. Thus $\A'$ has a complete description of the map in the subspace; and as a result, has necessarily a greater success probability than $\A$. \begin{equation}
Pr[1\leftarrow\Gnn{\UqPUF}{\qUni}(\lambda, \A)]\leq Pr[1\leftarrow\Gnn{\UqPUF}{\qUni}(\lambda, \A')]
\end{equation}
Therefore from now on throughout the proof, we calculate the success probability of $\A'$ who has full knowledge of the subspace, and we bound the success probability of $\A$ via $\A'$. We also note that the adversary cannot enhance their knowledge of the subspace by entangling their local system to the challenges of the learning phase since the reduced density matrix of the challenge/response entangled state lies in the same subspace $\Hild$ and $\Hildout$. Hereby, upper-bounding the success probability of $\A$ with the success probability of $\A'$ who has the full knowledge of the subspace we have also included the possible entangled queries. 

Now, we partition the set of all the challenges into two parts: the challenges that are completely orthogonal to the $\Hild$ subspace, and the rest of the challenges that have non-zero overlap with $\Hild$. We denote the subspace of all the states orthogonal to $\Hild$ as $\Hildperp$. We analyse the average success probability of $\A'$ in terms of the following partial probabilities:
\begin{equation}
    \underset{\ket{\psi}\in\Hildperp}{Pr}[F(\ket{\omega}, \kpo) \geq \delta] \quad \text{and } \quad \underset{\ket{\psi}\not\in\Hildperp}{Pr}[F(\ket{\omega}, \kpo) \geq \delta].
\end{equation}
We denote $F(\ket{\omega}, \kpo)$ as $F_{\omega}$ for simplicity.
Since the probability of $\ket{\psi}$ belonging to any particular subset is independent of the adversary's learnt queries, the success probability of $\A'$ can be written as:
\begin{equation}
\begin{split}
        Pr[1\leftarrow\Gnn{\UqPUF}{\qUni}(\lambda, \A')] & = \underset{\ket{\psi}\in\Hildperp}{Pr}[F_{\omega} \geq \delta]\times Pr[\ket{\psi} \in \Hildperp] \\
        & + \underset{\ket{\psi}\not\in\Hildperp}{Pr}[F_{\omega} \geq \delta]\times Pr[\ket{\psi} \not\in \Hildperp]
\end{split}        
\end{equation}
where $Pr[\ket{\psi} \in \Hildperp] = 1-Pr[\ket{\psi} \not\in \Hildperp]$ denotes the probability of the randomly selected $\ket{\psi}$ being projected into the subspace of $\Hildperp$ or in other words, have zero support in $\Hild$. From \lemref{lem:qpuf-avg-projection-prob}, we know that this probability for any subspace, is equal to the ratio of the dimensions. Here $\Hildperp$ is a $D-d$ dimensional subspace, thus $Pr[\ket{\psi} \in \Hildperp] = \frac{D-d}{D}$ and respectively $Pr[\ket{\psi} \not\in \Hildperp] = \frac{d}{D}$. Also the probability is upper-bounded by the cases that the adversary can always win the game for $\ket{\psi} \not\in \Hildperp$\footnote{This is one of the main reasons that our obtained upper-bound for the universal unforgeability is not tight, as this assumes the cases where adversary always wins with probability 1 if the state has any non-zero overlap with the sample subspace. Nevertheless, this upper-bound is enough for our purpose to show the unforgeability. Yet, obtaining tight upper-bounds for universal unforgeability is an interesting open question.}. So, we have,
\begin{equation}\label{eq:qpuf-unf-proof-prob-total}
    Pr[1\leftarrow\Gnn{\UqPUF}{\qUni}(\lambda, \A')] \leq \underset{\ket{\psi}\in\Hildperp}{Pr}[F_{\omega} \geq \delta]\times(\frac{D-d}{D}) + \frac{d}{D}
\end{equation}
Finally, the only remaining term to be calculated is $\underset{\ket{\psi}\in\Hildperp}{Pr}[F_{\omega} \geq \delta]$.

\noindent We write the expansion of $\ket{\psi}\in\HilD$ in an orthonormal basis for $\HilD$ as $\ket{\psi} = \sum^D_{i=1}c_i\ket{e_i}$. For any $\ket{\psi}\in\Hildperp$, the set of $\{\ket{e_i}\}^D_{i=1}$ can be the a union of the bases of $\Hild$, \emph{i.e.} $\{\ket{e^{in}_i}\}^d_{i=1}$ and the bases of $\Hildperp$, \emph{i.e.} $\{\ket{e'_i}\}^{D}_{i=d+1}$. Note that any state in $\Hildperp$ is orthogonal to all the $\ket{e^{in}_i}$ states. Thus, we can rewrite as follows
\begin{equation}
\ket{\psi} = \sum^d_{i=1}c^{in}_i\ket{e^{in}_i} + \sum^D_{i=d+1}c'_i\ket{e'_i}
\end{equation}

\noindent Recall the case of interest is when $\ket{\psi}\in\Hildperp$, and , $\mbraket{\psi}{e^{in}_i} = 0$. As a result, $c^{in}_i = 0$ and we have,
\begin{equation}
\ket{\psi} = \sum^D_{i=d+1}c'_i\ket{e'_i}
\end{equation}
Similarly for the output state $\kpo = \sum^d_{i=1}c^{out}_i\ket{e^{out}_i} + \sum^D_{i=d+1}\alpha_i\ket{b_i}$, as the unitary preserves the inner product, $c^{out}_i = \mbraket{e^{out}_i}{\po} = \bra{e^{in}_i}U^{\dagger}U\ket{\psi} = \mbraket{e^{in}_i}{\psi} = 0$, and the correct output state can be written as
\begin{equation}
\kpo = \sum^D_{i=d+1}\alpha_i\ket{b_i} 
\end{equation}
where $\{\ket{b_i}\}^{D-d}_{i=1}$ are a set of bases for $\Hildperpo$. 

\noindent Finally, the adversary $\A'$ can produce a forgery written as follows
\begin{equation}
    \ket{\omega} = \sum^d_{i=1}\beta_i\ket{e^{out}_i} + \sum^D_{i=d+1}\gamma_i\ket{q_i} 
\end{equation}
where the first part is spanned by the basis of learnt output subspace and the second part has been produced in $\Hildperpo$ with $\{\ket{q_i}\}^{D-d}_{i=1}$ being a set of bases for $\Hildperpo$. Based on unitarity argued above, the first part of the state $\ket{\omega}$ always gives a $0$ fidelity, and for $\A'$ to optimise the probability all $\beta_i$ should be zero. WHich makes $\sum^{D-d}_{i=1}\gamma_i\ket{q_i} \in \Hildperpo$ where the normalization condition is $\sum^{D-d}_{i=1}|\gamma_i|^2 = 1$. 

Since $\ket{\psi}$ is an unknown state selected uniformly at random and independent of the adversary, there are infinite choices for a set of bases orthogonal to $\{\ket{e^{out}_i}\}^d_{i=1}$, there is no unique way for $\A'$ choose or obtain the rest of the bases to complete the set. As a result, the choice of the $\ket{q_i}$ bases are also independent of $\ket{e'_i}$ or $\ket{b_i}$. In other words, knowing a matching pair of $(\ket{q_i},\ket{b_i})$ increases the dimension of the known subspace by one meaning the adversary has more information than it is assumed to have.

So, for each new challenge, $\A'$ produces a state $\ket{\omega} = \sum^{D-d}_{i=1}\gamma_i\ket{q_i}$ with a totally independent choice of bases. Without loss of generality we can fix the bases $\ket{q_i}$ for different $\ket{\omega}$. To calculate the success probability of $\A'$, we calculate the fidelity averaging over all the possible choices of $\ket{\psi}$. The unitary transformation also preserves the distribution, in this case Haar distribution. This leads to a uniform distribution of all the possible $\kpo$. As a result, the average probability taken over all possible $\ket\psi$ is equal to the average probability over all possible $\kpo$,
\begin{equation}
    \underset{\ket{\psi}\in\Hildperp}{Pr}[F_{\omega} \geq \delta] = \underset{\kpo\in\Hildperpo}{Pr}[F_{\omega} \geq \delta].
\end{equation}
We now show that $\A'$ also needs to output $\ket{\omega}$ according to the uniform Haar distribution to win the game in the average case with the highest probability. Let $\A'$ output the states according to a probability distribution $\mathfrak{D}$ which is not uniform. Then, by repeating the experiment asymptotically many times, the correct response $\kpo$ covers the whole $\Hildperpo$ while $\ket{\omega}$ covers a subspace of $\Hildperpo$. This decreases the average success probability of $\A'$. So, the best strategy for $\A'$ is to generate the states $\ket{\omega}$ such that they span the whole $\Hildperpo$, \emph{i.e.} generating them according to the symmetric Haar uniform distribution. 

Based on the above argument, and the fact that all the $\ket{\omega}$s are produced independently, we show that the average fidelity over all the $\kpo$ is equivalent to the average fidelity over all the $\ket{\omega}$. There are different methods for calculating the average fidelity over Hilbert spaces \cite{zyczkowski_average_2005}, a common approach is to integrate over the symmetric measure such as Haar. In our case, the average fidelity can be formulated as $
    \underset{\kpo\in\Hildperpo}{\int}|\mbraket{\omega}{\po_x}|^2d\mu_x $
where $d\mu$ is the Haar measure based on which the reference state has been parameterized. Note that $\ket{\omega}$ can be different for any new challenge. Now we rewrite the above average with the new parameters:
\begin{equation}
\begin{split}
    \underset{\kpo\in\Hildperpo}{\int} F(\ket\omega,\ket{\psi_x^{out}})d\mu_x  & =  \underset{\kpo\in\Hildperpo}{\int}|\mbraket{\omega}{\po_x}|^2d\mu_x \\ & = \underset{\kpo\in\Hildperpo}{\int}|\sum^{D-d}_{i=1}\overline{\gamma_i}\mbraket{q_i}{\po_x}|^2d\mu_x \\
    & = \underset{\kpo\in\Hildperpo}{\int}|\sum^{D-d}_{i=1}\overline{\gamma_{i_{x}}}\mbraket{q_i}{\po}|^2d\mu_x \\ & = \underset{\ket{\omega}\in\Hildperpo}{\int}|\mbraket{\omega_x}{\po}|^2d\mu_x \\
    & = \underset{\ket\omega\in\Hildperpo}{\int} F(\ket{\omega_x},\ket{\psi^{out}})d\mu_x
\end{split}
\end{equation}

We used the fact that fidelity is a symmetric function of two states and the measure of integral is the same for both cases where either of $\kpo$ or $\ket{\omega}$ are smoothly parametrized according to the symmetric measure. We use this equality for averaging all the possible outputs for one $\kpo$. We wanted to calculate the probability of this average fidelity being greater than $\delta$. To this end, we first calculate more generally, the probability that the average fidelity is non-zero. since we have 
$
    \underset{\ket{\omega}\in\Hildperpo}{Pr}[F_{\omega} \neq 0] + \underset{\ket{\omega}\in\Hildperpo}{Pr}[F_{\omega} = 0] = 1,
$  
we calculate the probability of the zero fidelity as follows,
\begin{equation}
\begin{split}
    \underset{\ket{\omega}\in\Hildperpo}{Pr}[F_{\omega} = 0] &= \underset{\ket{\omega}\in\Hildperpo}{Pr}[|\mbraket{\omega}{\po}|^2 = 0] \\ & = Pr[\int|\sum^{D-d}_{i=1}\overline{\gamma_{i_{x}}}\mbraket{q_i}{\po}|^2d\mu_x = 0] \\
    & = \underset{x}{Pr}[(\sum^{D-d}_{i,j=1}\overline{\gamma_{i_{x}}}\alpha_j\mbraket{q_{i_{x}}}{b_j})^2 = 0]
\end{split}
\end{equation}
Based on the Cauchy–Schwarz inequality we obtain the following inequality:
\begin{equation}
    [\sum^{D-d}_{i,j=1}\overline{\gamma_{i_{x}}}\alpha_j\mbraket{q_{i}}{b_j}]^2 \geq  \sum^{D-d}_{i,j=1}|\overline{\gamma_{i_{x}}}\alpha_j|^2|\mbraket{q_{i}}{b_j}|^2
\end{equation}
where, 
\begin{equation}
\begin{split}
    \sum^{D-d}_{i,j=1}|\overline{\gamma_{i_{x}}}\alpha_j|^2|\mbraket{q_{i}}{b_j}|^2 & =
     \sum^{D-d}_{i,j=1}|\overline{\gamma_{i_{x}}}\alpha_j|^2|\mbraket{q_{i}}{b_j}\mbraket{b_j}{q_{i}}| = \sum^{D-d}_{i,j=1}|\overline{\gamma_{i_{x}}}\alpha_j|^2|\bra{q_{i}}\Pi_j\ket{q_{i}}|
\end{split}
\end{equation}
Overall, we have, 
\begin{equation}
\underset{\ket{\omega}\in\Hildperpo}{Pr}[F_{\omega} = 0] \geq \underset{x}{Pr}[\sum^{D-d}_{i,j=1}|\overline{\gamma_{i_{x}}}\alpha_j|^2|\bra{q_{i}}\Pi_j\ket{q_{i}}|=0]
\end{equation}

While the RHS is the probability of $\ket\omega$ being projected into the orthogonal subspace of a space that only includes $\kpo$ averaging over all the projectors. We use again \lemref{lem:qpuf-avg-projection-prob}. Here the dimension of the orthogonal subspace is equal to $D-d-1$, since the target subspace is one-dimensional and thus dimension of the orthogonal subspace needs to be subtracted by 1. We then have,
\begin{equation}
\begin{split}
    \underset{\ket{\omega}\in\Hildperpo}{Pr}[F_{\omega} = 0] & \geq
    \underset{x}{Pr}[(\sum^{D-d}_{i,j=1}|\overline{\gamma_{i_{x}}}\alpha_j|^2|\bra{q_{i}}\Pi_j\ket{q_{i}}|) = 0] \\
    & \geq \frac{D-d-1}{D-d}
\end{split}
\end{equation}
And as a result, 
\begin{equation}
    \underset{\kpo\in\Hildperpo}{Pr}[F_{\omega} \neq 0] = \underset{\kpo\in\Hildperpo}{Pr}[|\mbraket{\omega}{\po}|^2 \neq 0] \leq \frac{1}{D-d}
\end{equation}    
Which holds for any non-zero $\delta$ as well. By substituting back into \eqref{eq:qpuf-unf-proof-prob-total}, we conclude that the success probability of $\A'$ is
\begin{equation}
    Pr[1\leftarrow\Gnn{\UqPUF}{\qUni}(\lambda, \A')] = \frac{1}{D-d}\times(\frac{D-d}{D}) + \frac{d}{D} = \frac{d+1}{D}
\end{equation}
And the success probability of $\A$ is also bounded by the same bound,
\begin{equation}
Pr[1\leftarrow\Gnn{\UqPUF}{\qUni}(\lambda, \A)] \leq \frac{d+1}{D}
\end{equation}
which completes the proof.
\end{proof}

Using this theorem that establishes a bound on the success probability of a QPT adversary in terms of fidelity, we can now prove a similar result for a general test algorithm.

\begin{thmbox}
\begin{theorem}\label{th:qpuf-universal-unf}
\textbf{(Any UqPUF  provides quantum universal unforgeability)} Let the test algorithm $\T$ be defined according to \defref{def:test} and satisfy the condition $\err(\kappa_1, \kappa_2) = \negl(\kappa_1, \kappa_2)$. Then any $\UqPUF=(\qPUFGen,\qPUFEval, \T)$ satisfies quantum universal unforgeability as for any QPT adversary, the following holds,
\begin{equation}
    Pr[1\leftarrow \Gnn{\UqPUF}{\qUni}(\lambda, \A)] = \negl(\lambda).    
\end{equation}
\end{theorem}
\end{thmbox}

\begin{proof}
Let $\ket\psi$ be the challenge chosen in the universal challenge phase. Also, let $\kpo$ and $\ket\omega$ be the correct output of the UqPUF and the forgery state of adversary $\A$, respectively. Also, we assume there exists one copy of $\ket{\omega}$, thus $\kappa_1 = 1$, but the challenger may have $\kappa_2$ copies stored for verification. The success probability of $\A$ in the game $\Gnn{\UqPUF}{\qUni}(\lambda,\A)$ is equal to the probability of the test algorithm in outputting 1:
\begin{equation}
    Pr[1\leftarrow \Gnn{\UqPUF}{\qUni}(\lambda, \A)] = Pr[1\leftarrow \T(\kpo^{\otimes \kappa_1}, \ket{\omega}^{\otimes \kappa_2})]
\end{equation}
We simplify the notation of $Pr[1 \leftarrow \T(\ket\omega^{\otimes \kappa_1}, \kpo^{\otimes \kappa_2})]$ by substituting with $Pr[1 \leftarrow \T]$. We also note that all the probabilities are being taken on average over the uniform choice of the challenge, though we omit the notation. To calculate this probability, we consider two independent cases that leads to output 1. We introduce the parameter $\delta$ as the threshold for $F(\ket\omega,\kpo)$ which helps us to write the $Pr[1 \leftarrow \T]$ as sum of two terms, \emph{i.e.} the probability of $\T$ outputting 1 while $F(\ket\omega,\kpo)\ge \delta$ and the probability of $\T$ outputting 1 while $F(\ket\omega,\kpo) < \delta$:
\begin{equation}
Pr[1 \leftarrow \T] = Pr[1 \leftarrow \T , F(\ket\omega,\kpo)\geq \delta] + Pr[1 \leftarrow \T , F(\ket\omega,\kpo) < \delta]
\end{equation}
Let $\delta = \negl(\lambda)$. We have,
\begin{equation}
\begin{split}
    Pr[1 \leftarrow \T] & = Pr[1 \leftarrow \T | F(\ket\omega,\kpo) \geq \negl(\lambda)] Pr[F(\ket\omega,\kpo) \geq \negl(\lambda)] \\ & + Pr[1 \leftarrow \T | F(\ket\omega,\kpo) < \negl(\lambda)] Pr[F(\ket\omega,\kpo) < \negl(\lambda)]
\end{split}
\end{equation}
From \thmref{th:qpuf-uni-unf-fid}, we concluded that 
\begin{equation}
Pr[F(\ket\omega,\kpo) \geq \negl(\lambda)] \leq \frac{d+1}{D}
\end{equation}
where $d$ is the dimension of the subspace spanned by the learnt queries and $D=2^n$ is the dimension of the domain and range Hilbert spaces and $n$ is the number of qubits in each input/output state. Since the adversary is QPT, the number of learnt queries and as a result the value of $d$ should be polynomial in $n$, \emph{i.e.} $d=poly(n)$. Also, according to \defref{def:test}, we have,
\begin{equation}
Pr[1 \leftarrow \T | F(\ket\omega,\kpo) < \negl(\lambda)] = Err(\kappa_1, \kappa_2)
\end{equation}
And,
\begin{equation}
Pr[1 \leftarrow \T | F(\ket\omega,\kpo) \geq \negl(\lambda)]\leq F(\ket\omega,\kpo)
\end{equation}
Considering the equality cases and due to the fact that $Pr[F(\ket\omega,\kpo) < \negl(\lambda)] = 1 - Pr[F(\ket\omega,\kpo) \geq \negl(\lambda)]$, the following equation is obtained
\begin{equation}
Pr[1 \leftarrow \T] = Err(\kappa_1, \kappa_2) (1-\frac{d+1}{D}) + \negl(\lambda) \frac{d+1}{D}
\end{equation}
Recall that $\err(\kappa_1, \kappa_2)=\negl(\kappa_1,\kappa_2)$, $d=poly(n)$ and $D=2^n$ and hence $\frac{d+1}{D}=\negl(n)$ and the probability that the test algorithm outputs 1 is computed as
\begin{equation}
\begin{split}
Pr[1 \leftarrow \T] & = \negl(\kappa_1,\kappa_2) (1-\negl(n)) + \negl(\lambda) \negl(n) \\ &= \negl(\kappa_1,\kappa_2) + \negl(\lambda) \negl(n) 
\end{split}    
\end{equation}
Let $\lambda=f(\kappa_1,\kappa_2,n)$, therefore we have 
\begin{equation}
Pr[1\leftarrow \Gnn{\UqPUF}{\qUni}(\lambda, \A)] = Pr[1 \leftarrow \T] = \negl(\lambda)
\end{equation}
and the proof is complete.
\end{proof}

We have shown that general UqPUFs together with a reasonably good quantum test algorithm, always satisfy universal unforgeability. We note that in deriving this result, we have mostly used the symmetries and geometry of Hilbert spaces and the randomness of the selected challenge according to the Haar measure, emphasising that in the quantum case, unlike the classical regime, the unpredictability of the qPUF can be proven given its unknownness or single-shot indistinguishability as a hardware assumption. One can also intuitively infer that any initially unknown unitary is hard to learn on average, given efficient-size oracle access.

\subsection{A note on the unforgeability of quantum PUFs with public database}\label{sec:qpuf-publicdb-pufs}
As we discussed in the previous section, the randomness of the challenge state, and the fact that an unknown single copy of it is available for the adversary, plays an important role in the unforgeability property of qPUFs. We specifically point out the close relation to the fundamental limitation of the adversary in copying a single unknown quantum state with high fidelity. It might appear that using the no-cloning property of the challenge state, is enough to provide universal unforgeability, and one might even be able to achieve unforgeability if the unitary transformation is fully or partially public. In fact, this was one of the core ideas in early proposals for using quantum transformation as physical unclonable functions \cite{skoric_quantum_2010,nikolopoulos_continuous-variable_2017}. In these works, it has been conjectured that an efficient adversary (QPT) is still incapable of providing a good estimation or guess for the response state, even if the unitary, or essentially the classical description of all the potential challenges are known to such adversaries. The given argument is simply the consequence of the challenge state being unknown and provided as a single copy, which presumably makes it hard for the adversary to determine the correct response. In this case, the best strategy would be to measure and estimate the challenge state, which will lead to a small success probability, and as a result, guessing the response state based on such measurements has accordingly low success probability. Making a PUF's database public is motivated since it can clear the need for securely storing big classical or quantum data.

Nevertheless, in the light of our new cryptanalysis tool, namely the \QE, we note that such observations and conjectures are not in general correct and only apply to specific attack models such as cases where the adversary is restricted to only prepare and measure single-qubit quantum states. Against a general QPT adversary, we show that no unitary qPUF with a public or partially public database can provide universal unforgeability. This new result is a direct byproduct of our cryptanalysis of quantum emulation in \chapref{chap:unf-tools}, thus we present it as the following corollary.

\begin{corrbox}
\begin{corollary}\label{cor:qpuf-no-unf-pdb} Let $\UqPUF=(\qPUFGen,\qPUFEval,\Td)$, be a unitary qPUF with unitary $U$ of dimension $D$. Let $\A$ be a QPT adversary and let $\mathcal{S} = (\Sin,\Sout)$ be an efficient-size ($polylog(D)$) sample set of $\UqPUF$ including challenge and response pairs respectively, known to $\A$. Let $\Hild$ be the subspace fully spanned by $\mathcal{S}$. For any challenge state $\ket{\psi}$ selected from any arbitrary distribution over $\Hild$, $\A$ can produce a state $\ket{\omega}$, with a very high fidelity compared to $U\ket{\psi}$. Therefore $\UqPUF$s in this setting do not provide universal unforgeability.\footnote{We note that the use of the term \emph{universal unforgeability} here is slightly informal and different from the universal unforgeability as formally defined in \gameref{game:qpuf-unf} since here the challenge is selected from an arbitrary distribution over a subspace rather than being selected from Haar measure over the full space. However, we deliberately use the same expression, as it captures the same notion as the universal unforgeability where the challenge is being selected by the challenger and not the adversary. Only here, the selection of the challenge state is from a different space and distribution.}
\end{corollary}
\end{corrbox}
\begin{proof}
Let the database include $q$ quantum input-output query pairs. Let $\A$ run a q-block \QE\ using $\mathcal{S}$. Since, $\ket{\psi}$ is fully spanned by $\mathcal{S}$ or alternatively by the full basis of $\Hild$, then the quantum emulation algorithm can emulate the output of $\ket{\psi}$, \emph{i.e.} $U_{\text{qPUF}}\ket{\psi}$ with an almost 1 fidelity according to \thmref{th:qe-fidel}. Thus this output passes any test algorithm with an overwhelming probability.
\end{proof}

As a result of the above corollary, it is fairly obvious that an efficient adversary can always successfully emulate the response of the UqPUF if the database is publicly known since they can locally build the set $\mathcal{S}$, from which the challenge state is selected, and win the universal unforgeability game with probability almost close to 1. Note that a universal quantum emulator is an efficient quantum algorithm, hence can be run by a QPT adversary. More importantly, the above result states that, in all cases where the adversary has a considerable amount of knowledge over an efficient subspace from which the challenge is selected, the quantum emulation attack can be performed on the unknown challenge,  leading to a high fidelity forgery state and breaking unforgeability, even though the challenge state is unclonable.

\section{Discussion and conclusions}\label{sec:qpuf-disc}
We have formally defined quantum physical unclonable functions, as a new notion of unclonability. We established the minimum requirements and conditions to be satisfied at a hardware level such that a unitary transformation qualifies as a qPUF. In doing so, we have also studied the connection between unknownness, physical unclonability and no-cloning of unitary transformations. We have then analysed the unforgeability of qPUFs as their property of interest, both for understanding them as cryptographic primitives and from the application point of view. We proved that even though no qPUF can be exponentially or existentially unforgeable, our proposed general notion of unitary qPUF, provided the unknownness, always satisfies universal unforgeability that has a close connection with the unlearnability of these primitives efficiently. We now briefly discuss the relationship between our proposal and other types of PUFs, as well as the open questions and direction for future works.

First, we briefly discuss the relevance of our framework and result for cPUFs. As mentioned before, input-output pairs of a cPUF are bit-strings. Most of the available cPUF structures use digital encoding as their inputs and outputs. As a result, they can easily be integrated with other functionalities in Integrated Circuits (ICs). Considering encoding of such bit strings in the computational basis of a Hilbert space, the cPUFs can be considered as special types of UqPUFs. Given a quantum oracle access to such PUFs, they can be studied under the \emph{quantum security model} as discussed in \chapref{chap:unf-tools}, queried with quantum states. In this model, our no-go result stating that no UqPUF provides quantum existential unforgeability can be extended to cPUFs, showing that they are also unable to provide this level of unforgeability for $\mu\neq 1$. 
 
Another interesting point of comparison is to compare the assumptions that lead to unforgeability for qPUFs and cPUFs. According to \cite{armknecht_towards_2016}, the min-entropy requirement (which imposes that the cPUF responses are linearly independent) is the main requirement of a cPUF which leads to existential unforgeability \cite{armknecht_towards_2016} against classical adversaries with no quantum access to the cPUF. However, this requirement cannot be achieved with most of the common cPUF structures as shown in \cite{ganji_strong_2016,ruhrmair_puf_2014,ruhrmair_modeling_2010,khalafalla_pufs_2019}. On the other hand, qPUFs only need the basic assumption on PUFs that let the behaviour of PUF be unknown to anyone \cite{ruhrmair_pufs_2014}; and under that assumption, they can achieve a slightly weaker notion of unforgeability, yet against much stronger adversaries with quantum capabilities.

It is worth mentioning that in theory, UqPUFs (and as a result, cPUFs in the quantum security model) can still achieve existential unforgeability for $\mu = 1$. Nevertheless, finding physical structures and systems that provide this level of unforgeability is still an open question. To the best of our knowledge, there is no study on the quantum security of cPUFs in the literature. We emphasise that given the speedy progress in quantum technology the investigation of the security of cPUFs against quantum adversaries is crucial. The security of silicon cPUFs and the other types of cPUFs that cannot be queried by quantum states can be explored in the \emph{post-quantum (or standard) security model} where the quantum adversary has only classical interaction with the primitive while they are equipped with a quantum computer. However, for the other types of cPUF structures like optical PUFs that can naturally be queried with quantum states, the security of cPUFs needs to be analysed in the quantum security model with quantum access to the cPUF oracle.

Another main category of PUFs is Quantum Read-out PUFs (QR-PUFs). Since they are also modelled via unitary transformations, they can be naturally compared to UqPUFs. The original definition of QR-PUFs considered quantumly-encoded challenge-response pairs. \cite{skoric_quantum_2010,skoric_quantum_2012}. The security of QR-PUF-based identification protocols has been investigated in specific and limited security models, such as prepare-and-resend adversaries in \cite{skoric_quantum_2010,skoric_quantum_2012,nikolopoulos_continuous-variable_2017,goorden_quantum-secure_2014,skoric_security_2013,nikolopoulos_continuous-variable_2018,fladung_intercept-resend_2019} where either the full unitary transformation or equivalently the classical description of QR-PUF responses for any known challenge, is assumed to be public knowledge. The security of such PUF-based protocols relies on the bounds for estimating an unknown quantum challenge sent by the verifier. 

Although our current framework as it is, will not be directly applicable to all sorts of protocols and scenarios in which QR-PUFs are defined and used due to specific sets of assumptions and adversarial models considered in these scenarios, we believe that QR-PUFs as a stand-alone primitive can be studied in our proposed framework. Following section~\ref{sec:qpuf-publicdb-pufs}, we discuss an extended class of such qPUFs which we call Public-Database PUFs (or PDB-PUFs). They include any PUF that can be queried with quantum (or quantumly encoded) challenges, producing quantum responses and are modelled by a publicly known unitary transformation or a public database equivalently. Our framework provides security notions against general and quantum adversaries in the standard game-based model. Hence we can also investigate the security of PDB-PUFs, by relaxing the unknownness condition for this class. \corrref{cor:qpuf-no-unf-pdb} shows that no PUF in this class can provide universal unforgeability (and existential unforgeability). Yet interestingly, the constructions proposed for such PUFs can be potential candidates for secure qPUFs (such as the optical qPUF presented in \cite{nikolopoulos_continuous-variable_2017}), by removing the assumption of the public database while ensuring that the challenge subspace is unknown to the adversary. It is worth mentioning that the feasibility of other quantum attacks with current technologies has been discussed in \cite{skoric_quantum_2010,skoric_quantum_2012,goorden_quantum-secure_2014,skoric_security_2013,nikolopoulos_continuous-variable_2018,fladung_intercept-resend_2019}. However, it remains an interesting inquiry whether the quantum emulator attack can also be demonstrated on NISQ quantum devices.

Furthermore, we note that given that our bounds for universal unforgeability are not tight, it leaves a space for exploring the possibility of universal unforgeability under more efficient challenge sets. We will show an example of this in the next chapter. Finding tight bounds for the unforgeability of qPUFs and more generally, quantum primitives is yet another attractive and challenging open problem. Later on, in \chapref{chap:application}, we will see that toolkits from quantum information theory such as entropic uncertainty relations and data processing inequalities are useful and powerful tools to prove similar results for specific constructions and we, therefore, suggest that the proposed problem and supervised learning can be both attacked using similar tools. Information-theoretic bounds for quantum vs classical machine learning has been also studied in~\cite{huang_information-theoretic_2021}. Considering the close relationship between unforgeability and learning problems, we believe this work can assist uncover tighter universal forgery bounds for qPUF and other general quantum primitives. 

An important complementary question that we left open is the design of concrete qPUF constructions based on the proposed formal framework. Developing sufficiently secure constructions for quantum PUF would be much more complicated than their classical counterparts as one needs to deal with many complications of the quantum world such as noise and decoherence. One of the important steps in this route is to study non-unitary qPUFs while relaxing the strong collision-resistance requirement to a weaker version. Such qPUFs will allow for more general noise models, and if proven to be secure, they will push the practical construction of qPUFs one step further towards experimental realisation. Another challenge in the way of industrialising qPUFs is the need for quantum memory for some of the qPUF-based protocols. It is an interesting question to what degree this resource can be reduced or even removed in different protocols. We will try to address this question to some extent in \chapref{chap:application}.

Finally, certification of qPUFs brings up compelling questions that are both of theoretical and practical interest. One example is to develop new efficient techniques for certifying the \emph{effective dimensionality} of quantum black-box primitives such as UqPUFs. These techniques will not only be beneficial certification tools for qPUFs in practice but can also be new toolkits for certification more generally.

\subsection{Subsequent works}
To round off this chapter, let us briefly mention a few related works which appeared after the completion of the work presented in this chapter. Firstly, in \cite{kumar_efficient_2021} a circuit-based construction has been proposed for qPUFs. The construction uses t-designs to accomplish the functionality of UqPUFs as well as showing the requirements of qPUFs are satisfied. Although this specific construction does not provide physical unclonability against the manufacturer due to the gate-based construction, it can be used against third-party adversaries (and not the manufacturer).

Another proposal for achieving physical unclonability via quantum devices is given in \cite{phalak_quantum_2021}. In this work, the concept of a Classical-Readout QPUF (CR-QPUF) has been introduced where a quantum device is queried classically. Such PUFs aimed to utilize the noisy behaviour of quantum devices as the main source of physical unclonability and to remove the requirement for a quantum memory in the qPUF-based authentication protocols. In the proposed protocol, the challenge is a classical description of a parameterized unitary, which runs on a quantum computer. Then, the mean value of the measurement outcome over the qubits in the computational basis has been taken to be the response. In the proposed protocols both challenges and responses are communicated over a classical channel. Even though this construction seems practically feasible and interesting, one can debate that considering the mean-value as the response, will most probably substantially reduce the security as a consequence of shrinking the response space to a high degree. Later in \cite{pappa_learning_2021}, the authors will show that this observation is correct and such PUFs can be efficiently learned using machine learning methods. Another appealing aspect of the work in \cite{pappa_learning_2021} is that it formalises the class of Classical Readout Quantum
PUFs (CR-QPUFs) using the statistical query (SQ) model that seems another promising approach to studying non-unitary PUFs with underlying quantum properties. The modelling attack proposed in this work explicitly shows the insufficiency of this class by successfully implementing the attack on the $\IBM$ quantum machine.

The research on authentication protocols using optical PUFs also continued in the two following works \cite{wang_authentication_2021,nikolopoulos_remote_2021} demonstrating experimental realisation of photonic PUFs in specific authentication protocols. 
\chapter{Connection Between Quantum Pseudorandomness and Quantum Hardware Assumptions} \label{chap:pr-connection}
\begin{chapquote}{Leonard Susskind}
``Unforeseen surprises are the rule in science, not the exception. Remember: Stuff happens.''
\end{chapquote}
\section{Introduction}\label{sec:connection-intro}
In the previous chapter, we have thoroughly studied the concept of physical unclonability, and quantum physical unclonable functions as hardware assumptions. We have also proved several results about their main cryptographic properties. Furthermore, in the course of the last two chapters, we have attempted to disclose the fundamental connection between the notions of unclonability and randomness. In this chapter, we turn into another stimulating key concept in cryptography, namely \emph{pseudorandomness}, and we show the relationship between this notion and physical unclonability, or even more generally, hardware assumptions.

Pseudorandomness is one of the most fundamental concepts in cryptography and complexity theory. In contrast to true randomness, it captures the notion of primitives that behave randomly to the computationally-bounded observers \cite{yao_theory_1982,shamir_generation_1983,blum_how_1984}. Pseudorandom objects like pseudorandom number generators (PRGs) and pseudorandom functions (PRFs) play a crucial role in designing classical symmetric key cryptographic protocols for secure communications \cite{goldreich_how_1986,hastad_pseudorandom_1999,luby_how_1988,rompel_one-way_1990}. These pseudorandom objects can be designed by exploiting the algebraic properties of families of keyed functions like keyed hash functions. Nevertheless, constructing these pseudorandom objects is challenging and usually relies on some computational assumptions. Recently Ji, Liu, and Song \cite{shacham_pseudorandom_2018} introduced the concept of quantum pseudorandomness as a quantum analogue of this concept by introducing pseudorandom quantum states (PRS, \defref{def:prelim-prs}) and pseudorandom unitaries (PRU, \defref{def:prelim-pru}). These are families of states or unitary transformations indistinguishable from Haar measure (true random measure) to any quantum computationally-bounded observer. Even though quantum pseudorandomness is a very new field of research, it has already found many applications in cryptography \cite{shacham_pseudorandom_2018,canteaut_efficient_2020,micciancio_scalable_2020,morimae_quantum_2021,ananth_cryptography_2022}, complexity theory \cite{kretschmer_quantum_2021,brandao_models_2021}, learning theory \cite{huang_quantum_2021}, and high energy physics \cite{bouland_computational_2019,kim_ghost_2020}. The existing PRS schemes are constructed under computational assumptions such as quantum-secure PRFs or quantum secure one-way functions~\cite{micciancio_scalable_2020,shacham_pseudorandom_2018}. An interesting question arises here: \emph{Whether quantum pseudorandomness can be achieved under different sets of assumptions, for instance, hardware assumptions?} 
In this chapter, we mainly try to address this question. Given our specific interest in qPUFs, as a well-defined hardware assumption, we mostly focus on them and for the first time, we show the construction of quantum pseudorandom unitaries from quantum PUFs and vice-versa. We also point out that in the classical world, the relationship between PUFs and pseudorandomness has also been studied~\cite{ruhrmair_foundations_2009} and it would be interesting to see if such a relationship also exists in the quantum setting.

Understanding this connection not only provides a deeper understanding of quantum pseudorandomness itself but also can substantially improve the construction of qPUFs and qPUF-based applications as well. Let us give an example. In the previous chapter, we have seen that a Haar-random family of unitaries can, by definition, be a family of secure qPUFs. Nonetheless, sampling Haar-random unitaries and states requires exponential resources~\cite{knill_approximation_1995,nakata_quantum_2021} and hence is experimentally challenging~\cite{carolan_universal_2015}. Moreover, the challenge distribution for universal unforgeability is required to also be Haar. If PRSs, can be used within the framework of universal unforgeability as a challenge set, a considerable improvement will occur in the practicality of any universally unforgeable scheme, including quantum PUFs.

In this work, we make substantial progress in the challenges mentioned above. Firstly, we show that PRS can replace the Haar-random assumption in the challenge state's selection for universal unforgeability. We further show that PRUs can be used as a viable candidate for qPUFs. This result provides yet another novel and efficient technique for constructing qPUFs.

Concerning our first question about alternative ways of constructing quantum pseudorandomness, we show that a qPUFs family can also be a family of PRUs. This, in turn, makes them special physical generators of pseudorandom quantum states.

Later, we give a novel construction of PRUs by exploring yet another hardware requirement, \emph{i.e.} the uniqueness property. This result is shown generally for any family of unitary matrices with a certain specified degree of uniqueness, not only qPUFs. And as long as the uniqueness property can be assumed at a hardware level, it relates a hardware assumption to quantum pseudorandomness. Informally, we prove that any family of unitary transformations over $d$-dimensional Hilbert space satisfying almost-maximal uniqueness in the diamond norm is also a PRU family for sufficiently large $d$. Hence any PUF family satisfying this degree of uniqueness, is also a PRU.

\begin{figure}
\includegraphics[scale=0.35]{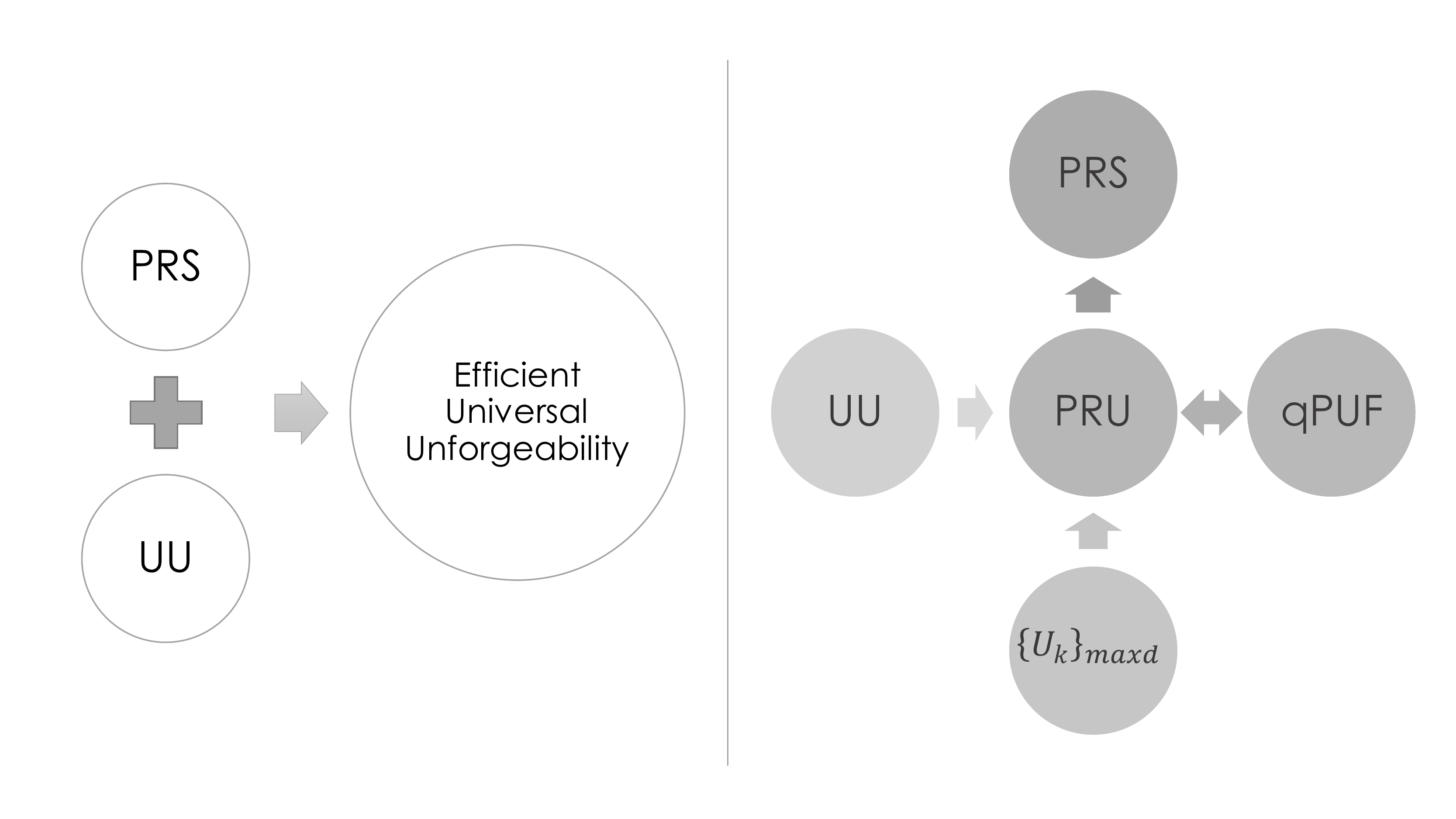}
    \centering
    \caption[Connection between different notions of quantum pseudorandomness, unknownness, and efficient universal unforgeability]{Pictorial summary of the results: The left-hand figure demonstrates \thmref{th:connection-efficientuu-prs} stating that universal unforgeability of unknown unitaries can be achieved efficiently using PRSs. The right figure depicts the relationship between unknown unitaries (UUs), quantum physical unclonable functions (qPUFs), pseudorandom unitaries (PRUs) and families of almost maximally-distanced unitaries ($\{U_k\}_{maxd}$) proved in \thmref{th:connection-pru-uu}, \thmref{th:connection-pru-unique}, \thmref{th:connection-practicaluu-pru}, and \thmref{th:connection-max-unique-pru}. It also shows that they can be used as generators for PRSs.}
    \label{fig:connection-results}
\end{figure}

Our investigation in this chapter helps establish a close connection between these two new fields and gives us novel insights into both physical unclonability and quantum pseudorandomness. A summary of our results is shown in \figref{fig:connection-results}. We are optimistic that the connections we foster here will enrich both fields.

\subsection{Structure of the chapter}
We begin the chapter with a question: Is it possible to have universal unforgeability with a PRS challenges set instead of a Haar-random state without losing any security guarantee? In Section \ref{sec:connection-unf-prs}, we give a positive answer to this question with a formal security proof. In section \ref{sec:connection-unf-pru-tdesign}, we show that one can construct a family of unknown unitaries from PRUs, which gives a potentially efficient proposal for constructing a family of qPUFs. Finally, in section \ref{sec:connection-pru-from-uu-unique}, we show that given hardware assumptions such as uniqueness and unknownness, one can achieve quantum pseudorandomness, which completes our picture.

\section{Efficient unforgeability with PRSs}\label{sec:connection-unf-prs}
In this section, we investigate the problem of \textit{universal unforgeability} with efficiently producible pseudorandom quantum states. As specified in the universal unforgeability framework in \chapref{chap:unf-tools}, the challenge states should be picked at random from Haar measure by the challenger. This is an important condition for the unforgeability of unknown unitary transformations. Since producing Haar random state is a challenging and resource-intensive task, to take the first step towards the realization of universally unforgeable schemes, we attempt to replace this condition with its computational equivalent, \emph{i.e.} the notion of PRS, introduced in \chapref{chap:prelim} (Section~\ref{sec:prelim-quanutm-pseudorandom}). We first relax this condition by defining a variant of the universal unforgeability game, namely quantum \emph{Efficient Universal Unforgeability (qEUU)} where the challenger picks the challenge states from a pseudorandom family of quantum states. Then we formally prove that unknown unitaries satisfy this notion of unforgeability. Furthermore, we briefly discuss how such pseudorandom quantum states can be efficiently generated using classical pseudorandom functions.

We define efficient universal unforgeability as follows:

\begin{defbox}
\begin{definition}[quantum Efficient Universal Unforgeability (qEUU)]\label{def:qunf-efficient}
Let Game $\Gnef$ be the same as \gameref{game:qpuf-unf}, except that in the challenge phase, the challenge states are being picked from the PRS family of states with a generation algorithm $G(k)$ with a key $k\in\K$, run in the setup phase. A primitive provides \emph{efficient quantum universal unforgeability} if the success probability of any QPT adversary $\A$ in winning the game $\Gnef$ is negligible in the security parameter $\lambda$,
\begin{equation}
Pr[1\leftarrow \Gnef(\lambda, \A)] = \negl(\lambda)
\end{equation}
\end{definition}
\end{defbox}

\noindent For the purpose of our proof, we also rewrite the pseudorandomness property of the PRS as a game which we formalized in the following:

\begin{gamebox}{PRS distinguishability game}
\begin{game}\label{game:prs} Let $\Hil$ be a Hilbert space and $\K$ the key space. The dimension of $\Hil$ and size of $\K$ depend on the security parameter $\lambda$. Let $\{\ket{\phi_k}\in S(\Hil)\}_{k\in\K}$ be a keyed family of quantum states with efficient generation algorithm $G(k) = \ket{\phi_k}$ on input $k$. We define the following distinguishability game between an adversary $\A$ and a challenger $\C$:
\begin{itemize}
    \item [] \textbf{Setup phase.} The challenger $\C$ selects $k \overset{\$}{\leftarrow} \K$ and $b \overset{\$}{\leftarrow} \{0,1\}$ at random.
    \item [] \textbf{Challenge phase.}
        \begin{itemize}
            \item If $b = 0$ (PRS world): $\C$ prepares $m$ copies of $\ket{\phi^0} = \ket{\phi_k}$ by running $G(k)$. 
            \item If $b = 1$ (Random world): $\C$ prepares $m$ copies of a Haar-random state $\ket{\phi^1} = \ket{\psi}$.
            \item $\C$ sends $\ket{\phi^b}^{\otimes m}$ to $\A$.
        \end{itemize} 
    \item [] \textbf{Guess phase.} $\A$ guesses $b$.
\end{itemize}
\end{game}
\end{gamebox}

\noindent We now establish our main result regarding efficient unforgeability of unknown unitary primitives. 

\begin{thmbox}
\begin{theorem}\label{th:connection-efficientuu-prs}
Any unitary transformation $U$ selected from a family of unknown unitaries satisfies \textbf{quantum efficient universal unforgeability} against any QPT adversary. 
\end{theorem}
\end{thmbox}

\begin{proof}
We prove this theorem by contraposition in a game-based setting. We want to show that starting from the assumption of pseudorandomness of PRS in the efficient universal unforgeability game, if there exists a QPT adversary who succeeds to win this game, with non-negligible probability, there will also exist an adversary who can efficiently distinguish between PRS and Haar random states, which is in contrast with the initial assumption and as a result show a contradiction. First, we need to specify the following games:
\begin{itemize}
    \item \texttt{Game 1}: This is the universal unforgeability game as specified in \gameref{game:unf-full}, with the only difference that the challenge state $\rho^* = \ket{\phi_{k^*}}\bra{\phi_{k^*}}$ is chosen from a PRS family.
    \item \texttt{Game 2}: This is the PRS distinguishability game as specified in \gameref{game:prs}.\footnote{One small remark is that in PRS game, the unitary is picked inside the game, while the universal unforgeability game takes the unitary as part of the primitive and hence applies to any selected unitary. However, since we will show that our result applies to \emph{all} unknown unitary matrices, it will also hold in the average case. Thus we drop this distinction in the course of the proof to avoid confusion.}
    \item \texttt{Game 3}: This is a variation of \gameref{game:prs}  where $\C$ in addition to initial resources, has also access to a publicly known and implementable unitary $U$. In the challenge phase, $\C$ does the following: Generates $m$ copies of $\ket{\phi^0} = \ket{\phi_k}$ using $G(k)$, or $m$ copies of Haar random states $\ket{\phi^1} = \ket{\psi}$ depending of $b$, then on each copies applies the public unitary $U$ and sends $(U\ket{\phi^b})^{\otimes m}$ to $\A$. The rest of the game is similar to \texttt{Game 2}. 
    \item \texttt{Game 4}: This game is similar to \texttt{Game 3}, except that $\C$ publicly chooses an $l$ and $l'$ such that $l + l' = m$ and sends $l$ copies of the generated state and $l'$ copies of the state after applying the unitary $U$, \emph{i.e.} sends $\ket{\phi^b}^{\otimes l} \otimes (U\ket{\phi^b})^{\otimes l'}$ to $\A$.  
    \item \texttt{Game 5}: This game is similar to \texttt{Game 4} except the public unitary has been replaced by an unknown unitary $\tilde{U}$ of the same dimension. Hence in this game, similar to \gameref{game:unf-full}, we also assume a learning phase for $\A$ before the challenge phase. The learning phase is as follows: $\A$ issues $q=poly(\lambda)$ queries $\{\rho_i\}^q_{i=1}$ to $\C$, on each query $\C$ generates $\rho^{out}_i = \tilde{U}\rho_i\tilde{U}^{\dagger}$ by applying the unitary on the query state and sends $\rho^{out}_i$ to $\A$. Then the rest of the game is similar to \texttt{Game 4} and at the end of the challenge phase $\A$ receives $\ket{\phi^b}^{\otimes l} \otimes (\tilde{U}\ket{\phi^b})^{\otimes l'}$
\end{itemize}

\begin{figure}[h!]
\includegraphics[scale=0.45]{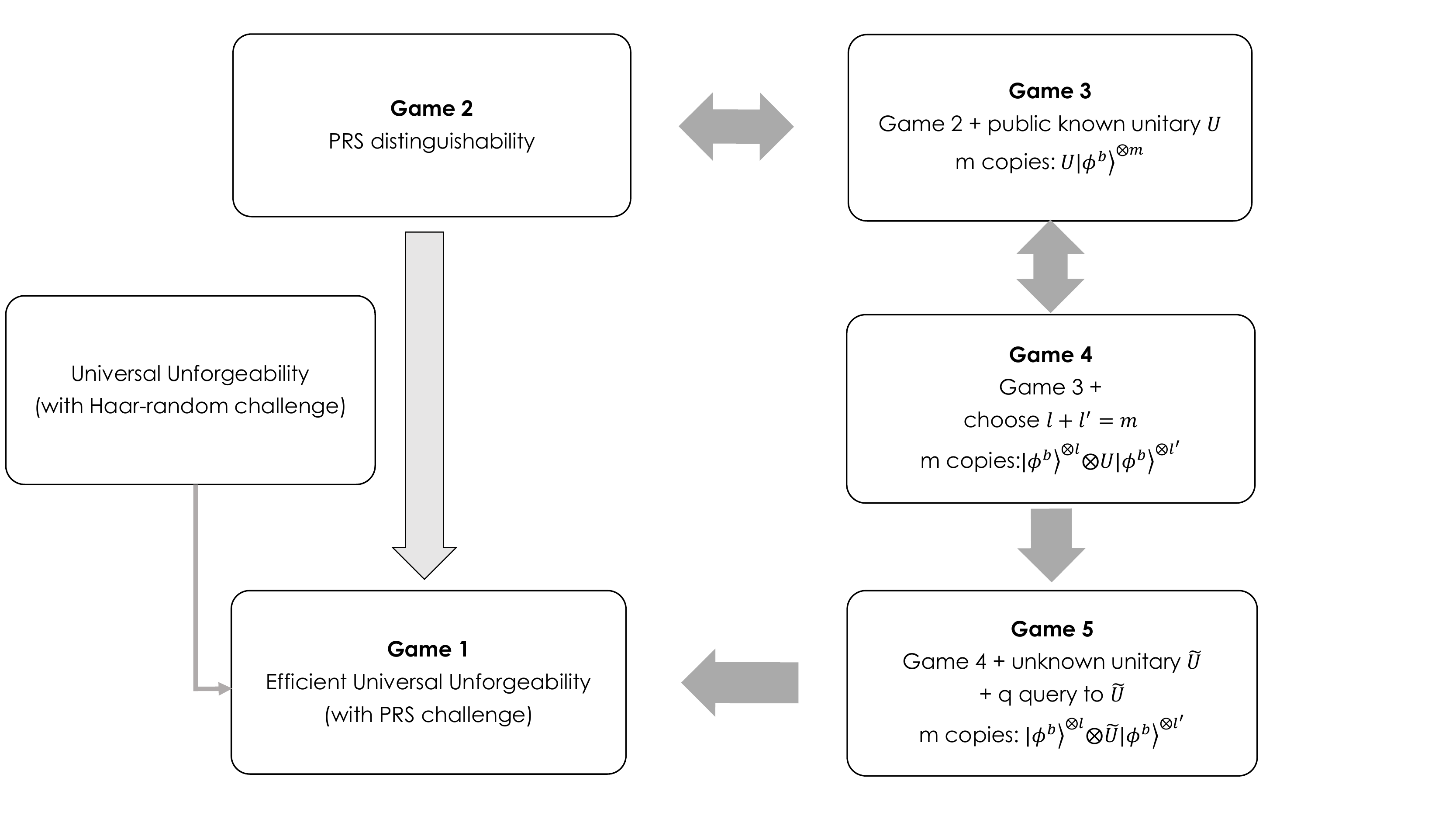}
    \centering
    \caption{Proof sketch of Theorem~\ref{th:connection-efficientuu-prs} with the intermediate games.}
    \label{fig:prs-unf-proof}
\end{figure}

Figure~\ref{fig:prs-unf-proof} illustrates the sketch of the proof. We first show that \texttt{Game 2}, \texttt{Game 3} and \texttt{Game 4} are equivalent. We note that unitary transformations are distance invariant and hence they also preserve the distribution of states, as a result applying a unitary to the state will not affect the distribution and the distinguishability of the quantum states, and as a result \texttt{Game 2} and \texttt{Game 3} are equivalent. Furthermore, in \texttt{Game 4}, since the unitary is public, $\A$ can either apply $U$ on the first $l$ copies $\ket{\phi^b}^{\otimes l}$ and end up with $m$ copies of $(U \ket{\phi^b})^{\otimes m}$ or alternatively apply $U^{\dagger}$ on the next $l'$ copies $(U\ket{\phi^b})^{\otimes l'}$ and get $m$ copies of $\ket{\phi^b}^{\otimes m}$, and hence be reduced to either \texttt{Game 2} or \texttt{Game 3}. As a result, we have
\begin{equation}
    \texttt{Game 2} \equiv \texttt{Game 3} \equiv \texttt{Game 4}
\end{equation}
Now we show that \texttt{Game 4} implies \texttt{Game 5} \emph{i.e.} if an adversary wins distinguishability in \texttt{Game 5} with probability $p$, they will also win in \texttt{Game 4} with the same probability.

The proof is straightforward as highlighted here. Let $\A$ be an adversary who wins \texttt{Game 5}, which means after the learning phase leading to a polynomial-size database of input-outputs of the unknown unitary $\tilde{U}$, and receiving $\ket{\phi^b}^{\otimes l} \otimes (\tilde{U}\ket{\phi^b})^{\otimes l'}$, they can guess $b$ with non-negligible probability better than random guess:
\begin{equation}
    \underset{\ket{\phi^b}}{Pr}[b \leftarrow \A(\ket{\phi^b}^{\otimes l} \otimes (\tilde{U}\ket{\phi^b})^{\otimes l'})] = \frac{1}{2} + \nonnegl(\lambda).
\end{equation}
Now let's assume an adversary $\A'$ who plays \texttt{Game 4} and has to guess $b$ by receiving the state $\ket{\phi^b}^{\otimes l} \otimes (U\ket{\phi^b})^{\otimes l'}$ can guess $b$ with same $l$ and $l'$ where $U$ is a public unitary. Now $\A'$ can run $\A$ as a subroutine and $\A'$ sends to $\A$ the response to the same learning phase states from $U$. Since $U$ is public $\A'$ can run it locally and produce the required queries. Then $\A'$ also sends the state $\ket{\phi^b}^{\otimes l} \otimes (U\ket{\phi^b})^{\otimes l'}$ to $\A$ and since $\A$ guesses the $b$ with a probability non-negligibly better than half, so does $\A'$. As a result, we have shown that:
\begin{equation}
    \texttt{Game 4} \Rightarrow \texttt{Game 5}
\end{equation}
Finally, we show that \texttt{Game 5} implies \texttt{Game 1}. By contradiction, we assume there exist an adversary $\A$ who wins the unforgeability game with non-negligible probability. Let $\tilde{U}$ be the unknown unitary and $\A$'s forgery state be $\ket{\omega}$ and let the challenge state of \texttt{Game 1} be a PRS state $\ket{\phi_k}$. We have:
\begin{equation}
\begin{split}
        Pr[1\leftarrow \Gnef(\lambda, \A)] & = \underset{k}{Pr}[1 \leftarrow \T(\ket{\omega}, (\tilde{U}\ket{\phi_k})^{\otimes \kappa})] \\
        & = \underset{k}{Pr}[F(\ket{\omega}, \tilde{U}\ket{\phi_k}) = \nonnegl(\lambda)] \\
        & = \nonnegl(\lambda).
\end{split}
\end{equation}
Now we construct an adversary $\A'$ playing an instance of \texttt{Game 5} where $l = 1$ and $l' = m - 1$. In the learning phase $\A$ interacts with the unknown unitary $\tilde{U}$ with the same learning phase states required for $\A$ and sends the query states $\{\rho^{out}_i\}^q_{i=1}$ together with the challenge state $\ket{\phi^b}$ to $\A$. Then $\A$ produces the forgery $\ket{\omega}$ as his guess for $\tilde{U}\ket{\phi^b}$. Now $\A'$ verifies $\ket{\omega}$ with the same test algorithm $\T$ where $\kappa = m - 1$, since $\A'$ has $m-1$ copies of $\tilde{U}\ket{\phi^b}$ to check with. Then $\A'$ outputs the same $b$ as outputted by the $\T$. The success probability of $\A'$ is as follows. If $b=0$, the state is a PRS and the contradiction assumption is satisfied. Hence $\A$'s forgery state will pass the test algorithm with high probability. On the other hand if $b=1$, the state has been picked from Haar measure and as a result of \thmref{th:qpuf-universal-unf}, the success probability of $\A$ winning the forgery game and producing a state to pass the test is negligible. Since guessing $b$ in Game 5 with probability better than random guess is equivalent to the difference between the success probability of $\A'$ in winning the game in the two different scenarios, we have:
\begin{equation}
\begin{split}
        & |\underset{k \leftarrow \K}{Pr}[\A'(\ket{\phi_k} \otimes (\tilde{U}\ket{\phi_k})^{\otimes m-1})=1] - \underset{\ket{\psi} \leftarrow \mu}{Pr}[\A'(\ket{\psi} \otimes (\tilde{U}\ket{\psi})^{\otimes m-1})=1]| \\
        & = |\underset{k \leftarrow \K}{Pr}[\A(\ket{\phi_k})=1] - \underset{\ket{\psi} \leftarrow \mu}{Pr}[\A(\ket{\psi})=1]| \\
        & = \nonnegl(\lambda) - \negl(\lambda) = \nonnegl(\lambda)
\end{split}
\end{equation}
Here, as a concrete example, we can consider the GSWAP to be the equality test and, we show how this check can efficiently be performed to show the gap and hence the implication of the two later games. Let us denote the adversary's purified forgery state as $\ket{\omega_b}$. According to \eqref{eq:gswap}, the probability of the GSWAP accepting this state given $m-1$ copies of reference state $\tilde{U}\ket{\phi^b}$, has the following relation with the fidelity of the forgery state:
\begin{equation}\label{eq:gswap-attack-test}
     \text{Pr}[\text{GSWAP accept}] = \frac{1}{m} + \frac{m-1}{m} F(\tilde{U}\ket{\phi^b}, \ket{\omega_b})^2
\end{equation}
Assuming $\A$ wins the unforgeability game for PRS state with non-negligible probability implies that this fidelity is a non-negligible value in the security parameter, hence $F(\tilde{U}\ket{\phi^0}, \ket{\omega_0}) = \delta = \nonnegl(\lambda)$. On the other hand, for Haar-random state this fidelity is always a negligible value and we have that $F(\tilde{U}\ket{\phi^1}, \ket{\omega_1}) = \negl(\lambda)$. As a result the difference between $\A$'s success probability in the two cases is as follows:
\begin{equation}
\begin{split}
        & |\underset{k \leftarrow \K}{Pr}[\A'(\ket{\phi_k} \otimes (\tilde{U}\ket{\phi_k})^{\otimes m-1})=1] - \underset{\ket{\psi} \leftarrow \mu}{Pr}[\A'(\ket{\psi} \otimes (\tilde{U}\ket{\psi})^{\otimes m-1})=1]| \\
        & = \frac{1}{m} + \frac{m-1}{m}F(\tilde{U}\ket{\phi^0}, \ket{\omega_0}) - \frac{1}{m} + \frac{m-1}{m}F(\tilde{U}\ket{\phi^1}, \ket{\omega_1}) \\
        & = \frac{m-1}{m}(\delta - \negl(\lambda)) \approx \frac{m-1}{m}\delta = \nonnegl(\lambda)
\end{split}
\end{equation}

As a result, we have shown that there exist a non-negligible gap and hence $\A'$ can also win the \texttt{Game 5}. In conclusion, we have shown the following relation:
\begin{equation}
    \texttt{Game 2} \equiv \texttt{Game 3} \equiv \texttt{Game 4} \Rightarrow \texttt{Game 5} \Rightarrow \texttt{Game 1}
\end{equation}
This means that an adversary winning the unforgeability game, with the challenge being picked from a PRS family, can also distinguish PRS states from Haar random states which is a contradiction and concludes the proof. 
\end{proof}

We have formally shown that PRS states are enough to achieve quantum universal unforgeability. For completeness let us briefly discuss the construction of these states with the existing proposals. Ji, Liu, and Song~\cite{shacham_pseudorandom_2018} propose several constructions for generating a PRS family using classical quantum-secure PRFs. Hence, they show that PRS can be constructed under the assumption that a quantum-secure one-way function exists. Another similar notion called Asymptotically Random State (ARS) has also been introduced in~\cite{hofheinz_pseudo_2019}. In both works, first, oracle access to a classical random function is given to efficiently construct a PRS, indistinguishable from Haar random states even for exponential adversaries. Then by relying on the existence of quantum-secure one-way functions, they replace the truly random function, with a post-quantum secure PRF to achieve security against polynomial adversaries. With this approach, one can construct computationally secure $n$-qubit PRS, which is also desired for the unforgeability security property. However, as discussed in~\cite{micciancio_scalable_2020}, these methods are not scalable and an n-qubit PRS generator cannot necessarily be employed to produce a random state for k-qubit where $k < n$. For these reasons, in~\cite{micciancio_scalable_2020} the authors introduce a scalable construction for PRSs which, unlike prior works, relies on randomising the amplitudes of the states instead of the phase. The authors use Gaussian sampling methods to efficiently achieve PRSs.

\section{From pseudorandom unitaries to UU and UqPUFs}\label{sec:connection-unf-pru-tdesign}
We prove that a family of unitaries satisfying the computational assumption of PRU is also a family of unknown unitary transformations. As a result of this implication, efficient constructions such as PRU or t-design can also satisfy the notion of universal unforgeability. Moreover, this result establishes for the first time, a link between a computational assumption of PRU with a hardware assumption such as unknownness.

\begin{thmbox}
\begin{theorem}\label{th:connection-pru-uu}
A family of PRUs, $\Uset = \{U_k\}_{k \in \K}$ is also a family of unknown unitary (UU).
\end{theorem}
\end{thmbox}
\begin{proof}
We prove this by contradiction. Let $\Uset$ be a family of PRUs but not a family of UU which means that there is a quantum polynomial-time (QPT) adversary $\A$ who can estimate the output of a randomly picked $U \leftarrow \Uset$ where $\Uset$ is a family of UU, on a state $\ket{\psi}$, non-negligible better than the output of a $U \leftarrow \mu$ picked from a Haar-random unitary $\mu$ over a $d$-dimensional Hilbert space. Thus for $\A$ the following holds: 
\begin{equation}
\begin{split}
        & |\underset{U \leftarrow \Uset}{Pr}[F(\A(\ket{\psi}),U\ket{\psi}) \geq \nonnegl(\lambda)] - \underset{U_{\mu} \leftarrow \mu}{Pr}[F(\A(\ket{\psi}),U_{\mu}\ket{\psi}) \geq \nonnegl(\lambda)]| \\
        & = \nonnegl(\lambda).
\end{split}
\end{equation}
Let $\A$' be a QPT adversary who aims to break the pseudorandomness property of $\Uset$ using $\A$, and works as follows:\\
\textit{$\A$' picks $\ket\psi$ as one of her chosen inputs in the learning phase of the pseudorandomness game. Then $\A$' also runs $\A$ internally on $\ket{\psi}$.}\\
From the previous equation, we know that $\A$ can estimate the output of $U \ket{\psi}$ better than $\U_{\mu}\ket{\psi}$ where $\U_{\mu}$ is a Haar random unitary, by a non-negligible value. Also by definition, we know that the probability that any QPT algorithm estimates the output of any Haar randomly given unitary, is negligible, as the response maps to any random state in the Hilbert space $\Hild$ with exponential distribution~\cite{dankert_exact_2009,nielsen_quantum_2010}. Thus the equation implies that:
\begin{equation}
        |\underset{U \leftarrow \Uset}{Pr}[F(\A(\ket{\psi}),U\ket{\psi}) \geq \nonnegl(\lambda)]| = \nonnegl(\lambda).
\end{equation}
Meaning that $\A$ can estimate the output with non-negligible fidelity if $U$ had been picked from the family. Now $\A$' runs a quantum equality test on $U \ket{\psi}$ obtained in the learning phase and $\A(\ket{\psi})$. In the case where $U$ is picked from the PRU family, the estimated output and the real output have non-negligible fidelity, and the test returns equality with a non-negligible probability. Otherwise, the test shows they are not equal, and $\A$' can conclude that the unitary has been picked from Haar unitaries. Thus for $\A$', we have:
\begin{equation}
    \underset{U \leftarrow \Uset}{Pr}[\A'^{U}(1^{\lambda})=1] - \underset{U_{\mu} \leftarrow \mu}{Pr}[\A'^{U_{\mu}}(1^{\lambda})=1]=\nonnegl(\lambda)
\end{equation}
Therefore we conclude the contradiction.
\end{proof}

We have shown that PRUs imply unknown unitaries, and combined with the results from the previous chapter, we conclude that PRUs make a set of universally unforgeable unitaries. Now we show that PRU can also be considered a qPUF family. To do this, we need to show that the PUF requirements given in \defref{def:qPUF} are satisfied. Since the $\delta_r$-Robustness and $\delta_c$-Collision Resistance are trivially satisfied by the unitarity, we only need to argue about the $\delta_u$-Uniqueness requirement.

\begin{thmbox}
\begin{theorem}\label{th:connection-pru-unique}
Let $\Uset = \{U_k\}_{k \in \K}$ be a family of PRUs, where each $U_i$ is a unitary matrix over a d-dimensional Hilbert space and is universally-unforgeable. Then there exist a $\delta_u = \nonnegl(\lambda) = \nonnegl(polylog(d))$ such that $\Uset$ satisfies $\delta_u$-uniqueness.
\end{theorem}
\end{thmbox}
\begin{proof}
We prove by contraposition and we assume that there exists no non-negligible $\delta_u$ to satisfy $\delta_u$-uniqueness. This means that for any two unitary $U_i$ and $U_j$ picked uniformly at random from $\Uset$, the two unitary are $\zeta$-close in the diamond norm with a high probability. Otherwise if there exist a minimum $\zeta_{min} = \nonnegl(\lambda)$ distance in diamond norm between any two unitaries we have already shown the $\delta_u$ exists. Hence we assume that we have the following condition:
\begin{equation}\label{eq:proof-contra-diamond}
    Pr[\parallel (U_i - U_j)_{i\neq j}\parallel_\diamond \leq \zeta] \geq 1 - \epsilon(\lambda)
\end{equation}
where both $\zeta$ and $\epsilon(\lambda)$ are negligible functions in the security parameter. Now we assume an adversary $\A$ wants to distinguish between $\Uset$ and the set of Haar-random unitaries. By assumption, we have that all the unitaries in $\Uset$ are universally unforgeable. So now we let $\A$ play the PRU game\footnote{Which is the indistinguishability game version of the pseudorandomness property, similar to \gameref{game:prs}} while running the universal unforgeability game as a distinguishing subroutine. Let $\C$ be the honest party picking at random a bit $b \in \{0,1\}$ where if $b=0$, a unitary $U$ is picked at random from $\Uset$ and we are in the PRU world and otherwise $U$ is picked from $\mu$ that denotes the set of Haar-random unitary matrices. Then $\A$ gets polynomial oracle access to the $U$ and after the interaction, needs to guess $b$. Now, since there exists an efficient public generation algorithm $Q$ for the PRU set, we let the adversary sample another unitary $U'$ from $Q$ locally and uniformly at random. According to the contraposition assumption give in \eqref{eq:proof-contra-diamond}, if $b=0$, with high probability these two unitaries are $\zeta$-close in the diamond norm, \emph{i.e.} $\parallel (U - U')\parallel_\diamond \leq \zeta$. Given this promise, the adversary performs the following strategy: $\A$ locally plays the universal unforgeability game on $U$, by picking a state $\ket{\psi}$ uniformly at random from Haar measure and querying it to $\C$ as part of the polynomial oracle interaction with $U$. $\A$ will receive $U\ket{\psi}$ and can ask for multiple copies of it so long as the total number of queries to the oracle remains polynomial. Now we also rely on the fact that since PRU has the efficient computation property, meaning that $\A$ can locally compute $U'\ket{\psi}$ to get multiple copies. Now $\A$'s strategy to win the unforgeability game is to output $U'\ket{\psi}$ as the forgery for $\ket{\psi}$.

Again in the case of $b=0$, since the two unitaries are negligibly close in the diamond norm with a high probability we have the following:
\begin{equation}
    Pr[\parallel (U - U')\parallel_\diamond \leq \zeta] \geq 1 - \epsilon \Rightarrow Pr[F(U\ket{\psi}, U'\ket{\psi}) \geq 1 - \zeta] \geq 1 - \epsilon
\end{equation}
This holds since the diamond norm is defined as a maximum over all density matrices. Therefore, if the two unitaries are very close in the diamond norm, their output over a random state is also very close on average. Thus, the adversary can run a local efficient verification test (for instance, a GSWAP test) between $U'\ket{\psi}$ and $U\ket{\psi}$ and use the output of the test as a distinguisher between pseudorandom and Haar-random world. If $b=0$, we have:
\begin{equation}
    Pr[F(U\ket{\psi}, U'\ket{\psi}) \geq 1 - \zeta] \geq 1 - \epsilon \Rightarrow Pr[1\leftarrow \Gnn{U'}{\qUni}(\lambda, \A)] = \nonnegl(\lambda)
\end{equation}
Hence $\A$ will win the game with a high probability. However, in the case of $b=1$ where $U$ is a Haar-random unitary, we can use a lemma in~\cite{kretschmer_quantum_2021}, that states for a fixed state $\ket{\phi} \in \Hild$ and a Haar-random state $\ket{\psi} \leftarrow \mu$, and any $\epsilon > 0$ we have:
\begin{equation}
    \underset{\ket{\psi} \leftarrow \mu}{Pr}[|\mbraket{\phi}{\psi}|^2 \geq \epsilon] \leq e^{-\epsilon d}
\end{equation}
This implies we can take $U'\ket{\psi} = \ket{\phi}$ to be the fixed state. Since $U$ is a Haar-random unitary then $U\ket{\psi}$ is also a Haar-random state and hence the probability that the fidelity $F(U\ket{\psi}, U'\ket{\psi})$ is a non-negligible value (with respect to $polylog(d)$) such as $ 1 - \zeta$, is exponentially low. Hence in case $b=1$, the probability that the adversary's state passes the verification is exponentially low. Hence using this strategy, there will always be a distinguisher that can distinguish between $\Uset$ and Haar-random unitaries \emph{i.e.}:
\begin{equation}
    \underset{U \leftarrow \Uset}{Pr}[\A'^{U}(1^{\lambda})=1] - \underset{U_{\mu} \leftarrow \mu}{Pr}[\A'^{U_{\mu}}(1^{\lambda})=1]=\nonnegl(\lambda)
\end{equation}
But this is in contrast with the assumption that $\Uset$ is a PRU. Hence we have reached a contradiction, and the proof is complete.
\end{proof}

\section{Pseudorandom unitaries and states from hardware assumptions}\label{sec:connection-pru-from-uu-unique}
As discussed earlier pseudorandom quantum states can be constructed under the assumption of qPRF or quantum one-way functions. Given the relationship that we have explored in the previous section between the unforgeability of qPUF and quantum pseudorandomness, here we ask whether it is possible to construct pseudorandom quantum states under a different set of assumptions? In this section, we discuss how one can achieve PRUs and PRSs under hardware assumptions on a family of unitary transformations. These hardware assumptions are generally discussed in the context of quantum PUFs, nevertheless, our results can be in general applied to any sets of unitaries with the given properties.

Let $\Uset = \{U_i\}^{\K}_{i=1}$ be a family of unitaries, where each $U_i$ is a unitary matrix over a d-dimensional Hilbert space. Let us bring a specific assumption offered by the physical nature of such unitaries. We want to use the above family as a PRU family or generators for PRS. As shown in~\cite{shacham_pseudorandom_2018}, if $\Uset$ is a PRU then it is also a generators for PRS states \emph{i.e.} $G(k) = U_k\ket{0} = \ket{\phi_k}$. To this end, we investigate the properties of a qPUF family that can be used to achieve pseudorandomness. In the last section, we have shown that PRU implies the notion of unknown unitary assumption, or in other words, single-shot unknownness. Now we explore the relation of PRUs and another notion of unknownness called \emph{practical unknownness} by Kumar et al.~\cite{kumar_efficient_2021}. This definition is better suited for t-design unitary sets constructions and is defined as follows:

\begin{defbox}
\begin{definition}[$\epsilon,t,d-$ Practical unkownnness~\cite{kumar_efficient_2021}]\label{def:pu} We say a unitary transformations $U$, from a set $\Uset \subseteq U(d)$\footnote{where $U(d)$ denotes the set of all unitary matrices over $d$-dimensional Hilbert space} is $(\epsilon,t,d)$- practically unknown if provided a bounded number $t \leq poly(\log_2 d)$ of queries $U\rho U^{\dagger}$, for any $\rho \in \Hild$, the probability that any $poly(\log_2 d)$-time adversary can perfectly distinguish $U$ from a Haar distributed unitary is upper bounded by $1/2(1 + 0.5 \epsilon)$. Here $0 < \epsilon < 1$, t are functions of $log_2 d$, and $\lim_{\log_2(d) \rightarrow \infty}\epsilon = 0$.
\end{definition}
\end{defbox}

For the sake of our proof, we need a variation of this definition which is for any polynomial number of queries in the security parameter. Hence, we define the following:

\begin{defbox}
\begin{definition}[$\epsilon,d-$ Practical unkownnness]\label{def:pu-poly} We say a unitary transformations $U$, from a set $\Uset \subseteq U(d)$ is $(\epsilon,d)$- practically unknown if it is $(\epsilon,t,d)$- practically unknown for any $t=poly(\lambda) = poly(\log d)$.
\end{definition}
\end{defbox}

Now we first show that the assumption of $\epsilon,d-$ Practical unkownness implies PRU.

\begin{thmbox}
\begin{theorem}\label{th:connection-practicaluu-pru}
A family of $(\epsilon,d)$- practically unknown unitaries where $\epsilon = \negl(\lambda)$ is a PRU family.
\end{theorem}
\end{thmbox}

\begin{proof}
We prove this by contraposition. Let $\Uset = \{U_k\}^{\K}_{i=1} \subseteq U(d)$ be a $(\epsilon,d)$- practically unknown family, that is not a PRU. This means that there exists a QPT adversary $\A$ for which we have the following after some $q=poly(\lambda) = poly(\log(d))$ queries to the unitary oracle:
\begin{equation}
    |\underset{k \leftarrow \K}{Pr}[\A^{U_k}(1^{\lambda})=1] - \underset{U \leftarrow \mu}{Pr}[\A^U(1^{\lambda})=1]| = \delta = \nonnegl(\lambda).
\end{equation}
Equivalently, we can say that if a unitary is randomly picked from either of the set $\Uset$ or a set of Haar-random distributed unitaries with a random bit $b$, the advantage of the adversary in guessing bit $b$ is a non-negligible function $\delta$ greater than $\frac{1}{2}$. If such an adversary exists, there also exists an adversary $\A'$ that querying the same $q$ states, can distinguish the $U_k \in \Uset$ from a Haar-random unitary with the following probability:
\begin{equation}
    Pr[\text{distinguish } U_k] \geq \frac{1}{2} + \delta
\end{equation}
On the other hand, if $\Uset$ is $(\epsilon,d)$-practically unknown this probability is equal to $\frac{1}{2}(1 + 0.5 \epsilon)$ where $\frac{\epsilon}{4}$ is a negligible function while as $\delta$ is non-negligible. Hence we reach a contradiction and the proof is complete.  
\end{proof}

We have shown that given the hardware assumption of practical unknownness, over a set of unitary transformations such as unitary qPUFs, one can get PRU and as a result generate PRSs by applying random elements of the set on the computational basis state. However, practical unknownnes is a stronger assumption than UU, and it is not surprising that it will lead to PRU. Now, we look at other properties of a qPUF family and see whether there exists a more interesting assumption under which pseudorandomness can be achieved.

One of the main requirements on a qPUF family is the \emph{uniqueness} property (Requirement~\ref{def:uniq}, \defref{def:qPUF}) that ensures any two qPUFs in the family are sufficiently distinguishable in the diamond norm. In what follows we show a family of unknown and (almost) maximally distinguishable unitary matrices, such as unitary qPUFs, also form a family of PRUs and are a generator for PRSs. 

\begin{thmbox}
\begin{theorem}\label{th:connection-max-unique-pru}
Let $\Uset_{\K} = \{U_k\}^{\K}_{k=1} \subseteq U(d)$ be a family of unitary transformation selected at random from a distribution $\chi_{\Uset}$ such that they satisfy almost maximal uniqueness \emph{i.e.} for any randomly picked pairs of unitary matrices from $\Uset_{\K}$, we have $\parallel (U_i - U_j)_{i\neq j}\parallel_\diamond = 2 - \epsilon$ where $\epsilon = \negl(\lambda)$, then for a sufficiently large $\K$ and $d$, the $\Uset_{\K}$ is also a PRU.
\end{theorem}
\end{thmbox}

\begin{proof}
We first show that if the maximum uniqueness is on average satisfied for any pairs of unitary matrices of $\Uset_{\K}$, then the distribution $\chi_{\Uset}$ converges to Haar measure in the limits of large $d$. We attempt to prove this convergence for a specific degree of uniqueness which is $2 - \epsilon$ where the maximum of the diamond norm is 2. The general proof idea is to show that the distribution of the eigenvalues of $2 - \epsilon$-distinguishable unitary matrices looks like the eigenvalue distribution of a Haar-random matrix. We use the toolkit from the random matrix theory introduced in \chapref{chap:prelim} (Section~\ref{sec:prelim-haar-random}) to show this statement. First, note that we have,
\begin{equation}
    \parallel (U_i - U_j)_{i\neq j}\parallel_\diamond = 2 - \epsilon = 2\sqrt{1 - \delta(U_i^{\dagger}U_j)^2}
\end{equation}
Where the $\delta(M) = \underset{\ket{\phi}}{min}|\bra{\phi}M\ket{\phi}|$ is the minimum of absolute value over the numerical range of the operator $M$. From the above equation we have:
\begin{equation}
     \delta(U_i^{\dagger}U_j)^2 = \epsilon - \frac{\epsilon^2}{4} \approx 0
\end{equation}
Since the diamond norm is unitary invariant, we can multiply all the unitaries of the family by a fixed unitary matrix which results in the set including the identity matrix $\mathcal{I}$, hence the above equation can be rewritten as:
\begin{equation}
     \delta(U'_k)^2 = \epsilon - \frac{\epsilon^2}{4}
\end{equation}
where the set of unitary matrices $U'$ is equivalent to the initial set up to a unitary transformation. Now let $\{e^{i\theta_1},\dots,e^{i\theta_d}\}$ be the eigenvalues of $U'_k$. The eigenvalues of a unitary matrix lie on a unit circle $\mathbb{S}^1 \subset \mathcal{C}$. As shown in~\cite{kumar_efficient_2021}, the following relation exists between the distribution of the eigenvalues of a general unitary matrix in an arc of size $\theta$, and the function $\delta(U)$:
\begin{equation}
     \delta(U'_k)^2 = \frac{1}{2} + \frac{1}{2} \cos{\theta}
\end{equation}
Where $\theta = \theta_j - \theta_k$ for pairs of eigenvalues $\{e^{i\theta_j},e^{i\theta_k}\}$. From the above equation we have:
\begin{equation}
     \theta = \theta_j - \theta_k = \arccos{(-1 + 2\epsilon - \frac{\epsilon^2}{2})} \approx \pi - \sqrt{\epsilon} + \dots
\end{equation}
Now we can use \thmref{th:prelim-wieand}. Let $N_{\theta}$ be a random variable that represents the number of eigenvalues in an arc of size $\theta$. Then we have the expectation value of this random variable for the given distribution where the $\theta = \pi - \epsilon'$, and $\epsilon' = \negl(\lambda)$, to be
\begin{equation}
     \mathbb{E}_d[N_{\theta}] = \frac{d\times\theta}{2\pi} = \frac{d}{2} - \frac{\epsilon'd}{2\pi}
\end{equation}
which is close to half of the total number of eigenvalues since the second term is always smaller than 1. This means that in the limit of large $d$, every diameter of the unit circle divide the circle into two areas that each on average includes half of the eigenvalues. Also the variance of the random variable $N_{\theta}$ will be: 
\begin{equation}
     Var(N_{\theta}) = \frac{1}{\pi^2}(\log(d) + 1 + \gamma + \log|2\sin(\frac{\pi - \epsilon'}{2})|) + o(1) \approx \frac{\log(d)}{\pi^2} + c' + o(1)
\end{equation}
where $\gamma \approx 0.577$ and $c' < 1$. Next, we calculate the probability that for our given distribution, there are more than half of the eigenvalues in each half of the circle denoted by an arc or size $\pi - \epsilon'$. Using the Chernoff bound we have:
\begin{equation}
     Pr[N_{\pi - \epsilon'} - \mathbb{E}_d[N_{\pi - \epsilon'}]| > x\mathbb{E}_d[N_{\pi - \epsilon'}]] \leq e^{-\frac{x^2}{2+x}\mathbb{E}_d[N_{\pi - \epsilon'}]}
\end{equation}
Here we want the $x\mathbb{E}_d[N_{\pi - \epsilon'}]$ to be equal to $\frac{d}{2}$, so we have $x = \frac{d/2}{d/2 - d\epsilon'/2\pi} = \frac{1}{1 - \epsilon'/\pi}$ and since the $x$ is a small value the above inequality can be used. Substituting this into the above equation we will have:
\begin{equation}
     Pr[N_{\pi - \epsilon'} - \mathbb{E}_d[N_{\pi - \epsilon'}]| > \frac{d}{2}] \leq e^{-\frac{(\frac{1}{1 - \epsilon'/\pi})^2}{2+\frac{1}{1 - \epsilon'/\pi}}\times(d/2 - \epsilon'd/2\pi)} \approx e^{-d/6}
\end{equation}
since $\epsilon'$ is negligible. This shows that with a very high probability, on every half of the unit circle, there exist half of the eigenvalues of the random matrix from our specified distribution. We conclude eigenvalues of a random unitary from the distribution $\chi_{\Uset}$ are uniformly distributed on the unit circle. Let us denote this uniform distribution on $\mathbb{S}^1$ by $\nu$. In order to compare the distribution of $\chi_{\Uset}$ with the Haar measure, we use the empirical spectral measure introduces in Section~\ref{sec:prelim-haar-random}.
We denote the empirical spectral distance of $\chi_{\Uset}$ as $\tilde{\mu}_{\chi}$ and for Haar measure we denote it as $\tilde{\mu}_{H}$. Since we have shown that the eigenvalues of matrices from $\chi_{\Uset}$ are distributed uniformly on $\mathbb{S}^1$, it is easy to see that $\mathbb{E}(\tilde{\mu}_{\chi}) = \nu$ and in the limit of large $d$ we have the convergence in probability $\tilde{\mu} \overset{d\rightarrow \infty}{\longrightarrow} \nu$. Now we use the \thmref{th:prelim-diaconis-shah} that implies the convergence of the empirical spectral measure of the set of unitaries picked from Haar measure to $\nu$, in the limit of large $d$. Having the these two convergence and the properties of the limit we can conclude that the empirical spectral measure for $\chi_{\Uset}$ converges to the one for Haar measure. Then we look at Kolmogorov distance of the eigenvalues of these two distributions. We rely on the result given in~\cite{meckes_random_2019} that shows the Kolmogorov distance between the distributions of eigenvalues of random unitary matrices is given by $d_K(\mu, \nu) = \underset{0\leq \theta < 2\pi}{sup} |\frac{N_{\theta}}{d} - \frac{\theta}{2\pi}|$ and specifically for Haar measure it is bounded by 
\begin{equation}
     d_K(\mu_{H}, \nu) \leq c \frac{\log(d)}{d}
\end{equation}
Where $c > 0$ is a universal constant. Given the fact that for the specific value of $\theta$ for the distribution of $\chi_{\Uset}$ the Kolmogorov distance $d_K(\mu_{\chi}, \nu)$ is of the order $\frac{1}{d}$ which is negligible and using the triangle inequality for the Kolmogorov distance we have
\begin{equation}
\begin{split}
         d_K(\mu_{H}, \mu_{\chi}) & \leq d_K(\mu_{H}, \nu) + d_K(\nu, \mu_{\chi})\\
         & \leq c \frac{\log(d)}{d} + \negl(\lambda)\\
         & \leq \negl(\lambda)
\end{split}
\end{equation}
Thus the distribution of the eigenvalues of the random matrices of $\chi_{\Uset}$ is negligibly close to the Haar measure. Also for any randomly picked matrix from each of these distributions, the eigenvalues are fixed. As a result, the convergence between the distribution of the eigenvalues of matrices leads to the fact that in the limit of large $d$, $\chi_{\Uset}$ converges to the Haar measure on the unitary set. 

Finally, we show that a polynomial time quantum adversary given a polynomial query to each unknown unitary $U_k$ cannot distinguish any member of this family from Haar measure. This is straightforward since the two distributions are asymptotically  close. Thus we have:
\begin{equation}
    |\underset{k \leftarrow \K}{Pr}[\A^{U_k}(1^{\lambda})=1] - \underset{U \leftarrow \mu}{Pr}[\A^U(1^{\lambda})=1]| = \negl(\lambda).
\end{equation}
And we have shown that the set $\Uset_{\K}$ is a PRU.
\end{proof}

\section{Discussion and conclusions}\label{sec:connection-conclusion}
We have explored in this chapter, the connection between quantum pseudorandomness and quantum hardware assumptions such as quantum physical unclonability. As one of the main cryptographic properties of quantum physical unclonable functions is the notion of universal unforgeability, we have inspected whether quantum pseudorandomness would be enough as a challenge sampling requirement, to achieve this level of unforgeability. We have formally proved that the answer to this question is positive. This result improves the practicality of qPUF-based constructions and protocols since it replaces the requirement of Haar-randomness on the challenge states, which is resourceful and experimentally challenging. We will articulate this improvement in the next chapter.

We have also established the link between the notions of \emph{unknownness} of unitary families and PRUs. We proved that any family of PRUs is also a family of unknown unitaries and, hence they could be a potential candidate for the construction of qPUF devices. This result complements the result of~\cite{kumar_efficient_2021} where they show t-designs can also satisfy a similar notion, namely practical unknownness, which leads to an efficient proposal for constructing quantum PUFs. 

Then we also looked at the problem of generating pseudorandom quantum states from hardware assumptions. Our results show that different physical assumptions proposed in the context of PUFs, such as uniqueness or practical unknownnes, can also imply quantum pseudorandomness. This result is of theoretical interest as it shows an alternative way of achieving quantum pseudorandomness which is different from current approaches based on post-quantum and computational assumptions. Apart from the cryptography perspective, having a different set of assumptions for PRSs and PRUs can find potential applications in physics where PRSs have been shown recently to be relevant in the AdS/CFT correspondence for the study of quantum gravity~\cite{bouland_computational_2019}. Nonetheless, due to arguments given in~\cite{bouland_computational_2019}, the proposed PRS constructions in~\cite{shacham_pseudorandom_2018}, are not directly applicable within this framework. Given our results in this chapter, a potential follow-up question would be whether a PRS state derived from physical unclonability assumptions can be used as an alternative solution. Another interesting future direction would be to further explore the relationship between unclonability and quantum pseudorandomness that has initially been discussed in~\cite{shacham_pseudorandom_2018}, relying upon our new results.

The final open problem that we would like to bring forward to conclude the chapter, is that of establishing concrete bounds on the randomness and pseudorandomness of unitary families, given different degrees of uniqueness or distinguishability (not negligibly close to perfect distinguishability). We believe this question has an interesting and non-trivial relationship to the study of t-design unitaries. A curious inquiry is whether the random matrix theory toolkit and the potential extension of our last result in the current chapter, regarding the relationship between distinguishability and pseudorandomness can also lead to novel constructions for t-designs.

\chapter{Applications of Quantum Physical Unclonable Functions} \label{chap:application}
\begin{chapquote}{Carl Sagan}
``If you wish to make an apple pie from scratch, you must first invent the universe.''
\end{chapquote}

\section{Introduction}\label{sec:application-intro}
In the last two chapters, we have studied quantum physical unclonable functions, both as theoretical objects and provably unforgeable hardware tokens, while also exploring the connection between physical unclonability and quantum pseudorandomness. Moving from foundations to applications, it is now time to introduce applications of qPUFs in quantum communication and quantum cryptography. This chapter is dedicated to the design and security analysis of protocols based on quantum PUFs. In the course of the chapter, we also attempt to move towards more efficient variants of the proposed protocols and make them more accessible for implementation. 

The recent advances in developing the quantum internet have enabled a broad range of applications from simple secure communications all the way to delegated quantum computation, with often no counterparts in classical networks  \cite{broadbent_quantum_2016,wehner_quantum_2018,fitzsimons_private_2017,veriqloud_quantum_2019, pirandola_advances_2020,diamanti_demonstrating_2019,dynes_cambridge_2019,kozlowski_designing_2020,caleffi_quantum_2018,cacciapuoti_quantum_2020,unruh_everlasting_2013}. For most of such applications, a key security feature is the ability of secure authentication which plays a central role in performing secure communications over untrusted channels \cite{alagic_quantum_2017, dulek_secure_2020, boneh_quantum-secure_2013}. The general term of \emph{authentication} encloses different definitions and levels depending on the strength of the security requirement and the nature of the subject of authentication, for instance, whether it is a message or an entity. Amongst different types of required security features, including confidentiality and authentication of data, mutual entity authentication is a crucial, yet sometimes neglected, aspect \cite{kang_controlled_2018,gollmann_what_1996}. Entity authentication also referred to as \emph{Identification}, is a method to prove the identity of one party called \emph{prover} to another party called \emph{verifier}.

The focus of this chapter is on secure identification as it is a central application of quantum communication, as well as a building block for many other applications of quantum networks. We aim to propose resource-efficient solutions for mutual entity authentication between two parties who can also be two nodes of a quantum network. We explore the advantages of quantum communication in achieving protocols with fewer assumptions or stronger security guarantees compared to their classical counterparts or existing solutions.

We consider both complementary scenarios where either the trusted verifier or a potentially malicious prover has limited resources. 
To better motivate the two scenarios, consider the quantum cloud service platforms that are commercially available today \cite{arute_quantum_2019,cross_ibm_2018,rigetti_welcome_nodate, bergholm_pennylane_2020, blinov_comparison_2021}. In the first setting, a client with a low quantum resource (such as the one defined in \cite{broadbent_universal_2009}) wishes to identify a high-resource quantum centre that they perhaps have had a previous contract with, before proceeding to access their platform and load its sensitive data. In the complimentary setting, the quantum cloud provider wishes to verify the identity of its customer possessing low quantum resources before providing them with access. This asymmetry between the verifier and the prover calls for `party resource-specific' identification protocols which exploit this asymmetry to enhance the efficiency. Another potential approach is include the mutual identification within one protocol which requires symmetrizing the parties as much as possible. We will explore both of these approaches via our proposals in this chapter.

Most of the typical classical solutions for authentication and identification rely on computational assumptions or a perfectly random key being securely shared between the two parties, or in some cases, both. Throughout this chapter, we replace these computational assumptions, or secure classical key sharing, with the hardware assumption of physical unclonability. The protocols have the structure of a symmetric-key protocol, although the \emph{key} here is some unclonable hardware. Another prominent aspect of our proposals is the employment of quantum communication. Similar to most functionalities and protocols, if one wishes for the security in the quantum era, there are usually two options available: either to go for the post-quantum alternatives and use assumptions that are believed to be hard for quantum computers, or to take advantage of the power of quantum information and quantum communication to attain quantum security. We focus on the second option here, for achieving provable security against quantum adversaries under minimal assumptions. We note that each of these approaches has its pros and cons, and the comparison between them is not the purpose of this chapter. We remark that the spirit of the works presented in this thesis is closer to exploiting the physical properties and fundamental limitations of quantum mechanics in the design of protocols.

First, we propose two entity authentication protocols based on the quantum PUFs that we have defined in \chapref{chap:qpuf}. Our first proposal is a secure qPUF-based device identification protocol which requires the prover to only have access to the valid qPUF device without requiring any quantum memory or quantum computational resource, while the verifier is required to possess a local quantum database and the ability to perform quantum operations. This covers the scenario presented before where a quantum cloud provider wants to identify its customer.

Our second proposal is a qPUF based protocol where the prover has a high computational resource, while, the verifier runs a purely classical algorithm, hence does not require performing quantum operations. This protocol can enable an \emph{almost classical} client, to identify a quantum server in a quantum network. Construction of this protocol has taken inspiration from the ideas of blind quantum computing \cite{broadbent_universal_2009} to introduce the idea of randomly placing trap quantum states in-between the valid states. This, coupled with the unknownness property of the qPUF device provides provable security against any QPT adversary. 
We also provide a comparison between the two protocols on different aspects and resources to get a better picture of their use-case in different scenarios.

Next, we exploit the result we have established in \chapref{chap:pr-connection}, to improve the efficiency of our protocols and make them more amenable to implementation, while formally proving that this step, on the way to practicality can be made with no compromise in the security guarantee.

Finally, in attempting to propose a yet more practical solution, we explore a different construction for PUFs, which although weaker than full qPUFs, can still enable secure quantum entity authentication while also being implementable with the technology and infrastructures that are available today. This new construction, called \emph{Hybrid PUF}, combines classical PUFs with quantum encoding and using some additional techniques from the world of classical hardware security, can lead to quantum-secure mutual identification that does not require quantum database or preparation of resourcefully complicated quantum states. The latest protocol we present has some further properties, such as the re-usability of challenge states during the protocol, which we will investigate in detail.

We believe that all these proposed protocols are just the start of the road for applications that utilize physical unclonability and quantum information since identification is essential yet quite a simple functionality. Our studies presented in this chapter show that there are still many applications to come using this newly introduced assumption.  

\subsection{Structure of the chapter}
In Section~\ref{sec:application-qpuf-id-protocols} we present our client-server qPUF-based identification protocols. The protocol with high-resource verifier has been introduced in Subsection~\ref{sec:application-hrv-protocol}, and the low-resource verifier protocol in Subsection~\ref{sec:application-lrv-protocol}. A generalisation of the second protocol is also discussed in Subsection~\ref{sec:application-lrv-general} and the comprehensive comparison between them is given in Subsection~\ref{sec:application-resource-comparison}.

In Section~\ref{sec:application-efficent-qpufid-prs} we use pseudorandom quantum states to reduce the assumptions and requirements for our proposed protocol and introduce a more efficient version.

Finally, Section~\ref{sec:application-hpuf-practical} focuses on presenting the Hybrid PUF construction as well as the identification protocol that is based on it and its security analysis. In Subsection~\ref{sec:application-hlpuf} the construction is given. In Subsection~\ref{sec:application-hlpuf} an enhanced version this construction called \emph{Hybrid Locked PUF} has been introduced which later is used within the identification protocol presented in Subsection~\ref{sec:application-hlpuf-protocol}. Subsection~\ref{sec:application-hlpuf-security} discussed the security analysis of HPUF, HLPUF and related protocol and Subsection~\ref{sec:application-hlpuf-reusability} investigates the challenge re-usability property.

\subsection{Related works}
The idea of taking advantage of quantum communication between the verifier and the prover in PUF-based identification protocols was first introduced by Skoric in \cite{skoric_quantum_2010} with the concept of \emph{quantum read-out of PUF (QR-PUF)}. The identification protocols based on this construction have been proposed in \cite{skoric_quantum_2010,skoric_quantum_2012,gianfelici_theoretical_2020,nikolopoulos_remote_2021}. The security of the majority of these protocols has been proved against limited types of attacks including intercept-resend \cite{skoric_quantum_2010,skoric_quantum_2012}, and Quantum Cloning \cite{yao_quantum_2016} attacks. The practical realization of this protocol was shown by Goorden et al. \cite{goorden_quantum-secure_2014,nikolopoulos_remote_2021}. 
In another work (also mentioned in \chapref{chap:qpuf}), Nikolopoulos and Diamanti introduce a different setup for QR-PUF-based identification protocols in which classical data is encoded to the continuous quadrature components of the quantized electromagnetic field of the probe \cite{nikolopoulos_continuous-variable_2017}. The security of this scheme has also been proved in \cite{nikolopoulos_continuous-variable_2018,fladung_intercept-resend_2019} against a bounded adversary who can only prepare and measure the quantum states.
The common feature of the mentioned protocols is full or partial knowledge of the unitary modelling of the QR-PUF. However, as thoroughly discussed in \ref{sec:qpuf-publicdb-pufs}, this extra information usually compromises the security and as a result, such protocols can only be proven secure against specific types of adversarial attacks. The main advantage of our qPUF-based proposals over the previous ones is their provable security against the most general form of attacks considering a QPT adversary. 

Related to our Hybrid construction, first we mention some classical constructions for classical PUFs such as \cite{gassend_silicon_2002,guajardo_fpga_2007,kim_cspan_2018}. The literature of classical PUFs, specifically regarding the implementation, is very vast and covering the full references to them is outside the scope of this chapter, but we refer the reader to \cite{maes_physically_2013} for a detailed review of the constructions of classical PUFs. We have also mentioned that most such classical PUFs are vulnerable to machine learning attacks. Some of the attacks on classical PUFs have been performed and studied in \cite{guneysu_gap_2015,becker_pitfalls_2015,delvaux_machine-learning_2019,ruhrmair_modeling_2010}. Furthermore, we also borrow an idea from classical hardware security literature, known as the lockdown technique (or as we call them, locking mechanism), that has been introduced by Yu et al.\cite{yu_lockdown_2016} as a proposal to prevent such machine learning attacks on classical PUFs.


\section{Quantum-secure identification protocols using quantum PUF}\label{sec:application-qpuf-id-protocols}
In this section, we aim to provide protocols for the task of identification, using quantum PUF and quantum communication as our main ingredients.
Intending to perform low-cost secure identification of the prover by the verifier using qPUF, we categorise the resources into three major segments. First is the `memory resource' which quantifies the type and amount of storage resources that a party requires. It can either be a classical memory that we label as low cost, or a quantum memory which is high cost since such a memory tends to be highly fragile and dissipative to the environment \cite{lvovsky_optical_2009}. Second is the `computing ability' resource which indicates the kind of operations a given party can perform. We denote a party with high computing ability as the one that can perform any bounded polynomial quantum circuit operations \cite{watrous_complexity_2003}, and a low ability party as the one that is restricted to generation and measurement of quantum states on a certain basis. And the third resource is the type and number of `communication rounds' required between the parties to establish identification. Often it is not possible to devise an identification scheme that minimises all the three types of resources for both the involved parties without compromising the underlying security. Hence, in this work, we propose two qPUF based identification schemes that achieve similar security guarantees but are vastly different in terms of the resource requirement for the involved parties. This allows the flexibility to deploy either of these schemes specific to each application.

The first protocol allows a low-resource party who has only access to the qPUF, to prove its identity to a high resource party with more quantum capabilities such as quantum computing capability and quantum memory. In the second protocol, we explore the other direction and try to minimize the resources on the verifier's side as much as possible. This leads to a novel qPUF-based protocol, which is different from the usual PUF-based protocols known in the literature. We give complete formal security proofs for each of the protocols. Then, we also provide a comprehensive comparison between the two proposed protocols in terms of our categorised resources.

But before introducing the protocols, let us give a general description of an identification protocol to provide a better intuition of the functionality we are trying to achieve.

\subsection{General description of device-based identification protocol}\label{sec:protocol}

An identification protocol, also called a device-authentication protocol, is run between a verifier and a prover. A verifier's task is to check the identity of the prover by identifying whether the prover is the correct owner of a valid device. Our setting assumes that the verifier and the prover having a valid device behave honestly. The security is provided against an adversary with limited access to the valid device\footnote{This the same QPT adversarial model we have described in the universal unforgeability game in \chapref{chap:qpuf}. This is a primitive-level access, while as in the level of our protocols we (usually) do not assume any bound on the adversary.}. The objective of the adversary is to successfully impersonate themselves as the valid owner of the device. Prior to providing the details of the construction of device identification protocols using qPUF, we describe a common structure in these protocols. Any such protocol consists of three sequential phases: \emph{setup phase} (or enrollment phase), \emph{identification phase} and \emph{verification phase} \cite{nikolopoulos_continuous-variable_2017,pappu_physical_2002,gianfelici_theoretical_2020}.

\begin{enumerate}
    \item \emph{Setup phase}: A setup phase is the beginning phase of the protocol. Here the verifier has the valid device (in this case a PUF/qPUF) and locally prepares a database consisting of multiple challenge and response pairs of this device. The challenges and responses, namely Challenge-Response pairs (CRPs) are stored in the verifier's local database. We assume that the verifier's quantum capabilities are restricted to quantum polynomial time, and a polynomial-size database. Once the local database is generated, the device is physically transferred to the prover over a public channel.
    
    \item \emph{Identification phase}: The setup phase is followed by the identification phase where the verifier sends one or multiple challenges, usually chosen at random, to the prover from the CRP database. The challenge(s) is sent over a public (quantum) channel to the prover. The prover who has the valid device obtains the responses to the received challenges by querying the device and obtaining the response. Then the prover sends either the response directly, or sends some classical or quantum information related to the response to the verifier. We note that qPUF-based identification protocols would mostly differ in this phase by varying the number of challenges sent to the prover and the type of information received by the verifier. 
    
    \item \textit{Verification phase}: In the verification phase, the verifier runs a quantum or classical verification algorithm on the information received from the prover. We denote that the verifier correctly identifies the prover if the verification algorithm outputs 1. Otherwise, it aborts. 
\end{enumerate}

The \textbf{Correctness} or \textbf{Completeness} of an identification protocol is defined as the success probability of an honest prover over multiple rounds of identification, in the absence of any adversary or noise, should be one. The \textbf{Soundness} of an identification protocol ensures that the success probability of any adversary (depending on the adversarial model) in passing the verification phase over the multiple rounds of identification, should be negligible in the security parameter. 

\subsection{Quantum identification protocol with high-resource verifier}\label{sec:application-hrv-protocol}
The identification protocol runs between a verifier and a prover. The verifier is tasked with correctly identifying the prover who owns the device. Our setting assumes that the verifier and device owner behave honestly. The security has been shown against an adversary (computationally bounded in the learning phase) willing to be identified as the valid device owner. We propose the construction of two identification protocols using qPUFs which provide exponential security against any QPT adversary. The qPUF used in this protocol is a UqPUF as defined in \defref{def:uqPUF}.

\begin{figure}[t]
\includegraphics[scale=0.6]{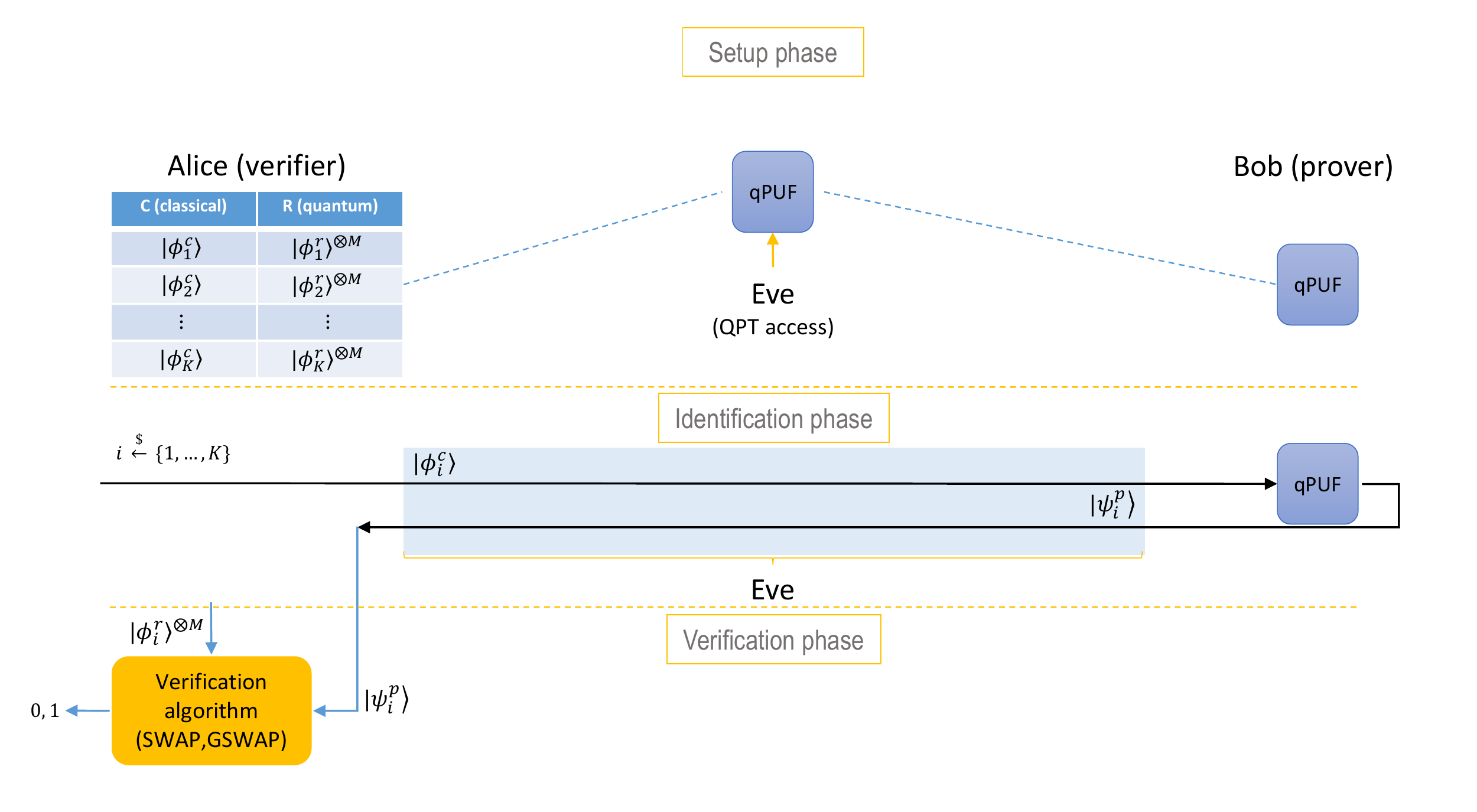}
\centering
\caption[qPUF-based identification protocol with high-resource verification]{qPUF-based identification protocol with high-resource verification between Alice (verifier) and Bob (prover) (\hrv). The protocol is divided into three sequential phases, \emph{setup phase}, \emph{identification phase}, and \emph{verification phase}. The protocol is analysed in the presence of a QPT adversary (Eve) which can gain information about the device during the \emph{setup phase}. In the last phase, verifier runs a quantum verification algorithm and outputs a classical bit `1' if prover's device is correctly identified. Otherwise, verifier outputs `0'. }
\label{fig:application-qid-p1}
\end{figure}

The first qPUF-based device identification protocol we propose is the quantum analogue of the standard PUF-based identification scheme between the verifier (Alice) and the prover (Bob) as shown in \figref{fig:application-qid-p1}. Before detailing the protocol, we list its salient features,
\begin{itemize}
    \item The prover is not required to have quantum memory, and computing ability resources\footnote{Here we note that the prover applies the qPUF transformation (the unitary) on the challenge states. Nevertheless, we do not consider this as computing ability of the prover, thus by no computing ability we refer to the fact that the prover does not need to run any extra quantum computations other than the physical interaction with the qPUF hardware.}, whereas the verifier is required to have high quantum memory and high computing ability resources (restricted to polynomial-size memory and QPT computation).
    \item The protocol requires a 2-way quantum communication link between the prover and verifier.
    \item The protocol has a quantum verification phase \emph{i.e.} the prover sends information in quantum states to the verifier who then performs a verification test to certify if the device is valid.
    \item The protocol provides perfect completeness and an exponentially-high security guarantee against any adversary with QPT resources. 
\end{itemize}

\subsubsection{Protocol description}
This protocol, referred as \hrv, is run between the verifier, and the prover and it is divided into three sequential phases,
\begin{protocol}[\hrv(K,N,M,D)] qPUF-based Identification protocol with high-resource verifier:\footnote{We drop the parameters $(K,N,M,D)$ from now on for simplicity whenever we refer to this protocol.}\\
\hrule
\begin{enumerate}
    \item \emph{Setup phase:}
            \begin{enumerate}
                \item Verifier has the qPUF device. 
                \item Verifier randomly picks $K \in \mathcal{O}(\text{poly} \log D)$ classical strings $\phi_i \in \{0,1\}^{\log D}$.
                \item Verifier selects and applies a Haar-random state generator operation denoted by the channel $\E_{prep}$ to locally create the corresponding quantum states in $\HilD$: $\phi_i \overset{\E_{prep}}{\rightarrow} \kc,\hspace{2mm} \forall i \in [K]$.
                \item Verifier queries the qPUF individually with each challenge $\kc$ a total of $M$ number of times to obtain $M$ copies of the response state $\kr$ and stores them in their local database $S \equiv \{\kc, \krm\}_{i=1}^{K}$. 
                \item Verifier publicly transfers the qPUF to prover.
            \end{enumerate}
            
            To be able to investigate the security in a strong and general setting, we do not assume the qPUF's transition to be done securely, in the sense that any QPT adversary (Eve) is allowed to query the qPUF during transition an $\mathcal{O}(\text{poly} \log D)$ number of times and thus build its local database. Due to the conditions on universal unforgeability of the qPUF, it is important that verifier picks the challenges $\kc \in S$ at random from a uniform distribution over the Hilbert space $\HilD$. This, in turn, implies that the encoding unitary operation $\E_{prep}$ is a Haar random unitary. We relax this condition in the upcoming section of this chapter.
    \item \emph{Identification phase}:
            \begin{enumerate}
                \item Verifier uniformly selects a challenge labelled ($i \xleftarrow{\$} [K]$), and sends the state $\kc$ over a public quantum channel to prover.
                \item Prover generates the output $\kp$ by querying to the qPUF device, the challenge received from the verifier.
                \item The output state $\kp$ is sent to verifier over a public quantum channel.
                \item This procedure is repeated with the same or different states a total of $R \leq K$ times\footnote{$R = M \times N$}. 
            \end{enumerate}

    \item \emph{Verification phase}:
            \begin{enumerate}
                \item Verifier runs a quantum equality test algorithm on the received response from the prover and the $M$ copies of the correct response that exists in the database. This algorithm is run for all the $R$ total number of CRP pairs.
                \item Verifier outputs `1' implying successful identification if the test algorithm returns `1' on all CRPs. Otherwise, outputs `0'.
            \end{enumerate}
\end{enumerate}
\hrule\vspace{3mm} 
Sections~\ref{sec:swapver} and \ref{sec:gswapver} describe the quantum verification algorithm run by the verifier. 
\end{protocol}

For this protocol, we define the security in terms of completeness and soundness properties. Completeness of {\hrv} protocol is the probability that verifier outputs `1' in the verification phase in absence of an adversary Eve. This implies that the verification algorithm must output `1' for all the $R$ rounds of the protocol with a probability that differs negligibly in the security parameter from 1,
\begin{equation}
     \text{Pr}[\text{Ver accept}_{\text{H}}] =  \text{Pr}\big[\prod_{i=1}^{R}\texttt{(qVer}(\kp, \kr) = 1)\big] = 1 - \negl(\lambda)
     \label{eq:complete1}
\end{equation}
where the subscript H denotes the honest device holder.

Soundness of the protocol is defined as the probability that a QPT adversary (Eve) passes the verification test. We say the {\hrv} is sound (or secure) if this probability is negligible in the security parameter:
\begin{equation}
    \text{Pr}[\text{Ver accept}_{\text{Eve}}] =  \text{Pr}\big[\prod_{i=1}^{R}\texttt{(qVer}(\rho_i, \kr) = 1)\big] =  \negl(\lambda)
    \label{eq:gensound1}
\end{equation}
where $\rho_{i}$ is the state sent by adversary in the $i$-th round.

Since our protocol is based on UqPUF, the verifier has no knowledge about the unitary of qPUF except the database $S$ which can be obtained by querying. Consequently, the responses stored in $S$ are unknown quantum states. This calls for quantum equality test verification algorithms to enable the verifier to validate the received states. We investigate the optimal one-sided error test, the SWAP test \cite{buhrman_quantum_2001} and the GSWAP \cite{chabaud_optimal_2018} as described in \chapref{chap:prelim}, Section~\ref{sec:prelim-swap-gswap}. 

\subsubsection{Verification with SWAP test} \label{sec:swapver}
The first proposal for verifier's \texttt{qVer} algorithm is the SWAP test and the identification protocol using this test is called {\hrvs}.
Its single run inputs one copy of each received state and verifier's response state and produces a binary outcome to determine the equality between two states. A single run, however, does not provide a low enough test error rate. To obtain an exponentially low rate, the test is repeated $M$ times for the same challenge state where $M$ is proportional to the inverse-log of the desired error probability. The error can be further decreased by choosing $N \leq K$ distinct challenge states such that the test is run for $R = N\times M$ number of times and the prover is successfully identified, only if he passes all the runs. In the next two theorems, we show that the SWAP-based test algorithm provides us with the desired completeness and soundness properties required in the protocol.

\begin{thmbox}
\begin{theorem}[\textbf{Completeness of \hrv\ with SWAP }]\label{th:application-hrv-swap-comp} In absence of an adversary Eve, the probability that the response state of an honest prover $\kp = U_{qPUF} \kc$, generated from the valid qPUF, passes all the $R$ {\normalfont SWAP} test runs is,
{\normalfont 
\begin{equation}
    \text{Pr}[\text{Ver accept}_{\text{H}}] =  \text{Pr}\big[\prod_{i=1}^{R}\text{(SWAP}(\kp, \kr) = 1)\big] = 1
\end{equation}}
\end{theorem}
\end{thmbox}
\begin{proof}
When verifier receives prover's response $\kp$ which is generated from the valid qPUF device for all the $i \in [R]$ copies of the challenge state, then $\kp = \kr$. This implies that $F(\kp,\kr) = 1$ for all $i \in [R]$. From \eqref{eq:swapaccept}, we see that,
\begin{equation}
\text{Pr}\big[(\text{SWAP}(\kp, \kr) = 1] = \frac{1}{2} + \frac{1}{2}F(\kp,\kr) = 1, \hspace{2mm} \forall i \in [R]  
\end{equation}
Since in the honest setting, the states received from prover over $R$ rounds are all valid qPUF pure states which are unentangled to each other, hence the SWAP tests for all the $R$ rounds are independent tests. This implies that,
\begin{equation}
\begin{split}
        \text{Pr}[\text{Ver accept}_{\text{H}}] & =  \text{Pr}\big[\prod_{i=1}^{R}\text{(SWAP}(\kp, \kr) = 1)\big] \\
        & = \prod_{i=1}^{R}\text{Pr}\big[\text{SWAP}(\kp, \kr) = 1\big] \\
        & = 1
\end{split}
\end{equation}
This completes the proof.
\end{proof}

To characterise the soundness, we bound Eve's success probability in passing the verification test \emph{i.e.} the probability that the state $\rho^R$ she sends to verifier passes all the $R$ runs of the SWAP test.
Even though the test runs are independent, if a generalised entangled state $\rho^R$ is sent by Eve, her success probability across the runs may no longer be the product of success probability of individual test runs. This implies that Eve's strategy might result in a higher success probability in some rounds based on the results of previous rounds. However, we show that since the $N$ distinct challenges being picked by verifier are all uniformly random, Eve does not gain anything by entangling the states across rounds corresponding to different challenges. To this end, we assume Eve can achieve optimal success probability by sending the state $ \bigotimes_{i=1}^{N}\rho_i^M$, where $\rho_i^M$ is a generalised state sent to $M$ runs of the SWAP test corresponding to the same challenge $\kc$.
Across these $j \in [M]$ runs corresponding to $\kc$, the state received by verifier is $\rho_{i,j} = \text{Tr}_{\{1\cdots M/j\}}(\rho_i^M)$, where $\rho_{i,j}$ is obtained by tracing out the M-1 instances $\{1,\cdots M/j \}$. Let $\rho_i^{\text{max}}$ be Eve's response state corresponding to challenge $\kc$, with the highest fidelity with the correct response, \emph{i.e.}

\begin{equation}\label{eq:application-maxrho}
   \forall{j}\in M \quad F(\rho_i^{\text{max}}, \kr) = \bra{\phi^r_i}\rho_i^{\text{max}}\kr \geqslant \bra{\phi^r_i}\rho_{i,j}\kr \hspace{2mm}
\end{equation}

Since the SWAP test success probability is directly proportional to the fidelity between the two input states, this implies that Eve can maximise her success probability by sending $M$ unentangled states $\rho_i^{\text{max}}$ to verifier instead of the generalised state $\rho_{i}^M$. The above equation \eqref{eq:application-maxrho} can be used to bound Eve's success probability in passing verifier's verification test,
\begin{equation}\label{eq:swapEve}
    \begin{split}
        \text{Pr}[\text{Ver accept}_{\text{Eve}}] &=  \text{Pr}\big[\prod_{i=1}^{R}\text{(SWAP}(\rho_i, \kr) = 1)\big] \\
        & = \prod_{i=1}^{N}\text{Pr}\big[\prod_{j=1}^{M}\text{(SWAP}(\rho_{i,j}, \kr) = 1)\big]\\
        &\leqslant \prod_{i=1}^{N}\prod_{j=1}^{M}\text{Pr}\big[\text{SWAP}(\rho_i^{\text{max}}, \kr) = 1\big]\\
        &\leqslant \prod_{i=1}^{N}\Big(\frac{1}{2} + \frac{1}{2}F_i \Big)^{M} = \epsilon
    \end{split}
\end{equation}
\noindent where $\rho_i = \text{Tr}_{\{1\cdots R/i\}}(\rho^R)$, and $F_i =  F(\rho_i^{\text{max}}, \kr)$.

Now using the fact that the qPUF device exhibits universal unforgeability against any QPT adversary (\thmref{th:qpuf-universal-unf}), we bound the success probability of Eve using the following theorem. 
\begin{thmbox}
\begin{theorem}[\textbf{Soundness of \hrv\ with SWAP }]\label{th:application-hrv-swap-sound} Let qPUF be a universally unforgeable UqPUF over $\HilD$. The success probability of any QPT adversary Eve, to pass the {\normalfont SWAP}-test based verification of the {\hrvs} protocol is at most $\epsilon$, given that there are $N$ different CRPs, each with $M$ copies. The $\epsilon$ is bounded as follows:
{\normalfont
     \begin{equation}
     \text{Pr}[\text{Ver accept}_{\text{Eve}}] \leqslant \epsilon \approx \mathcal{O}(\frac{1}{2^{NM}})  \end{equation}
     }
\end{theorem}
\end{thmbox}
\begin{proof}
From \eqref{eq:swapEve}, we see that the optimal strategy of Eve is to produce the response states $\rho_i^{\text{max}}$ which maximises the fidelity $F_i$ for each CRP $(\kc, \krm)$. We provided an upper bound on the fidelity when Eve has polynomial access to the qPUF in \thmref{th:qpuf-uni-unf-fid} stating that the fidelity $F_i$ is bounded as,
\begin{equation}\label{eq:application-swap-unfor}
    \text{Pr}[F_i \geqslant \delta] \leqslant \frac{d + 1}{D} 
\end{equation}
for any $\delta > 0$. Here $d = poly(\lambda) = poly\log(D)$ is the dimension of subspace that Eve has learnt from $\HilD$. For $D = 2^d$, this implies that the maximum fidelity state that Eve can create on average is non-orthogonal to the valid response state $\kr$ with a negligible probability $\approx \mathcal{O}(2^{-d})$. Hence $F_i = \delta \rightarrow 0$ with overwhelming probability. This bound holds   true for all distinct CRPs labelled by $i \in [N]$.

Thus from \eqref{eq:swapEve} and \eqref{eq:application-swap-unfor}, the probability that Eve passes verifier's SWAP based verification test is,
\begin{equation}
\begin{split}
    \text{Pr}[\text{Ver accept}_{\text{Eve}}] &\leqslant \prod_{i=1}^{N}\Big(\frac{1}{2} + \frac{1}{2}F_i \Big)^{M}\\ 
    &\leqslant  \prod_{i=1}^{N}\Big(\frac{1}{2} + \frac{1}{2}\delta \Big)^{M} \\
    &\approx \mathcal{O}(\frac{1}{2^{NM}}) = \negl(\lambda) 
\end{split}
\end{equation}

Note that here we also take into account the adaptive strategy of the adversary. That is even by assuming the previous rounds are added as extra states to Eve's learning phase, the dimension of the subspace $d$ will remain polynomial in $\lambda$. This completes the proof.
\end{proof}

The bound indicated above shows that one can achieve an exponentially secure qPUF-based identification using SWAP test based verification protocol with just a single challenge state \emph{i.e.} $N = 1$ and repeated for $M$ instances. However, non-ideal cases would make identification with different challenge states necessary. Hence we provide a general recipe involving multiple distinct challenges each running for multiple instances. Our protocol requires $R = N \times M$ number of rounds and uses $T = 2R$ number of communicated states. 

\subsubsection{Verification with GSWAP test} \label{sec:gswapver}

The second proposal for verifier's \texttt{qVer} algorithm is the GSWAP test and the identification protocol using this test is called {\hrvg}. Its single run requires one copy of the received state and $M$ copies of verifier's response state and produces a binary outcome to determine the equality between two states with a polynomial one-sided error \emph{i.e.} $\propto 1/M$. To boost the security to exponentially low error with a polynomial number of copies, the verifier first runs the challenge phase with $R = N \subset K$ distinct challenge states, then uses the GSAWP test as \texttt{qVer} algorithm to test the equality. To this end, she consumes $N$ received response states and $N\times M$ numbers of valid response states in her database.
In the next two theorems, we show that GSWAP based test algorithm provides us with the desired completeness and soundness properties required in the protocol.

\begin{thmbox}
\begin{theorem}[\textbf{Completeness of \hrv\ with GSWAP }]\label{th:application-hrv-gswap-comp} In absence of an adversary Eve, the probability that the response state of an honest prover, $\kp = U_{qPUF}\kc$ generated from the valid UqPUF passes all the $R = N$ test runs is,
{\normalfont 
\begin{equation}
\text{Pr}[\text{Ver accept}_{\text{H}}] =  \text{Pr}\big[\prod_{i=1}^{N}\text{(GSWAP}(\kp, \krm) = 1)\big] = 1    
\end{equation}
 }
\end{theorem}
\end{thmbox}
\begin{proof}
When verifier receives prover's response $\kp$ which is generated from the valid qPUF device for all the $i \in [R]$ copies of the challenge state, then $\kp = \kr$. This implies that $F(\kp,\kr) = 1$ for all $i \in [R]$. From Eq~\ref{eq:gswap}, we see that,
\begin{equation}
\text{Pr}\big[(\text{GSWAP}(\kp, \krm) = 1] = \frac{1}{M+1} + \frac{M}{M+1}F(\kp,\kr) = 1, \hspace{2mm} \forall i \in [N]  
\end{equation}
Since in the honest setting, the states received from prover over $R$ rounds are all valid qPUF pure states which are unentangled to each other, hence the GSWAP tests for all the $R$ rounds are independent tests. This implies that,
\begin{equation}
\begin{split}
    \text{Pr}[\text{Ver accept}_{\text{H}}] & =  \text{Pr}\big[\prod_{i=1}^{N}\text{(GSWAP}(\kp, \kr) = 1)\big] \\
    & = \prod_{i=1}^{N}\text{Pr}\big[\text{GSWAP}(\kp, \kr) = 1\big] \\
    & = 1
\end{split}
\end{equation}
This completes the proof.
\end{proof}
To characterise the soundness, we bound Eve's success probability in simultaneously passing the $N$ runs of GSWAP test when she sends the generalised entangled state $\rho^N$ to verifier. Similar to the argument provided for SWAP test soundness, Eve does not gain anything by entangling the states across different test runs. Thus Eve's probability in passing the verification test by sending the state $ \bigotimes_{i=1}^{N}\rho_i$ is the same as that for a generalised state $\rho^N$, where $\rho_i$ is the state sent to the instance of GSWAP test corresponding to the same challenge $\kc$. As a result, Eve's optimal success probability can be expressed as a product of individual GSWAP instance success probability,
\begin{equation}
    \begin{split}
        \text{Pr}[\text{Ver accept}_{\text{Eve}}] &=  \text{Pr}\big[\prod_{i=1}^{N}\text{(GSWAP}(\rho_i, \krm) = 1)\big] \\
        & = \prod_{i=1}^{N}\text{Pr}\big[\text{GSWAP}(\rho_i, \krm) = 1\big]\\
        &\leqslant \prod_{i=1}^{N}\Big(\frac{1}{M+1} + \frac{M}{M+1}F_i \Big) = \epsilon
    \end{split}
    \label{eq:gswapEve}
\end{equation}

where $F_i =  F(\rho_i, \kr)$ is the fidelity between Eve's state and the valid qPUF response state for the $i$-th round.

\begin{thmbox}
\begin{theorem}[\textbf{Soundness of \hrv\ with GSWAP }]\label{th:application-hrv-gswap-sound} Let qPUF be a universally unforgeable UqPUF over $\HilD$. The success probability of any QPT adversary Eve, to pass the {\normalfont GSWAP}-test based verification of the {\hrvg} protocol is at most $\epsilon$, given that there are $N$ different CRPs, each with $M$ copies. The $\epsilon$ is bounded as follows:
{\normalfont 
\begin{equation}
    \text{Pr}[\text{Ver accept}_{\text{Eve}}] \leqslant \epsilon \approx \mathcal{O}\big(\frac{1}{(M+1)^{N}}\big)
\end{equation}
}
\end{theorem}
\end{thmbox}
\begin{proof}
From \eqref{eq:gswapEve}, we see that the optimal strategy of Eve is to produce the response states $\rho_i$ which maximises the fidelity $F_i$ for each CRP $(\kc, \krm)$. We utilise the same universal unforgeability result to bound the fidelity $F_i$ with which Eve can produce the states $\rho_i$,
\begin{equation}\label{eq:application-gs-unfor}
    \text{Pr}[F_i \geqslant \delta] \leqslant \frac{d + 1}{D} 
\end{equation}
for any $\delta > 0$. Here $d = poly(\lambda) = poly\log(D)$ is the dimension of subspace that Eve has learnt from $\HilD$. For $D = 2^d$, this implies that the maximum fidelity state that Eve can create on average is non-orthogonal to the valid response state $\kr$ with a negligible probability $\approx \mathcal{O}(2^{-d})$. Hence $F_i = \delta \rightarrow 0$ with overwhelming probability. This bound holds true for all distinct CRPs labelled by $i \in [N]$.

Thus from Eq~\ref{eq:gswapEve} and \ref{eq:application-gs-unfor}, the probability that Eve passes verifier's GSWAP based verification test is,
\begin{equation}
\begin{split}
    \text{Pr}[\text{Ver accept}_{\text{Eve}}] &\leqslant \prod_{i=1}^{N}\Big(\frac{1}{M+1} + \frac{M}{M+1}F_i \Big)\\ 
    &\leqslant  \prod_{i=1}^{N}\Big(\frac{1}{M+1} + \frac{M}{M+1}\delta \Big) \approx \mathcal{O}\big(\frac{1}{(M+1)^{N}}\big) = \negl(\lambda) 
\end{split}
\end{equation}

We have also taken into account the adaptive strategy of Eve since our security is analysed for the most general attack strategy. This completes the proof.
\end{proof}

The last equation shows that to achieve an exponentially secure qPUF based identification using the GSWAP based verification protocol with only a polynomial sized register $S$, the protocol needs to be repeated for multiple $N$ instances. Our protocol requires $R = N$ number of communication rounds and uses $T = 2R$ number of communicated states.

\subsection{Quantum identification protocol with low-resource verifier}\label{sec:application-lrv-protocol}
Our second protocol enables a weak verifier to identify a quantum server prover in the network. We achieve this by delegating the equality testing to the prover thus effectively removing the quantum computational requirement on the verifier. While this might look like it could facilitate a malicious Eve to fool the weak verifier easily, we demonstrate due to the unforgeability of qPUF that the security is not affected. Before describing the details, we list the salient features of our protocol,
\begin{itemize}
    \item The protocol requires the prover to hold quantum computing capability, whereas the verifier is just required to have quantum memory and no quantum computing resources during the identification and verification phase\footnote{The state preparation phase happens in the setup phase of the protocol and it is a common property of all qPUF-based protocols. Here, we are mainly interested in the computing ability in the verification phase, which is the major difference between such protocols since verifying quantum states is a challenging task.} (restricted memory and computation).
    \item The protocol requires a one-way quantum communication link directed from the verifier to the prover. The prover to the verifier directed link is a classical channel.
    \item The protocol has a classical verification phase \emph{i.e.} the prover locally performs the verification test and sends the classical information to the verifier.
    \item The protocol provides perfect completeness and an exponentially-high security guarantee against any adversary with QPT resources.
\end{itemize}

\begin{figure}[ht!]
\includegraphics[scale=0.56]{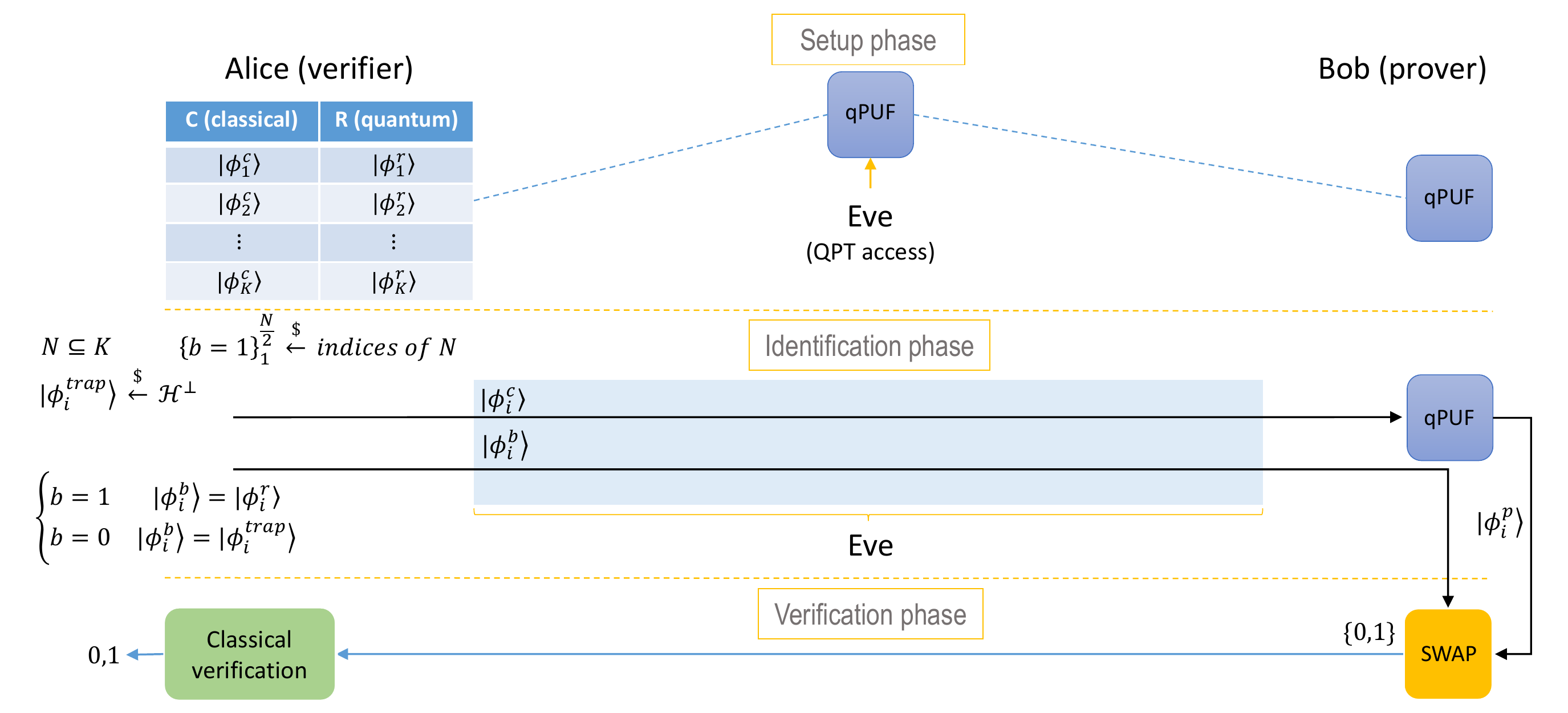}
\centering
\caption[qPUF-based identification protocol with low-resource verifier and classical verification]{qPUF-based identification protocol with low-resource verification between Alice (verifier) and Bob (prover) (\lrv). The protocol is divided into three sequential phases, \emph{setup phase}, \emph{identification phase} and \emph{verification phase}. In the identification phase, Alice randomly picks a subset $N \subseteq K$ of challenges which are sent to Bob. She also employs a trap based scheme where she sends either the correct response state of the challenges or the trap states which are states orthogonal to the valid response states. Bob performs the SWAP-test verification and sends the classical bits back to Alice. Alice finally performs a classical verification to check.}  
\label{fig:application-qid-p2}
\end{figure}

\subsubsection{Protocol description}

This protocol is run between a verifier, a prover in three sequential phases,

\begin{protocol}[\lrv(K,N,D)]
qPUF-based identification protocol with low resource verifier and classical verification algorithm:\footnote{We drop the parameters $(K,N,D)$ from now on for simplicity whenever we refer to this protocol.}
\vspace{3mm}\hrule
\begin{enumerate}
    \item \emph{Setup phase}:
            \begin{enumerate}
                \item Verifier has the qPUF device. 
                \item Verifier randomly picks $K \in \mathcal{O}(\text{poly} \log D)$ classical strings $\phi_i \in \{0,1\}^{\log D}$.
                \item Verifier selects and applies a Haar-random state generator operation denoted by the channel $\E$ to locally create the corresponding quantum states in $\HilD$: $\phi_i \overset{\E}{\rightarrow} \kc,\hspace{2mm} \forall i \in [K]$.
                \item Verifier queries the qPUF individually with each quantum challenge $\kc$ to obtain the response state $\kr$.
                \item Verifier creates states $\ket{\phi_i^{\perp}}$ orthogonal to $\kc$ and queries the qPUF device with them to obtain the trap states labelled as $\ket{\phi_i^{\text{trap}}}$. The unitary property of qPUF device ensures that $\langle \phi_i^{\text{trap}}|\phi_i^r\rangle = 0$. 
                \item Verifier creates a local database $S \equiv \{\kc,\{ \kr, \ket{\phi_i^{\text{trap}}}\}\}$ for all $i \in [K]$. Thus the $S$ registers stores the challenge state $\kp$ and the corresponding valid response state and the trap state which is orthogonal to the response state.
                \item Verifier publicly transfers the qPUF to prover.
            \end{enumerate}
            The transition is non-secure and Eve is allowed $\mathcal{O}(\text{poly} \log D)$ query access to the qPUF to build her own local database. 
    \item \emph{Identification phase}:
            \begin{enumerate}
                \item Verifier randomly selects a subset $N \subseteq K$ different challenges $\kc$ and sends them over a public channel to prover.
                \item Verifier randomly selects $N/2$ positions, marks them $b = 1$ and sends the valid response states $\ket{\phi_i^1} = \kr$ to prover. On the remaining $N/2$ positions, marked as $b = 0$, the verifier sends the trap states $\ket{\phi_i^0} = \ket{\phi_i^{\text{trap}}}$.
            \end{enumerate}
    \item \emph{Verification phase}:
            \begin{enumerate}
                \item Prover queries the qPUF device with the challenge states received from verifier to generate the response states $\kp$ for all $i \in [N]$. 
                \item Prover performs a quantum equality test algorithm by performing a SWAP test between $\kp$ and the response state $\ket{\phi_i^b}$ received from the verifier. This algorithm is repeated for all the $N$ distinct challenges.   
                \item Prover labels the outcome of $N$ instances of the SWAP test algorithm by $s_i \in \{0,1\}$ and sends them over a classical channel to verifier.
                \item Verifier runs a classical verification algorithm \texttt{cVer($s_1,...,s_N$)} and outputs `1' implying that prover's qPUF device has been successfully identified, and outputs `0' otherwise. 
            \end{enumerate}
\end{enumerate}
\hrule\vspace{3mm}
\end{protocol}

\figref{fig:application-qid-p2} shows the qPUF based identification protocol with low-resource verification denoted as {\lrv}. 
For the {\lrv} protocol, completeness is the probability that Verifier's verification algorithm $\texttt{cVer}$ returns an outcome `1' in absence of Eve. Ideally we require completeness to differ negligibly from 1,
\begin{equation}\label{eq:application-complete2}
     \text{Pr}[\text{Ver accept}_{\text{H}}] =  \text{Pr}\big[\texttt{cVer}(S_N) = 1\big] = 1 - \negl(\lambda)
\end{equation}
where $\lambda$ is the security parameter. 

Soundness of the protocol is the probability that \texttt{cVer} returns an outcome `1' in presence of Eve. For security, we require the soundness to be negligible in $\lambda$,
\begin{equation}\label{eq:application-gensound2}
    \text{Pr}[\text{Ver accept}_{\text{Eve}}] =  \text{Pr}\big[\texttt{cVer}(S_N) = 1\big] =  \negl(\lambda)
\end{equation}
We investigate the security of our protocol when the prover uses the SWAP test and the verifier uses the classical verification algorithm $\texttt{cVer}$. We remark that the prover can alternatively use GSWAP testing to generate the outcomes, however, this would require the verifier to send multiple copies of the same challenge state to the prover, thus incurring higher resources on the verifier's side.

\subsubsection{\texttt{cVer} algorithm}
The main ingredient of verification is the \texttt{cVer} classical test algorithm employed by the verifier to certify whether the prover's device has been identified.  As described in \algoref{alg:cver}, 
\texttt{cVer} receives an $N$-bit binary string $S_N$ as input. The algorithm is divided into two tests. \texttt{test1} first checks whether in the $N/2$ positions marked as $b = 1$, \emph{i.e.} the positions where the verifier had sent a valid qPUF response state to the prover if the corresponding bits in $S_N$ are all 0.

If this test succeeds, then the algorithm proceeds to \texttt{test2} which is a test on the positions where the verifier had sent the trap states to the prover. If on these positions, the expected number of bits in $S_N$ which are 0 lie between $\{\kappa\frac{N}{2} - \delta_{er}, \kappa\frac{N}{2} + \delta_{er}\}$, then \texttt{cVer} algorithm outputs `1' indicating that the device has been identified. Here $\kappa\frac{N}{2}$ is the expected number of bits in $b=1$ positions with outcome `0' that prover would obtain after the equality test algorithm measurement, in absence of any adversary Eve. In our case when the prover uses the SWAP test, $\kappa = 0.5$. Here, $\delta_{er}$ accounts for the statistical error in the measurement.

\begin{algorithm}[ht!]
\SetAlgoLined
\textbf{Description:} Let $S_N = \{0,1\}^N$ be the input $N$-bit string. Let $P=\{i_k\}^{N/2}_{k=1}$ be the set of indices showing the rounds of the protocol where $b=1$. Algorithm consists of two tests, \texttt{test1} and \texttt{test2} as follows:\\
   \texttt{test1:}\\
   \ForAll{$i$ in P}{
    \If{$s_{i} = 0$}{
        $count\gets count+1$\;
    }}
    \eIf{$count = \frac{N}{2}$}{
        \Return 1\;
    }{\Return 0\;}

\hrulefill\\
\texttt{test2:}\\
    \eIf{\texttt{test1} = 0}{
        \Return 0\;
    }
    {
    \ForAll{$i$ not in P}{
    \If{$s_{i} = 1$}{
        $count\gets count+1$\;
    }}
    \eIf{$\lvert count - \delta\frac{N}{2} \rvert \leqslant \delta_{er}$}{
        \Return 1\;
    }{\Return 0\;}
    }
 \caption{\texttt{cVer} algorithm}\label{alg:cver}
\end{algorithm}

\subsubsection{Verification using SWAP test and \texttt{cVer} algorithm}
Here we explicitly describe and calculate the completeness and soundness probabilities of the \lrv~protocol which employs the verification algorithm involving the prover's SWAP test, followed by the verifier's \texttt{cVer} algorithm. This allows the verifier to efficiently identify the valid qPUF device even though the SWAP test algorithm has been delegated to the prover. A single instance of the prover's SWAP test requires a single copy of the response state received from the verifier (either the valid qPUF response state or the trap state) and the response state that the prover generates by querying the verifier's challenge state in his qPUF device. To obtain a desired low enough error rate in the verification algorithm, the SWAP test is performed on $N$ distinct instances of the received response state and response state generated by prover by querying distinct challenges states. The responses of the SWAP test instances are classical bits. Thus the $N$ bit binary classical outcome string is sent to the verifier who employs the algorithm \texttt{cVer} described in \algoref{alg:cver}.
An identification protocol performed using $N$ distinct challenge states consumes a combined total of $2N$ copies of the received state and the response state generated by the verifier. In the next two sections, we show that SWAP based test algorithm provides us with the desired completeness and soundness properties required in the protocol.

\begin{thmbox}
\begin{theorem}[\textbf{\texttt{cVer} Completeness}]\label{th:application-cver-comp} In absence of an adversary Eve, the probability that the $N$-bit string $S_N = \{s_1,...,s_N\}$ sent by prover, passes the {\normalfont \texttt{cVer}($S_N$)} algorithm is,
{\normalfont 
\begin{equation}
\text{Pr}[\text{Ver accept}_{\text{H}}] =  \text{Pr}\big[\texttt{cVer}(S_N) = 1\big] = 1 - 2e^{-N/4}     
\end{equation}
}
\end{theorem}
\end{thmbox}

\begin{proof}
To prove this theorem, we separately analyse the $N/2$ positions where verifer sends the valid qPUF response state to prover (marked as $b =1$), and the remaining positions where she sends the trap state (marked as $b = 0$),
\begin{enumerate}
    \item $b = 1$ positions: When prover prepares the response state $\kp$ by querying her qPUF device with verifer's challenge state $\kc$, then prover's generated response state is equal to verifer's response state sent to prover, \emph{i.e.} $\kr = \kp$. This implies that $F(\kp,\kr) = 1$ for all $i \in [N]$ marked $b=1$. From \eqref{eq:swapaccept}, we see that,
    \begin{equation}
    \text{Pr}\big[\text{SWAP}(\kp, \kr) = 1] = \frac{1}{2} + \frac{1}{2}F(\kp,\kr) = 1, \hspace{2mm}   
    \end{equation}
    Note that $[\text{SWAP}(\kp, \kr) = 1]$ corresponds to the classical outcome 0. This implies that $s_i = 0$ for all $i \in [N]$ marked $b=1$ with certainty. Thus when verifer employs the \texttt{cVer} algorithm, prover always achieves a $count = N/2$ in the \texttt{test1} and thus passes it with certainty,
    \begin{equation}
        \text{Pr}[\texttt{test1 pass}] = 1
    \end{equation}    
    \item $b = 0$ positions: These positions correspond to verifer sending the trap states $\ket{\phi_i^{\text{trap}}}$ to prover such that prover's generated response state $\kp$ is orthogonal to the trap state. In other words, $F(\kp, \ket{\phi_i^{\text{trap}}}) = 0$ for all $i \in [N]$ marked $b=0$. This implies that,
    \begin{equation}
    \text{Pr}\big[\text{SWAP}(\kp, \ket{\phi_i^{\text{trap}}}) = 1] = \frac{1}{2} + \frac{1}{2}F(\kp,\ket{\phi_i^{\text{trap}}}) = \frac{1}{2}, \hspace{2mm}
    \end{equation}
    Thus, half of the $N/2$ positions would produce the classical outcome 1 on average. When verifer employs \texttt{test2} of the \texttt{cVer} algorithm, $\mathbb{E}[count] = N/4$. Using the Chernoff-Hoeffding inequality \cite{mitzenmacher_probability_2017}, for any constant $\delta_{er} > 0$,
    \begin{equation}
        \text{Pr}[\texttt{test2 pass}] = \text{Pr}\Big[\Big\lvert count - \frac{N}{4}\Big\rvert \leqslant \delta_{er}\Big] \geqslant 1 - 2e^{-N\delta_{er}^2}
    \end{equation}
\end{enumerate}
From the above results and using the fact that $\delta_{er} = 0.5$ for SWAP test based algorithm,
\begin{equation}
    \begin{split}
        \text{Pr}[\text{Ver accept}_{\text{H}}] &=  \text{Pr}\big[\texttt{cVer}(s_1,\cdots, s_N) = 1\big]\\
        &= \text{Pr}\big[\texttt{test1 pass} \wedge \texttt{test2 pass}\big]\\
        &= \text{Pr}[\texttt{test1 pass}]\cdot \text{Pr}[\texttt{test2 pass}]\\
        &\geqslant 1 - 2e^{-N/4}
    \end{split}
\end{equation}
This completes the proof.
\end{proof}

The next section details the soundness proof of the {\lrv} protocol.

\subsubsection{Soundness of {\lrv} protocol}\label{sec:application-lrv-soundness}
To characterise the soundness, we bound Eve's success probability in passing the \texttt{cVer} test. Since the verification test is reduced to a classical test, we consider the soundness in the presence of two types of Eve. The first is a \emph{classical Eve} who does not process any quantum resources. The second is a \emph{quantum Eve}, which possesses QPT memory and computing capability. We separately analyse the security against both types of Eve and prove that \emph{quantum Eve} gains only an exponentially-small advantage compared to the \emph{classical Eve}, thus reducing the security to analysing only the classical adversary. We show that since the verification test is classical, the only way for a \emph{quantum Eve} to succeed better than a \emph{classical Eve} is to succeed at guessing the trap positions better than a random guess of \emph{classical Eve}. We utilise the unforgeability property of qPUF to prove that a \emph{quantum Eve} can have only negligible advantage in guessing the trap positions compared to a \emph{classical Eve}, thus enabling the reduction.\\ 

\noindent\textbf{(I) Security against classical adversary}
\begin{thmbox}
\begin{theorem}[\textbf{Soundness against classical Eve}]\label{th:application-cver-cl-attack} The probability that any classical PPT adversary (Eve) produces an $N$-bit string $S_N = \{s_1,...,s_N\}$ which passes the {\normalfont \texttt{cVer}} algorithm is bounded as,
{\normalfont 
\begin{equation}
 \text{Pr}[\text{Ver accept}_{\text{Eve}}] =  \text{Pr}\big[\texttt{cVer}(S_N) = 1\big] \leqslant \mathcal{O}(2^{-N})      
\end{equation}
}
\end{theorem}
\end{thmbox}
\begin{proof}
First, we remark that any classical Eve's strategy to produce a valid $N$-bit string $S_N$ can be divided into two categories,

\begin{enumerate}
    \item \textbf{Independent guessing strategy:} Under this strategy, Eve tries to independently guess each bit of the string $S_N$ that would pass the verifier's \texttt{cVer} algorithm. This also relates to the strategy of independently finding valid responses and trap positions.

    \item \textbf{Global strategy:} Here, Eve's strategy is to output a string $S_N$ using the global properties of the \texttt{cVer}, such that it passes the verification test with maximum probability. In contrast to the previous strategy, the probability of outputting each bit $s_i$ is not necessarily independent of the global strategy.  
\end{enumerate}

We calculate the optimal success probability of Eve in both cases and show that by optimizing for both the strategies, we obtain a higher success probability for Eve in the optimal global strategy scenario. Although, the two strategies converge in the limit of large $N$. Hence we bound Eve's success probability by the optimal global strategy.\\

\noindent \textbf{1. Independent guessing strategy:} Under this strategy, Eve independently guesses each bit with the probability,

\begin{equation}
    Pr[s_i = 0] = \alpha, \quad Pr[s_i = 1] = 1 - \alpha
\end{equation}
where $\alpha \in [0,1]$.

We denote the resulting string generated by Eve's strategy as $S_{id} = \{s_1,\cdots, s_N\}$. In order for $S_{id}$ to pass the \texttt{cVer} verification algorithm, it must simultaneously pass the \texttt{test1} and \texttt{test2}. Since Eve's strategy is guessing each bit independently, hence the probability for her to pass the \texttt{test1} and \texttt{test2} are independent. Let us look at the probability of passing the \texttt{test1} (which corresponds to checking the $N/2$ positions marked $b = 1$,

\begin{equation}
        \text{Pr}[\texttt{test1 pass}] = \text{Pr}[s_{p_1} = 0]\times\cdots\times \text{Pr}[s_{p_{\frac{N}{2}}} = 0] = \alpha^{\frac{N}{2}}
\end{equation}
where $p_i$ correspond to the $b=1$ marked positions. 

If Eve's generated string passes \texttt{test1}, then verifer runs the \texttt{test2} to check if \emph{count}, which is the number of bits that are 1 in the remaining $N/2$ bits marked with $b=0$, lies within the interval $\big \lvert count - \frac{N}{4} \big\rvert \leqslant \delta_{er}$. Eve succeeds in passing this test with the probability, 
\begin{equation}
\begin{split}
    \text{Pr}[\texttt{test2 pass}] & = \sum_{x= N/4 - \delta_{er}}^{N/4 + \delta_{er}}(1-\alpha)^{x}\alpha^{\frac{N}{2} - x}\times {N/2 \choose x} \\ & \approx (2\delta_{er}+1)(1-\alpha)^{\frac{N}{4}}\alpha^{\frac{N}{4}}\times {N/2 \choose N/4} 
\end{split}
\end{equation}
where the approximation holds since we assume that $\delta_{er} \ll N$. From the above results, we see that the probability that Eve's string $S_{id}$ passes the \texttt{cVer} verification algorithm is,
\begin{equation}
\begin{split}
    \text{Pr}[\text{Ver Accept}_{\text{Eve},\alpha}] & = \text{Pr}[\texttt{test1 pass}_{\alpha}]\cdot \text{Pr}[\texttt{test2 pass}_{\alpha}] \\
    & \approx (2\delta_{er}+1)\alpha^{\frac{3N}{4}}(1-\alpha)^{\frac{N}{4}}\times {N/2\choose N/4}
\end{split}        
\end{equation}
This is Eve's acceptance probability for a given $\alpha$. An optimal strategy for Eve is to find the optimal value of $\alpha$ that maximises the acceptance probability. This corresponds to, 

\begin{equation}
        \frac{\partial}{\partial \alpha}\text{Pr}[\text{Ver Accept}_{\text{Eve},\alpha}] \Rightarrow \frac{\partial}{\partial \alpha}(\alpha^{\frac{3N}{4}}(1-\alpha)^{\frac{N}{4}}) = 0 \Rightarrow \alpha = \frac{3}{4}
\end{equation}

Thus the maximum acceptance probability of Eve using an independent guessing strategy is:
\begin{equation}
     \text{Pr}[\text{Ver Accept}_{\text{Eve}}] = (2\delta_{er} + 1) \frac{3^{\frac{3N}{4}}}{2^{2N}}\times {N/2\choose N/4} \approx \mathcal{O}(2^{-N})
\end{equation} \\

\noindent\textbf{2. Global strategy:} The second category of Eve's strategy is to guess the $N$ bit string which passes the \texttt{cVer} test algorithm with maximum probability. Here, Eve is not restricted to choosing each bit independently.  
To find the optimal global strategy we look at the \texttt{test1} and \texttt{test2} algorithms and extract essential properties that can be leveraged by Eve to pass the verification test. We note that
\begin{itemize}
    \item Since the good and trap response positions corresponding to $b = 0$ and 1 are chosen uniformly randomly by verifer, hence verifer does not have any information on the index set $P$ corresponding to $b = 1$ (thus no information on $b=0$ positions too).

    \item Eve knows the statistics of 0's and 1's in the desired string to pass the \texttt{cVer}. For example, a string must have a minimum of $\approx 3N/4$ bits which are $0$, otherwise, the string necessarily fails the \texttt{test1} or \texttt{test2} or both.
\end{itemize}
Based on the above facts, any global strategy for Eve should consist of optimizing the number of 0's and 1's to pass both verification tests.

Before considering the optimal global attack strategy, we give an example of a specific (non-optimal) attack strategy to provide intuition on the kind of strategies that Eve can adopt here.\\

\noindent \textit{Example of a global strategy:} The first global strategy that one might think of is to try to guess $P$, since passing the \texttt{test1} reduces to finding the strings that have bits `0' is all the $p_i$ positions \emph{i.e.} positions marked $b=1$. If Eve successfully manages to guess the $b=1$ positions, then she has a deterministic strategy of winning the \texttt{test2}, since she also knows the $b=0$ trap positions. Across these positions she can deterministically assign the bits such that the \emph{count} of the number of 1 bits lie within the interval $\big \lvert count - \frac{N}{4} \big\rvert \leqslant \delta_{er}$.
  
We denotes Eve's generated string with this strategy to be $S_g$. Hence the probability of $S_g$ passing \texttt{test1} is equal to correctly guessing the $\frac{N}{2}$ positions marked $b=1$,
\begin{equation}
\text{Pr}[\texttt{test1 pass}_{S_g}] = \text{Pr}[\text{guess $b=1$ positions}] = {N\choose N/2}^{-1}
\end{equation}

Once this test passes, then test2 passes with certainty. Now the probability of passing the \texttt{cVer} verification algorithm is, 

\begin{equation}
\begin{split}
    \text{Pr}[\text{Ver accept}_{\text{Eve},S_g}] &= \text{Pr}[\texttt{test1 pass}_{S_g}\wedge \texttt{test2 pass}_{S_g}]\\
    &= \text{Pr}[\texttt{test1 pass}_{S_g}]\cdot \text{Pr}[\texttt{test2 pass}_{S_g}|\texttt{test1 pass}_{S_g}] \\
    &= {N\choose N/2}^{-1}\cdot 1\\
    &\leqslant N^{-\frac{N}{2}} 
\end{split}
\end{equation}
We show that this global strategy is not optimal and Eve can design an optimal global strategy by properly utilising the part the second part of the information.

First, we argue that maximising the number of 0's will necessarily increase the success probability of passing \texttt{test1}. Let us assume that Eve sends an all `0' string $S_g$ to the verifier. Since \texttt{test1} checks only if in the $b=1$ marked positions are 0, so $S_g$ will always pass the first test. However, this string necessarily fails the \texttt{test2} since the \emph{count} for this test is $N/2$ which is much higher than the tolerated limit. 
 
Thus there always exists a global strategy with an optimal number of bits (number of 1) in $S_g$ in the case of $\delta_{er}=0$, or more precisely a strategy that allows the flexibility of having a set of values for the number of `1' bits that the \texttt{test2} tolerates in case of $\delta_{er}\neq 0$. \\

\noindent \textbf{Optimal global strategy}: We say that an optimal global strategy $\E_{gop}$ is the one that outputs a string $S_{gop}$  with $c_1$ number of 1 bits, where $c_1 \in m_{valid} = \{\frac{N}{4} - \delta_{er}, \dots,  \frac{N}{4} + \delta_{er}\}$.\\

\noindent \textbf{Optimality argument}: We prove the optimality of our test by the contradiction argument. Let us assume that there is a strategy $\E_{g}$ different from above which produces a string $S_g$ that succeeds with the verification acceptance probability higher than $S_{gop}$. Now, either all the strings that $\E_g$ outputs have $c_1$ number of 1 bits, where $c_1$ lies within the optimal boundary $ m_{valid}$. In this case $\E_g$ falls within the $\E_{gop}$ strategy set. Or, there is at least one string that $\E_g$ outputs with $c_1$ number of 1 bits such that $c_1 \not\in m_{valid}$. In this case, that string will necessarily fail \texttt{test2}, even if it passes \texttt{test1}. This is because for the strategy $\E_g \not\in \E_{gop}$ to pass, the bits in $S_g$ which are 1 must necessarily appear in the positions marked $b=0$ (trap positions). And since the number of 1 bits $c_1 \not\in m_{valid}$, this implies it will fail the \texttt{test2}. Thus, $\text{Pr}[\text{Ver Accept}_{\text{Eve},\E_g \not\in \E_{gop}}] = 0$.

Note that the condition of $c_1 \in m_{valid}$ is necessary but not a sufficient condition for passing the verification algorithm $\texttt{cVer}$ \emph{i.e.} any string with with $c_1 \not\in m_{valid}$, will always fail but not all strings with $c_1 \in m_{valid}$ will always pass the verification. Thus we can define the largest possible set of potentially valid strings which Eve needs to choose from to maximise her acceptance probability. As a result, we can define the optimal strategy $\E_{gop}$'s event space to be ${N\choose c_1}$. This is the set of all strings with the number of bits $c_1 \in m_{valid}$. We can now find the optimal global probability which is the probability that both the tests of \texttt{cVer} pass,

\begin{equation}
\begin{split}
    \text{Pr}[\text{Ver accept}_{\text{Eve}, S_{gop}}] &= \text{Pr}[\texttt{test1 pass}_{S_{gop}}\wedge \texttt{test2 pass}_{S_{gop}}]\\
    &= \text{Pr}[\texttt{test1 pass}_{S_{gop}}]\cdot \text{Pr}[\texttt{test2 pass}_{S_{gop}}|\texttt{test1 pass}_{S_{gop}}] 
\end{split}
\end{equation}
To calculate $\text{Pr}[\texttt{test1 pass}_{S_{gop}}]$, we need to find the number of strings $S_{gop}$ from the whole set of strings $\{0,1\}^N$ with $c_1 \in m_{valid}$ bits and which passes the first test. In other words, the string $S_g$ must have bits 0 in all the $b=1$ marked positions and the bits  1 in the $b=0$ marked positions.

Thus there are $N/2$ positions out $N$ where the bits 1 can be placed without the \texttt{test1} getting rejected.

For a specific $c_1$, the total number of such strings is equal to the possible ways of distributing $c_1$ objects (1's) in $N/2$ positions:
\begin{equation}
\#\text{(correct strings)} = {N/2 \choose c_1}    
\end{equation}

If one of these `correct strings' is picked, it will necessarily also satisfy the condition of the second test. Hence the conditional probability is $$\text{Pr}[\texttt{test2 pass}_{S_{gop}}|\texttt{test1 pass}_{S_{gop}}]  = 1$$.
And the probability of passing the first test is,
\begin{equation}
    \text{Pr}[\texttt{test1 pass}_{S_{gop}}] = {N/2 \choose c_1}\Big/{N \choose c_1}
\end{equation}

The above \texttt{test1} passing probability is for a single $c_1 \in m_{valid}$. Summing over the probabilities of all the accepted $c_1$ ,
\begin{equation}
\text{Pr}[\texttt{test1 pass}_{S_{gop}}] = \sum_{c_1 \in m_{valid}} \frac{{\frac{N}{2}\choose c_1}}{{N\choose c_1}} = \sum^{\delta_{er}}_{k=-\delta_{er}} \frac{{\frac{N}{2}\choose \frac{N}{4} + k}}{{N\choose \frac{N}{4} + k}} = \frac{(\frac{N}{2})!}{N!} \sum^{\delta_{er}}_{k=-\delta_{er}} \frac{(\frac{3N}{4} - k)!}{(\frac{N}{4} - k)!}
\end{equation}

In the limit $\delta_{er} \ll N$, the sum will converge,
\begin{equation}
\text{Pr}[\texttt{test1 pass}_{S_{gop}}] = (2\delta_{er} + 1)\cdot \frac{(\frac{N}{2})!(\frac{3N}{4})!}{N!(\frac{N}{4})!}    
\end{equation}

From the above equations, the probability that Eve passes the \texttt{cVer} algorithm  using the global strategy,

\begin{equation}
\begin{split}
\text{Pr}[\text{Ver accept}_{\text{Eve},S_{gop}}] 
&= \text{Pr}[\texttt{test1 pass}_{S_{gop}}]\cdot \text{Pr}[\texttt{test2 pass}_{S_{gop}}|\texttt{test1 pass}_{S_{gop}}] \\
&= (2\delta_{er} + 1)\times \frac{(\frac{N}{2})!(\frac{3N}{4})!}{N!(\frac{N}{4})!}\cdot 1 \\ 
&\leqslant \mathcal{O}(N^{-N/2})
\end{split}
\end{equation}

\noindent\textbf{3. Probability comparison of Independent guessing strategy and Global strategy:} To find the optimal classical attack, we compare the two categories of the attack strategies of Eve.

We fix the accepted tolerance value $\delta_{er}=1$ for the comparison. The same result holds for other fixed $\delta_{er}$ values. \figref{fig:application-prob-compare} shows the behaviour of the acceptance probabilities of Eve in the independent guessing strategy and global strategy as an increasing function of the string length $N$.

\begin{figure}[h!]
\includegraphics[scale=0.7]{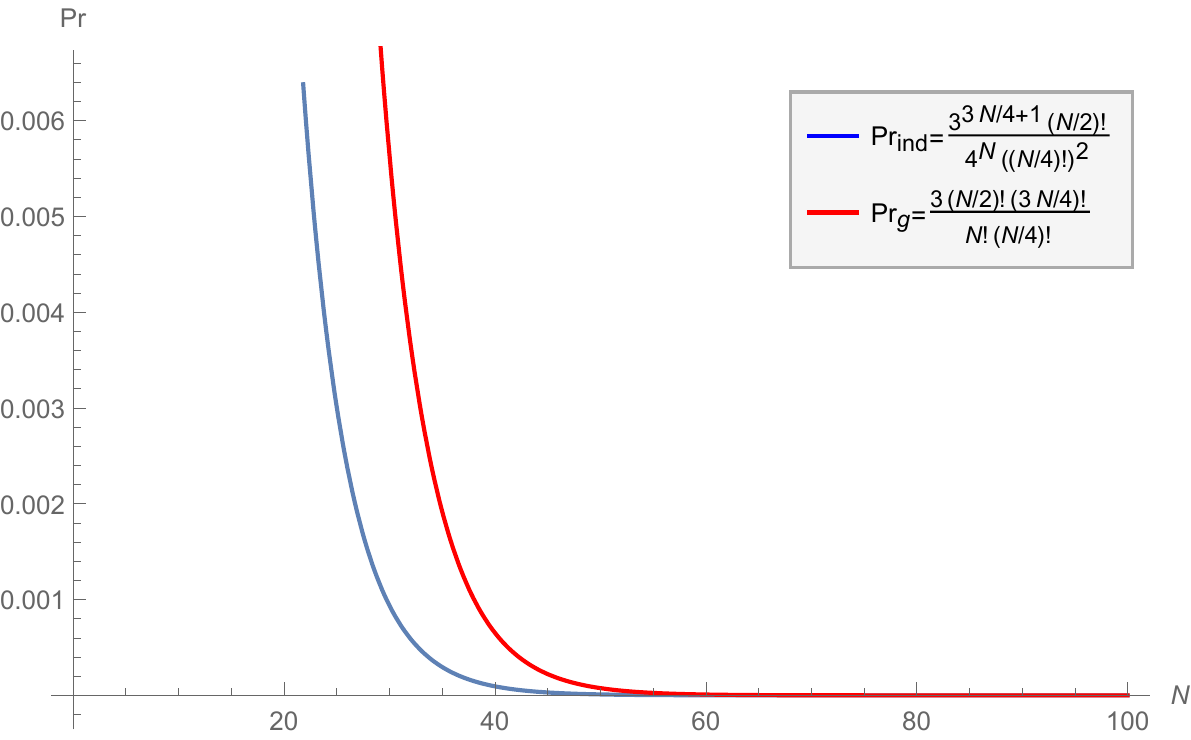}
\centering
\caption[Probability comparison for different classical adversarial strategies against \lrv]{Comparison of the acceptance probabilities of a classical adversary (Eve) in the independent guessing strategy (in blue) and global strategy (in red) as a decreasing function of the string length $N$ for the tolerance value $\delta_{er}=1$}
\label{fig:application-prob-compare}
\end{figure}

From the simulation, we infer that the two strategies have an inverse exponential form as expected. Also, they both converge for large enough $N$ values. This also confirms the fact that the optimal strategy lies in finding the correct number of 1's in the string and the difference comes from our approximation in using the frequency interpretation of the probabilities in the smaller N. Using Stirling's approximation $n! \approx \sqrt{2n\pi}(\frac{n}{e})^n$ one can check that $\frac{1}{{N\choose \frac{N}{4}}} \approx (\frac{4}{3^{3/4}})^{-N}$ which gives exactly the same bound as the independent guessing strategy. Although, in small $N$ the global strategy is slightly better.
Finally, we use Stirling's approximation ${2n\choose n} \approx \frac{2^{2n}}{\sqrt{\pi n}}$ to obtain the common factor of both probabilities we can bound the adversary's optimal success probability as,
\begin{equation}
\text{Pr}[\text{Ver Accept}_{\text{Eve}}] \approx \frac{3^{3N/4}}{2^{2N}}\times\frac{2^{N/2}}{\sqrt{\frac{\pi N}{4}}} = \frac{2}{\sqrt{N\pi}}(\frac{2^6}{{3^3}})^{-N/4} \approx \mathcal{O}(2^{-N}) \quad \text{for large enough N}    
\end{equation}
\noindent This completes the proof.
\end{proof}

\noindent\textbf{(II) Security against quantum adversary}\\

\noindent We now investigate the soundness property of the protocol against QPT Eve by modelling Eve's strategy with a completely positive trace preserving (CPTP) map that takes as input the target challenge $\kc$, the unknown state $\kb$, and ancilla qubits and outputs the classical bits which are sent to verifier for verification. This map utilises the database information created by Eve during the qPUF transition. A QPT Eve's strategy can be divided into two categories,
\begin{enumerate}
    \item \textbf{Collective attack strategy}: Eve applies an independent CPTP map on each of the $N$ rounds.
    \item \textbf{Coherent attack strategy}: Eve applies a CPTP map on the combined N distinct challenge and their corresponding response states that the verifier sends to the prover.
\end{enumerate}
A collective strategy is a special case of Eve's coherent strategy. However, we show that independence in choosing the trap states by verifier reduces the coherent strategy to the collective strategy by Eve. We analyse the collective security first and then give a reduction of the coherent strategy to the collective strategy.\\

\begin{figure}[h!]
\includegraphics[scale=0.7]{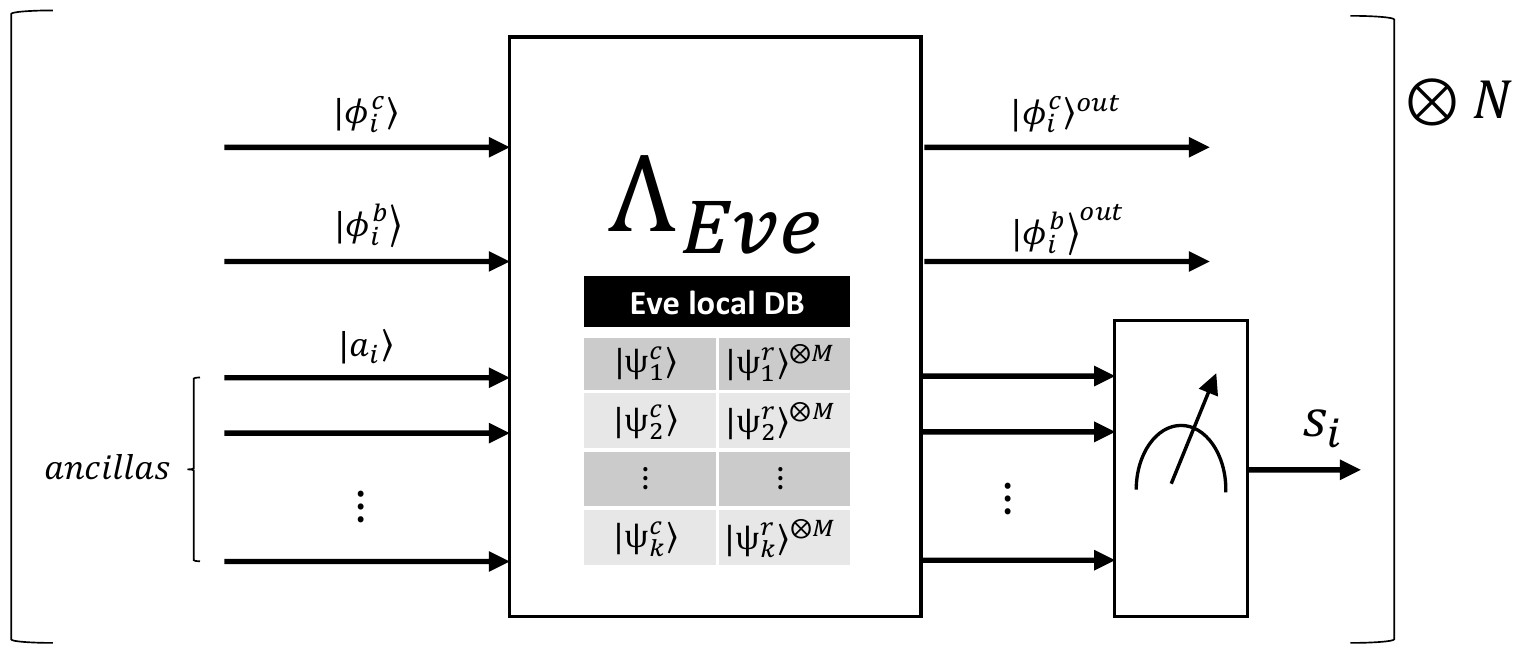}
\centering
\caption[Quantum collective attack strategy on {\lrv}]{Quantum collective attack strategy performed by Eve on {\lrv} protocol by applying the same local-database-dependent CPTP map on each round of the challenge and response state $\kc$ and $\kb$ respectively. The output of the single instance of the map is a bit $S_i$.}
\label{fig:application-quantum-adv-col}
\end{figure}

\noindent\textbf{1. Collective strategy:} Under this strategy, Eve optimises over all the CPTP maps that input verifier's states $\kc$ and $\kb$ and outputs a single bit $s_i$ to maximise the acceptance probability. 
\figref{fig:application-quantum-adv-col} shows Eve performing a general collective strategy. 

We denote Eve's quantum map to be, 
\begin{equation}
    \Lambda_{Eve} \equiv \bigotimes_{i=1}^{N}\Lambda_i \quad.
\end{equation}
Contrary to the classical Eve who is unable to figure out the trap positions in any round with a probability higher than half, a QPT Eve, by leveraging her local database information, could be expected to do better than a random guess.
More formally, we say that the {\lrv} protocol is secure against any QPT Eve that performs a CPTP map $\Lambda_i$ on the states $\kc, \kb$ for each $i \in [N]$ if the resulting success probability of correctly guessing the bit $b$ for each position differs negligibly in the security parameter from half.

We need a small toolkit, which is the abstraction of an ideal test in a single instance case (when one is provided with a single copy of one quantum state and multiple copies of the other state), in terms of fidelity. This definition is very similar to \defref{def:test-delta} in \chapref{chap:qpuf} where we remove the quantifier $\delta$ for simplicity:
\begin{defbox}
\begin{definition}[Single Instance Ideal Test Algorithm]\label{def:application-ideal-test} We call a test algorithm according to \defref{def:test}, a $\Ti$ test algorithm when one is provided a single copy of the state $\rho$ and multiple copies of the state $\ket{\psi}$ (or vice-versa) with fidelity $F(\rho, \ket{\psi}\bra{\psi})$ the test responds as follows:
\end{definition}
\begin{equation}
\Ti := \mathrm{Pr}[1 \leftarrow \Ti(\rho, \ket{\psi}\bra{\psi})] = F(\rho, \ket{\psi}\bra{\psi})
\end{equation}
\end{defbox}

Now we can show the security of \lrv~against QPT adversaries with collective attack strategies with the following theorem:

\begin{thmbox}
\begin{theorem}[\textbf{Security against collective attack}] \label{th:application-lrv-colattack} The success probability of any QPT adversary Eve, in correctly guessing whether $\kb = \kr$ for each $i \in [N]$ differs negligibly from half,
\begin{equation}
    \text{Pr}[b \leftarrow \Lambda_i(\kc,\kb)] \leqslant \frac{1}{2} + \mathcal{O}(2^{- d}) \hspace{2mm} \forall i \in [N]
\end{equation}
where \normalfont{$d = \mathcal{O}(\text{poly} \log D)$} is the size of Eve's database and qPUF is in $\mathcal{H}^{D}$.
\end{theorem}
\end{thmbox}
\begin{proof}
First, we use the symmetry of the problem to restrict ourselves to cases where $b=1$. We prove the theorem by contradiction \emph{i.e.}, suppose there exists an algorithm $W$ that wins the quantum security game for each index $i \in [N]$ with a probability non-negligibly better than a random guess. In other words, $W = 1$ if the index $b$ is correctly guessed, and $W = 0$ otherwise. Let $f(\lambda) \geqslant 0$ be a non-negligible function of the security parameter. 
The joint probabilities for all collective possible values of $b$ and $W$ can be written as,
\begin{equation}
\begin{split}
        & \text{Pr}[W=1, b=1] = \frac{1}{4} + f(\lambda)\quad \text{Pr}[W=1, b=0] = \frac{1}{4} - f(\lambda) \\
        & \text{Pr}[W=0, b=0] = \frac{1}{4} + f(\lambda)\quad \text{Pr}[W=0, b=1] = \frac{1}{4} - f(\lambda)
\end{split}
\end{equation}
where the joint probabilities are higher when $W$ correctly guesses $b$, and is lower otherwise. From this, we can define the following conditional probability of winning for cases where $b=1$ as follows:
\[Pr[W=1 | b=1] = \frac{Pr[W=1, b=1]}{Pr[b=1]} = \frac{1}{2} + f'(\lambda)\]
Where $f' = 2f$ is again a non-negligible function in the security parameter $\lambda$. This is the same probability of winning when $b = 0$ \emph{i.e.} $\text{Pr}[W = 0|b = 0]$.

Now we show that the success probability of Eve in successfully guessing whether $\kb = \kr$ reduces to finding a CPTP map $\Lambda_i$ which performs an optimal quantum test to distinguish the response state $\kb$ with the reference state $\ke$. As Eve has no access to the actual response $\kr$, the reference state $\ke$ should be generated within the $\Lambda_i$ itself. Thus without loss of generality, any attack map $\Lambda_i$, consists of two parts. The first part uses a generator algorithm \texttt{gen} to generate a reference state $\ke$, or more generally a mixed state $\rho_e$ by using the local database and the input challenge state $\kc$, and the second part performs a test algorithm $\T$ on $\kb$ and $\rho_e$,

\begin{equation}
    \Lambda_i \equiv \T(\kb, \rho_e \leftarrow \texttt{gen}(DB,\kc))
\end{equation}

where $DB$ is the local database of Eve generated in the  \emph{setup phase}. To further provide the capability to Eve, we assume that her test $\T$ is an optimal test equality test algorithm also referred to as the ideal test algorithm in \defref{def:application-ideal-test}, \emph{i.e.} $\T = \Ti$. Note that $\Ti$ is the optimal test allowed by quantum mechanics where the probability of succeeding in the equality test is proportional to the square of the fidelity distance of the two states. Now we state the following contraposition:
Let us assume that there exists a winning algorithm $W$ running $\Lambda = \Ti(\kb, \rho_e)$ such that,
\begin{equation}
\text{Pr}[1 \leftarrow \Lambda(\kc,\kb)| b = 1] \leqslant \frac{1}{2} + \nonnegl(\lambda)   
\label{eq:nonnegl}
\end{equation}

From \defref{def:application-ideal-test}, we see that $\Ti$ outputs $1$ with probability $p = F(\kb, \rho_e)$. In other words,
\begin{equation}
     \text{Pr}[1 \leftarrow \Lambda(\kc,\kb)| b = 1] = \text{Pr}[1 \leftarrow \Ti] = F(\kb, \rho_e) \leqslant \frac{1}{2} + \nonnegl(\lambda)
\end{equation}

This implies that if an algorithm $W$ exists for Eve, then she is able to generate the state $\rho_e$ with non-negligible fidelity with the valid qPUF response (for b=1), and similarly with trap states (for b = 0). And this would hold for all $i \in [N]$. 
But this contrasts with the universal unforgeability of the qPUF which states that the success probability of any QPT adversary having polynomial-size access to the qPUF is bounded as $\frac{d_e+1}{D}$ (\thmref{th:qpuf-uni-unf-fid}) where $d_e = poly(\lambda) = \text{poly} \log(D)$ is the dimension of subspace that Eve has learnt from $\HilD$. Thus such $\Lambda$ cannot exist even with the most efficient test $\Ti$. This concludes the proof.
\end{proof}

\begin{figure}[h!]
\includegraphics[scale=0.7]{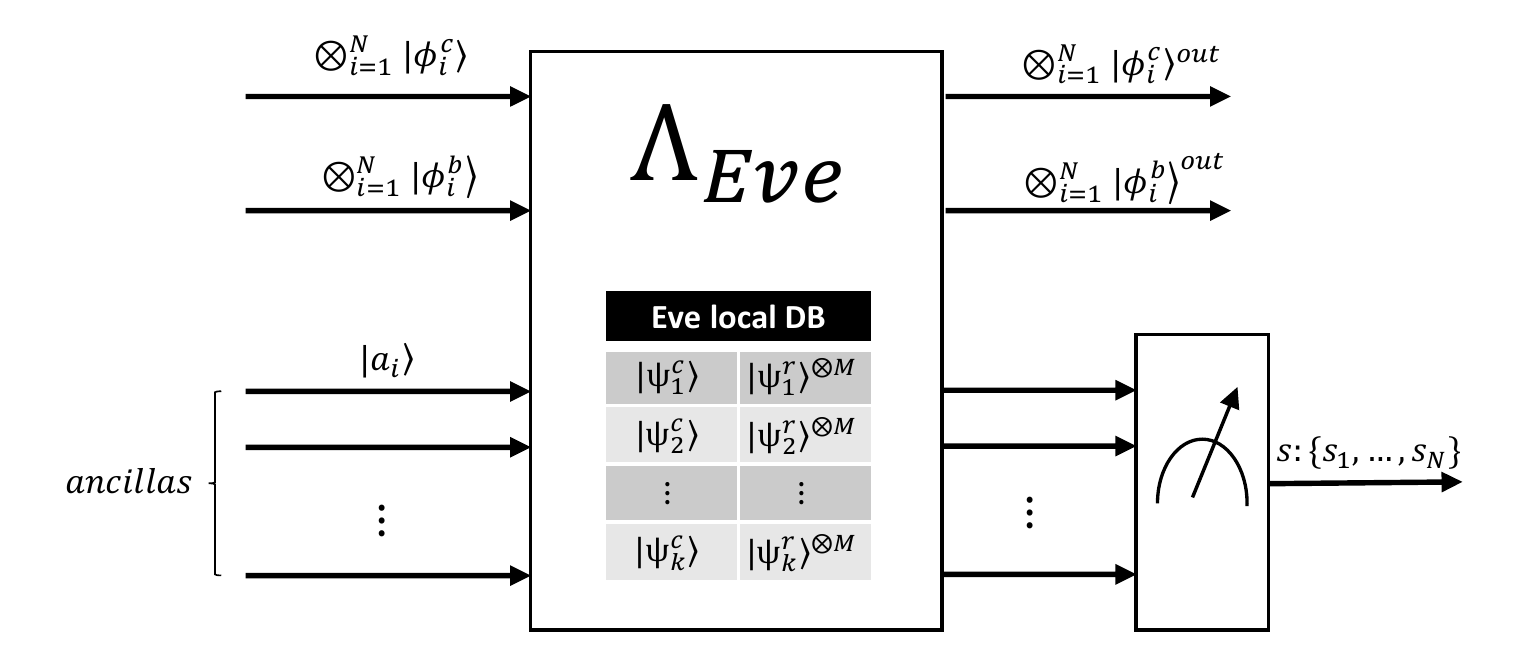}
\centering
\caption[Quantum coherent attack strategy on {\lrv}]{Quantum coherent attack strategy performed by Eve on {\lrv} protocol by applying the general local-database-dependent, CPTP map on the combined $N$ challenges and response states $\kc$ and $\kb$ respectively. The output is the $N$ bit string $s: \{ s_1,\cdots,s_N\}$.}
\label{fig:application-quantm-adv-coh}
\end{figure}

\noindent\textbf{2. Coherent Strategy:} The collective strategy is restricted to Eve applying individual unentangled maps in each round. A more generalised strategy, the coherent strategy, involves applying a CPTP map collectively on all the rounds thus potentially leveraging entanglement capabilities across rounds. Such a strategy takes as input the $N$ challenge states $\otimes_{i=1}^{N}\kc$, the $N$ response state $\otimes_{i=1}^{N}\kb$ and the ancilla qubits, and outputs a $N$ bit string $S_N$ which is sent to verifier for verification.  \figref{fig:application-quantm-adv-coh} depicts this strategy.  Eve's objective is to produce the $S_N$ which maximises the \texttt{cVer} passing probability. We denote Eve's quantum map to be,
\begin{equation}
    \Lambda_{Eve} \equiv \Lambda^{N}
\end{equation}

We say that the {\lrv} protocol is secure against any QPT Eve who performs the map $\Lambda^N$ if the resulting success probability of correctly guessing the $b$ value for all the $N$ positions is negligibly small in the security parameter.

\begin{thmbox}
\begin{theorem}[\textbf{Security against coherent attack}]\label{th:application-lrv-cohattack} The success probability of any QPT adversary Eve, in correctly guessing the $b$ values for all the $N$ positions, denoted by $[b_1,\cdots,b_N]$ is,
{\normalfont 
\begin{equation}
    \text{Pr}[\textbf{b} \leftarrow \Lambda^N(\ket{\phi^{\textbf{c}}},\ket{\phi^{\textbf{b}}})] \leqslant \Big(\frac{1}{2} + \mathcal{O}(2^{-d})\Big)^N 
\end{equation}
}
where $\textbf{b}:[b_1,\cdots,b_N]$ are the bits corresponding to correct $b$ values, $\ket{\phi^{\textbf{c}}} = \otimes_{i=1}^{N}\kc$, $\ket{\phi^{\textbf{b}}} = \otimes_{i=1}^{N}\kb$ and \normalfont{$d = \mathcal{O}(\text{poly} \log D)$} is the size of Eve's database.
\end{theorem}
\end{thmbox}

\begin{proof}
To prove this theorem, we notice that Eve applies a generalised map $\Lambda^N$ on the challenge and the response states of verifer to be able to correctly distinguish whether the response states are $\kb = \kr$ for all $i \in [N]$. Thus the probability to correctly guess $\textbf{b}$ reduces to Eve applying a CPTP map $\Lambda^N$ to perform an optimal test to distinguish the response state $\ket{\phi^{\textbf{b}}}$ with her reference  state $\rho_e^N$, where $\rho_e^N$ is the generalised entangled state. Thus without loss of generality, any attack map $\Lambda^N$, consists of two parts. The first part uses a generator algorithm $gen_N$ to generate a reference state $\rho_e^N$ by using the local database and the input challenge state $\ket{\phi^{\textbf{c}}}$, and the second part performs a test algorithm $\T$ on $\ket{\phi^{\textbf{b}}}$ and $\rho_e^N$,
\begin{equation}
    \Lambda^N = \T(\ket{\phi^{\textbf{b}}}, \rho_e^N \leftarrow gen(DB,\ket{\phi^{\textbf{c}}}))
\end{equation}
where $DB$ is the local database of Eve generated in the  \emph{setup phase}. Similar to the collective strategy proof, we assume Eve's testing algorithm $\T$ is the optimal test equality test algorithm, also referred as ideal test algorithm in  \defref{def:application-ideal-test}, \emph{i.e.} $\T = \Ti$. Here $\Ti$ again relates to the fidelity distance between the two states,
\begin{equation}
     \text{Pr}[1 \leftarrow \Lambda^N(\ket{\phi^{\textbf{c}}},\ket{\phi^{\textbf{b}}})] = \text{Pr}[1 \leftarrow \Ti] = F(\ket{\phi^{\textbf{b}}}, \rho_e^N) 
\end{equation}
Since each $b$ across the $N$ positions are chosen independently and randomly, this implies at entangling the map across different rounds does not help Eve in any way. Thus to correctly guess the $b$ values for all the $N$ positions, the optimal attack strategy of Eve is to generate the reference state $\rho_{max}^{\otimes N}$, such that,
\begin{equation}
    \forall{i \in [N]} \quad F(\rho^{\text{max}}, \kr) = \bra{\phi^b_i}\rho^{\text{max}}\kb \geqslant \bra{\phi^b_i}\rho_{i}\kb \hspace{2mm}  
\end{equation}
where $\rho_{i} = \text{Tr}_{\{1\cdots N/i\}}(\rho_e^N)$, \emph{i.e.} $\rho_{i}$ is obtained by tracing out the N-1 instances $\{1,\cdots N/i \}$. 

This further implies that attack map $\Lambda^N$ is reduced to $\Lambda_{ind}^{\otimes N}$, where the map $\Lambda_{ind}^{\otimes N}$ involves a generator algorithm that produces the state $\rho^{max}$ which maximises the average fidelity with verifier's response state across all the $N$ rounds. This implies that,
\begin{equation}
    \begin{split}
        \text{Pr}[\{b_1,\cdots,b_N\} \leftarrow \Lambda^N(\ket{\phi^{\textbf{c}}},\ket{\phi^{\textbf{b}}})] &=  \prod_{i=1}^{N}\text{Pr}[b_i \leftarrow \Lambda_{ind}(\kc,\kb)] \\
        &\leqslant \Big(\frac{1}{2} + \negl(\lambda)\Big)^N 
    \end{split}
\end{equation}
where we used the result of \thmref{th:application-lrv-colattack} after the reduction from coherent to the collective attack. This completes the proof.
\end{proof}

\noindent\textbf{(III) Comparing Classical and Quantum Strategies:}
Using the above \thmref{th:application-lrv-colattack} and \thmref{th:application-lrv-cohattack} we show that a QPT Eve does not have any non-negligible advantage in passing the \texttt{cVer} verification test compared to the purely classical Eve. Thus, we can bound the success probability of a general QPT Eve which the success probability of the classical Eve from the \thmref{th:application-cver-cl-attack},
\begin{equation}
\text{Pr}[\text{Ver accept}_{\text{QPT Eve}}]  \leqslant \text{Pr}[\text{Ver accept}_{\text{Classical Eve}}] + \mathcal{O}(2^{-N}) \approx \mathcal{O}(2^{-N})
\end{equation}

\subsection{Generalisation of low-resource protocol to arbitrary distribution of traps}\label{sec:application-lrv-general}
In the original {\lrv} protocol, verifier randomly picks half of the $N/2$ positions, and marks them $b = 1$. The rest is marked $b = 0$.  Here, even though an adversary Eve does not know the locations of valid qPUF response states and the trap states, she knows that half of the positions are traps. In this section, we generalise the {\lrv} protocol, to further hide the number of traps from Eve. This is done with the hope that hiding the number of trap and good response states could further decrease the probability of Eve passing the \texttt{cVer} test, especially against a fully classical Eve who only uses the statistics information to attack the protocol. Here verifier chooses an arbitrary number of trap positions. In other words, she randomly pics a value $p \in [0,1]$, then randomly picks $pN$ locations out of $N$ and marks them $b = 1$ (valid response states). The rest of $(1-p)N$ positions are assigned $b = 0$ (trap positions). 
One can observe that the protocol on the prover's side does not depend on this value $p$ hence verifier is not required to make the $p$ value public. 
We note that $b = 1$ positions must all have bits valued 0, and $b = 0$ positions must have half bits valued 0 and the rest are valued 1 (assuming $\delta_{er} = 0$ for simplicity) if the $N$ bits have to pass the classical verification algorithm \texttt{cVer}. Now, upon running the {\lrv} protocol, there are in total $N(1+p)/2$ number of  0 bits and $N(1-p)/2$ number of `1' bits in the desired bit-string $S_N$ which can pass the verification. 
Changing the tolerance value $\delta_{er}$ will not affect the result as we have seen in the previous section that by having a $\delta_{er}$ much smaller than $N$ the probability only multiplies to a constant factor.
We follow the same argument as in the proof of \thmref{th:application-cver-cl-attack}, for finding the optimal success probability of Eve generating successful bit-strings for the new classical verification. We say that the optimal strategies are the ones where their string space consists of exactly $c_1$ bits that are 1, where here $c_1=N(1-p)/2$. For the specific case of $p=0.5$, we have proven the optimality of such strategies. Hence in this specific case, we can refer to the same proof. In the generalised setting, the $p$ value is unknown, and as a result, $c_1$ is unknown to Eve as well. Therefore the overall winning probability of Eve will depend on first guessing the correct values of $c_1$ and then the probability of such strings passing both tests. Also, we know that the probability of any strings with incorrect $c_1$ is necessarily 0, hence we can write the probability that Eve passes the verification test as follows,
\begin{equation}
\begin{split}
\text{Pr}[\text{Ver accept}_{\text{Eve}}] & = \text{Pr}[\text{guess } c_1]\times \text{Pr}[\text{Ver accept}_{\text{Eve},S_{gop}} | c_1=\frac{N(1-p)}{2}] \\
& = \text{Pr}[\text{guess } c_1]\times \frac{{N-Np\choose\frac{N-Np}{2}}}{{N\choose\frac{N-Np}{2}}}    
\end{split}
\end{equation}

Let us assume that verifer, in order to maximize the randomness over the correct choice of $c_1$, picks $p$ completely uniformly from $[0,1]$. In this case, the number of trap responses can be any number between 0 (for $p=1$) and N (for $p=0$). Consequently, $c_1 \in \{0, 1,\dots, \frac{N}{2}\}$ and if any of these values occur with equal probability, then Eve can guess $c_1$ with the following probability:
\[\text{Pr}[\text{guess } c_1] = \frac{1}{\frac{N}{2} + 1}\]

Now one can calculate the average wining probability of Eve over p:
\begin{equation}
\underset{p}{\text{Pr}[\text{Ver accept}_{\text{Eve}}]} = \int_{0}^{1} \frac{2}{N + 2}\frac{(N-Np)!(\frac{N+Np}{2})!}{N!(\frac{N-Np}{2})!} dp    
\end{equation}

Now, we approximate the above integral as follows:

\begin{thmbox}
\begin{theorem}[Average success probability convergence with arbitrary distribution of traps]\label{th:applications-generaltrap-prob-integral}
Let $p$ be the probability of choosing correct responses in \lrv~protocol. Then the average wining probability of Eve over $p$, approximately converges as follows:
\begin{equation}
\underset{p}{\text{Pr}[\text{Ver accept}_{\text{Eve}}]} \approx \overline{\text{Pr}_{win}} = \frac{2}{N + 2} \sum^{N}_{k=0} \frac{(N-k)!(\frac{N+k}{2})!}{N!(\frac{N-k}{2})!} \approx \frac{6}{N(N+2)} = \mathcal{O}(\frac{1}{N^2})    
\end{equation}
\end{theorem}
\end{thmbox}
\begin{proof}
We approximate the following integral for the average probability that Eve wins the classical verification by performing the optimal classical strategy when $p$ is chosen to be a uniform distribution.
\[\underset{p}{\text{Pr}[\text{Ver accept}_{\text{Eve}}]} = \int_{0}^{1} \frac{2}{N + 2}\frac{(N-Np)!(\frac{N+Np}{2})!}{N!(\frac{N-Np}{2})!} dp\]

We choose $NP=k$ thus we have $Ndp=dk$ and we can rewrite the integral as:
\[\underset{p}{\text{Pr}[\text{Ver accept}_{\text{Eve}}]} =  \frac{2}{N(N + 2)}\int_{0}^{N}\frac{(N-k)!(\frac{N+k}{2})!}{N!(\frac{N-k}{2})!} dk\]

Now we can approximate the integral for discrete $k \in \{0,1,\dots,N\}$. Hence we have:
\[\underset{p}{\text{Pr}[\text{Ver accept}_{\text{Eve}}]} \approx \overline{\text{Pr}[\text{Ver accept}_{\text{Eve}}]} = \frac{2}{N(N + 2)} \sum^{N}_{k=0} \frac{(N-k)!(\frac{N+k}{2})!}{N!(\frac{N-k}{2})!}\]

The above series can be opened further as:

\begin{equation}
\begin{split}
    \sum^{N}_{k=0} \frac{(N-k)!(\frac{N+k}{2})!}{N!(\frac{N-k}{2})!} & = 1 + \frac{(N-1)!}{N!}\times\frac{(\frac{N}{2}+\frac{1}{2})!}{(\frac{N}{2}-\frac{1}{2})!} + \frac{(N-2)!}{N!}\times\frac{(\frac{N}{2}+1)!}{(\frac{N}{2}-1)!} + \dots + 1 \\
    & = 1 + \frac{1}{N}\times\frac{(\frac{N}{2} + \frac{1}{2})\cancel{(\frac{N}{2} - \frac{1}{2})!}}{\cancel{(\frac{N}{2} - \frac{1}{2})!}} + \frac{1}{N(N-1)}\times\frac{(\frac{N}{2}+1)(\frac{N}{2})\cancel{(\frac{N}{2}-1)!}}{\cancel{(\frac{N}{2}-1)!}} + \dots + 1 \\
    & \underset{N \gg 1}{\approx} 2 + \frac{\frac{N}{2}}{N} + \frac{(\frac{N}{2})^2}{N^2} + \frac{(\frac{N}{2})^3}{N^3} + \dots \\
    & = 2 + \sum^{N-1}_{i=1}(\frac{1}{2})^i \approx 2 + (1 - 2^{1-N}) \approx 3 
\end{split}
\end{equation}

where the sum has been approximated for large $N$. Thus we can write the average probability in the limit of large $N$ as follows,
\begin{equation}
\underset{p}{\text{Pr}[\text{Ver accept}_{\text{Eve}}]} \approx \overline{\text{Pr}[\text{Ver accept}_{\text{Eve}}]} = \frac{6}{N(N+2)}    
\end{equation}
This concludes the proof.
\end{proof}

This means that by choosing $p$ form a uniform distribution, the average success probability of the adversary becomes polynomially small in $N$ which reduces the security of the protocol to polynomial. This may seem a surprising result although the reason is that the probability function for $p=0$ and $p=1$ is 1.
On the other hand, from the security result for $p=\frac{1}{2}$, we know that the probability function's behaviour can be inverse exponential. This gives rise to the interesting question of whether one can find a boundary for $p$ in which $\text{Pr}[\text{Ver accept}_{\text{Eve}}]$ is negligible. Before addressing this problem, it is worth mentioning that by hiding $p$, one can hope the protocol's security to be boosted by at most a polynomial factor ($\frac{1}{\mathcal{O}(N)}$) as Eve's probability of guessing the correct $c_1$ depends only on the different number of 1's in the string that results from different choices of $p$. Even though for large $N$ this polynomial factor can be ignored, assuming that the verifier has a good choice of $p$ which leads to exponential security, in relatively smaller $N$ the hiding can practically boost the security of the identification.

Now to be able to analyse the $\text{Pr}[\text{Ver accept}_{\text{Eve}}]$, we rewrite the factorials with Gamma function and we define $z=\frac{N-Np}{2}$ where $z \in \{0,1,\dots,\frac{N}{2}\}$. Considering that $\Gamma(z+1)=z\Gamma(z)$, the probability is,
\begin{equation}
\begin{split}
\text{Pr}[\text{Ver accept}_{\text{Eve}}] & = \frac{(N-Np)!(\frac{N+Np}{2})!}{N!(\frac{N-Np}{2})!} = \frac{\Gamma(2z+1)\Gamma(N-z+1)}{N!\Gamma(z+1)} \\
& = \frac{2}{N!}\times\frac{\Gamma(2z)\Gamma(N-z+1)}{\Gamma(z)}
\end{split}
\end{equation}

Using properties of Gamma functions we have that $\frac{\Gamma(2z)}{\Gamma(z)} = \frac{2^{2z-1}}{\sqrt{\pi}}\Gamma(z+\frac{1}{2})$. Thus we can simplify the function to be:
\begin{equation}
\begin{split}
\text{Pr}[\text{Ver accept}_{\text{Eve}}]&  = \frac{2}{\sqrt{\pi}}\times\frac{2^{2z-1}}{N!}\Gamma(z+\frac{1}{2})\Gamma(N-z+1) \\
& \approx \frac{2^{2z-1}}{N!}\Gamma(z+\frac{1}{2})\Gamma(N-z+1)  
\end{split}
\end{equation}

\begin{figure}[t]
    \includegraphics[width=1\textwidth]{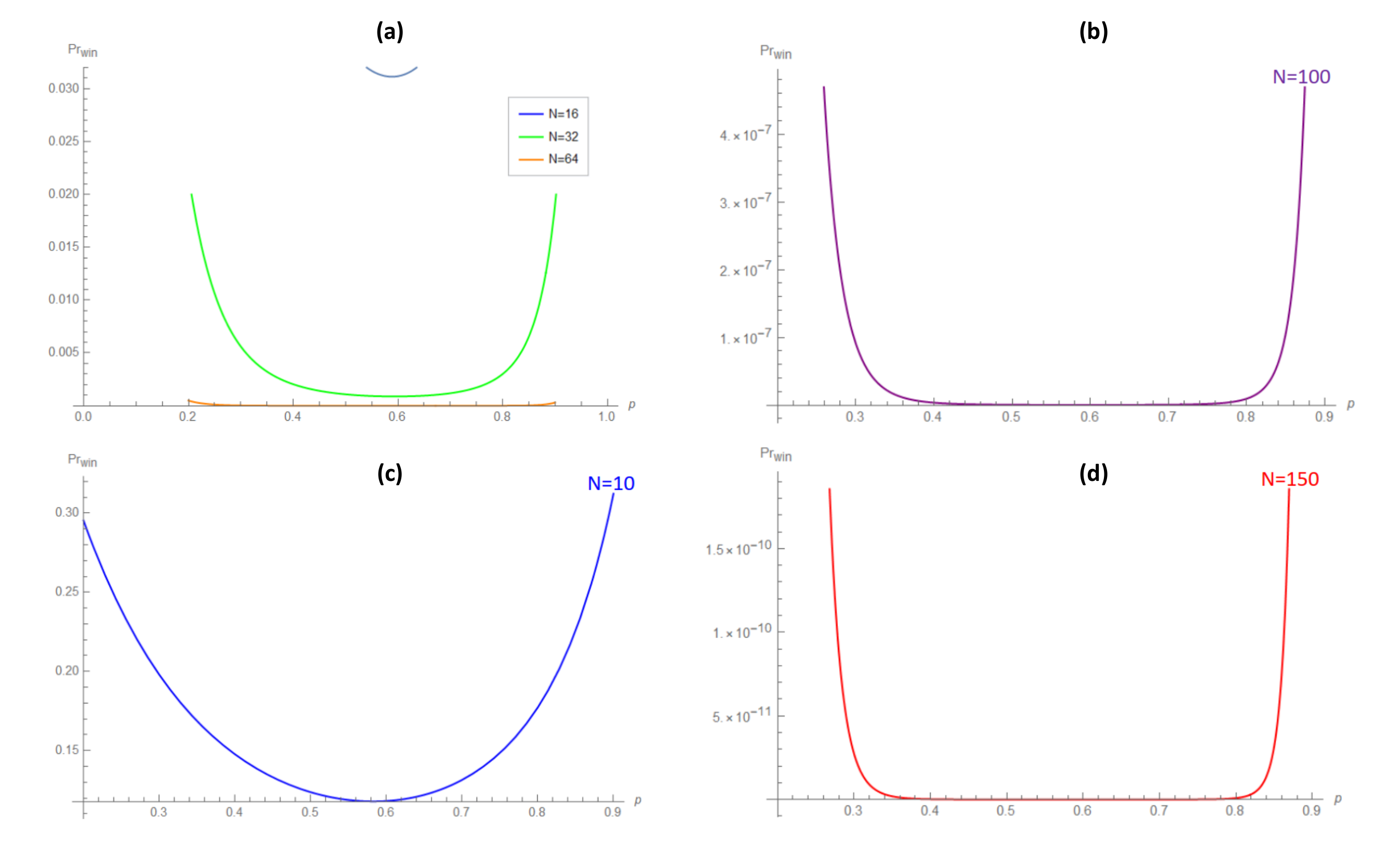}
    \caption{Behaviour of Eve's success probability $\text{Pr}[\text{Ver accept}_{\text{Eve}}]$ as a function of $p$ (corresponding to number of valid qPUF responses), for different values of $N$.}\label{fig:application-prob-generalp-n}
\end{figure}

For a large enough fixed $N$, the factor $\frac{2^{2z-1}}{N!} \ll 1$. However it is an increasing function in $z$ and $\Gamma(z+\frac{1}{2})\Gamma(N-z+1)$ is a large factor which quickly decreases with $z$. Also at the beginning and the end of the period where $z=0, z=\frac{N}{2}$, the probability is 1, and it reduces to a small value for certain $z$. Thus we deduce that the function will necessarily have a minimum for any $N$. The \figref{fig:application-prob-generalp-n}, different $\text{Pr}[\text{Ver accept}_{\text{Eve}}]$ for different $N$ has been shown. We have renormalised the probabilities as a function of $p$ to be able to compare them. As we can see, the function for all the different values of $N$ falls exponentially in a minimum region where there are the desirable values of $p$. As $N$ grows, the range of desirable $p$ expands, which can be found in the top right plot where we compare the probability for $N=16, N=32$ and $N=64$. Also by comparing the probability range for $N=10, N=100, N=150$ one can see how the exponential security is achieved for a $p$ which has been chosen in the \emph{good} region. This specification of the success probability would be useful for the verifier to be able to optimise the protocol based on her resources. Moreover, the freedom of choosing traps according to desired distribution, conditioning that it bounds the value of $p$ to the minimum region, enables the protocol to be useful in other scenarios.

\subsection{Resource comparison of protocols}\label{sec:application-resource-comparison}
The two proposed qPUF-based identification protocols differ a great deal in terms of the type and amount of resources available to the concerned parties. We divide the resources into three categories: quantum memory, quantum computing ability, and the number of communication rounds required to achieve identification. Here, quantum memory is quantified by the number of quantum states stored in a register, and the computing ability resource is quantified in terms of the number of quantum gates required to implement a specific quantum circuit.

\begin{table}[h!]
\centering
\resizebox{\textwidth}{!}{
\begin{tabular}{|c|c|c|c|c|c|c|c|c|}
\hline
Protocol & \multicolumn{2}{l|}{Security} & \multicolumn{2}{l|}{Quantum Memory} & \multicolumn{2}{l|}{Verification computing ability} & \multicolumn{2}{l|}{Communication round} \\ \hline
  &                    &  & Verifier & Prover & Verifier & Prover & Quantum & Classical \\ \hline
{hrv-id-swap} & \multirow{3}{*}{$\epsilon$} &  = $2^{-MN}$ & $\log 1/\epsilon$      & 0     & $poly \log D$       &   0   & $\log 1/\epsilon$      & 0        \\ \cline{1-1} \cline{3-9} 
{hrv-id-gswap} &                    &  = $(M+1)^{-N}$ & $\frac{M}{\log M+1}\log 1/\epsilon$       & 0     & $poly \log MD$       & 0     & $\frac{1}{\log M+1}\log 1/\epsilon$      & 0        \\ \cline{1-1} \cline{3-9} 
{lrv-id} &                    & = $2^{-N}$ & $\log 1/\epsilon$       & 0     & 0       & $poly \log D$     & $\log 1/\epsilon$      & 1        \\ \hline
\end{tabular}}
\caption{Comparison of different qPUF-based identification protocols in terms of security ($\text{Pr}[\text{Ver accept}_{\text{Eve}}] = \epsilon$) against any QPT adversary and the three resource categories of the verifier and the prover: quantum memory, computing ability and number of communication rounds. Here all the resources are in $\mathcal{O}(.)$. All our proposed protocols exhibit $\epsilon$  exponential security with polynomial sized resource $\mathcal{O}(\log 1/\epsilon)$ memory/communication and $\mathcal{O}(poly \log D)$ computing ability in both the parties. Here $D$ is the size of qPUF.}
\label{table:comp}
\end{table}

\tableref{table:comp} compares the resources of the two protocols that we have introduced. For a fair comparison between the above protocols, we fix the maximum acceptance probability for any QPT adversary, $\text{Pr}[\text{Ver accept}_{\text{Eve}}]$, to be $\epsilon$, and compute the number of resources required to achieve that desired acceptance probability. In all the protocols, we assume that during one identification, $N$ copies of different states, each with $M$ identical copies are used. For the specific case of {\lrv} protocol, $M = 1$. For the {\hrvs} protocol, where the quantum verification is via the SWAP test circuit, the adversary's acceptance probability is $\epsilon = \mathcal{O}(2^{-MN})$. In this protocol, the verifier requires $MN = \mathcal{O}(\log 1/\epsilon)$ size quantum memory and computing ability of $\mathcal{O}(poly \log D)$ quantum gates, where $D$ is the size of qPUF. The prover, on the other hand, requires no quantum memory and computing ability. The number of communication rounds required to achieve the desired security is $MN = \mathcal{O}(\log 1/\epsilon)$. The protocol {\hrvg}, where the quantum verification is via the GSWAP test circuit, the adversary's acceptance probability is $\epsilon = \mathcal{O}((M+1)^{-N})$. In this protocol, the verifier requires $MN = \mathcal{O}(\frac{M}{\log M+1} \log 1/\epsilon)$ size quantum memory and a computing ability of $\mathcal{O}(poly \log MD)$ quantum gates. Similar to {\hrvs}, the prover requires no quantum memory and computing ability. The number of communication rounds required to achieve the desired security is $N = \mathcal{O}(\frac{1}{\log M+1} \log 1/\epsilon)$. Thus for large $M$ values, the verifier's quantum memory requirement is less while using SWAP compared to GSWAP, but the number of communication rounds is higher using the SWAP test.

\begin{figure}[t]
\includegraphics[width=0.98\textwidth]{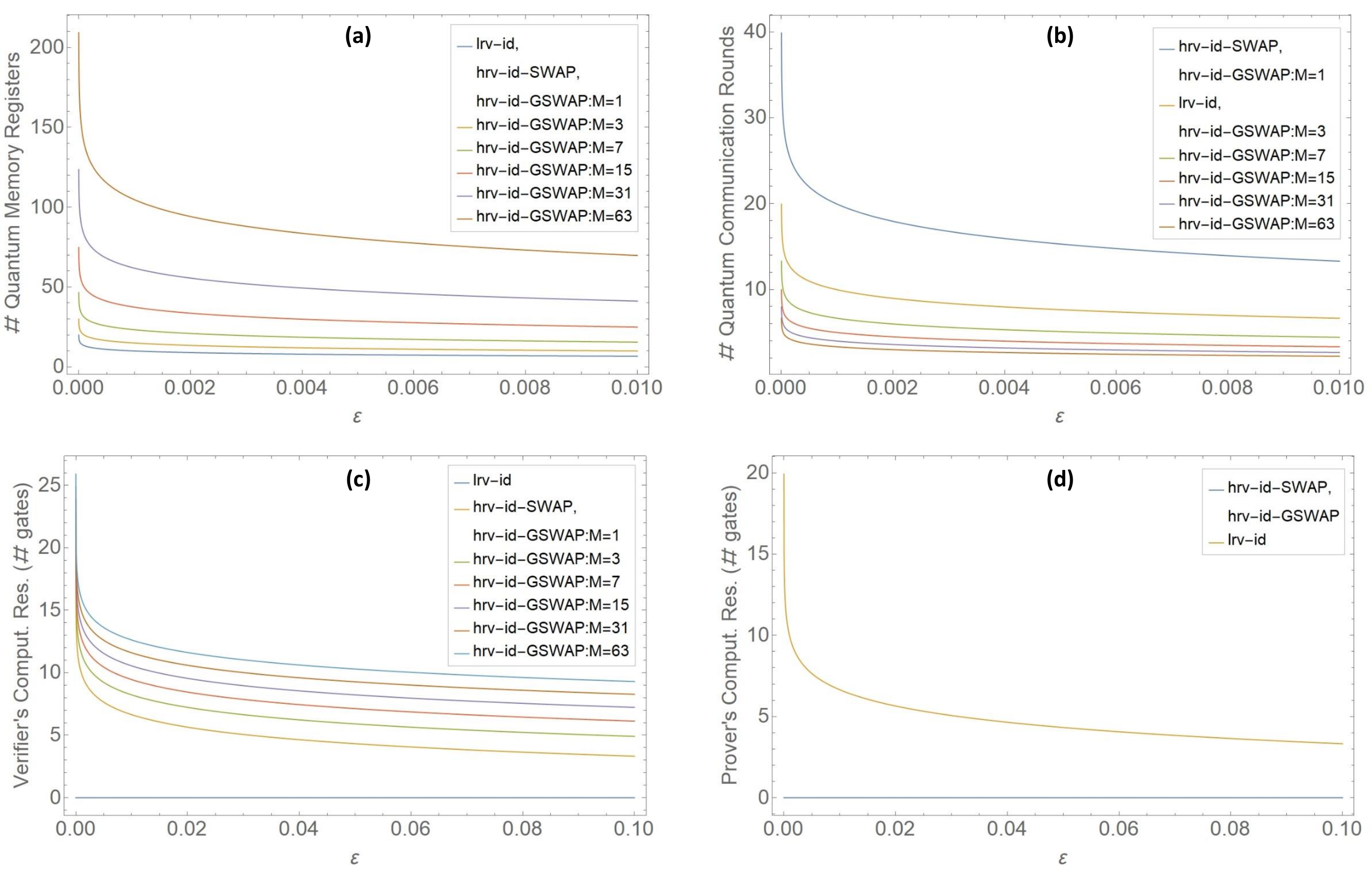}
\centering
\caption[Comparison of the resources required by the prover and verifier in the three qPUF-based identification protocols (\hrvs, \hrvg, and \lrv)]{Comparison of the resources required by the prover and verifier in the three qPUF-based identification protocols (\hrvs, \hrvg, and \lrv) for varying security values $\epsilon$. We choose the $\epsilon$ to range from 
$10^{-6}$ and $10^{-2}$ for the top row and between $10^{-6}$ and $10^{-1}$ for the bottom row. 
Plot~top left compares the verifier's quantum memory resource vs $\epsilon$ for the three protocols. The plot shows that the least memory requirement is minimum in {\hrvs} and {\lrv} protocols while it increases by increasing the number of local copies $M$ required in the GSWAP test for {\hrvg} protocol. We note that the prover's memory requirement is 0 in all three protocols. Plot~top right similarly compares the number of quantum communication rounds in the three protocols. The requirement is minimum in the {\lrv} while it increases with $M$ in the {\hrvg}. The communication round in {\hrv} is double compared to the {\lrv} requirement to indicate the two-way quantum communication instead of one way in the latter. 
Plots~bottom left and bottom right compares the computational resource vs $\epsilon$ for the verifier and prover respectively. Here we have taken $D = 1/\epsilon$ for comparison. 
}  
\label{fig:all-resources}
\end{figure}

Now for the {\lrv} protocol, the protocol with the low-resource verifier, the adversary's acceptance probability is $\epsilon = \mathcal{O}(2^{-N})$. In this protocol, the verifier requires $N = \mathcal{O}(\log 1/\epsilon)$ size quantum memory. Since the verifier performs classical verification, hence she does not require a quantum computing ability. The prover here requires no quantum memory but since he performs the SWAP test circuit, his computing ability is required to be $\mathcal{O}(poly \log D)$. The number of quantum communication rounds required to achieve the desired security is $N = \mathcal{O}(\log 1/\epsilon)$. This protocol also requires a single round of classical communication transmitting $N$ bits.

\figref{fig:all-resources} demonstrates the graphical comparison of different resources among the three qPUF-based identification protocols. The plots show a tradeoff in resources between different protocols to achieve the desired success probability of $\epsilon$. We choose the $\epsilon$ to range from $10^{-6}$ to $10^{-1}$. Since the computing ability resource depends on the qPUF size $D$, we choose $D = 1/\epsilon$ for comparison. 

\begin{figure}[t]
\includegraphics[width=0.98\textwidth]{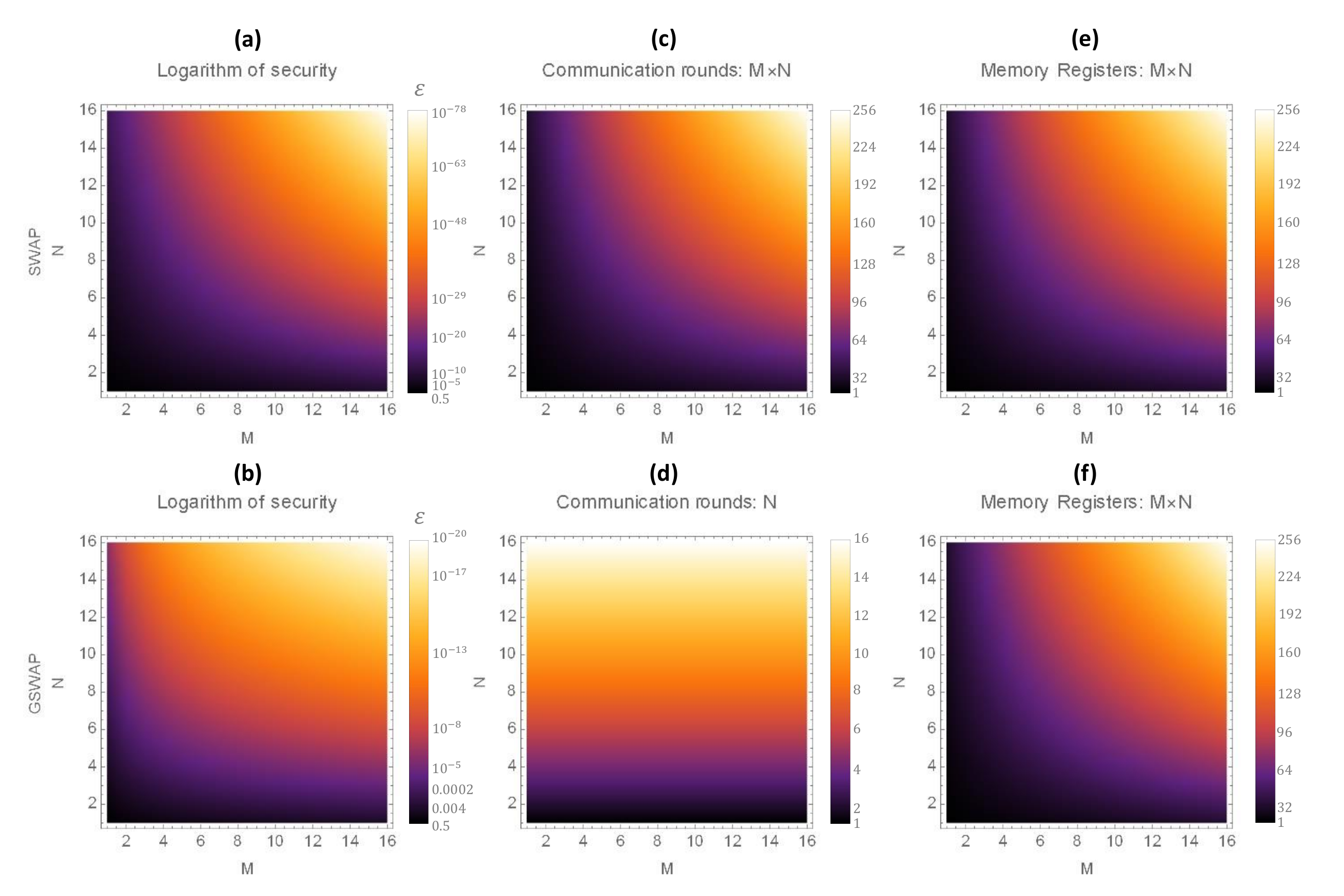}
\centering
\caption[Comparison of verification based on SWAP and GSWAP for identification protocols]{Comparison of verification based on SWAP and GSWAP for identification protocols. The top row is associated with SWAP and the button row with GSWAP. The x-axis of the plots are all $M$ (the number of local copies) and the y-axes are all $N$ (the number of different states) and the security, quantum memory and quantum communications have been shown with a colour spectrum. The left column shows the security $\epsilon$ where we have $\epsilon = 2^{-MN}$ for SWAP and $\epsilon = (M+1)^{-N}$ for GSWAP, in a logarithmic scale for more visibility. The middle column shows the required communication where we see that for GSWAP the communication rounds are independent of $M$ and only linearly growing with $N$ while for SWAP the communication rounds grow also linear by increasing the number of local copies. The right column shows the memory which has been fixed for both SWAP and GSWAP to $M\times N$. The comparison between security and communication plots shows a trade-off between SWAP and GSWAP as the quantum verification algorithm.}  
\label{fig:swap-gswap-density}
\end{figure}

We identify that the difference in resources primarily comes about due to the different requirements of SWAP and GSWAP tests. To illustrate this graphically, we provide density plots in
\figref{fig:swap-gswap-density} to showcase the trade-off between the success probability $\epsilon$ and the memory and communication round resources required for different $M$ ad $N$'s for protocols based on SWAP vs GSWAP tests. 

\section{Towards more efficient qPUF-based identification protocols}\label{sec:application-efficent-qpufid-prs}
In this section, we show that using our results form \chapref{chap:pr-connection}, we can make the protocols we presented in the previous section yet more efficient. We have seen how these identification protocols exploit the unforgeability of qPUFs, to achieve exponential security against QPT adversaries in a polynomial number of rounds. We have already discussed that these protocols are resource-efficient in many aspects, but one of the main practical challenges in implementing them is to sample the challenge states at random from the Haar measure. We have seen that this requirement is crucial for achieving unforgeability for qPUFs. Nevertheless, \thmref{th:connection-efficientuu-prs} showed that universal unforgeability can still be achieved with the same security guarantee if the states are sampled from a PRS family instead of Haar-random states. Here, we show that the qPUF-based identification protocols can also achieve exponential security using PRS. This transition to a more efficient sampling of challenge states brings us one step closer to the practical implementations of quantum identification protocols with exponential security against powerful quantum adversaries and leads to promising solutions to the problem of untrusted manufacturers.

We start with \hrv, and we introduce a computationally efficient variation of this protocol which we call \emph{Efficient-hrv-id} protocol, by replacing the Haar-random challenges with pseudorandom quantum states in the setup phase as follows:

\begin{protocol}[Efficient-hrv-id] Efficient version of \hrv~protocol using pseudorandom challenges:
\vspace{3mm}\hrule
\begin{enumerate}
    \item \emph{Setup Phase}:
            \begin{enumerate}
                \item Verifier has the qPUF device with unitary evaluation $U$.
                \item Verifier has also access to a family of PRS $\{\ket{\phi_{k}}\in S(\Hild)\}_{k\in\K}$ and randomly picks $Q \in \mathcal{O}(\text{poly} \log d)$ of them as the challenge states.
                \item Verifier queries the $U$ individually with each challenge $\ket{\phi_{k}}$ a total of $M$ number of times to obtain $M$ copies of the response state $\ket{\phi^r_{k}}$ and stores them in their local database $S$. 
                \item The verifier transfers the $U$ to Prover.
            \end{enumerate}
    \item \emph{Identification phase}: same as \hrv.
    \item \emph{Verification phase}: same as \hrv.
\end{enumerate}
\hrule\vspace{3mm}
\end{protocol}
The following statement which is a corollary of the previous results shows that the \emph{Efficient-hrv-id} protocol is also exponentially secure against QPT adversaries with the same security bounds.

\begin{corrbox}
\begin{corollary}\label{th:connection-eff-hrv-sound} Let $U$ be a UqPUF over $\HilD$. The success probability of any QPT adversary to pass the {\normalfont SWAP}-test or {\normalfont GSWAP}-test verification of the \emph{Efficient-hrv-id} is at most $\epsilon$, given that there are $N$ different CRPs, each with $M$ copies. The $\epsilon$ is bounded as follows for each verification:
\begin{equation}
         \text{Pr}[\text{Ver accept}_{\A}] \leqslant \epsilon \quad \quad \epsilon_{\text{SWAP}} \approx \mathcal{O}(\frac{1}{2^{NM}}) \quad \quad \epsilon_{\text{GSWAP}} \approx \mathcal{O}\big(\frac{1}{(M+1)^{N}}\big)
\end{equation}
\end{corollary}
\end{corrbox}
\begin{proof}
First, \thmref{th:connection-efficientuu-prs} that states any unknown unitary satisfies efficient universal unforgeability where the challenge states are selected from a PRS family. Then we can directly use the result states in \thmref{th:application-hrv-swap-sound} and \thmref{th:application-hrv-gswap-sound} using the SWAP and GSWAP test which shows the same security bound in the number of rounds and copies of challenge-response pairs.
\end{proof}

Similarly, we can improve the efficiency of the second protocol, namely \lrv~ by substituting the Haar random states with PRS as follows:

\begin{protocol}[Efficient-lrv-id] Efficient version of \lrv~protocol using pseudorandom challenges
\vspace{3mm}\hrule
\begin{enumerate}
    \item \emph{Efficient-lrv-id Setup Phase}:
            \begin{enumerate}
                \item Verifier has the qPUF device with unitary evaluation $U$.
                \item Verifier has also access to a family of PRS $\{\ket{\phi_{k}}\in S(\Hild)\}_{k\in\K}$ and randomly picks $Q \in \mathcal{O}(\text{poly} \log d)$ of them as the challenge states.
                \item Verifier queries the $U$ individually with each challenge $\ket{\phi_{k}}$ a total of $M$ number of times to obtain $M$ copies of the response state $\ket{\phi^r_{k}}$ and stores them in their local database $S$. 
                \item Verifier selects states $\ket{\phi^{\perp}}$ orthogonal to the selected challenge's subspace and queries the $U$ with them to obtain the trap states labelled as $\ket{\phi^{\text{trap}}}$. The unitary property ensures that $\langle \phi^{\text{trap}}|\phi_{k}^r\rangle = 0$.
                \item The verifier transfers the $U$ to Prover.
            \end{enumerate}
    \item \emph{Identification phase}: same as \lrv.
    \item \emph{Verification phase}: same as \lrv.
\end{enumerate}
\hrule\vspace{3mm}
\end{protocol}

Again using the proof techniques presented in the previous sections, and our results from \chapref{chap:pr-connection}, we show that the \emph{Efficient lrv-id protocol} satisfies exponential security against QPT adversary both under the coherent and collective attack models.

\begin{corrbox}
\begin{corollary}\label{th:application-eff-lrv-sound} Let $U$ be a UqPUF over $\HilD$. The success probability of a QPT adversary $\A$ to pass the verification of the \emph{Efficient lrv-id protocol} is at most $\epsilon$, in $N$ rounds. The $\epsilon$ is bounded as follows:
\begin{equation}
         \text{Pr}[\text{Ver accept}_{\A}] \leqslant \epsilon \quad \quad \epsilon \approx \mathcal{O}(\frac{1}{2^{N}})
\end{equation}
\end{corollary}
\end{corrbox}
\begin{proof}
First, we specify that we can directly use the result of \thmref{th:application-cver-cl-attack} which bounds the success probability of a classical adversary in passing the classical verification algorithm. Then the success probability against a quantum adversary with the collective and coherent attack is defined as the advantage of the quantum adversary over that classical adversary in guessing the trap states, using all the side information obtained from the $U$ in the learning phase. 
We use \thmref{th:connection-efficientuu-prs} that states any unknown unitary satisfy efficient universal unforgeability with PRS challenge states. Next, the conditions of \thmref{th:application-lrv-colattack} and \thmref{th:application-lrv-cohattack} are satisfied and we can directly use those results which gives the following bounds.
\begin{equation}
    \text{Pr}[b \leftarrow \Lambda_{\A}] \leqslant \frac{1}{2} + \mathcal{O}(2^{-N})
\end{equation}
Where $\Lambda_{\A}$ denotes any map that $\A$ uses to distinguish the traps states.
Finally, putting all the above results together we have
\begin{equation}
\text{Pr}[\text{Ver accept}_{\A}]  \leqslant \epsilon = \text{Pr}[\text{Ver accept}_{\text{Classical Adv}}] + \mathcal{O}(2^{-N}) \approx \mathcal{O}(2^{-N}) 
\end{equation}
This concludes the soundness proof of \emph{Efficient lrv-id protocol}.
\end{proof}

\section{Hybrid PUF: A practical solution}\label{sec:application-hpuf-practical}
As our last contribution towards the practical realisation of qPUF-based applications, we propose a more implementation-friendly construction called Hybrid PUF (HPUF). This new type of PUF, as opposed to quantum PUFs which exploit quantum randomness, uses a classical PUF as a weak source of randomness and enhances it using quantum communication, hence the name \emph{Hybrid PUF}.

Let us recall the main implementation challenges of qPUF-based identification protocols and see how our proposal for a hybrid construction can overcome these challenges. The first challenge is the implementation of the UqPUF itself, which requires either sampling unitaries from Haar-measure or equivalently a family of PRU or UU. As mentioned in \chapref{chap:qpuf}, one of the promising candidates for qPUFs is optical devices. Nevertheless, in using them as qPUFs the main challenge will be to certify the unknownness property and dimension of the unitary since these parameters are directly related to the security of qPUFs. On the other hand, the literature on classical PUFs is rich, and there is a multitude of constructions available based on several different hardware technologies \cite{gassend_silicon_2002,guajardo_fpga_2007,kim_cspan_2018,maes_physically_2013}. Although all of those constructions can be manufactured quite easily and they provide unique and inexpensive hardware fingerprints, they all suffer from the lack of enough randomness and as a result, do not provide satisfactory unpredictability. Thus, most of the existing CPUF constructions are vulnerable against the machine learning modeling-based attacks \cite{guneysu_gap_2015,becker_pitfalls_2015,delvaux_machine-learning_2019,ruhrmair_modeling_2010}. In these types of attacks, the attacker first collects a sufficient number of CRPs by adaptively querying the PUF. Then the collected data is used to derive a numerical model that mimics the behaviour of the PUF, using the tools and techniques from machine learning. The central idea behind the Hybrid PUF is to use a classical PUF as an embedded hardware module that is easy to implement but does not offer suitable security, to construct a secure hardware token that uses commercially available quantum communication tools and provides sophisticated security guarantees. At the same time, we aim for a technologically available construction to overcome the manufacturing obstacle of a secure PUF that uses quantum CRPs.

The second major challenge regarding qPUF-based identification protocols becoming widely available today or in the near term is the fact that the verifier needs to store the CRP database on a quantum memory. Although there has been significant progress in the implementation of quantum memories in the recent years \cite{lvovsky_optical_2009,wang_efficient_2019,bradley_ten-qubit_2019,gyongyosi_optimizing_2020,lago-rivera_telecom-heralded_2021,bouillard_quantum_2019,wallucks_quantum_2020,dennis_topological_2002}, storing large quantum states for a considerable time is still infeasible given today's technology. The hybrid construction can solve this problem by fully removing the quantum memory requirement. Since an HPUF encodes classical responses in separable single-qubit states, verifying the response states can be done more easily using the underlying classical information from the CPUF, and as we will show, having a classical database will suffice to verify an HPUF.

Finally, through studying this construction, we will also address a long-standing open problem in the field of PUF-based identification, which is the re-usability of challenge-response pairs stored in the database. One significant drawback of the PUF-based authentication protocols is that the server cannot use the same challenge multiple times to authenticate a client due to man-in-the-middle attacks. There is no way to avoid this limitation for classical PUFs. However, in this section, we show that due to the entropic uncertainty principle of quantum information theory, with our given construction, the server can reuse a challenge as long as they have been successfully authenticated by the client using that challenge in the previous rounds, overcoming this problem and proving for the first time, the challenge re-usability of PUF-based application.

In this section, we give a construction for HPUF based on \emph{conjugate coding}, which enhances the security of classical PUFs against a weaker class of quantum adversaries (as opposed to adaptive QPT adversaries that we have been considering so far). Then by exploiting a technique known as \emph{Locking mechanism}, we boost the security of this construction to our usual adaptive QPT adversaries, therefore presenting the concept of \emph{Hybrid Locked PUF} (HLPUF) as our candidate for a secure and efficiently implementable quantum hardware token. We then show a secure HLPUF-based identification protocol. Finally, we formally prove the challenge re-usability property of this protocol.

But first, we introduce the theoretical model that we adopt for the CPUF which is going to be used as the building block of our construction. 

\subsection{CPUF model}
We have introduced classical PUFs in \chapref{chap:qpuf}. Here we provide some additional technical tools and definitions that we need to introduce our Hybrid construction. Classical PUFs are usually defined with probabilistic functions, due to their inherent physical randomness. Here we also consider them as probabilistic functions.

A classical PUF can be modelled as a probabilistic function $f:\R \times\X\rightarrow\Y$ where $\X$ is the input space, $\Y$ is the output space of $f$ and $\R$ is the identifier. As defined in \ref{sec:qpuf-classic-puf}, the creation of a classical PUF is formally expressed by invoking a manufacturing process $f\leftarrow\gen_{C}(\lambda)$, where $\lambda$ is the security parameter. Here $f$ is the evaluation algorithm of CPUF which needs to satisfy the requirements of robustness, collision-resistance and uniqueness as defined before \cite{armknecht_towards_2016}. For a fixed input $x \in \X$, and a random coin (or key) $R \leftarrow \R$, we denote the probability distribution of the output random variable $f(x) := f(R,x)$ over all $ y \in \Y$ as,
\begin{equation}
    p^f_x(y) := Pr[f(x) = y|x] = \sum_{r:f(r,x) = y} Pr[R = r].
\end{equation}

Now, let us define a $\emph{parameterised randomness}$ definition for the classical PUF $f$ as follows:

\begin{defbox}
\begin{definition}[$p$-Randomness]\label{def:p-randomness}
We define the $p$-randomness of a classical PUF $f:\R \times \X\rightarrow\Y$ as \begin{equation}
    p := \max_{\substack{x \in \X \\ y \in \Y}} p^f_x(y) = \max_{\substack{x \in \X \\ y \in \Y}} \underset{R}{Pr}[f(R,x) = y].
\end{equation}
\end{definition}
\end{defbox}

We use this definition to characterize the quality of a CPUF with a quantitative measure of its randomness. This parameter has some relation to a property of min-entropy for classical PUFs but is formally defined differently. Nevertheless, for technical purposes, we choose to use this definition.

\subsection{Construction for Hybrid PUF}\label{sec:application-hpuf}
For our construction, we start with a classical PUF with a certain amount of randomness characterized by the $p$-randomness value we defined in \defref{def:p-randomness}. We construct a new PUF that is the combination of a classical PUF, and a \emph{quantum encoder}, which encodes the output of the CPUF into non-orthogonal quantum states. The output qubits are the response of the PUF and will be sent through the quantum communication channel. We refer to the entire system, \emph{i.e.} the combination of CPUF and the quantum encoding, as \emph{Hybrid PUF (HPUF)}.  
A HPUF receives a \emph{classical} challenge and produces a \emph{quantum} response. In \constref{cons:hpuf_1} we give a simple design of a HPUF based on conjugate coding \cite{wiesner_conjugate_1983}.

\begin{constbox}
\begin{construction}[Hybrid PUF]
\label{cons:hpuf_1}
Suppose $f:\{0,1\}^n \rightarrow \{0,1\}^{4m}$ be a classical PUF, that maps an $n$-bit string $x_i \in \{0,1\}^n$ to an $4m$-bit string output $y_i \in \{0,1\}^{4m}$. We denote the $j$-th bit of $y_i$ as $y_{i,j} \in \{0,1\}$. From the $4m$-bit string, we prepare the set of $2m$-tuples $\{(y_{i,(2j-1)}, y_{i,2j})\}_{1\leq j \leq 2m}$. The hybrid PUF encodes each of the tuples $(y_{i,(2j-1)}, y_{i,2j})$ into a single qubit \emph{BB84 states}, $\ket{\psi^{i,j}}$. The exact expression of the encoding is defined in the following way,
\begin{equation}
    \ket{\psi^{i,j}_{out}}\bra{\psi^{i,j}_{out}} := 
    \begin{cases}
        \ket{0}\bra{0}~&(y_{i,(2j-1)}, y_{i,2j})= (0,0)\\
        \ket{1}\bra{1}~&(y_{i,(2j-1)}, y_{i,2j})= (1,0)\\
        \ket{+}\bra{+}~&(y_{i,(2j-1)}, y_{i,2j})= (0,1)\\
        \ket{-}\bra{-}~&(y_{i,(2j-1)}, y_{i,2j})= (1,1)
    \end{cases}
\end{equation}
For any $x_i \in \{0,1\}^n$, the mapping of the HPUF $\E_{f}:\{0,1\}^{n} \rightarrow (\Hil^{2})^{\otimes 2m}$ is defined as follows.
\begin{equation}
    \label{mqp:hpuf1}
    x_i \rightarrow \ket{\psi^i_{out}}\bra{\psi^i_{out}}~~(\text{or}~ \ket{\psi_{f(x_i)}}\bra{\psi_{f(x_i)}})
\end{equation}
where $\ket{\psi^i_{out}}\bra{\psi^i_{out}}=\bigotimes_{j=1}^{2m} \ket{\psi^{i,j}_{out}}\bra{\psi^{i,j}_{out}}$.
\end{construction}
\end{constbox}

Intuitively, if the adversary wants to extract the information about the $i,2j$-th bit out of the classical PUF corresponding to a challenge $x_i$, they need to guess whether the state is prepared in $Z = \{\ket{0},\ket{1}\}$ basis or in $X = \{\ket{+}, \ket{-}\}$ basis, then knowing the encoded bit. In Section~\ref{sec:application-hlpuf-security}, we estimate the success probability of a (weak) adversary in winning the universal unforgeability game for the HPUF as a function of the number of required queries.

Another remark here is that HPUF can be considered and studied within the quantum PUF framework that we have defined in \chapref{chap:qpuf}. However, one should consider it as a non-unitary qPUF since it includes a classical pre-processing and state preparation which can be described by CPTP maps but is not necessarily a unitary. Nevertheless, we treat the HPUF as a new type of PUF and prove its unforgeability in a stand-alone manner. Moreover, we require that the CPUF inside the construction satisfies the robustness and collision resistance requirements, as have also been defined for the qPUFs. If these requirements are satisfied by the CPUF, the HPUF will also deliver them (to the same degree) since the encoding part of the construction is fully deterministic. Finally, we do not investigate the uniqueness property of the HPUF here as we use it as a single device within the other construction and protocol that we will present, which only assume the robustness, collision resistance and $p$-randomness from the classical PUF.

\subsection{Hybrid Locked PUF}\label{sec:application-hlpuf}
As discussed before, most classical PUFs are vulnerable to machine learning attacks. To perform such attacks, the adversary needs to get access to many CRPs of the PUF to use this data for training and obtaining a model for CPUF. One common scenario that makes such attacks viable is when an adversary intercepts the communication channel between the verifier and prover during an identification protocol and pretends to be the verifier. Then the adversary can send their favourite queries as challenges to the prover, who will provide the adversary with the correct response. In this way, an adversary can build a local database, even during the identification phase. To address this issue, a technique has been introduced in the literature of classical PUFs known as \emph{Lockdown technique} (or locking mechanism) \cite{yu_lockdown_2016} that upper-bounds adversary’s capability in querying CRPs by converting the adaptive adversary into a weak one. We recall from Section~\ref{sec:unf-weak-adaptive-adv} that a \emph{weak adversary}, in contrast to an \emph{adaptive adversary} who queries the oracle or device with their chosen and potentially adaptive queries, has only access to a random set of challenges and responses (or input-output queries) that are selected at random by an honest party. we recall that this is equivalent to the random-message attack model as we have also discussed in Section~\ref{sec:prelim-adversarial-models}.

The central idea of the locking mechanism is that the prover (client) can also identify the verifier (server) during the identification. This mutual identification prevents an adversary from querying the PUF arbitrarily. One method is that the verifier sends part of the response along with the challenge so that the prover having access to the PUF device can check if the challenge has really come from the server or from an adversary who is trying to increase their information on the PUF device.
We adopt the idea of the locking mechanism and we apply it on a HPUF that leads to our next construction, namely~\constref{cons:hlpuf}. We refer to this construction as Hybrid Locked PUFs (HLPUFs). First, we devide the output of the HPUF $\E_f : \{0,1\}^n \rightarrow (\Hil^{2})^{\otimes 2m}$ corresponding to a classical PUF $f:\{0,1\}^n \rightarrow \{0,1\}^{4m}$ into two separate parts. The first part contains the first $m$ qubits, and the second half contains the last $m$ qubits of the outcome of HPUF. Note that, the first $m$ qubits of the HPUF's outcome is generated from the first $2m$ bits outcome of the corresponding classical PUF $f$. For any challenge $x \in \{0,1\}^n$ we can write the outcome of the classical PUF as $f(x) = f_1(x)||f_2(x)$, where the mapping $f_1 :\{0,1\}^n \rightarrow \{0,1\}^{2m}$ denotes the first $2m$ bits of $f$ and $f_2 :\{0,1\}^n \rightarrow \{0,1\}^{2m}$ denotes the last $2m$ bits of $f$. Similarly, we can rewrite the HPUF $\E_f$ as a tensor product of two mappings $\E_{f_1}:\{0,1\}^n \rightarrow (\Hil^{2})^{\otimes m}$, and $\E_{f_2}:\{0,1\}^n \rightarrow (\Hil^{2})^{\otimes m}$, where for any challenge $x \in \{0,1\}^n$, $\E_{f_1}(x)$ denotes the first $m$ qubits of $\E_f(x)$, and $\E_{f_2}(x)$ denotes the last $m$ qubits of $\E_f(x)$.

The hybrid locked PUF, takes the classical input $x_i$ and a quantum state $\tilde \rho_1$ and produces the second half of the response of the hybrid PUF,  $\ket{\psi_{f_2(x_i)}}\bra{\psi_{f_2(x_i)}}$, as an output if $\tilde \rho_1$ is equal to the first half of the output of the hybrid PUF $\ket{\psi_{f_1(x_i)}}\bra{\psi_{f_1(x_i)}}$. The construction is shown in \figref{fig:qhlpuf}. We formalise it as follows.

\begin{constbox}
\begin{construction}[HLPUF]
\label{cons:hlpuf}
Suppose we have a hybrid PUF $\E_f$ where $f:\{0,1\}^n \rightarrow \{0,1\}^{4m}$ is a CPUF. The mapping of the HLPUF $\E^{L}_{f}:\din\times \Hil^{d_{out_{1}}}\rightarrow \Hil^{d_{out_{2}}}\otimes\Hil^{\perp}$ corresponding to a hybrid PUF $\E$ is defined as follows:
\begin{equation}
    \label{eq:qlpuf_2}
    (x_i,\tilde \rho_1) \rightarrow \begin{cases}
    & \hspace{-0.15in}\ket{\psi_{f_2(x_i)}}\bra{\psi_{f_2(x_i)}}~\text{if } \texttt{Ver}(\ket{\psi_{f_1(x_i)}}\bra{\psi_{f_1(x_i)}},\tilde \rho_1) = 1\\
    & \perp \hspace{2cm}\text{otherwise.}
    \end{cases}
\end{equation}
where $\texttt{Ver}(.,.)$ is verification algorithm that checks the equality of the first half of the response based on the classical response $y^1_i$.
\end{construction}
\end{constbox}
More precisely, $\texttt{Ver}(.,.)$ is specified by measuring each qubit of the incoming quantum state with corresponding basis according to $\{y_{i,2j}\}_{1\leq j\leq 2m}$ of response $y_i$ and check the equality $\texttt{Equal}(y_{i,2j}, \tilde y_{i,2j})_{1\leq j\leq 2m}$ in our construction.

\begin{figure}[H]
    \centering \small
    \begin{tikzpicture}[
    node distance = 3mm, every node/.style = {rectangle, rounded corners, align=center}]
    \node (puf) [draw, inner xsep=5mm, inner ysep=3mm] {HPUF $\E_f$};
    \node [coordinate, right=1.8cm of puf] (ADL){};
    \node (first) [below=1cm of ADL, draw, inner xsep=3mm, inner ysep=3mm] {$\texttt{Ver}(\ket{\psi_{f_1(x_i)}}\bra{\psi_{f_1(x_i)}},\tilde \rho_1)$};
    \node (second) [right=3.8cm of puf,draw, inner xsep=3mm, inner ysep=3mm] {Output};
    \node[name=outer1, dashed, fit=(puf) (first) (second), draw, inner xsep=2.5mm, inner ysep=3mm] {};
    \node (input_1)[left=0.5cm of puf] {$x_i$};
    \node (output_1) [right=0.4cm of puf, inner ysep=1mm] {$\ket{\psi_{f_1(x_i)}}\bra{\psi_{f_1(x_i)}}_1$};
    \node (input_2)[left=2.6cm of first] {$\tilde \rho_1$};

    \node (output_2) [right=0.5cm of second] {$\ket{\psi_{f_2(x_i)}}\bra{\psi_{f_2(x_i)}}_2/\perp$};
    \draw[-latex'] (input_1) -- (puf); 
    \draw[-latex'] (puf) -- (output_1); 
    \draw[-latex'] (output_1) -- (second); 
    \draw[-latex'] (output_1.south -| first.north) -- (first); 
    \draw[-latex'] (input_2) -- (first);
    \draw[-latex'] (second) -- (output_2);

    \draw[-latex'] (first) -| node[pos=0.7,right,font=\footnotesize] {b} (second);
    \end{tikzpicture}
    \caption{Hybrid Locked PUF (HLPUF) $\E^L_f$ with \constref{cons:hlpuf}}
    \label{fig:qhlpuf}
\end{figure}

In \cite{chakraborty_quantum_2021} the possibility of exploiting the lockdown technique for quantum PUFs has also been investigated where we have developed the mathematical model for it. First of all, it is interesting to see whether the lockdown technique can enable to reduce the adversarial power in the quantum case as well. Moreover, this applicated is well-motivated, especially for the weaker types of qPUF that we have mentioned before, such as QR-PUFs, since arbitrarily querying the PUF with multiple copies allows for quantum emulation attacks (discussed in \chapref{chap:qpuf}). One possible way to protect the PUF from such sophisticated attacks is to use the locking technique. However, similar to the HPUF setting, a central component of the locking mechanism is the verification subroutine. In the quantum case, this verification consists of testing the equality of unknown quantum states. We show a no-go result stating that due to the entanglement that can be generated by the unknown unitary over the subsystems of the response, the verification of such subsystems in a way that can be used for a locking mechanism is impossible, unless in very limited cases. Nevertheless, we avoid presenting the details of this result here, and we conclude with this brief mention not to make the section unnecessarily long. We refer the reader to the main paper for more information.


\subsection{Quantum identification protocol using Hybrid Locked PUF}\label{sec:application-hlpuf-protocol}
After introducing the HLPUF construction, we describe an identification protocol based on it. The description of the protocol is given in~\protoref{prot:hlpuf}. Note that we use $\tilde{\rho}_1$ and $\tilde{\rho}_2$ to denote the quantum state received by the prover/verifier respectively. 

\begin{protocol}\label{prot:hlpuf}[HLPUF-based Authentication] An authentication protocol based on HLPUF construction.
\vspace{3mm}\hrule
    \begin{enumerate}
        \item \textbf{Set-up:}
        \begin{enumerate}
            \item The Prover $\Pv$ equips a Hybrid Locked PUF: $\E_{f}^L$ with HPUF $\E_{f}: \{0,1\}^{n} \rightarrow (\Hil^{2})^{\otimes 2m}$ constructed upon a classical PUF $f:\X \rightarrow \Y$. Here, the classical PUF $f$ maps an $n$-bit string $x_i \in \{0,1\}^n$ to an $4m$-bit string output $y_i \in \{0,1\}^{4m}$.
            \item The Verifier $\V$ has a classical database $D:= \{(x_i,y_i)\}_{i=1}^{d}$ with all $d$ CRPs of $f$, as well as the necessary quantum devices for preparing and measuring quantum states. 
        \end{enumerate}
        \item \textbf{Authentication:}
        \begin{enumerate}
            \item \label{a} $\V$ randomly chooses a CRP $(x_i,y_i)$ and splits the response equally into two partitions $y_i=f_1(x_i)||f_2(x_i)=y_i^1||y_i^2$ with length $2m$.
            \item \label{b} $\V$ then encodes the first partition of response into $\ket{\psi_{f_1(x_i)}}\bra{\psi_{f_1(x_i)}}:=\bigotimes_{j=1}^m \ket{\psi^{i,j}_{f_1(x_i)}}\bra{\psi^{i,j}_{f_1(x_i)}}$ and issues the joint state $(x_i,\ket{\psi_{f_1(x_i)}}\bra{\psi_{f_1(x_i)}})$ to the client. 
            \item \label{c} $\Pv$ receives the joint state $(x_i,\tilde \rho_1)$ and queries Hybrid Locked PUF $\E_{f}^L$. If the verification algorithm $\texttt{Ver}(\ket{\psi_{f_1(x_i)}}\bra{\psi_{f_1(x_i)}},\tilde \rho_1)\geq 1-\epsilon(\lambda)$ with negligible $\epsilon(\lambda)$, $\Pv$ obtains $\ket{\psi_{f_2(x_i)}}\bra{\psi_{f_2(x_i)}}:=\bigotimes_{j=1}^m \ket{\psi^{i,j}_{f_2(x_i)}}\bra{\psi^{i,j}_{f_2(x_i)}}$ from $\E_{f}^L$ and sends back to $\V$. Otherwise, the authentication aborts.
            \item \label{d} $\V$ receives the quantum state $\tilde{\rho}_2$ and performs the the verification algorithm $\texttt{Ver}(.,.)$. If the verification  $\texttt{Ver}(\ket{\psi_{f_2(x_i)}}\bra{\psi_{f_2(x_i)}},\tilde \rho_2)\geq 1-\epsilon(\lambda)$ with negligible $\epsilon(\lambda)$, the authentication passes. Otherwise it aborts.
        \end{enumerate}
    \end{enumerate}
\hrule\vspace{3mm} 
\end{protocol}

We note that this protocol requires only a classical database, but two-way quantum communication. Nevertheless, the quantum states used in this protocol are easy to prepare and measure given the infrastructures that already exist for QKD and the current stage of a quantum internet \cite{wehner_quantum_2018}.

\subsection{Security analysis}\label{sec:application-hlpuf-security}
Now, we give a comprehensive security analysis of the proposed protocol. We will prove the security step-by-step. First, we show that using hybrid construction will exponentially improve security compared to CPUF. More precisely, it will exponentially decrease the success probability of a \emph{weak} quantum adversary in the universal unforgeability game, compared to a classical PUF with the same number queries in the learning phase. For this part, we will use the weak quantum adversarial model and the respective variant of the universal unforgeability game as defined in Section~\ref{sec:unf-weak-adaptive-adv}. This result shows how much quantum communication can improve the security of a weaker classical PUF against quantum adversaries. Using this improvement, we propose an efficient and secure construction using existing classical PUFs. Then, we analyse the completeness and security of the HLPUF-based device authentication protocol and show that given that the intrinsic classical PUF is not fully broken against a weak quantum adversary, the HLPUF-based protocol will be secure against a QPT adaptive adversary. The HLPUF-based protocol, in fact, provides mutual authentication of both prover/client and verifier/server due to the specific construction of the quantum lock. However, in our security analysis, we only formally prove the authentication of the prover to the verifier, and the lock has only been shown to reduce the adversary's capability. Nevertheless, since the verification mechanism is similar on both sides, the alternative side of authentication can be proven similarly. Let us first formalise our assumption on the underlying CPUF:\\

\noindent\textbf{Assumptions on the CPUFs}\\
For the security analysis of our constructions we consider the following assumptions of the CPUFs $f:\{0,1\}^n \rightarrow \{0,1\}^{4m}$.
\begin{enumerate}
    \item For any input $x \in \{0,1\}^n$ the probability distributions of the $4m$ output bits $f(x)_1, \ldots , f(x)_{4m}$ are independent and identically distributed (i.i.d)\footnote{This assumption is not strictly required in practice, for the HLPUF construction to be secure, as our simulation results show in~\cite{chakraborty_quantum_2021}. It is mostly required for our theoretical bounds and even so, we parameterised the deviation from perfect randomness or identical distribution with the randomness parameter of CPUF, although we require that the encoded qubits are independent.}.
    \item The output distributions $\{p^f_x(y)\}_{y \in \{0,1\}^{4m}}$ for all the inputs $x$ are independent and identically distributed (i.i.d).
\end{enumerate}

\subsubsection{Universal Unforgeability of HPUF}\label{sec:application-hpuf-security-proofs}
Intuitively the security of our HPUF comes from the indistinguishability property of the non-orthogonal quantum states. First we show that the HPUFs are at least as secure as the underlying CPUFs.

\begin{thmbox}
\begin{theorem}\label{th:cpuf_vs_hpuf}
Let $f:\{0,1\}^n \rightarrow \{0,1\}^{2m}$ be a classical PUF. If there is no QPT weak adversary who can win the universal unforgeability game for CPUF with a non-negligible probability in the security parameter, then the HPUF constructed from $f$ according to \constref{cons:hlpuf}, is also universally unforgeable. 
\end{theorem}
\end{thmbox}
\begin{proof}
We show the contrapositive statement that if you can break HPUF you can also break underlying CPUF. Here we give the proof for $m=1$, and it can easily be generalised for any arbitrary integer $m > 0$. Suppose for the HPUF, a $q$-query weak-adversary win the unforgeability game with a non-negligible probability $P(m=1,p,q)$. This implies, given a database of $q$ random challenge response from the HPUF, the adversary can produce $\ket{\psi_{f(x^*)}}$ corresponding to a random challenge $x^* \in \{0,1\}^n$ with a non-negligible probability $P(m=1,p,q)$. Note that, for the deterministic adversarial strategy, the adversary can produce multiple copies of the forged state $\ket{\psi_{\tilde f(x^*)}}$ for a random challenge $x^*$. For the random adversaries we can produce the multiple copies of the same forged state $\ket{\psi_{\tilde f(x^*)}}$ just by fixing the internal randomness parameter of the adversarial strategy. Hence, both the random and deterministic adversary can produce multiple copies of the forged state $\ket{\psi_{\tilde f(x^*)}}$ for a random challenge $x^*$. From the multiple (say $K$) such copies of $\ket{\psi_{\tilde f(x^*)}}$, the adversary will extract $\tilde{f}(x^*)$ using the following strategy.
\begin{algorithm}
    \caption{Algorithm to Forge CPUF from HPUF}\label{alg:cap}
    \begin{algorithmic}
    \Require{$K \geq 2$-copies of the forged state $\ket{\psi_{\tilde f(x^*)}}$}
    \State{ - Measure the $1$-st copy of the state $\ket{\psi_{\tilde f(x^*)}}$ in $\{\ket{0},\ket{1}\}$-basis.}
    \State{ - Let $z_1 \in \{0,1\}$ be the measurement outcome.}\\
    
    \For{$i=2$; $i \leq (K-1)$; $i++$}{
    \State{Measure the $i$-th copy of the state $\ket{\psi_{\tilde f(x^*)}}$ in $\{\ket{0},\ket{1}\}$-basis.}
    \State{Let $z_i \in \{0,1\}$ be the measurement outcome.}\\
    \If{$z_i \neq z_{i-1}$}{
        \State \textbf{break}
          \Comment{Implies $\ket{\psi_{\tilde f(x^*)}} \in \{\ket{+},\ket{-}\}$.}
         }
    }
    \noindent\If{$i=K$}{
        \State \textbf{Return }$\tilde f(x^*) = (0,z_i)$} \\
        \Else{
        \State Measure the $i+1$-th copy in $\{\ket{+},\ket{-}\}$-basis.
        \State Let $z_{i+1}$ be the measurement outcome.
        \State \textbf{Return} $\tilde f(x^*) = (1,z_{i+1})$.
        }
    \end{algorithmic}
\end{algorithm}

If $\ket{\psi_{f(x^*)}} = \ket{\psi_{\tilde f(x*)}} \in \{\ket{0},\ket{1}\}$ then in \algoref{alg:cap} all the measurement outcomes $z_i$ (for $1\leq i \leq K$) would be the same, and $\tilde f(x^*) = f(x^*)$. However, if $\ket{\psi_{f(x^*)}} = \ket{\psi_{\tilde f(x*)}} \in \{\ket{+},\ket{-}\}$ then we $\tilde f(x^*) \neq f(x^*)$ if and only if all the measurement outcomes $z_i$ are equal ($1 \leq i \leq K$). This happens with probability $\frac{1}{2^K}$. Therefore, we get

\begin{equation}
\label{eq:forge_f_psi_x}
    \Pr_{x^*}[\tilde f(x^*) = f(x^*)| |\psi_{f(x^*)}\rangle = |\psi_{\tilde f(x*)}\rangle] \geq 1-\frac{1}{2^K}.
\end{equation}

If the adversary successfully forges the HPUF with a non-negligible probability $P(m=1,p,q)$ then from \eqref{eq:forge_f_psi_x} we get that the adversary manages the CPUF with probability at least $P(m=1,p,q) = 1-\frac{1}{2^K}$, that is an overwhelming probability. Therefore, if an adversary successfully wins the unforgeability game for the HPUF with a non-negligible probability, then using the same forging strategy it can also win the unforgeability game for the corresponding CPUF with a non-negligible probability. This implies, that if no QPT weak adversary can win the universal unforgeability game with a non-negligible probability for the CPUF then no QPT adversary can win the universal unforgeability game with a non-negligible probability for the corresponding HPUF. This concludes the proof.
\end{proof}

The above theorem is an intuitive result that shows HPUF is stronger or at least as strong as the underlying CPUF. Although we want to prove a more powerful and explicit statement regarding HPUFs by quantifying how much the hybrid construction will boost the security. In fact, we want to show that one can construct a secure unforgeable HPUF against a quantum adversary even if the underlying CPUF is breakable (with a certain probability) against the classical forger. To this end, we compare the success probability of a QPT adversary in breaking the HPUF in the universal unforgeability game, with the success probability of the adversary who breaks the CPUF with a certain non-negligible probability, in a fixed query setting. This, allows us to show that some of the weak and considerably broken CPUFs can still be used to construct an asymptotically secure HPUF against stronger quantum adversaries since the quantum encoding drastically decreases the success probability. Before giving our main theorem, we need to prove two lemmas. In the first one, we give an upper bound on the adversary's guessing probability of the response $f(x_i)$ corresponding to a challenge $x_i$ and a single copy of the quantum response state $\ket{\psi_{f(x_i)}}$.

\begin{lembox}
\begin{lemma}\label{lem:application-guess_prob}
Let $f:\{0,1\}^n \rightarrow \{0,1\}^{4m}$ be a CPUF with the following property,
\begin{equation}
    \forall~~x_i\in \{0,1\}^n, \forall~~1\leq j \leq 4m,~~p^{f}_{x_i}(y_{i,j} = 0) =\frac{1}{2}+\delta_r, 
\end{equation}
with a biased distribution $p=\frac{1}{2}+\delta_r$ where $0 \leq \delta_r \leq \frac{1}{2}$, and $\E_f$ be a HPUF corresponding to $f$ that we construct using \constref{cons:hpuf_1}. Let a quantum adversary $\A$ extract the value $y_{i,(2j-1)}$ out of $(y_{i,(2j-1)},y_{i,2j})$ from quantum state $\ket{\psi^{i,j}_{out}}\bra{\psi^{i,j}_{out}}$ corresponding to a random challenge $x_i$. If all the output bits of the CPUF are independent and identically distributed, then for any quantum adversary $\A$, and  $\forall~x_i\in \{0,1\}^n$ then,
\begin{align}
    \nonumber
    p_{\guess} := \Pr[\A&(x_i, \ket{\psi^{i,j}_{out}}\bra{\psi^{i,j}_{out}})= y_{i,(2j-1)}] \\
    \nonumber &\leq p(1+\sqrt{p^2+(1-p)^2})\\
    &\leq p(1+\sqrt{2}p)
\end{align}
\end{lemma}
\end{lembox}
\begin{proof}
According to \constref{cons:hlpuf}, for a given $x_i$, we use the $2j$-th bit $y_{i,2j} \in \{0,1\}$ of the outcome of the CPUF to choose the basis (either $\{\ket{0},\ket{1}\}$-basis or $\{\ket{+},\ket{-}\}$-basis) of the $j$-th qubit output of the HPUF. Further we use the $y_{i,(2j-1)}\in \{0,1\}$ to choose a state from the chosen basis. Here, if $y_{i,(2j-1)} = 0$ then from an adversarial point of view, the output state is $\rho_0=(\frac{1}{2}+\delta_r)\ket{0}\bra{0} + (\frac{1}{2}-\delta_r)\ket{+}\bra{+}$. Similarly, if $y_{i,(2j-1)} = 1$ then from an adversarial point of view, the output state is $\rho_1=(\frac{1}{2}+\delta_r)\ket{1}\bra{1} + (\frac{1}{2}-\delta_r)\ket{-}\bra{-}$. For the adversary, the probability of correctly guessing $y_{i,(2j-1)}$ is the same as distinguishing the two states $\rho_0,\rho_1$. Here $Pr[\A(x_i, \ket{\psi^{i,j}_{out}}\bra{\psi^{i,j}_{out}})= y_{i,(2j-1)}]$ denotes the optimal probability of guessing the basis correctly. From the Holevo-Helstorm bound \cite{holevo_statistical_1973} (see Section~\ref{sec:prelim-distinguish-quantum-test}) we get,
\begin{align}
\nonumber
    \Pr[\A(x_i, \ket{\psi^{i,j}_{out}}\bra{\psi^{i,j}_{out}})= y_{i,(2j-1)}] & \leq p[1+\max_{E} \tr[E(\rho_0 - \rho_1)]] \\
    \nonumber &= p[1 + \frac{1}{2}\norm{\rho_0- \rho_1}_1].\\
    \nonumber &= p(1+\sqrt{p^2+(1-p)^2})\\
    &\leq p(1+\sqrt{2}p)
\end{align}
This concludes the proof.
\end{proof}

The next lemma shows that the adversary needs to extract the classical information $f(x)$ that is encoded in the quantum state $\ket{\psi_{f(x)}}$ for the forgery of the HPUFs. This will be a key step in our proof, since using this lemma we can put bounds on the maximum amount of information the adversary can extract from the overall response state using quantum information tools we have presented in the preliminaries, in Section~\ref{sec:prelim-uncert}.

\begin{lembox}
\begin{lemma}\label{lem:application-database-forgery}
Let $\ket{D_q} = \bigotimes_{i=1}^q \left(\ket{x_i}_C \otimes \ket{\psi_{f(x_i)}}_R \right)$ denotes the adversary's database of $q$ random CRPs that are generated from a HPUF $\E_f:\{0,1\}^n \rightarrow (\Hil^2)^{\otimes m}$. Let $E(D_q)$ denote the optimal measurement strategy for forging the HPUF with probability $p_{\forge}$ using the database $D_q$, then the following measure-then-forge strategy can optimally forge the HPUF with the same probability $p_{\forge}$.
\begin{itemize}
    \item Adversary extracts the classical encoding $\{f(x_i)\}_{1 \leq i \leq q}$ from $\ket{D_q}$. Let $\{\tilde f(x_i)\}_{1 \leq i \leq q}$ denotes the extracted classical string.
    \item The QPT adversary applies a forging strategy using the extracted data set $\{\tilde f(x_i)\}_{1 \leq i \leq q}$.
\end{itemize}
\end{lemma}
\end{lembox}
\begin{proof}
For a successful forgery, the adversary needs to win the universal unforgeability game that we define in \gameref{game:uni-unf-weak-adv} in \chapref{chap:unf-tools}. This implies, using the measurement strategy $E(D_q)$ the adversary needs to produce a quantum state $\ket{\psi_{f(x^*)}}$ corresponding to a challenge $x^* \in_R \{0,1\}^n$ that is chosen uniformly at random. Without loss of generality we can write the measurement strategy as a POVM with two outcomes $E(D_q) = \{E_{\forge}(D_q,x^*),E_{\fail}(D_q,x^*)\}$, where $E_{\forge}(D_q,x^*), E_{\fail}(D_q,x^*)$ denote the measurement operators corresponding to the successful forgery and the failure forgery respectively. Therefore, we can write the successful forging probability $p_{\forge}$ as follows.
\begin{equation}
    p_{\forge} = \tr[E_{\forge}(D_q,x^*)\rho_{D_q}^{x^*}],
\end{equation}
where $\rho_{D_q}^{x^*} := \ket{D_q}\bra{D_q} \otimes \ket{x^*}\bra{x^*}\otimes \ket{0^m}_{out}\bra{0^m}$. Here the $out$ register would contain the forged state. If we write $E_{\forge}(D_q,x^*) = M^{\dagger}_{\forge}(D_q,x^*)M_{\forge}(D_q,x^*)$, then we can rewrite the post-measurement state corresponding to the successful forgery as follows:
\begin{equation}\label{eq:post_state}
    \frac{M_{\forge}(D_q,x^*)\ket{D_q} \otimes \ket{x^*} \otimes \ket{0^{m'}}_{out}}{\sqrt{p_{\forge}}} 
    = \frac{\ket{\tilde D_q}_R \otimes \ket{x^*} \otimes \ket{\psi_{f(x^*)}}_{out} \otimes \ket{\tilde a}_{out}}{\sqrt{p_{\forge}}},
\end{equation}
where $|\tilde D_q\rangle_R$ denotes the post-measurement database state, and $\ket{\tilde a}_{out}$ is the post-measurement state of the ancillary system which is a $(m'-m)$ dimensional state while as $\ket{\psi_{f(x^*)}}_{out}$ is $m$ dimensional. As $\bigotimes_{i=1}^q|x_i\rangle_C$ is a classical state, in the rest of the proof we don't write them in the expressions. 

Using the \emph{Neimark's theorem}\footnote{The version of Neimark's theorem, is similar but more genral than the one we introduced in Section~\ref{sec:prelim-distinguish-quantum-test}.} we can replace the POVM measurement strategy $E(D_q)$ with the combination of a unitary acting on an extended system including an ancilla $\ket{anc}_A$, followed by a projective measurement. Let us denote the unitary as $U^{x^*}_{D_q}$ which couples the input state $\ket{D_q} \otimes \ket{0^{m'}}_{out}$ with the ancillary system $|anc\rangle_A$, and let $\{|v\rangle\}$ be the basis on which the projective measurement is applied to the ancilla. We first rewrite the impact of the unitary $U^{x^*}_{D_q}$ on the input state:
\begin{align}\label{eq:application-proof-neimark-unit}\nonumber
    U^{x^*}_{D_q}\left(\bigotimes_{i=1}^q \ket{\psi_{f(x_i)}}_R \otimes \ket{0}_{out} \otimes \ket{anc}_A\right)
    & = U^{x^*}_{D_q}\left( \ket{\Psi^q_{f}}_R \otimes \ket{0}_{out} \otimes \ket{anc}_A\right) \\ 
    & = \sum_{v} \sqrt{p_v} \ket{\Psi^q_v}_R \otimes \ket{\tilde \psi_v}_{out} \otimes \ket{v}_{A}.
\end{align}
where in the second line we have rewritten everything after applying the unitary in the $\{\ket{v}\}$-basis. Now, the adversary performs a projective measurement on the state \eqref{eq:application-proof-neimark-unit} in this basis. Suppose for the correct forgery, the ancilla is projected into the $\ket{v_{\forge}}_A$ state. Therefore we can rewrite the expression of $p_{\forge}$ as follows:
\begin{equation}
    p_{\forge} = \sum_{v: v = v_{\forge}} p_v |\mbraket{v_{\forge}}{v}|^2.
\end{equation}
Overall, following this strategy, the purification of the adversary's post-measurement state with an optimal POVM measurement, can be written as the following:
\begin{equation}\label{eq:application-post-state-neimark}
    \frac{\ket{\tilde D_q}_R \otimes \ket{x^*} \otimes \ket{\psi_{f(x^*)}}_{out} \otimes \ket{v_{\forge}}_A}{\sqrt{p_{\forge}}}, 
\end{equation}
where $\ket{\tilde D_q}$ denotes the post-measurement database state. Note that, due to Neimark's theorem the post-measurement database states in Equation \eqref{eq:post_state}, and \eqref{eq:application-post-state-neimark} are the same, if the same ancillary systems has been assumed after the purification and POVM, \emph{i.e.} if $\ket{v_{\forge}}_A = \ket{\tilde a}_{out}$.

Now, let us use the unitary $U^{x^*}_{D_q}$ and the measurement basis $\{\ket{v}\}$ to construct a \emph{measure-then-forge} strategy. As the unitary $U^{x^*}_{D_q}$ only depends on the input $x^*$ and $D_q$, we can rewrite it in the basis that is diagonalised with respect to the states $\{\ket{\Psi^q_v,v}\}_v$.
For the post-measurement state $\ket{v_{\forge}}$, of the ancilla, the adversary applies $U^{x,x^*}_{D_q,\Psi^q_{\forge},v_{\forge}}$ on the $\ket{0}_{out}$ register. Note that, the adversary doesn't have any information about the $\{f(x_i)\}_{1 \leq i \leq q}$ before measuring the ancillary sub-system in the $\{\ket{v}\}$-basis. Hence, the measurement basis $\{\ket{v}\}$ choice only depends on the classical challenges $x_i$'s and $x^*$. Therefore, the adversary can use the same information to find the $\{\ket{v}\}$-basis, and first performs the measurement on the $RA$ register in $\{\ket{\Psi^q_v,v}\}$-basis, and obtains the state $\ket{\Psi^q_{\forge},v_{\forge}}$ with the same probability $p_{\forge}$. After the measurement, the adversary applies the unitary $U^{x^*}_{D_q,\Psi^q_{\forge},v_{\forge}}$ on $\ket{0}_{out}$, and get the forged state $\ket{\psi_{f(x^*)}}$. Therefore, with this strategy the adversary also win the unforgeability game with the probability $p_{\forge}$.

Note that, there always exists a unitary $U$ such that $U(\bigotimes_{i=1}^q\ket{\tilde{f}(x_i)}) \otimes \ket{anc} = \ket{\Psi^q_{\forge},v_{\forge}}$, where $\tilde f(x_i)$ denotes the extracted information about $f(x_i)$'s from the encoded database $\ket{D_q}$. Therefore, from any generalised measurement strategy $E(D_q)$ we can construct a strategy for the measure-then-forge protocol that can win the universal unforgeability game with the same probability $p_{\forge}$. This concludes the proof.
\end{proof}

\lemref{lem:application-database-forgery} suggests that an optimal strategy of the adversary including general POVM strategies, is equivalent to optimally extracting the classical information from the database (state $\ket{D_q}$) and then performing the most optimal forgery strategy on the measurement results. In general, if the extracted classical information $\{\tilde f(x_i)\}_{1 \leq i \leq q}$ from the database state $\ket{D_q}$ is very far from the original encoded string $\{f(x_i)\}_{1 \leq i \leq q}$ then the forgery will not perform well and the overall forgery attack will have a low probability. Based on this, we now bound the overall amount of information that can be extracted from the outputs of HPUF, using the information quantities such as min-entropy and we will show how much the quantum encoding will contribute to reducing the success probability.

\begin{thmbox}
\begin{theorem}
\label{th:hpuf-security-prob}
Let  $f:\{0,1\}^n \rightarrow \{0,1\}^{4m}$ be a classical PUF with $p$-randomness, where $p = \frac{1}{2}+\delta_r$. Let $p^{\text{classic}}_{\forge}(m,p,q)$ denote the optimal success probability of any $q$-query weak quantum adversary to wins the universal unforgeability game for the CPUF $f$. Then a $q$-query weak quantum adversary can win the universal unforgeability game for the HPUF $\E_f$ at most the following probability
\begin{equation}
    p_{\forge}^{\quantum}(m,p,q) = p_{\forge}^{\classical}(m,p,q) \times (p(1 + \sqrt{2}p))^{2mq}
\end{equation}
\end{theorem}
\end{thmbox}
\begin{proof}
We want to quantify the success probability of the QPT adversary in attacking HPUF, in comparison with the QPT adversary who attacks the classical PUF with a fixed number of queries. Let $\A_c$ be the QPT adversary attempting to forge CPUF, where they produce a classical forgery $f(x^*)$ for a randomly selected challenge $x^*$, from a classical database consisting of $q$ pairs of $\{(x_i,f(x_i))\}^q_{i=1}$ input-outputs of CPUF with probability $p^{\text{classic}}_{\forge}(m,p,q)$. Note that in general the success probability, is a function of the CPUF's randomness parameter $p$, the output size $m$ and the number of queries $q$. Let $\A_h$ be a quantum adversary who plays the unforgeability game against the HPUF. $\A_h$ has access to $q$ queries of HPUF included in the database state $\ket{D_q}$. The goal of $\A_h$ is to produce a valid forgery $\ket{\psi_{f(x^*)}}$ for a random challenge $x^*$. Let $p_{\forge}^{\quantum}$ be the optimal success probability of any adversary in successfully doing so.

Now, for the purpose of the proof, we introduce another intermediate quantum adversary $\B$ who plays the same version of the unforgeability game as $\A_h$, although has access to a combined database of $\A_h$ and $\A_c$ \emph{i.e.} the triplet $\{(x_i, f(x_i), \ket{\psi_{f(x_i)}})\}^q_{i=1}$, and will output a valid forgery of the form $\ket{\psi_{f(x^*)}}$ for a random challenge $x^*$. We show that the adversaries $\A_c$ and $\B$ are equivalent in the success probability up to a negligible factor. First, note that $\B$ is obviously at least as strong as $\A_c$ since has an extended database, and can simply ignore the quantum encoded detest and run $\A_c$ as a subroutine so we have $p_{\B}(m,p,q) \geq p^{\text{classic}}_{\forge}(m,p,q)$. But on the other hand, $\A_c$ can also locally construct the third column of the database $\ket{\psi_{f(x_i)}}$ easily and run $\B$ as a subroutine. Then $\A_c$ can use this to produce polynomial many copies of $\ket{\psi_{f(x^*)}}$ and measure them with the optimal POVM measurement and get $f(x^*)$ with very high probability (See the proof of \thmref{th:cpuf_vs_hpuf}). Also note that by definition of the unforgeability game, $\B$ produces the forgery with non-negligible success probability, meaning that the $\ket{\psi_{f(x^*)}}$ is close to the actual BB84 encoding of $f(x^*)$ (in terms of fidelity). Thus $\B$ has at most a negligible advantage over $\A_c$ and we can conclude: $p_{\B}(m,p,q) \approx p^{\text{classic}}_{\forge}(m,p,q)$.

Now we have two adversaries who produce quantum state as a forgery and we can compare the success probability of $\A_h$ with $\B$, which are both QPT adversaries producing the same quantum forgery while having access to different input databases since $\B$ has the underlying classical information $f(x_i)$ for each encoded quantum query, while $\A_h$ has only access to the encoded states in the form of $\ket{\psi_{f(x_i)}}$. We compare the success probability of these two adversaries by comparing the accessible amount of information via entropy inequalities. First we note that according to \lemref{lem:application-database-forgery}, the optimal forgery for a quantum adversary includes optimally extracting the classical information then applying the forgery (which is equivalent to the classical forgery on the classical database). This exactly quantifies the relation between the success probabilities of $\A_h$ with $\B$ which will give us the reduction to the problem of extracting information from the quantum-classical database of  $\A_h$.

To do so, we reformulate the forging probability (the success probability of the adversary in the unforgeability game) in terms of quantum processing. We consider $\B$ as a CPTP map over a database of size $N$, denoted as $\rho^{\B^N}$, where each input $\rho^{\B}_i = \ket{x_i}\bra{x_i} \otimes \ket{f(x_i)}\bra{f(x_i)} \otimes \ket{\psi_{f(x_i)}}\bra{\psi_{f(x_i)}}$ is a classical-quantum state. Adversary $\A_h$ can be defined directly from $\B$ through another CPTP map which we denote by $\Lambda_h$. For each record we have $\rho^{\A_h}_i = \Lambda_h (\rho^{\B}_i)$. Assuming the queries to be i.i.d and the way we have defined these two adversaries, The overall action of the CPTP map $\A_h$ is captured by the density matrix $\rho^{\A_h^N} =  \Lambda_h^{\otimes N}(\rho^{\B^N})$. Also, let $F$ represent the random variable of getting the correct output of the PUF, over the uniform choice of the input $x^*$, by processing the given input database.

We now use the inequality for conditional min-entropy to relate the above success probability. Using \lemref{lem:application-database-forgery} stating that the optimal strategy is equivalent to extracting the underlying classical information (optimal state discrimination) and then run an optimal forgery algorithm that depends on the extracted information. Thus we can write the success probability of $\A_h$ in terms of the extraction probability as follows:
\begin{equation}
    p_{\forge}^{\quantum}(m,p,q) = p_{\extract} \times p_{\B}(m,p,q) = 2^{- H_{min}(F|C^N)}
\end{equation}
where $C^N$ denotes $\rho^{\A_h^N}$ as the full quantum system of $\A_h$ and also let $C$ denotes $\rho^{\A_h}$ which is the single-qubit database of $\A_h$. We can rewrite the success probability of forgery for adversary $\B$, in terms of the min-entropy as follows:
\begin{equation}
    p_{\B}(m,p,q) = p^{\B}_{\text{forge}} = 2^{- H_{min}(F|B^N)}
\end{equation}
Where $B^N$ represent the full system of $\B^N$ \emph{i.e.} $\rho^{\B^N}$ for short. 
Now for the single-qubit database, the following relation holds:
\begin{equation}
    H_{min}(F|C) \geq H_{min}(\tilde{Y}|C) + H_{min}(F|\tilde{Y})
\end{equation}
Where $\tilde{Y}$ denotes the random variable of the estimated output $Y = f(X)$ which is the 2-bit fraction of the output of CPUF. We also note that $H_{min}(F|\tilde{Y}) \geq H_{min}(F|B)$, and the equality is when the $\tilde{Y}$ is arbitrarily close to $Y$. As a result we have\footnote{This inequality in fact gives and improvement to the data processing inequality (introduced in Section~\ref{sec:prelim-uncert}) for our specific case. Since according to the data-processing inequality we have that the entropy min-entropy increases via any CPTP channel acting on the joint state of the system, which results in $H_{min}(F|B) \leq H_{min}(F|C)$ which denotes that the success probability of adversary $\B$ who has access to additional classical output is higher than $\A_h$. The new inequality quantifies the bound on the difference.}:
\begin{equation}
    H_{min}(F|C) \geq H_{min}(\tilde{Y}|C) + H_{min}(F|B)
\end{equation}
Extending to $N$-fold database we have:
\begin{equation}\label{eq:optimal-forgery-Bn-Cn}
    H_{min}(F|C^N) \geq H_{min}(\tilde{Y}^N|C^N) + H_{min}(F|B^N)
\end{equation}
Next, we need to relate the first term of the right-hand side, which denoted the min-entropy of extracting information from all $N$ given qubits to the min-entropy quantity $H_{min}(\tilde{Y}|C)$. For that, we use the quantum-classical AEP that we have introduced in \thmref{th:prelim-quantum-aep} as follows:
\begin{equation}
   H^{\epsilon}_{min}(\tilde{Y}^N|C^N) \geq N (H(\rho_{\tilde{Y}C}) - H(\rho_C)) - N\eta
\end{equation}
where $\eta := (2H_{max}(\rho_F)+3)\sqrt{\frac{\log(\frac{1}{\epsilon})}{N} + 1}$, is a function of the smoothing parameter $\epsilon$ and $N$. Here we select the smoothing parameter $\epsilon$ such that $N\eta$ becomes a negligible function in the security parameter. Given that $H(\tilde{Y}|C) = H(\rho_{\tilde{Y}C}) - H(\rho_C)$ and the fact that $H_{min}(\tilde{Y}|C) \leq H(\tilde{Y}|C)$, we have:
\begin{equation}
   H^{\epsilon}_{min}(\tilde{Y}|C^N) \geq N H_{min}(\tilde{Y}|C)
\end{equation}
By substituting the above inequality back into~\ref{eq:optimal-forgery-Bn-Cn}, we can conclude the following:
\begin{equation}
   H_{min}(F|C^N) \geq N H_{min}(\tilde{Y}|C) + H_{min}(F|B^N)
\end{equation}
The final step is to determine $N$. We note that there exist $q$ number of i.i.d queries in each database, but each query itself consists of a $2m$ number of qubits in tensor product form. Although the effective size or the amount of information included in these $2m$ qubits depends on the bias of the PUF. In general, using quantum data compression inequality, the effective size of such $2m$ tensor product states is given by $2m S(\rho_f)$, where $S(\rho_f)$ is the von-Neumann entropy of each encoded state $\rho_f$ given as follows concerning the HPUF construction and the PUF bias:
\begin{equation}
   \rho_f = (\frac{1}{2} + \delta_r)^2\ket{0}\bra{0} + (\frac{1}{4} - \delta_r^2)(\ket{1}\bra{1} + \ket{+}\bra{+}) + (\frac{1}{2} - \delta_r)^2 \ket{-}\bra{-}
\end{equation}
We can then calculate $S(\rho_f)$ which gives the following result while we are discarding $\mathcal{O}(\delta^3)$ and higher:
\begin{equation}
   S(\rho_f) = 1 - (\frac{1}{2} - \delta_r)\log(1 - 2\delta_r) - (\frac{1}{2} + \delta_r)\log(1 + 2\delta_r)
\end{equation}
Let us call $g(\delta_r) = (\frac{1}{2} - \delta_r)\log(1 - 2\delta_r) + (\frac{1}{2} + \delta_r)\log(1 + 2\delta_r)$
Thus we have $N \approx 2mq - 2mqg(\delta_r)$. Also, for small enough values of $\delta_r$, we have $g(\delta_r) \approx 0$. Let us also denote $H_{min}(\tilde{Y}|C) = -\log(p^1_{\extract})$. Where $p^1_{\extract}$ is the probability of extracting the classical information from a single qubit. We can then conclude the following relations between the success probabilities:
\begin{equation}
    p_{\forge}^{\quantum}(m,p,q) \leq p_{\B}(m,p,q)\times (p^1_{\extract})^{2mq - 2mq g(\delta_r)}
\end{equation}
For our given construction $p^1_{\extract} = p_{\text{guess}} = p(1 + \sqrt{2}p)$ according to the optimal discrimination probability given in \lemref{lem:application-guess_prob}. Thus we can rewrite the above equation as:
\begin{equation}
\begin{split}
    p_{\forge}^{\quantum}(m,p,q) & \leq p_{\forge}^{\classical}(m,p,q) \times (p(1 + \sqrt{2}p))^{2mq - 2mq g(\delta_r)} \\
    & \approx  p_{\forge}^{\classical}(m,p,q) \times (p(1 + \sqrt{2}p))^{2mq}
\end{split}
\end{equation}
which is the bound we wanted to prove, and we have also used that $p_{\forge}^{\classical}(m,p,q) \approx p_{\B}(m,p,q)$. As a final remark, we note that the optimal probability is a function of the number of queries, thus we can show that the optimal overall probability is achieved given the adversary optimises on the number of queries used for extracting the information for the forgery. Analysing the upper bound of the probability as a function of $q$, one can see that the first term is a non-decreasing function of $q$ while the second term is always strictly decreasing with $q$. As a result, the combined function has necessary an extremum over $q$, which we denote by $q_{opt}$. Assuming the two cases where the given number of queries is smaller or larger than $q_{opt}$, we have the following bounds:
\begin{equation}
\begin{split}
    & p_{\forge}^{\quantum} \leq p_{\forge}^{\classical}(m,p,q) \times (p_{\text{guess}})^{2mq (1 - g(\delta_r))} \quad \quad \quad \quad q < q_{opt} \\
    & p_{\forge}^{\quantum} \leq p_{\forge}^{\classical}(m,p,q_{opt})\times (p_{\text{guess}})^{2mq_{opt} (1 - g(\delta_r))} \quad \quad q \geq q_{opt}
\end{split}
\end{equation}
Summarizing the above cased and given that $g(\delta_r)$ is small we have:
\begin{equation}
    p_{\forge}^{\quantum} \leq \sup_{q} [p_{\forge}^{\classical}(m,p,q) \times (p_{\text{guess}})^{2mq}]
\end{equation}
which concludes the proof.\footnote{In the latest version of the paper \cite{chakraborty_quantum_2021}, we have given an alternative version of this proof which does not use the AEP, and instead relies on giving the bound on the forgery probability over a noisy database where the bound is slightly different although very similar. The exponential decay in the probability happens in both cases. In that proof, one no longer needs to give an optimality argument over the number of queries. However, since proving the result, in this way seemed more intuitive and used a rather nice quantum information tools, we have decided to present this proof in the thesis. The reader can refer to the paper for the other result.}
\end{proof}

Finally, let us present the following corollary that ensures the universal unforgeability of an HPUF constructed from a CPUF that does not provide suitable security, yet is not totally broken with overwhelming probability.

\begin{corrbox}
\begin{corollary}\label{cor:hpuf}
Let the success probability of any QPT weak-adversary in the universal unforgeability game with a CPUF $f:\{0,1\}^n \rightarrow \{0,1\}^{4m}$ with $p$-randomness, be at most $p^{\text{classic}}_{\forge}$, where $0 \leq p^{\text{classic}}_{\forge} \leq 1 - \text{non-negl}(2m)$. Then, the success probability of any QPT adversary in the universal unforgeability game for the HPUF $\E_f$, is at most $\epsilon(2m)$, which is a negligible function in the security parameter. Hence such HPUFs are universally unforgeable.
\end{corollary}
\end{corrbox}
\begin{proof}
This directly follows from \thmref{th:hpuf-security-prob} where $p^{\text{classic}}_{\forge} = p_{\forge}^{\classical}(m,p,q)$ for any $q = poly(m)$ is a value between $0$ and $1$, and not negligibly close to $1$. As shown in the proof of \thmref{th:hpuf-security-prob} the second term of the probability, becomes negligibly small (in $2m$) and hence the overall probability becomes a negligible function $\epsilon(2m)$.
\end{proof}

\subsubsection{Universal Unforgeability of HLPUF}
So far, we have analysed the security of the HPUFs only against weak adversaries. In the next theorem, we show that if the HPUF is secure against weak adversaries, then using the locking mechanism, we can make the HLPUF secure against adaptive adversaries.

\begin{thmbox}
\begin{theorem}\label{th:application-hpuf_hlpuf-adaptive-secure}
Let $\E_f: \{0,1\}^{n}  \rightarrow (\Hil^{2})^{\otimes m} \otimes (\Hil^{2})^{\otimes m}$ be a hybrid PUF that we construct from a classical PUF $f:\{0,1\}^n \rightarrow \{0,1\}^{2m} \times \{0,1\}^{2m}$ and let $\E^L_{f}: \{0,1\}^{n} \times (\Hil^{2})^{\otimes m}\rightarrow (\Hil^{2})^{\otimes m}$ denotes the HLPUF that we construct from $\E_f$ using the \constref{cons:hlpuf}.
If $\E_f = \E_{f_1} \otimes \E_{f_2}$ and if each of the mappings $\E_{f_1},\E_{f_2}$ has $(\epsilon, m)$-universal unforgeability against the $q$-query weak adversaries, then the corresponding HLPUF $\E^L_f$ is $(\epsilon, m)$-secure against the $q$-query adaptive adversaries.
\end{theorem}
\end{thmbox}

\begin{proof}
At the $i$-th round, the HLPUF $\E^L_f$ receives the queries of the form $(x_i, \tilde \rho_1)$, where the classical string $x_i \in \{0,1\}^n$, and $\tilde \rho_1 \in (\Hil^{2})^{\otimes m}$. The HLPUF returns $\E_{f_2}(x_i)$ if $\ver(\tilde \rho^i_{i}, \E_{f_1}(x_i)) = 1$, otherwise it returns an abort state $\ket{{\perp}}\bra{\perp}$ corresponding to $\perp$. Hence, to avoid getting state $\ket{\perp}$ from the HLPUF, the adaptive adversaries $\Aad$ need to produce a query of the form $(x_i, \E_{f_1}(x_i))$. As the adversary doesn't have any direct access to the mapping $\E_{f_1}$, the only way it can get any information about $\E_{f_1}(x_i)$ by intercepting the challenges that are sent by the server to the client. Suppose that the adaptive adversary has access to a set of $q$ queries $X_{[q]} := \{X_i\}_{1\leq i \leq q}$ and the corresponding responses $\Psi_{[q]} := \{\E_{f_1}(x_i)\}_{1\leq i \leq q}$. Here each $X_i$ follows a uniform distribution over the challenge set $\{0,1\}^n$. Hence, for the mapping $\E_{f_1}$ the power of the adaptive adversary reduces to the power of a weak adversary. As $\E_{f_1}$ has the universal unforgeability property against any $q$-query weak adversary, hence we get, for any random challenge $X \not \in X_{[q]}$,

\begin{equation}\label{eq:unforge_e1_1}
    \Pr_{X,X_{[q]}}[1 \leftarrow \Gea(\Aad, m,X,X_{[q]})] =\Pr_{X,X_{[q]}}[1 \leftarrow \Gea(\Ana, m, X, X_{[q]})]\leq\epsilon(m).
\end{equation}

This implies, that using the set of challenges $X_{[q]}$ and responses $\Psi_{[q]}$ the adversary cannot produce the response corresponding to a random challenge $X \not \in X_{[q]}$. 
Suppose from the query set $X_{[q]}$ and the responses, the adaptive adversary successfully generates a set $X'_{[q']}$ of $q'$ adaptive queries, and corresponding responses $\Psi_{[q']}$ for the HLPUF $\E^L_f$. Without any loss of generality we assume that for all of the queries $X'_i \in X'_{[q']}$ the HLPUF returns a non-abort state.  We assume that the adaptive adversary wins the universal unforgeability game using the query set $ X_{\text{ad}} = X_{[q]} \cap X'_{[q']}$. This implies,

\begin{equation}\label{eq:assump_el}
    \Pr_{X,X^{\E_L}_{[q]_{\text{ad}}}}[1 \leftarrow \Gel(\Aad, m, X, X_{\text{ad}})] \geq \text{non-negl}(m).
\end{equation}

\noindent From the HLPUF \constref{cons:hlpuf}, we get that winning the universal unforgeability game with the HLPUF $\E^L_{f}$ implies winning the universal unforgeability with $\E_{f_2}$. Hence, we can rewrite \eqref{eq:assump_el} in the following way,

\begin{equation}
    \label{eq:assump_e2}
    \Pr_{X,X_{\text{ad}}}[1 \leftarrow \Geb(\Aad, m, X, X_{\text{ad}})] \geq  \text{non-negl}(m).
\end{equation}

Note that, if the adaptive adversary manages to get non-abort outcomes from the HLPUF corresponding to all $X'_i \in X_{\text{ad}}$ then from the \constref{cons:hlpuf} we get, $1 \leftarrow \Gea(\Aad, m, X'_i, X_{\text{ad}})$. Due to the unforgeability assumption of Equation \eqref{eq:unforge_e1_1} we have,

\begin{equation}\label{eq:assump_e1_2}
    \Pr_{X,X_{[q]}}[1 \leftarrow \Gea(\Ana, m,X,X_{[q]})] = \Pr_{X,X_{\text{ad}}}[1 \leftarrow \Gea(\Aad, m, X, X_{\text{ad}})] \leq   \epsilon(m).
\end{equation}

Note that, the main difference between adaptive and weak adversaries lies in the choice of the query set. If we fix the query set $X_{\text{ad}}$, then both adaptive $\Aad$ and a weak adversary can extract the same amount of information from the responses corresponding to the query set $X_{\text{ad}}$. Therefore, their winning probability of the universal unforgeability game becomes equivalent. This implies, we can rewrite Equation \eqref{eq:assump_e1_2} in the following way,

\begin{equation}
    \Pr_{X,X_{\text{ad}}}[1 \leftarrow \Gea(\Aad, m, X, X_{\text{ad}})] \label{eq:assump_e1_3} = \Pr_{X,X_{\text{ad}}}[1 \leftarrow \Gea(\Ana, m, X, X_{\text{ad}})]\leq \epsilon(m).
\end{equation}

By combining Equation \eqref{eq:assump_e1_2} and Equation \eqref{eq:assump_e1_3} we get, both the random variables $X_{[q]}$ and $X_{\text{ad}}$ are equivalent. 
From the universal unforgeability property of the PUF $\E_{f_2}$ against any $q$-query weak adversary, we get

\begin{equation}\label{eq:unforge_e2_1}
     \Pr_{X,X_{[q]}}[1 \leftarrow \Geb(\Ana, m, X, X_{[q]})] \leq  \epsilon(m).
\end{equation}
As both of the random variables $X_{[q]}$ and $X_{\text{ad}}$ are equivalent, so we get, 

\begin{equation}\label{eq:final_thm_2}
    \begin{split}
        & \Pr_{X,X_{[q]}}[1 \leftarrow \Geb(\Ana, m, X, X_{[q]})] \\
        & = \Pr_{X,X_{\text{ad}}}[1 \leftarrow \Geb(\Ana, m, X, X_{\text{ad}})]\\
        &= \Pr_{X,X_{\text{ad}}}[1 \leftarrow \Geb(\Aad, m, X, X_{\text{ad}})] \leq \epsilon(m). 
    \end{split}
\end{equation}

The second equality follows from the fact that for a fixed query set $X_{\text{ad}}$ the adaptive adversary $\Aad$ and weak adversary $\Ana$ becomes equivalent. Note that, only one of \eqref{eq:assump_e2} and \eqref{eq:final_thm_2}  is true. The \eqref{eq:final_thm_2} is true because of the unforgeability of $\E_{f_2}$. Hence, our assumption of \eqref{eq:assump_e2} is wrong. Therefore, \eqref{eq:assump_el} is also not true. Hence, with conclude our proof by contradiction.
\end{proof}

Apart from the theoretical results provided in this section, we have also simulated the design of HPUF constructions with underlying silicon CPUFs instantiated by \emph{pypuf} \cite{wisiol_nils-wisiolpypuf_2021} which is a python-based emulator that features different existing CPUFs. Furthermore, we simulate the situation where an adversary acquires classical challenges and quantum-encoded responses from HPUF and converts the response into classical bit string by measurement behaviour. The adversary then performs some machine learning-based attacks with CRPs to reproduce a model that accurately predicts enough the behaviours of underlying CPUF. Such an adversary possibly forges the HPUF given (exponentially in the security parameter) many CRPs. Our simulation results assist to demonstrate the exponential gap in the security between CPUF and HPUF in a regime outside the polynomial-size database and for classical PUFs that are commercially available. Since the simulations have not been done by the author, we have excluded them from this chapter. However, we refer the reader to \cite{chakraborty_quantum_2021} for the full work, including the simulation results.

\subsubsection{Security of the HLPUF-based authentication protocol:}
We now have all the elements to be able to prove the completeness and security (or soundness) of our HLPUF-based authentication protocol. Firstly, we define the completeness and security property for \protoref{prot:hlpuf}. Then, in \thmref{thm:HLPUF_auth} we will prove that they are satisfied. We start with the completeness:

\begin{defbox}
\begin{definition}[Completeness of HLPUF-based Authentication~\protoref{prot:hlpuf}]\label{def:comp_lpuf}
We say the HLPUF-based authentication \protoref{prot:hlpuf} satisfies completeness if in the absence of any adversary, an honest verifier and prover generating $\ket{\psi_{f_1(x_i)}}\bra{\psi_{f_1(x_i)}}$ and $\ket{\psi_{f_2(x_i)}}\bra{\psi_{f_2(x_i)}}$ with a valid HLPUF for any selected challenge $x_i$, can pass the verification algorithms with overwhelming probability:
\begin{equation}
   Pr[\texttt{Ver}(\ket{\psi_{f_1(x_i)}}\bra{\psi_{f_1(x_i)}},\tilde \rho_1) =\texttt{Ver}(\ket{\psi_{f_2(x_i)}}\bra{\psi_{f_2(x_i)}}, \tilde{\rho}_2)= 1] \geq 1 - \epsilon(\lambda)
\end{equation}
\end{definition}
\end{defbox}

We also define the security of the protocol, in relation to the universal unforgeability game as follows: 

\begin{defbox}
\begin{definition}[Security of the HLPUF-based Authentication~\protoref{prot:hlpuf}] 
\label{def:secu_lpuf}
We say the HLPUF-based authentication \protoref{prot:hlpuf} is secure if the success probability of any QPT adaptive adversary $\Aad$ in winning the universal unforgeability game to forge an output of HLPUF according to \constref{cons:hlpuf}, for any randomly selected challenge of the form $\tilde{c} = (x,\ket{\psi_{f_1(x)}}\bra{\psi_{f_1(x)}})$ is at most negligible in the security parameter:
\begin{equation}
    Pr[1\leftarrow \Gnhl(\Aad, \lambda)] \leq \epsilon(\lambda)
\end{equation}
where the verification algorithm of the universal unforgeability game checks the adversary's output $\sigma_2$, with the output of the HLPUF, $\ket{\psi_{f_2(x)}}\bra{\psi_{f_2(x)}}$.
\end{definition}
\end{defbox}

Through the following theorem, we can see that \protoref{prot:hlpuf} satisfies both completeness and security according to the above definitions.

\begin{thmbox}
\begin{theorem}
\label{thm:HLPUF_auth}
If the HLPUF $\E^L_f$ is constructed from a hybrid PUF $\E_f$ using \constref{cons:hlpuf}, then the HLPUF-based authentication \protoref{prot:hlpuf} satisfies both the completeness and security conditions.
\end{theorem}
\end{thmbox}

\begin{proof}
In \protoref{prot:hlpuf} with hybrid PUF $\E_f = \E_{f_1} \otimes \E_{f_2}$, the verifier(server) chooses the classical input $x_i\in\X$, encodes the quantum state corresponding to $2m$ bits of $f_1(x_i)$ and issues the joint state to the prover(client). If there is no adversary, the prover receives the joint state and queries $\E_f$ with $x_i$ and $\tilde \rho_1$, where $\tilde \rho_1=\E_{f_1}(x_i)=\ket{\psi_{f_1(x_i)}}\bra{\psi_{f_1(x_i)}}$ for the first $m$ qubits of $\E_{f}(x_i)$. Hence we have:
\begin{equation}
    \label{eq:thm_cs_1}
    Pr\left[\texttt{Ver}(\ket{\psi_{f_1(x_i)}}\bra{\psi_{f_1(x_i)}},\tilde \rho_1)=1\right] = 1  
\end{equation}

On the prover's side, since the verification algorithm of HLPUF $\E_f^L$ always passes with $\texttt{Ver}(\ket{\psi_{f_1(x_i)}}\bra{\psi_{f_1(x_i)}},\tilde \rho_1)=1$, and they return the quantum state $\E_{f_2}(x_i)=\ket{\psi_{f_2(x_i)}}\bra{\psi_{f_2(x_i)}}$ corresponding to $2m$ bits of $f_2(x_i)$ to the verifier. Without the presence of adversary, the verifier always receives the state with $\tilde{\rho}_2=\ket{\psi_{f_2(x_i)}}\bra{\psi_{f_2(x_i)}}$, and we obtain the equation similarly to \eqref{eq:thm_cs_1}. Therefore, we can say the HLPUF-based authentication protocol satisfies the completeness condition with
\begin{equation}
    \label{eq:thm_cs_2}
    \text{Pr}\left[\texttt{Ver}(\ket{\psi_{f_1(x_i)}}\bra{\psi_{f_1(x_i)}},\tilde \rho_1)=\texttt{Ver}(\ket{\psi_{f_2(x_i)}}\bra{\psi_{f_2(x_i)}}, \tilde{\rho}_2)= 1\right] = 1
\end{equation}

On the other hand, for the security property, we rely on \thmref{th:application-hpuf_hlpuf-adaptive-secure} that the HLPUF $\E_f^L$ is $(\epsilon,m)$-secure against any QPT adaptive adversary (a $q$-query adaptive adversary for any $q$ polynomial in the security parameter). For both $\E_{f_1}$ and $\E_{f_2}$ of HPUF $\E_f$, we show that the power of an adaptive adversary can be reduced to the power of a weak adversary, due to the locking mechanism. Also since $\E_{f_1}$ has the universal unforgeability against a weak adversary by definition, for any adaptive query of the form $(x_i, \sigma_1)$ that an adaptive adversary issues to the HLPUF, the following applies: 
\begin{equation}
    \label{eq:thm_cs_3}
    \text{Pr}\left[\texttt{Ver}(\ket{\psi_{f_1(x_i)}}\bra{\psi_{f_1(x_i)}}, \sigma_1)= 1\right] \leq \epsilon(m)
\end{equation}
Where $\ket{\psi_{f_1(x_i)}}\bra{\psi_{f_1(x_i)}}$ is the correct response constructed from CPUF according to HPUF construction. Thus the power of the adaptive adversary reduces to the power of weak adversary and we have:
\begin{equation}
    Pr[1\leftarrow \Gnhl(\Aad, m)] \approx Pr[1\leftarrow \Gnhl(\Ana, m)]
\end{equation}
Now given the fact that the adaptive adversary cannot boost from the weak-learning phase to the HPUF, producing a forgery $\sigma_2$ for the HLPUF that passes the verification $\texttt{Ver}(\ket{\psi_{f_2(x_i)}}\bra{\psi_{f_2(x_i)}}, \sigma_2)$, reduces to forging the HPUF $\E_{f_2}$. Again by assumption, $\E_{f_2}$ has the universal unforgeability against weak adversary, hence we have:
\begin{equation}
    \label{eq:thm_cs_4}
    Pr[1\leftarrow \Gnhl(\Aad, m)] = Pr[1\leftarrow \Gnh(\Ana, m)] \leq \epsilon(m)
\end{equation}
This concludes the proof.
\end{proof}

Therefore, we have shown that \protoref{prot:hlpuf}, under certain reasonable assumptions on the underlying classical PUF, is correct and achieves suitable security against QPT adversaries.

\subsection{Challenge re-usability}
\label{sec:application-hlpuf-reusability}
In any PUF-based protocol relying on the classical communication of challenges and responses of the PUF, each challenge can only be used once as the adversary can simply copy and record the challenges and responses and have a perfect copy of the challenger's database which later they can use to falsely identify themselves. This is an important limitation of the classical PUFs~\cite{suh_physical_2007,herder_physical_2014}. Quantum communication can solve this issue due to the unclonability of quantum states. In this section, we discuss how our hybrid construction can allow for challenge states to be used several times during the authentication, under the circumstances of previous successful authentication rounds. This property will resolve an important practical issue as the challenger can avoid storing a big database or renewing the database of challenge responses frequently.

First, we need to clarify the conditions under which the challenge can be reused. We assume the challenger's database to only include $q$ number of challenge-response pairs such that $q$ is polynomial in the security parameter. We also need to recall that in our hybrid construction, the challenges are still being sent as \emph{classical} bit-strings over the public channel, hence the adversary, after polynomial rounds of communication, can have the same challenge set as the server's database. Due to this fact, we should emphasize that the adversary does not get any physical access to the internal classical PUF in the HLPUF construction during the authentication and no query can be directly issued to the CPUF by the adversary. This condition is satisfied using our locking mechanism. Thus, the adversary has access to the following information: a pre-learnt polynomial-size local database of challenge-responses of the CPUF, a set of classical challenges used during the protocol, and the set of quantum states that encode either the first or second half of the response, in the BB84 states.

It is a straightforward observation that the challenges for which the verification test has failed should never be used again. A trivial attack, in this case, would be that the adversary intercepts the communication and stores the response state, and later when the same challenge has been queried again, will re-send the stored correct response state to pass the verification. As a result, all the challenges in the failed rounds should be discarded.

Nonetheless, we argue that in the events of successful authentication, the challenges can be re-used. Here, by successful identification, we mean that the received response state passes the verification on the client and server sides and the prover identifies an honest party. Even though the events of false identification of an adversary, is still possible (for example, if the challenge is the same as one of the challenges that previously exists in the adversary's local database), the unforgeability of PUF and our security proof for the hybrid construction, ensures that these events occur only with negligible probability. 

We are thus interested in the eavesdropping attacks by the adversary on the first and second half of the response states that are of the form $\ket{\psi_{f_1(x_i)}}\bra{\psi_{f_1(x_i)}}=\bigotimes_{j=1}^m \ket{\psi^{i,j}_{f_1(x_i)}}\bra{\psi^{i,j}_{f_1(x_i)}}$ and $\ket{\psi_{f_2(x_i)}}\bra{\psi_{f_2(x_i)}}=\bigotimes_{j=1}^m \ket{\psi^{i,j}_{f_2(x_i)}}\bra{\psi^{i,j}_{f_2(x_i)}}$
Note that eavesdropping on the states which encode the first part of the response will lead to breaking the locking mechanism while eavesdropping on the second half will lead to an attack on the identification. Without loss of generality, we only consider one of the cases where the adversary wants to eavesdrop on the first (or second) half to break the protocol in the upcoming rounds where the challenge is re-used. The arguments will hold equivalently for both cases since the states and verification are symmetric.

Given all these considerations, the challenge re-usability problem will reduce to the optimal probability of the eavesdropping attack on state  $\ket{\psi_{f_1(x_i)}}\bra{\psi_{f_1(x_i)}}$ which is in fact $m$ qubit states encoded in conjugate basis same as BB84 states. In the most general case, the adversary can perform any arbitrary quantum operation on the state $\bigotimes_{j=1}^m |\psi^{i,j}_{f_1(x_i)}\rangle\langle\psi^{i,j}_{f_1(x_i)}|$ or separately on each qubit state $\ket{\psi^{i,j}_{f_1(x_i)}}$, together with a local ancillary system and sends a partial state of this larger state to the verifier to pass the verification test, and keep the local state to extract the encoded response bits. Let $\rho_{SEC}$ be the joint state of the server, the eavesdropper and the client. Since the states used in the protocol are from Mutually Unbiased Basis (MUB) states \emph{i.e.} from either $Z = \{\ket{0}, \ket{1}\}$ or $X = \{\ket{+}, \ket{-}\}$, in order to show the optimal attack, we can rely on the entropy uncertainty relations that have been used for the security proof of QKD. The measurements for verification are also performed in the $\{Z,X\}$ bases accordingly. We use the entropy uncertainty relations from~\cite{coles_entropic_2017} where the security criteria for QKD have been given in terms of the conditional entropy for MUBs measurements. Using these results we show that the entropy of Eve in guessing the correct classical bits for the response is very high if the state sent to the verification algorithm passes the verification with a high probability. Intuitively this is due to the uncertainty that exists related to the commutation relation between $X$ and $Z$ operators in quantum mechanics. Hence we conclude that the success probability of Eve in extracting information from the encoded halves of the response is relatively low. Also, we show that this uncertainty increases linearly with $m$ similar to the number of rounds for QKD. This argument results in the following theorem. In proving this theorem, we have used the entropic uncertainty relation introduced in \chapref{chap:prelim}, Section~\ref{sec:prelim-uncert}.

\begin{thmbox}
\begin{theorem}\label{th:chreuse-unc-formal}
In \protoref{prot:hlpuf}, let $x$ be a challenge and $(y_{1},\dots,y_{2m})$ be the response of a classical PUF used within the HPUF construction, with randomness bias $p = (\frac{1}{2} + \delta_r)^{2m}$ over the classical responses. If the verification algorithm for a state $\tilde{\rho}$ passes with probability $1 - \epsilon(m)$, then Eve's conditional min-entropy $H_{min}^{Eve}$ in terms of von Neumann entropy over the verifier/prover's (server/client) classical response, satisfies the following inequality:
\begin{equation}
  H_{min}^{Eve} = H_{min}(S^m|ER^m) \geq m - \epsilon(m)  
\end{equation}
\end{theorem}
\end{thmbox}

\begin{proof}
We prove this theorem based on the first half of the state used in \protoref{prot:hlpuf}, \emph{i.e.} the state $\ket{\psi_{f_1(x_i)}}\bra{\psi_{f_1(x_i)}}=\bigotimes_{j=1}^m |\psi^{i,j}_{f_1(x_i)}\rangle\langle\psi^{i,j}_{f_1(x_i)}|$ that is being sent by the verifier/server denoted by (S) and received and measured by the prover/client denoted by (C). Nevertheless, we note that the same proof applies for the second state due to the symmetry of the states and the protocol. 

Let $R^m = (R_1,\dots,R_m)$ be the randomness bitstring showing the choice of the basis encoding of the response, $S^m = (S_1,\dots,S_m)$ be the server's bit encoded in the $R^m$ bases. Note that both $R^m$ and $S^m$ are produced according to the bitstring $(y_{1},\dots,y_{2m})$ which is the first half of the response of CPUF to a given challenge $x$. Also, let $C^m = (C_1,\dots,C_m)$ be the client's correct bit string. We denote the arbitrary joint state of three systems by $\rho_{S^m E C^m}$ where $E$ denotes any arbitrary quantum system held by the eavesdropper. Now, let the the Client's measurement outcomes, after the verification be $\tilde{Y}^m = (\tilde{Y_1},\dots,\tilde{Y_m})$ which shows the estimated bits by the Client. Now we can write the tripartite uncertainty principle, in terms of the von Neumann entropy, for MUB measurements and MUB states as follows:
\begin{equation}\label{eq:uncert-tripartite}
    H(X_1 X_2 Z_3 X_4 \dots X_{m-1} Z_m | E) + H(Z_1 Z_2 X_3 Z_4 \dots Z_{m-1} X_m | C) \geq \log_2 (\frac{1}{c})^m
\end{equation}
where $c = \max_{x,z} c_{xz}$ and $c_{xz} = \parallel \sqrt{M^x}\sqrt{N^z} \parallel^2$ for an arbitrary POVM sets $M = \{M^x\}_x$ and $N = \{N^z\}_z$. We note that if the CPUF creates perfect random bitstring for $R^m$ then states are perfect MUB states and $c = \frac{1}{2}$. Nonetheless we consider a weaker CPUF with a biased distribution of $p = (\frac{1}{2} + \delta_r)^{2m}$ in creating $0$s and $1$s in the response. Hence, we can translate this imperfectness into a disturbance in the measurement bases. Let $M^0 = \ket{0}\bra{0}$ and $M^1 = \ket{1}\bra{1}$ be the usual measurement in the computational basis but let the $N$ measurements be a slightly shifted version of the measurements in the $X$ basis. Consider the following states:
\begin{equation}
\begin{split}
        & \ket{\psi_N} = \sqrt{\frac{1}{2} + \delta_r}\ket{0} + \sqrt{\frac{1}{2} - \delta_r}\ket{1} \\
        & \ket{\psi^{\perp}_N} = \sqrt{\frac{1}{2} - \delta_r}\ket{0} - \sqrt{\frac{1}{2} + \delta_r}\ket{1} 
\end{split}
\end{equation}
We define the new $N$ projective operators according to the following states as $N^0 = \ket{\psi_N}\bra{\psi_N}$ and $N^1 = \ket{\psi^{\perp}_N}\bra{\psi^{\perp}_N}$. Now we calculate the operator norm for all the pairs of measurements and we have:
\begin{equation}
\begin{split}
        &  \parallel\sqrt{M^0}\sqrt{N^0}\parallel^2 = \frac{1}{2} + \delta_r, \quad  \parallel\sqrt{M^0}\sqrt{N^1}\parallel^2 = \frac{1}{2} - \delta_r\\
        &  \parallel\sqrt{M^1}\sqrt{N^0}\parallel^2 = \frac{1}{2} - \delta_r, \quad  \parallel\sqrt{M^1}\sqrt{N^1}\parallel^2 = \frac{1}{2} + \delta_r
\end{split}
\end{equation}
Thus we conclude that $c = \frac{1}{2} + \delta_r$ and the Equation~\eqref{eq:uncert-tripartite} can be re-written as follows:
\begin{equation}
    H(X_1 X_2 Z_3 X_4 \dots X_{m-1} Z_m | E) + H(Z_1 Z_2 X_3 Z_4 \dots Z_{m-1} X_m | C) \geq m  - m\log_2 (1 + 2\delta_r)
\end{equation}
Now, we use the data processing inequality~\cite{coles_entropic_2017}, we have got the following security criteria that show Eve's uncertainty (in terms of the von Neumann entropy) of the actual response bits $S^m$:
\begin{equation}
    H(S^m | ER^m) + H(S^m | \tilde{Y}^m) \geq m  - m\log_2 (1 + 2\delta_r)
\end{equation}

We can get the same inequality in terms of smooth min and max entropy \cite{coles_entropic_2017,tomamichel_uncertainty_2011} (see Section~\ref{sec:prelim-uncert}), which is more appropriate for ensuring the security in the finite size, for min and max entropy we equivalently have:

\begin{equation}\label{eq:sec-criteria-uncert}
    H^{\epsilon}_{min}(S^m | ER^m) \geq m  - H^{\epsilon}_{max}(S^m | \tilde{Y}^m) - m\log_2 (1 + 2\delta_r)
\end{equation}
To calculate the above bound we need to find the bound on the second term of the right-hand side, \emph{i.e.} $H^{\epsilon}_{max}(S^m | \tilde{Y}^m)$. Here we use another result from~\cite{tomamichel_uncertainty_2011} where it states that for any bitstring $X$ of $n$ bit and the respective measurement outcome $X'$, which at most a fraction $\zeta$ of them disagree according to the performed statistical test, then the smooth max entropy is bounded as follows:
\begin{equation}
    H^{\epsilon}_{max}(X|X') \leq nh(\zeta)
\end{equation}
where $h(.)$ denotes the classical binary Shannon entropy. Now we can use this result and our assumption of successful verification together. Given the assumption that the verification is passed with a probability $1 - \epsilon(m)$, and the verification algorithm consists of measuring the states in the $Z$ and $X$ bases, we can conclude that the final bits differ in at most a fraction $\zeta = \epsilon(m)$ where $\epsilon(m)$ is a negligible function. As a result, we have:
\begin{equation}\label{eq:max-entropy}
    H^{\epsilon}_{max}(S^m | \tilde{Y}^m) \leq mh(\zeta) \approx m \epsilon(m)
\end{equation}
Putting Equations~\eqref{eq:sec-criteria-uncert} and~\eqref{eq:max-entropy} together, we have:
\begin{equation}
    H^{\epsilon}_{min}(S^m | ER^m) \geq m  - m \epsilon(m) - m\log_2 (1 + 2\delta_r)
\end{equation}
On the right-hand side of the above inequality, the second term is still a negligible function and the third term depends on the CPUF bias probability distribution. We assume the CPUF satisfies $p$-randomness, as defined in the \defref{def:p-randomness}, thus the $\delta_r$ is a small value and hence the term $(1 + 2\delta_r)$ is negligibly close to $1$, which means that the third term, is negligibly close to $0$ in the security parameter which is $m$. Finally, we conclude that:
\begin{equation}
    H^{Eve}_{min} = H^{\epsilon}_{min}(S^m | ER^m) \geq m  - \epsilon'(m)
\end{equation}
where $\epsilon'(m)$ is a negligible function and the proof is complete.
\end{proof}

Let us see how the abouve information-theoretic bound can be used to prove the challenge-reusability of the \protoref{prot:hlpuf}. First, define the re-usability in relation with the unforgeability game and then using \thmref{th:chreuse-unc-formal}, we prove the challenge re-usability of our protocol.

\begin{defbox}
\begin{definition}[Challenge ($k$-)re-usability in the universal unforgeability game]
Let $\Gnre(\lambda, \A, x_{k+1})$ be a special instance of the universal unforgeability game, where a challenge $x$, picked uniformly at random by the challenger, has been previously used $k$ times. We are interested in the events where the same challenge is used in the $(k+1)$-th round, which we denote by $x_{k+1}$. We say the challenge $x$ is \textit{($k$-)re-usable} if the success probability of any QPT adversary in winning $\Gnre(\lambda, \A, x_{k+1})$, i.e, in forging message $x_{k+1}$, is negligible in the security parameter:
\begin{equation}
    p_{forge}(\A, x_{k+1}) = Pr[1 \leftarrow \Gnre(\lambda, \A, x_{k+1})] \leq \epsilon(\lambda)
\end{equation}
\end{definition}
\end{defbox}

\begin{thmbox}
\begin{theorem}[Challenge re-usability of HLPUF-based Authentication \protoref{prot:hlpuf}] A challenge $x$ can be reused $k$ times during the \protoref{prot:hlpuf} as long as the received respective response $\sigma$ for each round passes the (client's or server's) verification with overwhelming probability. In other words, given the successful verification, the success probability of any quantum adversary in passing the $(k+1)$-th round with the same challenge $x$ is bounded as follows:
\begin{equation}
    p_{forge}(\A, x_{k+1}) \leq k 2^{-m} \approx \epsilon(m).
\end{equation}
\end{theorem}
\end{thmbox}

\begin{proof}
To prove this theorem, we use \thmref{th:chreuse-unc-formal} directly. First, we assume that $x$ has been used one time before in a previous round. Given the assumption that the verification is passed with probability $1 - \epsilon(m)$, and this theorem, we conclude that the uncertainty of the adversary in guessing the encoded response of the HLPUF is larger than $m - \epsilon(m)$. In our case, the joint quantum state between the server and the adversary is a classical-quantum state (server has the classical description of $f(x)$, and the adversary has the quantum state $\ket{\psi_{f(x)}}$). For such states, Eve's uncertainty, $H_{min}^{Eve}$ is same as $-\log p_{guess}^{Eve}$, where $p_{guess}^{Eve}$ is Eve's guessing probability of the classical information encoded in the quantum state \cite{konig_operational_2009}. Therefore,
\begin{equation}
\begin{split}
    p_{guess}^{Eve} & = 2^{-H_{min}^{Eve}}\\
    & \leq 2^{-m + \epsilon(m)}
\end{split}
\end{equation}
This probability is negligible in the security parameter, which means that after performing any arbitrary quantum operations, the adversary's local state includes at most, a negligible amount of information on the response of $x$, each round that the state $x$ is reused. Now, we can use the union bound (See Preliminaries, \ref{sec:}) to show that this success probability only linearly scales with $k$:
\begin{equation}
    p_{guess}^{Eve,k} = Pr(\bigcup^k_{i=1} E_{guess}^i) \leq \sum^k_{i=1} p(E_{guess}^i) \approx k2^{-m}
\end{equation}
where $E_{guess}^i$ are the events where Eve correctly guesses the response and where $p(E_{guess}^i)=(p_{guess}^{Eve})^i$ is the success probability of Eve in guessing in the $i$-th round. Finally, let the success probability of an adversary in the universal unforgeability game for the HLPUF be upper-bounded by $\epsilon_1(m)$ which is a negligible function in the security parameter since we assume that the HLPUF satisfies the universal unforgeability. This is the same as the success probability of the adversary in passing the verification for a new challenge, chosen at random from the database. Now in the $(k+1)$-th round, where the same $x$ is reused, the success probability is at most boosted by the guessing probability over the previous $k$-th rounds, hence we will have:
\begin{equation}
    p_{forge}(\A, x_{k+1}) \leq \epsilon_1(m) + k2^{-m} = \epsilon(m)
\end{equation}
As long as $k$ is polynomial in the security parameter, the second term is also a negligible function and since the sum of two negligible probabilities will be also negligible. This concludes the proof.
\end{proof}

\section{Discussion and conclusions}
We have proposed three different identification protocols based on quantum PUFs and Hybrid classical-quantum PUFs which provide exponential (or relatively exponential) security against any QPT adversary by exploiting physical unclonability (both in the quantum and classical sense) as a hardware assumption instead of the usual cryptographic assumptions. The first two protocols use full quantum PUFs and the last one combines a classical PUF with quantum encoding to give rise to our hybrid construction. Our primary classification in the first two protocols has come about from the practical scenarios in a network, \emph{i.e.} parties with varying capabilities should be able to run a secure identification protocol. The first protocol, \hrv, is proposed to be suited more in the mobile-like device settings \emph{i.e.} provers having low resources would want their device to be correctly identified by a high resource verifier. This protocol can be used as a subroutine for many other applications, specifically in a quantum internet or a quantum network with \emph{star-like} architecture \cite{chen_field_2009,pirandola_physics_2016,linke_experimental_2017}, where a server (central node) can use this protocol to identify each of the clients. Also, since the identification protocol requires multi-round communication between the prover and the verifier, we have proposed efficient quantum equality-testing verification to reduce the communication overhead requirement.

Our second protocol, \lrv, is suited to the quantum verification setting \emph{i.e.} a low-resource and classical-like verifier would want to verify the identity of a high resource quantum device, like a quantum cloud server. The advantage of this protocol is that a purely classical verification algorithm is sufficient to verify the prover's device with provable security. {\lrv} is based on the idea of trapification, where the verifier inserts random trap states in between the communication rounds which facilitates a secure delegation of the quantum testing to the prover. This allows the verifier to simply run a classical algorithm on the quantum test outcomes to perform successful identification. We have also shown an extension of the {\lrv} protocol that generalises it to an arbitrary distribution of traps instead of randomly inserting them in half of the positions as proposed in the current version of the protocol. With this generalisation on hiding the trap distribution, one hopes for further enhancement in security against a QPT adversary. We draw non-trivial conclusions from this generalisation, including the worsening of security to polynomial in the number of communication rounds (instead of exponential as our current protocol) when the number of trap positions is chosen uniformly over the total positions. We also remark that some distributions provide a polynomial enhancement over the current exponential security bound, thus justifying the need for hiding the number of trap positions. We also believe that this generalisation provides a potential use case of a similar protocol for a certain degree of verification and certification of quantum devices using embedded quantum PUFs. We see this direction as an engaging future subject of study since qPUFs provide natural and physical randomness, which combined with more enhanced trapification techniques, can potentially lead to a new class quantum verification/certification protocols.

An important future direction regarding the practicality of the presented protocols is to study the effect of noise and robustness of the protocols under honest noise. So far, the protocols have been studied in the noiseless setting, and the requirements of verification are strictly rigid (for instance, all rounds of verification have to be successfully passed), however, both protocols can be made more robust by relaxing some of these requirements and allowing for a tolerance parameter while remaining secure. We leave the study of the robustness and security properties of the protocol in the noisy setting for future work. 

Then, we have used our result regarding efficient universal unforgeability, to reduce the Haar-random sampling of our proposed protocols to pseudorandom quantum states that are efficient to generate, as a result, introducing a more implementation-friendly version of our qPUF-based protocols. Quantum pseudorandomness is yet a very young field of research, and we believe the advancement in this field can assist the qPUF constructions and the protocols based on them to become more and more practical.

Finally, in our latest proposal, we have exploited one of the main sources of security in the quantum information world, namely the concept of conjugate coding \cite{broadbent_quantum_2016}, to propose a new construction that uses an internal (and almost weak) classical PUF and enhances its unpredictability to a high degree using this quantum encoding. Nonetheless, this enhancement is only against non-adaptive adversaries. For security against our usual QPT adversary that is in general adaptive, we use an additional technique, namely the locking mechanism that provides us with our HLPUF construction used in our proposed authentication protocol. An important property of the new construction is the combination of classical challenges and quantum responses, which harnesses the power of quantum information over an untrusted quantum channel while the verifier does not need to store the responses quantumly, which fully removes the quantum-memory requirement. As a result, the implementation of hybrid PUF is practical nowadays with quantum communication technology. Another advantage of the HLPUF-based protocol is that each challenge-response pair used for a successful authentication round can be used several times for authentication due to the unclonability and other fundamental quantum mechanical properties of the response quantum states. Therefore, with our solution, a server can continue the client authentication protocol for a longer period without exhausting its CRP database. This result overcomes the fundamental drawbacks of the existing classical PUF-based authentication protocols and offers a novel use case, not only for our construction but also for quantum communication in general. 

However, there are several thought-provoking questions, yet to be explored. In bounding the success probability of a QPT adversary against HPUF in \thmref{th:hpuf-security-prob}, we have established a connection between unforgeability and the learnability of a classical function (in our case CPUF's evaluation function) from a quantumly encoded random set of data, using quantum informatics approaches. Despite the current proof being specific to our construction, we believe that most of the techniques we have used are fairly general and can be used to formally establish a link between cryptographic properties and learning problems, using quantum information theory. Furthermore, if the same result can be generalised to bound the success probability of a QPT adversary respective to a classical PPT adversary, for learning a classical function from a general quantum encoding, then one can relate the problem to the open question of the advantage of \emph{supervised learning} with quantum models \cite{schuld_effect_2021,servedio_equivalences_2004}. The use of quantum information theory in learning theory has recently led to influential results regarding the comparison between classical and quantum machine learning \cite{huang_information-theoretic_2021}. Therefore, we conjecture that further expanding this connection to cryptography and quantum-information-based cryptography can provide new insights into these very challenging and exciting problems.


We finally discuss the experimental requirement of our HLPUF-based proposal. We note that by selecting the quantum encoding to be BB84 states, the protocol can be implemented with resources similar to quantum key distribution. QKD technology is one of the most mature quantum technologies. The long-distance QKD networks are already implemented and used in many different countries like the USA, UK, China, EU, Japan, \cite{sasaki_field_2011,stucki_long-term_2011,poppe_outline_2008,wang_field_2014,courtland_chinas_2016} etc. Many commercially available QKD infrastructures provide almost $300$kb/s secret key rate over optical fibre links of length $120$km \cite{frohlich_long-distance_2017}. Moreover, the availability of the mature QKD on-chip technology \cite{sibson_chip-based_2017,semenenko_chip-based_2020,bunandar_metropolitan_2018} makes all the proposed constructions in this work implementable inside the IoT devices. Given all these available technologies, our proposal can solve almost all of the shortcomings of the device authentication problem. To further study the feasibility and practicality of hybrid PUF constructions, an important future direction would be toward the experimental implementation of our proposal and the HLPUF-based authentication protocol. Furthermore, we believe that due to the matching of required resources and the strong security guarantee of our protocol, it can be easily incorporated with the QKD itself as a promising solution for providing the authenticated channel that is required for QKD \cite{scarani_security_2009}. Hence an immediate and important future research direction would be the usage of the protocol for the task of message authentication and the composition of this protocol with QKD. Furthermore, since QKD has been proven composable secure \cite{ben-or_universal_2005,renner_security_2008,tomamichel_largely_2017,leverrier_composable_2015}, an important future direction would be to study the security of the proposed protocol within the existing composable frameworks such as universal composability framework \cite{canetti_universally_2001} or abstract cryptography framework \cite{maurer_abstract_2005,maurer_constructive_2012}.

\chapter{Variational Quantum Cloning: A New Cryptanalysis Toolkit}\label{chap:varqlone}
\begin{chapquote}{Robert Ford, Westworld (S1.Ep2: Chestnut)}
``Everything in this world is magic, except to the magician.''
\end{chapquote}
\section{Introduction}\label{sec:varqlone-intro}
We have set out on a long journey to understand new aspects of `Unclonability', from a foundational point of view in discovering its relationship to randomness and learnability, all the way to introducing new applications in quantum cryptography. The three past chapters navigate around the concept of physical unclonability. In this chapter, we come back to the more familiar notion of quantum unclonability, that is, the no-cloning theorem and unclonability of quantum states.

In \chapref{chap:prelim} (Section \ref{sec:prelim-cloning}), we have covered the no-cloning theorem and the concept of approximate cloning. We have seen that it is both possible to create \emph{imperfect} copies of unknown quantum states (approximate cloning) or to have a quantum operation that only \emph{sometimes} gives you two perfectly similar copies of general quantum states (probabilistic cloning). Among these two categories, approximate cloning is particularly interesting for us since it somehow matches the idea that we have pursued in this thesis in understanding the fundamental relation between unclonability, learnability and the level of \emph{unknownness}. Intuitively we focus on the amount of information that exists prior to performing the cloning mechanism, about the entity that one aims to clone. As we have seen in Section~\ref{sec:unclone-unknown} and Section~\ref{sec:unclone-different-learning}. This specific prior information leads to different classes of cloners with the ability to clone the particular family of states corresponding to that information, where the quality (or more technically, fidelity) of these clones is higher than the universal cloner. To roughly summarise this argument, the more you know (or learn) about the states you want to copy, the better you can copy them. Having this inherent relation in mind, now we move to the field of quantum cryptography, where no-cloning is at the heart of the security of many quantum protocols (QKD, coin-flipping, verifiable blind quantum computing, etc.). Here, the following question arises:
\begin{center}
\emph{`In what ways the ability to clone a specific family of states with partial prior information can affect the security of quantum protocols?'}    
\end{center}

Despite the fact that the study of approximate cloning was born many years ago with the remarkable discovery of \emph{Buzeh and Hillery}~\cite{buzek_quantum_1997}, and despite the existence of a rich literature on the subject, there are still very limited classes of states that are known how to be cloned approximately with optimal fidelity \cite{scarani_quantum_2005, fan_quantum_2014}. More importantly, even for some of these known classes of cloning, the unitary circuits (or circuit decompositions) of these cloners are not known. Having these circuits and unitaries explicitly is not only important for practical applications of quantum cloning machines, but also a very relevant problem to quantum compilation. Notably, in touching on quantum cryptanalysis, this `prior information' becomes much more general and can include cases where our knowledge about the cloning machines (for the specific problem of interest) is narrow. Moreover, if cloning based on a specific family of the state is to be used as an attack model, the complexity of the circuit and technological feasibility of performing those cloners will be considerable factors, which once again calls for being able to have an explicit form of the cloning machines.

Quantum cloning has been previously considered as an attack model for some quantum protocols such as QKD. It turns out that in some cases, these types of attacks are in fact, optimal~\cite{scarani_quantum_2005}. In cases where they are not, cloning provides a means to determine lower bounds on the strategies of an adversary~\cite{xiong_general_2012}. However, \emph{implementing} such cloning-based attacks might be non-trivial in practice due to the difficulties mentioned above. Also, the effect of decoherence and errors in NISQ devices makes the production of high-quality clones out of reach for the adversary, which limits the power of practical quantum cryptanalysis in the NISQ era. On the other hand, there has been much interest in implementing quantum cloning and cryptographic attacks on protocols via specific and tailored experiments (for example~\cite{lamas-linares_experimental_2002,fiurasek_optical_2003,chen_experimental_2007, bartkiewicz_experimental_2013,bartkiewicz_experimental_2017}), but these may not be easily reconfigurable or generalizable to other scenarios. In summary, finding and constructing quantum cloning circuits for preparing high fidelity clones on NISQ hardware is challenging.

All the arguments given above, motivate us to seek a new approach to efficiently produce optimal cloning machines and their circuits for specific classes of states. We are in particular interested in the ones that are implementable efficiently on NISQ devices. We also note that existing approaches in the literature are by no means optimal if one wants to generalise this question in order to target applications, especially targeting applications in cryptanalysis given the everyday-expanding variety of quantum protocols. 
The known analytical approaches proceed as follows: Firstly, the most general unitary for the cloning machine has been considered. Secondly, by imposing the symmetries and conditions on the specified family of states, the output fidelity of the machine has been optimised, irrespective of the characterisation of the unitary. Thirdly, one should try to find a unitary matrix that achieves that maximum fidelity, which is often a non-trivial and challenging task. Some inspiring alternative approaches have also been given for the case of equatorial states\footnote{We recall that we have defined these states in \chapref{chap:prelim}, Section~\ref{sec:prelim-cloning-phase-cov}} \cite{cerf_cloning_2002}, which exploit the symmetries in a much more intriguing way than the previous formalism for cloning. However, these approaches do not seem generalisable to other classes of cloning.

The approach we propose here takes a very different path. We start with the idea of \emph{letting a quantum machine learn how to clone} a specific given family of quantum states. We give a novel algorithm: `Variational Quantum Cloning' ($\VQC$) which uses quantum machine learning (QML) \cite{wecker_progress_2015, biamonte_quantum_2017, kopczyk_quantum_2018, schuld_machine_2018, schuld_prospects_2018} techniques to \emph{learn} how to clone quantum states in an end-to-end manner. $\VQC$ is made possible by recent advances and techniques in the field of \emph{variational} quantum algorithms (VQAs)~\cite{mcclean_theory_2016,biamonte_universal_2021,endo_hybrid_2021, wecker_progress_2015,cerezo_variational_2021}. VQAs are intentionally tailored to be useful on NISQ devices, which are limited in scale and noisy to implement `coherent' algorithms with speedups, such as factoring large prime numbers~\cite{shor_algorithms_1994}. However, such devices are capable of performing tasks which cannot be simulated by any classical device in reasonable time~\cite{arute_quantum_2019,zhong_quantum_2020,centrone_experimental_2021}. This motivates the search for dedicated applications for a topic of likely practical relevance. 

Variational quantum algorithms have been proposed and used for various applications, including quantum chemistry~\cite{peruzzo_variational_2014} and combinatorial optimization~\cite{farhi_quantum_2014}. The core quantum component is typically a parameterised quantum circuit (PQC)~\cite{du_expressive_2020} (as mentioned in Section~\ref{sec:prelim-vqc}). When VQAs are applied to machine-learning problems, they have come to be seen as quantum neural networks (QNNs)~\cite{benedetti_parameterized_2019, killoran_continuous-variable_2019}. This is because they can achieve many of the same tasks as classical neural networks, \cite{mitarai_quantum_2018, grant_hierarchical_2018} and can outperform them in certain cases~\cite{wright_capacity_2020, coyle_born_2020, cong_quantum_2019}. Furthermore, machine learning techniques, both quantum~\cite{morales_variational_2018,khatri_quantum-assisted_2019,jones_robust_2022,heya_variational_2018,bravo-prieto_variational_2020,xu_variational_2021,huang_near-term_2021} and classical~\cite{krenn_automated_2016, melnikov_active_2018, odriscoll_hybrid_2019, nichols_designing_2019, wallnofer_machine_2020} have proven to be useful in \emph{discovering} and providing insights into quantum algorithms and subroutines. This line of study even extends to the foundations of quantum mechanics. We refer the reader to this paper \cite{arrasmith_variational_2019} about variational consistent histories.

$\VQC$ is different from other variational algorithms in that it can be viewed as the first step into a new area of applications, \emph{variational quantum cryptanalysis}. Specifically, by using QML techniques to learn to clone quantum states, $\VQC$ can discover unique ways to attack quantum protocols, in particular those whose underlying security can be reduced to quantum cloning. Furthermore, in developing such techniques more generally, we can determine the relationship between classical machine learning and deep learning, with classical cryptography~\cite{ateniese_hacking_2015,maghrebi_breaking_2016,papernot_towards_2016,alani_applications_2019}.

We believe this new approach for approximate cloning is reasonably general and can be used to deepen and widen our understanding of approximate cloning. However, as a concrete case study and proof of concept of our new approach to cryptanalysis, we map two well-known families of approximate cloning machines, \emph{i.e.} \emph{phase-covariant cloning} \cite{brus_phase-covariant_2000} and \emph{fixed-overlap state-dependent cloning} \cite{brus_optimal_1998,brus_optimal_1999} to two well-known families of quantum protocols, namely \emph{quantum key distribution} and \emph{quantum coin flipping} respectively \cite{mayers_unconditionally_2018,aharonov_quantum_2000}. The latter is probably more exciting since, to the best of our knowledge, this is the first time such a connection has been revealed in the case of state-dependent cloning. One can also see this work as an attempt to uncover the core ingredient of security of different quantum protocols from the perspective of the family of states they rely on.

As part of our research in developing this algorithm, we define suitable cost functions which depend on the symmetries used in the cloning problem, and we use them to prove theoretical guarantees for them (including notions of faithfulness~\cite{khatri_quantum-assisted_2019}). As we have discussed in Section \ref{sec:prelim-vqc} since VQAs are heuristic techniques, usually being able to provide such theoretical guarantees is rare in the field (also one of the reasons we have started the chapter with 'everything in this world is magic, except to the magician'!) and this work is one of the few exceptions.   

Finally, to underline the practical potential of our approach we implemented it on the Rigetti \computerfont{Aspen} quantum computer and show how $\VQC$ can learn to clone states with a higher fidelity on this device than previously known `analytic' quantum circuits, highlighting the flexibility of our approach. Furthermore, the nature of $\VQC$ allows us to improve cloning fidelities generically, on quantum computers available through the cloud~\cite{larose_overview_2019}, without requiring significant tweaking and custom-built experimental hardware.

\subsection{Structure of the chapter}
First, in Section~\ref{sec:varqlone-cloning-crypt} we introduce how different classes of cloning can be used as a cryptanalysis toolkit to attack protocols that benefit from using certain classes of states. For our case study we introduce cloning attacks on the BB84 protocol in \ref{sec:varqlone-phasecov-crypt} in relation to phase covariant cloning, and two different quantum coin-flipping protocols in Section~\ref{sec:varqlone-statedep-cloning-crypt} in connection to state-dependent cloning. We present our cloning-based attacks based on an optimal cloner which we then replace with the cloning machines learned through our variational algorithm. In Section~\ref{sec:varqlone-algorithm} we introduce some theoretical aspects and specifications of $\VQC$ such as cost functions (Section~\ref{sec:varqlone-algorithm-cost-function}), gradient of the proposed cost functions (Section~\ref{sec:varqlone-algorithm-gredient}) and more importantly, theoretical guarantees for our algorithm based on the proposed cost function (Section~\ref{sec:varqlone-algorithm-cost-guarantee}). Finally, in Section~\ref{sec:varqlone-numerics} we present both simulation and experimental results based on $\VQC$, including the circuits found for our specified problems and the probability analysis of the attacks based on them.

\section{Quantum cryptanalysis based on different classes of cloning}\label{sec:varqlone-cloning-crypt}
In the first part of this chapter, and before introducing our machine learning algorithm, we want to establish the theoretical ground for our cryptanalysis based on different types of cloning with \emph{partial prior information} about the states. Here, we only look at cryptanalysis using approximate quantum cloning, and we do not study the use-cases and relevance of probabilistic cloners. Using cloning as an attack strategy has been considered previously only in limited settings, for instance, regarding the QKD protocol, where it has been shown that there exists an optimal cloning-based attack on the BB84 protocol \cite{scarani_quantum_2005}. Nevertheless, here we take a more methodologic approach. Keeping in mind that approximate cloning is categorised into different classes characterised by \emph{specific} families of states, we look for quantum protocols that use the same state families for which we have a cloning machine. 

Among different types of cloning, universal cloning does not seem very thrilling for this purpose, as it is state-agnostic. Moreover, as we have seen, restricting the class of states that one wants to clone,\emph{i.e.} having prior information about the states leads to higher fidelity clones. In cryptanalysis, we ever optimise over all the adversarial strategies, which in this case includes using the best cloning strategy given the protocol's characteristics.

We choose two main families of approximate cloning for our purpose, namely \emph{phase-covariant cloning} (see Section~\ref{sec:prelim-cloning-phase-cov}) and \emph{fixed-overlap state-dependent cloning} (see Section~\ref{sec:prelim-state-dep-cloning}). For the former, our target example protocol is QKD, where we will explain the optimal cloning-based attack. This analysis serves as an important tool as we use the calculations later for our variational cloning attacks. The latter class of approximate cloning, on the other hand, has never been used before for the purpose of cryptanalysis to the best of our knowledge. We notice that the specific family of states used in this type of cloning matches the class of states used in quantum coin-flipping. Hence, we introduce, for the first time, cloning-based attacks on two different quantum coin-flipping protocols.

Finally, we emphasise that the purpose of this study is mostly the illustrations of applications of variational cloning based techniques, for practical cryptanalysis.\footnote{To clarify, let us take QKD as an example. This protocol has been proven information-theoretic secure, so we do not intend to break it using variational attacks as it seems pointless. But rather QKD will serve as an example of phase-covariant cloning attacks in general. Moreover, as we will see it is not the case for all the coin-flipping protocols we study in this chapter.} Nonetheless, the attacks we present in this section either saturate the optimal bounds, where there exists a rigorous security proof or even lead to a complete novel security attack, in the absence of such strong proof.  
\subsection{Cryptanalysis based on phase-covariant cloning}\label{sec:varqlone-phasecov-crypt}
Let us begin with phase-covariant cloning and the quantum key distribution protocols by focusing on the BB84 protocol~\cite{bennett_quantum_1984,bennett_quantum_2014}. In this protocol, one honest party, Alice, sends single-qubit states in two orthogonal bases (for instance, the eigenstates of the Pauli $\mathsf{X}$ and Pauli $\mathsf{Y}$ matrices, $\ket{\pm}$ and $\ket{\pm i}$) to a second honest party, Bob, via a quantum channel that is susceptible to an eavesdropping adversary, Eve. Eve's goal is to extract the secret information exchanged between Alice and Bob, encoded in the states. It turns out that the optimal `individual' (or incoherent) Eve's attack~\cite{scarani_quantum_2005} on this protocol is given by cloning so-called \emph{phase-covariant}~\cite{brus_phase-covariant_2000} states of the form:
\begin{equation} \label{eq:x_y_plane_states_phase_cov}
    \ket{\psi_{xy}(\eta)} = \frac{1}{\sqrt{2}}\left(\ket{0} + e^{i\eta}\ket{1}\right)
\end{equation}
For these states, some \emph{analytic} circuits are given in~\cite{buzek_quantum_1997, fan_quantum_2001, fan_quantum_2014}.
For this family of states, Eve can construct a cloning machine with fidelity $F_{\mathsf{L},  \text{opt}}^{\text{PC}, \mathrm{E}} \approx 0.85$.

There are different families of attacks considered on such protocols. The simplest attack by Eve is a so-called ‘incoherent’ or individual attack, where
Eve attacks each quantum state individually before the reconciliation phase of the protocol. The security of this protocol relies on the information-theoretic bounds on the information shared between Alice and Bob as compared to the information that Eve was able to extract from the key. In the incoherent attacks, the security condition states that a secret key can only be extracted as long as the amount of Eve's information is less than what Bob has received. Thus, one key parameter in the protocol is what is called the \emph{critical error rate}, $\Dcrit$, which defines the threshold above which Alice and Bob abort the protocol and conclude that the channel is insecure. 

For incoherent attacks, the optimal error rate for the ideal incoherent attack is $\Dcrit^{\text{incoh}} = 1 - F_{\mathsf{L},  \text{opt}}^{\text{PC}, \mathrm{E}} \approx 14.6\%$~\cite{scarani_quantum_2005}.

However, as discussed in \cite{scarani_quantum_2005}, this is not the best way of analysing the cloning-based attacks against this protocol, since it does not allow for a comparison between a cloning machine that uses the ancilla, and one that does not. The importance of this comparison is that the cloning machine with ancillary inputs may provide Eve with extra information about both parties. A more appropriate way of calculating the key rate which generalises the strategies is via the Holevo quantity, denoted as $\chi$ which is defined as follows from von Neumann entropy:
\begin{equation}\label{eq:var-qkd-holevo-quantity}
    \chi(Q:E) := S(\rho_E) - \frac{1}{2}S(\rho^0_E) - \frac{1}{2}S(\rho^1_E)
\end{equation}
In~\eqref{eq:var-qkd-holevo-quantity}, $\rho_E$ denotes the mixed state of Eve over all of the combinations of Alice's choice of input, and $\rho^0_E$ and $\rho^1_E$ denote the states of Eve for the random variables that encode $0$ and $1$ in the protocol respectively.

Combining mutual information with Holevo quantity gives a concrete and simple method for calculating the key rate in QKD protocols. The key rate for QKD can be defined as follows:
\begin{equation} \label{eq:var-qkd-key-rate}
    R = I(A:B) - \min\{\chi(A:E_Q), \chi(B:E_Q)\}
\end{equation}
where $I(A:B)$ denotes the mutual information between Alice and Bob and the index $Q$ denotes that Eve may employ general quantum strategies. Intuitively, \eqref{eq:var-qkd-key-rate} states that no key can be extracted at $R=0$ which is when Alice and Bob's mutual information is the minimum value between Alice and Eve and Bob and Eve. At any point after that, Eve has increased the correlation to the key, to the point that the key is compromised.


For calculating the quantity $\Dcrit$, one needs to calculate the Holevo quantity for Eve, set $R = 0$, compute the mutual information, $I(A : B) = 1−H(\Dcrit)$ and finally solve the resulting equation for $\Dcrit$. For the ideal incoherent attack, this value is again proven to be $\Dcrit^{\text{incoh}} \approx 14.6\%$. We go back to this calculation in Section~\ref{sec:varqlone-numerical-results-phase-cov} where we will show that the cloning transformations we learn using our variational cloning algorithm give an approximately close critical error rate while being experimentally superior to the ideal proposed circuits.

\subsection{Cryptanalysis based on state-dependent cloning}\label{sec:varqlone-statedep-cloning-crypt}
Now, let us examine the class of states used in \emph{state-dependent} cloning, which are states with fixed and known overlap. Non-orthogonal quantum states are among the elements that are often present in quantum cryptography since they can encode information that is not easily decodable for an adversary who does not know the basis. The most famous example is, of course, conjugate coding~\cite{wiesner_conjugate_1983}. But despite the generality of the state-dependent cloning framework, it is surprising that this type of cloning machine has not been studied as a concrete attack model, and to the best of our knowledge, this is the first time we use state-dependent cloning as a cryptanalysis tool. A concrete example of the protocols that exploit such states is the family of \emph{quantum coin-flipping} protocols. Coin-flipping is a cryptographic task where two mutually distrustful parties, who are usually spatially separated and want to agree on a common random bit (see Section~\ref{sec:prelim-coin-flip} for more details about coin-flipping). Classical and quantum coin-flipping have a vast literature, but here for our case study, we focus on two specific \emph{strong quantum coin-flipping} protocols. The two protocols we consider are that of Mayers et al.~\cite{mayers_unconditionally_1999}, and that of Aharonov et al.~\cite{aharonov_quantum_2000}. First, let us introduce the common aspect of these protocols as well as the states used in the protocols.

\subsubsection{Quantum coin flipping states}\label{sec:varqlone-coinflip-states}
Let us first introduce quantum coin flipping in more detail. The task of the quantum coin flipping is similar to the classical one, only the parties can have quantum capabilities. We say the coin is `biased' when one outcome is more likely to occur than the other, for example, with the following probabilities:
\begin{equation}\label{eq:coin_flipping_epsilon_biased}
    \begin{split}
    \Pr(y = 0) = 1/2 + \epsilon\\
    \Pr(y = 1) = 1/2 -  \epsilon
    \end{split}
\end{equation}
where $y$ is the output bit. The above coin is an $\epsilon$-biased coin with a bias towards the outcome $0$. In contrast, a fair coin would correspond to $\epsilon=0$.

\begin{figure}[t]
    \centering
    \includegraphics[width=0.45\columnwidth, height = 0.45\columnwidth]{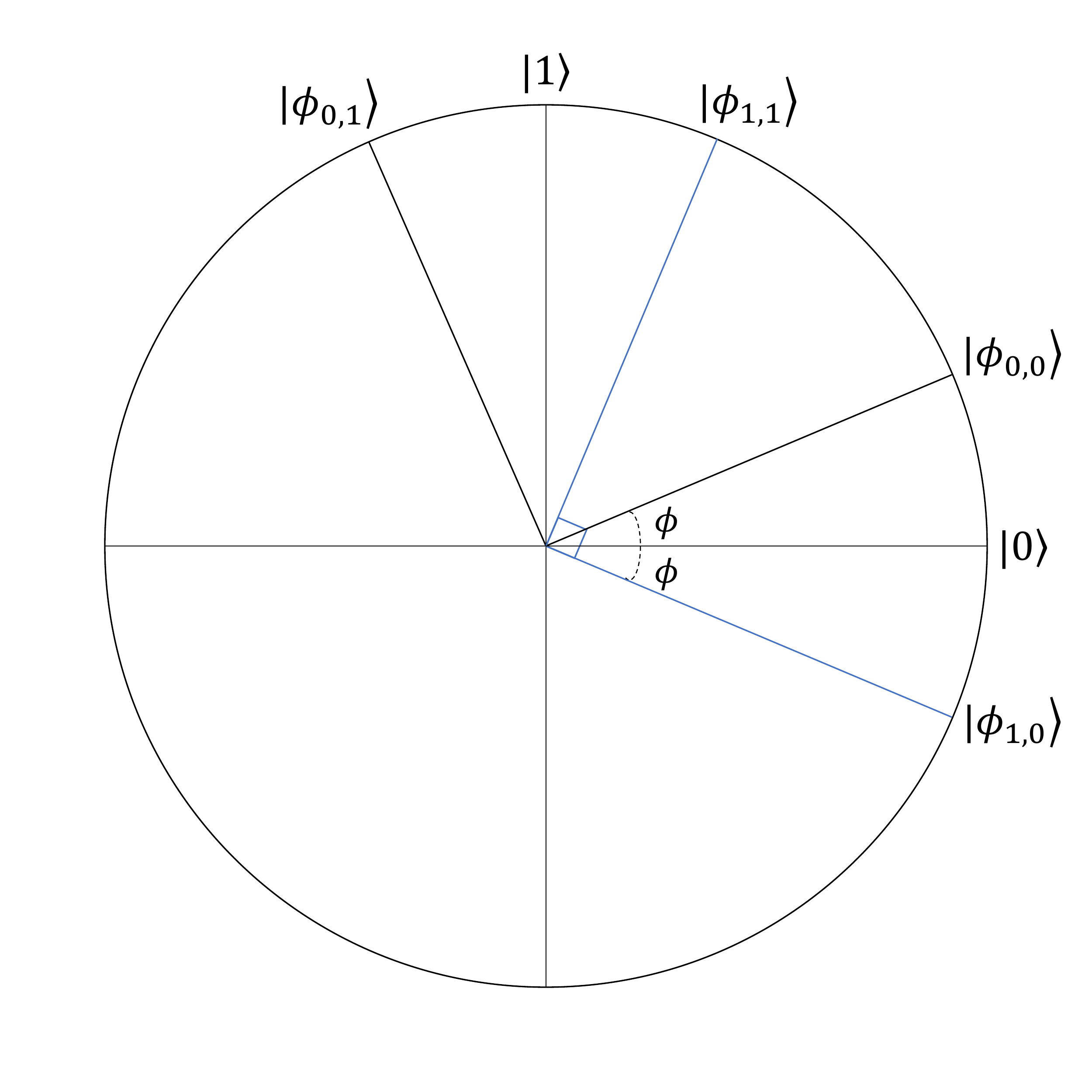}
    \caption{States used for quantum coin-flipping. The first bit represents the \emph{basis}, while the other represents one of the two orthogonal states.}
    \label{fig:varqlone-coinflipstates}
\end{figure}

We recall from Section~\ref{sec:prelim-coin-flip} that it is impossible in an information-theoretic manner, to achieve a perfectly secure (with zero bias $\epsilon=0$) strong coin-flipping protocol in both the classical and quantum setting~\cite{blum_coin_1983,lo_why_1998,mayers_unconditionally_1999}. Several protocols have been proposed for $\epsilon$-biased strong coin flipping~\cite{mayers_unconditionally_1999, aharonov_quantum_2000, bennett_quantum_2014, berlin_fair_2009}, the states used by these protocols share a common structure. Here we introduce a more general form of these states, which will be useful for our purpose. The following set of states (illustrated in \figref{fig:varqlone-coinflipstates}) have been used in the protocols that we will investigate:
\begin{equation}\label{eq:coinflip-st}
    \ket{\phi_{x, a}} = 
    \begin{cases}
    \ket{\phi_{x,0}} = \cos\phi\ket{0} + (-1)^x\sin\phi\ket{1} \\
    \ket{\phi_{x,1}} =  \sin\phi\ket{0} + (-1)^{x \oplus 1}\cos\phi\ket{1}
  \end{cases} 
\end{equation}
where $x \in \{0,1\}$, and the angle $\phi$ determines the overlap between the pairs of states.

Additionally, these protocols share the following shared structure: One of the parties, or sometimes both, will encode some random classical bits into the above states and then exchange some classical/quantum information as part of the protocol. The attack of the malicious party who is trying to bias the coin is then, reduced to the ability to learn the encoded classical bit, from the state (or in some cases, to prepare states deviating from the perfect ones). To see why this is the case, we need to look at the impossibility of the classical coin-flipping task~\cite{blum_coin_1983}. The intuitive reason behind this impossibility lies in the \emph{order} or asymmetry between the two parties. In other words, since the outcome has to be determined after a certain number of communication rounds in the protocol consisting of sending some messages, one can always find a message such that, before the message is sent, the outcome is not yet determined, but once the message has been sent it becomes determined. This means that the party who receives this `extra' information first can always bias the protocol. This limitation cannot be overcome in the classical world. Therefore, there is no value of $\epsilon < 0.5$ for which, the protocol can be secure. However, using non-orthogonal quantum states gives the parties the ability to `hide' their choice of the random bit before the other party also flips a coin. Therefore the existence of quantum coin-flipping protocols with a certain degree of bias is closely related to the fact that an adversary, is fundamentally bounded by quantum mechanics, in performing the task of distinguishing non-orthogonal quantum states.

Now, going back to our selected protocols, we have the two following cases:
\begin{enumerate}
    \item The protocol of Mayers \textit{et. al.}~\cite{mayers_unconditionally_1999} (denoted by $\mathcal{P}_1$) in which the states, $\{\ket{\phi_{0,0}}, \ket{\phi_{1,0}}\}$ are used (which have a fixed overlap $s = \cos\left(2\phi\right)$).
    
    \item The protocol of Aharonov \textit{et. al.}~\cite{aharonov_quantum_2000} (denoted by $\mathcal{P}_2$), which uses the full set of states, \emph{i.e.}\@ $\{\ket{\phi_{x, a}}\}$.
\end{enumerate} 

This set of states is conveniently related through a reparameterisation of the angle $\phi$~\cite{bru_approximate_2006}, which makes them easier to deal with mathematically. 

In general for the security analysis of strong quantum coin-flipping protocols, one considers both cases where Alice or Bob are being dishonest. Here, for simplicity of comparison and since our goal is to demonstrate cloning-based attacks, we only focus on a dishonest Bob who tries to bias the bit by cloning the non-orthogonal states sent by Alice. 

In the following two subsections, the biases are computed assuming access to the \emph{ideal} cloning machine (\emph{i.e.} the one which clones the input states with the optimal, analytical fidelities). Later, we compare these ideal biases with those achievable using the quantum cloning machines learned by our variational quantum cloner.

\subsubsection{Cloning attack on 2-state quantum coin flipping protocol}\label{sec:varqlone-2state-coinflip}

The Mayers' protocol was incidentally one of the first protocols proposed for strong quantum coin-flipping. Here, Alice utilizes the states\footnote{Since the value of the overlap is the only relevant quantity, the different parameterisation of these states compared to the ones in \eqref{eq:coinflip-st} does not make a difference for our purposes. However, we note that explicit cloning unitary would be different in both cases.} $\ket{\phi_0} := \ket{\phi_{0,0}}$ and $\ket{\phi_1} := \ket{\phi_{1, 0}}$ such that the angle between them is $\phi := \frac{\pi}{18} \implies s := \cos(\frac{\pi}{9})$. In the following, we describe the general version of the protocol with $k$ rounds (of quantum communication). We also discuss the proposed attack in more detail and prove the relevant theorems for fixed-overlap state-dependent cloning attacks.

\begin{protocol}[$\mathcal{P}_1$ with $k$ rounds]\label{prot:mayers_general_krounds}
During the protocol, each party sends multiple copies of either $\ket{\phi_0}\otimes\ket{\phi_1}$ or $\ket{\phi_1}\otimes\ket{\phi_0}$
\vspace{3mm}\hrule
\begin{enumerate}
    \item Alice and Bob now choose $k$ random bits, $\{a_1, \dots, a_k\}$ and $\{b_1, \dots, b_k\}$ respectively. The final bit is now equal to the \computerfont{XOR} of input bits over all $k$ rounds \emph{i.e.},
\begin{equation}
x = \bigoplus_j a_j  \oplus \bigoplus_j b_j   
\end{equation}
    \item Each round has $n$ steps identified by $i$: In each round $j = 1,\dots, k$ of the protocol, and for every step $i=1, \dots, n$ within each round, Alice uniformly picks a random bit $c_{i,j}$ and sends the state $\ket{\phi_c^{i,j}} := \ket{\phi_{c_{i,j}}} \otimes\ket{\phi_{\overline{c_{i,j}}})}$ to Bob (where $\overline{c_{i,j}}$ denotes the complement of $c_{i,j}$).
    
    \item Bob uniformly picks a random bit $d_{i,j}$ and sends the state $\ket{\phi_d^{i,j}} := \ket{\phi_{d_{i,j}}} \otimes\ket{\phi_{\overline{d_{i,j}}}}$ to Alice.\footnote{Note that if $c_{i,j}$ and $d_{i,j}$ are chosen independently of $a_j$ and $b_j$, no information about the primary bits has been transferred.}
    
    \item For each $j$ and $i$, Alice announces the value $a_j\oplus c_{i,j}$.
    
    \item\label{bobreturns} If $a_j\oplus c_{i,j} = 0$, Bob returns the second state of the pair $(i,j)$ back to Alice, and sends the first state otherwise.
    
    \item Bob announces $b_j\oplus d_{i,j}$.
    
    \item Alice returns one of the states back to Bob accordingly (similar to step \ref{bobreturns}).
    
    \item $a$ and $b$ are announced by both sides.
    
    \item Alice measures the remaining states with the projectors, $(E_{b}, E^{\perp}_{b})$ and the returned states by Bob with $(E_{\overline{a}}, E^{\perp}_{\overline{a}})$ (\eqref{eq:mayers_povm}). She aborts the protocol if the measurement result corresponds to $\perp$, and declares Bob as being dishonest\footnote{The similar verification measurement is performed by Bob to verify Alice, however here we skip that part since we are only interested in half of the protocol, \emph{i.e.} dishonest Bob}.
\end{enumerate}
\hrule\vspace{3mm}
\end{protocol}

Considering steps (5) to (7) of the protocol, we argue that it is sufficient to only consider a single round in the protocol from the point of view of a cloning attack. This is because a dishonest Bob can bias the protocol if he learns about Alice's bit $a_j$ (for any choice of $j$), which he can do by guessing $c_{i,j}$ with probability better than $1/2$. With this knowledge, Bob only needs to announce a single false $b_j \oplus d_{i,j}$ to cheat, and so this strategy can be deferred to the final round~\cite{mayers_unconditionally_1999}. Hence a single round of the protocol is sufficient for analysis, and we herein drop the $j$ index. 

In the last phase of the protocol, after $a$ and $b$ are announced by both sides (so $x$ can be computed by both sides), Alice performs the measurements $(E_{b}, E^{\perp}_{b})$ and $(E_{\overline{a}}, E^{\perp}_{\overline{a}})$ on the remaining states. (as defined in \eqref{eq:mayers_povm}) for checking whether Bob has cheated or not. In this sense, the use of quantum states in this protocol is purely for cheat-detection.
\begin{align}\label{eq:mayers_povm}
 E_l &= \ket{\phi_l}\bra{\phi_l}^{\otimes n} \\
     E_l^\perp &= \mathds{1} - \ket{\phi_l}\bra{\phi_l}^{\otimes n}, \qquad l \in\{0, 1\}
\end{align}

\noindent\textbf{A Cloning Attack on \texorpdfstring{$\mathcal{P}_1$}{}:} Next, we present the explicit attack and calculation that can be implemented by Bob on $\mathcal{P}_1$. Without loss of generality, we assume that Bob wishes to bias the bit towards $x = 0$. For clarity, we give the attack for when Alice only sends one copy of the state ($n=1$), but we discuss the general case later:

\begin{attack}[Cloning Attack on $\mathcal{P}_1$ with $k=1$]\label{attack:mayers-attack-1-round}
The goal is to bias the bit towards $0$, \emph{i.e.}\@ $p(x=0) > 1/2$
\vspace{3mm}\hrule\vspace{3mm}
\noindent\textit{Inputs.} Random bit for Alice ($a \overset{\$}{\leftarrow} \{0, 1\}$) and Bob ($b \overset{\$}{\leftarrow} \{0, 1\}$). Bob receives a state $\ket{\phi_c^i}$ from Alice.\\
\textit{The attack:}
\begin{enumerate}
\item for $i = 1, \dots, n$: 
  \begin{enumerate}
    \item \textbf{Step 1:} Alice announces $a \oplus c_i$. If $a \oplus c_i = 0$, Bob sends the second qubit of $\ket{\phi_c^i}$ to Alice, otherwise he sends the first qubit. 
    \item \textbf{Step 2:} Bob runs a $1\rightarrow 2$ state-dependent cloner on the qubit he has to return to Alice, producing $2$ approximate clones. He sends her one clone and keeps the other.
    \item \textbf{Step 3:} Bob runs an optimal state discrimination on the remaining qubit (and any other auxiliary output of the cloner, if exists), and finds $c_1$ with a maximum success probability $P^{\opt}_{\mathrm{disc}, \mathcal{P}_1}$. He then guesses a bit $a'$ such that $P_{\mathrm{succ}, \mathcal{P}_1}(a' = a) := P^{\opt}_{\mathrm{disc}, \mathcal{P}_1}$.
    \item \textbf{Step 4:} If $a'\oplus b = 0$ he continues the protocol honestly and announces $b \oplus d_1$, otherwise he announces $a' \oplus d_1$. The remaining qubit on Alice's side is $\ket{\phi^i_{a}}$.
  \end{enumerate}
\end{enumerate}
\hrule
\end{attack}

Now, we find the success probability of the above attack:

\begin{thmbox}
\begin{theorem}\label{thm:mayers_attack_bias_probability}[Bias of ideal cloning attack on $\mathcal{P}_1$]
Bob can achieve a bias of $\epsilon \approx 0.27$ using an ideal state-dependent cloning attack on the protocol $\mathcal{P}_1$ using a single copy of Alice's state.
\end{theorem}
\end{thmbox}

\begin{proof}
As mentioned in the previous section, the final measurements performed by Alice on her remaining $n$ states, plus the $n$ states returned to her by Bob allow her to detect his nefarious behaviour. If he performed a cloning attack, the $\perp$ outcomes would be detected by Alice with some probability. We must compute both probabilities: the probability of guessing the value of Alice's bit $a$ (by guessing the value of the bit $c_1$), and the probability of being detected by Alice. This would provide us with Bob's final success probability in cheating, hence the bias probability. 

At the start of the attack, Bob has a product state of either $\ket{\phi_0}\otimes\ket{\phi_1}$ or $\ket{\phi_1}\otimes\ket{\phi_0}$ (but he does not know which). After the announcement stage, depending on Alice's announced bit, Bob proceeds to clone one of the qubits, sends one copy to Alice and keeps the other to himself. Without loss of generality, we assume that Alice's announced bit is $0$. In this case, at this point of the attack, he has one of the following pairs: $\ket{\phi_0}\bra{\phi_0}\otimes\rho^1_c$ or $\ket{\phi_1}\bra{\phi_1}\otimes\rho^0_c$, where $\rho^1_c$ and $\rho^0_c$ are leftover clones (the second state of the cloner together with any existing ancillary systems) for $\ket{\phi_1}$ and $\ket{\phi_0}$ respectively. 

Bob must now discriminate between the following density matrices:
\begin{align}\label{eq:mayers_bob_pairs_discriminate}
   \rho_1 &= \ket{\phi_0}\bra{\phi_0}\otimes\ket{\phi_1}\bra{\phi_1}\\ \textrm{ and }  \qquad \rho_2 &= \ket{\phi_1}\bra{\phi_1}\otimes\rho^0_c
\end{align}

Alternatively, if Alice announced $a\oplus c_i = 1$, he would have:
\begin{align}\label{eq:mayers_bob_pairs_discriminate_alternate}
    \rho_1 &= \ket{\phi_1}\bra{\phi_1}\otimes\ket{\phi_0}\bra{\phi_0}, \\ \textrm{ and }  \qquad  \rho_2 &= \ket{\phi_0}\bra{\phi_0}\otimes\rho^1_c
\end{align}
In either case, we have that the minimum discrimination error for two density matrices is given by the Holevo-Helstrom bound~\cite{holevo_statistical_1973,helstrom_quantum_1969} (also see Section~\ref{sec:prelim-distinguish-quantum-test} for more information) bound as follows\footnote{This also is because we assume a symmetric cloning machine for both $\ket{\phi_0}$ and $\ket{\phi_1}$. If this is not the case, the guessing probability is instead the average of the discrimination probabilities of both cases.}:
\begin{equation}\label{eq:varqlone-helstrom}
    P^{\opt}_{\mathrm{disc}} = \frac{1}{2} + \frac{1}{4}\norm{\rho_1 - \rho_2}_{\Tr} = \frac{1}{2} + \frac{1}{2}\dtr(\rho_1, \rho_2)
\end{equation}
The ideal symmetric cloning machine for these states will have an output of the form:
\begin{equation}\label{eq:reduced_local_state_mayers_attack}
      \rho_c = \alpha\ket{\phi_0}\bra{\phi_0} + \beta \ket{\phi_1}\bra{\phi_1}
     + \gamma (\ket{\phi_0}\bra{\phi_1} + \ket{\phi_1}\bra{\phi_0})  
\end{equation} 
where $\alpha, \beta$ and $\gamma$ are functions of the overlap $s = \mbraket{\phi_0}{\phi_1} = \cos{\frac{\pi}{9}}$. Now, using \eqref{eq:mayers_bob_pairs_discriminate}, $\rho_2$ can be written as follows:
\begin{equation}\label{eq:reduced_local_state_rho2_mayers_attack}
\begin{split}
      \rho_2 = & \alpha\ket{\phi_1}\bra{\phi_1}\otimes\ket{\phi_0}\bra{\phi_0} + \beta  \ket{\phi_1}\bra{\phi_1}\otimes\ket{\phi_1}\bra{\phi_1} \\
     & + \gamma ( \ket{\phi_1}\bra{\phi_1}\otimes\ket{\phi_0}\bra{\phi_1} +  \ket{\phi_1}\bra{\phi_1}\otimes\ket{\phi_1}\bra{\phi_0})  
\end{split}     
\end{equation} 
Finally, by plugging in the values of the coefficients in \eqref{eq:reduced_local_state_mayers_attack} for the optimal local cloning machine~\cite{brus_optimal_1998} and finding the eigenvalues of $\sigma := (\rho_1 - \rho_2)$, we can calculate the corresponding value for \eqref{eq:varqlone-helstrom}, and recover the following minimum error probability:
\begin{equation}\label{eq:mayers-cloning-min-error}
P_{\mathrm{fail}, \mathcal{P}_1} = P^{\mathrm{er}}_{\mathrm{disc}, \mathcal{P}_1} = 1 - P^{\opt}_{\mathrm{disc}, \mathcal{P}_1} \approx 0.214
\end{equation}

\noindent This means that Bob can successfully guess $c_1$ with ${P}^{1}_{\textrm{succ}, \mathcal{P}_1} = 78.5\%$ probability.

Now we look at the probability of a cheating Bob being detected by Alice. We note that whenever Bob guesses $a$ successfully, the measurements $(E_{b}, E^{\perp}_{b})$ will be passed with probability $1$, hence we use $(E_{\overline{a}}, E^{\perp}_{\overline{a}})$ where the states sent by Bob will be measured. Using \eqref{eq:prelim-local_optimal_non_ortho_fidelity_1to2} (in Section~\ref{sec:prelim-state-dep-cloning}) with the value of overlap $s = \cos\left(\pi/9\right)$, the optimal fidelity is $F_{\mathsf{L}} \approx 0.997$ and so the probability of Bob getting caught is at most $1\%$. Putting this together with Bob's guessing probability for $a$ gives his overall success probability of $77.5\%$.

This implies that Bob is able to successfully create a bias of $\epsilon \approx 0.775 - 0.5 = 0.275$.
\end{proof}

We also have the following corollary, for a general number $n$ of exchanged states, which shows the protocol can be completely broken and Bob can enforce an arbitrary bias:

\begin{corrbox}
\begin{corollary}
The probability of Bob successfully guessing Alice's bit $a$, over $n$ rounds and from all $n$ copies of the received stated, has the property:
    \begin{equation} \label{eqn:bob_guess_prob_n_rounds_mayers_appendix}
        \lim\limits_{n\rightarrow \infty}{P}^{n}_{\mathrm{succ}, \mathcal{P}_1} = 1
    \end{equation}
\end{corollary}
\end{corrbox}

\begin{proof}
If Bob repeats the above \attref{attack:mayers-attack-1-round} over all $n$ copies, he will guess $n$ different bits $\{a'_{i}\}_{i=1}^n$. He can then take a majority vote and announce $b$ such that $a^* \oplus b = 0$, where we denote $a^*$ as the bit he guesses in at least $\frac{n}{2} + 1$ of the rounds.

If $n$ is even, he may have guessed $a'$ to be $0$ and $1$ an equal number of times. In this case, the attack becomes indecisive and Bob is forced to guess at random. Hence we separate the success probability for even and odd $n$ as follows: 
\begin{equation}\label{eq:mayers_attack_guess_probability}
    P^n_{\textrm{succ}, \mathcal{P}_1} = 
    \begin{cases}
    \sum\limits_{k=\frac{n+1}{2}}^{n} \binom{n}{k}(1 - P_{\mathrm{fail}})^k P_{\mathrm{fail}}^{n-k} &n  \text{ odd,} \\
     \sum\limits_{k=\frac{n}{2} + 1}^{n} \binom{n}{k}(1 - P_{\mathrm{fail}})^k P_{\mathrm{fail}}^{n-k} + \frac{1}{2}\binom{n}{n/2}(1 - P_{\mathrm{fail}})^{\frac{n}{2}} P_{\mathrm{fail}}^{\frac{n}{2}}  & n \text{ even} \\
  \end{cases} 
\end{equation}
By substituting the value of $P_{\mathrm{fail}}$ one can see that the function is uniformly increasing with $n$ so $\lim\limits_{n\rightarrow \infty}{P}^{n}_{\textrm{succ}, \mathcal{P}_1} = 1$\footnote{Although, as Bob's success probability in guessing correctly increases with $n$, the probability of his cheating strategy getting detected by Alice will also increase, yet does not converge to 1 as fast. We also note that this strategy is independent of $k$, the number of different bits used during the protocol.}. This concludes the proof.
\end{proof}

\subsubsection{Cloning attack on 4-state quantum coin flipping protocol}\label{sec:varqlone-4state-coinflip}

Another class of coin-flipping protocols are those which require all the four states in~\eqref{eq:coinflip-st}. One such protocol was proposed by Aharonov \textit{et al.}~\cite{aharonov_quantum_2000}, where the optimal $\phi$ is set as $\frac{\pi}{8}$ \emph{i.e.} resulting in the following states:

\begin{equation}\label{eq:aharonov_coinflip_states}
    \ket{\phi_{x, a}} = 
    \begin{cases}
    \ket{{\frac{\pi}{8}}_{x,0}} = \cos\left( \frac{\pi}{8} \right)\ket{0} + (-1)^x\sin\left( \frac{\pi}{8} \right)\ket{1} \\
    \ket{{\frac{\pi}{8}}_{x,1}} =  \sin\left( \frac{\pi}{8} \right)\ket{0} + (-1)^{x \oplus 1}\cos\left( \frac{\pi}{8} \right)\ket{1}
  \end{cases} 
\end{equation}

In protocols of this form, Alice encodes her bit as `basis information' of the family of states. More specifically, her random bit is encoded in the state $\ket{\phi_{x, a}}$. For instance, we can take $\{\ket{\phi_{0,0}}, \ket{\phi_{1, 0}}\}$ to encode the bit $a = 0$; and $\{\ket{\phi_{0, 1}}, \ket{\phi_{1,1}}\}$ to encode $a = 1$. The goal again is to produce a final `coin flip' $y = a\oplus b$, while ensuring that no party has biased the bit $y$. A similar protocol has also been proposed using BB84 states~\cite{bennett_quantum_2014} where $\ket{\phi_{0,0}} := \ket{0}, \ket{\phi_{0,1}} := \ket{1}, \ket{\phi_{1,0}} := \ket{+}$ and $\ket{\phi_{1,1}} := \ket{-}$. In this case, the states (as well as some protocol steps) are different but the angle between them is the same as with the states in $\mathcal{P}_2$. A fault-tolerant version of $\mathcal{P}_2$ has also been proposed in Ref.~\cite{berlin_fair_2009}, which uses a generalized angle as in~\eqref{eq:coinflip-st}.

\begin{protocol}[$\mathcal{P}_2$ (Aharonov's coin flipping)]\label{prot:aharonov-coinflip}
The protocol uses all four possible states from \eqref{eq:aharonov_coinflip_states}.
\vspace{3mm}\hrule 
\begin{enumerate}
    \item Alice selects two random bits $a  \overset{\$}{\leftarrow} \{0,1\}$ and $x  \overset{\$}{\leftarrow} \{0,1\}$.
    \item Alice sends one of the states, $\ket{\phi_{x, a}}$ to Bob.
    
    \item Bob selects his random bit $b  \overset{\$}{\leftarrow} \{0,1\}$ and sends to Alice.
    
    \item One of two following things happens:
    \begin{enumerate}
        \item (either) Alice will send the bits $x$ and $a$ to Bob, who measures the qubit on a suitable basis to check if Alice was honest.
        \item (or) Bob is asked to return the qubit $\ket{\phi_{x, a}}$ to Alice, who measures it and verifies if it is correct.
    \end{enumerate}
    
    \item If no party declares cheating, the final output bit, will be $c = a \oplus b$.
\end{enumerate}
\hrule \vspace{3mm}
\end{protocol}

Now, we can discuss the cheating strategies of each of the players. Examples of the cheating strategies for Alice include incorrect preparation of $\ket{\phi_{x, a}}$ and giving Bob the wrong information about $(x, a)$, or Bob trying to determine the bits $x, a$ from $\ket{\phi_{x, a}}$ before Alice has revealed them classically. We again focus only on Bob's strategies here to use cloning arguments. We note that the information-theoretic achievable bias of $\epsilon = 0.42$ proven in Ref.~\cite{aharonov_quantum_2000} applies only to Alice's strategy since she has greater control of the protocol (she prepares the original state). In general, with a cloning based attack strategy, Bob will be able to achieve a lower bias, as we show next. As mentioned above, Bob randomly selects his own bit $b$ and sends it to Alice. He then builds a QCM to clone all 4 states in \eqref{eq:aharonov_coinflip_states}.

We next sketch the two cloning attacks on Bob's side of $\mathcal{P}_2$. Again, as with the protocol, $\mathcal{P}_1$, Bob can cheat using as much information as he can gain about $a$ and again, once Bob has performed the cloning, his strategy boils down to the problem of state discrimination. In both attacks, Bob will use a state-dependent cloning machine.

In the first attack model (which we denote {\RNum{1}} - see \figref{fig:aharonov_1to2_cloning_fidelities_variational_plus_attack_models}(a) in Section~\ref{sec:varqlone-numerical-results-4-state-coinflip} where we introduce the variational cloning version of the attack) Bob measures \emph{all} the qubits outputted from the cloner to guess $(x, a)$. As such, it is the \emph{global} fidelity that will be the relevant quantity. This strategy would be useful in the first possible challenge in the protocol, where Bob is not required to send anything back to Alice. We will discuss how using cloning in this type of attack can also reduce practical resources for Bob from a general POVM to projective measurements, which may be of independent interest. The main attack here boils down to Bob measuring the global output state from his QCM using the projectors $\{\ket{v}\bra{v}, \ket{v^\perp}\bra{v^\perp}\}$, and from this measurement, determines $a$. These projectors are constructed explicitly relative to the input states using the Neumark theorem~\cite{bae_quantum_2015} (see Section~\ref{sec:prelim-distinguish-quantum-test}).

The second attack model (which we denote {\RNum{2}} - see \figref{fig:aharonov_1to2_cloning_fidelities_variational_plus_attack_models}(a) in Section~\ref{sec:varqlone-numerical-results-4-state-coinflip}) is instead a \emph{local} attack and as such will depend on the optimal local fidelity. It may also be more relevant in the scenario where Bob is required to return a quantum state to Alice. We note that Bob could also apply a global attack in this scenario but we do not consider this possibility here to give more interesting and distinct examples. In what follows we explain the attacks in detail. For simplicity, we compute a bias assuming he does not return a state to Alice thus the bias will be equivalent to his discrimination probability. The analysis could be tweaked to take a detection probability for Alice into account as well. In this scenario, Bob again applies the QCM, but now he only uses one of the clones to perform state discrimination (given by the \emph{Discriminator} in \figref{fig:aharonov_1to2_cloning_fidelities_variational_plus_attack_models}(a)).\\


\noindent\textbf{Attack} \textbf{\RNum{1}} \textbf{on $\mathcal{P}_2$:}\\
We note that attack {\RNum{1}}, is a $4$ state \emph{global} attack on $\mathcal{P}_2$ and that this attack model (\emph{i.e.} based on cloning) can be considered a constructive way of implementing the optimal discrimination strategy of the states Alice is to send. To bias the bit, Bob needs to discriminate between the four pure states in \eqref{eq:coinflip-st} or equivalently between the ensembles of states encoding $a=\{0,1\}$, where the optimal discrimination is done via a set of POVM measurements. 

However, by implementing a cloning based attack, we can simplify the implementation of optimal discrimination strategies.
This is because the symmetric state-dependent cloner (which is a unitary) has the interesting feature that for either case ($a=0$ or $a=1$), the cloner's output is a pure state in the $2$-qubit Hilbert space. As such, the states (after going through the QCM) can be optimally discriminated via a set of projective measurements $\{P_v, P_{v^{\perp}}\}$, rather than general POVMs. This may not seem very important at this stage, but later we will see that it will relate a theoretical bound to an implementational attack strategy. Especially when we introduce the variational cloner that can learn to optimally clone these states efficiently, which, in turn, will assist the benchmarking of existing protocols. Let us now establish this bound and prove it. 

\begin{thmbox}
\begin{theorem}\label{th:aharonov_attack_I_bias_probability}[Ideal Cloning Attack ({\RNum{1}}) Bias on $\mathcal{P}_2$]
Using a cloning attack on the protocol, $\mathcal{P}_2$, (in attack model I) Bob can achieve a bias:
\begin{equation}
    \epsilon^{\mathrm{I}}_{\mathcal{P}_2, \mathrm{ideal}} \approx 0.35
\end{equation}
\end{theorem}
\end{thmbox}

\begin{proof}
The attack involves the global output state of the cloning machine. For this attack we can use the fixed overlap $1 \rightarrow 2$ cloner with the global fidelity given by \eqref{eq:prelim-global_optimal_non_ortho_fidelity_1to2}:
\begin{equation}\label{eq:optimal_fidelity_g_1_2}
    F^{{\mathrm{FO}, \textrm{opt}}}_{\mathsf{G}}(1,2) = \frac{1}{2}\left( 1 + s^{3} + \sqrt{1-s^{2}}\sqrt{1-s^{4}} \right) \approx 0.983
\end{equation}
where $s=\sin(2\phi) = \cos(\frac{\pi}{4})$ for $\mathcal{P}_2$. Also alternatively we can use the 4-state cloner which clones the two states with a fixed overlap plus their orthogonal set. For both of these cloners, we are interested in the global state of the cloner which we denote as $\ket{\psi_{x, a}^{1\rightarrow 2}}$ for an input state $\ket{\phi_{x, a}}$.

In order for Bob to guess $a$ he must discriminate between $\ket{\phi_{0, 0}}$ (encoding $a=0$) and $\ket{\phi_{1, 1}}$ (encoding $a=1$) or alternatively the pair of states $\{\ket{\phi_{0, 1}}, \ket{\phi_{1, 0}}\}$. This is due to the pairs $\{\ket{\phi_{0, 0}}, \ket{\phi_{0, 1}}\}$ being orthogonal and $\{\ket{\phi_{1, 0}}, \ket{\phi_{1, 1}}\}$ both encode $a=0$, so the only choice is to discriminate between $\ket{\phi_{0, 0}}$ and $\ket{\phi_{1, 1}}$. Due to the symmetry and without an ancilla, the cloner preserves the overlap between each pairs \emph{i.e.}\@ $\mbraket{\psi_{0, 0}^{1\rightarrow 2}}{\psi_{1, 1}^{1\rightarrow 2}} = \mbraket{\phi_{0,0}}{\phi_{1,1}} = s$ (we also have $\mbraket{\psi_{0, 1}^{1\rightarrow 2}}{\psi_{1, 0}^{1\rightarrow 2}} = s$).

Now we select the projective measurements $P_v=\ket{v}\bra{v}$ and $P_{v^{\perp}}=\ket{v^{\perp}}\bra{v^{\perp}}$ such that $\mbraket{v}{v^{\perp}} = 0$. One can show that the discrimination probability is optimal when $\ket{v}$ and $\ket{v^{\perp}}$ are symmetric with respect to the target states according to the Neumark theorem. We have that $\mbraket{v}{v^{\perp}} = 0$ so $2\theta + 2\phi = \frac{\pi}{2} \Rightarrow \theta = \frac{\pi}{4} - \phi$. Finally, writing the cloner's states for $\{\ket{\psi_{0, 0}^{1\rightarrow 2}}, \ket{\psi_{1,1}^{1\rightarrow 2}}\}$ in the basis $\{\ket{v}, \ket{v^{\perp}}\}$ gives:
\begin{equation}\label{eq:coin_flip_output_cloner_states_in_v}
    \begin{split}
    & \ket{\psi_{0, 0}^{1\rightarrow 2}} = \cos\left(
    \frac{\pi}{4} - \phi\right)\ket{v} + \sin\left(
    \frac{\pi}{4} - \phi\right)\ket{v^{\perp}},\\
    & \ket{\psi_{1, 1}^{1\rightarrow 2}} = \cos\left(\frac{\pi}{4} - \phi\right)\ket{v} - \sin\left(\frac{\pi}{4} - \phi\right)\ket{v^{\perp}}
    \end{split}
\end{equation}
where it can be checked that $\mbraket{\psi_{0, 0}^{1\rightarrow 2}}{\psi_{1, 1}^{1\rightarrow 2}} = \cos\left(\frac{\pi}{2} - 2\phi\right) = \sin\left(2\phi\right) = s$. Hence $\ket{v}$ and $\ket{v^{\perp}}$ can be explicitly derived. Note that these bases are also symmetric with respect to the other pair i.e $\{\ket{\psi_{0, 1}^{1\rightarrow 2}}, \ket{\psi_{1,0}^{1\rightarrow 2}}\}$. Finally, the success probability of this measurement is then given by:
\begin{equation}\label{eq:coin_flip_global_disc_probability}
P^{\opt, \mathrm{I}}_{\mathrm{disc}, \mathcal{P}_2} = \frac{1}{2} + \frac{1}{2}\mbraket{\psi_{0, 0}^{1\rightarrow 2}}{\psi_{1, 1}^{1\rightarrow 2}} =   \frac{1}{2} + \frac{1}{2}\sin{2\phi} = 0.853
\end{equation}
which is the maximum cheating probability for Bob. From this, we derive the bias as:
\begin{equation}
   \epsilon^{\mathrm{I}}_{\mathcal{P}_2, \mathrm{ideal}} =  P^{\opt, \mathrm{I}}_{\mathrm{disc}, \mathcal{P}_2} -   \frac{1}{2} = 0.353
\end{equation}
which completes the proof.
\end{proof}

\noindent\textbf{Attack} \textbf{\RNum{2}} \textbf{on $\mathcal{P}_2$:}\\

\noindent Finally, we consider a second attack model (attack {\RNum{2}}) on the protocol, $\mathcal{P}_2$, which is in the form of a `local' attack. Here, we further consider two scenarios:
\begin{enumerate}
    \item A cloning machine which is able to clone \emph{all} $4$ states $\ket{\phi_{0, 0}}, \ket{\phi_{1, 1}}$ \emph{and} $\ket{\phi_{0, 1}}, \ket{\phi_{1, 0}}$,
    \item A cloning machine tailored for only the two states, $\ket{\phi_{0, 0}}$ and $\ket{\phi_{1, 1}}$ (which Bob needs to discriminate between).
\end{enumerate}

We focus on the former scenario since it connects in a more clear way to the $\VQC$ clone fidelities, while the second scenario enables a stronger attack (in the ideal scenario).\\

\noindent \textbf{\underline{Scenario 1:}}\\

\noindent In this case, we can compute an exact discrimination probability, but it will result in a non-optimal attack (smaller success probability compared to the second one).

\begin{thmbox}
\begin{theorem}\label{th:aharonov_4state_attack_II_bias_probability}[Ideal Cloning Attack ({\RNum{2}}) Bias on $\mathcal{P}_2$ in Scenario $1$.]
Using a cloning attack on the protocol $\mathcal{P}_2$, (in attack model {\RNum{2}} with $4$-states) Bob can achieve the following bias:
\begin{equation}\label{eqn:attack_4_state_aharonov_success_probability_exact_appendix}
    \epsilon^{\mathrm{II}}_{\mathcal{P}_2, \mathrm{ideal}} = 0.25
\end{equation}
\end{theorem}
\end{thmbox}

\begin{proof}
Considering the 4 states to be in the $\mathsf{X} - \mathsf{Z}$ plane of the Bloch sphere, the density matrices of each state can be represented as:
\begin{equation} \label{eq:4state-density-bloch}
\rho_{ij} = \frac{1}{2}(\mathds{1} + m_{ij}^x\sigma_x + m_{ij}^z\sigma_z)
\end{equation}
where $\sigma_x$ and $\sigma_z$ are Pauli matrices and $m_{ij}^x$ and $m_{ij}^z$ are $3$ dimensional vectors given by:
\begin{equation} \label{eq:4state-cloner-vectors-bloch}
\begin{split}
    & m_{00} := [\sin(2\phi), 0, \cos(2\phi)]\\
    & m_{01} := [-\sin(2\phi), 0, -\cos(2\phi)]\\
    & m_{10} := [-\sin(2\phi), 0, \cos(2\phi)]\\
    & m_{11} := [\sin(2\phi), 0, -\cos(2\phi)]
\end{split}
\end{equation}
After the cloning (in the ideal case), the density matrix of each clone will become:

\begin{equation} \label{eqn:4state-cloner-density-bloch}
    \rho_{ij}^c = \frac{1}{2}(\mathds{1} + \eta_x m_{ij}^x\sigma_x + \eta_z m_{ij}^z\sigma_z)
\end{equation}

where $\eta_x$ and $\eta_z$ are the shrinking factors in each direction given as follows:

\begin{equation} \label{eq:4state-cloner-shrinking-factors}
\eta_x = \sin^2(2\phi)\sqrt{\frac{1}{\sin^4(2\phi) + \cos^4(2\phi)}}, \quad \quad 
\eta_z = \cos^2(2\phi)\sqrt{\frac{1}{\sin^4(2\phi) + \cos^4(2\phi)}}
\end{equation}

For the states used in $\mathcal{P}_2$, we have $\phi = \frac{\pi}{8}$ and hence $\eta_x = \eta_z := \eta =  \frac{1}{\sqrt{2}}$. Again, we can reduce the problem to the discrimination probability between the two ensembles encoding $a=0$ and $a=1$ in \eqref{eqn:ensembles}. Let us define $\rho^c$ to be the output clone that Bob chooses to use ($c\in \{1, 2\}$). We have:
\begin{equation}\label{eq:4state-cloner-final-probability}
\begin{split}
     P^{\opt, \mathrm{II}}_{\mathrm{disc}, \mathcal{P}_2} &= \frac{1}{2} + \frac{1}{4}\left|\left|\rho_{(a=0)} - \rho_{(a=1)}\right|\right|_{\Tr}\\
     &= \frac{1}{2} + \frac{1}{4}\left|\left|\frac{1}{2}\left[(\rho_{00}^c - \rho_{11}^c) + (\rho_{10}^c - \rho_{01}^c\right]\right|\right|_{\Tr} \\
     &= \frac{1}{2} + \frac{1}{4}\left|\left|\frac{\eta}{4}((m_{00}^x - m_{11}^x + m_{10}^x - m_{01}^x)\sigma_x + (m_{00}^z - m_{11}^z + m_{10}^z - m_{01}^z)\sigma_z \right|\right|_{\Tr} \\
       &= \frac{1}{2} + \frac{\eta\cos(2\phi)}{4}\left|\left|\sigma_z \right|\right|_{\Tr} \\
     &= \frac{1}{2} + \frac{\eta\cos(2\phi)}{2} = \frac{3}{4}
\end{split}     
\end{equation}
Computing the bias in the same way as above completes the proof.
\end{proof}

\noindent \textbf{\underline{Scenario 2:}}\\

\noindent Here, we give a bound on the success probabilities of Bob in terms of the local fidelities of the QCM where the cloning machine is only tailored to clone two fixed-overlap states. We rely on the fact that Bob can discriminate between the two ensembles of states (for $a=0$, $a=1$) with equal probabilities. 

\begin{thmbox}
\begin{theorem}\label{th:aharonov_2state_attack_II_bias_probability}
The optimal discrimination probability for a cloning attack on the protocol $P_2$, (in attack model II, with $2$ states) is:
\begin{equation} \label{eq:attack_2_state_aharonov_success_probability_bound}
     0.619 \leq P^{\opt, \mathrm{II}}_{\mathrm{disc}, \mathcal{P}_2} \leq 0.823
\end{equation}
\end{theorem}
\end{thmbox}

\begin{proof}
For each of the input states, $\ket{\phi_{i,j}}$ in \eqref{eq:aharonov_coinflip_states}, we denote  $\rho_{ij}^c$ to be a clone outputted from the QCM. Due to symmetry, we only need to consider one of the two output clones. We can now write the effective states for each encoding ($a=0, a=1$) as:
\begin{equation}\label{eqn:ensembles}
\rho_{(a=0)} := \frac{1}{2}(\rho_{00}^c + \rho_{10}^c), \qquad \qquad \rho_{(a=1)} := \frac{1}{2}(\rho_{01}^c + \rho_{11}^c)
\end{equation}
Dealing with these two states is sufficient since it can be shown that discriminating between these two density matrices, is equivalent to discriminating between the entire set of $4$ states in \eqref{eq:coinflip-st}.

Again, we use the discrimination probability from the Holevo-Helstrom bound:
\begin{equation}\label{eqn:opt-helstrom-clone}
 P^{\opt,  \mathrm{II}}_{\mathrm{disc}, \mathcal{P}_2} := P^{\opt}_{\mathrm{disc}}(\rho_{(a=0)},\rho_{(a=1)}) := \frac{1}{2} + \frac{1}{2}\dtr(\rho_{(a=0)},\rho_{(a=1)})
\end{equation}
Now, we have:
\begin{equation}\label{eqn:opt-trace-distance}
\begin{split}
   \dtr(\rho_{(a=0)},\rho_{(a=1)}) & = \frac{1}{2}\left|\left|\rho_{(a=0)} - \rho_{(a=1)}\right|\right|_{\Tr} \\
   & = \frac{1}{2}\left|\left|\frac{1}{2}(\rho_{00}^c - \rho_{11}^c) + \frac{1}{2}(\rho_{10}^c - \rho_{01}^c)\right|\right|_{\Tr} \\
   & \leq \frac{1}{4}\left|\left|(\rho_{00}^c - \rho_{11}^c)\right|\right|_{\Tr} + \left|\left|(\rho_{10}^c - \rho_{01}^c)\right|\right|_{\Tr} \\
   & \leq \frac{1}{2} \left[\dtr(\rho_{00}^c, \rho_{11}^c) + \dtr(\rho_{01}^c,\rho_{10}^c)\right]\\ \\
   \implies    P^{\opt}_{\mathrm{disc}}(\rho_{(a=0)},\rho_{(a=1)}) &\leq  \frac{1}{2} (P^{\opt}_{\mathrm{disc}}(\rho_{00}^c,\rho_{11}^c) + P^{\opt}_{\mathrm{disc}}(\rho_{01}^c,\rho_{10}^c))  \\
   &= P^{\opt}_{\mathrm{disc}}(\rho_{00}^c,\rho_{11}^c) 
\end{split}
\end{equation}

\noindent The last equality follows since for both ensembles, $\{\ket{\phi_{0,0}},\ket{\phi_{1,1}}\}$ and $\{\ket{\phi_{0,1}},\ket{\phi_{1,0}}\}$, we have that their output clones having equal discrimination probability:
\begin{equation}
    P^{\opt}_{\mathrm{disc}}(\rho_{00}^c,\rho_{11}^c)  = P^{\opt}_{\mathrm{disc}}(\rho_{01}^c,\rho_{10}^c) 
\end{equation}
This is because the QCM is symmetric, and depends only on the overlap of the states (we have in both cases $\mbraket{\phi_{00}}{\phi_{11}} = \mbraket{\phi_{01}}{\phi_{10}} = \sin(2\phi)$).

Furthermore, since the cloning machine can only lower the discrimination probability between two states, we have:
\begin{equation}
   P^{\opt}_{\mathrm{disc}}(\rho_{00}^c,\rho_{11}^c) \leq P^{\opt}_{\mathrm{disc}}(\rho_{00}^c,\ket{\phi_{1,1}}\bra{\phi_{1,1}}) =: \overline{P^{\opt}_{\mathrm{disc}}}
\end{equation}
Now, using the relationship between fidelity and the trace distance (\eqref{}), we have the following bounds:
\begin{equation}\label{eqn:opt_disc_probability_helstrom}
\frac{1}{2}+\frac{1}{2}\left(1 - \sqrt{\bra{\phi_{1,1}}\rho^c_{00}\ket{\phi_{1,1}}}\right) \le \overline{P^{\opt}_{\mathrm{disc}}} \le \frac{1}{2} + \frac{1}{2}\sqrt{1 - \bra{\phi_{1,1}}\rho^c_{00}\ket{\phi_{1,1}}}
\end{equation}
By plugging this inequality in the observed density matrix for the output clone, we can find this discrimination probability.

As in the previous section, the output density matrix from the QCM for an output clone can be written as \eqref{eq:reduced_local_state_mayers_attack}:
\begin{equation}
   \rho^c_{00} = \alpha\ket{\phi_{0,0}}\bra{\phi_{0,0}} + \beta \ket{\phi_{1,1}}\bra{\phi_{1,1}} +
\gamma (\ket{\phi_{0,0}}\bra{\phi_{1,1}}+ \ket{\phi_{1,1}}\bra{\phi_{0,0}}) 
\end{equation}

Hence the output state has a local fidelity, $F_{\Lbs} = \bra{\phi_{0,0}}\rho^c_{00}\ket{\phi_{0,0}} = \alpha + s^2\beta + s\gamma$. On the other hand, we have $F(\rho^c_{00}, \ket{\phi_{1, 1}}\bra{\phi_{1, 1}}) = \bra{\phi_{1, 1}}\rho^c_{00}\ket{\phi_{1, 1}} = s^2\alpha + \beta + s\gamma$. Combining these two, we then have:
\begin{equation}
F(\rho^c_{00}, \ket{\phi_{1, 1}}\bra{\phi_{1, 1}}) = F_{\Lbs} + (s^2 - 1)(\alpha - \beta)
\end{equation}
Plugging in $F_{\Lbs}$ from \eqref{eq:prelim-local_optimal_non_ortho_fidelity_1to2}, and $\alpha - \beta = \sqrt{\frac{1-s^2}{1-s^4}}$ (for an optimal state-dependent cloner), we get:
\begin{equation} \label{eqn:attack_2_state_aharonov_success_probability_bound_not_filled_appendix}
    \frac{1}{2} + \frac{1}{2}\left[1 - \sqrt{F_{\Lbs} + (s^2 - 1) \sqrt{\frac{1-s^2}{1-s^4}}}\right] \leq P^{\opt, \textrm{II}}_{\mathrm{disc}, \mathcal{P}_2} \leq \frac{1}{2} + \frac{1}{2}\sqrt{1 - F_{\Lbs} - (s^2 - 1) \sqrt{\frac{1-s^2}{1-s^4}}}
\end{equation}
To complete the proof, we use $F_{\Lbs} \approx 0.989$ and $s = 1/\sqrt{2}$ which gives the numerical discrimination probabilities above.
\end{proof}

\section{Variational Quantum Cloner: specifications of the algorithm}\label{sec:varqlone-algorithm}
Now we introduce our machine learning algorithm \emph{Variational Quantum Cloner} or $\VQC$ that given a specific family of states, learns the circuit that optimally clones that family. We recall that our motivation is to find short-depth circuits to clone a given family of states, and also use this toolkit to investigate the family of states where the optimal figure of merit is unknown.

$\VQC$ is a variational quantum algorithm with similar core parts to other VQAs. We have given an overview of such techniques in the preliminaries (Section \ref{sec:prelim-vqc}). In particular, this variational method uses a parameterised state, denoted by $\rho_{\paramtheta}$, typically prepared by some short-depth parameterised unitary on some initial state $\rho_{\paramtheta} := U(\paramtheta)\ket{0}\bra{0}U^{\dagger}(\paramtheta)$. The parameters are then optimized by minimizing (or maximizing) a \emph{cost function}, typically a function of $k$ observable measurements on $\rho_{\paramtheta}$, $\mathsf{O}_k$. This resembles a classical neural network, and indeed, techniques and ideas from classical machine learning can be borrowed and adapted to our setting. Nevertheless, we need to develop the core ingredients such as differentiable cost functions for gradient-descent based optimisation and theoretical guarantees on these cost functions specific to our problem. Additionally, for all the results we give here we use the gradient-descent-based optimizer (as discussed in Section~\ref{sec:prelim-vqc-opt}) \cite{kingma_adam_2017} with our cost functions.

We also note that other than the core theoretical subjects discussed in this section, the machine learning algorithm itself as well as the codes and simulations have not been developed and run by the author and therefore have been excluded from this thesis. In this section, we only present theoretical results where the author has contributed, such as defining suitable cost functions for the cloning problem and providing theoretical guarantees on them. For more details on the algorithm and related information, we refer the reader to the paper \cite{coyle_progress_2022}.

\subsection{Cost functions}\label{sec:varqlone-algorithm-cost-function}
In this section, we propose several cost functions for our problems and discuss their advantages and differences. Primarily, we propose the so-called `local' cost functions of the following functional form:
\begin{align}\label{eq:local_cost}
    \Cbs_{\mathsf{loc}}^{M\rightarrow N}(\paramtheta) &:= \mathop{\mathbb{E}}_{\substack{\ket{\psi} \in S}} f(\mathsf{O}^{\psi}_{\Lbs}, \rho_{\paramtheta}, M, N)
\end{align}
Where $\ket{\psi}$ denotes target state of the cloning machine, $S$ denotes the set of states to be cloned, and $M$ and $N$ are the number of copies for the input states and output clones respectively.
Choosing $f$, or $f_{\sq}$ (denoting the squared cost function) as follows:
\begin{equation} \label{eqn:f_function_squared_cost}
    f_{\sq} := \sum\limits_{i=1}^N (1-F^i_{\Lbs}(\paramtheta))^2
    + \sum\limits_{i<j}^N (F^i_{\Lbs}(\paramtheta)-F^j_{\Lbs}(\paramtheta))^2
\end{equation}
results in what for brevity we refer to as the \emph{squared} cost function, a generalization of the cost also proposed in \cite{jasek_experimental_2019}. Here  $F^j_{\Lbs}(\boldsymbol{\theta}) := F_{\Lbs}(\ket{\psi}\bra{\psi}, \rho^j_{\boldsymbol{\theta}})$ is the local fidelity of the parameterized state relative to output clone $j$. This is generated using the observable $\mathsf{O}^{\psi}_{\sq} = \ket{\psi}\bra{\psi}$ for the specific instance of state to be cloned from the set, $\ket{\psi} \in S$. As such, we define $\Cbs_{\sq}^{M\rightarrow N}(\paramtheta) := \mathop{\mathbb{E}}_{\substack{\ket{\psi} \in S}} \left[f_{\sq}\right]$.

Let us now elaborate a bit on the alternative choices of cost functions. The second local cost, which we call the \emph{linear} local cost or `local cost' again for brevity, is given by:
\begin{equation}\label{eq:local_cost_full_supp}
    \Cbs_{\Lbs}^{M\rightarrow N}(\paramtheta) := \mathop{\mathbb{E}}_{\substack{\ket{\psi} \in S}} \left[\Cbs^{\psi}_{\Lbs}(\paramtheta) \right] :=  \mathop{\mathbb{E}}_{\substack{\ket{\psi} \in S}} \left[\Tr( \mathsf{O}^{\psi}_{\Lbs} \rho_{\paramtheta})\right], \quad
    \mathsf{O}^{\psi}_{\Lbs}  := \mathds{1} - \frac{1}{N}\sum\limits_{j=1}^N\ket{\psi}\bra{\psi}_j \otimes \mathds{1}_{\Bar{j}}
\end{equation}
where $\ket{\psi} \in S$ is state to be cloned.

\noindent These first two cost functions, are related only in that they are both functions of \emph{local} observables, or in other words, the local fidelities. The third cost, on the other hand, is fundamentally different compared to the other two proposals, since it captures the global fidelity \emph{i.e.} uses global observables, and as such, we refer to it as the `\emph{global cost}':
\begin{equation}\label{eq:global_cost_full_supp}
    \Cbs_{\Gbs}^{M\rightarrow N}(\paramtheta) := \mathop{\mathbb{E}}_{\substack{\ket{\psi} \in S}} \left[\Tr(\mathsf{O}^{\psi}_{\Gbs}\rho_{\paramtheta})\right], \qquad
     \mathsf{O}^{\psi}_{\Gbs} := \mathds{1} - \ket{\psi}\bra{\psi}^{\otimes N}
\end{equation}
The second local cost, and our global cost functions are adapted from the literature on variational algorithms~\cite{larose_variational_2019,cerezo_cost_2021,khatri_quantum-assisted_2019,sharma_noise_2020}. For compactness, we will drop the superscript $M \rightarrow N$ when the meaning is clear from context. 

Now, we motivate our choices for the above cost functions. For \eqref{eq:local_cost}, if we restrict to the special case of $1\rightarrow 2$ cloning (\emph{i.e.}\@ we have only two output parties, $j\in \{B, E\}$), and remove the expectation value over states, we recover the cost function used in Ref.~\cite{jasek_experimental_2019}. A useful feature of this cost is that symmetry is explicitly enforced by the difference term $(F_i(\paramtheta) - F_j(\paramtheta))^2$. 

In contrast, the local and global cost functions are inspired by other variational algorithm in the literature~\cite{larose_variational_2019,cerezo_cost_2021,khatri_quantum-assisted_2019, sharma_noise_2020} where their properties have been extensively studied, particularly in relation to the phenomenon of `barren plateaus'~\cite{mcclean_barren_2018, cerezo_cost_2021}. Since we have not covered the topic of barren plateaus in \chapref{chap:prelim}, we will give a brief description here. Barren plateaus is a phenomenon where the gredient-based optimisation in the quantum landscape ends up with no interesting search directions to go. It has been demonstrated that hardware efficient $\Ansatze$ are untrainable using a global cost function similar to the one given in \eqref{eq:global_cost_full_supp}, since they have exponentially vanishing gradients~\cite{schuld_evaluating_2019}, often leading to a barren plateau . In contrast, local cost functions (\eqref{eq:local_cost_full_supp}, \eqref{eq:local_cost}) are shown to be efficiently trainable with $\mathcal{O}(\log N)$ depth hardware efficient $\Ansatze$~\cite{cerezo_cost_2021} (see Section~\ref{sec:prelim-qml-ansatz} for more details about different types of \Ansatze).

We also remark that typically global cost functions are usually more favourable from the point of view of \emph{operational meaning}. For example in variational compilation~\cite{khatri_quantum-assisted_2019}, this cost function compares the closeness of two global unitaries. In this respect, local cost functions are usually used as a proxy to optimize a global cost function. 

In our case, the nature of quantum cloning allows $\VQC$ local cost functions to have immediate operational meaning, illustrated through the following example (using the local cost, \eqref{eq:local_cost_full_supp}) for $1\rightarrow 2$ cloning:
\begin{align*}
    \Cbs^{\psi}_{\Lbs}(\paramtheta) &= \Tr\left[\left(\mathds{1} - \frac{1}{2}\sum\limits_{j=1}^2\ket{\psi}\bra{\psi}_j \otimes \mathds{1}_{\Bar{j}}\right) \rho_{\paramtheta}\right]\\
     \implies  \Cbs_{\Lbs}(\paramtheta)  &= 1 - \frac{1}{2}\mathbb{E}\left[F_{\Lbs}\left(\ket{\psi}\bra{\psi}, \rho_{\paramtheta}^1\right) + F_{\Lbs}\left(\ket{\psi}\bra{\psi}, \rho_{\paramtheta}^2\right)\right]
\end{align*}
where $\mathbb{E}[F_{\Lbs}]$ is the average fidelity~\cite{scarani_quantum_2005} over the possible input states. The final expression of $\Cbs_{\Lbs}(\paramtheta)$ in the above equation follows from the expression of fidelity when one of the states is pure. Similarly, the global cost function relates to the global fidelity of the output state concerning the input state(s).

\subsection{Cost function gradients}\label{sec:varqlone-algorithm-gredient}
In the gradient-descent-based optimization approach which we use for developing our algorithm, we require efficient computation of the gradients. Here, we derive the analytic gradients for our cost functions. We use the local cost function, \eqref{eq:local_cost} as an explicit example and the derivations for the other cost functions follow straightforwardly.
As a reminder, the squared cost is given by:
\begin{align}\label{eq:squared_local_cost_mton_gradient}
    \Cbs_{\sq}^{M\rightarrow N}(\paramtheta) := \mathop{\mathbb{E}}_{\substack{\ket{\psi} \in S}}\left[ \sum\limits_{i=1}^N (1-F^i_{\Lbs}(\paramtheta))^2\right. 
    \left.+ \sum\limits_{i<j}^N (F^i_{\Lbs}(\paramtheta)-F^j_{\Lbs}(\paramtheta))^2\right] 
\end{align}
where the expectation is taken over the set of states with uniform distribution. For example, in the phase-covariant cloner of the states \eqref{eq:x_y_plane_states_phase_cov}, the parameter $\eta$ is sampled uniformly from the interval $[0, 2\pi)$.

Now, the derivative of \eqref{eq:squared_local_cost_mton_gradient}, with respect to a single parameter, $\theta_l$ is given by:
\begin{align}\label{eq:squared_local_cost_mton_gradient_calc}
\small
    \frac{\partial \Cbs_{\sq}(\boldsymbol{\theta})}{\partial \theta_l} =
    2\mathop{\mathbb{E}}_{\substack{\ket{\psi} \in S}}\left[\sum\limits_{i=1}^N (1-F^i_{\Lbs}(\paramtheta)) \left[-\frac{\partial  F^i_{\Lbs}(\boldsymbol{\theta})}{\partial \theta_l}\right]+\sum\limits_{i<j}^N (F^i_{\Lbs}(\paramtheta)-F^j_{\Lbs}(\paramtheta)) \left[\frac{\partial  F^i_{\Lbs}(\boldsymbol{\theta})}{\partial \theta_l}  - \frac{\partial  F^j_{\Lbs}(\boldsymbol{\theta})}{\partial \theta_l} \right]\right]
\end{align}
We can rewrite the expression for the fidelity of the $j^{th}$ clone as:
\begin{align}
    F^{j}_{\Lbs}(\boldsymbol{\theta}) = \bra{\psi}\rho_j(\boldsymbol{\theta})\ket{\psi} = \Tr\left[\ket{\psi}\bra{\psi}\rho_{j}\right] =  \Tr\left[\ket{\psi}\bra{\psi}\Tr_{\bar{j}}\left(U(\boldsymbol{\theta}) \rho_{\text{init}}U(\boldsymbol{\theta})^{\dagger} \right)\right] 
\end{align}

Using the linearity of the trace, the derivative of the fidelities with respect to the parameters, $\theta_l$, can be computed as:
\begin{align}
    \frac{\partial   F^{j}_{\Lbs}(\boldsymbol{\theta})}{\partial \theta_l} =   \Tr\left[\ket{\psi}\bra{\psi}\Tr_{\bar{j}}\left(\frac{\partial  U(\boldsymbol{\theta})\rho_{\text{init}}U(\boldsymbol{\theta})^{\dagger}}{\partial \theta_l} \right)\right] 
\end{align}
Using the \emph{parameter shift rule} (\thmref{th:perlim-param-shift-rule} from Section~\ref{sec:prelim-vqc-opt}) technique the explicit expression of the cost function's gradients can be calculated. The calculation has been given in \appref{app:vqc-gradient-calc}.

\subsection{Cost function guarantees}\label{sec:varqlone-algorithm-cost-guarantee}
One of the interesting problems in the area of theoretical machine learning is showing theoretical guarantees for the cost function, \emph{i.e.} achieving the cost minimum indicates a solution to the problem in question~\cite{khatri_quantum-assisted_2019, bravo-prieto_variational_2020}. This property is known as \emph{faithfulness}.

For our approximate quantum cloning problem, due to the information-theoretic limits, the above costs cannot have a minimum at $0$, but instead at some finite positive value (say $\Cbs^{\textrm{opt}}_{\Lbs}$ for the local cost).

Despite this, we can still derive certain theoretical guarantees about them. Specifically, we consider notions of \emph{strong} and \emph{weak} faithfulness, relative to the learner's error in our solution. Our goal is to provide statements about the \emph{generalization performance} of the cost functions, by considering how close are the states we output by our cloning machine, to those which would be outputted from the `\emph{optimal}' cloner, relative to some metrics. In the following, we denote $\rho_{\mathrm{opt}}^{\psi, j}$ ($\rho^{\psi, j}_{\paramtheta}$) to be the optimal ($\VQC$ learned) reduced state for qubit $j$, for a particular input state $\ket{\psi}$. If the superscript $j$ is not present, we mean the global state of all clones. Let us give the definitions of faithfulness.
\begin{defbox}
\begin{definition}[Strong Faithfulness]
A cloning cost function, $\Cbs$, is strongly faithful if for all $\ket{\psi} \in S$, optimising the closeness in cost function implies the the optimally close  states \emph{i.e.}:
    \begin{equation}\label{eq:strongly_faithful_cost_defn}
        \Cbs(\paramtheta) = \Cbs^{\mathrm{opt}} \implies \rho_{\paramtheta}^\psi = \rho_{\opt}^{\psi} \qquad \forall \ket{\psi} \in S
    \end{equation}
    where $\Cbs^{\mathrm{opt}}$ is the minimum value achievable (allowed by quantum mechanics) for the cost $\Cbs$, and $S$ is the given set of states to be cloned.
\end{definition}
\end{defbox}
\begin{defbox}
\begin{definition}[$\epsilon$-Weak Local Faithfulness]
A local cloning cost function, $\Cbs_{\Lbs}$, is $\epsilon$-weakly faithful if for all $\ket{\psi} \in S$ and for all the \emph{local} clones, the closeness of local cost function to its optimal value implies the closeness of local clone states \emph{i.e.}:
    \begin{equation}\label{eq:weakly_faithful_local_cost_defn}
        |\Cbs_{\Lbs}(\paramtheta) - \Cbs_{\Lbs}^{\mathrm{opt}}| \leqslant \epsilon \implies D(\rho_{\paramtheta}^{\psi, j}, \rho_{\opt}^{\psi, j}) \leqslant f(\epsilon), \qquad \forall \ket{\psi} \in S, \forall j 
    \end{equation}
    where $D( \cdot, \cdot)$ is a chosen metric in the Hilbert space between the two states and $f$ is a polynomial function of $\epsilon$.
\end{definition}
\end{defbox}
\begin{defbox}
\begin{definition}[$\epsilon$-Weak Global Faithfulness]
    A global cloning cost function, $\Cbs_{\Gbs}$, is $\epsilon$-weakly faithful if for all $\ket{\psi} \in S$ the closeness of global cost function to its optimal value implies the closeness of global optimal state \emph{i.e.}:
    \begin{equation}\label{eqn:weakly_faithful_global_cost_defn}
        |\Cbs_{\Gbs}(\paramtheta) - \Cbs_{\Gbs}^{\mathrm{opt}}| \leqslant \epsilon \implies D(\rho_{\paramtheta}^{\psi}, \rho_{\opt}^{\psi}) \leqslant f(\epsilon) \ \ \forall \ket{\psi} \in S
    \end{equation}
\end{definition}
\end{defbox}
One could also define local and global versions of the strong faithfulness, but this is less attractive as it is included in the other case. Thus we do not focus on it here. Let us begin by examining the squared local cost function. For this case, we will provide the most extensive analysis, and faithfulness proofs for the other cost functions can be derived using similar methods.\\

\noindent\textbf{Squared Cost Function}\\

\noindent First, we start with the squared cost function which we rewrite as:
\begin{equation} \label{eq:squared_local_cost_mton_supp_redo}
        \Cbs_{\sq}^{M\rightarrow N}(\paramtheta) = \frac{1}{\mathcal{N}}\int_{S}\left[ \sum\limits_{j=1}^N (1-F_i(\paramtheta))^2 + \sum\limits_{i<j}^N (F_i(\paramtheta)-F_j(\paramtheta))^2\right]d \psi
\end{equation}
where the expectation of a fidelity $F_i$ over the states in distribution $S$ is defined as $\mathbb{E}[F_i] = \frac{1}{\mathcal{N}}\int_{S}F_i\cdot d\psi$, with the normalisation condition being $\mathcal{N} = \int_{S}d\psi$. For qubit states, if the normalisation is over the entire Bloch sphere in $SU(2)$, then $\mathcal{N} = 4\pi$. For notation simplicity, we herein denote the $\Cbs_{\sq}^{M\rightarrow N}(\paramtheta)$ as $\Cbs_{\sq}(\paramtheta)$. We begin with a proof of the fact that the cost function is \emph{strongly} faithful.

\begin{thmbox}
\begin{theorem} \label{thm:squared_local_cost_squared_strong_faithful_local_appendix}[Strong faithfulness of the squared cost function]
The squared local cost function is locally strongly faithful, \emph{i.e.}:
    \begin{equation}\label{eq:squared_cost_function_locally_faithful}
        \Cbs_{\sq}(\paramtheta) = \Cbs_{\sq}^{\mathrm{opt}} \implies \rho_{\paramtheta}^{\psi, j} = \rho_{\opt}^{\psi, j} \qquad \forall \ket{\psi} \in S, \forall j \in [N]
    \end{equation}
\end{theorem}
\end{thmbox}

\begin{proof}
The cost function $\Cbs_{\textrm{sq}}(\paramtheta)$ achieves a minimum at the joint maximum of $\mathbb{E} [F_i(\paramtheta)]$ for all $i \in [N]$. In symmetric $M \rightarrow N$ cloning, the expectation value of all the $N$ output fidelities peak at $F_i = F_{\opt}$ for all input states $\ket{\psi}$. This corresponds to a unique optimal joint state $\rho_{\opt}^{\psi, j} = U_{\opt}\ket{\psi^{\otimes M}, 0^{\otimes N-M}}\bra{\psi^{\otimes M}, 0^{\otimes N-M}}U_{\opt}^{\dagger}$ for each $\ket{\psi} \in S$ and for any $j \in [N]$, where $U_{\opt}$ is the unitary producing the the optimal state. Since the joint optimal state and the corresponding fidelities are unique for all input states in the distribution, we conclude that the cost function achieves a minimum under precisely the unique condition \emph{i.e.}\@ $\mathbb{E}[F_j(\paramtheta)] = F_{\opt}$ for all $j \in [N]$. This condition implies that, 
\begin{equation}\label{Eq:strong_faithful_condition}
    \rho^{\psi,j}_{\paramtheta} = \rho^{\psi,j}_{\opt}, \qquad \forall \ket{\psi} \in S, \forall j \in [N]
\end{equation}
We note that since $F_{\opt}$ is the same for all the reduced states $j \in [N]$, this implies that the optimal reduced states are all the same for a given $\ket\psi \in S$.  Thus \eqref{Eq:strong_faithful_condition} provides the necessary guarantee that minimizing the cost function over the parameter space, results in the corresponding circuit's output, being equal to the optimal cloned state for all the inputs.  
\end{proof}

\noindent Now, we take the weaker notion of faithfulness into account. Computing the exact fidelities of the output states requires an infinite number of copies. In reality, we run the iteration only a finite number of times and thus, our cost function can only reach the optimal cost up to some precision. This is also relevant when running the circuit on devices in the NISQ era which would inherently introduce noise in the system. Thus, we can only hope to minimise the cost function up to some precision of the optimal cost. This statement can be formalised via the following lemma: 

\begin{lembox}
\begin{lemma}\label{lemma:trace_bound_squared_cost}
Suppose the cost function is $\epsilon$-close to the optimal cost in symmetric cloning
\begin{equation}
    \Cbs_{\mathrm{sq}}(\paramtheta) - \Cbs^{\opt}_{\mathrm{sq}} \leqslant \epsilon
    \label{eq:cost_to_epsilon_squared}
\end{equation}
Then we have,
\begin{equation}\label{eq:tracecloseness_squared_local_appendix}
        \Tr\left[(\rho_{\opt}^{\psi, j} - \rho^{\psi, j}_{\paramtheta})\ket{\psi}\bra{\psi}\right] \leqslant \frac{\mathcal{N}\epsilon}{2(1 - F_{\opt})}, \qquad \forall \ket{\psi} \in S, \forall j \in [N]
\end{equation}
\end{lemma}
\end{lembox}

\begin{proof}
In $M \rightarrow N$ symmetric cloning, the optimal cost function value is achieved when each output clone achieves the fidelity $F_{\opt}$.  Thus, using  \eqref{eq:local_cost} (or \eqref{eq:squared_local_cost_mton_gradient}), the optimal cost function value is given by,
\begin{equation}
    \Cbs^{\opt}_{\sq} = N \cdot (1 - F_{\opt})^2
\end{equation}
The optimal cost function is achieved when all output clones have the same fidelity. Therefore, as we begin to minimize the cost $\Cbs_{\mathrm{sq}}(\paramtheta)$, all the output clones start to produce states with approximately the same fidelity. This is explicitly enforced by taking the limit $\epsilon \rightarrow 0$, in which case the difference terms of \eqref{eq:squared_local_cost_mton_gradient} vanish. Thus, the cost function explicitly enforces the symmetry property. Let us assume $\epsilon \rightarrow 0$, and consider the quantity $ \Cbs_{\sq}(\paramtheta) - \Cbs^{\opt}_{\sq}$:
\begin{equation}     \label{eq:costtotrace_squared}
    \begin{split}
        \Cbs_{\sq}(\paramtheta) - \Cbs^{\opt}_{\sq} &=
        \frac{1}{\mathcal{N}}\int_{S}\left[ \sum\limits_{i}^N (1-F_i(\paramtheta))^2 + \sum\limits_{i<j}^N  (F_i(\paramtheta)-F_j(\paramtheta))^2\right]d \psi  - N\cdot (1 - F_{\opt})^2\\
        &\underset{\epsilon \rightarrow 0}{\approx} \frac{1}{\mathcal{N}}\int_{S}\left[ \sum\limits_{j}^N (1-F_j(\paramtheta))^2  - N\cdot (1 - F_{\opt})^2 \right]d \psi  \\
    &\approx \frac{1}{\mathcal{N}}\int_{S}\left[ \sum\limits_{j}^N (F_{\opt} - F_j(\paramtheta))(2 - F_{\opt} - F_j(\paramtheta)) \right]d \psi  \\
     &\geqslant \frac{2(1 - F_{\opt})}{\mathcal{N}}\int_{S}\left[ \sum\limits_{j}^N (F_{\opt} - F_j(\paramtheta)) \right]d \psi  \\
        &= \frac{2(1 - F_{\opt})}{\mathcal{N}}\left[ \sum\limits_{j}^N \int_{S}\text{Tr}[(\rho_{\opt}^{\psi, j} - \rho^{\psi, j}_{\paramtheta})\ket{\psi}\bra{\psi}]d\psi \right]
    \end{split}
\end{equation}
The second line follows since $F_{\opt}$ is the same for each input state $\ket{\psi}$. Utilizing the inequality in \eqref{eq:cost_to_epsilon_squared} and \eqref{eq:costtotrace_squared}, we obtain,
\begin{equation}
\begin{split}
       &  \sum\limits_{j}^N \int_{S}\text{Tr}\left[(\rho_{\opt}^{\psi, j} - \rho^{\psi, j}_{\paramtheta})\ket{\psi}\bra{\psi}\right]d\psi  \leqslant \frac{\mathcal{N}\epsilon}{2(1 - F_{\opt})} \\
       & \implies \text{Tr}\left[(\rho_{\opt}^{\psi, j} - \rho^{\psi, j}_{\paramtheta})\ket{\psi}\bra{\psi}\right] \leqslant \frac{\mathcal{N}\epsilon}{2(1 - F_{\opt})}, \hspace{3mm} \forall \ket{\psi} \in S, \forall j \in [N]
\label{eq:tracecloseness_squared_local}
\end{split}
\end{equation}
This concludes the proof.
\end{proof}
\noindent The above inequality allows us to quantify the closeness of the state produced by $\VQC$ and the unique optimal clone for any $\ket{\psi} \in S$. We quantify this closeness of the states in a popular distance measure in quantum information, namely the Fubini-Study (or Bures angle) distance between two quantum states (introduced in Section~\ref{sec:prelim-distances}). Using the above lemma, we can prove the following two theorems for the squared local cost function:
\begin{thmbox}
\begin{theorem}\label{thm:squared_local_cost_squared_FS_weak_faithful}[Weak faithfulness of the local squared const function]
The squared cost function as defined in \eqref{eq:local_cost}, is $\epsilon$-weakly faithful with respect to the Bures angle $\dba$ (or alternatively Fubini-distance measure $\dfs$).
In other words, if the squared cost function is $\epsilon$-close to its minimum, \emph{i.e.}\@:
\begin{equation}\label{eq:squared_cost_to_epsilon}
    \Cbs_{\mathrm{sq}}(\paramtheta) - \Cbs^{\mathrm{opt}}_{\mathrm{sq}} \leqslant \epsilon
\end{equation}
where $\Cbs^{\opt}_{\mathrm{sq}} := \underset{\paramtheta}{\textrm{min}}\sum\limits_{i}^N (1-F_i(\paramtheta))^2 + \sum\limits_{i<j}^N  (F_i(\paramtheta)-F_j(\paramtheta))^2 = N(1-F_{\mathrm{opt}})^2$ is the optimal theoretical cost using fidelities produced by the ideal \emph{symmetric} cloning machine, then the following holds:
\begin{equation}\label{eqn:fubini_study_bound_squared}
    \dba(\rho^{\psi, j}_{\paramtheta}, \rho_{\opt}^{\psi, j}) \leqslant \frac{\mathcal{N}}{2(1 - F_{\mathrm{opt}})\sin(F_{\mathrm{opt}})}\cdot \epsilon := f_1(\epsilon),   \qquad \forall \ket{\psi} \in S, \forall j \in [N]
\end{equation}
\end{theorem}
\end{thmbox}

\begin{proof}
To prove this theorem, we revisit and rewrite the Bures angle from \eqref{eq:prelim-bures-angle} (see Section~\ref{sec:prelim-distances}):
\begin{equation} \label{eq:fubini_study_defn}
    \dba(\rho,\sigma) = \arccos{\sqrt{F(\rho, \sigma)}} = \text{arccos}\hspace{1mm} \bra{\phi}\tau\rangle
\end{equation}
where $\ket{\phi}$ and $\ket{\tau}$ are the purifications of $\rho$ and $\sigma$ respectively which maximize the overlap. We note that $\dba(\rho,\sigma)$ lies in the interval $[0,\pi/2]$, with the value $\pi/2$ corresponding to the unique solution $\rho = \sigma$. Since this distance is a metric,  it obeys the triangle's inequality, \emph{i.e.}, for any three states $\rho, \sigma$ and $\delta$,
\begin{equation}
     \dba(\rho,\sigma) \leqslant  \dba(\rho,\delta) +  \dba(\sigma,\delta)
     \label{eq:triangle-inequality}
\end{equation}

Rewriting the result of \lemref{lemma:trace_bound_squared_cost} in terms of fidelity for each $\ket{\psi} \in S$ and correspondingly in terms of Bures distance using \eqref{eq:fubini_study_defn} is, 
\begin{equation}\label{eq:fidelityFS_squared_local}
\begin{split}
    & F(\rho_{\opt}^{\psi, j}, \ket{\psi}) - F(\rho^{\psi,j}_{\paramtheta}, \ket{\psi}) \leqslant \epsilon' \\\\
    & \implies  \cos^2(\dba(\rho_{\opt}^{\psi, j}, \ket{\psi})) -  \cos^2(\dba(\rho^{\psi,j}_{\paramtheta}, \ket{\psi})) \leqslant \epsilon' 
\end{split}    
\end{equation}
where $\epsilon' = \mathcal{N}\epsilon/2(1 - F_{\opt})$. 
Let us denote $D_{\pm}^{\psi} = \dba(\rho_{\opt}^{\psi, j}, \ket{\psi}) \pm \dba(\rho^{\psi,j}_{\paramtheta}, \ket{\psi})$
This inequality in \eqref{eq:fidelityFS_squared_local} can be further rewritten as,
\begin{equation}
\small
\begin{split}
    \cos(\dba(\rho_{\opt}^{\psi, j}, \ket{\psi})) -  \cos(\dba(\rho^{\psi,j}_{\paramtheta}, \ket{\psi})) &\leqslant \frac{\epsilon'}{\cos(\dba(\rho_{\opt}^{\psi, j}, \ket{\psi})) +  \cos(\dba(\rho^{\psi,j}_{\paramtheta}, \ket{\psi}))} \\
    \cos(\dba(\rho_{\opt}^{\psi, j}, \ket{\psi})) -  \cos(\dba(\rho^{\psi,j}_{\paramtheta}, \ket{\psi})) &\lessapprox \frac{\epsilon'}{2\cos(\dba(\rho_{\opt}^{\psi, j}, \ket{\psi}))} \\
    2\sin\left(\frac{D^{\psi}_{+}}{2}\right)\sin\left(\frac{\textrm{D}^{\psi}_{-}}{2}\right) &\leqslant \frac{\epsilon'}{2\cos(\dba(\rho_{\opt}^{\psi, j}, \ket{\psi}))} \\
    \implies \textrm{D}^{\psi}_{-} &\leqslant \frac{\epsilon'}{\sin(\dba(\rho_{\opt}^{\psi, j}, \ket{\psi}))}  = \frac{\mathcal{N}\epsilon}{2(1 - F_{\opt})\sin(F_{\opt})}
\end{split}
\label{eq:cosinerelation}
\end{equation}
where we have used the approximations that in the limit $\epsilon \rightarrow 0$, $\dba(\rho_{\opt}^{\psi, j}, \ket{\psi}) \approx \dba(\rho^{\psi,j}_{\paramtheta}, \ket{\psi})$ and the trigonometric identities $\cos\left( x - y \right) = 2\sin \left(\frac{x+y}{2}\right)\sin \left(\frac{x-y}{2}\right)$, and $\sin 2x = 2\sin x \cos x$. 

Further, using the Fubini-Study metric triangle's inequality on the set of states $\{\rho_{\opt}^{\psi, j}, \rho^{\psi,j}_{\paramtheta}, \ket{\psi}\}$
results in,
\begin{equation}
 \dba(\rho^{\psi,j}_{\paramtheta}, \ket{\psi}) \leqslant \dba(\rho_{\opt}^{\psi, j}, \ket{\psi}) + \dba(\rho^{\psi,j}_{\paramtheta}, \rho_{\opt}^{\psi, j}) 
\end{equation}
Combining the above inequality and \eqref{eq:cosinerelation} results in,
\begin{equation} \label{eqn:squared_cost_fubini_study_bound_appendix}
    \dba(\rho^{\psi,j}_{\paramtheta}, \rho_{\opt}^{\psi, j}) \leqslant \frac{\mathcal{N}}{2(1 - F_{\opt})\sin(F_{\opt})}\cdot \epsilon, \hspace{3mm} \forall \ket{\psi} \in S
\end{equation}
This bounds the closeness of the trained output state and the optimal output state as a function of $\epsilon$.
\end{proof}

A similar result can be derived relative to trace distance instead of the Bures/Fubini-Study distance. However, we avoid presenting the result here, since it is very similar in nature. Instead, we refer the reader to \cite{coyle_progress_2022} for the faithfulness result using the trace distance.\\\\
\noindent\textbf{Local Cost Function}\\\\
Next, we prove analogous results for the local cost function, defined for $M \rightarrow N$ cloning. We rewrite the cost function with an average integral form over the set $S$:
\begin{equation} \label{eq:local_cost_full_suppforproof}  
    \Cbs_{\Lbs}(\paramtheta) := \mathbb{E}\left[1 - \frac{1}{N}\left(\sum\limits_{j=1}^{N} F_j({\paramtheta})\right)\right] = 1 - \frac{1}{N\mathcal{N}}\int_{S} \sum\limits_{j=1}^{N} F_j({\paramtheta}) d\psi
\end{equation}
where $\mathcal{N} = \int_{S}d\psi$ is the normalisation condition. 
As above, we can show this cost function also exhibits strong faithfulness:

\begin{thmbox}
\begin{theorem}[Strong faithfulness of the local cost function] \label{thm:squared_local_cost_local_FS_strong_faithful}
    The local squared cost function is locally strongly faithful:
    \begin{equation}\label{eqn:strongly_local_faithful_local_cost_defn_appendix}
        \Cbs_{\Lbs}(\paramtheta) = \Cbs_{\Lbs}^{\mathrm{opt}} \implies \rho_{\paramtheta}^{\psi, j} = \rho_{\opt}^{\psi, j} \qquad \forall \ket{\psi} \in S, \forall j \in [N]
    \end{equation}
\end{theorem}
\end{thmbox}

\begin{proof}
Similar to the faithfulness arguments of the squared cost function, one can immediately see that the cost function $\Cbs_{\Lbs}(\paramtheta)$ achieves a unique minimum at the joint maximum of $\mathbb{E}[F_j(\paramtheta)]$ for all $j \in [N]$. Thus, the minimum of $\Cbs_{\Lbs}(\paramtheta)$ corresponds to the unique optimal joint state with its unique local reduced states $\rho_{\opt}^{\psi, j}$ for each $j \in [N]$, and for each input state $\ket{\psi} \in S$. 
Thus the cost function achieves a minimum under precisely the unique condition \emph{i.e.} the output state is equal to the optimal clone state.
\end{proof}

\noindent Now, we can also prove analogous versions of weak faithfulness. Many of the steps in the proof follow similarly to the squared cost derivations above, so we omit them for brevity where possible. As above, we first have the following lemma:

\begin{lembox}
\begin{lemma} \label{lemma:trace_bound_local_cost}
Suppose the cost function is $\epsilon$-close to the optimal cost in symmetric cloning
\begin{equation}
    \Cbs_{\Lbs}(\paramtheta) - \Cbs^{\opt}_{\Lbs} \leqslant \epsilon
    \label{eqn:costtoepsilon_local_appendix}
\end{equation}
where we assume $ \lim_{\epsilon \rightarrow 0} |\mathbb{E}[F_i(\paramtheta)] - \mathbb{E}[F_j(\paramtheta)]| \rightarrow 0, \forall i, j$, and therefore $\Cbs_{\opt} := 1-F_{\opt}$. Then,
\begin{equation}\label{eq:tracecloseness_local_cost_appendix}
        \Tr[(\rho_{\opt}^{\psi, j} - \rho^{\psi, j}_{\paramtheta})\ket{\psi}\bra{\psi}] \leqslant \mathcal{N}\epsilon, \qquad \forall \ket{\psi} \in S, \forall j \in [N]
\end{equation}
\end{lemma}
\end{lembox}

The proof of \lemref{lemma:trace_bound_local_cost} follows almost identically to \lemref{lemma:trace_bound_squared_cost}, but with the exception that we can write $\Cbs_{\Lbs}(\paramtheta) - \Cbs^{\opt}_{\Lbs} = \mathbb{E}\left(F_{\opt} - F(\paramtheta)\right)$ in the symmetric case, assuming $F_i(\paramtheta) \approx F_j(\paramtheta)$,  $\forall i \neq j \in [N]$. Thus we skip the proof and we show the weak faithfulness in the following theorem:

\begin{thmbox}
\begin{theorem} \label{thm:local_cost_FS_weak_faithful}
The local cost function, \eqref{eq:local_cost_full_supp}, is $\epsilon$-weakly faithful with respect to $\dba$
\begin{equation}  \label{eq:cost_to_epsilon_local}
    \Cbs_{\Lbs}(\paramtheta) - \Cbs_{\Lbs}^{\opt} \leqslant \epsilon
\end{equation}
Then the following holds:
\begin{equation}     \label{eq:fubini_study_bound_local}
    \dba(\rho^{\psi,j}_{\paramtheta}, \rho^{\psi, j}_{\opt}) \leqslant \frac{\mathcal{N}\epsilon}{\sin(F_{\opt})} =: f_2(\epsilon), \hspace{3mm} \forall \ket{\psi} \in S, \forall j \in [N]
\end{equation}
where $\Cbs_{\Lbs}^{\opt} := 1 - F_{\mathrm{opt}}$
\end{theorem}
\end{thmbox}

\begin{proof}
We rewrite the \eqref{eq:tracecloseness_local_cost_appendix} in terms of the Bures angle,
\begin{equation}\label{eq:fidelityFS_local}
\begin{split}
    & F(\rho_{\opt}^{\psi, j}, \ket{\psi}) - F(\rho^{\psi,j}_{\paramtheta}, \ket{\psi}) \leqslant \mathcal{N}\epsilon \\\\
    & \implies  \cos^2(\dba(\rho_{\opt}^{\psi, j}, \ket{\psi})) -  \cos^2(\dba(\rho^{\psi,j}_{\paramtheta}, \ket{\psi})) \leqslant \mathcal{N}\epsilon 
\end{split}    
\end{equation}
Following the derivation in the squared cost function section, we obtain the Bures angle/Fubini-Study closeness as,
\begin{equation}
    \dba(\rho^{\psi,j}_{\paramtheta}, \rho_{\opt}^{\psi, j}) \leqslant \frac{\mathcal{N}\epsilon}{\sin(F_{\opt})}, \qquad \forall \ket{\psi} \in S, \forall j \in [N]
    \label{eq:FSbound-standard-local}
\end{equation}
This concludes the proof.
\end{proof}

\noindent\textbf{Global Cost Function}\\\\
\noindent Finally, we show in the next theorems that the global cost function exhibits similar notions of faithfulness:

\begin{thmbox}
\begin{theorem} \label{thm:global_cost_strong_faithful}[Strong faithfulness of the global cost function]
    The global cost function is globally strongly faithful, meaning the following implication holds for all the states $\ket{\psi} \in S$:
    \begin{equation}\label{eq:strongly_faithful_cost_defn}
        \Cbs_{\Gbs}(\paramtheta) = \Cbs^{\mathrm{opt}}_{\Gbs} \implies \rho_{\paramtheta}^\psi = \rho_{\opt}^{\psi} \qquad \forall \ket{\psi} \in S
    \end{equation}
\end{theorem}
\end{thmbox}
\begin{proof}
The global cost function $\Cbs_{\Gbs}(\paramtheta)$ achieves the minimum value $\Cbs^{\opt}_{\Gbs}$ at a unique point corresponding to $\mathbb{E}[F_{\Gbs}(\paramtheta)] = F_{\Gbs}^{\opt}$, where $F_{\Gbs}^{\opt}$ corresponds to the fidelity term for $\Cbs^{\mathrm{opt}}_{\Gbs}$. This corresponds to the unique global clone state $\rho^{\psi}_{\opt}$. Thus the cost function, achieves a unique minimum under precisely the unique condition \emph{i.e.} the output global state is equal to the optimal clone state for all inputs in the distribution.\end{proof}
Now, we provide a statement of weak faithfulness that is much more relevant in the practical implementation of the cloning scheme using global optimization.

\begin{lembox}
\begin{lemma} \label{lemma:global-state-closeness}
Suppose the cost function is $\epsilon$-close to the optimal cost in symmetric cloning
\begin{equation}
    C_{\Gbs}(\paramtheta) - C^{\opt}_{\Gbs} \leq \epsilon
    \label{eq:costtoepsilon_global_trace}
\end{equation}

where $\Cbs^{\opt}_{\Gbs} := 1 - F_{\Gbs}^{\opt}$. Then,

\begin{equation}\label{eq:tracecloseness}
        \Tr\left[(\rho^{\psi}_{\opt} - \rho^{\psi}_{\paramtheta})\ket{\psi}^{\otimes 2}\bra{\psi}^{\otimes 2}\right] \leqslant \mathcal{N}\epsilon, \qquad \forall \ket{\psi} \in S
\end{equation}
\end{lemma}
\end{lembox}

\begin{proof}
The proof follows identically to \lemref{lemma:trace_bound_local_cost} but with the exception that $C_{\Gbs}(\paramtheta) - C^{\opt}_{\Gbs} = \mathbb{E}[ F_{\Gbs}^{\opt} - F_{\Gbs}(\paramtheta)]$.
\end{proof}

Finally, we have the following theorem regarding the weak faithfulness of the global cost function:

\begin{thmbox}
\begin{theorem}\label{thm:trace_FS_bound_global}[Weak faithfulness of the global cost function]
Suppose the cost function is $\epsilon$-close to the optimal cost in symmetric cloning
\begin{equation}\label{eq:costtoepsilon_global_FS}
    \Cbs_{\Gbs}(\paramtheta) - \Cbs^{\opt}_{\Gbs} \leq \epsilon
\end{equation}
where $\Cbs^{\opt}_{\Gbs} := 1 - F_{\Gbs}^{\opt}$. Then,
\begin{equation}     \label{eqn:global_fubini_study_bound_squared_appendix}
    \dba(\rho^{\psi}_{\paramtheta}, \rho^{\psi}_{\opt}) \leqslant \frac{\mathcal{N}\epsilon}{\sin(F_{\Gbs}^{\opt})} =: f_3(\epsilon),   \qquad \forall \ket{\psi} \in S
\end{equation}
\end{theorem}
\end{thmbox}

\begin{proof}
The proof follows along the same lines as the proof of closeness of the Bures angl/Fubini-Study distance for local cost function provided in \thmref{thm:local_cost_FS_weak_faithful}.
\end{proof}
%

\subsubsection{Global versus Local Faithfulness} \label{sec:global_v_local_faithfulness}
This section explores the relationship between local and global cost function optimization for different cloners (universal, phase-covariant, etc.). In particular, we address the question of whether optimizing a cloner with a local or a global cost function also achieves an optimal solution relative to the other cost (operational meaning). If the answer is affirmative, we can use whichever cost exhibits the most desirable qualities and be confident they will achieve the same results. If not, we must be more careful as the choice may not lead to the optimal behaviour we desire and so will be application dependent.

We note that this relationship only manifests in \emph{symmetric} cloning since there is no possibility to enforce asymmetry in the global cost function. The tradeoff between local and global faithfulness turns out to be subtle when dealing with cloning problems and is in contrast to similar studies in analogous variational algorithm literature \cite{kempe_complexity_2006,cerezo_cost_2021}. To begin, we have the following theorem:

\begin{thmbox}
\begin{theorem}\label{thm:relationship-local-global}
For the general case of $M \rightarrow N$ cloning, the global cost function $\Cbs_{\Gbs}(\paramtheta)$ and the local cost function $\Cbs_{\Lbs}(\paramtheta)$ satisfy the inequality,
\begin{equation}
 \Cbs_{\Lbs}(\paramtheta) \leqslant \Cbs_{\Gbs}(\paramtheta) \leqslant N\cdot\Cbs_{\Lbs}(\paramtheta)  
\end{equation}
\end{theorem}
\end{thmbox}

\begin{proof}
We first prove the first part of the inequality,
\begin{equation}\label{eq:global-local}
\begin{split}
    \Cbs_{\Gbs}(\paramtheta) - \Cbs_{\Lbs}(\paramtheta) &= \frac{1}{\mathcal{N}}\int_{S}\textrm{Tr}((\mathsf{O}_{\Gbs}^{\psi} - \mathsf{O}_{\Lbs}^{\psi})\rho_{\paramtheta}^{\psi})d\psi \\
    &= \frac{1}{\mathcal{N}N}\int_{S}\textrm{Tr}\left(\left(
    \sum\limits_{j=1}^N\left(\ket{\psi}\bra{\psi}_j \otimes \mathds{1}_{\Bar{j}} - \ket{\psi}\bra{\psi}_1 \otimes \cdots \ket{\psi}\bra{\psi}_N \right)\right)\rho^{\psi}_{\paramtheta}\right) \geq 0\\
    &\qquad\implies \Cbs_{\Gbs}(\paramtheta) \geq \Cbs_{\Lbs}(\paramtheta)
\end{split}   
\end{equation}
where $\mathsf{O}_{\Lbs}^{\psi}$ is defined in \eqref{eq:local_cost_full_supp}, and the inequality in the second line holds due to the following
\begin{equation}
 \sum\limits_{j=1}^N\left(\ket{\psi}\bra{\psi}_j \otimes \mathds{1}_{\Bar{j}} - \ket{\psi}\bra{\psi}_1 \otimes \cdots \ket{\psi}\bra{\psi}_N \right) =  \sum\limits_{j=1}^N\ket{\psi}\bra{\psi}_j \otimes (\mathds{1}_{\Bar{j}} - \ket{\psi}\bra{\psi}_{\Bar{j}}) \geq 0, \quad \forall \ket{\psi} \in S   
\end{equation}

For the second part of the inequality, we consider the operator $N \mathsf{O}_{\Lbs}^{\psi} - \mathsf{O}_{\Gbs}^{\psi}$,
\begin{equation}
     \begin{split}
         N \mathsf{O}_{\Lbs}^{\psi} - \mathsf{O}_{\Gbs}^{\psi} &= (N - 1)\mathds{1} - \sum_{j=1}^{N}\left(\ket{\psi}\bra{\psi}_j \otimes \mathds{1}_{\Bar{j}}\right) + \ket{\psi}\bra{\psi}_1 \otimes \cdots \ket{\psi}\bra{\psi}_N \\
         &= \sum_{j = 1}^{N-1}\left(\mathds{1}_j \otimes \mathds{1}_{\Bar{j}} - \ket{\psi}\bra{\psi}_j \otimes \mathds{1}_{\Bar{j}}\right) - \ket{\psi}\bra{\psi}_{N} \otimes \mathds{1}_{\Bar{N}} + \ket{\psi}\bra{\psi}_1 \otimes \cdots \ket{\psi}\bra{\psi}_N \\
         &= \sum_{j = 1}^{N-1}\left((\mathds{1} - \ket{\psi}\bra{\psi})_{j}) \otimes \mathds{1}_{\Bar{j}}\right) - \bigotimes_{j=1}^{N-1}(\mathds{1} - \ket{\psi}\bra{\psi})_j) \otimes \ket{\psi}\bra{\psi}_N \\
         &= (\mathds{1} - \ket{\psi}\bra{\psi})_{1}\otimes \left(\mathds{1}_{\Bar{1}} - \bigotimes_{j=2}^{N-1}(\mathds{1} -  \ket{\psi}\bra{\psi}_{j}) \otimes \ket{\psi}\bra{\psi}_N\right) \\
         & + \sum_{j = 2}^{N-1}\left((\mathds{1} - \ket{\psi}\bra{\psi})_{j}) \otimes \mathds{1}_{\Bar{j}}\right) \\
         &\geq 0
     \end{split}
\end{equation}   
where the second last line is positive because each individual operator is positive for all $\ket{\psi} \in S$.
\end{proof}
A similar inequality was proven in the work of  \cite{bravo-prieto_variational_2020}.
But interestingly, the inequality proven in \thmref{thm:relationship-local-global} (unlike in \cite{bravo-prieto_variational_2020}) does not allow us make statements about the similarity of individual clones from the closeness of the global cost function and vice versa. This can be seen as follows:
\begin{equation}
\begin{split}
    \Cbs_{\Gbs}(\paramtheta) - \Cbs_{\Gbs}^{\opt} \leqslant \epsilon &\implies  \Cbs_{\Lbs}(\paramtheta) - \Cbs_{\Lbs}^{\opt} \leqslant \epsilon - (\Cbs_{\Gbs}(\paramtheta) - \Cbs_{\Lbs}(\paramtheta)) + (\Cbs_{\Lbs}^{\opt} - \Cbs_{\Gbs}^{\opt})\\ 
    &\implies \Cbs_{\Lbs}(\paramtheta) - \Cbs_{\Lbs}^{\opt} \leqslant \epsilon + (\Cbs_{\Lbs}^{\opt} - \Cbs_{\Gbs}^{\opt}) \\ 
    & \qquad \nRightarrow \Cbs_{\Lbs}(\paramtheta) - \Cbs_{\Lbs}^{\opt} \leqslant \epsilon
\end{split}    
\end{equation}
Here we have used the result of \thmref{thm:relationship-local-global} that $\Cbs_{\Gbs}(\paramtheta) \geq \Cbs_{\Lbs}(\paramtheta)$ and we note that $\Cbs_{\Lbs}^{\opt} - \Cbs_{\Gbs}^{\opt} \neq 0$ for all the $M \rightarrow N$ cloning. In particular, for $1 \rightarrow 2$ cloning, $\Cbs_{\Lbs}^{\opt} = 5/6$, while $\Cbs_{\Gbs}^{\opt} = 2/3$. This is due to the non-vanishing property of these cost functions, and highlights the subtlety of the case in hand.

While we are unable to leverage generic inequalities for our purpose, based on the cost functions, we can make statements in \emph{specific} cases. In other words, by restricting the cloning problem to a specific input set of states, we can guarantee that optimizing \emph{globally} will be sufficient to also optimize \emph{local} figures of merit. 
 
In particular, in the following, we establish these strong and weak faithfulness guarantees for the \emph{special cases} of universal and phase-covariant cloning by analyzing problem-specific features.
\begin{thmbox}
\begin{theorem}\label{thm:local_clones_from_global_cost_universal_universal}
The \emph{global} cost function is \emph{locally} strongly faithful for a universal symmetric cloner, i.e,:
\begin{equation}\label{eqn:strongly_faithful_global_cost_local_clone_strong_universal_appendix}
    \Cbs_{\Gbs}(\paramtheta) = \Cbs^{\mathrm{opt}}_{\Gbs} \iff \rho_{\paramtheta}^{\psi, j} = \rho_{\opt}^{\psi, j} \qquad \forall \ket{\psi} \in \mathcal{H}, \forall j \in \{1,\dots,N\}
\end{equation}
\end{theorem}
\end{thmbox}
\begin{proof}
In the symmetric universal case, $\Cbs^{\opt}_{\Lbs}$ has a unique minimum when, each local fidelity saturates:
\begin{align} \label{eqn:univ-localfid}
    F_{\Lbs}^{\opt} = \frac{M(N+2) + N - M}{N(M+2)}
\end{align}
achieved by local reduced states $\{\rho_{\opt}^{\psi, j}\}_{j=1}^N$. Now, it has been shown that the optimal global fidelity $F_{\Gbs}$ that can be reached~\cite{buzek_quantum_1997, scarani_quantum_2005} is,
\begin{equation}\label{eq:univ-globalfid}
    F_{\Gbs}^{\opt} = \frac{N!(M+1)!}{M!(N+1)!}
\end{equation}
which also is the corresponding unique minimum value for $\Cbs^{\opt}_{\Gbs}$, achieved by some global state $\rho_\opt^\psi$.

Finally, it was proven in \cite{werner_optimal_1998, keyl_optimal_1999} that the cloner which achieves one of these bounds is unique and also saturates the other bound, and therefore must also achieve the unique minimum of both global and local cost functions, $\Cbs^{\opt}_{\Gbs}$ and $\Cbs^{\opt}_{\Lbs}$. Hence, the local states which optimize $\Cbs^{\opt}_{\Lbs}$  must be the reduced density matrices of the global state which optimizes $\Cbs^{\opt}_{\Gbs}$ and so:
\begin{equation}    \label{eq:local-global-state-closeness}
    \rho_{\opt}^{\psi, j} := \Tr_{\Bar{j}}(\rho_{\opt}^{\psi}), \ \ \forall j
\end{equation}
Thus for a universal cloner, the cost function with respect to both local and global fidelities will converge to the same minimum.
\end{proof}

Now, before proving an analogous statement in the case of phase-covariant cloning, we first need the following lemma (we return to the notation of $B, E$ and $E^*$ for clarity):

\begin{lembox}
\begin{lemma}\label{lemma:phase-covariant-clone-uniqueness}
For any $1\rightarrow 2$ phase-covariant cloning machine which takes states $\ket{0}_{B}\otimes\ket{\psi}_{E}$ and an ancillary qubit $\ket{A}_{E^*}$ as input, where $\ket{\psi} := \frac{1}{\sqrt{2}}(\ket{0} + e^{i\theta}\ket{1})$, and outputs a 3-qubit state $\ket{\Psi_{BEE^*}}$ in the following form:
\begin{equation}\label{eq:phasecov-tripartite-output}
\begin{split}
    \ket{\Psi_{BEE^*}} = & \frac{1}{2} [\ket{0,0} + e^{i\phi}(\sin\eta\ket{0,1} + \cos\eta\ket{1,0}))\ket{0}_{E^*}
    \\ & + e^{i\phi}\ket{1,1} + (\cos\eta\ket{0,1} + \sin\eta\ket{1,0})\ket{1}_{E^*}]
\end{split}
\end{equation}
 the global and local fidelities are simultaneously maximized at $\eta = \frac{\pi}{4}$ where $0\leq \eta \leq \frac{\pi}{2}$ is the `shrinking factor'. 
\end{lemma}
\end{lembox}

\begin{proof}
To prove this, we follow the formalism that was adopted by Cerf \textit{et. al.}\cite{cerf_cloning_2002}. This uses the fact that a symmetric phase-covariant cloner induces a mapping of the following form~\cite{scarani_quantum_2005}:
\begin{align} \label{eq:phasecov-map}
    \begin{split}
        & \ket{0}\ket{0}\ket{0} \rightarrow \ket{0}\ket{0}\ket{0} \\
        & \ket{1}\ket{0}\ket{0} \rightarrow (\sin\eta\ket{0}\ket{1} + \cos\eta\ket{1}\ket{0})\ket{0} \\
        & \ket{0}\ket{1}\ket{1} \rightarrow (\cos\eta\ket{0}\ket{1} + \sin\eta\ket{1}\ket{0})\ket{1} \\
        & \ket{1}\ket{1}\ket{1} \rightarrow \ket{1}\ket{1}\ket{1}
    \end{split}
\end{align}
Next, we calculate the global state by tracing out the ancillary state to get $\rho_{\Gbs}^{\opt}$:
\begin{align} \label{eq:phasecov-global-density}
    \rho_{\Gbs}^{\opt} = \textrm{Tr}_{E^*}(\ket{\Psi_{BEE^*}}\bra{\Psi_{BEE^*}}) = \ket{\Phi_1}\bra{\Phi_1} + \ket{\Phi_2}\bra{\Phi_2}
\end{align}
where,
\begin{align}
    & \ket{\Phi_1} := \frac{1}{2}\left[\ket{0,0} + e^{i\phi}(\sin\eta\ket{0,1} + \cos\eta\ket{1,0})\right], \\
    & \ket{\Phi_2} := \frac{1}{2}\left[e^{i\phi}\ket{1,1} + (\cos\eta\ket{0,1} + \sin\eta\ket{1,0})\right]
\end{align}
\noindent Hence the global fidelity can be computed as:
\begin{align} \label{eq:phasecov_global_fid}
    F_{\Gbs}^{\opt} = \Tr(\ket{\psi}\bra{\psi}^{\otimes 2}\rho_{\Gbs}^{\opt}) = |\braket{\psi^{\otimes 2}}{\Phi_1}|^2 +|\braket{\psi^{\otimes 2}}{\Phi_2}|^2 = \frac{1}{8}(1 + \sin\eta + \cos\eta)^2
\end{align}
Now, optimising $F_{\Gbs}^{\opt}$ with respect to $\eta$, we see that $F_{\Gbs}^{\opt}$ has only one extremum value between $[0,\frac{\pi}{2}]$ specifically at $\eta=\frac{\pi}{4}$. We can also see that the local fidelity is also achieved for the same $\eta$ and is equal to:
\begin{align} \label{eq:phasecov_local_fid}
    F_{\Lbs}^{\opt} = \frac{1}{2}\left(1 + \frac{\sqrt{2}}{2}\right)
\end{align}
which is the upper bound for local fidelity of the phase-covariant cloner.
\end{proof}

With \lemref{lemma:phase-covariant-clone-uniqueness} established, we can next prove:

\begin{thmbox}
\begin{theorem}\label{thm:local_clones_from_global_cost_phase-covariant}
The global cost function is locally strongly faithful for phase-covariant symmetric cloner, \emph{i.e.}:
\begin{equation}\label{eq:strongly_faithful_global_cost_local_clone_strong_phase_covariant}
    \Cbs_{\Gbs}(\paramtheta) = \Cbs^{\mathrm{opt}}_{\Gbs} \iff \rho_{\paramtheta}^{\psi, j} = \rho_{\opt}^{\psi, j} \qquad \forall \ket{\psi} \in S, \forall j \in \{B, E\}
\end{equation}
where $S$ is the distribution corresponding to phase-covariant cloning.  
\end{theorem}
\end{thmbox}

\begin{proof}
We have shown in \lemref{lemma:phase-covariant-clone-uniqueness} that the global and local fidelities of a phase-covariant cloner are both achieved with a cloning transformation of the form in \eqref{eq:phasecov-map}. Applying this transformation unitary to $\ket{\psi}\ket{\Phi^+}_{BE}$ (where $\ket{\Phi^+}_{BE}$ is a Bell state) leads to Cerf's formalism for cloning. Furthermore, we can observe that due to the symmetry of the problem, this transformation is unique (up to global phases) and so any optimal cloner must achieve it.

Furthermore, one can check that the ideal circuit in \figref{fig:learned_vs_ideal_circ_on_hw}(b) does indeed produce an output in the form of \eqref{eq:phasecov-tripartite-output} once the preparation angles have been set for phase-covariant cloning. By a similar argument to the above, we can see that a variational cloning machine which achieves an optimal cost function value, \emph{i.e.} $\Cbs_{\Gbs}(\paramtheta) = \Cbs^{\mathrm{opt}}_{\Gbs}$ much also saturate the optimal cloning fidelities. Furthermore, by the uniqueness of the above transformation (\eqref{eq:phasecov-map}) we also have that the local states of $\VQC$ are the same as the optimal transformation, which completes the proof.
\end{proof}

\subsection{Summary of other specifications}\label{sec:varqlone-algorithm-other-specs}
In this section for completeness , we give an overview of some of the other specifications of the algorithm. We will not go in-depth to prove them as they are neither the author's contribution nor directly relevant to the main topic of this thesis. Nevertheless, it will contribute to understanding the algorithm for potential future applications.

First, we start with the choice of Ansatz (see Section~\ref{sec:prelim-qml-ansatz}). A key element in variational algorithms is the choice of Ansatz that is used in parameterized quantum circuits. The primary Ansatz we choose is one with a \emph{variable} structure. This allows us to learn cloning circuits in an end-to-end manner. The idea is to optimize over both the continuous parameters of a quantum circuit, but also over the gates within the circuit itself, which come from a discrete set. The goal is to solve the following optimization problem~\cite{li_quantum_2020}:
\begin{equation} \label{eq:variable_structure_ansatz_optimisation_problem}
    (\paramtheta^{*}, \boldsymbol{g}^{*}) = \argmin_{\paramtheta, \boldsymbol{g} \in \mathcal{G}} \Cbs(\paramtheta, \boldsymbol{g})
\end{equation}
where $\mathcal{G}$ denotes the gate set. Such variable-structure $\Ansatze$ approaches can be broadly dubbed as the \emph{Quantum Architecture Search} (QAS) \cite{zhang_differentiable_2021} similar to \emph{Neural Architecture Search} (NAS) in classical ML \cite{yao_taking_2019, liu_darts_2018}. Approaches to QAS have appeared in many forms~\cite{cincio_learning_2018, grimsley_adaptive_2019,ostaszewski_structure_2021,chivilikhin_mog-vqe_2020, li_quantum_2020,pirhooshyaran_quantum_2021}. In this work, $\mathcal{G}$ is a gateset \emph{pool}, from which a particular sequence $\boldsymbol{g}$ is chosen. As a summary, to solve this problem, we iterate over $\boldsymbol{g}$, swap out gates, and re-optimize the parameters $\paramtheta$, until a minimum of the cost, $\Cbs(\paramtheta^{*}, \boldsymbol{g}^{*})$ is found. This is a combination of a discrete and continuous optimization problem, where the discrete parameters are the indices of the gates in $\boldsymbol{g}$ (\emph{i.e.}\@, the circuit structure), and the continuous parameters are represented by $\paramtheta$. Each time the circuit structure is changed (a subset of gates are altered), the continuous parameters are re-optimized, as in \cite{cincio_learning_2018}.  Variations of this approach have been proposed in \cite{du_quantum_2020,li_quantum_2020} which could be easily incorporated, and we leave such investigation to future work. For the results shown for $1\rightarrow 2$ cloning phase-covariant states, we use the following three qubit gate pool:

\begin{multline} \label{eq:phase_covariant_cloning_gateset}
    \mathcal{G}_{\textrm{PC}} := \left\{\right. \mathsf{R}^2_{z}(\theta), \mathsf{R}^3_{z}(\theta), \mathsf{R}^4_{z}(\theta),
    \mathsf{R}^2_{x}(\theta), \mathsf{R}^3_{x}(\theta), \mathsf{R}^4_{x}(\theta), \\
    \mathsf{R}^2_{y}(\theta), \mathsf{R}^3_{y}(\theta), \mathsf{R}^4_{y}(\theta), 
    \CZ_{2, 3}, \CZ_{3, 4}, \CZ_{2, 4}\left.\right\}
\end{multline}

In order to attack protocol $\mathcal{P}_1$ using $1\rightarrow 2$ state dependent cloning, we use the following pool:

\begin{equation} \label{eq:mayers_1to2_state_dependent_cloning_gateset}
    \mathcal{G}_{\mathcal{P}_1^{1\rightarrow 2}} := \left\{
    \mathsf{R}^i_{j}(\theta), \CZ_{2, 3}, \CZ_{3, 4} \right\} \quad \forall i \in \{2, 3, 4\}, \forall j \in \{x, y, z\}
\end{equation}

where $\mathsf{R}_j^i$ indicates the $j^{th}$ Pauli rotation with angle $\theta$ acting on the $i^{th}$ qubit and $\CZ$ is the controlled-$\mathsf{Z}$ gate. In both cases, we use the qubits indexed $2, 3$ and $4$ in an \computerfont{Aspen-8} sublattice. Note that in the latter case, we allow only a linear, nearest-neighbour (NN) connectivity, which removes the need for inserting $\SWAP$ gates by the quantum compiler. For more detailed specifications about the algorithm via supplementary numerical results, we refer to \cite{coyle_progress_2022}.

Next, we talk about the sample complexity of $\VQC$. We discussed that $\VQC$ requires classical minimisation of one of the cost functions $\Cbs(\paramtheta) :=\{\Cbs_{\textrm{sq}}(\paramtheta), \Cbs_{\Lbs}(\paramtheta), \Cbs_{\Gbs}(\paramtheta) \}$ to achieve the optimal cost value. To do so, we must be able to efficiently evaluate the cost function of choice. In our case, this can be achieved via a method that allows the computation of the fidelity between quantum states. This estimation can be done either via $\SWAP$ test, which is a powerful toolkit that we have used many times so far in this thesis; or via computing an estimator for the true cost $\Cbs(\paramtheta) = \mathop{\mathbb{E}}_{\substack{\ket{\psi} \in S}}[\Cbs^{\psi}(\paramtheta)]$ using $K$ different states sampled from $S$. Estimating the overlap in the mentioned way is sufficient for our purposes since this coincides with the fidelity when at least one of the states is a pure state:
\begin{equation}\label{eq:fidelity_equals_state_overlap_one_pure_state}
    F(\ket{\psi}\bra{\psi}, \rho) = \bra{\psi}\rho\ket{\psi} = \Tr(\ket{\psi}\bra{\psi} \rho)
\end{equation}
Since VQAs are heuristic algorithms, there are no guarantees on the number of training iterations over $\paramtheta$ to converge to $\Cbs^{\opt}$. However, one can at least provide guarantees on the number of samples required to estimate the cost, for a particular instance of the parameters. Since this is a necessary subroutine in the algorithm, it must be efficient. It can be shown that the number of samples $L \times K$, where $K$ is the number of distinct states $\ket{\psi}$ sampled uniformly at random from the distribution $S$, and $L$ is the number of copies of each input state, required to estimate the cost function $\Cbs(\paramtheta)$ up to $\epsilon'$-additive error with a success probability $\delta$ is,
\begin{equation}\label{eqn:number-of-samples}
    L \times K =  \mathcal{O}\left(\frac{1}{\epsilon'^2}\log \frac{2}{\delta}\right)
\end{equation}
We refer to \cite{coyle_progress_2022} for the proof.

Finally, we can examine $\VQC$ for the existence of barren plateaus. We prove in \cite{coyle_progress_2022}, that the local cost function that we have presented does not exhibit barren plateaus for a sufficiently shallow alternating layered Ansatz, \emph{i.e.} $U(\paramtheta)$ contains blocks $W$, acting on alternating pairs of qubits~\cite{cerezo_cost_2021}.

\section{Practical cryptanalysis based on the numerical results of VarQlone}\label{sec:varqlone-numerics}
At last, we present the numerical and experimental results of VarQlone for the cryptographic problems that we have introduced in Section~\ref{sec:varqlone-cloning-crypt}. Let us start with the phase covariant case study.

\subsection{Variational phase-covariant cloning}\label{sec:varqlone-numerical-results-phase-cov}
\begin{figure}[ht]
    \centering
\includegraphics[width=1.1\columnwidth]{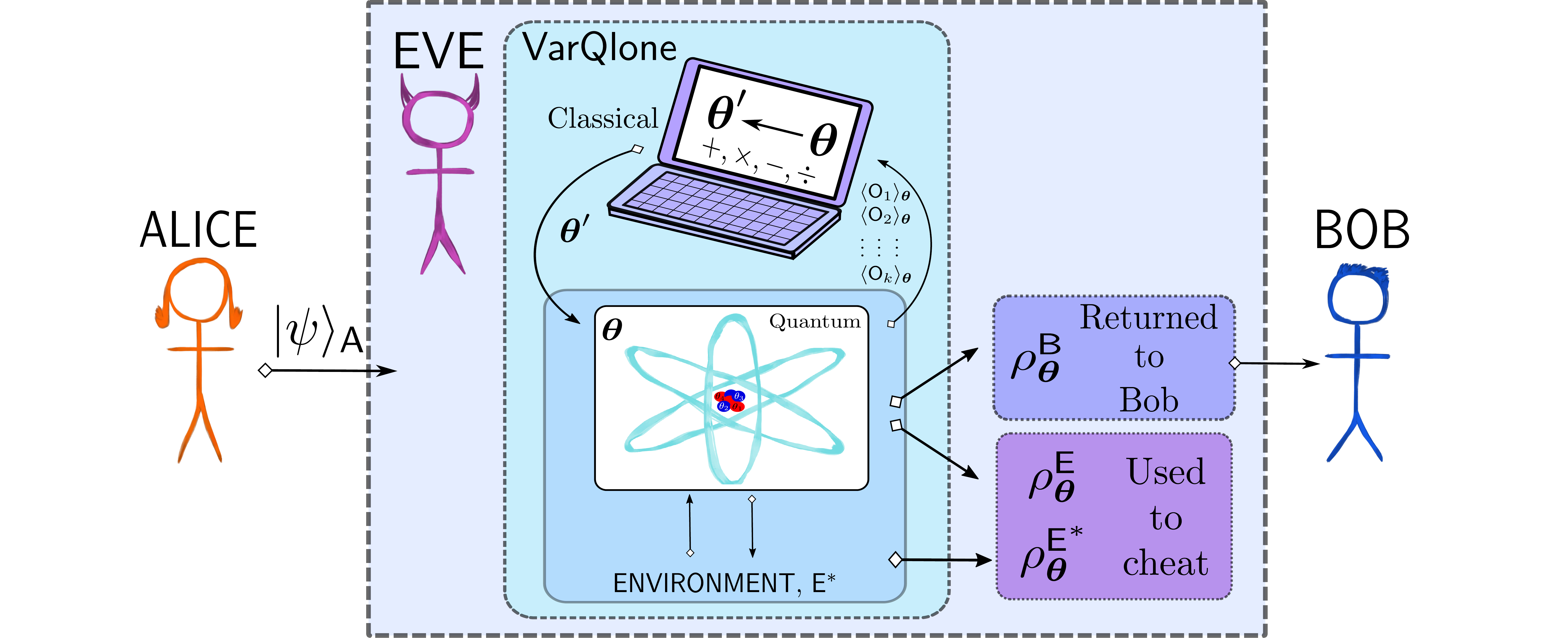}
    \caption[Cartoon overview of $\VQC$ in a cryptographic attack]{
    Cartoon overview of $\VQC$ in a cryptographic attack. Here an adversary Eve, $E$, implements a $1\rightarrow 2$ cloning attack on states used in a quantum protocol (for example QKD) between Alice and Bob. Eve intercepts the states sent by Alice $\ket{\psi}_A$ and may interact with an ancillary `environment', $E^*$. This interaction is trained (an optimal parameter setting $\paramtheta$ is found) by Eve to optimally produce clones, $\rho^B_{\paramtheta}, \rho^E_{\paramtheta}$. In order to attack the protocol, Eve will return $\rho^{B}_{\paramtheta}$ to Bob and use the rest (her clone, $\rho^{E}_{\paramtheta}$ plus the remaining environment state, $\rho^{E^*}_{\paramtheta}$) to cheat. The training procedure consists of using a classical computer to optimize the quantum parameters, via a cost function. The cost is a function of $k$ observables, $\mathsf{O}_k$, measured from the output states, which are designed to extract fidelities of the states to compare against the ideal state.
    }
    \label{fig:variational_algorithm_vqc}
\end{figure}

We start with the attack we have presented in Section~\ref{sec:varqlone-phasecov-crypt} to attack the BB84 protocol, while here we replace the theoretical cloner with $\VQC$. \figref{fig:variational_algorithm_vqc} demonstrates at a high level, how $\VQC$ is inserted into an attack (on QKD or similar protocols).

Here $\VQC$ has $K$ layers in the ansatz, in each layer there is a fixed structure. For simplicity, we choose each layer to have parameterised single-qubit rotations, $\RY(\theta)$, and nearest neighbour $\CZ$ gates. Our primary target is $1 \leftarrow 2$ cloning, so we use 3 qubits and therefore we have 2 $\CZ$ gates per layer. Not surprisingly, in the experiment, we observe convergence to the minimum as the number of layers increases, saturating at $K = 3$.

\begin{figure}[ht]
\centering
    \includegraphics[width=1\textwidth]{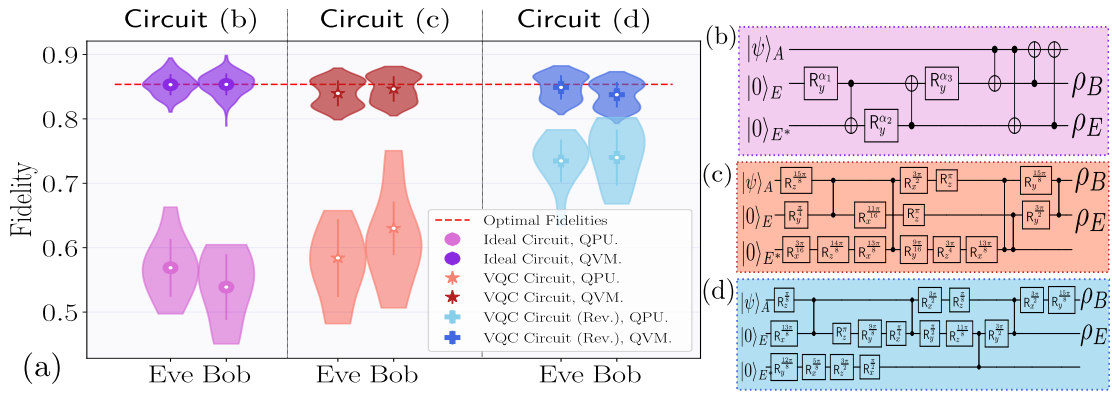}
    \caption[Variational Quantum Cloning implemented on phase-covariant states using three qubits of the Rigetti \computerfont{Aspen-8} chip (QPU), plus simulated results (QVM)]{Variational Quantum Cloning implemented on phase-covariant states using three qubits of the Rigetti \computerfont{Aspen-8} chip (QPU), plus simulated results (QVM). Violin plots in (a) show the cloning fidelities, for Bob and Eve, found using each of the circuits shown in (b)--(d) respectively. Shown in red is the maximal possible fidelity for this problem. (b) is the ideal circuit with clones appearing in registers $2$ and $3$. (c) shows the structure-learned circuit for the same scenario, using one less entangling gate. (d) demonstrates the effect of allowing clones to appear in registers $1$ and $2$. In the latter case, only four (nearest-neighbour) entangling gates are used, demonstrating a significant boost in performance on the  QPU.}
    \label{fig:learned_vs_ideal_circ_on_hw}
\end{figure}

Now we show a proof of principle implementation of our methods for phase-covariant cloner. The results of this can be seen in \figref{fig:learned_vs_ideal_circ_on_hw}.

Let us begin by describing some problem parameters. Firstly, we allow 3 qubits (2 output clones plus 1 ancilla) in the circuit. Next, we give the $\VQC$ the fully connected (FC) gateset pools introduced in \eqref{eq:phase_covariant_cloning_gateset}. Let us now analyse the two candidate resulting circuits of $\VQC$ in \figref{fig:learned_vs_ideal_circ_on_hw}(c,d) in comparison with the optimal `analytic' circuit \figref{fig:learned_vs_ideal_circ_on_hw}(b) introduced in \cite{buzek_quantum_1997,fan_quantum_2014}.

Firstly, we note that all three circuits approximately saturate the optimal bound for phase-covariant cloning ($F_L = 0.85$) when simulated (\emph{i.e.} without quantum noise). But we notice that the ideal circuit in \figref{fig:learned_vs_ideal_circ_on_hw}(b) suffers degradation in performance when implemented on the QPU since it requires $6$ entangling gates as it is attempting to transfer the information across the circuit. Furthermore, since the \computerfont{Aspen-8} chip does not have any $3$ qubit loops in its topology, it is necessary for the compiler to insert $\SWAP$ gates.

Next, we compare the ideal circuit to two examples learned by $\VQC$. Firstly, we force the qubit clones to appear in registers $2$ and $3$, (demonstrated in \figref{fig:learned_vs_ideal_circ_on_hw}(c)) exactly as in \figref{fig:learned_vs_ideal_circ_on_hw}(b). Secondly, we allow the clones to appear instead in registers $1$ and $2$ (demonstrated in \figref{fig:learned_vs_ideal_circ_on_hw}(d) - The circuit labeled `Rev.' (`Reverse').) The ability to make such a subtle change demonstrates clearly the advantage of our flexible approach. We notice that the restriction imposed in \figref{fig:learned_vs_ideal_circ_on_hw}(c) results in only slightly improved performance over the ideal. However, by allowing the clones to appear in registers $1$ and $2$, $\VQC$ can find much more conservative circuits, having fewer entangling gates, and are directly implementable on a linear topology. This gives a significant improvement in the cloning fidelities, of about $15\%$ when the circuit is run on the QPU, as observed in \figref{fig:learned_vs_ideal_circ_on_hw}(a). For all results shown using a variable structure $\Ansatz$, we use the \computerfont{forest-benchmarking} library~\cite{combes_forest_2019} to reconstruct the output density matrix in order to mitigate the effect of quantum noise.

Finally, we can calculate the success probability of a real implementable attack on BB84 using today's quantum hardware as a result of $\VQC$. We specifically analyse the
performance of one of these $\VQC$-learned circuits (Circuit(c)) in such an attack. The reason for this choice is that while the circuit in \figref{fig:learned_vs_ideal_circ_on_hw}(d) achieves higher fidelities on the Aspen hardware, it does not actually make use of the ancillary qubit (one can observe that the sequence of gates acting on it, is approximately an identity gate). We will do so by computing the corresponding critical error rate, $\Dcrit$, using \eqref{eq:qkd-key-rate} as discussed in Section \ref{sec:varqlone-phasecov-crypt}. for the BB84 protocol run in $X-Y$ Pauli basis. First, we compute the resulting mixed states outputted over all input states of the cloning machine, for each basis state: $\{\ket{+}, \ket{-}, \ket{+ i}, \ket{- i}\}$ so $\rho_E$ is given by:
\begin{equation}\label{eq:eve_state_varqlone_attack}
    \rho_E := \frac{1}{4} (\rho_E^{+} + \rho_E^{-} + \rho_E^{+i} + \rho_E^{-i})
\end{equation}

Similarly, $\rho_E^0 , \rho_E^1$ in \eqref{eq:qkd-holevo-quantity} are the mixed states encoding the random bit $0$ (corresponding to $\{\ket{+}, \ket{+ i}\}$) and bit $1$ (corresponding to $\{\ket{-}, \ket{- i}\}$), so are given by:

\begin{equation}\label{eq:eve_state_varqlone_attack_0_and_1}
    \rho_E^0 := \frac{1}{2} (\rho_E^{+} + \rho_E^{+i}), \quad\quad \rho_E^1 := \frac{1}{2} (\rho_E^{-} + \rho_E^{-i})
\end{equation}

Calculating the minimum Holevo quantity $\chi_{min}$ for the above density matrices outputted by the circuit in \figref{fig:learned_vs_ideal_circ_on_hw}(c) gives the following:

\begin{equation}\label{eq:dcrit_varqlone_attack_calc}
\begin{split}
    & 1 - H(\Dcrit) - \chi_{min} = 0 \\
    & \implies 1 - \chi_{min} + (\Dcrit \log_2(\Dcrit) + (1 - \Dcrit)\log_2(1 - \Dcrit))) = 0 \\
    & \implies \Dcrit = 15.8\%.
\end{split}    
\end{equation}

Recalling the optimal bound for the individual attack, one can see that the $\Dcrit$ obtained by the result of $\VQC$ is very close to that bound. Nevertheless, as pointed out in \cite{scarani_quantum_2005,ferenczi_symmetries_2012}, the same bound can be reached by a collective attack (where Eve defers all the measurements until the end of the reconciliation phase and applies a general strategy to all collected states) so long as the individual quantum operations are still given by the optimal phase-covariant cloner. Thus, the $\VQC$ learned circuits can be used to perform collective attacks and almost saturate the optimal collective bound.

Finally, one may observe that Circuit (d) in \figref{fig:learned_vs_ideal_circ_on_hw} achieves an even higher fidelity on the actual hardware, but it does so without using the ancilla to reduce the circuit
depth. Therefore, it is a more suited and non-trivial circuit for purely performing phase covariant cloning.

\subsection{Variational state-dependent cloning}\label{sec:varqlone-numerical-results-statedep}
In this section, we present the results of $\VQC$ when learning to clone the states used in the two coin-flipping protocols described in Section~\ref{sec:varqlone-2state-coinflip} and Section~\ref{sec:varqlone-4state-coinflip}. Firstly, we focus on the states used in the original protocol, $\mathcal{P}_1$ for $1\rightarrow 2$ cloning, and then move to the 4 state protocol, $\mathcal{P}_2$. In the latter, we also extend from $1\rightarrow 2$ cloning to $1\rightarrow 3$ and $2\rightarrow 4$. These extensions will allow us to probe certain features of $\VQC$, in particular explicit symmetry in the cost functions. In all cases, we use the variable structure Ansatz, and once a suitable candidate has been found, the solution is manually further optimised. The learned circuits that are used to produce the figures  and results in this section, are given in \figref{fig:mayers_1to2_cloning_vqc_circuit} and \figref{fig:aharonov_1to2_1to3_2to4_circuits}.

\subsubsection{Variational cloning attack on 2-state quantum coin-flipping}\label{sec:varqlone-numerical-results-2-state-coinflip}
As a reminder, the two states used in this protocol are:
\begin{align} \label{eqn:mayers_states_explicit_numerical}
    \ket{\phi_0} := \ket{\phi_{0, 0}} = \cos\left(\frac{\pi}{18}\right)\ket{0} + \sin\left(\frac{\pi}{18}\right)\ket{1}\\
    \ket{\phi_1} := \ket{\phi_{0, 1}} = \cos\left(\frac{\pi}{18}\right)\ket{0} - \sin\left(\frac{\pi}{18}\right)\ket{1}
\end{align}

\begin{figure}[ht]
\centering
    \includegraphics[width=1\textwidth]{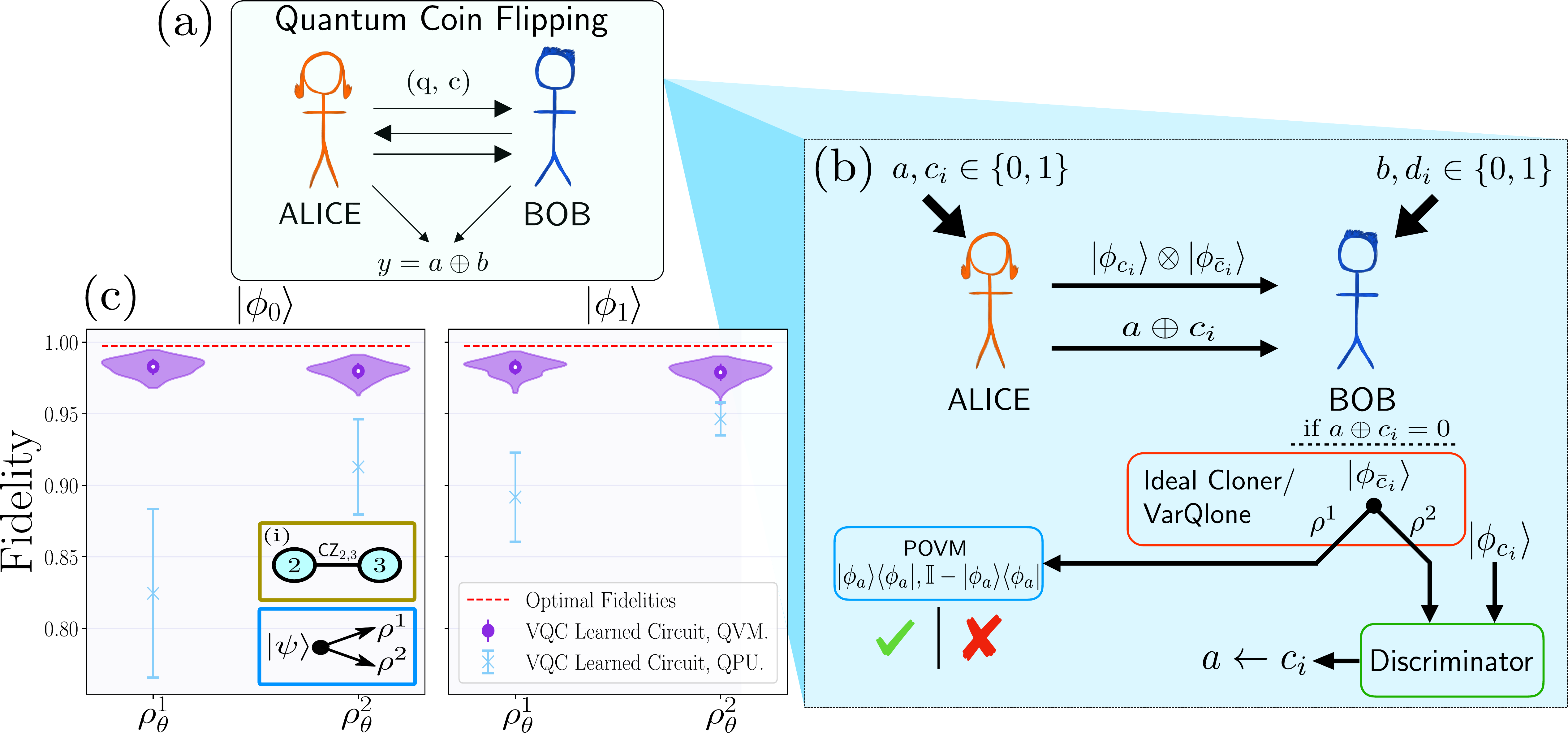}
    \caption[Overview of cloning-based attack on the protocol of Mayers \textit{et. al.}~\cite{mayers_unconditionally_1999}, plus corresponding numerical results for $\VQC$]{Overview of cloning-based attack on the protocol of Mayers \textit{et. al.}~\cite{mayers_unconditionally_1999}, plus corresponding numerical results for $\VQC$. (a) Cartoon of coin flipping protocols, Alice and Bob send quantum (q) and/or classical (c) information to agree on a final `coin flip' bit, $y$. (b) The relevant part of the protocol of Mayers \textit{et. al.}, $\mathcal{P}_1$, plus a cloning based attack on Bob's side. He builds a cloning machine using $\VQC$ to produce two clones of Alice's sent states, one of which he returns, and the other is used to guess Alice's input bit, $a$. (c) Fidelities of each output clone, $\rho^j_{\theta}$ achieved using $\VQC$ when ($1\rightarrow 2$) cloning the family of states used in, $\mathcal{P}_1$.  In the left (right) panel, $\ket{\phi_0}$ ($\ket{\phi_1}$) is. Figure shows both simulated (QVM - purple circles) and on Rigetti hardware (QPU - blue crosses). For the QVM (QPU) results, 256 (5) samples of each state are used to generate statistics. Violin plots show complete distribution of outcomes and error bars show the means and standard deviations. Inset (i) shows the two qubits of the \computerfont{Aspen-8} chip which were used, with the allowed connectivity of a $\CZ$ between them. Note an ancilla was also allowed, but $\VQC$ chose not to use it in this example.}
    \label{fig:mayers_1to2_cloning_fidelities_variational}
\end{figure}

The cloning-based attacks and the obtained fidelities achieved by the $\VQC$ learned circuit can be seen in \figref{fig:mayers_1to2_cloning_fidelities_variational} where we use the gate pool \eqref{eq:mayers_1to2_state_dependent_cloning_gateset} which allows a linear entangling connectivity.

In the above figure, a deviation from the optimal fidelity can be seen, even in the simulated case. We believe this is mostly due to tomographic errors in reconstructing the cloned states. Before analysing the results for our given attack, let us also have a look at the circuits learned by $\VQC$ for the task of state-dependent cloning with fixed-overlap. 

\begin{figure}
    \centering
    \includegraphics[width=0.8\columnwidth]{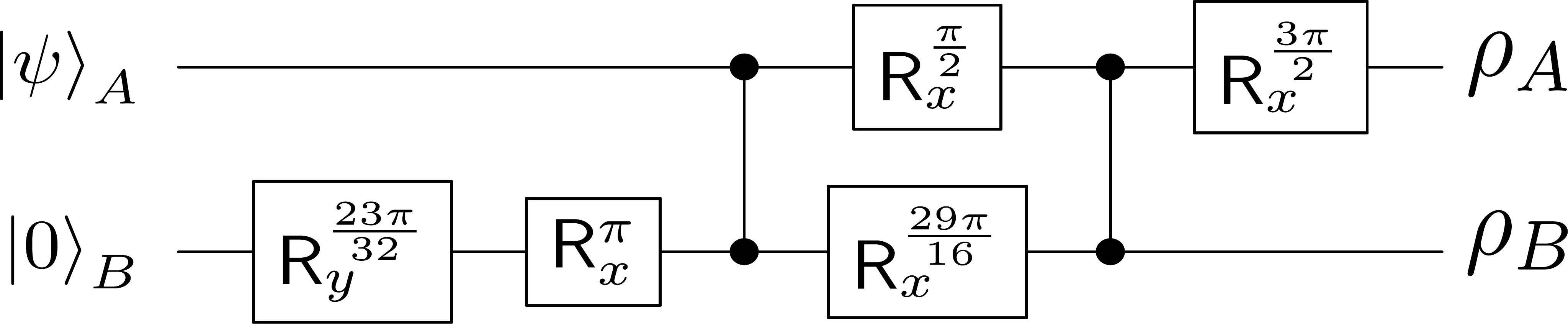}
    \caption[Circuit learned by $\VQC$ in to clone states, $\ket{\phi_0}, \ket{\phi_1}$, with an overlap $s=\cos\left(\pi/9\right)$ in the protocol, $\mathcal{P}_1$]{Circuit learned by $\VQC$ in to clone states, $\ket{\phi_0}, \ket{\phi_1}$, with an overlap $s=\cos\left(\pi/9\right)$ in the protocol $\mathcal{P}_1$. For example, $\rho_A$ is the clone sent back to Alice, while $\rho_B$ is kept by Bob.}
    \label{fig:mayers_1to2_cloning_vqc_circuit}
\end{figure}

\figref{fig:mayers_1to2_cloning_vqc_circuit} shows the circuit used to achieve the fidelities in the attack on $\mathcal{P}^1$. In training, we still allowed an ancilla to aid the cloning, but the example in \figref{fig:mayers_1to2_cloning_vqc_circuit} did not make use of it (in other words, $\VQC$ only applied gates which were equivalent to applying identity on the ancilla), so we remove it to improve hardware performance. This repeats the behaviour seen for the circuits learned in phase-covariant cloning. We mention again, that some of the learned circuits did make use of the ancilla with similar performance. This mimics the behaviour seen in the previous example of phase-covariant cloning. As such, we only use the two qubits shown in the inset (i) of the figure when running on the QPU to improve performance.

Now, we proceed with calculating the success probability of the attack on $\mathcal{P}_1$ given the above experimental results. For illustration, let us return to the example in~\eqref{eq:mayers_bob_pairs_discriminate}, where instead the cloned state is now produced from our $\VQC$ circuit, $\rho^0_c\rightarrow \rho^0_{\VQC}$.

\begin{thmbox}
\begin{theorem} \label{thm:vqc_bias_mayers_protocol}[$\VQC$ Attack Bias on $\mathcal{P}_1$]
Bob can achieve a bias of $\epsilon \approx 0.29$ using a state-dependent $\VQC$ attack on the protocol $\mathcal{P}_1$, with a single copy of Alice's state.
\end{theorem}
\end{thmbox}
\begin{proof}
For the proof, we compute the success probability in the same way as in \thmref{thm:mayers_attack_bias_probability}, as follows:
\begin{equation}\label{eq:mayers_vqc_clone0}
\begin{split}
     P^{\VQC}_{\mathrm{succ}, \mathcal{P}_1} = \frac{1}{2}+\frac{1}{4}\Tr|\rho_1 - \ket{\phi_1}\bra{\phi_1}\otimes  \rho^0_{\VQC}| \approx 0.804
\end{split}
\end{equation}
Here $\rho_1= \ket{\phi_0}\bra{\phi_0}\otimes\ket{\phi_1}\bra{\phi_1}$ (similar to the case in \eqref{eq:mayers_bob_pairs_discriminate}). Here, we have a higher probability for Bob to correctly guess Alice's bit $a$, but correspondingly, the detection (by Alice) probability is higher than in the ideal case, due to a lower local fidelity of $F^{\VQC}_{\Lbs} = 0.985$.
\end{proof}

\subsubsection{Variational cloning attack on 4-state quantum coin-flipping}\label{sec:varqlone-numerical-results-4-state-coinflip}

For the attacks on $\mathcal{P}_2$ using $\VQC$, again we recall the family of states:
\begin{equation}\label{eq:aharonov_coinflip_states}
    \ket{\phi_{x, a}} = 
    \begin{cases}
    \ket{{\frac{\pi}{8}}_{x,0}} = \cos\left( \frac{\pi}{8} \right)\ket{0} + (-1)^x\sin\left( \frac{\pi}{8} \right)\ket{1} \\
    \ket{{\frac{\pi}{8}}_{x,1}} =  \sin\left( \frac{\pi}{8} \right)\ket{0} + (-1)^{x \oplus 1}\cos\left( \frac{\pi}{8} \right)\ket{1}
  \end{cases} 
\end{equation}

Here first we mention the attack that uses $1 \rightarrow 2$ cloning similar to the ones discussed in Section~\ref{sec:varqlone-4state-coinflip}. But then we generalise the result to $1 \rightarrow 3$ and $2 \rightarrow 4$ cloning as well. One interesting aspect of the result given here is that there is no explicit analytical circuit known so far prior to our results, for these types of state-dependent cloning.\\

\noindent\textbf{$1 \rightarrow 2$ Cloning.}\\

\begin{figure}[ht]
    \centering
        \includegraphics[width=1\textwidth]{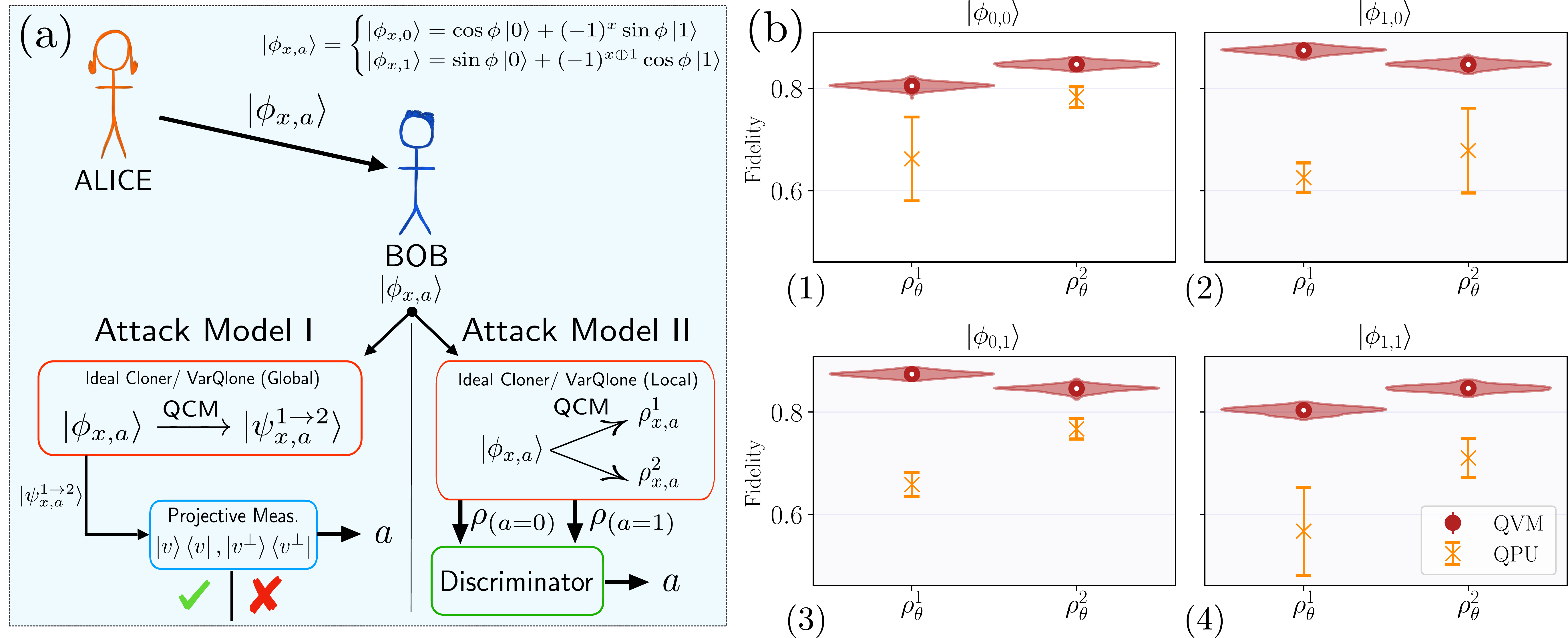}
    \caption[Cloning attacks and numerical results for the protocol, $\mathcal{P}_2$]{Cloning attacks and numerical results for the protocol, $\mathcal{P}_2$.(a) The two cloning based attacks we consider. In attack model {\RNum{1}} (left), Bob measures both output states with a set of fixed projective measurements, defined relative to the cloner output states, $\ket{\psi^{1\rightarrow 2}}_{a, x}$ and guesses Alice's bit, $a$. In attack model {\RNum{2}}, Bob keeps one clone for either testing Alice later or to send back the deposit qubit requested by Alice. He uses then the other local clone to discriminate and guess $a$. (b) The fidelities achieved cloning the each state, $\{\ket{\phi_{x, a}}\}$ used in $\mathcal{P}_2$ with $\VQC$. These numerics relate to scenario $1$ from attack model {\RNum{2}}. Each panel (1-4) shows both simulated (QVM - red circles) and on Rigetti hardware (QPU - orange crosses). We indicate the fidelities of each clone received by Alice and Bob. For the QVM (QPU) results, 256 (3) samples of each state are used to generate statistics. Violin plots show the complete distribution of outcomes and error bars show the means and standard deviations. Inset (i) shows the connectivity we allow in $\VQC$ for this example.}
    \label{fig:aharonov_1to2_cloning_fidelities_variational_plus_attack_models}
\end{figure}

\noindent Firstly, we use the same gate set and subset of the \computerfont{Aspen-8} lattice $(\mathcal{G}_{\mathcal{P}_2^{1\rightarrow 2}} = \mathcal{G}_{\mathcal{P}_1^{1\rightarrow 2}})$. We use the local cost, \eqref{eq:local_cost_full_supp}, to train the model, with a sequence length of $35$ gates. The attack model and the numerical results are shown in \figref{fig:aharonov_1to2_cloning_fidelities_variational_plus_attack_models}. Specifically, in part (b) the results for both QVM (simulation) and QPU (experiment) are given. We note that the solution exhibits some small degree of asymmetry in the output states, due to the form of the local cost function. This asymmetry is particularly pronounced as we scale the problem size and aim to produce $N$ output clones, which we further discuss in the next section.

Now, we can relate the performance of the $\VQC$ cloner to the attacks discussed in Section~\ref{sec:varqlone-4state-coinflip}. We do this by explicitly analyzing the output states produced in the circuits used to achieve fidelities shown in \figref{fig:aharonov_1to2_cloning_fidelities_variational_plus_attack_models}(b) and following the derivation in Section~\ref{sec:varqlone-4state-coinflip}, we show in \thmref{th:aharonov_attack_I_bias_probability_vqc} and \thmref{th:aharonov_4state_attack_II_bias_probability_vqc}:

\begin{thmbox}
\begin{theorem}\label{th:aharonov_attack_I_bias_probability_vqc}[$\VQC$ Cloning Attack ({\RNum{1}}) Bias on $\mathcal{P}_2$]
Using a cloning attack on the protocol $\mathcal{P}_2$, (in attack model {\RNum{1}}) Bob can achieve a bias:
\begin{equation}
    \epsilon^{\mathrm{I}}_{\mathcal{P}_2, \VQC} \approx 0.345
\end{equation}
\end{theorem}
\end{thmbox}

Similarly, we have the bias which can be achieved with attack {\RNum{2}}:
\begin{thmbox}
\begin{theorem}\label{th:aharonov_4state_attack_II_bias_probability_vqc}[$\VQC$ Cloning Attack ({\RNum{2}}) Bias on $\mathcal{P}_2$]
Using a cloning attack on the protocol $\mathcal{P}_2$, (in attack model {\RNum{2}}) Bob can achieve a bias:
\begin{equation}\label{eq:attack_2_aharonov_success_probability_bound_real}
    \epsilon^{\mathrm{II}}_{\mathcal{P}_2, \VQC} = 0.241
\end{equation}
\end{theorem}
\end{thmbox}

The small variation between these results and the ideal biases proved in \thmref{th:aharonov_attack_I_bias_probability} and \thmref{th:aharonov_4state_attack_II_bias_probability} is primarily due to the small degree of asymmetry induced by the heuristics of $\VQC$. However, we emphasize that these biases can now be achieved via constructive attacks on the hardware.\\

\begin{figure}
    \begin{center}
    \includegraphics[width=1\textwidth]{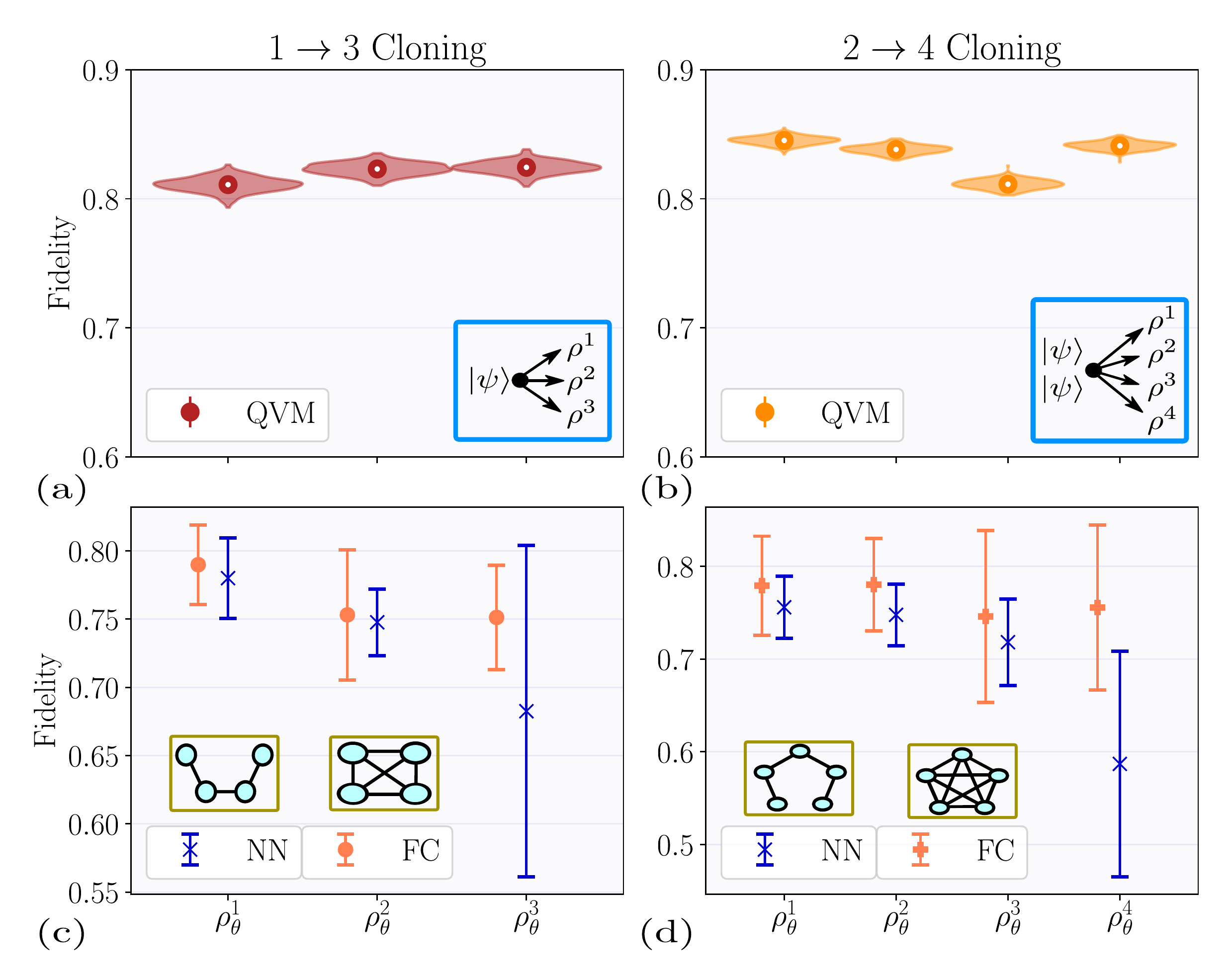}
    \caption[Clone fidelities for optimal circuits learned by $\VQC$ for (a) $1\rightarrow 3$ and (b) $2 \rightarrow 4$ cloning of the states used in the coin-flipping protocols]{Clone fidelities for optimal circuits learned by $\VQC$ for (a) $1\rightarrow 3$ and (b) $2 \rightarrow 4$ cloning of the states used in the coin-flipping protocol of~\cite{aharonov_quantum_2000} \textit{et. al.}\@. Mean and standard deviations of $256$ samples are shown (violin plots show the full distribution of fidelities), where the fidelities are computed using tomography only on the Rigetti QVM. In both cases, $\VQC$ is able to achieve average fidelities $> 80\%$. (c-d) shows the mean and standard deviation of the optimal fidelities found by $\VQC$ over 15 independent runs ($15$ random initial structures, $\boldsymbol{g}$) for the nearest neighbour (NN - purple) versus (d) fully connected (FC - pink) entanglement connectivity allowed in the variable structure Ansatz for $1\rightarrow 3$ and $2 \rightarrow 4$ cloning of $\mathcal{P}_2$ states. Insets of (c-d) shown corresponding allowed $\CZ$ gates in each example. 
    }
    \label{fig:1to3_2to4_aharonov_optimal_fidelities_plus_nn_vs_fc}
    \end{center}
\end{figure}

\noindent \textbf{$1 \rightarrow 3$ and $2\rightarrow 4$ Cloning.}\\\\
\noindent Finally, we extend the above analysis to the more general scenario of $M\rightarrow N$ cloning, taking $M=1, 2$ and $N=3, 4$. The result for $1 \rightarrow 3$ and $2\rightarrow 4$ are illustrated in \figref{fig:1to3_2to4_aharonov_optimal_fidelities_plus_nn_vs_fc}.

These examples are illustrative since they demonstrate the strengths of the squared local cost function  in~\eqref{eq:local_cost} over the local cost function in~\eqref{eq:local_cost_full_supp}. In particular, we find that the local cost function does not enforce symmetry strongly enough in the output clones and using only the local cost function, suboptimal solutions are found. We particularly observed this in the example of $2\rightarrow 4$ cloning, where $\VQC$ tended to take a shortcut by allowing one of the input states to fly through the circuit (resulting in nearly $100\%$ fidelity for that clone). It then attempts to perform $1\rightarrow 3$ cloning with the remaining input state. By strongly enforcing symmetry in the output clones using the squared cost, this can be avoided.

We also test two connectivities in these examples, a fully connected (FC) and the nearest neighbour (NN) architecture as allowed by the following gate sets:
\begin{equation}\label{eq:aharonov_1to3_state_dependent_cloning_gateset_NN}
    \mathcal{G}^{\textnormal{NN}}_{\mathcal{P}_2^{1\rightarrow 3}} = \{\mathsf{R}^i_{z}(\theta),  \mathsf{R}^i_{x}(\theta), \mathsf{R}^i_{y}(\theta), 
    \CZ_{2, 3}, \CZ_{3, 4}, \CZ_{4, 5}\} \quad
    \forall i \in \{2, 3, 4, 5\}
\end{equation}

\noindent and 
\begin{equation}\label{eq:aharonov_1to3_state_dependent_cloning_gateset_FC}
\begin{split}
    & \mathcal{G}^{\textnormal{FC}}_{\mathcal{P}_2^{1\rightarrow 3}} = \{\mathsf{R}^i_{z}(\theta),  \mathsf{R}^i_{x}(\theta), \mathsf{R}^i_{y}(\theta), \CZ_{2, 3},  \CZ_{2, 4},  \CZ_{2, 5}, 
    \CZ_{3, 4}, \CZ_{3, 5},  \CZ_{4, 5}\} \\
    & \forall i \in \{2, 3, 4, 5\}
    \end{split}
\end{equation}

Note, that for $1\rightarrow 3$ ($2\rightarrow 4$) cloning, we actually use $4$ ($5$) qubits, with one being an ancilla. The results of these experiments are also given in \figref{fig:1to3_2to4_aharonov_optimal_fidelities_plus_nn_vs_fc}.

\begin{figure}
    \centering
    \includegraphics[width=0.8\textwidth]{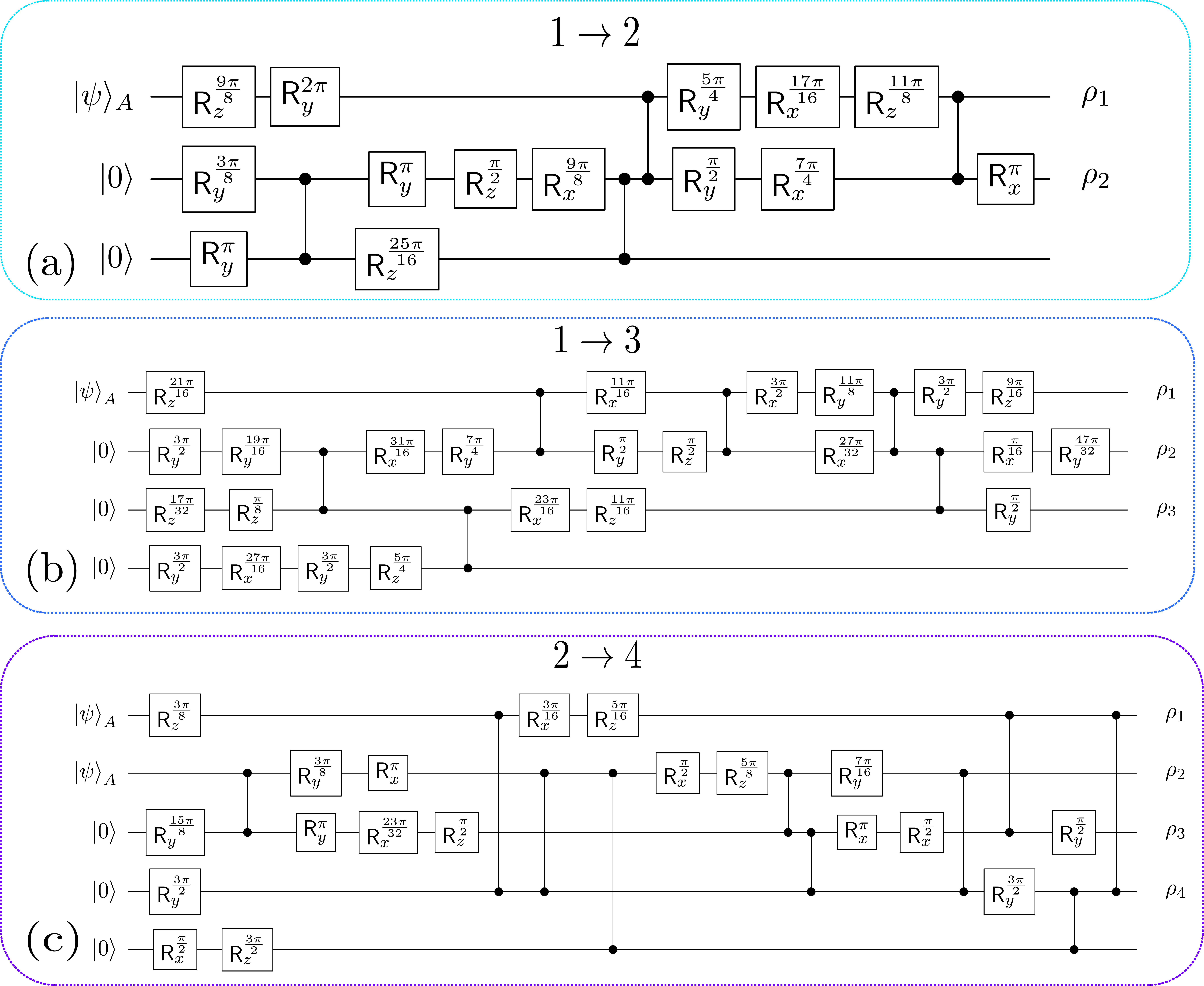}
    \caption[Circuits learned by $\VQC$ to clone states from the protocol, $\mathcal{P}_2$]{Circuits learned by $\VQC$ to clone states from the protocol, $\mathcal{P}_2$ for (a) $1\rightarrow 2$, (b) $1\rightarrow 3$ and (c) $2 \rightarrow 4$ cloning. These specific circuits produce the fidelities in \figref{fig:aharonov_1to2_cloning_fidelities_variational_plus_attack_models}(b) for $1\rightarrow 2$, (using the local cost function), and in \figref{fig:1to3_2to4_aharonov_optimal_fidelities_plus_nn_vs_fc} for $1\rightarrow 3$ and $2 \rightarrow 4$ (using the squared cost function). We allow an ancilla for all circuits, and $\rho_k$ indicates the qubit which will be the $k^{th}$ output clone.}
    \label{fig:aharonov_1to2_1to3_2to4_circuits}
\end{figure}

Finally, we give the the circuits learned by $\VQC$ and approximately clone all four states in \eqref{eq:aharonov_coinflip_states} in the protocol, $\mathcal{P}_2$, for $1\rightarrow 2, 1 \rightarrow 3$ and $2 \rightarrow 4$ cloning in \figref{fig:aharonov_1to2_1to3_2to4_circuits}. These are the specific circuits used to produce the fidelities in \figref{fig:aharonov_1to2_cloning_fidelities_variational_plus_attack_models}(b) and \figref{fig:1to3_2to4_aharonov_optimal_fidelities_plus_nn_vs_fc}.

\section{Discussion and conclusions}\label{sec:varqlone-conclusion}
We have shown, throughout this chapter, yet another face of \emph{unclonability}, not only as a core ingredient for quantum cryptanalysis but also with roots in foundational questions of quantum mechanics. Our attempts in this chapter have given partial answers to the following fundamental question: `How do we construct efficient, flexible, and noise-tolerant circuits to perform approximate cloning? and `How this ability will impact the security of real-life quantum protocols?' This latter question is especially pertinent in the current NISQ era, where the search for beneficial applications on small-scale noisy quantum devices remains at the forefront. On the other hand, this is an important and relevant question from a quantum communication perspective, given the existing gaps between the real implementations of quantum protocols and the proven theoretical results. In this work, exploring the exciting era between cryptanalysis and quantum machine learning, we have proposed our variational quantum cloner ($\VQC$), a cloning device that utilizes the capability of short-depth quantum circuits and the power of classical computation to learn the ability to clone specific set of states. This brings into view a whole new domain of performing realistic implementation of attacks on quantum cryptographic systems. We note, however, that in order to fully implement realistic and practical attacks, one must consider all aspects of the protocol environment, including, for example, the input and output mechanisms to the quantum cloner. Incorporating $\VQC$ into the full analysis of the experimental implementation of quantum protocols, for example as in~\cite{bartkiewicz_experimental_2013,bartkiewicz_experimental_2017}, is a fruitful avenue for future work.

We remark that our work opens new frontiers for analyzing quantum cryptographic schemes using quantum machine learning. In particular, this is applicable to secure quantum communication schemes which are becoming increasingly relevant in the quantum internet era.

We also note that one of the applications of our work is to find new cloning circuits that perform better on specific hardware. We specify that even though the states used in our examples and experiments are of low dimensions, in which case the emulator or the state vector machines will also provide the required result for cryptanalysis, using the VQA in $\VQC$ provides a circuit with better fidelity than the optimal theoretical (or emulated) circuits when run on real hardware, as we have seen in \figref{fig:learned_vs_ideal_circ_on_hw}. This point establishes a unique use-case of quantum machine learning techniques for a problem which is quantum in nature.

We also believe that the tools we have developed in this study can be used to clone a new family of states with partial prior information, which leads to expanding our fundamental understanding of approximate cloning and unclonability in general. Therefore we conclude that finding new classes of cloners and their circuits is a potential use-case of our work and a new appealing future research direction.

\chapter{Conclusion} \label{chap:conclusion}
\begin{chapquote}{Cheshire Cat - Alice in Wonderland}
``Imagination is the only weapon in the war with reality.''
\end{chapquote}

\noindent This thesis started with questions about unclonability in quantum mechanics and its role in quantum cryptanalysis. We were also seeking to grasp a deeper insight into the capacities of a quantum entity whose purpose is to attack the quantum and classical cryptosystems, specifically given the recent advancement in theoretical and experimental verges of the research on quantum technologies. In doing so, we have joined paths with various concepts such as unforgeability, unknownness, pseudorandomness, learnability, physical unclonability and variational algorithms, each of which has helped us to uncover a new connection to unclonability and let us to this point to conclude the thesis with a brief summary of this tortuous road. Hoping that the reader is not too weary by now, we will also discuss a general outlook and future direction. 

In \chapref{chap:unf-tools} we studied a more general notion of unclonability with a new perspective that related the unknownness of quantum states and processes to their unclonability. Following this direction, we correspondingly discussed the relation to learnability and eventually to a related cryptographic notion: unforgeability. We then devised a framework in which the quantum unforgeability of cryptographic primitives can be studied, regardless of being classical or quantum. This game-based security framework has been our cryptographic handbook in the majority of our security proofs throughout the thesis.

In \chapref{chap:qpuf}, we met a new kind of unclonability, the physical unclonability, which we have formally defined as a mathematical concept in the quantum world. We managed to thoroughly study the \emph{unpredictability} of some physical devices with the physical unclonability properties in terms of their unforgeability. Thus the notion of quantum physical unclonable functions has helped us to formalise some of our intuitions about unforgeability, unclonability and learnability, which we have put forward earlier. We have shown that the unforgeability of quantum PUFs is a provable consequence of their unknownness as hardware assumptions, which is unlike classical PUFs as they usually require assuming requirements that are morally equivalent to their unpredictability. Another point worth mentioning is the midst of proving the unforgeability of the unitary qPUF family, we apprehended the crucial role of quantum randomness. We have followed this thread in \chapref{chap:pr-connection} where we studied the computational or cryptographic counterpart of quantum randomness, \emph{i.e.}, quantum pseudorandomness. The investigation of quantum pseudorandomness concerning physical unclonability has uncovered interesting facts, some of which we conjecture and hope to be of interest in different areas of physics and cryptography.

As the thesis title includes ``...: from foundations to applications'', one would expect that this road reaches `Applications' at some point! That point was \chapref{chap:application}, where we search for applications of quantum physical unclonable function. We provide several proposals that demonstrate the applicability and relevance of this notion in designing a new genre of quantum protocols: secure quantum protocols based on hardware assumptions. Even though our proposed protocols aim for rather non-complex functionality, they are significant as a building block for other more complicated protocols and functionalities. This factor is particularly relevant if one takes a modular and composable view over quantum protocols, which we believe should be the next era in quantum protocol design and security analysis.\footnote{Historically, quantum protocols have not been designed with this mindset. One reason perhaps, is that they have been developed separately and by very different communities (physics, cryptography, math). However, we have adopted this kind of modular view in the development of \emph{quantum protocol zoo} \cite{veriqloud_quantum_2019} where we have gathered and studied different quantum protocols and showed their composition into simpler subroutines and building blocks. This composition and modularity are beneficial in designing advanced functionalities, as well as security proofs (Especially in composable frameworks such as universal composability or abstract cryptography). We have excluded our contributions on this topic from the thesis to keep it more coherent, though we believed it would be worth a short remark, in the conclusion.} 

Finally, we aimed to utilise recent developments in quantum computing for the purpose of cryptanalysis. One of the most recent tools and topics of research in this area is quantum machine learning and variational quantum algorithms. As a result, in the last chapter (\chapref{chap:varqlone}), we turned to another type of application: practical cryptanalysis using variational algorithms. However, since we could not resist a gaze into foundations, this chapter also includes questions and contributions regarding foundations and, more specifically, approximate quantum cloning. Our attempts led to the design of $\VQC$, our variational quantum cloner, using which we have performed a cloning-based security analysis on different quantum protocols. Although we have demonstrated specific case studies, we argue that the foremost importance of this contribution is not the particular examples we have investigated, but rather the new method and mentality that it uses for cryptanalysis. We believe this method can be used in a handful of scenarios, and most importantly, for protocols and cryptosystems for which we do not have full security proof while having such a tool can provide valuable insight. Yet another significance of this work is its compatibility with NISQ devices since it would allow hardware-efficient and high-quality cloning circuits according to the available hardware.

At the end of each chapter, we have discussed the potential future direction, and remaining questions in each of the topics, which we do not intend to repeat here. Instead, we discuss a broader outlook and prospective research direction in this field.

I believe one of the most exciting realms to step into, as also probably mentioned several times in the thesis, is the relationship between cryptography and learning theory in the quantum world. One evident reason is that learning theory offers powerful tools, both concretely theoretical and heuristic, which is maybe unconventional (compared to the approaches used in what I call `hard-core cryptography') but exciting approaches towards cryptanalysis. Although the two fields have quite different \emph{languages}, it appears to me that in many cases, they talk about closely related concepts, maybe from different perspectives. Thus, an idea that hopefully, this thesis has managed to convey to some extent is that there might be some level of correspondence between the two fields that could be capturable inside a new framework (maybe too naively and ambitiously). This sort of generalisation is, in my humble opinion, more feasible with quantum systems since; first, its mathematical framework is enough to include classical cases, and second, it inherently encompasses \emph{physics} or the \emph{actual systems} into the picture, while it is often neglected in cryptography or classical learning theory. Additionally, the recent works regarding learning different properties of the quantum states, or the relationship between quantum information and quantum machine learning, are an indication of this potential. An example of a concrete question I can propose here is proving tight bounds for quantum unforgeability using these tools, which would be in turn of interest in terms of learnability. 

Next, let me go back to my other favourite research areas: physics and foundations. The history of quantum cryptography shows how physics can influence (and has influenced) cryptography. But can cryptography do the same? Can powerful mathematical techniques and frameworks of cryptography help us to better understand nature? Especially the cryptography that has been already armed with physical phenomena such as unclonability and entanglement. To this end, a deep understanding of concepts such as unclonability would become handy since it is both central to physics and cryptography. Yet another concept that we have discussed in this thesis and can be considered in this regime is quantum pseudorandomness. Computational randomness is a cryptographic concept, however, as we have discussed in \chapref{chap:pr-connection}, pseudorandom quantum states have already been of interest for fundamental physics and quantum gravity. One cannot help oneself to wonder if perhaps there is a deeper level to this ``rabbit hole'' (which would be a very convenient term if we were to study black holes, for instance). Maybe this is motivating enough for `Alice's in the future to continue the study of quantum pseudorandomness and quantum physical unclonability in this context.

And finally, applications! Quantum hardware security (or quantum hardware cryptography) is a very young\footnote{I believe it did not really exist as a concrete field of research when I started my PhD, despite the fact that there have been several works in this area.}, yet promising field in terms of application. In this field, the ultimate goal is to exploit the unique properties of quantum hardware to reduce or remove computational assumptions or resource-intensive machinery and achieve secure and efficient quantum protocols on this ground. This goal, however exciting, and despite our attempts in this area, is admittedly still not too close to reality. Regarding qPUFs, many open questions and potential extensions exist, among which I can mention the realisation of efficient and secure quantum PUFs, and certifying existing hardware to satisfy the criteria of qPUF as the most influential ones. Both are challenging and potentially intriguing problems that can help close the gap between the theoretical security analysis and the hardware implementation of a quantum PUF. Even considering our proposed hybrid PUF construction, which gives a significant practicality improvement, analysis of noise and several experimental aspects has remained untouched. A more general remark on the field is that PUF is not the only potential subject in hardware security, and the discovery and study of other existing hardware assumptions in the quantum setting can be remarkably fruitful.

I conclude this chapter and this thesis with a slightly less scientific and more personal point, as I let my doubtful inner scientist elaborate. Despite the considerable recent progress in building quantum computers, it is still a possibility that either the noisy nature of these systems or our technological limitations in other ways will defeat us in the conquest to achieve large scale and fault-tolerant quantum computers (in which case there will be no need to protect ourselves against them). Also, despite the strong complexity theory evidence, it is still a possibility for quantum computation to be proven to have no advantages over classical. Even more drastically, we might find ourselves in a situation where quantum mechanics turns out to be insufficiently correct or severely incomplete (which is a possibility every scientist concerning any scientific theory should be prepared for, even if one develops affection for the beauty of a theory like quantum mechanics). In the unlikely event of any of these happening in the future, I am aware that it will hugely affect the validity and relevance of this thesis. Yet this thesis and my works during the period of my PhD have still been a (hopefully meaningful) attempt toward understanding, and if any piece of this attempt will ever create any tiny bit of imagination, curiosity or excitement in anyone, I can hope that all this effort has not been in vain, as ``Imagination is the only weapon in the war with reality.''

\appendix
\chapter{Additional proofs and derivations}
\section{Proof of \thmref{th:qe-fins} in \chapref{chap:unf-tools}}\label{app:qe-final-state-proof}
Here we give the full proof of \thmref{th:qe-fins} as follows:
\begin{proof}
We prove the theorem by induction. For the first block ($K=1$), according to \eqref{eq:qe-recur} and letting $\ket{\chi_0} = \ket{\psi}$ we have:
\begin{equation}
    \ket{\chi_1}= \frac{1}{2}[(\mathbb{I} - R(\phi_r))\ket{\psi}\ket{0} + R(\phi_i)(\mathbb{I} + R(\phi_r))\ket{\psi}\ket{1}]    
\end{equation}
where the term $\mathbb{I} - R(\phi_r) = 2\ket{\phi_r}\bra{\phi_r}$ projects the previous state to $\ket{\phi_r}$ with the coefficient $\mbraket{\phi_r}{\psi}$ and the term $R(\phi_i)(\mathbb{I} + R(\phi_r))$ is equal to:
\begin{equation}\label{eq:qe2term}
    R(\phi_i)(\mathbb{I} + R(\phi_r)) = 2[\mathbb{I} - \ket{\phi_r}\bra{\phi_r} - 2\ket{\phi_i}\bra{\phi_i} + 2\bra{\phi_i}\phi_r\rangle \ket{\phi_i}\bra{\phi_r}].
\end{equation}
Thus, the final relation between all the parameters in the first block is as follows.
\begin{equation}
\begin{split}
    \ket{\chi_1} = \mbraket{\phi_r}{\psi} \ket{\phi_r}\ket{0} + \ket{\psi}\ket{1} & - \mbraket{\phi_r}{\psi} \ket{\phi_r}\ket{1}\\
    & -2\mbraket{\phi_1}{\psi} \ket{\phi_1}\ket{1}
+2\mbraket{\phi_r}{\psi}\mbraket{\phi_r}{\phi_1} \ket{\phi_1}\ket{1}
\end{split}
\end{equation}

As can be seen, it satisfies the form of \eqref{eq:qe-fins} where the first sum is zero and in the second sum $g_{10}=-1, g_{11}=+1$, $l'_{10} = l'_{11} = 1$, $x'_{10} = z'_{10} = 0, y'_{10} = 1$, $x'_{11} = z'_{11} = 1$ and $y'_{11} = 0$.

Now we write $\ket{\chi_{K}}$ according to recursive relation of \eqref{eq:qe-recur}. We assume $\ket{\chi_{K-1}}$ is written in form of \eqref{eq:qe-fins} and show $\ket{\chi_{K}}$ also satisfies this equation.
\begin{equation}
\begin{split}
    \ket{\chi_K} = & \mbraket{\phi_r}{\chi_{K-1}} \ket{\phi_r}\ket{0} + \ket{\chi_{K-1}}\ket{1} - \mbraket{\phi_r}{\chi_{K-1}} \ket{\phi_r}\ket{1} -2\mbraket{\phi_K}{\chi_{K-1}} \ket{\phi_K}\ket{1} \\
    & +2\mbraket{\phi_r}{\chi_{K-1}}\mbraket{\phi_r}{\phi_K} \ket{\phi_K}\ket{1}
\end{split}
\end{equation}
By substituting $\ket{\chi_{K-1}}$ with its equivalent based on \eqref{eq:qe-fins}, we calculate each term in the above formula. Note that the coefficient in the third term is the same as the first one with a minus sign, and the ancillary state for the first term is $\ket{0}$ while for the third term is $\ket{1}$. Thus, we only show the details of the calculation for the first term:
\begin{equation}
\begin{split}
    \mbraket{\phi_r}&{\chi_{K-1}} \ket{\phi_r}\ket{0} = \\ &
    \mbraket{\phi_r}{\psi}\ket{\phi_r}\ket{0}^{\otimes K} + \mbraket{\phi_r}{\psi}\ket{\phi_r}\ket{1}^{\otimes K-1}\ket{0} - \mbraket{\phi_r}{\psi}\ket{\phi_r}\ket{1}^{\otimes K-1}\ket{0} + \\
    & + \sum^{K-1}_{i=1}\sum^{i}_{j=0} [f_{ij} 2^{l_{ij}} |\mbraket{\phi_r}{\psi}|^{x_{ij}} |\mbraket{\phi_i}{\psi}|^{y_{ij}} |\mbraket{\phi_r}{\phi_i}|^{z_{ij}}]\ket{\phi_r}\ket{q_{anc}(i,j)}\ket{0} \\
    & + \sum^{K-1}_{i=1}\sum^{i}_{j=0} [g_{ij} 2^{l'_{ij}} |\mbraket{\phi_r}{\psi}|^{x'_{ij}} |\mbraket{\phi_i}{\psi}|^{y'_{ij}} |\mbraket{\phi_r}{\phi_i}|^{z'_{ij}+1}]\ket{\phi_i}\ket{q'_{anc}(i,j)}\ket{0}.
\end{split}
\end{equation}
The second term is calculated as follows:
\begin{equation}
\begin{split}
    \ket{\chi_{K-1}}\ket{1} &= \mbraket{\phi_r}{\psi}\ket{0}^{\otimes K-1}\ket{1} + \ket{\psi}\ket{1}^{\otimes K} -\mbraket{\phi_r}{\psi}\ket{\phi_r}\ket{1}^{\otimes K} + \\
    & + \sum^{K-1}_{i=1}\sum^{i}_{j=0} [f_{ij} 2^{l_{ij}} |\mbraket{\phi_r}{\psi}|^{x_{ij}} |\mbraket{\phi_i}{\psi}|^{y_{ij}} |\mbraket{\phi_r}{\phi_i}|^{z_{ij}}]\ket{\phi_r}\ket{q_{anc}(i,j)}\ket{1} \\
    & + \sum^{K-1}_{i=1}\sum^{i}_{j=0} [g_{ij} 2^{l'_{ij}} |\mbraket{\phi_r}{\psi}|^{x'_{ij}} |\mbraket{\phi_i}{\psi}|^{y'_{ij}} |\mbraket{\phi_r}{\phi_i}|^{z'_{ij}}]\ket{\phi_i}\ket{q'_{anc}(i,j)}\ket{1}.
\end{split}
\end{equation}
The forth term $-2\mbraket{\phi_K}{\chi_{K-1}}\ket{\phi_K}\ket{1}$ has the coefficient $-2\mbraket{\phi_K}{\chi_{K-1}}$, which produces the same sigma terms while only  $l'_{i,j}, x'_{i,j}, y'_{i,j}$ and $z'_{i,j}$ are increased by one. The fifth term $2\mbraket{\phi_r}{\chi_{K-1}}\mbraket{\phi_r}{\phi_K}\ket{\phi_K}\ket{1}$ has the coefficient  $2\mbraket{\phi_r}{\chi_{K-1}}\mbraket{\phi_r}{\phi_K}$ and similarly produces the same sigma terms where $l_{i,j}$, $x_{i,j}$, $y_{i,j}$ and $z_{i,j}$ are increased by one (Note that the $\mbraket{\phi_r}{\phi_K}$ is itself one of the terms of the sigma). Finally by putting all these terms together, \eqref{eq:qe-fins} is obtained which completes the proof.
\end{proof}
\section{Proof of \thmref{th:suf-uuf} in \chapref{chap:unf-tools}}\label{app:proof-suf-uuf}
\begin{proof}
To show this implication we will show that if a QPT adversary $\A$ can win in \uuf, then $\A$ can also win against \sufm. Although for simplicity we restrict the proof for the case of $\mu=1$ and the generalisation to any $\mu$ is straightforward from the hierarchy of the definition for different $\mu$ shown in \thmref{th:mu-smaller-stronger}. Also, we recall that \suf\ and \euf\ are equivalent. Let $\A$ play the game $\GCM{\F}{\qUni}(\lambda, \A)$ by picking a set of learning phase state $\{\ket{\phi_i}\}^K_{i=1}$. Let the dimension of the unitary oracle $\eO$ be $D = 2^n$ and let the subspace of $\sigma_{in}$ be of dimension $d=poly(n)$. If $\A$ wins the game, then the average probability of $\A$ generating an acceptable output for any $x \in \M$ picked uniformly at random by $\C$ is non-negligible:
\begin{equation}
    Pr[1\leftarrow \GCM{\F}{\qUni}(\lambda, \A)] = \underset{x \in \M}{Pr}[1\leftarrow \A(x)] = \nonnegl(\lambda).
\end{equation}
where $\underset{x \in \M}{Pr}[1\leftarrow \A(x)]$ denotes the success probability of the adversary winning the game for input $x$. Now to be able to translate this game to the \suf\ game, first, we need to make sure that the set of states that $\A$ picks the challenge from them, satisfies the distinguishability condition for $\mu=1$ i.e. they are orthogonal to all the learning phase states. Let $\M'$ be the set of all the challenges with no overlap with any of the learning phase states $\rho^{in}_i$. Then we can rewrite the average success probability as follows:
\begin{equation}
\begin{split}
    \underset{x \in \M}{Pr}[1\leftarrow \A(x)] & = \underset{x \in \M'}{Pr}[1\leftarrow \A(x)]Pr[x \in \M'] + \underset{x \not\in \M'}{Pr}[1\leftarrow \A(x)]Pr[x \not\in \M'] \\
    & = \nonnegl(\lambda).
\end{split}
\end{equation}
since the dimension of the subspace that $\sigma_{in}$ spans is $d$ and it is polynomial with respect to the size of $\M$ then $\frac{|\M'|}{|\M|} \approx 1$. Hence $Pr[x \in \M'] \approx 1$ but $Pr[x \not\in \M'] = 1-Pr[x \in \M'] = \negl(\lambda)$. As a result the second term will be negligible and for the whole expression to become non-negligible, the following should hold:
\begin{equation}
\underset{x \in \M'}{Pr}[1\leftarrow \A(x)] = \nonnegl(\lambda).
\end{equation}
Now let $\A'$ be an adversary who wants to win the game $\GCM{\F}{\qSel, \mu}(\lambda, \A')$ by using $\A$. As $\A'$ picks the challenge of their choice, we will show that there is a strategy for $\A'$ to win the game relying on the average success probability of $\A$ being non-negligible over $\M'$. But also as $\A'$ is a QPT, we will show there exist a poly size subspace of $\M'$ in which $\A'$ will win with non-negligible probability. First we assume that $\M'$ is partitioned into $K$ different subsets (or subspaces) $S_i$ with equal size (or dimension in the quantum case) $|S_1|=\dots =|S_K|= l = poly(\lambda)$. Note that this partitioning is only for simplicity and any random partitioning of $\M'$ into the equal size subspace will be enough for our purpose. Now let $\A'$ pick one of the subsets of message space which consists of picking one of the $S_i$ with probability $\frac{1}{K}$. We want to show that if $\A'$ picks the $S_i$ at random and calls $\A$ on that $S_i$ the probability that in the picked subspace the following condition holds is non-negligible:
\begin{equation}\label{eq:suf-uuf-subset-prob}
\underset{x \in S_i}{Pr}[1\leftarrow \A(x)] = \nonnegl(\lambda)
\end{equation}
If this is the case, then by the definition of the average probability there exists at least one $x^*$ for which the $Pr[1\leftarrow \A(x^*)] = \nonnegl(\lambda)$ and hence the $\A'$ has won the game with a non-negligible probability. Thus we need to find the number of the success probability of $\A'$ picking a desirable subset. This probability is given by:
\begin{equation}
    Pr_{succ} = \frac{\#(S_i: \underset{x \in S_i}{Pr}[1\leftarrow \A(x)] = \nonnegl(\lambda))}{K} = \frac{Q}{K}
\end{equation}
where $Q$ denotes the number of subsets $S_i$ which satisfy the condition and $K = O(|\M'|)$. We then only need to show that $\frac{Q}{K}$ is non-negligible in the security parameter. For simplicity let us replace average probability of $\A$ in wining the game over $\M'$, with the expected value of wining probability of $\A$ over all the different elements of $\M'$ i.e.
\begin{equation}
\underset{x \in \M'}{Pr}[1\leftarrow \A(x)] = \nonnegl(\lambda) \Rightarrow \underset{\M'}{\mathbb{E}}[\A(x)] = \nonnegl(\lambda)
\end{equation}
Then we rewrite the expectation value in terms of all the subsets of $\M'$. As $\M' = S_1 \cup S_2 \cup \dots \cup S_K$, we have:
\begin{equation}
    \underset{\M'}{\mathbb{E}}[\A(x)] = \frac{1}{K}\sum^K_{i=1} \mathbb{E}_i = \nonnegl(\lambda)
\end{equation}
where $\mathbb{E}_i = \underset{S_i}{\mathbb{E}}[\A(x)]$. We then rearrange all the $\mathbb{E}_i$ descending such that the $Q$th term shows the last smallest $\mathbb{E}_i$ for which the condition is satisfied. Hence we have:
\begin{equation}
    \underset{\M'}{\mathbb{E}}[\A(x)] = \frac{1}{K}\sum^Q_{i=1} \mathbb{E}_i + \frac{1}{K}\sum^K_{i=Q+1} \mathbb{E}_i = \nonnegl(\lambda)
\end{equation}
The above equality holds if at least one of the two sums is non-negligible. If the first sum is non-negligible we have:
\begin{equation}
    \frac{1}{K}\sum^Q_{i=1} \mathbb{E}_i \geq \frac{Q\mathbb{E}_Q}{K}
\end{equation}
As $\mathbb{E}_i$s have been ordered and $\mathbb{E}_Q$ is the smallest one which is still non-negligible. Then we can conclude that:
\begin{equation}
    \frac{Q}{K} = \nonnegl(\lambda)
\end{equation}
which is what we wanted to show. The second case is when the first sum is negligible and the second sum needs to be non-negligible for the equality to hold. Similar to the previous case due to the descending ordering, we have: 
\begin{equation}
    \frac{1}{K}\sum^K_{i=Q+1} \mathbb{E}_i \leq \frac{(K-Q)\mathbb{E}_{Q+1}}{K}
\end{equation}
But followed by our assumption the $\mathbb{E}_{Q+1}$ is itself negligible and $0 < \frac{K-Q}{K} < 1$, thus this sum can never converge to a non-negligible function of $\lambda$. Hence we conclude that necessarily the first sum, and as a result, $\frac{Q}{K}$ is non-negligible. Thus we have shown the equation~\ref{eq:suf-uuf-subset-prob}, and there exists a strategy for $\A'$ to win the game by calling $\A$. This concludes that \suf (\sufm) implies \uuf\ and the proof is complete.
\end{proof}

\section{Proof of \thmref{th:1guf-bu} in \chapref{chap:unf-tools}}\label{app:proof-1gu-bu-equivalent}
\begin{proof}
We show that \euf\ implies BU and vice versa. First, we show that if a scheme is not BU unforgeable against a QPT adversary then it is not \euf\ unforgeable either. Let $\A$ be a QPT adversary who forges a scheme $\F = (\ES, \E, \V)$ with message set $\M = \{0,1\}^n$ in the BU definition. Following the formal definition of BU provided in \defref{def:prelim-bu}, $\A$ selects an $\epsilon$ for which the blinded region $\Be$ is created by selecting each $m \in \M$ at random with an $\epsilon$-related probability. Then there exists a non-empty set $\Be$ for which $\A$ interacts with the blinded oracle associated with it and outputs a pair $(m^*,t^*)$ where $t^* = f(m^*)$ (where $f$ is the classical function of the evaluation $\E$, for instance a \textit{MAC}(.)) such that $\V = Ver_k(m^*,t^*) = acc$, and also the $m^* \in \Be$ with non-negligible probability in $\lambda = poly(n)$. By the definition of the blinding oracle, $\A$ receives a $\ket{\perp}$ for any of the computational basis that is in the blinded region. As a result, we can write $\A$'s input and output queries as follows:
\begin{equation}
\begin{split}
    & \ket{\phi_i} = \sum_{m_i \not\in \Be} \alpha_i \ket{m_i, y_i} + \sum_{\overline{m}_j \in \Be} \beta_j \ket{\overline{m}_j, y_j} \\
    & \ket{\phi^{out}_i} = \sum_{m_i \not\in \Be} \alpha_i \ket{m_i, y_i\oplus f(m_i)} + \sum_{\overline{m}_j \in \Be} \beta_j \ket{\overline{m}_j, y_j \oplus \perp}
\end{split}
\end{equation}
Now assuming the quantum encoding of the challenge $m^* \in \Be$ to be $\ket{m^*, 0}$ and the tag/output to be $\ket{m^*, t^*} = \ket{m^*, f(m^*)}$, we can see that $\mbraket{m^*, t^*}{\phi^{out}_i} = 0$ since $m^*$ will have no overlap with the first part of the superposition, and also to the second part due to the blinding. Now, we show that there exists a unitary non-blinding oracle that generates equivalent queries for this scenario. Let $\Ue$ be the unitary evaluation oracle such that $\ket{m^*, t^*} = \Ue\ket{m^*, 0}$, and similarly for all the queries. Due to the unitarity, we have that $\mbraket{m^*, t^*}{\phi^{out}_i} = \bra{m^*, 0}\Ue^{\dagger}\Ue\ket{\phi_i} = 0$. Thus there will also exist an adversary $\A'$ with equivalent queries except that the target forgery will be always orthogonal to the selected challenge. Hence for this adversary, the condition of \euf\ is satisfied. Then by calling $\A$, the adversary $\A'$ can generate an output state $\ket{m^*, t^*}$ that passes the test algorithm with also non-negligible probability. Hence we have shown that \euf\ implies \bu. 

To prove the other way of implication we need to show whenever there is an attack on \euf, then there will also be an attack on \bu\ definition and hence the scheme is also \bu\ insecure. This time we consider $\A$ to be a QPT adversary who wins \euf\ by selecting a challenge state $\ket{m^*, y}$ where the $m^*$ is the classical challenge and $y$ is the ancillary register, and querying a set of states $\{\ket{\phi_i}\}^{q}_{i=1}$ s.t. $\forall \ket{\phi_i}: \mbraket{m^*}{\phi_i} = 0$ and $q=poly(n)$. Then by definition, $\A$ can output a $\ket{m^*, t^*} = \Ue\ket{m^*, y}$ that passes the test algorithm with non-negligible probability. Now an adversary $\A'$ calls $\A$ to win the \bu\ with non-negligible probability.

At this stage we recall the \thmref{th:prelim-bu} and we show that an $\A'$ satisfies the conditions of this theorem. Let us write the learning phase queries in the computational basis as follows:
\begin{equation}
    \ket{\phi^{out}_i} = \sum^{d}_{j=1} \alpha_{i,j} \ket{b_j}
\end{equation}
where $\{\ket{b_j}\}^{d}_{j=1}$ is the set of computational bases spanning the effective learning phase subspace. Now we create a non-empty set $R$ by selecting each $x \in \M$ as follows
\begin{equation}
    R = \{x \in \M : \ket{x} \neq \ket{b_j}_X \}
\end{equation}
Where $\ket{b_j}_X$ denotes the input register of the full basis. 
Note that $R$ will always be non-empty as the basis set will only cover a polynomial-size subspace of the whole Hilbert space of messages. Moreover, since $\A'$ includes $\A$ and $m^*$ has no overlap with any of the input queries, it will also have no overlap with the input register of the output queries. As a result, $R$ has at least one element. Hence the set of all input elements that have non-zero overlap with the queries and the elements included in $R$ have no intersection. This shows that $supp(\A)\cap R = \emptyset$ if the support is defined for the oracle $\Ora_f$ for a fixed randomly picked classical function $f$ (or key $k$) during the game. Thus we also have $supp(\A')\cap R = \emptyset$ and $m^* \in R$. Nevertheless, in~\cite{alagic_quantum-access-secure_2020} it has been mentioned that the support is taken to be the union of the support of all the queries over the choice of the function. In this case, we can also redefine our set and the queries of $\A'$ such that it satisfies the condition of the theorem respectively. We take the set $R'$ to only include one element which is the forgery message $m^*$. As in the \euf\ the function (or the key for the keyed functions) is selected at random in the setup phase, the success probability of $\A$ is inherently taken over the choice of the function. Then $\A'$ queries all the queries of $\A$ for any randomly selected $f$ during the experiment. For any other functions, excludes any queries for which the support will include $m^*$. Now we can see that $\A'$ can output a valid pair $(m^*, t^*)$ by measuring $\ket{m^*, t^*}$ in the computational basis with probability 1 while $supp(\A')\cap R' = \emptyset$ and $m^* \in R'$. Hence $\A'$ breaks the \bu\ unforgeability and we have shown that \bu\ implies \euf. This mutual implication shows that these definitions are equivalent and the proof is complete.
\end{proof}
\section{Alternative model for an adaptive quantum adversary}\label{app:alt-adaptive-adv}
In this appendix, we introduce an alternative way for capturing full quantum adaptive adversaries. Here we also consider QPT adversaries who have $q$-query access to the evaluation function of a primitive $\F$, namely $\E$ where $q$ is polynomial in the security parameter. An adaptive adversary can choose and issue any arbitrary query which could also depend on the previous responses received from the black-box oracle. An adaptive quantum adversary is likely to consume the quantum state of the response to be able to pick the next query adaptively. Hence modeling the post-query database of an adaptive quantum adversary is more challenging. In what follows we give a $q$-query mathematical model for adaptive adversaries.

\begin{defbox}
\begin{definition}\label{def:adap-full-quant}
Let $q$ be a positive integer, and $\E : \Hil^{d_{\inp}} \rightarrow \Hil^{d_{\out}}$ be a quantum evaluation.  
We model a probabilistic adversary as a CPTP map 
$\A : \R \times (\Hil^{d_{\inp}})^{\otimes q} \otimes (\Hil^{d_{\out}})^{\otimes q} \rightarrow (\Hil^{d_{\inp}})$. Such an adversary is called an \textbf{adaptive} adversary $\A_{ad}$ if for all random coin $r \in \R$ and for any $\bigotimes_{i=1}^q \rho^{\inp}_i \in (\Hil^{d_{\inp}})^{\otimes q}$ and for $\bigotimes_{i=1}^q  \rho^{\out}_i \in (\Hil^{d_{\out}})^{\otimes q}$ (where $\rho^{\out}_i :=\E(\rho^{\inp}_i)$), the mapping $\bigotimes_{i=1}^q (\rho^{\inp}_i\otimes \rho^{\out}_i) \rightarrow \A^{r}_{ad}(\bigotimes_{i=1}^q (\rho^{\inp}_i\otimes \rho^{\out}_i))$ is dependent on the  $\rho^{\inp}_1\otimes \rho^{\out}_1, \ldots , \rho^{\inp}_q\otimes\rho^{\out}_q$;
\end{definition}
\end{defbox}

Intuitively, an adaptive adversary $\A : \R \times (\Hil^{d_{\inp}})^{\otimes q} \otimes (\Hil^{d_{\out}})^{\otimes q} \rightarrow (\Hil^{d_{\inp}})$ captures the strategy to choose the query input $\rho^{\inp}_{q+1} \in \Hil^{d_{\inp}}$ to $\E$. The adversary can use these query response pairs to predict the output of $\E$. 
We call the pair $(\bigotimes_{i=1} ^q \rho^{\inp}_i,\bigotimes_{i=1} ^q \rho^{\out}_i)$ that is generated after the $q$-round of interaction between an adversary $\A$ and $\E$, as a transcript. Note, that the transcripts depend on the choice of the random coins of $\A$.

However, since this model is more complicated to work with, we use our usual notation used in \chapref{chap:unf-tools}.

\section{Proof of \thmref{th:unf-uni-aua} in \chapref{chap:unf-tools}}\label{app:uni-adaptive}
\begin{proof}
Let $\A$ be the QPT adversary playing the game $\GCM{\F}{\uuf-aua, \mu}(\lambda, \A)$ and running the algorithm described in \algoref{alg:aua-uuf}.

\begin{algorithm}[ht!]
\SetAlgoLined
\begin{itemize}
    \item {\bf First learning phase:} $\nul$
    \item {\bf Challenge phase:}
        \begin{itemize}
            \item prepare qubit $\ket{0}_a$
            \item receive $\ket{\psi_m}$ as a challenge
        \end{itemize}
    \item {\bf Second learning phase:}
    \begin{itemize}
        \item $\ket{\Psi}_{ca} = CNOT_{c,a}(\ket{\psi_m}\ket{0})$\footnote{The subscript $c$ denotes the challenge and the subscript $a$ denotes the adversary's qubit.} 
        \item query register $c$ ($\A$ sends the challenge part of the entangled system, $\rho_c$ as a query.)
        \item receive $\Ue\rho_c\Ue^{\dagger}$ or $(\Ue\otimes\mathcal{I})\ket{\Psi}_{ca}$
    \end{itemize}
  \item {\bf Guess phase:}
    \begin{itemize}
        \item $\ket{\psi_m^{out}}\otimes\ket{\pm} \leftarrow Measure(\ket{\Psi}_{ca}, \{\ket{\pm}\}$)
        \item \textbf{if} {$\ket{\pm} = \ket{+}$} {\setlength\itemindent{25pt} \item \textbf{output:} $\ket{t} = \ket{\psi_m^{out}}$}
        \item \textbf{else} {\setlength\itemindent{25pt} \item \textbf{output:} $\ket{t} = CZ^{\otimes n-1}(\ket{\psi_m^{out}})$}
        \item $Measure(\ket{\Psi}_{ca}, \{\ket{\pm}\}$ outputs the result of the measurement.
  \end{itemize}
\end{itemize}
\caption{aua attack on \uuf}\label{alg:aua-uuf}
\end{algorithm}

$\A$ does not issue any query during the first learning phase. Then $\A$ receives an unknown challenge state $\ket{\psi_m} = \sum_{i=1}^{D} \alpha_i \ket{b_i}$ where $\{\ket{b_i}\}_{i=1}^{D}$ is a set of complete orthonormal bases for $\HilD$. Now, $\A$ prepares state $\ket{0}$ and performs a CNOT gate on the first qubit of the unknown challenge state and the ancillary qubit ($\ket{0}$) with the control qubit on the challenge state. We can assume the order of the bases is such that in the first half, the first qubit is $\ket{0}$ and in the second half the first qubit is $\ket{1}$. Then the output entangled state is
\begin{equation*}
\ket{\Psi}_{ca} = \sum_{i=1}^{D/2} \alpha_i \ket{b_i}_{c}\otimes\ket{0}_{a} + \sum_{i=\frac{D}{2} + 1}^{D} \alpha_i \ket{b_i}_{c}\otimes\ket{1}_{a}
\end{equation*}
Now we can compute the final state of the two systems after the second learning phase which is:
\begin{equation*}
\ket{\Psi^{out}}_{ca} = \sum_{i=1}^{D/2} \alpha_i (\Ue\otimes\mathbb{I})(\ket{b_i}_{c}\otimes\ket{0}_{a}) + \sum_{i=\frac{D}{2} + 1}^{D} \alpha_i (\Ue\otimes\mathbb{I})(\ket{b_i}_{c}\otimes\ket{1}_{a}).
\end{equation*}
By rewriting the first qubit in the $\ket{+}$ basis we have
\begin{equation*}
\ket{\psi_m^{out}} = [\Ue(\sum_{i=1}^{D} \alpha_i \ket{b_i}_{c})]\frac{\ket{+}}{\sqrt{2}} + [\Ue(\sum_{i=1}^{D/2} \alpha_i \ket{b_i}_{c} - \sum_{i=\frac{D}{2} + 1}^{D} \alpha_i \ket{b_i}_{c})]\frac{\ket{-}}{\sqrt{2}}.
\end{equation*}
Then, the adversary measures his local qubit in the $\{\ket{+}, \ket{-}\}$ bases. If he obtains $\ket{+}$, the state collapses to $\Ue(\sum_{i=1}^{D} \alpha_i \ket{b_i}_{c}) = \Ue\ket{\psi_m}$ that is the desired state with fidelity 1. If the output of the measurement is $\ket{-}$, half of the terms have a minus sign. In this case, $\A$ applies a controlled-Z gate on the second half of the state to obtain again $\Ue\ket{\psi_m}$. As a result, for any $\kappa_1$ and $\kappa_2$, we have:
\begin{equation*}
    Pr[1\leftarrow \GCM{\F}{\qUni-aua, \mu}(\lambda, \A)] = Pr[1\leftarrow\T((\Ue\ket{\psi_m})^{\otimes\kappa_1}, \ket{t}^{\otimes\kappa_2})] = 1.
\end{equation*}
Now to complete the proof, we show that the $\mu$-distinguishability is satisfied on average.
We need to calculate the reduced density matrix of this state and compare it with the density matrix $\rho_{\psi}=\ket{\psi}\bra{\psi}$ in terms of the Uhlmann's fidelity. The reduced density matrix of the challenge state can be calculated as follows:
\begin{equation*}
\begin{split}
\rho_c = Tr_{a}[\ket{\psi}\bra{\psi}_{ca}] = & \sum^{D}_{i=1}|\alpha_i|^2\ket{b_i}\bra{b_i} + \sum^{\frac{D}{2}}_{i=j=1}\sum^D_{j\neq i,j=\frac{D}{2}+1}\overline{\alpha_i}\alpha_j\ket{b_i}\bra{b_j} + \\ & \sum^{D}_{i=\frac{D}{2}+1}\sum^{\frac{D}{2}}_{j\neq i, j=1}\overline{\alpha_i}\alpha_j\ket{b_i}\bra{b_j}
\end{split}
\end{equation*}
where $Tr_{a}$ denoted the partial trace taken over the adversary's sub-system. And the first sum shows the diagonal terms of the density matrix. As it can be seen these density matrices are different in half of the non-diagonal terms with the $\rho_{\psi}$. According to the Uhlmann's fidelity definition in the preliminary, and the fact that $\ket{\psi}$ is a pure state the fidelity reduce to:
\begin{equation*}
F(\rho_{\psi},\rho_c)=[Tr(\sqrt{{\sqrt{\rho_{\psi}}}\rho_c {\sqrt{\rho_{\psi}}}})]^{2} = \bra{\psi}\rho_c\ket{\psi} = \sum^D_{i=1}|\alpha_i|^2\bra{b_i}\rho_c\ket{b_i}.
\end{equation*}
By substituting the $\rho_c$ from above, the result will be as follows:
\begin{equation*}
F(\rho_{\psi},\rho_c)= \sum^D_{i=1}|\alpha_i|^4 + \sum^{\frac{D}{2}}_{i=1}\sum^{D}_{j=\frac{D}{2}+1}2|\alpha_i\alpha_j|^2 = 1 - \sum^{\frac{D(D-1)}{4}}_{i=1}2|\gamma_i|^2
\end{equation*}
where $|\gamma_i|^2$ denoted the square of a quarter of the non-diagonal elements of $\rho_{\psi}$. This is a positive value and on average over all the state $\ket{\psi}$, non-negligible compared to the dimensionality of the state. Hence:
\[F(\rho_{\psi},\rho_c) \leq 1 - \nonnegl(\lambda)\]
and the distinguishability condition is satisfied and the proof is complete.
\end{proof}

\section{Calculation of the cost function's gradient for $\VQC$}\label{app:vqc-gradient-calc}
In this appendix we calculate the gradient of the cost functions we have introduced in Section~\ref{sec:varqlone-algorithm-gredient} of \chapref{chap:varqlone}, using the techniques introduced in Section~\ref{sec:prelim-vqc-opt}.

We remind the squared cost:
\begin{align}\label{eq:app-squared_local_cost_mton_gradient}
    \Cbs_{\sq}^{M\rightarrow N}(\paramtheta) := \mathop{\mathbb{E}}_{\substack{\ket{\psi} \in S}}\left[ \sum\limits_{i=1}^N (1-F^i_{\Lbs}(\paramtheta))^2\right. 
    \left.+ \sum\limits_{i<j}^N (F^i_{\Lbs}(\paramtheta)-F^j_{\Lbs}(\paramtheta))^2\right] 
\end{align}
and its partial derivative from \eqref{eq:squared_local_cost_mton_gradient_calc}
\begin{align}\label{eq:app-squared_local_cost_mton_gradient_calc}
\small
    \frac{\partial \Cbs_{\sq}(\boldsymbol{\theta})}{\partial \theta_l} =
    2\mathop{\mathbb{E}}_{\substack{\ket{\psi} \in S}}\left[\sum\limits_{i=1}^N (1-F^i_{\Lbs}(\paramtheta)) \left[-\frac{\partial  F^i_{\Lbs}(\boldsymbol{\theta})}{\partial \theta_l}\right]+\sum\limits_{i<j}^N (F^i_{\Lbs}(\paramtheta)-F^j_{\Lbs}(\paramtheta)) \left[\frac{\partial  F^i_{\Lbs}(\boldsymbol{\theta})}{\partial \theta_l}  - \frac{\partial  F^j_{\Lbs}(\boldsymbol{\theta})}{\partial \theta_l} \right]\right]
\end{align}
Now we assume that each $U(\paramtheta) := U(\theta_d)U(\theta_{d - 1}) \dots U(\theta_{1})$ is composed of unitary gates of the form: $U(\theta_l) = \exp\left(-i\theta_l\Sigma_l\right)$, where $\Sigma_l^2= \mathds{1}$ (for example, a tensor product of Pauli operators). We can use the \emph{parameter shift rule} from \thmref{th:perlim-param-shift-rule}, we get:
\begin{align}
    \frac{\partial  U(\boldsymbol{\theta})\rho_{\text{init}}U(\boldsymbol{\theta})^{\dagger}}{\partial \theta_l} = U^{l+\frac{\pi}{2}}(\boldsymbol{\theta})\rho_{\text{init}}(U(\boldsymbol{\theta})^{l+\frac{\pi}{2}})^{\dagger} -  U^{l-\frac{\pi}{2}}(\boldsymbol{\theta})\rho_{\text{init}}(U(\boldsymbol{\theta})^{l-\frac{\pi}{2}})^{\dagger}
\end{align}
Where the notation, $U^{l\pm\frac{\pi}{2}}$, indicates the $l^{th}$ parameter has been shifted by $\pm \frac{\pi}{2}$, i.e.\@ $U^{l\pm\frac{\pi}{2}} := U(\theta_d)U(\theta_{d - 1}) \dots U(\theta_l\pm \pi / 2) \dots  U(\theta_{1})$. We get:
\begin{equation}
\begin{split}
    \frac{\partial  F^{j}_{\Lbs}(\boldsymbol{\theta})}{\partial \theta_l} &=   \Tr\left[\ket{\psi}\bra{\psi}\Tr_{\bar{j}}\left(U^{l+\frac{\pi}{2}}(\boldsymbol{\theta})\rho_{\text{init}}(U(\boldsymbol{\theta})^{l+\frac{\pi}{2}})^{\dagger} \right)\right] \\
    & - \Tr\left[\ket{\psi}\bra{\psi}\Tr_{\bar{j}}\left(U^{l-\frac{\pi}{2}}(\paramtheta)\rho_{\text{init}}(U(\boldsymbol{\theta})^{l-\frac{\pi}{2}})^{\dagger} \right)\right]\\\\
        \implies \frac{\partial   F^{ j}_{\Lbs}(\boldsymbol{\theta})}{\partial \theta_l} &=   \Tr\left[\ket{\psi}\bra{\psi}\rho_j^{l+\frac{\pi}{2}}(\boldsymbol{\theta})\right]
    -   \Tr\left[\ket{\psi}\bra{\psi}\rho_j^{l-\frac{\pi}{2}}(\boldsymbol{\theta})\right]
    \\
    & = F^{(j, l+\frac{\pi}{2})}_{\Lbs}(\boldsymbol{\theta}) - F^{(j, l-\frac{\pi}{2})}_{\Lbs}(\boldsymbol{\theta})
\end{split}     
\end{equation}
where we define $F^{(l\pm \frac{\pi}{2})}_j(\boldsymbol{\theta}) \coloneqq \bra{\psi}\rho_j^{l\pm\frac{\pi}{2}}(\boldsymbol{\theta})\ket{\psi}$ the fidelity of the $j^{th}$ clone, when prepared using a unitary whose $l^{th}$ parameter is shifted by $\pm \frac{\pi}{2}$, with respect to a target input state, $\ket{\psi}$. 

Plugging this into \eqref{eq:app-squared_local_cost_mton_gradient_calc}, we get:
\begin{align}\label{eq:analytic_squared_grad_mton}
    \frac{\partial \Cbs_{\sq}(\boldsymbol{\theta})}{\partial \theta_l} =
    \mathop{2\mathbb{E}}_{\substack{\ket{\psi} \in S}} & \left[
     \sum\limits_{i<j}^N (F^i_{\Lbs}-F^j_{\Lbs}) \left[F^{(i, l+\frac{\pi}{2})}_{\Lbs} - F^{(i, l-\frac{\pi}{2})}_{\Lbs} - F^{(j, l+\frac{\pi}{2})}_{\Lbs} + F^{(j, l-\frac{\pi}{2})}_{\Lbs} \right]\right. \\
     & - \left.\sum\limits_{i=1}^N (1-F^i_{\Lbs})\left[F^{(i, l+\frac{\pi}{2})}_{\Lbs} - F^{(i, l-\frac{\pi}{2})}_{\Lbs}\right] \right]
\end{align}

Using the same method, we can also derive the gradient of the local cost, \eqref{eq:local_cost_full_supp} with $N$ output clones as:
\begin{equation}\label{eq:gradient_local_cost_full}
    \frac{\partial \Cbs_{\Lbs}(\paramtheta)}{\partial \theta_l} = \mathbb{E}\left(\sum_{i=1}^N \left[F_{\Lbs}^{i, l-\pi/2} -  F_{\Lbs}^{i, l+\pi/2} \right]\right)
\end{equation}
Finally, similar techniques result in the analytical expression of the gradient of the global cost function:
\begin{equation}\label{eqn:gradient_global_cost_full}
    \frac{\partial \Cbs_{\Gbs}(\paramtheta)}{\partial \theta_l} = \mathbb{E}\left(F_{\Gbs}^{l-\pi/2} - F_{\Gbs}^{l+\pi/2}\right)    
\end{equation}
where $F_{\Gbs}:= F(\ket{\psi}\bra{\psi}^{\otimes N}, \rho_{\paramtheta})$ is the global fidelity between the parameterised output state and an $N$-fold tensor product of input states to be cloned.

\bibliographystyle{alpha}

\singlespace


\bibliography{thesis-refs}

\end{document}